%% file: KdV_documenatation.tex
\documentclass[10pt]{report}

\usepackage{amsmath,amsfonts,amsthm,amscd,amssymb,graphicx}
\usepackage{epstopdf}
\usepackage[usenas,dvipsnames]{xcolor}
\usepackage{hyperref}

\title{STABLAB Documentation for KdV}

\author{\sc \small Blake Barker}

\begin{document}

\maketitle

\tableofcontents

% Include all of the other pieces of your thesis:

\include{prelude_kdv}
\include{new_commands}

% begin{place code}

\include{interval_arithmetic_single}

\include{interval_arithmetic_all}

% end{place code}

\end{document}

%% file: prelude_kdv.tex
%
% Introduction
%

\chapter{Introduction}

We document the STABLAB code used to study the stability of traveling-wave solutions of the Kuramoto-Sivashinsky equation in the Korteweg-de Vries limit. 

%% file: new_commands.tex
% new commands

%
% The numbers
%

\newcommand{\R}{\mathbb{R}}
\newcommand{\C}{\mathbb{C}}
\newcommand{\Z}{\mathbb{Z}}
\newcommand{\N}{\mathbb{N}}
\newcommand{\Q}{\mathbb{Q}}

%
% math commands
%

\newcommand{\ti}{\times}
\newcommand{\grad}{\triangledown}
\newcommand{\mat}[1]{\begin{pmatrix}#1\end{pmatrix}}
\newcommand{\imply}[1]{\Rightarrow}
\newcommand{\sech}{\mathrm{sech}}
\newcommand{\pd}[2]{\frac{\partial #1}{\partial #2}}

% 
% text formatting commands
%

\newcommand{\n}{\newline}
\newcommand{\qut}[1]{\textquotedblleft #1\textquotedblright}

%
% equation modes
%

\newcommand{\be}{\begin{equation}}
\newcommand{\ee}{\end{equation}}
\newcommand{\eq}[2]{\begin{equation}\begin{split}#1\end{split}#2\end{equation}}
\newcommand{\eqn}[2]{\begin{equation}\begin{split}#1\end{split}#2\notag\end{equation}}
\newcommand{\eqnb}[2]{\begin{equation}\left\{\begin{split}#1\end{split}\right.#2\notag\end{equation}}
\newcommand{\eqbn}[2]{\begin{equation}\left\{\begin{split}#1\end{split}\right.#2\notag\end{equation}}
\newcommand{\eqb}[2]{\begin{equation}\left\{\begin{split}#1\end{split}\right.#2\end{equation}}

%
% Greek letters
%

\newcommand{\eps}{\varepsilon}
\newcommand{\del}{\Delta}

%% file: interval_arithmetic_single.tex
\color{Black}\chapter{Study of single wave}

\color{Black}\section{bound\_numer.m}

\color{Green}\color{BrickRed}\color{NavyBlue}\-\ function\-\ \color{BrickRed}\-\ [M\_psi,M\_x,M\_q]\-\ =\-\ bound\_numer(dm,rho\_psi,rho\_x,rho\_q,a\_q,b\_q,a\_psi,b\_psi,ntilde)\color{Green}
$\\$
$\\$
$\\$\color{Green}$\%$
$\\$\color{Green}$\%$\-\ constants\-\ 
$\\$\color{Green}$\%$
$\\$
$\\$
$\\$\color{BrickRed}min\_abs\_q\-\ =\-\ (a\_q+b\_q)/2\-\ -(b\_q-a\_q)*(rho\_q+1/rho\_q)/4;\color{Green}
$\\$
$\\$
$\\$\color{Green}$\%$
$\\$\color{Green}$\%$\-\ bound\-\ for\-\ interpolation\-\ in\-\ psi\-\ ------------------------------------------------------------
$\\$\color{Green}$\%$
$\\$
$\\$
$\\$\color{Green}$\%$\-\ top\-\ of\-\ ellipse\-\ \color{Black}$E_{\rho_{\psi}}$ \color{Green}
$\\$\color{BrickRed}max\_imag\_x\-\ =\-\ 0;\color{Green}
$\\$\color{BrickRed}max\_real\_psi\-\ =\-\ (1+(rho\_psi+1/rho\_psi)/2)/2;\color{Green}
$\\$\color{BrickRed}max\_abs\_q\-\ =\-\ sup(real(b\_q));\color{Green}
$\\$\color{BrickRed}prodq\-\ =\-\ abs(log(a\_q)/2);\color{Green}
$\\$
$\\$
$\\$\color{BrickRed}max\_xi\-\ =\-\ bound\_xi(a\_psi,b\_psi,rho\_psi,a\_q,b\_q,1,ntilde);\color{Green}
$\\$\color{BrickRed}max\_xi\_der\-\ =\-\ bound\_xi\_der...\color{Green}
$\\$\color{BrickRed}\-\ \-\ \-\ \-\ (max\_abs\_q,min\_abs\_q,a\_q,b\_q,max\_real\_psi,rho\_q,rho\_psi,a\_psi,b\_psi,ntilde);\color{Green}
$\\$
$\\$
$\\$\color{BrickRed}M\_psi\-\ =\-\ 2*local\_bounds(dm,max\_imag\_x,max\_real\_psi,max\_abs\_q,max\_xi,max\_xi\_der,prodq);\color{Green}
$\\$
$\\$
$\\$\color{Green}$\%$
$\\$\color{Green}$\%$\-\ bound\-\ for\-\ interpolation\-\ in\-\ x\-\ ------------------------------------------------------------
$\\$\color{Green}$\%$
$\\$
$\\$
$\\$\color{Green}$\%$\-\ top\-\ of\-\ ellipse\-\ \color{Black}$E_{\rho_{\beta}}$ \color{Green}
$\\$\color{BrickRed}max\_imag\_x\-\ =\-\ (rho\_x-1/rho\_x)/2;\color{Green}
$\\$\color{BrickRed}max\_real\_psi\-\ =\-\ 1;\color{Green}
$\\$\color{Green}$\%$\-\ max\_abs\_q\-\ the\-\ same
$\\$\color{Green}$\%$\-\ prodq\-\ the\-\ same
$\\$\color{Green}$\%$\-\ max\_xi\-\ the\-\ same
$\\$\color{Green}$\%$\-\ max\_xi\_der\-\ the\-\ same
$\\$
$\\$
$\\$\color{BrickRed}M\_x\-\ =\-\ local\_bounds(dm,max\_imag\_x,max\_real\_psi,max\_abs\_q,max\_xi,max\_xi\_der,prodq);\color{Green}
$\\$
$\\$
$\\$\color{Green}$\%$
$\\$\color{Green}$\%$\-\ bound\-\ for\-\ interpolation\-\ in\-\ q\-\ ------------------------------------------------------------
$\\$\color{Green}$\%$
$\\$
$\\$
$\\$\color{Green}$\%$\-\ top\-\ of\-\ ellipse\-\ \color{Black}$E_{\rho_{\beta}}$ \color{Green}
$\\$\color{BrickRed}max\_imag\_x\-\ =\-\ 0;\color{Green}
$\\$\color{BrickRed}max\_real\_psi\-\ =\-\ 1;\color{Green}
$\\$\color{BrickRed}max\_abs\_q\-\ =\-\ (a\_q+b\_q)/2\-\ +\-\ ((b\_q-a\_q)/4)*(rho\_q+1/rho\_q);\color{Green}
$\\$
$\\$
$\\$\color{BrickRed}max\_xi\-\ =\-\ bound\_xi(a\_psi,b\_psi,rho\_psi,a\_q,b\_q,1,ntilde);\color{Green}
$\\$\color{BrickRed}max\_xi\_der\-\ =\-\ bound\_xi\_der...\color{Green}
$\\$\color{BrickRed}\-\ \-\ \-\ \-\ (max\_abs\_q,min\_abs\_q,a\_q,b\_q,max\_real\_psi,rho\_q,rho\_psi,a\_psi,b\_psi,ntilde);\color{Green}
$\\$
$\\$
$\\$\color{BrickRed}min\_abs\_q\-\ =\-\ (a\_q+b\_q)/2\-\ -(b\_q-a\_q)*(rho\_q+1/rho\_q)/4;\color{Green}
$\\$\color{BrickRed}prodq\-\ =\-\ abs(log(min\_abs\_q)/2);\color{Green}
$\\$
$\\$
$\\$\color{BrickRed}M\_q\-\ =\-\ 2*local\_bounds(dm,max\_imag\_x,max\_real\_psi,max\_abs\_q,max\_xi,max\_xi\_der,prodq);\color{Green}
$\\$
$\\$
$\\$\color{Green}$\%$--------------------------------------------------------------------------------------------------
$\\$\color{Green}$\%$\-\ local\_bounds
$\\$\color{Green}$\%$--------------------------------------------------------------------------------------------------
$\\$
$\\$
$\\$\color{BrickRed}\color{NavyBlue}\-\ function\-\ \color{BrickRed}\-\ out\-\ =\-\ local\_bounds(dm,max\_imag\_x,max\_real\_psi,max\_abs\_q,xi,xi\_der,prodq)\color{Green}
$\\$
$\\$
$\\$\color{BrickRed}pie\-\ =\-\ nm('pi');\color{Green}
$\\$\color{BrickRed}prod\-\ =\-\ pie/2;\color{Green}
$\\$
$\\$
$\\$\color{Green}$\%$\-\ bound\-\ on\-\ \color{Black} $\vartheta_1^{(m)}(\frac{\pi}{2\omega}(\omega x \pm i\omega'))$  \color{Green}
$\\$\color{BrickRed}temp\-\ =\-\ bound\_theta1\_m(max\_imag\_x,max\_real\_psi,max\_abs\_q,4);\color{Green}
$\\$\color{BrickRed}n0\-\ =\-\ temp(1);\-\ \color{Green}
$\\$\color{BrickRed}n1\-\ =\-\ temp(2);\-\ \color{Green}$\%$\-\ first\-\ derivatve
$\\$\color{BrickRed}n2\-\ =\-\ temp(3);\-\ \color{Green}$\%$\-\ second\-\ derivative
$\\$\color{BrickRed}n3\-\ =\-\ temp(4);\-\ \color{Green}$\%$\-\ third\-\ derivative
$\\$\color{BrickRed}n4\-\ =\-\ temp(5);\color{Green}
$\\$
$\\$
$\\$\color{Green}$\%$\-\ bound\-\ on\-\ \color{Black} $\vartheta_1^{(m)}(\frac{\pi}{2\omega}
\color{Black} (\omega x \pm i\omega'))$  \color{Green}
$\\$\color{BrickRed}max\_real\_psi\-\ =\-\ 0;\color{Green}
$\\$\color{BrickRed}temp\-\ =\-\ bound\_theta1\_m(max\_imag\_x,max\_real\_psi,max\_abs\_q,4);\color{Green}
$\\$\color{BrickRed}d1\-\ =\-\ temp(2);\-\ \color{Green}$\%$\-\ first\-\ derivative
$\\$\color{BrickRed}d2\-\ =\-\ temp(3);\-\ \color{Green}$\%$\-\ second\-\ derivative
$\\$\color{BrickRed}d3\-\ =\-\ temp(4);\-\ \color{Green}$\%$\-\ third\-\ derivative
$\\$\color{BrickRed}d4\-\ =\-\ temp(5);\color{Green}
$\\$
$\\$
$\\$\color{Green}$\%$\-\ bound\-\ on\-\ w(x),\-\ the\-\ conjugate\-\ of\-\ w(x),\-\ and\-\ their\-\ derivatives
$\\$\color{Green}$\%$\-\ on\-\ the\-\ ellipse\-\ \color{Black}$E_{\rho_x}$ \color{Green}
$\\$
$\\$
$\\$\color{Green}$\%$\-\ w(x)
$\\$\color{BrickRed}w0\-\ =\-\ n0\verb|^|2/dm\verb|^|2;\color{Green}
$\\$
$\\$
$\\$\color{Green}$\%$\-\ w'(x)
$\\$\color{BrickRed}w1\-\ =\-\ 2*prod*(\-\ n0*n1/dm\verb|^|2\-\ +\-\ w0*d1/dm\-\ );\color{Green}
$\\$
$\\$
$\\$\color{Green}$\%$\-\ w''(x)
$\\$\color{BrickRed}w2\-\ =\-\ 4*prod\verb|^|2*(n1\verb|^|2/dm\verb|^|2+n0*n2/dm\verb|^|2+2*n0*n1*d1/dm\verb|^|3+w1*d1/dm+w0*d2/dm+w0*d1\verb|^|2/dm\verb|^|2);\color{Green}
$\\$
$\\$
$\\$\color{Green}$\%$\-\ w'''(x)
$\\$\color{BrickRed}w3\-\ =\-\ 8*prod\verb|^|3*(\-\ 2*n1*n2/dm\verb|^|2+2*n1\verb|^|2*d1/dm\verb|^|3+n1*n2/dm\verb|^|2+n0*n3/dm\verb|^|2+2*n0*n2*d1/dm\verb|^|3+...\color{Green}
$\\$\color{BrickRed}\-\ \-\ \-\ \-\ 2*n1\verb|^|2*d1/dm\verb|^|3+2*n0*n2*d1/dm\verb|^|3+2*n0*n1*d2/dm\verb|^|3+6*n0*n1*d1\verb|^|2/dm\verb|^|4+w2*d1/dm\-\ +\-\ w1*d2/dm+...\color{Green}
$\\$\color{BrickRed}\-\ \-\ \-\ \-\ w1*d1\verb|^|2/dm\verb|^|2+w1*d2/dm+\-\ w0*d3/dm+w0*d1*d2/dm\verb|^|2+w1*d1\verb|^|2/dm\verb|^|2+2*w0*d1*d2/dm\verb|^|2+...\color{Green}
$\\$\color{BrickRed}\-\ \-\ \-\ \-\ 2*w0*d1\verb|^|3/dm\verb|^|3);\color{Green}
$\\$
$\\$
$\\$\color{Green}$\%$\-\ bound\-\ on\-\ derivative\-\ of\-\ w(x)\-\ with\-\ respect\-\ to\-\ psi
$\\$\color{BrickRed}w0\_psi\-\ =\-\ 2*prodq*(n0*n1/dm\verb|^|2);\color{Green}
$\\$
$\\$
$\\$\color{Green}$\%$\-\ derivative\-\ of\-\ w'(x)\-\ with\-\ respect\-\ to\-\ psi
$\\$\color{BrickRed}w1\_psi\-\ =\-\ 2*prodq\verb|^|2*(n0*n2/dm\verb|^|2+n1\verb|^|2/dm\verb|^|2+w0\_psi*d1/dm);\color{Green}
$\\$
$\\$
$\\$\color{Green}$\%$\-\ bound\-\ on\-\ derivative\-\ of\-\ w''(x)\-\ with\-\ respect\-\ to\-\ psi
$\\$\color{Green}$\%$\-\ |w''(x)|\-\ $<$\-\ 4*prod\verb|^|2*(\-\ (2*n0*n1*d1)/dm\verb|^|3+\-\ (n1\verb|^|2+n0*n2+w0*d1\verb|^|2)/dm\verb|^|2\-\ +\-\ (w1*d1+\-\ w0*d2)/dm\-\ );
$\\$\color{BrickRed}w2\_psi\-\ =\-\ 4*prodq\verb|^|3*(\-\ (2*n1\verb|^|2*d1+2*n0*n2*d1+2*n0*n1*d2)/dm\verb|^|3\-\ +\-\ ...\color{Green}
$\\$\color{BrickRed}\-\ \-\ \-\ \-\ (2*n1*n2+n1*n2+n0*n3+w0\_psi*d1\verb|^|2+\-\ w0*2*d1*d2)/dm\verb|^|2\-\ +\-\ ...\color{Green}
$\\$\color{BrickRed}\-\ \-\ \-\ \-\ (w1\_psi*d1+w1*d2+w0\_psi*d2+w0*d3)/dm);\color{Green}
$\\$
$\\$
$\\$\color{Green}$\%$\-\ bound\-\ on\-\ derivative\-\ of\-\ w'''(x)\-\ with\-\ respect\-\ to\-\ psi
$\\$\color{BrickRed}w3\_psi\-\ =\-\ 8*prodq\verb|^|4*(...\color{Green}
$\\$\color{BrickRed}\-\ \-\ \-\ \-\ 2*n2*n2/dm\verb|^|2+2*n1*n3/dm\verb|^|2\-\ +\-\ ...\-\ \color{Green}$\%$
$\\$\color{BrickRed}\-\ \-\ \-\ \-\ 4*n1*n2*d1/dm\verb|^|3+2*n1\verb|^|2*d2/dm\verb|^|3+\-\ ...\color{Green}$\%$
$\\$\color{BrickRed}\-\ \-\ \-\ \-\ n2\verb|^|2/dm\verb|^|2+\-\ n1*n3/dm\verb|^|2\-\ +\-\ ...\-\ \color{Green}$\%$
$\\$\color{BrickRed}\-\ \-\ \-\ \-\ n1*n3/dm\verb|^|2+n0*n4/dm\verb|^|2\-\ +\-\ ...\-\ \color{Green}$\%$
$\\$\color{BrickRed}\-\ \-\ \-\ \-\ 2*n1*n2*d1/dm\verb|^|3\-\ +\-\ 2*n0*n3*d1/dm\verb|^|3\-\ +\-\ 2*n0*n2*d2/dm\verb|^|3\-\ +...\-\ \color{Green}$\%$
$\\$\color{BrickRed}\-\ \-\ \-\ \-\ 4*n1*n2*d1/dm\verb|^|3\-\ +\-\ 2*n1\verb|^|2*d2/dm\verb|^|3\-\ +\-\ ...\-\ \color{Green}$\%$
$\\$\color{BrickRed}\-\ \-\ \-\ \-\ 2*n1*n2*d1/dm\verb|^|3\-\ +\-\ 2*n0*n3*d1/dm\verb|^|3\-\ +\-\ 2*n0*n2*d2/dm\verb|^|3\-\ +\-\ ...\-\ \color{Green}$\%$
$\\$\color{BrickRed}\-\ \-\ \-\ \-\ 2*n1\verb|^|2*d2/dm\verb|^|3\-\ +\-\ 2*n0*n2*d2/dm\verb|^|3+2*n0*n1*d3/dm\verb|^|3+\-\ ...\-\ \color{Green}$\%$
$\\$\color{BrickRed}\-\ \-\ \-\ \-\ 6*n1\verb|^|2*d1\verb|^|2/dm\verb|^|4\-\ +\-\ 6*n0*n2*d1\verb|^|2/dm\verb|^|4+6*n0*n1*2*d1*d2/dm\verb|^|4\-\ +\-\ ...\-\ \color{Green}$\%$
$\\$\color{BrickRed}\-\ \-\ \-\ \-\ w2\_psi*d1/dm\-\ +\-\ w2*d2/dm\-\ +...\-\ \color{Green}$\%$
$\\$\color{BrickRed}\-\ \-\ \-\ \-\ w1\_psi*d2/dm\-\ +\-\ w1*d3/dm+\-\ ...\-\ \color{Green}$\%$
$\\$\color{BrickRed}\-\ \-\ \-\ \-\ w1\_psi*d1\verb|^|2/dm\verb|^|2\-\ +\-\ w1*2*d1*d2/dm\verb|^|2\-\ +\-\ ...\-\ \color{Green}$\%$
$\\$\color{BrickRed}\-\ \-\ \-\ \-\ w1\_psi*d2/dm\-\ +\-\ w1*d3/dm\-\ +\-\ ...\-\ \color{Green}$\%$
$\\$\color{BrickRed}\-\ \-\ \-\ \-\ w0\_psi*d3/dm\-\ +\-\ w0*d4/dm\-\ +\-\ ...\-\ \color{Green}$\%$
$\\$\color{BrickRed}\-\ \-\ \-\ \-\ w0\_psi*d1*d2/dm\verb|^|2\-\ +\-\ w0*d2*d2/dm\verb|^|2\-\ +\-\ w0*d1*d3/dm\verb|^|2\-\ +\-\ ...\-\ \color{Green}$\%$
$\\$\color{BrickRed}\-\ \-\ \-\ \-\ w1\_psi*d1\verb|^|2/dm\verb|^|2\-\ +\-\ w1*2*d1*d2/dm\verb|^|2\-\ +\-\ ...\-\ \color{Green}$\%$
$\\$\color{BrickRed}\-\ \-\ \-\ \-\ 2*w0\_psi*d1*d2/dm\verb|^|2\-\ +\-\ 2*w0*d2*d2/dm\verb|^|2\-\ +\-\ 2*w0*d1*d3/dm\verb|^|2\-\ +\-\ ...\color{Green}$\%$
$\\$\color{BrickRed}\-\ \-\ \-\ \-\ 2*w0\_psi*d1\verb|^|3/dm\verb|^|3\-\ +\-\ 2*w0*3*d1\verb|^|2*d2/dm\verb|^|3\-\ ...\-\ \color{Green}$\%$
$\\$\color{BrickRed}\-\ \-\ \-\ \-\ );\color{Green}
$\\$
$\\$
$\\$\color{BrickRed}xi2\-\ =\-\ xi*xi;\color{Green}
$\\$\color{BrickRed}xi3\-\ =\-\ xi2*xi;\color{Green}
$\\$
$\\$
$\\$\color{Green}$\%$\-\ derivatives\-\ of\-\ v\-\ with\-\ respect\-\ to\-\ x
$\\$\color{BrickRed}v1\-\ =\-\ w1+\-\ xi*w0;\color{Green}
$\\$\color{BrickRed}v2\-\ =\-\ w2+2*xi*w1+xi2*w0;\color{Green}
$\\$\color{BrickRed}v3\-\ =\-\ w3+3*xi*w2+3*xi2*w1+xi3*w0;\color{Green}
$\\$
$\\$
$\\$\color{Green}$\%$\-\ derivatives\-\ with\-\ respect\-\ to\-\ psi
$\\$\color{BrickRed}v1\_psi\-\ =\-\ w1\_psi+xi\_der*w0+xi*w0\_psi;\color{Green}
$\\$\color{BrickRed}v2\_psi\-\ =\-\ w2\_psi+2*xi\_der*w1+2*xi*w1\_psi+2*xi*xi\_der*w0+xi2*w0\_psi;\color{Green}
$\\$\color{BrickRed}v3\_psi\-\ =\-\ w3\_psi+3*xi\_der*w2+3*xi*w2\_psi+6*xi*xi\_der*w1...\color{Green}
$\\$\color{BrickRed}\-\ \-\ \-\ \-\ \-\ \-\ \-\ \-\ \-\ \-\ \-\ \-\ \-\ \-\ \-\ \-\ +3*xi2*w1\_psi\-\ +\-\ 3*xi2*xi\_der*w0+xi3*w0\_psi;\color{Green}
$\\$\color{BrickRed}\-\ \-\ \-\ \-\ \-\ \color{Green}
$\\$\color{Green}$\%$\-\ bound\-\ on\-\ functions\-\ to\-\ interpoate\-\ \-\ \-\ \-\ \-\ \-\ \-\ \-\ \-\ 
$\\$\color{BrickRed}f1\-\ =\-\ v1*v2;\color{Green}
$\\$\color{BrickRed}f2\-\ =\-\ v3*v2;\color{Green}
$\\$\color{BrickRed}g\-\ =\-\ w0*v1;\color{Green}
$\\$\color{BrickRed}\-\ \-\ \-\ \-\ \-\ \-\ \-\ \-\ \-\ \-\ \-\ \-\ \color{Green}
$\\$\color{Green}$\%$\-\ bound\-\ on\-\ derivative\-\ with\-\ respect\-\ to\-\ psi\-\ of\-\ 
$\\$\color{Green}$\%$\-\ functions\-\ to\-\ interpolate
$\\$\color{BrickRed}f1\_psi\-\ =\-\ v1\_psi*v2+v1*v2\_psi;\color{Green}
$\\$\color{BrickRed}f2\_psi\-\ =\-\ v3\_psi*v2+v3*v2\_psi;\color{Green}
$\\$\color{BrickRed}g\_psi\-\ =\-\ w0\_psi*v1+w0*v1\_psi;\color{Green}
$\\$
$\\$
$\\$\color{Green}$\%$\-\ here\-\ =\-\ f1(round(0.5*length(f1)))
$\\$
$\\$
$\\$\color{Green}$\%$\-\ bound\-\ on\-\ all\-\ sub\-\ bounds
$\\$\color{BrickRed}out\-\ =\-\ nm(max(sup([f1,f2,g,f1\_psi,f2\_psi,g\_psi])));\color{Green}
$\\$
$\\$
$\\$\color{Black}\section{bound\_numer\_unstable.m}

\color{Green}\color{BrickRed}\color{NavyBlue}\-\ function\-\ \color{BrickRed}\-\ M\_x\-\ =\-\ bound\_numer\_unstable(dm,rho\_x,a\_q,b\_q)\color{Green}
$\\$
$\\$
$\\$\color{Green}$\%$
$\\$\color{Green}$\%$\-\ bound\-\ for\-\ interpolation\-\ in\-\ x\-\ ------------------------------------------------------------
$\\$\color{Green}$\%$
$\\$
$\\$
$\\$\color{BrickRed}pie\-\ =\-\ nm('pi');\color{Green}
$\\$\color{Green}$\%$\-\ top\-\ of\-\ ellipse\-\ \color{Black}$E_{\rho_{\beta}}$ \color{Green}
$\\$\color{BrickRed}max\_abs\_q\-\ =\-\ sup(real(b\_q));\color{Green}
$\\$\color{BrickRed}prodq\-\ =\-\ abs(log(a\_q)/2);\color{Green}
$\\$\color{BrickRed}max\_imag\_x\-\ =\-\ (rho\_x-1/rho\_x)/2;\color{Green}
$\\$\color{BrickRed}max\_real\_psi\-\ =\-\ 0;\color{Green}
$\\$
$\\$
$\\$\color{BrickRed}max\_xi\-\ =\-\ (2*pie).*((1-a\_q)./(1+a\_q))+4*pie*(1./(1-b\_q.\verb|^|2)).*(b\_q./(1-b\_q));\color{Green}
$\\$
$\\$
$\\$\color{BrickRed}max\_xi\_der\-\ =\-\ -4*pie*log(a\_q)*(1/(1-b\_q\verb|^|2))*(b\_q/(1-b\_q)\verb|^|2);\color{Green}
$\\$
$\\$
$\\$\color{BrickRed}M\_x\-\ =\-\ local\_bounds(dm,max\_imag\_x,max\_real\_psi,max\_abs\_q,max\_xi,max\_xi\_der,prodq);\color{Green}
$\\$
$\\$
$\\$\color{Green}$\%$--------------------------------------------------------------------------------------------------
$\\$\color{Green}$\%$\-\ local\_bounds
$\\$\color{Green}$\%$--------------------------------------------------------------------------------------------------
$\\$
$\\$
$\\$\color{BrickRed}\color{NavyBlue}\-\ function\-\ \color{BrickRed}\-\ out\-\ =\-\ local\_bounds(dm,max\_imag\_x,max\_real\_psi,max\_abs\_q,xi,xi\_der,prodq)\color{Green}
$\\$
$\\$
$\\$\color{BrickRed}pie\-\ =\-\ nm('pi');\color{Green}
$\\$\color{BrickRed}prod\-\ =\-\ pie/2;\color{Green}
$\\$
$\\$
$\\$\color{Green}$\%$\-\ bound\-\ on\-\ \color{Black} $\vartheta_1^{(m)}(\frac{\pi}{2\omega}(\omega x \pm i\omega'))$  \color{Green}
$\\$\color{BrickRed}temp\-\ =\-\ bound\_theta1\_m(max\_imag\_x,max\_real\_psi,max\_abs\_q,4);\color{Green}
$\\$\color{BrickRed}n0\-\ =\-\ temp(1);\-\ \color{Green}
$\\$\color{BrickRed}n1\-\ =\-\ temp(2);\-\ \color{Green}$\%$\-\ first\-\ derivatve
$\\$\color{BrickRed}n2\-\ =\-\ temp(3);\-\ \color{Green}$\%$\-\ second\-\ derivative
$\\$\color{BrickRed}n3\-\ =\-\ temp(4);\-\ \color{Green}$\%$\-\ third\-\ derivative
$\\$\color{BrickRed}n4\-\ =\-\ temp(5);\color{Green}
$\\$
$\\$
$\\$\color{Green}$\%$\-\ bound\-\ on\-\ \color{Black} $\vartheta_1^{(m)}(\frac{\pi}{2\omega}
\color{Black} (\omega x \pm i\omega'))$  \color{Green}
$\\$\color{BrickRed}max\_real\_psi\-\ =\-\ 0;\color{Green}
$\\$\color{BrickRed}temp\-\ =\-\ bound\_theta1\_m(max\_imag\_x,max\_real\_psi,max\_abs\_q,4);\color{Green}
$\\$\color{BrickRed}d1\-\ =\-\ temp(2);\-\ \color{Green}$\%$\-\ first\-\ derivative
$\\$\color{BrickRed}d2\-\ =\-\ temp(3);\-\ \color{Green}$\%$\-\ second\-\ derivative
$\\$\color{BrickRed}d3\-\ =\-\ temp(4);\-\ \color{Green}$\%$\-\ third\-\ derivative
$\\$\color{BrickRed}d4\-\ =\-\ temp(5);\color{Green}
$\\$
$\\$
$\\$\color{Green}$\%$\-\ bound\-\ on\-\ w(x),\-\ the\-\ conjugate\-\ of\-\ w(x),\-\ and\-\ their\-\ derivatives
$\\$\color{Green}$\%$\-\ on\-\ the\-\ ellipse\-\ \color{Black}$E_{\rho_x}$ \color{Green}
$\\$
$\\$
$\\$\color{Green}$\%$\-\ w(x)
$\\$\color{BrickRed}w0\-\ =\-\ n0\verb|^|2/dm\verb|^|2;\color{Green}
$\\$
$\\$
$\\$\color{Green}$\%$\-\ w'(x)
$\\$\color{BrickRed}w1\-\ =\-\ 2*prod*(\-\ n0*n1/dm\verb|^|2\-\ +\-\ w0*d1/dm\-\ );\color{Green}
$\\$
$\\$
$\\$\color{Green}$\%$\-\ w''(x)
$\\$\color{BrickRed}w2\-\ =\-\ 4*prod\verb|^|2*(n1\verb|^|2/dm\verb|^|2+n0*n2/dm\verb|^|2+2*n0*n1*d1/dm\verb|^|3+w1*d1/dm+w0*d2/dm+w0*d1\verb|^|2/dm\verb|^|2);\color{Green}
$\\$
$\\$
$\\$\color{Green}$\%$\-\ w'''(x)
$\\$\color{BrickRed}w3\-\ =\-\ 8*prod\verb|^|3*(\-\ 2*n1*n2/dm\verb|^|2+2*n1\verb|^|2*d1/dm\verb|^|3+n1*n2/dm\verb|^|2+n0*n3/dm\verb|^|2+2*n0*n2*d1/dm\verb|^|3+...\color{Green}
$\\$\color{BrickRed}\-\ \-\ \-\ \-\ 2*n1\verb|^|2*d1/dm\verb|^|3+2*n0*n2*d1/dm\verb|^|3+2*n0*n1*d2/dm\verb|^|3+6*n0*n1*d1\verb|^|2/dm\verb|^|4+w2*d1/dm\-\ +\-\ w1*d2/dm+...\color{Green}
$\\$\color{BrickRed}\-\ \-\ \-\ \-\ w1*d1\verb|^|2/dm\verb|^|2+w1*d2/dm+\-\ w0*d3/dm+w0*d1*d2/dm\verb|^|2+w1*d1\verb|^|2/dm\verb|^|2+2*w0*d1*d2/dm\verb|^|2+...\color{Green}
$\\$\color{BrickRed}\-\ \-\ \-\ \-\ 2*w0*d1\verb|^|3/dm\verb|^|3);\color{Green}
$\\$
$\\$
$\\$\color{Green}$\%$\-\ bound\-\ on\-\ derivative\-\ of\-\ w(x)\-\ with\-\ respect\-\ to\-\ psi
$\\$\color{BrickRed}w0\_psi\-\ =\-\ 2*prodq*(n0*n1/dm\verb|^|2);\color{Green}
$\\$
$\\$
$\\$\color{Green}$\%$\-\ derivative\-\ of\-\ w'(x)\-\ with\-\ respect\-\ to\-\ psi
$\\$\color{BrickRed}w1\_psi\-\ =\-\ 2*prodq\verb|^|2*(n0*n2/dm\verb|^|2+n1\verb|^|2/dm\verb|^|2+w0\_psi*d1/dm);\color{Green}
$\\$
$\\$
$\\$\color{Green}$\%$\-\ bound\-\ on\-\ derivative\-\ of\-\ w''(x)\-\ with\-\ respect\-\ to\-\ psi
$\\$\color{Green}$\%$\-\ |w''(x)|\-\ $<$\-\ 4*prod\verb|^|2*(\-\ (2*n0*n1*d1)/dm\verb|^|3+\-\ (n1\verb|^|2+n0*n2+w0*d1\verb|^|2)/dm\verb|^|2\-\ +\-\ (w1*d1+\-\ w0*d2)/dm\-\ );
$\\$\color{BrickRed}w2\_psi\-\ =\-\ 4*prodq\verb|^|3*(\-\ (2*n1\verb|^|2*d1+2*n0*n2*d1+2*n0*n1*d2)/dm\verb|^|3\-\ +\-\ ...\color{Green}
$\\$\color{BrickRed}\-\ \-\ \-\ \-\ (2*n1*n2+n1*n2+n0*n3+w0\_psi*d1\verb|^|2+\-\ w0*2*d1*d2)/dm\verb|^|2\-\ +\-\ ...\color{Green}
$\\$\color{BrickRed}\-\ \-\ \-\ \-\ (w1\_psi*d1+w1*d2+w0\_psi*d2+w0*d3)/dm);\color{Green}
$\\$
$\\$
$\\$\color{Green}$\%$\-\ bound\-\ on\-\ derivative\-\ of\-\ w'''(x)\-\ with\-\ respect\-\ to\-\ psi
$\\$\color{BrickRed}w3\_psi\-\ =\-\ 8*prodq\verb|^|4*(...\color{Green}
$\\$\color{BrickRed}\-\ \-\ \-\ \-\ 2*n2*n2/dm\verb|^|2+2*n1*n3/dm\verb|^|2\-\ +\-\ ...\-\ \color{Green}$\%$
$\\$\color{BrickRed}\-\ \-\ \-\ \-\ 4*n1*n2*d1/dm\verb|^|3+2*n1\verb|^|2*d2/dm\verb|^|3+\-\ ...\color{Green}$\%$
$\\$\color{BrickRed}\-\ \-\ \-\ \-\ n2\verb|^|2/dm\verb|^|2+\-\ n1*n3/dm\verb|^|2\-\ +\-\ ...\-\ \color{Green}$\%$
$\\$\color{BrickRed}\-\ \-\ \-\ \-\ n1*n3/dm\verb|^|2+n0*n4/dm\verb|^|2\-\ +\-\ ...\-\ \color{Green}$\%$
$\\$\color{BrickRed}\-\ \-\ \-\ \-\ 2*n1*n2*d1/dm\verb|^|3\-\ +\-\ 2*n0*n3*d1/dm\verb|^|3\-\ +\-\ 2*n0*n2*d2/dm\verb|^|3\-\ +...\-\ \color{Green}$\%$
$\\$\color{BrickRed}\-\ \-\ \-\ \-\ 4*n1*n2*d1/dm\verb|^|3\-\ +\-\ 2*n1\verb|^|2*d2/dm\verb|^|3\-\ +\-\ ...\-\ \color{Green}$\%$
$\\$\color{BrickRed}\-\ \-\ \-\ \-\ 2*n1*n2*d1/dm\verb|^|3\-\ +\-\ 2*n0*n3*d1/dm\verb|^|3\-\ +\-\ 2*n0*n2*d2/dm\verb|^|3\-\ +\-\ ...\-\ \color{Green}$\%$
$\\$\color{BrickRed}\-\ \-\ \-\ \-\ 2*n1\verb|^|2*d2/dm\verb|^|3\-\ +\-\ 2*n0*n2*d2/dm\verb|^|3+2*n0*n1*d3/dm\verb|^|3+\-\ ...\-\ \color{Green}$\%$
$\\$\color{BrickRed}\-\ \-\ \-\ \-\ 6*n1\verb|^|2*d1\verb|^|2/dm\verb|^|4\-\ +\-\ 6*n0*n2*d1\verb|^|2/dm\verb|^|4+6*n0*n1*2*d1*d2/dm\verb|^|4\-\ +\-\ ...\-\ \color{Green}$\%$
$\\$\color{BrickRed}\-\ \-\ \-\ \-\ w2\_psi*d1/dm\-\ +\-\ w2*d2/dm\-\ +...\-\ \color{Green}$\%$
$\\$\color{BrickRed}\-\ \-\ \-\ \-\ w1\_psi*d2/dm\-\ +\-\ w1*d3/dm+\-\ ...\-\ \color{Green}$\%$
$\\$\color{BrickRed}\-\ \-\ \-\ \-\ w1\_psi*d1\verb|^|2/dm\verb|^|2\-\ +\-\ w1*2*d1*d2/dm\verb|^|2\-\ +\-\ ...\-\ \color{Green}$\%$
$\\$\color{BrickRed}\-\ \-\ \-\ \-\ w1\_psi*d2/dm\-\ +\-\ w1*d3/dm\-\ +\-\ ...\-\ \color{Green}$\%$
$\\$\color{BrickRed}\-\ \-\ \-\ \-\ w0\_psi*d3/dm\-\ +\-\ w0*d4/dm\-\ +\-\ ...\-\ \color{Green}$\%$
$\\$\color{BrickRed}\-\ \-\ \-\ \-\ w0\_psi*d1*d2/dm\verb|^|2\-\ +\-\ w0*d2*d2/dm\verb|^|2\-\ +\-\ w0*d1*d3/dm\verb|^|2\-\ +\-\ ...\-\ \color{Green}$\%$
$\\$\color{BrickRed}\-\ \-\ \-\ \-\ w1\_psi*d1\verb|^|2/dm\verb|^|2\-\ +\-\ w1*2*d1*d2/dm\verb|^|2\-\ +\-\ ...\-\ \color{Green}$\%$
$\\$\color{BrickRed}\-\ \-\ \-\ \-\ 2*w0\_psi*d1*d2/dm\verb|^|2\-\ +\-\ 2*w0*d2*d2/dm\verb|^|2\-\ +\-\ 2*w0*d1*d3/dm\verb|^|2\-\ +\-\ ...\color{Green}$\%$
$\\$\color{BrickRed}\-\ \-\ \-\ \-\ 2*w0\_psi*d1\verb|^|3/dm\verb|^|3\-\ +\-\ 2*w0*3*d1\verb|^|2*d2/dm\verb|^|3\-\ ...\-\ \color{Green}$\%$
$\\$\color{BrickRed}\-\ \-\ \-\ \-\ );\color{Green}
$\\$
$\\$
$\\$\color{BrickRed}xi2\-\ =\-\ xi*xi;\color{Green}
$\\$\color{BrickRed}xi3\-\ =\-\ xi2*xi;\color{Green}
$\\$
$\\$
$\\$\color{Green}$\%$\-\ derivatives\-\ of\-\ v\-\ with\-\ respect\-\ to\-\ x
$\\$\color{BrickRed}v1\-\ =\-\ w1+\-\ xi*w0;\color{Green}
$\\$\color{BrickRed}v2\-\ =\-\ w2+2*xi*w1+xi2*w0;\color{Green}
$\\$\color{BrickRed}v3\-\ =\-\ w3+3*xi*w2+3*xi2*w1+xi3*w0;\color{Green}
$\\$
$\\$
$\\$\color{Green}$\%$\-\ derivatives\-\ with\-\ respect\-\ to\-\ psi
$\\$\color{BrickRed}v1\_psi\-\ =\-\ w1\_psi+xi\_der*w0+xi*w0\_psi;\color{Green}
$\\$\color{BrickRed}v2\_psi\-\ =\-\ w2\_psi+2*xi\_der*w1+2*xi*w1\_psi+2*xi*xi\_der*w0+xi2*w0\_psi;\color{Green}
$\\$\color{BrickRed}v3\_psi\-\ =\-\ w3\_psi+3*xi\_der*w2+3*xi*w2\_psi+6*xi*xi\_der*w1...\color{Green}
$\\$\color{BrickRed}\-\ \-\ \-\ \-\ \-\ \-\ \-\ \-\ \-\ \-\ \-\ \-\ \-\ \-\ \-\ \-\ +3*xi2*w1\_psi\-\ +\-\ 3*xi2*xi\_der*w0+xi3*w0\_psi;\color{Green}
$\\$\color{BrickRed}\-\ \-\ \-\ \-\ \-\ \color{Green}
$\\$\color{Green}$\%$\-\ bound\-\ on\-\ functions\-\ to\-\ interpoate\-\ \-\ \-\ \-\ \-\ \-\ \-\ \-\ \-\ 
$\\$\color{BrickRed}f1\-\ =\-\ v1*v2;\color{Green}
$\\$\color{BrickRed}f2\-\ =\-\ v3*v2;\color{Green}
$\\$\color{BrickRed}g\-\ =\-\ w0*v1;\color{Green}
$\\$\color{BrickRed}\-\ \-\ \-\ \-\ \-\ \-\ \-\ \-\ \-\ \-\ \-\ \-\ \color{Green}
$\\$\color{Green}$\%$\-\ bound\-\ on\-\ derivative\-\ with\-\ respect\-\ to\-\ psi\-\ of\-\ 
$\\$\color{Green}$\%$\-\ functions\-\ to\-\ interpolate
$\\$\color{BrickRed}f1\_psi\-\ =\-\ v1\_psi*v2+v1*v2\_psi;\color{Green}
$\\$\color{BrickRed}f2\_psi\-\ =\-\ v3\_psi*v2+v3*v2\_psi;\color{Green}
$\\$\color{BrickRed}g\_psi\-\ =\-\ w0\_psi*v1+w0*v1\_psi;\color{Green}
$\\$
$\\$
$\\$\color{Green}$\%$\-\ bound\-\ on\-\ all\-\ sub\-\ bounds
$\\$\color{BrickRed}out\-\ =\-\ nm(max(sup([f1,f2,g,f1\_psi,f2\_psi,g\_psi])));\color{Green}
$\\$\color{Black}\section{bound\_sub\_integrals.m}

\color{Green}\color{BrickRed}\color{NavyBlue}\-\ function\-\ \color{BrickRed}\-\ [M\_psi,M\_x,M\_q]\-\ =\-\ bound\_sub\_integrals(dm,rho\_psi,rho\_x,rho\_q,a\_q,b\_q)\color{Green}
$\\$
$\\$
$\\$\color{Green}$\%$\-\ constants
$\\$\color{BrickRed}pie\-\ =\-\ nm('pi');\color{Green}
$\\$\color{BrickRed}prod\-\ =\-\ pie/2;\color{Green}
$\\$
$\\$
$\\$\color{Green}$\%$\-\ -----------------------------------------------------------
$\\$\color{Green}$\%$\-\ bound\-\ for\-\ interpolation\-\ in\-\ psi
$\\$\color{Green}$\%$\-\ -----------------------------------------------------------
$\\$
$\\$
$\\$\color{Green}$\%$\-\ top\-\ of\-\ ellipse\-\ \color{Black}$E_{\rho_x}$ \color{Green}
$\\$\color{BrickRed}max\_imag\_x\-\ =\-\ 0;\color{Green}
$\\$\color{BrickRed}max\_real\_psi\-\ =\-\ (1+(rho\_psi+1/rho\_psi)/2)/2;\color{Green}
$\\$\color{BrickRed}max\_abs\_q\-\ =\-\ sup(real(b\_q));\color{Green}
$\\$
$\\$
$\\$\color{BrickRed}vec\-\ =\-\ bound(prod,dm,max\_imag\_x,max\_real\_psi,max\_abs\_q);\color{Green}
$\\$
$\\$
$\\$\color{Green}$\%$\-\ largest\-\ bound\-\ on\-\ all\-\ the\-\ sub\-\ integrands\-\ when\-\ alpha\-\ =\-\ \-\ i\-\ beta
$\\$\color{BrickRed}M\_psi\-\ =\-\ sup(2*nm(max(vec)));\color{Green}
$\\$
$\\$
$\\$\color{Green}$\%$\-\ -----------------------------------------------------------
$\\$\color{Green}$\%$\-\ bound\-\ for\-\ interpolation\-\ in\-\ x
$\\$\color{Green}$\%$\-\ -----------------------------------------------------------
$\\$
$\\$
$\\$\color{Green}$\%$\-\ top\-\ of\-\ ellipse\-\ \color{Black}$E_{\rho_x}$ \color{Green}
$\\$\color{BrickRed}max\_imag\_x\-\ =\-\ (rho\_x-1/rho\_x)/2;\color{Green}
$\\$\color{BrickRed}max\_real\_psi\-\ =\-\ 1;\color{Green}
$\\$\color{BrickRed}max\_abs\_q\-\ =\-\ sup(real(b\_q));\color{Green}
$\\$
$\\$
$\\$\color{BrickRed}vec\-\ =\-\ bound(prod,dm,max\_imag\_x,max\_real\_psi,max\_abs\_q);\color{Green}
$\\$
$\\$
$\\$\color{Green}$\%$\-\ largest\-\ bound\-\ on\-\ all\-\ the\-\ sub\-\ integrands\-\ when\-\ alpha\-\ =\-\ \-\ i\-\ beta
$\\$\color{BrickRed}M\_x\-\ =\-\ max(vec);\color{Green}
$\\$
$\\$
$\\$\color{Green}$\%$\-\ -----------------------------------------------------------
$\\$\color{Green}$\%$\-\ bound\-\ for\-\ interpolation\-\ in\-\ q
$\\$\color{Green}$\%$\-\ -----------------------------------------------------------
$\\$
$\\$
$\\$\color{Green}$\%$\-\ top\-\ of\-\ ellipse\-\ \color{Black}$E_{\rho_x}$ \color{Green}
$\\$\color{BrickRed}max\_imag\_x\-\ =\-\ 0;\color{Green}
$\\$\color{BrickRed}max\_real\_psi\-\ =\-\ 1;\color{Green}
$\\$\color{BrickRed}max\_abs\_q\-\ =\-\ (a\_q+b\_q)/2\-\ +\-\ ((b\_q-a\_q)/4)*(rho\_q+1/rho\_q);\color{Green}
$\\$
$\\$
$\\$\color{BrickRed}vec\-\ =\-\ bound(prod,dm,max\_imag\_x,max\_real\_psi,max\_abs\_q);\color{Green}
$\\$
$\\$
$\\$\color{Green}$\%$\-\ largest\-\ bound\-\ on\-\ all\-\ the\-\ sub\-\ integrands\-\ when\-\ alpha\-\ =\-\ \-\ i\-\ beta
$\\$\color{BrickRed}M\_q\-\ =\-\ sup(2*nm(max(vec)));\color{Green}
$\\$
$\\$
$\\$\color{Green}$\%$\-\ -----------------------------------------------------------
$\\$\color{Green}$\%$\-\ function\-\ for\-\ bounds
$\\$\color{Green}$\%$\-\ -----------------------------------------------------------
$\\$\color{BrickRed}\color{NavyBlue}\-\ function\-\ \color{BrickRed}\-\ vec\-\ =\-\ bound(prod,dm,max\_imag\_x,max\_real\_psi,max\_abs\_q)\color{Green}
$\\$
$\\$
$\\$\color{Green}$\%$\-\ bound\-\ on\-\ \color{Black} $\vartheta_1^{(m)}(\frac{\pi}{2\omega}(\omega x \pm i\omega'))$  \color{Green}
$\\$\color{BrickRed}temp\-\ =\-\ bound\_theta1\_m(max\_imag\_x,max\_real\_psi,max\_abs\_q,4);\color{Green}
$\\$
$\\$
$\\$\color{BrickRed}n0\-\ =\-\ temp(1);\-\ \color{Green}
$\\$\color{BrickRed}n1\-\ =\-\ temp(2);\-\ \color{Green}$\%$\-\ first\-\ derivatve
$\\$\color{BrickRed}n2\-\ =\-\ temp(3);\-\ \color{Green}$\%$\-\ second\-\ derivative
$\\$\color{BrickRed}n3\-\ =\-\ temp(4);\-\ \color{Green}$\%$\-\ third\-\ derivative
$\\$
$\\$
$\\$\color{Green}$\%$\-\ bound\-\ on\-\ \color{Black} $\vartheta_1^{(m)}(\frac{\pi}{2\omega}
\color{Black} (\omega x \pm i\omega'+n\omega + i\beta))$  \color{Green}
$\\$\color{BrickRed}max\_real\_psi\-\ =\-\ 0;\color{Green}
$\\$\color{BrickRed}temp\-\ =\-\ bound\_theta1\_m(max\_imag\_x,max\_real\_psi,max\_abs\_q,4);\color{Green}
$\\$\color{BrickRed}d1\-\ =\-\ temp(2);\-\ \color{Green}$\%$\-\ first\-\ derivative
$\\$\color{BrickRed}d2\-\ =\-\ temp(3);\-\ \color{Green}$\%$\-\ second\-\ derivative
$\\$\color{BrickRed}d3\-\ =\-\ temp(4);\-\ \color{Green}$\%$\-\ third\-\ derivative
$\\$
$\\$
$\\$\color{Green}$\%$\-\ w(x)
$\\$\color{BrickRed}w0\-\ =\-\ n0\verb|^|2/dm\verb|^|2;\color{Green}
$\\$
$\\$
$\\$\color{Green}$\%$\-\ w'(x)
$\\$\color{BrickRed}w1\-\ =\-\ 2*prod*(\-\ n0*n1/dm\verb|^|2\-\ +\-\ w0*d1/dm\-\ );\color{Green}
$\\$
$\\$
$\\$\color{Green}$\%$\-\ w''(x)
$\\$\color{BrickRed}w2\-\ =\-\ 4*prod\verb|^|2*(n1\verb|^|2/dm\verb|^|2+n0*n2/dm\verb|^|2+2*n0*n1*d1/dm\verb|^|3+w1*d1/dm+w0*d2/dm+w0*d1\verb|^|2/dm\verb|^|2);\color{Green}
$\\$
$\\$
$\\$\color{Green}$\%$\-\ w'''(x)
$\\$\color{BrickRed}w3\-\ =\-\ 8*prod\verb|^|3*(\-\ 2*n1*n2/dm\verb|^|2+2*n1\verb|^|2*d1/dm\verb|^|3+n1*n2/dm\verb|^|2+n0*n3/dm\verb|^|2+2*n0*n2*d1/dm\verb|^|3+...\color{Green}
$\\$\color{BrickRed}\-\ \-\ \-\ \-\ 2*n1\verb|^|2*d1/dm\verb|^|3+2*n0*n2*d1/dm\verb|^|3+2*n0*n1*d2/dm\verb|^|3+6*n0*n1*d1\verb|^|2/dm\verb|^|4+w2*d1/dm\-\ +\-\ w1*d2/dm+...\color{Green}
$\\$\color{BrickRed}\-\ \-\ \-\ \-\ w1*d1\verb|^|2/dm\verb|^|2+w1*d2/dm+\-\ w0*d3/dm+w0*d1*d2/dm\verb|^|2+w1*d1\verb|^|2/dm\verb|^|2+2*w0*d1*d2/dm\verb|^|2+...\color{Green}
$\\$\color{BrickRed}\-\ \-\ \-\ \-\ 2*w0*d1\verb|^|3/dm\verb|^|3);\color{Green}
$\\$
$\\$
$\\$\color{Green}$\%$\-\ bounds\-\ on\-\ the\-\ 10\-\ sub\-\ integrands
$\\$
$\\$
$\\$\color{BrickRed}vec\-\ =\-\ [\-\ w1*w2;\-\ ...\color{Green}
$\\$\color{BrickRed}\-\ \-\ \-\ \-\ \-\ \-\ \-\ \-\ w0*w2+2*w1*w1;\-\ ...\color{Green}
$\\$\color{BrickRed}\-\ \-\ \-\ \-\ \-\ \-\ \-\ \-\ w1*w0+2*w0*w1;\-\ ...\color{Green}
$\\$\color{BrickRed}\-\ \-\ \-\ \-\ \-\ \-\ \-\ \-\ w0*w0;\-\ ...\color{Green}
$\\$\color{BrickRed}\-\ \-\ \-\ \-\ \-\ \-\ \-\ \-\ w3*w2;\-\ ...\color{Green}
$\\$\color{BrickRed}\-\ \-\ \-\ \-\ \-\ \-\ \-\ \-\ 3*w2*w2+2*w3*w1;\color{Green}
$\\$\color{BrickRed}\-\ \-\ \-\ \-\ \-\ \-\ \-\ \-\ w3*w0+6*w2*w1+3*w1*w2;\color{Green}
$\\$\color{BrickRed}\-\ \-\ \-\ \-\ \-\ \-\ \-\ \-\ 3*w2*w0+6*w1*w1+w0*w2;\color{Green}
$\\$\color{BrickRed}\-\ \-\ \-\ \-\ \-\ \-\ \-\ \-\ 3*w1*w0+2*w0*w1;\color{Green}
$\\$\color{BrickRed}\-\ \-\ \-\ \-\ \-\ \-\ \-\ \-\ w0*w0];\color{Green}
$\\$\color{BrickRed}\-\ \-\ \-\ \-\ \-\ \-\ \-\ \-\ \color{Green}
$\\$\color{Green}$\%$\-\ bound\-\ on\-\ 10\-\ sub\-\ integrands
$\\$\color{BrickRed}vec\-\ =\-\ sup(vec);\color{Green}
$\\$
$\\$
$\\$
$\\$
$\\$
$\\$
$\\$\color{Black}\section{bound\_theta1\_m.m}

\color{Green}\color{BrickRed}\color{NavyBlue}\-\ function\-\ \color{BrickRed}\-\ out\-\ =\-\ bound\_theta1\_m(max\_imag\_x,max\_real\_psi,max\_abs\_q,m)\color{Green}
$\\$
$\\$
$\\$
$\\$\color{Black}
The first Jacobi Theta function is given by the series,
\eqn{
\vartheta_1(z)&= 2\sum_{n=0}^{\infty} (-1)^n q^{(n+1/2)^2}\sin((2n+1)z),
}{}
and its $m$th derivative is given by,
\eqn{
\vartheta_1^{(m)}(z)&= 2\sum_{n=0}^{\infty}(-1)^n q^{(n+1/2)^2}(2n+1)^m f((2n+1)z),
}{}
where $f(\cdot) = \sin(\cdot)$ if $m\equiv 0 \mod 4$, $f(\cdot) = \cos(\cdot)$ if $m \equiv 1 \mod 4$,
$f(\cdot) = -\sin(\cdot)$ if $m\equiv 2 \mod 4$, and $f(\cdot) = -\cos(\cdot)$ if $m\equiv 3 \mod 4$.

$\\$
Now for $m \geq 0$, 
\eqn{
\left|(-1)^{n}q^{(n+1/2)^2}(2n+1)^m f((2n+1)z)\right| & \leq \left| q^{(n+1/2)^2}(2n+1)^m e^{(2n+1)|\Im(z)|}\right|\\
&\leq \left| q^{N^2+1/2}e^{(2N+1)|\Im(z)|}(2N+1)^m\right|q^n,
}{}
for $n\geq N$ with $N$ sufficiently large that
\eqn{
\left| q^{n^2+1/4}e^{(2n+1)|\Im(z)|}(2n+1)^m\right| & \leq \left| q^{N^2+1/4}e^{(2N+1)|\Im(z)|}(2N+1)^m\right|, 
}{}
whenever $n\geq N$. To determine how large $N$ must be, we define,
\eqn{
g(x):= q^{x^2+1/4}e^{(2x+1)|\Im(z)|}(2x+1)^m.
}{}
We will take $N$ large enough that $g'(x) < 0$ whenever $x\geq N$. If $m=0$, then
\eqn{
g'(x) = 2q^{x^2+1/4}e^{(2x+1)|\Im(z)|}\left( x\log(q)+|\Im(z)|\right),
}{}
and we see that 
\eqn{
N &> -\frac{|\Im(z)|}{\log(q)}
}{}
suffices. If $M > 0$,
\eqn{
g'(x) = 2q^{x^2+1/4}e^{(2x+1)|Im(z)|}(2x+1)^{m-1}\left((x\log(q)+|\Im(z)|)(2x+1)+m\right).
}{}
From 
\eqn{
(x\log(q)+|\Im(z)|)(2x+1)+m < 0,
}{}
we find that 
\eqn{
N > -\frac{2|\Im(z)|+\log(q) + \sqrt{(2|\Im(z)|+\log(q))^2-8\log(q)(|\Im(z)|+m)}}{4\log(q)}
}{}
suffices. For such an $N$, the error of the summation truncation is
\eqn{
q^{N^2+1/4}e^{(2N+1)|\Im(z)|}(2N+1)^m \frac{q^N}{1-q}.
}{}

$\\$
\color{Green}
$\\$
$\\$
$\\$
$\\$\color{Green}$\%$$\%$\-\ constants
$\\$\color{BrickRed}pie\-\ =\-\ nm('pi');\color{Green}
$\\$\color{BrickRed}q\-\ =\-\ max\_abs\_q;\color{Green}
$\\$\color{Green}$\%$\-\ \color{Black}Find max of $\vartheta_1(\frac{\pi}{2\omega}(\omega x \pm i\omega' +n\omega +i\beta))$ on 
\color{Black} the ellipse $E_{\rho}$. \color{Green}
$\\$
$\\$
$\\$\color{Green}$\%$\-\ abz\-\ =\-\ pie*(omega*max\_imag\_x+omega\_prime+max\_real\_beta)/(2*omega);
$\\$\color{BrickRed}abz\-\ =\-\ sup(pie*max\_imag\_x\-\ +\-\ abs(log(max\_abs\_q))*(1+max\_real\_psi))/2;\color{Green}
$\\$
$\\$
$\\$\color{BrickRed}qlog\-\ =\-\ log(q);\color{Green}
$\\$\color{BrickRed}one\_fourth\-\ =\-\ nm(1)/4;\color{Green}
$\\$\color{BrickRed}one\_half\-\ =\-\ nm(1)/2;\color{Green}
$\\$\color{BrickRed}con\-\ =\-\ q\verb|^|one\_fourth*exp(abz);\color{Green}
$\\$
$\\$
$\\$\color{Green}$\%$$\%$
$\\$
$\\$
$\\$\color{Green}$\%$\-\ find\-\ N\-\ large\-\ enough\-\ that\-\ 
$\\$\color{Green}$\%$\-\ \color{Black}$f(x):= q^{x^2+1/4}e^{(2x+1)|Im(z)|}(2x+1)^m$ \color{Green}
$\\$\color{Green}$\%$\-\ is\-\ decreasing\-\ for\-\ \color{Black}$ x \geq N$. \color{Green}
$\\$\color{BrickRed}\color{NavyBlue}\-\ if\-\ \color{BrickRed}\-\ m\-\ ==\-\ 0\color{Green}
$\\$\color{BrickRed}\-\ \-\ \-\ \-\ N\-\ =\-\ ceil(sup(-abz/qlog));\color{Green}
$\\$\color{BrickRed}\color{NavyBlue}\-\ else\-\ \color{BrickRed}\color{Green}
$\\$\color{BrickRed}\-\ \-\ \-\ \-\ c1\-\ =\-\ 2*abz+qlog;\color{Green}
$\\$\color{BrickRed}\-\ \-\ \-\ \-\ disc\-\ =\-\ c1\verb|^|2-8*qlog*(abz+m);\color{Green}
$\\$\color{BrickRed}\-\ \-\ \-\ \-\ c2\-\ =\-\ -c1/(4*qlog);\color{Green}
$\\$\color{BrickRed}\-\ \-\ \-\ \-\ c3\-\ =\-\ disc/(4*qlog);\color{Green}
$\\$\color{BrickRed}\-\ \-\ \-\ \-\ N\-\ =\-\ ceil(sup(c2-c3));\color{Green}
$\\$\color{BrickRed}\color{NavyBlue}\-\ end\-\ \color{BrickRed}\color{Green}
$\\$
$\\$
$\\$\color{Green}$\%$\-\ compute\-\ theta(z,q)\-\ and\-\ its\-\ derivatives\-\ with\-\ partial\-\ sum
$\\$\color{Green}$\%$\-\ \color{Black}$\vartheta_1^{(m)}(z)= 2\sum_{n=0}^{N-1}(-1)^n q^{(n+1/2)^2}(2n+1)^m f((2n+1)z)$ \color{Green}
$\\$\color{BrickRed}out\-\ =\-\ nm(zeros(m+1,1));\color{Green}
$\\$\color{BrickRed}\color{NavyBlue}\-\ for\-\ \color{BrickRed}\-\ n\-\ =\-\ 0:N-1\color{Green}
$\\$\color{BrickRed}\-\ \-\ \-\ \-\ prod\-\ =\-\ 2*n+1;\color{Green}
$\\$\color{BrickRed}\-\ \-\ \-\ \-\ zprod\-\ =\-\ prod*abz;\color{Green}
$\\$\color{BrickRed}\-\ \-\ \-\ \-\ ezprod\-\ =\-\ exp(zprod);\color{Green}
$\\$\color{BrickRed}\-\ \-\ \-\ \-\ gen\-\ =\-\ ezprod*q\verb|^|((n+one\_half)\verb|^|2);\color{Green}
$\\$\color{BrickRed}\-\ \-\ \-\ \-\ \color{NavyBlue}\-\ for\-\ \color{BrickRed}\-\ ind\-\ =\-\ 1:m+1\color{Green}
$\\$\color{BrickRed}\-\ \-\ \-\ \-\ \-\ \-\ \-\ \-\ out(ind)\-\ =\-\ out(ind)\-\ +\-\ gen*prod\verb|^|(ind-1);\color{Green}
$\\$\color{BrickRed}\-\ \-\ \-\ \-\ \color{NavyBlue}\-\ end\-\ \color{BrickRed}\color{Green}
$\\$\color{BrickRed}\color{NavyBlue}\-\ end\-\ \color{BrickRed}\color{Green}
$\\$
$\\$
$\\$\color{Green}$\%$\-\ truncation\-\ error
$\\$\color{Green}$\%$\-\ \color{Black}$q^{N^2+1/4}e^{(2N+1)|\Im(z)|}(2N+1)^m \frac{q^N}{1-q}$ \color{Green}
$\\$\color{Green}$\%$\-\ note\-\ that\-\ con\-\ =\-\ \color{Black}$q^{1/4}e^{|\Im(z)|}$ \color{Green}
$\\$\color{BrickRed}\color{NavyBlue}\-\ for\-\ \color{BrickRed}\-\ ind\-\ =\-\ 0:m\color{Green}
$\\$\color{BrickRed}\-\ \-\ \-\ \-\ out(ind+1)\-\ =\-\ out(ind+1)\-\ +\-\ con*q\verb|^|(N\verb|^|2)*exp(2*abz*N)*(2*N+1)\verb|^|ind*q\verb|^|N/(1-q);\color{Green}
$\\$\color{BrickRed}\color{NavyBlue}\-\ end\-\ \color{BrickRed}\color{Green}
$\\$
$\\$
$\\$\color{BrickRed}out\-\ =\-\ 2*out;\color{Green}
$\\$\color{BrickRed}out\-\ =\-\ sup(out);\color{Green}
$\\$\color{Black}\section{bound\_xi.m}

\color{Green}\color{BrickRed}\color{NavyBlue}\-\ function\-\ \color{BrickRed}\-\ out\-\ =\-\ bound\_xi(a\_psi,b\_psi,rho\_psi,a\_q,b\_q,rho\_q,ntilde)\color{Green}
$\\$
$\\$
$\\$\color{Green}$\%$\-\ latex\-\ description
$\\$
$\\$\color{Black}

$\\$
Now 
\eqn{
\xi(\alpha) &= 2i\left(\zeta(\alpha)-\frac{\alpha}{\omega}\zeta(\omega)\right)\\
&= 2i\left(\frac{\pi}{2\omega}\cot\left(\frac{\pi \alpha}{2\omega}\right)+\frac{2\pi}{\omega}
\sum_{k=1}^{\infty} \frac{q^{2k}}{1-q^{2k}}\sin\left(\frac{k\pi \alpha}{\omega}\right)\right).
}{}

$\\$
When $\alpha = \omega + i\psi \omega'$ we have 
\eq{
\xi(\omega + i\psi \omega')&= 2i\left( \frac{\pi}{2\omega}\cot\left(\frac{\pi(\omega + i\psi \omega')}{2\omega}\right) 
+ \frac{2\pi}{\omega} \sum_{k=1}^{\infty} \frac{q^{2k}}{1-q^{2k}}
\sin\left(\frac{k\pi (\omega + i\psi \omega')}{\omega}\right)\right)\\
&= 2i\left(-\frac{\pi}{2\omega} \tan\left( \frac{i \pi \psi \omega'}{2\omega}\right) 
+ \frac{2\pi}{\omega} \sum_{k=1}^{\infty}
\frac{q^{2k}}{1-q^{2k}}(-1)^k \sin\left(\frac{ik\pi \psi \omega'}{\omega}\right)\right).
}{\label{efdef}}

$\\$
When $\alpha = i\psi \omega'$ we have,
\eqn{
\xi(i\psi \omega')&= 2i\left(\frac{\pi}{2\omega}\cot\left(\frac{\pi(i\psi \omega')}{2\omega}\right) 
+ \frac{2\pi}{\omega} \sum_{k=1}^{\infty} \frac{q^{2k}}{1-q^{2k}}
\sin\left(\frac{k\pi ( i\psi \omega')}{\omega}\right)\right).
}{\label{efdef2}}

$\\$

$\\$
Note that if either $\psi \in [0,1]$ and $0<|q|< 1$ with $q\in \C$, or if $q\in [q_a,q_b]\subset (0,1)$, 
$|\Re(\psi)| < 2$ and
 $|\Im(\psi)|< \frac{\pi}{|\log(q_a)|}$, then
\eqref{efdef} is analytic in that region. If either $\psi \in (0,1]$ and $0<|q|< 1$ with
 $q\in \C$, or if $q\in [q_a,q_b]\subset (0,1)$, $|\Re(\psi)| < 2$  and
 $|\psi|>0$, then
\eqref{efdef2} is analytic. 

$\\$
Note that $\sin\left(\frac{i\pi \psi \omega'k}{\omega}\right) = \frac{1}{2i}\left(q^{\psi k}-q^{-\psi k}\right)$
 so that in both cases the infinite sum is bounded by 
\eqn{
2\sum_{k=1}^{\infty} \frac{q_0^{\gamma k}}{1-q_0^{2k}}&\leq
\frac{2}{1-q_0^2}\sum_{k=1}^{\infty} q_0^{\gamma k}\\
&\leq \frac{2 q_0^{\gamma }}{(1-q_0^2)(1-q_0^{\gamma})}.
}{}
where $|q|\leq q_0 < 1$ and $\gamma := 2-|\Re(\psi)|$.

$\\$

$\\$
Let $M$ be a bound on 
 $\tan\left( \frac{-i\psi\log(q)}{2}\right)$ or $\cot\left( \frac{-i\psi\log(q)}{2}\right)$, depending on the
 choice of $\alpha$. Then

$\\$
\eqn{
\left|\omega \xi(\tilde n + i\psi \omega')\right| &
 \leq M\pi+\frac{8\pi q_0^{\gamma }}{(1-q_0^2)(1-q_0^{\gamma})}.
}{}

$\\$
\color{Green}
$\\$
$\\$\color{Green}$\%$$\%$
$\\$
$\\$
$\\$\color{BrickRed}pie\-\ =\-\ nm('pi');\color{Green}
$\\$
$\\$
$\\$\color{Green}$\%$
$\\$\color{Green}$\%$\-\ get\-\ bound\-\ on\-\ tan\-\ or\-\ cot
$\\$\color{Green}$\%$
$\\$
$\\$
$\\$\color{BrickRed}theta\-\ =\-\ nm(linspace(0,sup(2*pie)));\color{Green}
$\\$\color{BrickRed}\color{NavyBlue}\-\ if\-\ \color{BrickRed}\-\ rho\_psi\-\ ==\-\ 1\color{Green}
$\\$\color{BrickRed}\-\ \-\ \-\ \-\ psivec\-\ =\-\ nm(linspace(inf(a\_psi),sup(b\_psi)));\color{Green}
$\\$\color{BrickRed}\color{NavyBlue}\-\ else\-\ \color{BrickRed}\color{Green}
$\\$\color{BrickRed}\-\ \-\ \-\ \-\ psivec\-\ =\-\ nm(1)/2+\-\ (rho\_psi*exp(1i*theta)+exp(-1i*theta)/rho\_psi)/4;\color{Green}
$\\$\color{BrickRed}\color{NavyBlue}\-\ end\-\ \color{BrickRed}\color{Green}
$\\$
$\\$
$\\$\color{BrickRed}\color{NavyBlue}\-\ if\-\ \color{BrickRed}\-\ rho\_q\-\ ==\-\ 1\color{Green}
$\\$\color{BrickRed}\-\ \-\ \-\ \-\ qvec\-\ =\-\ nm(linspace(inf(a\_q),sup(b\_q)));\color{Green}
$\\$\color{BrickRed}\color{NavyBlue}\-\ else\-\ \color{BrickRed}\color{Green}
$\\$\color{BrickRed}\-\ \-\ \-\ \-\ qvec\-\ =\-\ (a\_q+b\_q)/2+(b\_q-a\_q)*(rho\_q*exp(1i*theta)+exp(-1i*theta)/rho\_q)/4;\color{Green}
$\\$\color{BrickRed}\color{NavyBlue}\-\ end\-\ \color{BrickRed}\color{Green}
$\\$
$\\$
$\\$\color{BrickRed}qv\-\ =\-\ nm(qvec(1:end-1),qvec(2:end));\color{Green}
$\\$\color{BrickRed}psiv\-\ =\-\ nm(psivec(1:end-1),psivec(2:end));\color{Green}
$\\$\color{BrickRed}\-\ \-\ \-\ \-\ \color{Green}
$\\$\color{BrickRed}\color{NavyBlue}\-\ if\-\ \color{BrickRed}\-\ ntilde\-\ ==\-\ 1\color{Green}
$\\$\color{BrickRed}\-\ \-\ \-\ \-\ \color{Green}$\%$\-\ get\-\ bound\-\ on\-\ tan(1i*pi*psi*omega\_prime/(2*omega))\-\ \-\ \-\ \-\ 
$\\$\color{BrickRed}\-\ \-\ \-\ \-\ temp\-\ =\-\ sup(abs(tan(-1i*psiv.'*log(qv)/2)));\color{Green}
$\\$\color{BrickRed}\color{NavyBlue}\-\ elseif\-\ \color{BrickRed}\-\ ntilde\-\ ==\-\ 0\color{Green}
$\\$\color{BrickRed}\-\ \-\ \-\ \-\ \color{Green}$\%$\-\ get\-\ bound\-\ on\-\ cot(1i*pi*psi*omega\_prime/(2*omega))\-\ \-\ \-\ \-\ 
$\\$\color{BrickRed}\-\ \-\ \-\ \-\ temp\-\ =\-\ sup(abs(cot(-1i*psiv.'*log(qv)/2)));\color{Green}
$\\$\color{BrickRed}\color{NavyBlue}\-\ end\-\ \color{BrickRed}\color{Green}
$\\$
$\\$
$\\$\color{BrickRed}\color{NavyBlue}\-\ if\-\ \color{BrickRed}\-\ sum(sum(isnan(temp)))\-\ $>$\-\ 0\color{Green}
$\\$\color{BrickRed}\-\ \-\ \-\ \-\ error('NaN\-\ present');\color{Green}
$\\$\color{BrickRed}\color{NavyBlue}\-\ end\-\ \color{BrickRed}\color{Green}
$\\$
$\\$
$\\$\color{BrickRed}M\-\ =\-\ max(max(temp));\color{Green}
$\\$
$\\$
$\\$\color{BrickRed}max\_abs\_real\_psi\-\ =\-\ (a\_psi\-\ +\-\ b\_psi)/2+(b\_psi-a\_psi)*(rho\_psi\-\ +1/rho\_psi)/4;\color{Green}
$\\$\color{BrickRed}max\_abs\_q\-\ =\-\ (a\_q\-\ +\-\ b\_q)/2+(b\_q-a\_q)*(rho\_q\-\ +1/rho\_q)/4;\color{Green}
$\\$
$\\$
$\\$\color{BrickRed}gamma\-\ =\-\ 2-max\_abs\_real\_psi;\color{Green}
$\\$\color{BrickRed}qg\-\ =\-\ max\_abs\_q\verb|^|gamma;\color{Green}
$\\$\color{BrickRed}out\-\ =\-\ pie*M+\-\ \-\ (8*pie*qg)/((1-max\_abs\_q\verb|^|2)*(1-qg));\color{Green}
$\\$
$\\$
$\\$
$\\$
$\\$
$\\$
$\\$
$\\$
$\\$
$\\$
$\\$
$\\$
$\\$
$\\$
$\\$\color{Black}\section{bound\_xi\_der.m}

\color{Green}\color{BrickRed}\color{NavyBlue}\-\ function\-\ \color{BrickRed}\-\ out\-\ =\-\ bound\_xi\_der(abs\_q,min\_abs\_q,a,b,max\_real\_psi,rho\_q,rho\_psi,a\_psi,b\_psi,ntilde)\color{Green}
$\\$
$\\$
$\\$
$\\$
$\\$
$\\$
$\\$\color{Green}$\%$\-\ latex\-\ comments
$\\$
$\\$
$\\$
$\\$\color{Black}
If $\tilde n = 1$, then from the q-series representation of the Weierstrass elliptic function, 
\eq{
\pd{}{\psi} \xi(\tilde n \omega + i\psi\omega') & = 2\omega'\left( \wp(\tilde\omega + i\psi\omega') + \frac{\zeta(\omega)}{\omega}\right)\\
&= 2\omega'\left( \left(\frac{\pi}{2\omega}\right)^2 \sec^2\left(\frac{i\pi \psi\omega'}{2\omega}\right) - \frac{2\pi^2}{\omega^2}
\sum_{k=1}^{\infty} (-1)^k\frac{kq^{2k}}{1-q^{2k}}\cos\left(\frac{ik \pi \psi\omega' }{\omega}\right)\right)\\
&= 2\omega'\left( \left(\frac{\pi}{2\omega}\right)^2 \sec^2\left(\frac{i\pi \psi\omega'}{2\omega}\right) - \frac{2\pi^2}{\omega^2}
\sum_{k=1}^{\infty} (-1)^k\frac{kq^{2k}}{1-q^{2k}}\left(\frac{q^{\psi k}+q^{-\psi k}}{2}\right)\right).
}{\label{efdef}}

$\\$
If $\tilde n = 0$, then from the q-series representation of the Weierstrass elliptic function, 
\eq{
\pd{}{\psi} \xi(\tilde n \omega + i\psi\omega') & = 2\omega'\left( \wp(\tilde\omega + i\psi\omega') + \frac{\zeta(\omega)}{\omega}\right)\\
&= 2\omega'\left( \left(\frac{\pi}{2\omega}\right)^2 \csc^2\left(\frac{i\pi \psi\omega'}{2\omega}\right) - \frac{2\pi^2}{\omega^2}
\sum_{k=1}^{\infty} \frac{kq^{2k}}{1-q^{2k}}\cos\left(\frac{ik \pi \psi\omega' }{\omega}\right)\right)\\
&= 2\omega'\left( \left(\frac{\pi}{2\omega}\right)^2 \csc^2\left(\frac{i\pi \psi\omega'}{2\omega}\right) - \frac{2\pi^2}{\omega^2}
\sum_{k=1}^{\infty} \frac{kq^{2k}}{1-q^{2k}}\left(\frac{q^{\psi k}-q^{-\psi k}}{2i}\right)\right).
}{\label{efdef2}}

$\\$

$\\$
Note that if either $\psi \in [0,1]$ is fixed and $|q|< 1$ with $q\in \C$, or if $q\in [q_a,q_b]\subset (0,1)$ is fixed and
 $|\Im(\psi)|< \frac{\pi}{|\log(q_a)|}$, then
\eqref{efdef} is analytic in that region. Note that if either $\psi \in [0,1]$ is fixed and $|q|< 1$ with $q\in \C$, or if $q\in [q_a,q_b]\subset (0,1)$ is fixed and
 $\Re(\psi)>0$, then
\eqref{efdef} is analytic in that region.

$\\$
Let $0\leq q_0 <1$, $\gamma \in \R$, 
 and define 
\eq{
f(x)&:= \sum_{k=0}^{\infty} \frac{1}{\gamma \log(q_0)} q_0^{\gamma k x}\\
&= \frac{1}{\gamma \log(q_0)} \frac{1}{1-q_0^{\gamma x}}.
}{}
Note that
\eq{
f'(x)& = \sum_{k=0}^{\infty}kq_0^{\gamma x}\\
&= \frac{q_0^{\gamma x}}{(1-q_0^{\gamma x})^2}.
}{}

$\\$
Then
\eq{
\left| \sum_{k=1}^{\infty}
(-1)^{\tilde nk} \frac{kq^{2k}}{1-q^{2k}}\left(q^{\psi k}+q^{-\psi k}\right)\right| &\leq
2\sum_{k=0}^{\infty} \frac{kq_0^{\gamma k}}{1-q_0^{2k}}\\
&\leq \frac{2}{1-q_0^2} \sum_{k=0}^{\infty} kq_0^{\gamma k}\\
&\leq \frac{2}{1-q_0^2}\frac{q_0^{\gamma}}{(1-q_0^{\gamma})^2},
}{}
where $|q|\leq q_0<1$ and $\gamma:= 2-\Re(\psi)$.

$\\$
\color{Green}
$\\$
$\\$\color{Green}$\%$$\%$
$\\$
$\\$
$\\$\color{Green}$\%$
$\\$\color{Green}$\%$\-\ constants
$\\$\color{Green}$\%$
$\\$
$\\$
$\\$\color{BrickRed}pie\-\ =\-\ nm('pi');\color{Green}
$\\$\color{BrickRed}gamma\-\ =\-\ 2-max\_real\_psi;\color{Green}
$\\$\color{BrickRed}qg\-\ =\-\ abs\_q\verb|^|gamma;\color{Green}
$\\$
$\\$
$\\$\color{Green}$\%$\-\ error\-\ check
$\\$\color{BrickRed}\color{NavyBlue}\-\ if\-\ \color{BrickRed}\-\ inf(gamma)\-\ $<$=\-\ 0\color{Green}
$\\$\color{BrickRed}\-\ \-\ \-\ \-\ error('max\_psi\-\ too\-\ big');\color{Green}
$\\$\color{BrickRed}\color{NavyBlue}\-\ end\-\ \color{BrickRed}\color{Green}
$\\$
$\\$
$\\$\color{Green}$\%$\-\ 
$\\$\color{Green}$\%$\-\ get\-\ bound\-\ on\-\ (sec/csc)(1i*pi*psi*omega\_prime/(2*omega))
$\\$\color{Green}$\%$
$\\$
$\\$
$\\$\color{BrickRed}pnts\-\ =\-\ 100;\color{Green}
$\\$\color{Green}$\%$\-\ bound\-\ on\-\ stadium\-\ in\-\ variable\-\ psi
$\\$\color{BrickRed}qvec\-\ =\-\ nm(linspace(inf(a),sup(b),pnts));\color{Green}
$\\$\color{BrickRed}theta\-\ =\-\ nm(linspace(0,sup(2*pie),pnts));\color{Green}
$\\$\color{BrickRed}psivec\-\ =\-\ (a\_psi+b\_psi)/2+\-\ (b\_psi-a\_psi)*(rho\_psi*exp(1i*theta)+exp(-1i*theta)/rho\_psi)/4;\color{Green}
$\\$\color{BrickRed}qv\-\ =\-\ nm(qvec(1:end-1),qvec(2:end));\color{Green}
$\\$\color{BrickRed}psiv\-\ =\-\ nm(psivec(1:end-1),psivec(2:end));\color{Green}
$\\$\color{BrickRed}\color{NavyBlue}\-\ if\-\ \color{BrickRed}\-\ ntilde\-\ ==\-\ 1\color{Green}
$\\$\color{BrickRed}\-\ \-\ \-\ \-\ temp\-\ =\-\ sup(abs(sec(-1i*psiv.'*log(qv)/2)));\color{Green}
$\\$\color{BrickRed}\color{NavyBlue}\-\ elseif\-\ \color{BrickRed}\-\ ntilde\-\ ==\-\ 0\color{Green}
$\\$\color{BrickRed}\-\ \-\ \-\ \-\ temp\-\ =\-\ sup(abs(csc(-1i*psiv.'*log(qv)/2)));\color{Green}
$\\$\color{BrickRed}\color{NavyBlue}\-\ end\-\ \color{BrickRed}\color{Green}
$\\$\color{BrickRed}\color{NavyBlue}\-\ if\-\ \color{BrickRed}\-\ sum(sum(isnan(temp)))\-\ $>$\-\ 0\color{Green}
$\\$\color{BrickRed}\-\ \-\ \-\ \-\ error('NaN\-\ present');\color{Green}
$\\$\color{BrickRed}\color{NavyBlue}\-\ end\-\ \color{BrickRed}\color{Green}
$\\$\color{BrickRed}max\_temp1\-\ =\-\ max(max(temp));\color{Green}
$\\$
$\\$
$\\$\color{Green}$\%$\-\ bound\-\ on\-\ stadium\-\ in\-\ variable\-\ q
$\\$\color{BrickRed}psivec\-\ =\-\ nm(linspace(a\_psi,b\_psi,pnts));\color{Green}
$\\$\color{BrickRed}theta\-\ =\-\ nm(linspace(0,sup(2*pie),pnts));\color{Green}
$\\$\color{BrickRed}qvec\-\ =\-\ (a+b)/2+(b-a)*(rho\_q*exp(1i*theta)+exp(-1i*theta)/rho\_q)/4;\color{Green}
$\\$\color{BrickRed}qv\-\ =\-\ nm(qvec(1:end-1),qvec(2:end));\color{Green}
$\\$\color{BrickRed}psiv\-\ =\-\ nm(psivec(1:end-1),psivec(2:end));\color{Green}
$\\$
$\\$
$\\$\color{BrickRed}\color{NavyBlue}\-\ if\-\ \color{BrickRed}\-\ ntilde\-\ ==\-\ 1\color{Green}
$\\$\color{BrickRed}\-\ \-\ \-\ \-\ temp\-\ =\-\ sup(abs(sec(-1i*psiv.'*log(qv)/2)));\color{Green}
$\\$\color{BrickRed}\color{NavyBlue}\-\ elseif\-\ \color{BrickRed}\-\ ntilde\-\ ==\-\ 0\color{Green}
$\\$\color{BrickRed}\-\ \-\ \-\ \-\ temp\-\ =\-\ sup(abs(csc(-1i*psiv.'*log(qv)/2)));\color{Green}
$\\$\color{BrickRed}\color{NavyBlue}\-\ end\-\ \color{BrickRed}\color{Green}
$\\$\color{BrickRed}\color{NavyBlue}\-\ if\-\ \color{BrickRed}\-\ sum(sum(isnan(temp)))\-\ $>$\-\ 0\color{Green}
$\\$\color{BrickRed}\-\ \-\ \-\ \-\ error('NaN\-\ present');\color{Green}
$\\$\color{BrickRed}\color{NavyBlue}\-\ end\-\ \color{BrickRed}\color{Green}
$\\$\color{BrickRed}max\_temp2\-\ =\-\ max(max(temp));\color{Green}
$\\$
$\\$
$\\$\color{Green}$\%$\-\ take\-\ maximum\-\ of\-\ two\-\ bounds
$\\$\color{BrickRed}temp\_bd\-\ =\-\ nm(max(max\_temp1,max\_temp2));\color{Green}
$\\$
$\\$
$\\$\color{Green}$\%$\-\ 
$\\$\color{Green}$\%$\-\ bound\-\ on\-\ infinite\-\ sum\-\ part
$\\$\color{Green}$\%$
$\\$
$\\$
$\\$\color{BrickRed}out\-\ =\-\ (2/(1-abs\_q\verb|^|2))*(qg/(1-qg)\verb|^|2);\color{Green}
$\\$
$\\$
$\\$\color{Green}$\%$
$\\$\color{Green}$\%$\-\ combine\-\ all\-\ parts
$\\$\color{Green}$\%$
$\\$
$\\$
$\\$\color{BrickRed}out\-\ =\-\ 2*abs(log(min\_abs\_q))*pie*(out\-\ +\-\ temp\_bd\verb|^|2/4);\color{Green}
$\\$
$\\$
$\\$
$\\$
$\\$
$\\$
$\\$
$\\$
$\\$
$\\$
$\\$
$\\$
$\\$
$\\$
$\\$
$\\$
$\\$
$\\$
$\\$\color{Black}\section{bound\_xi\_der\_n1.m}

\color{Green}\color{BrickRed}\color{NavyBlue}\-\ function\-\ \color{BrickRed}\-\ out\-\ =\-\ bound\_xi\_der\_n1(abs\_q,min\_abs\_q,a,b,max\_real\_psi,rho\_q,rho\_psi)\color{Green}
$\\$\color{Green}$\%$\-\ out\-\ =\-\ bound\_xi\_der\_n1(psi,rho\_beta,omega,omega\_prime)
$\\$\color{Green}$\%$
$\\$\color{Green}$\%$\-\ Find\-\ an\-\ upper\-\ bound\-\ on\-\ the\-\ derivative\-\ of\-\ xi\-\ with\-\ respect\-\ to\-\ beta
$\\$\color{Green}$\%$\-\ when\-\ alpha\-\ =\-\ omega\-\ +\-\ i\-\ beta
$\\$
$\\$
$\\$\color{Green}$\%$\-\ latex\-\ comments
$\\$
$\\$\color{Black}

$\\$
Now 
\eqn{
\pd{}{\beta} \xi(\omega + i\beta) & = 2\left( \wp(\omega + i\beta) + \frac{\zeta(\omega)}{\omega}\right)\\
&= 2\left( \left(\frac{\pi}{2\omega}\right)^2 \sec^2\left(\frac{i\pi \beta}{2\omega}\right) - \frac{2\pi^2}{\omega^2}
\sum_{k=1}^{\infty} (-1)^k\frac{kq^{2k}}{1-q^{2k}}\cos\left(\frac{ik \pi \beta }{\omega}\right)\right).
}{}

$\\$
Then if $\beta = \psi \omega'$,
\eq{
\omega \pd{}{\psi}\xi(\omega + i\psi \omega')&=
 -2\log(q)\left(\frac{\pi}{4}\sec^2\left(\frac{i\pi \psi \omega'}{2\omega}\right)
-\pi \sum_{k=1}^{\infty} (-1)^k \frac{kq^{2k}}{1-q^{2k}}\left(q^{\psi k}+q^{-\psi k}\right)\right)\\
&= -2\log(q)\left(\frac{\pi\sec^2\left(-i\psi \log(q)/2 \right)}{4} -\pi \sum_{k=1}^{\infty}
(-1)^k \frac{kq^{2k}}{1-q^{2k}}\left(q^{\psi k}+q^{-\psi k}\right)\right).
}{\label{efdef}}
Note that if either $\psi \in [0,1]$ and $0<|q|< 1$ with $q\in \C$ or if $q\in [q_a,q_b]\subset (0,1)$ and
 $|\Im(\psi)|< \frac{\pi}{|\log(q_a)|}$, then
\eqref{efdef} is analytic in that region. 

$\\$
Let $0\leq q_0 <1$, $\gamma \in \R$, 
 and define 
\eq{
f(x)&:= \sum_{k=0}^{\infty} \frac{1}{\gamma \log(q_0)} q_0^{\gamma k x}\\
&= \frac{1}{\gamma \log(q_0)} \frac{1}{1-q_0^{\gamma x}}.
}{}
Note that
\eq{
f'(x)& = \sum_{k=0}^{\infty}kq_0^{\gamma x}\\
&= \frac{q_0^{\gamma x}}{(1-q_0^{\gamma x})^2}.
}{}

$\\$
Then in the above regions,
\eq{
\left| \sum_{k=1}^{\infty}
(-1)^k \frac{kq^{2k}}{1-q^{2k}}\left(q^{\psi k}+q^{-\psi k}\right)\right| &\leq
2\sum_{k=0}^{\infty} \frac{kq_0^{\gamma k}}{1-q_0^{2k}}\\
&\leq \frac{2}{1-q_0^2} \sum_{k=0}^{\infty} kq_0^{\gamma k}\\
&\leq \frac{2}{1-q_0^2}\frac{q_0^{\gamma}}{(1-q_0^{\gamma})^2},
}{}
where $|q|\leq q_0<1$ and $\gamma:= 2-\Re(\psi)$.

$\\$

$\\$
\color{Green}
$\\$
$\\$\color{Green}$\%$$\%$
$\\$
$\\$
$\\$\color{Green}$\%$
$\\$\color{Green}$\%$\-\ constants
$\\$\color{Green}$\%$
$\\$
$\\$
$\\$\color{BrickRed}pie\-\ =\-\ nm('pi');\color{Green}
$\\$\color{BrickRed}gamma\-\ =\-\ 2-max\_real\_psi;\color{Green}
$\\$\color{BrickRed}qg\-\ =\-\ abs\_q\verb|^|gamma;\color{Green}
$\\$
$\\$
$\\$\color{Green}$\%$\-\ error\-\ check
$\\$\color{BrickRed}\color{NavyBlue}\-\ if\-\ \color{BrickRed}\-\ inf(gamma)\-\ $<$=\-\ 0\color{Green}
$\\$\color{BrickRed}\-\ \-\ \-\ \-\ error('max\_psi\-\ too\-\ big');\color{Green}
$\\$\color{BrickRed}\color{NavyBlue}\-\ end\-\ \color{BrickRed}\color{Green}
$\\$
$\\$
$\\$\color{Green}$\%$\-\ 
$\\$\color{Green}$\%$\-\ get\-\ bound\-\ on\-\ sec(1i*pi*psi*omega\_prime/(2*omega))
$\\$\color{Green}$\%$
$\\$
$\\$
$\\$\color{BrickRed}qvec\-\ =\-\ nm(linspace(inf(a),sup(b)));\color{Green}
$\\$\color{BrickRed}theta\-\ =\-\ nm(linspace(0,sup(2*pie)));\color{Green}
$\\$\color{BrickRed}psivec\-\ =\-\ nm(1)/2+\-\ (rho\_psi*exp(1i*theta)+exp(-1i*theta)/rho\_psi)/4;\color{Green}
$\\$\color{BrickRed}qv\-\ =\-\ nm(qvec(1:end-1),qvec(2:end));\color{Green}
$\\$\color{BrickRed}psiv\-\ =\-\ nm(psivec(1:end-1),psivec(2:end));\color{Green}
$\\$\color{BrickRed}temp\-\ =\-\ sup(abs(sec(-1i*psiv.'*log(qv)/2)));\color{Green}
$\\$\color{BrickRed}\color{NavyBlue}\-\ if\-\ \color{BrickRed}\-\ sum(sum(isnan(temp)))\-\ $>$\-\ 0\color{Green}
$\\$\color{BrickRed}\-\ \-\ \-\ \-\ error('NaN\-\ present');\color{Green}
$\\$\color{BrickRed}\color{NavyBlue}\-\ end\-\ \color{BrickRed}\color{Green}
$\\$\color{BrickRed}max\_sec1\-\ =\-\ max(max(temp));\color{Green}
$\\$
$\\$
$\\$\color{BrickRed}psivec\-\ =\-\ nm(linspace(0,1));\color{Green}
$\\$\color{BrickRed}theta\-\ =\-\ nm(linspace(0,sup(2*pie)));\color{Green}
$\\$\color{BrickRed}qvec\-\ =\-\ (a+b)/2+(b-a)*(rho\_q*exp(1i*theta)+exp(-1i*theta)/rho\_q)/4;\color{Green}
$\\$\color{BrickRed}qv\-\ =\-\ nm(qvec(1:end-1),qvec(2:end));\color{Green}
$\\$\color{BrickRed}psiv\-\ =\-\ nm(psivec(1:end-1),psivec(2:end));\color{Green}
$\\$\color{BrickRed}temp\-\ =\-\ sup(abs(tan(-1i*psiv.'*log(qv)/2)));\color{Green}
$\\$\color{BrickRed}\color{NavyBlue}\-\ if\-\ \color{BrickRed}\-\ sum(sum(isnan(temp)))\-\ $>$\-\ 0\color{Green}
$\\$\color{BrickRed}\-\ \-\ \-\ \-\ error('NaN\-\ present');\color{Green}
$\\$\color{BrickRed}\color{NavyBlue}\-\ end\-\ \color{BrickRed}\color{Green}
$\\$\color{BrickRed}max\_sec2\-\ =\-\ max(max(temp));\color{Green}
$\\$
$\\$
$\\$\color{BrickRed}sec\_bd\-\ =\-\ nm(max(max\_sec1,max\_sec2));\color{Green}
$\\$
$\\$
$\\$\color{Green}$\%$\-\ 
$\\$\color{Green}$\%$\-\ bound\-\ on\-\ infinite\-\ sum\-\ part
$\\$\color{Green}$\%$
$\\$
$\\$
$\\$\color{BrickRed}out\-\ =\-\ (2/(1-abs\_q\verb|^|2))*(qg/(1-qg)\verb|^|2);\color{Green}
$\\$
$\\$
$\\$\color{Green}$\%$
$\\$\color{Green}$\%$\-\ combine\-\ all\-\ parts
$\\$\color{Green}$\%$
$\\$
$\\$
$\\$\color{BrickRed}out\-\ =\-\ 2*abs(log(min\_abs\_q))*pie*(out\-\ +\-\ sec\_bd\verb|^|2/4);\color{Green}
$\\$
$\\$
$\\$
$\\$
$\\$
$\\$
$\\$
$\\$
$\\$
$\\$
$\\$
$\\$
$\\$
$\\$
$\\$
$\\$
$\\$
$\\$
$\\$\color{Black}\section{bound\_xi\_n1.m}

\color{Green}\color{BrickRed}\color{NavyBlue}\-\ function\-\ \color{BrickRed}\-\ out\-\ =\-\ bound\_xi\_n1(q,rho\_psi,max\_tan)\color{Green}
$\\$
$\\$
$\\$\color{Green}$\%$\-\ latex\-\ description
$\\$
$\\$\color{Black}
Now 
\eqn{
\xi(\omega + i\beta)&= \frac{\pi}{2\omega}\cot\left(\frac{\pi(\omega + i\beta)}{2\omega}\right) 
+ \frac{2\pi}{\omega} \sum_{k=1}^{\infty} \frac{q^{2k}}{1-q^{2k}}
\sin\left(\frac{k\pi (\omega + i\beta)}{\omega}\right)\\
&= -\frac{\pi}{2\omega} \tan\left( \frac{i \pi \beta}{2\omega}\right) 
+ \frac{2\pi}{\omega} \sum_{k=1}^{\infty}
\frac{q^{2k}}{1-q^{2k}}(-1)^k \sin\left(\frac{ik\pi \beta}{\omega}\right).
}{}

$\\$
Note that 
\eqn{
|\tan(z)| &= \left| \frac{e^{2iz}-1}{i(e^{2iz}+1)}\right|\\
&\leq \frac{1+\left|e^{2iz}\right|}{\left| 1 + e^{2iz}\right|}\\
&\leq \frac{1 + e^{-2\Im(z)}}{\left| 1 + e^{2iz}\right|}.
}{}

$\\$
Let $x,y\in\R$ such that $2iz = x+iy$, and suppose that $|y|\leq \psi \pi$ where $0\leq \psi < 1$. Now
\eqn{
\left| 1 + e^{x+iy}\right|^2 &= 1+2e^x\cos(y)+e^{2x}\\
&\geq 1 + 2e^{x}\cos(\psi \pi)+e^{2x}.
}{}
By simple calculus we find that $1+2e^x\cos(\psi \pi) + e^{2x}$ is 
bounded below by 1 if $\psi \leq \frac{1}{2}$, and is bounded below by $\sin(\psi \pi)$
if $\frac{1}{2}< \psi < 1$.

$\\$
Then
\eqn{
|\tan(z)| & \leq \frac{1 + e^{-2\Im(z)}}{\sin(\psi \pi)}
}{}
if $\frac{1}{2} < \psi < 1$, and
\eqn{
|\tan(z)| &\leq 1 +e^{-2\Im(z)}
}{}
if $0\leq \psi \leq \frac{1}{2}$.

$\\$
Note that 
\eqn{
\left| \sum_{k=1}^{\infty} \frac{q^{2k}}{1-q^{2k}}(-1)^k\sin(kz)\right| & \leq
 \sum_{k=1}^{\infty} \frac{q^{2k}}{1-q^{2k}}e^{k|\Im(z)|}\\
&= \sum_{k=1}^{\infty} \frac{\left(e^{|\Im(z)|-2\pi \omega'/\omega} \right)^k}{1-q^{2k}}\\
& \leq \frac{e^{|\Im(z)|-2\pi \omega'/\omega}}{(1-q^2)(1-e^{|\Im(z)|-2\pi \omega'/\omega})},
}{}
so long as $|\Im(z)|-2\pi \omega'/\omega < 0$.

$\\$
Then if $\rho_{\beta} < 3+2\sqrt{2}$, we have 
\eqn{
|\xi(\omega + i\beta)| &\leq \frac{\pi (1+ e^{V})}{2\omega \sin(\psi \pi)}
+ \frac{2\pi e^Q}{\omega (1-q^2)(1-e^{Q})},
}{}
where 
\eqn{
Q:= \frac{\pi \omega'}{2\omega}\left( 1+\frac{1}{2}\left( \rho + \frac{1}{\rho}\right) \right) -
\frac{2\pi \omega'}{\omega},
}{}
and 
\eqn{
V = -2\left(\frac{\pi \omega'}{4\omega}\right)\left( 1 - \frac{1}{2}\left(\rho + \frac{1}{\rho}\right)\right).
}{}

$\\$
\color{Green}
$\\$
$\\$
$\\$
$\\$\color{BrickRed}pie\-\ =\-\ nm('pi');\color{Green}
$\\$
$\\$
$\\$\color{Green}$\%$\-\ c1\-\ =\-\ pie*omega\_prime/omega;
$\\$\color{BrickRed}c1\-\ =\-\ -log(q);\color{Green}
$\\$\color{BrickRed}c2\-\ =\-\ (rho\_psi\-\ +\-\ 1/rho\_psi)/2;\color{Green}
$\\$
$\\$
$\\$\color{BrickRed}Q\-\ =\-\ (c1/2)*(1+c2)-2*c1;\color{Green}
$\\$
$\\$
$\\$\color{BrickRed}out\-\ =\-\ (pie/2)*max\_tan\-\ +\-\ (2*pie)*exp(Q)/((1-q\verb|^|2)*(1-exp(Q)));\color{Green}
$\\$\color{BrickRed}out\-\ =\-\ nm(sup(out));\color{Green}
$\\$
$\\$
$\\$
$\\$
$\\$\color{Black}\section{cf\_biv\_cheby.m}

\color{Green}\color{BrickRed}\color{NavyBlue}\-\ function\-\ \color{BrickRed}\-\ cf\-\ =\-\ cf\_biv\_cheby(m,n,num\_funs,fun)\color{Green}
$\\$\color{Green}$\%$\-\ function\-\ cf\_padau(m,n,fun)
$\\$\color{Green}$\%$
$\\$\color{Green}$\%$\-\ Returns\-\ the\-\ coefficients\-\ for\-\ bivariate\-\ Chebyshev\-\ interpolation
$\\$
$\\$
$\\$
$\\$\color{Black}
Let $f(x,y):[-1,1]\times [-1,1] \to \mathbb{C}$. Let $T_i(x)$, $T_j(y)$ be
Chebyshev polynomials of the first kind. This function returns a matrix 
cf of coefficients to the bivariate interpolation polynomial, 
$p(x,y) = \sum_{i=0}^m\sum_{j=0}^n c_{i,j} T_i(x)T_j(y)$,
where $c_{i,j}$ is the $ith+1$, $jth+1$ row and column of the matrix cf.
\color{Green}
$\\$
$\\$
$\\$
$\\$\color{Green}$\%$\-\ constants
$\\$\color{BrickRed}pie\-\ =\-\ nm('pi');\color{Green}
$\\$\color{BrickRed}c\_1\-\ =\-\ pie/(2*(m+1));\color{Green}
$\\$\color{BrickRed}c\_2\-\ =\-\ pie/(2*(n+1));\color{Green}
$\\$
$\\$
$\\$\color{Green}$\%$\-\ theta\-\ for\-\ x\-\ and\-\ y
$\\$\color{BrickRed}theta\_xr\-\ =\-\ c\_1*(2*(0:1:m)+1);\color{Green}
$\\$\color{BrickRed}theta\_ys\-\ =\-\ c\_2*(2*(0:1:n)+1);\color{Green}
$\\$
$\\$
$\\$\color{Green}$\%$\-\ x\-\ and\-\ y\-\ interpolation\-\ points
$\\$\color{BrickRed}xr\-\ =\-\ cos(theta\_xr);\color{Green}
$\\$\color{BrickRed}ys\-\ =\-\ cos(theta\_ys);\-\ \color{Green}
$\\$
$\\$
$\\$\color{Green}$\%$\-\ evaluate\-\ function\-\ at\-\ grid\-\ points
$\\$\color{BrickRed}f\_xy\-\ =\-\ nm(zeros(m+1,n+1,num\_funs));\color{Green}
$\\$
$\\$
$\\$\color{BrickRed}fun2\-\ =\-\ \@(x)(fun(x,ys));\color{Green}
$\\$\color{BrickRed}\color{NavyBlue}\-\ parfor\-\ \color{BrickRed}\-\ r\-\ =\-\ 1:m+1\color{Green}
$\\$\color{BrickRed}\-\ \-\ \-\ \-\ \-\ \-\ \-\ \color{Green}
$\\$\color{BrickRed}\-\ \-\ \-\ \-\ \color{Green}$\%$\-\ 
$\\$\color{BrickRed}\-\ \-\ \-\ \-\ \color{Green}$\%$\-\ start\-\ intlab\-\ in\-\ matlab\-\ workers\-\ when\-\ in\-\ parallel
$\\$\color{BrickRed}\-\ \-\ \-\ \-\ \color{Green}$\%$\-\ 
$\\$
$\\$
$\\$\color{BrickRed}\-\ \-\ \-\ \-\ clc;\color{Green}
$\\$\color{BrickRed}\-\ \-\ \-\ \-\ curr\_dir\-\ =\-\ cd;\color{Green}
$\\$\color{BrickRed}\-\ \-\ \-\ \-\ \color{Green}$\%$$\%$\-\ startup\-\ commands
$\\$\color{BrickRed}\-\ \-\ \-\ \-\ cd('..');\color{Green}
$\\$\color{BrickRed}\-\ \-\ \-\ \-\ cd('..');\color{Green}
$\\$\color{BrickRed}\-\ \-\ \-\ \-\ startup('intlab','','start\-\ matlabpool','off');\color{Green}
$\\$\color{BrickRed}\-\ \-\ \-\ \-\ cd(curr\_dir);\color{Green}
$\\$
$\\$
$\\$\color{BrickRed}\-\ \-\ \-\ \-\ \color{Green}$\%$
$\\$\color{BrickRed}\-\ \-\ \-\ \-\ \color{Green}$\%$\-\ get\-\ function\-\ values
$\\$\color{BrickRed}\-\ \-\ \-\ \-\ \color{Green}$\%$
$\\$\color{BrickRed}\-\ \-\ \-\ \-\ \color{Green}
$\\$\color{BrickRed}\-\ \-\ \-\ \-\ f\_xy(r,:,:)\-\ =\-\ fun2(xr(r));\color{Green}
$\\$\color{BrickRed}\-\ \-\ \-\ \-\ \-\ \-\ \-\ \-\ \color{Green}
$\\$\color{BrickRed}\color{NavyBlue}\-\ end\-\ \color{BrickRed}\color{Green}
$\\$
$\\$
$\\$\color{Green}$\%$\-\ get\-\ Chebyshev\-\ polynomials\-\ evaluated\-\ at\-\ points;
$\\$\color{BrickRed}Tx\-\ =\-\ cos((0:1:m).'*theta\_xr);\color{Green}
$\\$\color{BrickRed}Ty\-\ =\-\ cos(theta\_ys.'*(0:1:n));\color{Green}
$\\$
$\\$
$\\$\color{BrickRed}cf\-\ =\-\ nm(zeros(m+1,n+1,num\_funs));\color{Green}
$\\$
$\\$
$\\$\color{BrickRed}\color{NavyBlue}\-\ for\-\ \color{BrickRed}\-\ j\-\ =\-\ 1:num\_funs\color{Green}
$\\$\color{BrickRed}\-\ \-\ \-\ \-\ cf(:,:,j)\-\ =\-\ (4/((m+1)*(n+1)))*Tx*f\_xy(:,:,j)*Ty;\color{Green}
$\\$\color{BrickRed}\-\ \-\ \-\ \-\ cf(:,1,j)\-\ =\-\ cf(:,1,j)/2;\color{Green}
$\\$\color{BrickRed}\-\ \-\ \-\ \-\ cf(1,:,j)\-\ =\-\ cf(1,:,j)/2;\color{Green}
$\\$\color{BrickRed}\color{NavyBlue}\-\ end\-\ \color{BrickRed}\color{Green}
$\\$
$\\$
$\\$\color{Black}\section{cf\_eval.m}

\color{Green}\color{BrickRed}\color{NavyBlue}\-\ function\-\ \color{BrickRed}\-\ out\-\ =\-\ cf\_eval(cf,x,y)\color{Green}
$\\$\color{Green}$\%$\-\ function\-\ out\-\ =\-\ cf\_eval(cf,x,y)
$\\$\color{Green}$\%$
$\\$\color{Green}$\%$\-\ Evaluates\-\ the\-\ two\-\ dimensional\-\ Chebyshev\-\ polynomial\-\ at\-\ (x,y)
$\\$
$\\$
$\\$
$\\$\color{Black}
Let $f(x,y):[-1,1]\times [-1,1] \to \mathbb{C}$. Let $T_i(x)$, $T_j(y)$ be
Chebyshev polynomials of the first kind. This function evaluates  
$p(x,y) = \sum_{i=0}^m\sum_{j=0}^n c_{i,j} T_i(x)T_j(y)$,
where $c_{i,j}$ is the $ith+1$, $jth+1$ row and column of the matrix cf.
\color{Green}
$\\$
$\\$
$\\$
$\\$\color{BrickRed}m\-\ =\-\ size(cf,1);\color{Green}
$\\$\color{BrickRed}b\-\ =\-\ nm(zeros(m,length(y)));\color{Green}
$\\$\color{BrickRed}\color{NavyBlue}\-\ for\-\ \color{BrickRed}\-\ i\-\ =\-\ 1:m\color{Green}
$\\$\color{BrickRed}\-\ \-\ \-\ \-\ \color{Green}$\%$\-\ evaluate\-\ \color{Black}$b_i:= \sum_{j=0}^n c_{i,j}T_j(y)$ \color{Green}
$\\$\color{BrickRed}\-\ \-\ \-\ \-\ b(i,:)\-\ =\-\ single\_cf(cf(i,:),y);\color{Green}
$\\$\color{BrickRed}\color{NavyBlue}\-\ end\-\ \color{BrickRed}\color{Green}
$\\$
$\\$
$\\$\color{Green}$\%$\-\ evaluate\-\ \color{Black}$\sum_{i=0}^m b_i T_i(x)$ \color{Green}
$\\$\color{BrickRed}out\-\ =\-\ single\_cf(b.',x);\color{Green}
$\\$
$\\$
$\\$\color{Green}$\%$------------------------------------------------------------
$\\$\color{Green}$\%$\-\ single\_cf(cf,z)
$\\$\color{Green}$\%$------------------------------------------------------------
$\\$\color{BrickRed}\color{NavyBlue}\-\ function\-\ \color{BrickRed}\-\ out\-\ =\-\ single\_cf(cf,z)\color{Green}
$\\$
$\\$
$\\$\color{Green}$\%$\-\ get\-\ theta
$\\$\color{BrickRed}theta\-\ =\-\ acos(z);\color{Green}
$\\$\color{Green}$\%$\-\ evalaute\-\ \color{Black}$ \sum_{j=0}^n c_j \cos(j\theta)$ \color{Green}
$\\$
$\\$
$\\$\color{BrickRed}out\-\ =\-\ cos(theta.'*(0:1:size(cf,2)-1))*cf.';\color{Green}
$\\$
$\\$
$\\$
$\\$
$\\$
$\\$
$\\$
$\\$
$\\$
$\\$
$\\$\color{Black}\section{distinct.m}

\color{Green}\color{BrickRed}clc;\-\ close\-\ all;\-\ beep\-\ off;\color{Green}
$\\$
$\\$
$\\$\color{BrickRed}curr\_dir\-\ =\-\ cd;\color{Green}
$\\$\color{BrickRed}cd('..');\color{Green}
$\\$\color{BrickRed}cd('..');\color{Green}
$\\$\color{BrickRed}startup('intlab','','start\-\ matlabpool','off');\color{Green}
$\\$\color{BrickRed}format\-\ long;\color{Green}
$\\$\color{BrickRed}clc;\color{Green}
$\\$\color{BrickRed}cd(curr\_dir);\color{Green}
$\\$\color{Green}$\%$\-\ Spectral\-\ stability\-\ of\-\ periodic\-\ wave\-\ trains\-\ of\-\ the\-\ Korteweg-de
$\\$\color{Green}$\%$\-\ Vries/Kuramoto-Sivashinsky\-\ equation\-\ in\-\ the\-\ Korteweg-de\-\ Vries\-\ limit
$\\$
$\\$
$\\$\color{Green}$\%$\-\ display\-\ type
$\\$\color{Green}$\%$\-\ intvalinit('DisplayMidRad');
$\\$\color{BrickRed}intvalinit('DisplayInfSup');\color{Green}
$\\$\color{Green}$\%$$\%$
$\\$
$\\$
$\\$
$\\$
$\\$\color{BrickRed}pie\-\ =\-\ nm('pi');\color{Green}
$\\$\color{BrickRed}k\-\ \-\ =\-\ nm('0.99');\color{Green}
$\\$
$\\$
$\\$\color{BrickRed}K\-\ =\-\ elliptic\_integral(k,1);\color{Green}
$\\$\color{BrickRed}E\-\ =\-\ elliptic\_integral(k,2);\color{Green}
$\\$\color{BrickRed}k2\-\ =\-\ k*k;\color{Green}
$\\$\color{BrickRed}c1\-\ =\-\ 1-k2;\color{Green}
$\\$\color{BrickRed}b1\-\ =\-\ k2*K/(E-K)\color{Green}
$\\$\color{BrickRed}b2\-\ =\-\ k2*c1*K/(c1*K-E)\color{Green}
$\\$\color{BrickRed}b3\-\ =\-\ c1*K/E\color{Green}
$\\$
$\\$
$\\$
$\\$
$\\$
$\\$
$\\$
$\\$
$\\$
$\\$
$\\$
$\\$
$\\$\color{Black}\section{instability.m}

\color{Green}\color{BrickRed}clear\-\ all;\-\ close\-\ all;\-\ beep\-\ off;\-\ clc;\-\ curr\_dir\-\ =\-\ cd;\color{Green}
$\\$\color{Green}$\%$$\%$\-\ startup\-\ commands
$\\$\color{BrickRed}cd('..');\color{Green}
$\\$\color{BrickRed}cd('..');\color{Green}
$\\$\color{BrickRed}startup('intlab','','start\-\ matlabpool','off');\color{Green}
$\\$\color{BrickRed}format\-\ long;\color{Green}
$\\$\color{BrickRed}clc;\color{Green}
$\\$\color{BrickRed}cd(curr\_dir);\color{Green}
$\\$\color{Green}$\%$\-\ Spectral\-\ stability\-\ of\-\ periodic\-\ wave\-\ trains\-\ of\-\ the\-\ Korteweg-de
$\\$\color{Green}$\%$\-\ Vries/Kuramoto-Sivashinsky\-\ equation\-\ in\-\ the\-\ Korteweg-de\-\ Vries\-\ limit
$\\$
$\\$
$\\$\color{Green}$\%$\-\ display\-\ type
$\\$\color{BrickRed}intvalinit('DisplayMidRad');\color{Green}
$\\$\color{Green}$\%$\-\ intvalinit('DisplayInfSup');
$\\$
$\\$
$\\$\color{BrickRed}pie\-\ =\-\ nm('pi');\color{Green}
$\\$\color{BrickRed}total\_time\-\ =\-\ tic;\color{Green}
$\\$
$\\$
$\\$\color{Green}$\%$$\%$
$\\$
$\\$
$\\$\color{Green}$\%$\-\ [dm,rho\_x,q\_min,q\_max]\-\ =\-\ lower\_bound\_unstable
$\\$\color{BrickRed}dm\-\ =\-\ nm('0.036927931316603');\color{Green}
$\\$\color{BrickRed}rho\_x\-\ =\-\ midrad(\-\ \-\ \-\ 5.46027719725235,\-\ \-\ 0.00000000000001);\color{Green}
$\\$\color{BrickRed}q\_min\-\ =\-\ nm('1e-7');\color{Green}
$\\$\color{BrickRed}q\_max\-\ =\-\ nm('0.4');\color{Green}
$\\$
$\\$
$\\$
$\\$
$\\$
$\\$\-\ \color{Black}
Note that $\rho_{\psi}$ must be chosen so that $\xi(\omega + i\psi \omega')$ does not have a pole.
The poles of $\xi$ are $z = 2m \omega + 2n \omega'$. Setting
$2m\omega + 2n\omega' = \omega + i\psi \omega'$ with $\psi = 1/2 + \tilde \psi/2$, we
find that $|\Im(\tilde \psi)|< -\frac{\pi}{\log(q)}$ is necessary and sufficient to ensure analyticity.
\color{Green}
$\\$
$\\$\color{Green}$\%$$\%$
$\\$
$\\$
$\\$\color{Green}$\%$\-\ find\-\ rho\_psi
$\\$\color{BrickRed}c\_psi\-\ =\-\ nm('0.9')*pie/abs(log(q\_max));\color{Green}
$\\$\color{BrickRed}rho\_psi\-\ =\-\ nm(inf(c\_psi\-\ +sqrt(c\_psi\verb|^|2+1)));\color{Green}
$\\$\color{BrickRed}d.rho\_psi\-\ =\-\ rho\_psi;\color{Green}
$\\$
$\\$
$\\$\color{Green}$\%$\-\ make\-\ sure\-\ rho\_psi\-\ is\-\ small\-\ enough\-\ that\-\ bounds
$\\$\color{Green}$\%$\-\ on\-\ xi(omega\-\ +\-\ 1i*psi*omega')\-\ are\-\ valid
$\\$\color{BrickRed}\color{NavyBlue}\-\ if\-\ \color{BrickRed}\-\ sup(rho\_psi)\-\ $>$=\-\ 3+sqrt(nm(2))\color{Green}
$\\$\color{BrickRed}\-\ \-\ \-\ \-\ rho\_psi\-\ =\-\ 3+sqrt(nm(2));\color{Green}
$\\$\color{BrickRed}\color{NavyBlue}\-\ end\-\ \color{BrickRed}\color{Green}
$\\$
$\\$
$\\$\color{BrickRed}\color{NavyBlue}\-\ if\-\ \color{BrickRed}\-\ sup(rho\_psi)\-\ $<$=\-\ inf(4-sqrt(nm(15)))\color{Green}
$\\$\color{BrickRed}\-\ \-\ \-\ \-\ error('problem');\color{Green}
$\\$\color{BrickRed}\color{NavyBlue}\-\ end\-\ \color{BrickRed}\color{Green}
$\\$\color{BrickRed}\-\ \color{Green}
$\\$\color{Green}$\%$
$\\$\color{Green}$\%$\-\ get\-\ bounds\-\ for\-\ alpha\-\ =\-\ omega\-\ +\-\ 1i*psi*omega'
$\\$\color{Green}$\%$
$\\$
$\\$
$\\$\color{BrickRed}time1\-\ =\-\ tic;\color{Green}
$\\$
$\\$
$\\$\color{BrickRed}M\_x\-\ =\-\ bound\_numer\_unstable(dm,rho\_x,q\_min,q\_max);\color{Green}
$\\$
$\\$
$\\$\color{BrickRed}abs\_tol\-\ =\-\ 1e-17;\color{Green}
$\\$\color{BrickRed}[N\_x,err\_x]\-\ =\-\ N\_nodes(rho\_x,M\_x,abs\_tol);\color{Green}
$\\$
$\\$
$\\$\color{Green}$\%$------------------------------------------------------------
$\\$\color{Green}$\%$\-\ get\-\ derivatives\-\ when\-\ ntilde\-\ =\-\ 1
$\\$\color{Green}$\%$------------------------------------------------------------
$\\$
$\\$
$\\$\color{BrickRed}ntilde\-\ =\-\ 1;\color{Green}
$\\$\color{BrickRed}psi\-\ =\-\ 1;\color{Green}
$\\$\color{BrickRed}fun\-\ =\@(q)(integrand\_numer(N\_x,err\_x,q,psi,ntilde));\color{Green}
$\\$
$\\$
$\\$\color{Green}$\%$\-\ k\-\ =\-\ nm('0.942202');
$\\$
$\\$
$\\$\color{BrickRed}k\-\ =\-\ nm('0.9','0.901');\color{Green}
$\\$
$\\$
$\\$\color{BrickRed}kappa\-\ =\-\ kappa\_of\_k(k);\color{Green}
$\\$\color{BrickRed}elipk\-\ =\-\ elliptic\_integral(k,1);\color{Green}
$\\$\color{BrickRed}elipk2\-\ =\-\ elliptic\_integral(sqrt(1-k\verb|^|2),1);\color{Green}
$\\$\color{BrickRed}omega\-\ =\-\ pie/kappa;\color{Green}
$\\$\color{Green}$\%$\-\ omega\_prime\-\ =\-\ elipk2*pie/(elipk*kappa);
$\\$\color{Green}$\%$\-\ X\-\ =\-\ 2*pie/kappa;
$\\$\color{Green}$\%$\-\ w1\-\ =\-\ omega;
$\\$\color{Green}$\%$\-\ w2\-\ =\-\ omega\_prime*1i;
$\\$\color{Green}$\%$\-\ eta\_omega\-\ =\-\ weierstrass\_eta1(w1,w2);
$\\$\color{BrickRed}q\-\ =\-\ exp(-pie*elipk2/elipk)\color{Green}
$\\$
$\\$
$\\$\color{BrickRed}vals\-\ =\-\ fun(q);\color{Green}
$\\$
$\\$
$\\$\color{BrickRed}numer\_psi\-\ =\-\ imag(vals(:,:,2)+vals(:,:,4)./omega.\verb|^|2)\color{Green}
$\\$\color{BrickRed}g\_psi\-\ =\-\ imag(vals(:,:,6))\color{Green}
$\\$
$\\$
$\\$\color{BrickRed}numer\_cond\-\ =\-\ any(inf(real(numer\_psi))$<$=0);\color{Green}
$\\$\color{BrickRed}denom\_cond\-\ =\-\ any(inf(real(g\_psi))$<$=0);\color{Green}
$\\$
$\\$
$\\$\color{BrickRed}\color{NavyBlue}\-\ if\-\ \color{BrickRed}\-\ (numer\_cond+denom\_cond)\-\ ==\-\ 0\color{Green}
$\\$\color{BrickRed}\-\ \-\ \-\ \-\ fprintf('instability\-\ shown\textbackslash n');\color{Green}
$\\$\color{BrickRed}\color{NavyBlue}\-\ else\-\ \color{BrickRed}\color{Green}
$\\$\color{BrickRed}\-\ \-\ \-\ \-\ error('instability\-\ not\-\ shown\textbackslash n');\color{Green}
$\\$\color{BrickRed}\color{NavyBlue}\-\ end\-\ \color{BrickRed}\color{Green}
$\\$
$\\$
$\\$
$\\$
$\\$
$\\$
$\\$
$\\$
$\\$
$\\$
$\\$
$\\$
$\\$
$\\$
$\\$
$\\$
$\\$\color{Black}\section{integrand\_numer.m}

\color{Green}\color{BrickRed}\color{NavyBlue}\-\ function\-\ \color{BrickRed}\-\ out\-\ =\-\ integrand\_numer(Nx,err,q,psi,ntilde)\color{Green}
$\\$
$\\$
$\\$\color{BrickRed}xi\-\ =\-\ xi\_q\_psi(q,psi,ntilde);\color{Green}
$\\$\color{BrickRed}xi\_der\-\ =\-\ xi\_der\_q\_psi(q,psi,ntilde);\color{Green}
$\\$
$\\$
$\\$\color{Green}$\%$\-\ hold\-\ on
$\\$\color{Green}$\%$\-\ plot(mid(psi),mid(xi),'.-k');
$\\$\color{Green}$\%$\-\ plot(mid(psi),mid(xi\_der),'.-b');
$\\$\color{Green}$\%$\-\ return
$\\$
$\\$
$\\$\color{Green}$\%$\-\ plot(mid(psi),mid(xi),'.k','MarkerSize',18);
$\\$
$\\$
$\\$\color{BrickRed}pie\-\ =\-\ nm('pi');\color{Green}
$\\$\color{BrickRed}half\-\ =\-\ nm(1)/2;\color{Green}
$\\$\color{BrickRed}theta\-\ =\-\ ((0:1:Nx)+half)*pie/(Nx+1);\-\ \color{Green}
$\\$\color{BrickRed}x\-\ =\-\ cos(theta);\color{Green}
$\\$
$\\$
$\\$\color{Green}$\%$\-\ alpha\-\ =\-\ 0
$\\$\color{BrickRed}J\-\ =\-\ theta\_vec(q,0,x,3,0);\color{Green}
$\\$\color{BrickRed}J\-\ =\-\ repmat(J,[1\-\ length(psi)]);\color{Green}
$\\$
$\\$
$\\$\color{BrickRed}E0\-\ =\-\ 1./J(:,:,1);\color{Green}
$\\$\color{BrickRed}E1\-\ =\-\ -J(:,:,2)./J(:,:,1).\verb|^|2;\color{Green}
$\\$\color{BrickRed}E2\-\ =\-\ 2*J(:,:,2).\verb|^|2./J(:,:,1).\verb|^|3-J(:,:,3)./J(:,:,1).\verb|^|2;\color{Green}
$\\$\color{BrickRed}E3\-\ =\-\ -6*J(:,:,2).\verb|^|3./J(:,:,1).\verb|^|4+6*J(:,:,2).*J(:,:,3)./J(:,:,1).\verb|^|3-J(:,:,4)./J(:,:,1).\verb|^|2;\color{Green}
$\\$
$\\$
$\\$\color{BrickRed}B0\-\ =\-\ E0.*E0;\color{Green}
$\\$\color{BrickRed}B1\-\ =\-\ 2*E0.*E1;\color{Green}
$\\$\color{BrickRed}B2\-\ =\-\ 2*(E0.*E2+E1.*E1);\color{Green}
$\\$\color{BrickRed}B3\-\ =\-\ 2*(E3.*E0+3*E1.*E2);\color{Green}
$\\$
$\\$
$\\$\color{Green}$\%$\-\ alpha\-\ \textbackslash neq\-\ 0
$\\$\color{BrickRed}L\-\ =\-\ theta\_vec(q,psi,x,4,ntilde);\color{Green}
$\\$
$\\$
$\\$\color{BrickRed}A0\-\ =\-\ L(:,:,1).*L(:,:,1);\color{Green}
$\\$\color{BrickRed}A1\-\ =\-\ 2*L(:,:,1).*L(:,:,2.);\color{Green}
$\\$\color{BrickRed}A2\-\ =\-\ 2*(L(:,:,1).*L(:,:,3)+L(:,:,2).*L(:,:,2));\color{Green}
$\\$\color{BrickRed}A3\-\ =\-\ 2*(L(:,:,4).*L(:,:,1)+3*L(:,:,2).*L(:,:,3));\color{Green}
$\\$\color{BrickRed}A4\-\ =\-\ 2*(L(:,:,1).*L(:,:,5)+4*L(:,:,2).*L(:,:,4)+3*L(:,:,3).*L(:,:,3));\color{Green}
$\\$
$\\$
$\\$\color{BrickRed}w0\-\ =\-\ A0.*B0;\color{Green}
$\\$\color{BrickRed}w1\-\ =\-\ A0.*B1+A1.*B0;\color{Green}
$\\$\color{BrickRed}w2\-\ =\-\ A0.*B2+2*A1.*B1+A2.*B0;\color{Green}
$\\$\color{BrickRed}w3\-\ =\-\ A0.*B3+3*A1.*B2+3*A2.*B1+\-\ A3.*B0;\color{Green}
$\\$
$\\$
$\\$\color{BrickRed}con\-\ =\-\ log(q)/(1i*pie);\color{Green}
$\\$\color{BrickRed}D0\-\ =\-\ con*(A1.*B0);\color{Green}
$\\$\color{BrickRed}D1\-\ =\-\ con*(A1.*B1+A2.*B0);\color{Green}
$\\$\color{BrickRed}D2\-\ =\-\ con*(A1.*B2+2*A2.*B1+A3.*B0);\color{Green}
$\\$\color{BrickRed}D3\-\ =\-\ con*(A1.*B3+3*A2.*B2+3*A3.*B1+\-\ A4.*B0);\color{Green}
$\\$
$\\$
$\\$\color{BrickRed}xicon\-\ =\-\ 1i*repmat(xi,length(x),1);\color{Green}
$\\$\color{BrickRed}xicon\_der\-\ =\-\ 1i*repmat(xi\_der,length(x),1);\-\ \color{Green}
$\\$
$\\$
$\\$\color{BrickRed}xicon2\-\ =\-\ xicon.*xicon;\color{Green}
$\\$\color{BrickRed}xicon3\-\ =\-\ xicon.*xicon2;\color{Green}
$\\$
$\\$
$\\$\color{BrickRed}v1\-\ =\-\ w1+xicon.*w0;\color{Green}
$\\$\color{BrickRed}v2\-\ =\-\ w2+2*xicon.*w1+xicon2.*w0;\color{Green}
$\\$\color{BrickRed}v3\-\ =\-\ w3+3*xicon.*w2+3*xicon2.*w1+xicon3.*w0;\color{Green}
$\\$
$\\$
$\\$\color{BrickRed}v1\_psi\-\ =\-\ D1\-\ +\-\ xicon\_der.*w0+xicon.*D0;\color{Green}
$\\$
$\\$
$\\$\color{BrickRed}v2\_psi\-\ =\-\ D2+2*xicon\_der.*w1+2*xicon.*D1+...\color{Green}
$\\$\color{BrickRed}\-\ \-\ \-\ \-\ 2*xicon.*xicon\_der.*w0+xicon2.*D0;\color{Green}
$\\$
$\\$
$\\$\color{BrickRed}v3\_psi\-\ =\-\ D3\-\ +\-\ 3*xicon\_der.*w2+3*xicon.*D2+6*xicon.*xicon\_der.*w1...\color{Green}
$\\$\color{BrickRed}\-\ \-\ \-\ \-\ +3*xicon2.*D1+3*xicon2.*xicon\_der.*w0+xicon3.*D0;\color{Green}
$\\$
$\\$
$\\$
$\\$
$\\$\color{BrickRed}f1\-\ =\-\ v1.*conj(v2);\color{Green}
$\\$\color{BrickRed}f1\_psi\-\ =\-\ v1\_psi.*conj(v2)+v1.*conj(v2\_psi);\color{Green}
$\\$
$\\$
$\\$\color{BrickRed}f2\-\ =\-\ v3.*conj(v2);\color{Green}
$\\$\color{BrickRed}f2\_psi\-\ =\-\ v3\_psi.*conj(v2)+v3.*conj(v2\_psi);\color{Green}
$\\$
$\\$
$\\$\color{BrickRed}g\-\ =\-\ w0.*conj(v1);\color{Green}
$\\$\color{BrickRed}g\_psi\-\ =\-\ D0.*conj(v1)+w0.*conj(v1\_psi);\color{Green}
$\\$
$\\$
$\\$\color{Green}$\%$\-\ Chebyshev\-\ polynomials\-\ evaluated\-\ at\-\ the\-\ points
$\\$\color{BrickRed}Tx\-\ =\-\ cos(theta.'*(0:2:Nx));\color{Green}
$\\$
$\\$
$\\$\color{BrickRed}cf1\-\ =\-\ 2*f1.'*Tx/(Nx+1);\color{Green}
$\\$\color{BrickRed}cf1(:,1)\-\ =\-\ cf1(:,1)/2;\color{Green}
$\\$\color{BrickRed}cf2\-\ =\-\ 2*f2.'*Tx/(Nx+1);\color{Green}
$\\$\color{BrickRed}cf2(:,1)\-\ =\-\ cf2(:,1)/2;\color{Green}
$\\$\color{BrickRed}cg\-\ =\-\ 2*g.'*Tx/(Nx+1);\color{Green}
$\\$\color{BrickRed}cg(:,1)\-\ =\-\ cg(:,1)/2;\color{Green}
$\\$
$\\$
$\\$\color{BrickRed}cf1\_psi\-\ =\-\ 2*f1\_psi.'*Tx/(Nx+1);\color{Green}
$\\$\color{BrickRed}cf1\_psi(:,1)\-\ =\-\ cf1\_psi(:,1)/2;\color{Green}
$\\$\color{BrickRed}cf2\_psi\-\ =\-\ 2*f2\_psi.'*Tx/(Nx+1);\color{Green}
$\\$\color{BrickRed}cf2\_psi(:,1)\-\ =\-\ cf2\_psi(:,1)/2;\color{Green}
$\\$\color{BrickRed}cg\_psi\-\ =\-\ 2*g\_psi.'*Tx/(Nx+1);\color{Green}
$\\$\color{BrickRed}cg\_psi(:,1)\-\ =\-\ cg\_psi(:,1)/2;\color{Green}
$\\$
$\\$
$\\$
$\\$
$\\$\color{BrickRed}out\-\ =\-\ nm(zeros(length(psi),1,6));\color{Green}
$\\$
$\\$
$\\$\color{BrickRed}out(:,1)\-\ =\-\ 2*cf1*(1./(1-(0:2:Nx).\verb|^|2)).';\color{Green}
$\\$\color{BrickRed}out(:,3)\-\ =\-\ 2*cf2*(1./(1-(0:2:Nx).\verb|^|2)).';\color{Green}
$\\$\color{BrickRed}out(:,5)\-\ =\-\ 2*cg*(1./(1-(0:2:Nx).\verb|^|2)).';\color{Green}
$\\$
$\\$
$\\$\color{BrickRed}out(:,2)\-\ =\-\ 2*cf1\_psi*(1./(1-(0:2:Nx).\verb|^|2)).';\color{Green}
$\\$\color{BrickRed}out(:,4)\-\ =\-\ 2*cf2\_psi*(1./(1-(0:2:Nx).\verb|^|2)).';\color{Green}
$\\$\color{BrickRed}out(:,6)\-\ =\-\ 2*cg\_psi*(1./(1-(0:2:Nx).\verb|^|2)).';\color{Green}
$\\$
$\\$
$\\$\color{Green}$\%$\-\ add\-\ integration\-\ (\-\ in\-\ x)\-\ error
$\\$\color{BrickRed}out\-\ =\-\ out\-\ +\-\ 2*(nm(-err,err)+1i*nm(-err,err));\color{Green}
$\\$
$\\$
$\\$
$\\$
$\\$
$\\$
$\\$
$\\$
$\\$
$\\$
$\\$
$\\$
$\\$
$\\$
$\\$
$\\$
$\\$\color{Black}\section{kappa\_of\_k.m}

\color{Green}\color{BrickRed}\color{NavyBlue}\-\ function\-\ \color{BrickRed}\-\ kappa\-\ =\-\ kappa\_of\_k(k)\color{Green}
$\\$\color{Green}$\%$\-\ function\-\ kappa\-\ =\-\ kappa\_of\_k(k)
$\\$\color{Green}$\%$
$\\$\color{Green}$\%$\-\ Returns\-\ kappa(k)\-\ 
$\\$
$\\$
$\\$
$\\$\color{Black}
From
\eqn{
\left(\frac{K(k)\mathcal{G}(k)}{\pi}\right)^2=
   \frac{7}{20}\frac{2(k^4-k^2+1)E(k)-(1-k^2)(2-k^2)K(k)}{(-2+3k^2+3k^4-2k^6)E(k)+(k^6+k^4-4k^2+2)K(k)}
}{}
we may determine $\kappa = \mathcal{G}(k)$.
\color{Green}
$\\$
$\\$
$\\$
$\\$\color{BrickRed}pie\-\ =\-\ nm('pi');\color{Green}
$\\$\color{Green}$\%$\-\ elliptic\-\ integrals
$\\$\color{BrickRed}K\-\ =\-\ elliptic\_integral(k,1);\color{Green}
$\\$\color{BrickRed}E\-\ =\-\ elliptic\_integral(k,2);\color{Green}
$\\$\color{Green}$\%$\-\ KmE\-\ =\-\ elliptic\_integral(k,3)
$\\$
$\\$
$\\$\color{Green}$\%$\-\ kappa(k)
$\\$\color{BrickRed}k2\-\ =\-\ k.*k;\color{Green}
$\\$\color{BrickRed}k4\-\ =\-\ k2.*k2;\color{Green}
$\\$\color{BrickRed}k6\-\ =\-\ k2.*k4;\color{Green}
$\\$\color{BrickRed}c1\-\ =\-\ 2*(k4-k2+1).*E-(1-k2).*(2-k2).*K;\color{Green}
$\\$\color{BrickRed}c2\-\ =\-\ (-2+3*k2+3*k4-2*k6).*E+(k6+k4-4*k2+2).*K;\color{Green}
$\\$\color{BrickRed}kappa\-\ =\-\ pie*sqrt(7*c1./(20*c2))./K;\color{Green}
$\\$
$\\$
$\\$
$\\$
$\\$
$\\$
$\\$\color{Black}\section{lambda\_xi.m}

\color{Green}\color{BrickRed}\color{NavyBlue}\-\ function\-\ \color{BrickRed}\-\ [f,fd,fdd,g,gd,gdd]\-\ =\-\ lambda\_xi(q,omega,omega\_prime,psi,ntilde)\color{Green}
$\\$\color{Green}$\%$
$\\$\color{Green}$\%$\-\ out\-\ =\-\ lambda\_xi(q,omega,ntilde,psi\_tilde)
$\\$\color{Green}$\%$
$\\$\color{Green}$\%$\-\ Returns\-\ in\-\ the\-\ first\-\ three\-\ components\-\ 
$\\$\color{Green}$\%$\-\ xi(ntilde*omega\_prime+1i*psi*omega\_prime)\-\ 
$\\$\color{Green}$\%$\-\ and\-\ its\-\ first\-\ two\-\ derviatives\-\ with\-\ respect\-\ to\-\ psi.\-\ Returns\-\ in\-\ the\-\ next\-\ three\-\ 
$\\$\color{Green}$\%$\-\ components\-\ 1i*c*lambda\_0(ntilde*omega+1i*psi*omega\_prime)\-\ where\-\ c\-\ is\-\ a\-\ real,\-\ nonzero\-\ constant.
$\\$
$\\$
$\\$
$\\$\color{Black}
Now
\eqn{
\wp'(z)&= -\frac{\sigma(2z)}{\sigma^4(z)}\\
\sigma(z) &= \frac{2\omega}{\pi}e^{\eta_1 z^2/2\omega}\vartheta_1(\pi z/2\omega)/\vartheta_1'(0)\\
\implies \wp'(z)&= -\frac{(\pi\vartheta_1'(0))^3}{8\omega^3}
\frac{\vartheta_1(\pi z/\omega)}{\vartheta_1^4(\pi z/2\omega)} }{}
\color{Green}
$\\$
$\\$\color{Green}$\%$$\%$
$\\$
$\\$
$\\$\color{Green}$\%$\-\ pi
$\\$\color{BrickRed}pie\-\ =\-\ nm('pi');\color{Green}
$\\$
$\\$
$\\$\color{Green}$\%$\-\ \color{Black}$\vartheta_1'(0)$ \color{Green}
$\\$\color{BrickRed}v0\-\ =\-\ theta\_vec\_z(q,0,1);\color{Green}
$\\$
$\\$
$\\$\color{Green}$\%$\-\ number\-\ of\-\ derivatives\-\ to\-\ take
$\\$\color{BrickRed}m\-\ =\-\ 2;\-\ \color{Green}
$\\$
$\\$
$\\$\color{Green}$\%$\-\ \color{Black}$\vartheta_1(\pi z/\omega)$ \color{Green}
$\\$\color{BrickRed}con\-\ =\-\ pie/(omega);\color{Green}
$\\$\color{BrickRed}z\-\ =\-\ con*(ntilde*omega+1i*psi*omega\_prime);\color{Green}
$\\$\color{BrickRed}vn\-\ =\-\ theta\_vec\_z(q,z,m);\color{Green}
$\\$
$\\$
$\\$\color{Green}$\%$\-\ \color{Black}$\vartheta_1(\pi z/(2\omega))$ \color{Green}
$\\$\color{BrickRed}con\-\ =\-\ pie/(2*omega);\color{Green}
$\\$\color{BrickRed}z\-\ =\-\ con*(ntilde*omega+1i*psi*omega\_prime);\color{Green}
$\\$\color{BrickRed}vd\-\ =\-\ theta\_vec\_z(q,z,m);\color{Green}
$\\$
$\\$
$\\$\color{Green}$\%$\-\ \color{Black}$\wp'(z)$ \color{Green}
$\\$\color{BrickRed}gdd\-\ =\-\ -v0(:,:,2).\verb|^|3*pie\verb|^|3.*vn(:,:,1)./(8*omega.\verb|^|3.*vd(:,:,1).\verb|^|4);\color{Green}
$\\$\color{Green}$\%$\-\ \color{Black}$\pd{^2}{\psi^2} \omega\zeta(\tilde n \omega + i\psi\omega')$ \color{Green}
$\\$\color{BrickRed}gdd\-\ =\-\ 2*1i*omega\_prime\verb|^|2*omega*gdd;\color{Green}
$\\$
$\\$
$\\$\color{Green}$\%$\-\ \color{Black}$ \xi(\tilde n + i\psi \omega')$ \color{Green}
$\\$\color{BrickRed}g\-\ =\-\ xi\_q\_psi(q,psi,ntilde);\color{Green}
$\\$\color{BrickRed}g\-\ =\-\ g.';\color{Green}
$\\$
$\\$
$\\$\color{Green}$\%$\-\ \color{Black}$\pd{}{\psi} \xi(\tilde n + i\psi \omega')$ \color{Green}
$\\$\color{BrickRed}gd\-\ =\-\ \-\ xi\_der\_q\_psi(q,psi,ntilde);\color{Green}
$\\$\color{BrickRed}gd\-\ =\-\ gd.';\color{Green}
$\\$
$\\$
$\\$\color{Green}$\%$\-\ auxiliary\-\ quantities
$\\$
$\\$
$\\$\color{BrickRed}A\-\ =\-\ ((pie/omega)*vd(:,:,1).\verb|^|4.*vn(:,:,2)-(2*pie/omega)*vn(:,:,1).*vd(:,:,1).\verb|^|3.*vd(:,:,2));\color{Green}
$\\$
$\\$
$\\$\color{BrickRed}Ad\-\ =\-\ (\-\ (4*pie/omega)*vd(:,:,1).\verb|^|3.*vd(:,:,2).*vn(:,:,2)*(pie/(2*omega))\-\ ...\color{Green}
$\\$\color{BrickRed}\-\ \-\ \-\ \-\ +\-\ (pie/omega)*vd(:,:,1).\verb|^|4.*vn(:,:,3)*(pie/omega)\-\ ...\color{Green}
$\\$\color{BrickRed}\-\ \-\ \-\ \-\ -\-\ (2*pie/omega)*vn(:,:,2).*(pie/omega).*vd(:,:,1).\verb|^|3.*vd(:,:,2)\-\ ...\color{Green}
$\\$\color{BrickRed}\-\ \-\ \-\ \-\ -(2*pie/omega)*vn(:,:,1).*3.*vd(:,:,1).\verb|^|2.*vd(:,:,2).*(pie/(2*omega)).*vd(:,:,2)\-\ ...\color{Green}
$\\$\color{BrickRed}\-\ \-\ \-\ \-\ -(2*pie/omega)*vn(:,:,1).*vd(:,:,1).\verb|^|3.*vd(:,:,3)*(pie/(2*omega)));\color{Green}
$\\$
$\\$
$\\$\color{BrickRed}B\-\ =\-\ vd(:,:,1).\verb|^|8;\color{Green}
$\\$
$\\$
$\\$\color{BrickRed}Bd\-\ =\-\ 8*vd(:,:,1).\verb|^|7.*vd(:,:,2)*(pie/(2*omega));\color{Green}
$\\$
$\\$
$\\$\color{Green}$\%$\-\ \color{Black} $ ic\lambda_0(\tilde n\omega+i\psi\omega') $  \color{Green}
$\\$\color{BrickRed}f\-\ =\-\ 1i*vn(:,:,1)./vd(:,:,1).\verb|^|4;\color{Green}
$\\$
$\\$
$\\$\color{Green}$\%$\-\ \color{Black} $\pd{}{\psi} ic\lambda_0(\tilde n\omega+i\psi\omega') $  \color{Green}
$\\$\color{BrickRed}fd\-\ =\-\ A./B;\color{Green}
$\\$\color{BrickRed}fd\-\ =\-\ 1i*fd*(1i*omega\_prime);\color{Green}
$\\$
$\\$
$\\$\color{Green}$\%$\-\ \color{Black} $\pd{^2}{\psi^2} ic\lambda_0(\tilde n\omega+i\psi\omega') $  \color{Green}
$\\$\color{BrickRed}fdd\-\ =\-\ (B.*Ad-A.*Bd)./B.\verb|^|2;\color{Green}
$\\$\color{BrickRed}fdd\-\ =\-\ 1i*fdd*(1i*omega\_prime)\verb|^|2;\color{Green}
$\\$
$\\$
$\\$
$\\$
$\\$
$\\$
$\\$\color{Black}\section{lower\_bound.m}

\color{Green}\color{BrickRed}\color{NavyBlue}\-\ function\-\ \color{BrickRed}\-\ [dm,rho\_x]\-\ =\-\ lower\_bound(q)\color{Green}
$\\$
$\\$
$\\$
$\\$
$\\$\color{BrickRed}frac\-\ =\-\ nm('0.9');\-\ \color{Green}$\%$\-\ strictly\-\ between\-\ 0\-\ and\-\ 1\-\ (for\-\ choosing\-\ rho\_x)
$\\$
$\\$
$\\$\color{BrickRed}dm\-\ =\-\ lower\_bound\_local(q,frac);\color{Green}
$\\$
$\\$
$\\$\color{BrickRed}pie\-\ =\-\ nm('pi');\color{Green}
$\\$\color{BrickRed}rho\_con\-\ =\-\ -frac*log(q)/pie;\color{Green}
$\\$\color{BrickRed}rho\_x\-\ =\-\ rho\_con+sqrt(rho\_con\verb|^|2+1);\color{Green}
$\\$
$\\$
$\\$\color{Green}$\%$------------------------------------------------------------
$\\$\color{Green}$\%$\-\ lower\_bound\_local
$\\$\color{Green}$\%$------------------------------------------------------------
$\\$\color{BrickRed}\color{NavyBlue}\-\ function\-\ \color{BrickRed}\-\ out\-\ =\-\ lower\_bound\_local(q,frac)\color{Green}
$\\$
$\\$
$\\$\color{BrickRed}steps\-\ =\-\ 8000;\color{Green}
$\\$\color{BrickRed}pie\-\ =\-\ nm('pi');\color{Green}
$\\$\color{Green}$\%$\-\ parity\-\ and\-\ mirror\-\ symmetry\-\ of\-\ \color{Black}$\vartheta_1$ \color{Green}only\-\ requries\-\ we\-\ go
$\\$\color{Green}$\%$\-\ to\-\ pi/2\-\ instead\-\ of\-\ 2\-\ pi.
$\\$\color{BrickRed}half\-\ =\-\ nm('0.5');\color{Green}
$\\$\color{BrickRed}con\-\ =\-\ (half*pie/steps);\-\ \color{Green}
$\\$
$\\$
$\\$\color{BrickRed}ind\-\ =\-\ 0:1:steps;\color{Green}
$\\$\color{BrickRed}theta\-\ =\-\ nm(ind(1:end-1)*con,ind(2:end)*con);\color{Green}
$\\$
$\\$
$\\$\color{BrickRed}psi\-\ =\-\ 0;\-\ ntilde\-\ =\-\ 0;\color{Green}
$\\$\color{BrickRed}fun\-\ =\-\ \@(q,x)(theta\_vec(q,psi,x,0,ntilde));\color{Green}
$\\$
$\\$
$\\$\color{BrickRed}rho\_x\-\ =\-\ get\_rho(q,frac);\color{Green}
$\\$
$\\$
$\\$\color{BrickRed}x\-\ =\-\ half*(rho\_x*exp(1i*theta)+exp(-1i*theta)/rho\_x);\color{Green}
$\\$
$\\$
$\\$\color{BrickRed}temp\-\ =\-\ abs(fun(q,x));\color{Green}
$\\$
$\\$
$\\$\color{BrickRed}theta(129)\color{Green}
$\\$\color{BrickRed}temp(129)\color{Green}
$\\$
$\\$
$\\$\color{BrickRed}out\-\ =\-\ min(inf(temp));\color{Green}
$\\$
$\\$
$\\$\color{Green}$\%$\-\ check\-\ if\-\ lower\-\ bound\-\ is\-\ not\-\ helpful
$\\$\color{BrickRed}\color{NavyBlue}\-\ if\-\ \color{BrickRed}\-\ out\-\ ==\-\ 0\color{Green}
$\\$\color{BrickRed}\-\ \-\ \-\ \-\ error('lower\-\ bound\-\ is\-\ 0');\color{Green}
$\\$\color{BrickRed}\color{NavyBlue}\-\ end\-\ \color{BrickRed}\color{Green}
$\\$
$\\$
$\\$\color{Green}$\%$\-\ check\-\ that\-\ out\-\ is\-\ not\-\ NaN
$\\$\color{BrickRed}\color{NavyBlue}\-\ if\-\ \color{BrickRed}\-\ ~(out\-\ $<$\-\ 10)\color{Green}
$\\$\color{BrickRed}\-\ \-\ \-\ \-\ error('error\-\ apparent')\color{Green}
$\\$\color{BrickRed}\color{NavyBlue}\-\ end\-\ \color{BrickRed}\color{Green}
$\\$
$\\$
$\\$
$\\$
$\\$\color{Green}$\%$------------------------------------------------------------
$\\$\color{Green}$\%$\-\ get\-\ rho
$\\$\color{Green}$\%$------------------------------------------------------------
$\\$\color{BrickRed}\color{NavyBlue}\-\ function\-\ \color{BrickRed}\-\ rho\_x\-\ =\-\ get\_rho(q,frac)\color{Green}
$\\$
$\\$
$\\$\color{Green}$\%$
$\\$\color{Green}$\%$$\%$\-\ compute\-\ rho\-\ for\-\ ellipse\-\ used\-\ in\-\ getting\-\ Chebyshev\-\ bounds\-\ for\-\ x\-\ variable
$\\$\color{Green}$\%$
$\\$
$\\$
$\\$
$\\$\color{Black}
To interpolate in $x\in[-1,1]$ we need a bound on the integrands. The integrands 
are analytic in and on an ellipise that does not intersect zeros of
$\vartheta_1(\pi(x\pm i\omega')/2)$. Now the zeros of $\vartheta_1$ are the set
$\{m\pi +n\pi i\omega'/\omega | m,n\in \N \}$ .
Then from
\eq{
\frac{\pi}{2\omega} \left(\omega x\pm i\omega'\right) &= m\pi + n\pi i \omega'/\omega
}{}
we find
\eq{
\Im(x)< \omega'/\omega.
}{}
is required.

$\\$
Now if $0<c<\omega'/\omega$ is the height of the top of the ellipse $E_{\rho}$, then
\eq{
\frac{1}{2} ( \rho - 1/\rho) = c,
}{}
then
\eq{
\rho = c + \sqrt{c^2+1}.
}{}

$\\$
\color{Green}
$\\$
$\\$
$\\$
$\\$\color{BrickRed}pie\-\ =\-\ nm('pi');\color{Green}
$\\$\color{BrickRed}rho\_con\-\ =\-\ -frac*log(q)/pie;\color{Green}
$\\$\color{BrickRed}rho\_x\-\ =\-\ rho\_con+sqrt(rho\_con\verb|^|2+1);\color{Green}
$\\$
$\\$
$\\$
$\\$
$\\$
$\\$
$\\$
$\\$
$\\$\color{Black}\section{lower\_bound\_unstable.m}

\color{Green}\color{BrickRed}\color{NavyBlue}\-\ function\-\ \color{BrickRed}\-\ [dm,rho\_x,q\_min\_out,q\_max\_out]\-\ =\-\ lower\_bound\_unstable\color{Green}
$\\$
$\\$
$\\$
$\\$
$\\$\color{BrickRed}temp\-\ =\-\ zeros(5,1);\color{Green}
$\\$\color{BrickRed}frac\-\ =\-\ nm('0.9');\-\ \color{Green}$\%$\-\ strictly\-\ between\-\ 0\-\ and\-\ 1\-\ (for\-\ choosing\-\ rho\_x)
$\\$
$\\$
$\\$\color{BrickRed}q\_min\-\ =\-\ nm('0.1');\color{Green}
$\\$\color{BrickRed}q\_max\-\ =\-\ nm('0.15');\color{Green}
$\\$\color{BrickRed}q\_max\_out\-\ =\-\ q\_max;\color{Green}
$\\$\color{BrickRed}num\_steps\-\ =\-\ 100;\color{Green}
$\\$\color{BrickRed}temp(1)\-\ =\-\ lower\_bound\_local(q\_min,q\_max,num\_steps,frac)\color{Green}
$\\$
$\\$
$\\$\color{BrickRed}q\_min\-\ =\-\ nm('0.01');\color{Green}
$\\$\color{BrickRed}q\_max\-\ =\-\ nm('0.1');\color{Green}
$\\$\color{BrickRed}num\_steps\-\ =\-\ 100;\color{Green}
$\\$\color{BrickRed}temp(2)\-\ =\-\ lower\_bound\_local(q\_min,q\_max,num\_steps,frac)\color{Green}
$\\$
$\\$
$\\$\color{BrickRed}q\_min\-\ =\-\ nm('0.001');\color{Green}
$\\$\color{BrickRed}q\_max\-\ =\-\ nm('0.01');\color{Green}
$\\$\color{BrickRed}num\_steps\-\ =\-\ 100;\color{Green}
$\\$\color{BrickRed}temp(3)\-\ =\-\ lower\_bound\_local(q\_min,q\_max,num\_steps,frac)\color{Green}
$\\$
$\\$
$\\$\color{BrickRed}q\_min\-\ =\-\ nm('0.0001');\color{Green}
$\\$\color{BrickRed}q\_max\-\ =\-\ nm('0.001');\color{Green}
$\\$\color{BrickRed}num\_steps\-\ =\-\ 100;\color{Green}
$\\$\color{BrickRed}temp(4)\-\ =\-\ lower\_bound\_local(q\_min,q\_max,num\_steps,frac)\color{Green}
$\\$
$\\$
$\\$\color{BrickRed}q\_min\-\ =\-\ nm('1e-7');\color{Green}
$\\$\color{BrickRed}q\_max\-\ =\-\ nm('0.0001');\color{Green}
$\\$\color{BrickRed}q\_min\_out\-\ =\-\ q\_min;\color{Green}
$\\$\color{BrickRed}num\_steps\-\ =\-\ 100;\color{Green}
$\\$\color{BrickRed}temp(5)\-\ =\-\ lower\_bound\_local(q\_min,q\_max,num\_steps,frac)\color{Green}
$\\$
$\\$
$\\$\color{BrickRed}dm\-\ =\-\ min(temp);\color{Green}
$\\$
$\\$
$\\$\color{BrickRed}pie\-\ =\-\ nm('pi');\color{Green}
$\\$\color{BrickRed}rho\_con\-\ =\-\ -frac*log(q\_max)/pie;\color{Green}
$\\$\color{BrickRed}rho\_x\-\ =\-\ rho\_con+sqrt(rho\_con\verb|^|2+1);\color{Green}
$\\$
$\\$
$\\$\color{Green}$\%$------------------------------------------------------------
$\\$\color{Green}$\%$\-\ lower\_bound\_local
$\\$\color{Green}$\%$------------------------------------------------------------
$\\$\color{BrickRed}\color{NavyBlue}\-\ function\-\ \color{BrickRed}\-\ out\-\ =\-\ lower\_bound\_local(q\_min,q\_max,num\_steps,frac)\color{Green}
$\\$
$\\$
$\\$
$\\$
$\\$\color{BrickRed}steps\-\ =\-\ 8000;\color{Green}
$\\$\color{BrickRed}pie\-\ =\-\ nm('pi');\color{Green}
$\\$\color{Green}$\%$\-\ parity\-\ and\-\ mirror\-\ symmetry\-\ of\-\ \color{Black}$\vartheta_1$ \color{Green}only\-\ requries\-\ we\-\ go
$\\$\color{Green}$\%$\-\ to\-\ pi/2\-\ instead\-\ of\-\ 2\-\ pi.
$\\$\color{BrickRed}half\-\ =\-\ nm('0.5');\color{Green}
$\\$\color{BrickRed}con\-\ =\-\ (half*pie/steps);\-\ \color{Green}
$\\$
$\\$
$\\$\color{BrickRed}ind\-\ =\-\ 0:1:steps;\color{Green}
$\\$\color{BrickRed}theta\-\ =\-\ nm(ind(1:end-1)*con,ind(2:end)*con);\color{Green}
$\\$
$\\$
$\\$\color{BrickRed}q\_del\-\ =\-\ (q\_max-q\_min)/num\_steps;\color{Green}
$\\$\color{BrickRed}psi\-\ =\-\ 0;\-\ ntilde\-\ =\-\ 0;\color{Green}
$\\$\color{BrickRed}fun\-\ =\-\ \@(q,x)(theta\_vec(q,psi,x,0,ntilde));\color{Green}
$\\$
$\\$
$\\$\color{BrickRed}out\-\ =\-\ 1000;\color{Green}
$\\$\color{BrickRed}\color{NavyBlue}\-\ for\-\ \color{BrickRed}\-\ j\-\ =\-\ 1:num\_steps-1\color{Green}
$\\$\color{BrickRed}\-\ \-\ \-\ \-\ \color{Green}
$\\$\color{BrickRed}\-\ \-\ \-\ \-\ q\-\ =\-\ q\_min\-\ +\-\ nm((j-1)*q\_del,j*q\_del);\color{Green}
$\\$\color{BrickRed}\-\ \-\ \-\ \-\ rho\_x\-\ =\-\ get\_rho(q,frac);\color{Green}
$\\$\color{BrickRed}\-\ \-\ \-\ \-\ \-\ \-\ \-\ \-\ \color{Green}
$\\$\color{BrickRed}\-\ \-\ \-\ \-\ x\-\ =\-\ half*(rho\_x*exp(1i*theta)+exp(-1i*theta)/rho\_x);\-\ \color{Green}
$\\$\color{BrickRed}\-\ \-\ \-\ \-\ out\-\ =\-\ min(out,\-\ min(inf(abs(fun(q,x)))));\color{Green}
$\\$
$\\$
$\\$\color{BrickRed}\color{NavyBlue}\-\ end\-\ \color{BrickRed}\color{Green}
$\\$
$\\$
$\\$\color{Green}$\%$\-\ check\-\ if\-\ lower\-\ bound\-\ is\-\ not\-\ helpful
$\\$\color{BrickRed}\color{NavyBlue}\-\ if\-\ \color{BrickRed}\-\ out\-\ ==\-\ 0\color{Green}
$\\$\color{BrickRed}\-\ \-\ \-\ \-\ error('lower\-\ bound\-\ is\-\ 0');\color{Green}
$\\$\color{BrickRed}\color{NavyBlue}\-\ end\-\ \color{BrickRed}\color{Green}
$\\$
$\\$
$\\$\color{Green}$\%$\-\ check\-\ that\-\ out\-\ is\-\ not\-\ NaN
$\\$\color{BrickRed}\color{NavyBlue}\-\ if\-\ \color{BrickRed}\-\ ~(out\-\ $<$\-\ 10)\color{Green}
$\\$\color{BrickRed}\-\ \-\ \-\ \-\ error('error\-\ apparent')\color{Green}
$\\$\color{BrickRed}\color{NavyBlue}\-\ end\-\ \color{BrickRed}\color{Green}
$\\$
$\\$
$\\$
$\\$
$\\$\color{Green}$\%$------------------------------------------------------------
$\\$\color{Green}$\%$\-\ get\-\ rho
$\\$\color{Green}$\%$------------------------------------------------------------
$\\$\color{BrickRed}\color{NavyBlue}\-\ function\-\ \color{BrickRed}\-\ rho\_x\-\ =\-\ get\_rho(q,frac)\color{Green}
$\\$
$\\$
$\\$\color{Green}$\%$
$\\$\color{Green}$\%$$\%$\-\ compute\-\ rho\-\ for\-\ ellipse\-\ used\-\ in\-\ getting\-\ Chebyshev\-\ bounds\-\ for\-\ x\-\ variable
$\\$\color{Green}$\%$
$\\$
$\\$
$\\$
$\\$\color{Black}
To interpolate in $x\in[-1,1]$ we need a bound on the integrands. The integrands 
are analytic in and on an ellipise that does not intersect zeros of
$\vartheta_1(\pi(x\pm i\omega')/2)$. Now the zeros of $\vartheta_1$ are the set
$\{m\pi +n\pi i\omega'/\omega | m,n\in \N \}$ .
Then from
\eq{
\frac{\pi}{2\omega} \left(\omega x\pm i\omega'\right) &= m\pi + n\pi i \omega'/\omega
}{}
we find
\eq{
\Im(x)< \omega'/\omega.
}{}
is required.

$\\$
Now if $0<c<\omega'/\omega$ is the height of the top of the ellipse $E_{\rho}$, then
\eq{
\frac{1}{2} ( \rho - 1/\rho) = c,
}{}
then
\eq{
\rho = c + \sqrt{c^2+1}.
}{}

$\\$
\color{Green}
$\\$
$\\$
$\\$
$\\$\color{BrickRed}pie\-\ =\-\ nm('pi');\color{Green}
$\\$\color{BrickRed}rho\_con\-\ =\-\ -frac*log(q)/pie;\color{Green}
$\\$\color{BrickRed}rho\_x\-\ =\-\ rho\_con+sqrt(rho\_con\verb|^|2+1);\color{Green}
$\\$
$\\$
$\\$
$\\$
$\\$
$\\$
$\\$
$\\$
$\\$\color{Black}\section{numer.m}

\color{Green}\color{BrickRed}\color{NavyBlue}\-\ function\-\ \color{BrickRed}\-\ out\-\ =\-\ numer(Nx,err,q,psi,ntilde)\color{Green}
$\\$\color{Green}$\%$\-\ out\-\ =\-\ numer(Nx,err,q,psi,ntilde)
$\\$\color{Green}$\%$
$\\$\color{Green}$\%$\-\ 
$\\$
$\\$
$\\$
$\\$\color{Black}

 Definition of $\lambda_1$

 \eq{
\lambda_1 
&= \frac{\int_{-1}^{1} \left[\frac{\partial}{\partial x}v(\omega x)+\frac{1}{\omega^2}\frac{\partial^3}{\partial x^3}v(\omega x)\right]\frac{\partial^2}{\partial x^2}\bar v(\omega x)dx}{\omega^2\int_{-1}^{1} v(\omega x)\frac{\partial}{\partial x}\bar v(\omega x)dx}.
}{\label{innerprod2}}

Now
\eq{
v(x) &= w(x)e^{\gamma x}\\
v'(x) & = w'(x) e^{\gamma x} + \gamma w(x)e^{\gamma x}\\
v''(x) & = w''(x) e^{\gamma x} + 2\gamma w'(x) e^{\gamma x} + \gamma^2 w(x) e^{\gamma x}\\
v'''(x) & = w'''(x) e^{\gamma x} +3\gamma w''(x) e^{\gamma x} + 3\gamma^2 w'(x) e^{\gamma x} + \gamma^3 w(x) e^{\gamma x}.
}{\notag}

Then
\eq{
v(\omega x) \bar v'(\omega x) &= w(\omega x) \bar w'(\omega x) + (i\xi) w(\omega x) \bar w(\omega x),\\
v(\omega x) \bar v''(\omega x) &= \sum_{n=0}^{3} \gamma^n c_n(x),\\
v'''(\omega x) \bar v''(\omega x) &= \sum_{n=0}^{5} \gamma^n h_n(x),
}{}
where

\eq{
c_0(x)&:= w'(\omega x) \bar w''(\omega x)\\
c_1(x)&:= w(\omega x) \bar w''(\omega x)-2w'(\omega x)\bar w'(\omega x)\\
c_2(x)&:= w'(\omega x)\bar w(\omega x) -2w(\omega x)\bar w'(\omega x)\\
c_3(x)&:= w(\omega x)\bar w(\omega x)\\
h_0(x)&:= w'''(\omega x)\bar w''(\omega x)\\
h_1(x)&:= 3w''(\omega x)\bar w''(\omega x)-2w'''(\omega x)\bar w'(\omega x)\\
h_2(x)&:= w'''(\omega x)\bar w(\omega x)-6w''(\omega x)\bar w'(\omega x)+3 w'(\omega x)\bar w''(\omega x)\\
h_3(x)&:= 3w''(\omega x)\bar w(\omega x)-6w'(\omega x)\bar w'(\omega x)+w(\omega x)\bar w''(\omega x)\\
h_4(x)&:= 3w'(\omega x)\bar w(\omega x)-2w(\omega x)\bar w'(\omega x)\\
h_5(x)&:= w(\omega x)\bar w(\omega x).
}{}

$\\$
\color{Green}
$\\$
$\\$\color{Green}$\%$$\%$
$\\$
$\\$
$\\$\color{Green}$\%$\-\ -----------------------------------------------------------
$\\$\color{Green}$\%$\-\ Get\-\ theta\-\ and\-\ derivatives
$\\$\color{Green}$\%$\-\ -----------------------------------------------------------
$\\$
$\\$
$\\$\color{Green}$\%$\-\ constants
$\\$\color{BrickRed}pie\-\ =\-\ nm('pi');\color{Green}
$\\$\color{BrickRed}half\-\ =\-\ nm(1)/2;\color{Green}
$\\$
$\\$
$\\$\color{Green}$\%$\-\ interpolation\-\ points\-\ in\-\ x\-\ for\-\ integration
$\\$\color{BrickRed}theta\-\ =\-\ ((0:1:Nx)+half)*pie/(Nx+1);\-\ \color{Green}
$\\$\color{BrickRed}x\-\ =\-\ cos(theta);\color{Green}
$\\$
$\\$
$\\$\color{Green}$\%$\-\ denominator\-\ term\-\ (alpha\-\ =\-\ 0)
$\\$\color{BrickRed}J\-\ =\-\ theta\_vec(q,0,x,3,0);\color{Green}
$\\$\color{BrickRed}J\-\ =\-\ repmat(J,[1\-\ length(psi)]);\color{Green}
$\\$\color{BrickRed}\-\ \-\ \color{Green}
$\\$\color{BrickRed}E0\-\ =\-\ 1./J(:,:,1);\color{Green}
$\\$\color{BrickRed}E1\-\ =\-\ -J(:,:,2)./J(:,:,1).\verb|^|2;\color{Green}
$\\$\color{BrickRed}E2\-\ =\-\ 2*J(:,:,2).\verb|^|2./J(:,:,1).\verb|^|3-J(:,:,3)./J(:,:,1).\verb|^|2;\color{Green}
$\\$\color{BrickRed}E3\-\ =\-\ -6*J(:,:,2).\verb|^|3./J(:,:,1).\verb|^|4+6*J(:,:,2).*J(:,:,3)./J(:,:,1).\verb|^|3-J(:,:,4)./J(:,:,1).\verb|^|2;\color{Green}
$\\$
$\\$
$\\$\color{BrickRed}B0\-\ =\-\ E0.*E0;\color{Green}
$\\$\color{BrickRed}B1\-\ =\-\ 2*E0.*E1;\color{Green}
$\\$\color{BrickRed}B2\-\ =\-\ 2*(E0.*E2+E1.*E1);\color{Green}
$\\$\color{BrickRed}B3\-\ =\-\ 2*(E3.*E0+3*E1.*E2);\color{Green}
$\\$
$\\$
$\\$\color{Green}$\%$\-\ numerator\-\ term\-\ (alpha\-\ =\-\ ntilde*omega\-\ +\-\ 1i*psi*omega')
$\\$\color{BrickRed}L\-\ =\-\ theta\_vec(q,psi,x,4,ntilde);\color{Green}
$\\$
$\\$
$\\$\color{BrickRed}A0\-\ =\-\ L(:,:,1).*L(:,:,1);\color{Green}
$\\$\color{BrickRed}A1\-\ =\-\ 2*L(:,:,1).*L(:,:,2.);\color{Green}
$\\$\color{BrickRed}A2\-\ =\-\ 2*(L(:,:,1).*L(:,:,3)+L(:,:,2).*L(:,:,2));\color{Green}
$\\$\color{BrickRed}A3\-\ =\-\ 2*(L(:,:,4).*L(:,:,1)+3*L(:,:,2).*L(:,:,3));\color{Green}
$\\$
$\\$
$\\$\color{BrickRed}w0\-\ =\-\ A0.*B0;\color{Green}
$\\$\color{BrickRed}w1\-\ =\-\ A0.*B1+A1.*B0;\color{Green}
$\\$\color{BrickRed}w2\-\ =\-\ A0.*B2+2*A1.*B1+A2.*B0;\color{Green}
$\\$\color{BrickRed}w3\-\ =\-\ A0.*B3+3*A1.*B2+3*A2.*B1+\-\ A3.*B0;\color{Green}
$\\$
$\\$
$\\$\color{Green}$\%$\-\ -----------------------------------------------------------
$\\$\color{Green}$\%$\-\ functions\-\ in\-\ expansions
$\\$\color{Green}$\%$\-\ -----------------------------------------------------------
$\\$\color{BrickRed}\-\ \color{Green}
$\\$\color{Green}$\%$\-\ \color{Black}$v(\omega x) \bar v''(\omega x) = \sum_{n=0}^{3} \gamma^n c_n(x)$ \color{Green}
$\\$\color{BrickRed}c0\-\ =\-\ w1.*conj(w2);\color{Green}
$\\$\color{BrickRed}c1\-\ =\-\ w0.*conj(w2)-2*w1.*conj(w1);\color{Green}
$\\$\color{BrickRed}c2\-\ =\-\ w1.*conj(w0)-2*w0.*conj(w1);\color{Green}
$\\$\color{BrickRed}c3\-\ =\-\ w0.*conj(w0);\color{Green}
$\\$
$\\$
$\\$\color{Green}$\%$\-\ \color{Black}$v'''(\omega x) \bar v''(\omega x) = \sum_{n=0}^{5} \gamma^n h_n(x)$ \color{Green}
$\\$\color{BrickRed}h0\-\ =\-\ w3.*conj(w2);\color{Green}
$\\$\color{BrickRed}h1\-\ =\-\ 3*w2.*conj(w2)-2*w3.*conj(w1);\color{Green}
$\\$\color{BrickRed}h2\-\ =\-\ w3.*conj(w0)-6*w2.*conj(w1)+3*w1.*conj(w2);\color{Green}
$\\$\color{BrickRed}h3\-\ =\-\ 3*w2.*conj(w0)-6*w1.*conj(w1)+w0.*conj(w2);\color{Green}
$\\$\color{BrickRed}h4\-\ =\-\ 3*w1.*conj(w0)-2*w0.*conj(w1);\color{Green}
$\\$\color{BrickRed}h5\-\ =\-\ w0.*conj(w0);\color{Green}
$\\$
$\\$
$\\$\color{Green}$\%$\-\ Chebyshev\-\ polynomials\-\ evaluated\-\ at\-\ the\-\ interpolation\-\ points
$\\$\color{BrickRed}Tx\-\ =\-\ cos(theta.'*(0:2:Nx));\color{Green}
$\\$
$\\$
$\\$\color{Green}$\%$\-\ Chebyshev\-\ coefficients
$\\$\color{BrickRed}cf\_c0\-\ =\-\ 2*c0.'*Tx/(Nx+1);\color{Green}
$\\$\color{BrickRed}cf\_c0(:,1)\-\ =\-\ cf\_c0(:,1)/2;\color{Green}
$\\$
$\\$
$\\$\color{BrickRed}cf\_c1\-\ =\-\ 2*c1.'*Tx/(Nx+1);\color{Green}
$\\$\color{BrickRed}cf\_c1(:,1)\-\ =\-\ cf\_c1(:,1)/2;\color{Green}
$\\$
$\\$
$\\$\color{BrickRed}cf\_c2\-\ =\-\ 2*c2.'*Tx/(Nx+1);\color{Green}
$\\$\color{BrickRed}cf\_c2(:,1)\-\ =\-\ cf\_c2(:,1)/2;\color{Green}
$\\$
$\\$
$\\$\color{BrickRed}cf\_c3\-\ =\-\ 2*c3.'*Tx/(Nx+1);\color{Green}
$\\$\color{BrickRed}cf\_c3(:,1)\-\ =\-\ cf\_c3(:,1)/2;\color{Green}
$\\$
$\\$
$\\$\color{BrickRed}cf\_h0\-\ =\-\ 2*h0.'*Tx/(Nx+1);\color{Green}
$\\$\color{BrickRed}cf\_h0(:,1)\-\ =\-\ cf\_h0(:,1)/2;\color{Green}
$\\$
$\\$
$\\$\color{BrickRed}cf\_h1\-\ =\-\ 2*h1.'*Tx/(Nx+1);\color{Green}
$\\$\color{BrickRed}cf\_h1(:,1)\-\ =\-\ cf\_h1(:,1)/2;\color{Green}
$\\$
$\\$
$\\$\color{BrickRed}cf\_h2\-\ =\-\ 2*h2.'*Tx/(Nx+1);\color{Green}
$\\$\color{BrickRed}cf\_h2(:,1)\-\ =\-\ cf\_h2(:,1)/2;\color{Green}
$\\$
$\\$
$\\$\color{BrickRed}cf\_h3\-\ =\-\ 2*h3.'*Tx/(Nx+1);\color{Green}
$\\$\color{BrickRed}cf\_h3(:,1)\-\ =\-\ cf\_h3(:,1)/2;\color{Green}
$\\$
$\\$
$\\$\color{BrickRed}cf\_h4\-\ =\-\ 2*h4.'*Tx/(Nx+1);\color{Green}
$\\$\color{BrickRed}cf\_h4(:,1)\-\ =\-\ cf\_h4(:,1)/2;\color{Green}
$\\$
$\\$
$\\$\color{BrickRed}cf\_h5\-\ =\-\ 2*h5.'*Tx/(Nx+1);\color{Green}
$\\$\color{BrickRed}cf\_h5(:,1)\-\ =\-\ cf\_h5(:,1)/2;\color{Green}
$\\$
$\\$
$\\$\color{BrickRed}out\-\ =\-\ nm(zeros(length(psi),1,10));\color{Green}
$\\$\color{BrickRed}out(:,1)\-\ =\-\ 2*cf\_c0*(1./(1-(0:2:Nx).\verb|^|2)).';\color{Green}
$\\$\color{BrickRed}out(:,2)\-\ =\-\ 2*cf\_c1*(1./(1-(0:2:Nx).\verb|^|2)).';\color{Green}
$\\$\color{BrickRed}out(:,3)\-\ =\-\ 2*cf\_c2*(1./(1-(0:2:Nx).\verb|^|2)).';\color{Green}
$\\$\color{BrickRed}out(:,4)\-\ =\-\ 2*cf\_c3*(1./(1-(0:2:Nx).\verb|^|2)).';\color{Green}
$\\$\color{BrickRed}out(:,5)\-\ =\-\ 2*cf\_h0*(1./(1-(0:2:Nx).\verb|^|2)).';\color{Green}
$\\$\color{BrickRed}out(:,6)\-\ =\-\ 2*cf\_h1*(1./(1-(0:2:Nx).\verb|^|2)).';\color{Green}
$\\$\color{BrickRed}out(:,7)\-\ =\-\ 2*cf\_h2*(1./(1-(0:2:Nx).\verb|^|2)).';\color{Green}
$\\$\color{BrickRed}out(:,8)\-\ =\-\ 2*cf\_h3*(1./(1-(0:2:Nx).\verb|^|2)).';\color{Green}
$\\$\color{BrickRed}out(:,9)\-\ =\-\ 2*cf\_h4*(1./(1-(0:2:Nx).\verb|^|2)).';\color{Green}
$\\$\color{BrickRed}out(:,10)\-\ =\-\ 2*cf\_h5*(1./(1-(0:2:Nx).\verb|^|2)).';\color{Green}
$\\$
$\\$
$\\$\color{Green}$\%$\-\ add\-\ integration\-\ (\-\ in\-\ x)\-\ error
$\\$\color{BrickRed}out\-\ =\-\ out\-\ +\-\ 2*(nm(-err,err)+1i*nm(-err,err));\color{Green}
$\\$
$\\$
$\\$
$\\$
$\\$\color{Black}\section{simplicity.m}

\color{Green}\color{BrickRed}clear\-\ all;\-\ close\-\ all;\-\ beep\-\ off;\-\ clc;\-\ curr\_dir\-\ =\-\ cd;\color{Green}
$\\$\color{Green}$\%$$\%$\-\ startup\-\ commands
$\\$\color{BrickRed}cd('..');\color{Green}
$\\$\color{BrickRed}cd('..');\color{Green}
$\\$\color{BrickRed}startup('intlab','','start\-\ matlabpool','off');\color{Green}
$\\$\color{BrickRed}format\-\ long;\color{Green}
$\\$\color{BrickRed}clc;\color{Green}
$\\$\color{BrickRed}cd(curr\_dir);\color{Green}
$\\$
$\\$
$\\$\color{Green}$\%$\-\ display\-\ type
$\\$\color{BrickRed}intvalinit('DisplayMidRad');\color{Green}
$\\$\color{Green}$\%$\-\ intvalinit('DisplayInfSup');
$\\$
$\\$
$\\$\color{Green}$\%$
$\\$\color{Green}$\%$$\%$\-\ parameters
$\\$\color{Green}$\%$
$\\$
$\\$
$\\$\color{BrickRed}k\-\ =\-\ nm('0.99');\color{Green}
$\\$
$\\$
$\\$\color{Green}$\%$
$\\$\color{Green}$\%$$\%$\-\ derived\-\ constants
$\\$\color{Green}$\%$
$\\$
$\\$
$\\$\color{BrickRed}p.k\-\ =\-\ k;\color{Green}
$\\$\color{BrickRed}pie\-\ =\-\ nm('pi');\color{Green}
$\\$\color{BrickRed}kappa\-\ =\-\ kappa\_of\_k(k);\color{Green}
$\\$\color{BrickRed}elipk\-\ =\-\ elliptic\_integral(k,1);\color{Green}
$\\$\color{BrickRed}elipk2\-\ =\-\ elliptic\_integral(sqrt(1-k\verb|^|2),1);\color{Green}
$\\$\color{BrickRed}omega\-\ =\-\ pie/kappa;\color{Green}
$\\$\color{BrickRed}omega\_prime\-\ =\-\ elipk2*pie/(elipk*kappa);\color{Green}
$\\$\color{BrickRed}X\-\ =\-\ 2*pie/kappa;\color{Green}
$\\$\color{BrickRed}w1\-\ =\-\ omega;\color{Green}
$\\$\color{BrickRed}w2\-\ =\-\ omega\_prime*1i;\color{Green}
$\\$\color{BrickRed}eta\_omega\-\ =\-\ weierstrass\_eta1(w1,w2);\color{Green}
$\\$\color{BrickRed}q\-\ =\-\ exp(-pie*elipk2/elipk);\color{Green}
$\\$
$\\$
$\\$\color{Green}$\%$
$\\$\color{Green}$\%$\-\ ----------------
$\\$\color{Green}$\%$
$\\$
$\\$
$\\$\color{BrickRed}ntilde\-\ =\-\ 1;\color{Green}
$\\$
$\\$
$\\$\color{BrickRed}num\-\ =\-\ 100;\color{Green}
$\\$\color{BrickRed}L\-\ =\-\ 50;\-\ \-\ \-\ \color{Green}$\%$\-\ \color{Black}$\beta_0 = 0.98$ \color{Green}\-\ 
$\\$\color{BrickRed}L2\-\ =\-\ 1;\-\ \-\ \-\ \color{Green}$\%$\-\ \color{Black}$\beta \in [0,\omega']$ \color{Green}
$\\$
$\\$
$\\$\color{BrickRed}psi\_x\-\ =\-\ nm(0:1:num)/(L2*num)+nm(L2-1)/L2;\color{Green}
$\\$\color{BrickRed}psi\_x\-\ =\-\ nm(psi\_x(1:end-1),psi\_x(2:end));\-\ \color{Green}$\%$\-\ make\-\ intervals
$\\$\color{BrickRed}psi\_y\-\ =\-\ nm(0:1:num)/(L*num)+nm(L-1)/L;\color{Green}
$\\$\color{BrickRed}psi\_y\-\ =\-\ nm(psi\_y(1:end-1),psi\_y(2:end));\-\ \color{Green}$\%$\-\ make\-\ intervals
$\\$\color{BrickRed}num\-\ =\-\ num\-\ -1;\color{Green}
$\\$
$\\$
$\\$\color{Green}$\%$\-\ get\-\ components\-\ of\-\ h(x,y)
$\\$\color{BrickRed}[f1,fd1,fdd1,g1,gd1,gdd1]\-\ =\-\ lambda\_xi(q,omega,omega\_prime,psi\_x,ntilde);\color{Green}
$\\$
$\\$
$\\$\color{BrickRed}F1\-\ =\-\ repmat(f1,1,num+1);\color{Green}
$\\$\color{BrickRed}Fd1\-\ =\-\ repmat(fd1,1,num+1);\color{Green}
$\\$\color{BrickRed}Fdd1\-\ =\-\ repmat(fdd1,1,num+1);\color{Green}
$\\$\color{BrickRed}G1\-\ =\-\ repmat(g1,1,num+1);\color{Green}
$\\$\color{BrickRed}Gd1\-\ =\-\ repmat(gd1,1,num+1);\color{Green}
$\\$\color{BrickRed}Gdd1\-\ =\-\ repmat(gdd1,1,num+1);\color{Green}
$\\$
$\\$
$\\$\color{Green}$\%$
$\\$\color{Green}$\%$\-\ ----------------
$\\$\color{Green}$\%$
$\\$
$\\$
$\\$\color{BrickRed}ntilde\-\ =\-\ 0;\color{Green}
$\\$
$\\$
$\\$\color{Green}$\%$\-\ get\-\ components\-\ of\-\ h(x,y)
$\\$\color{BrickRed}[f2,fd2,fdd2,g2,gd2,gdd2]\-\ =\-\ lambda\_xi(q,omega,omega\_prime,psi\_y,ntilde);\color{Green}
$\\$\color{BrickRed}f2\-\ =\-\ f2.';\-\ fd2\-\ =\-\ fd2.';\-\ fdd2\-\ =\-\ fdd2.';\-\ g2\-\ =\-\ g2.';\-\ gd2\-\ =\-\ gd2.';\-\ gdd2\-\ =\-\ gdd2.';\color{Green}
$\\$
$\\$
$\\$\color{BrickRed}F2\-\ =\-\ repmat(f2,num+1,1);\color{Green}
$\\$\color{BrickRed}Fd2\-\ =\-\ repmat(fd2,num+1,1);\color{Green}
$\\$\color{BrickRed}Fdd2\-\ =\-\ repmat(fdd2,num+1,1);\color{Green}
$\\$\color{BrickRed}G2\-\ =\-\ repmat(g2,num+1,1);\color{Green}
$\\$\color{BrickRed}Gd2\-\ =\-\ repmat(gd2,num+1,1);\color{Green}
$\\$\color{BrickRed}Gdd2\-\ =\-\ repmat(gdd2,num+1,1);\color{Green}
$\\$
$\\$
$\\$\color{Green}$\%$
$\\$\color{Green}$\%$\-\ ----------------
$\\$\color{Green}$\%$
$\\$
$\\$
$\\$\color{Green}$\%$\-\ \color{Black}Show that $\pd{}{\psi} \xi(\omega+i\psi\omega') > 0$ for $\psi \in [0,1]$. \color{Green}
$\\$\color{BrickRed}min\_gd1\-\ =\-\ min(inf(real(gd1)));\color{Green}
$\\$\color{BrickRed}fprintf('\textbackslash n\textbackslash nThe\-\ minimum\-\ of\-\ xi(omega\-\ +\-\ 1i*beta)\-\ \color{NavyBlue}\-\ for\-\ \color{BrickRed}\-\ beta\-\ in\-\ [0,omega'']\-\ is:\-\ $\%$16.16g\textbackslash n\textbackslash n',min\_gd1);\color{Green}
$\\$
$\\$
$\\$\color{BrickRed}con\-\ =\-\ 1;\color{Green}
$\\$
$\\$
$\\$\color{BrickRed}T\-\ =\-\ con*(F1+F2);\color{Green}
$\\$\color{BrickRed}Tx\-\ =\-\ con*(Fd1);\color{Green}
$\\$\color{BrickRed}Txx\-\ =\-\ con*(Fdd1);\color{Green}
$\\$\color{Green}$\%$\-\ Txy\-\ =\-\ 0
$\\$\color{BrickRed}Ty\-\ =\-\ con*(Fd2);\color{Green}
$\\$\color{BrickRed}Tyy\-\ =\-\ con*(Fdd2);\color{Green}
$\\$
$\\$
$\\$\color{BrickRed}S\-\ =\-\ G1+G2-2*pie;\color{Green}
$\\$\color{BrickRed}Sx\-\ =\-\ Gd1;\color{Green}
$\\$\color{Green}$\%$\-\ Sxy\-\ =\-\ 0
$\\$\color{BrickRed}Sxx\-\ =\-\ Gdd1;\color{Green}
$\\$\color{BrickRed}Sy\-\ =\-\ Gd2;\color{Green}
$\\$\color{BrickRed}Syy\-\ =\-\ Gdd2;\color{Green}
$\\$
$\\$
$\\$\color{BrickRed}h\-\ =\-\ T.\verb|^|2+S.\verb|^|2;\color{Green}
$\\$\color{BrickRed}hx\-\ =\-\ 2*T.*Tx+2*S.*Sx;\color{Green}
$\\$\color{BrickRed}hxx\-\ =\-\ 2*Tx.\verb|^|2+2*T.*Txx+2*Sx.\verb|^|2+2*S.*Sxx;\color{Green}
$\\$\color{BrickRed}hy\-\ =\-\ 2*T.*Ty+2*S.*Sy;\color{Green}
$\\$\color{BrickRed}hyy\-\ =\-\ 2*Ty.\verb|^|2+2*T.*Tyy+2*Sy.\verb|^|2+2*S.*Syy;\color{Green}
$\\$\color{BrickRed}hxy\-\ =\-\ 2*Ty.*Tx+2*Sy.*Sx;\color{Green}
$\\$
$\\$
$\\$\color{Green}$\%$Del
$\\$
$\\$
$\\$\color{BrickRed}Del\-\ =\-\ (hxx.*hyy-hxy.\verb|^|2);\color{Green}
$\\$\color{BrickRed}Del\-\ =\-\ inf(real(Del));\color{Green}
$\\$
$\\$
$\\$\color{BrickRed}\color{NavyBlue}\-\ if\-\ \color{BrickRed}\-\ \-\ abs(sum(sum(isfinite(Del)-1)))\-\ $>$\-\ 0\color{Green}
$\\$\color{BrickRed}\-\ \-\ \-\ \-\ error('Del\-\ is\-\ not\-\ finite');\color{Green}
$\\$\color{BrickRed}\color{NavyBlue}\-\ end\-\ \color{BrickRed}\color{Green}
$\\$
$\\$
$\\$\color{BrickRed}min\_Del\-\ =\-\ min(min(Del));\color{Green}
$\\$\color{BrickRed}fprintf('\textbackslash n\textbackslash nThe\-\ minimum\-\ of\-\ hxx*hyy-hxy\verb|^|2:\-\ $\%$16.16g\textbackslash n',min\_Del);\color{Green}
$\\$
$\\$
$\\$\color{Green}$\%$\-\ hxx
$\\$
$\\$
$\\$\color{BrickRed}hxx\-\ =\-\ inf(real(hxx));\color{Green}
$\\$
$\\$
$\\$\color{BrickRed}\color{NavyBlue}\-\ if\-\ \color{BrickRed}\-\ \-\ abs(sum(sum(isfinite(hxx)-1)))\-\ $>$\-\ 0\color{Green}
$\\$\color{BrickRed}\-\ \-\ \-\ \-\ error('hxx\-\ is\-\ not\-\ finite');\color{Green}
$\\$\color{BrickRed}\color{NavyBlue}\-\ end\-\ \color{BrickRed}\color{Green}
$\\$
$\\$
$\\$\color{BrickRed}min\_hxx\-\ =\-\ min(min(hxx));\color{Green}
$\\$\color{BrickRed}fprintf('The\-\ minimum\-\ of\-\ hxx:\-\ $\%$16.16g\textbackslash n',min\_hxx);\color{Green}
$\\$
$\\$
$\\$
$\\$
$\\$\color{BrickRed}temp1\-\ =\-\ -4*gdd1/(2*1i*omega\_prime\verb|^|2*omega);\color{Green}
$\\$\color{BrickRed}max\_gdd1\-\ =\-\ max(sup(abs(imag(temp1))));\color{Green}
$\\$
$\\$
$\\$\color{BrickRed}temp2\-\ =\-\ -4*gdd2(1)/(2*1i*omega\_prime\verb|^|2*omega);\color{Green}
$\\$\color{BrickRed}gdd1\_one\-\ =\-\ inf(imag(temp2));\color{Green}
$\\$
$\\$
$\\$\color{BrickRed}fprintf('Max\-\ abs(lambda\_0(omega+1i*psi*omega\_prime)):\-\ $\%$16.16g\textbackslash n',max\_gdd1);\color{Green}
$\\$\color{BrickRed}fprintf('Inf\-\ of\-\ lambda\_0(1i*psi\_0*omega\_prime):\-\ $\%$16.16g\textbackslash n',gdd1\_one);\color{Green}
$\\$\color{BrickRed}fprintf('Diff:\-\ $\%$16.16g\textbackslash n',inf(nm(gdd1\_one)-nm(max\_gdd1)))\color{Green}
$\\$\color{BrickRed}fprintf('Inf\-\ 3*omega*pi/X:\-\ $\%$16.16g\textbackslash n',\-\ inf(3*pie/2));\color{Green}
$\\$\color{BrickRed}fprintf('Sup\-\ of\-\ xi(1i*psi\_0*omega\_prime:\-\ $\%$16.16g\textbackslash n',\-\ sup(real(g2(1))));\color{Green}
$\\$\color{BrickRed}fprintf('Diff\-\ of\-\ last\-\ two:\-\ $\%$16.16g\textbackslash n\textbackslash n',\-\ -inf(g2(1)-3*pie/2));\color{Green}
$\\$
$\\$
$\\$\color{Green}$\%$\-\ throw\-\ errors\-\ if\-\ we\-\ do\-\ not\-\ verify\-\ the\-\ necessary\-\ properties
$\\$
$\\$
$\\$\color{BrickRed}\color{NavyBlue}\-\ if\-\ \color{BrickRed}\-\ min\_Del\-\ $<$=0\color{Green}
$\\$\color{BrickRed}\-\ \-\ \-\ \-\ error('Del\-\ not\-\ positive');\color{Green}
$\\$\color{BrickRed}\color{NavyBlue}\-\ end\-\ \color{BrickRed}\color{Green}
$\\$
$\\$
$\\$\color{BrickRed}\color{NavyBlue}\-\ if\-\ \color{BrickRed}\-\ min\_hxx\-\ $<$=\-\ 0\color{Green}
$\\$\color{BrickRed}\-\ \-\ \-\ \-\ error('hxx\-\ not\-\ positive');\color{Green}
$\\$\color{BrickRed}\color{NavyBlue}\-\ end\-\ \color{BrickRed}\color{Green}
$\\$
$\\$
$\\$\color{BrickRed}\color{NavyBlue}\-\ if\-\ \color{BrickRed}\-\ inf(nm(gdd1\_one)-nm(max\_gdd1))\-\ $<$=\-\ 0\color{Green}
$\\$\color{BrickRed}\-\ \-\ \-\ \-\ error('condition\-\ fails')\color{Green}
$\\$\color{BrickRed}\color{NavyBlue}\-\ end\-\ \color{BrickRed}\color{Green}
$\\$
$\\$
$\\$\color{BrickRed}\color{NavyBlue}\-\ if\-\ \color{BrickRed}\-\ -inf(g2(1)-3*pie/2)\-\ $<$=\-\ 0\color{Green}
$\\$\color{BrickRed}\-\ \-\ \-\ \-\ error('condition\-\ fails')\color{Green}
$\\$\color{BrickRed}\color{NavyBlue}\-\ end\-\ \color{BrickRed}\color{Green}
$\\$
$\\$
$\\$\color{Black}\section{stability.m}

\color{Green}\color{BrickRed}clear\-\ all;\-\ close\-\ all;\-\ beep\-\ off;\-\ clc;\-\ curr\_dir\-\ =\-\ cd;\color{Green}
$\\$\color{Green}$\%$$\%$\-\ startup\-\ commands
$\\$\color{BrickRed}cd('..');\color{Green}
$\\$\color{BrickRed}cd('..');\color{Green}
$\\$\color{BrickRed}startup('intlab','','start\-\ matlabpool','off');\color{Green}
$\\$\color{BrickRed}format\-\ long;\color{Green}
$\\$\color{BrickRed}clc;\color{Green}
$\\$\color{BrickRed}cd(curr\_dir);\color{Green}
$\\$\color{Green}$\%$\-\ Spectral\-\ stability\-\ of\-\ periodic\-\ wave\-\ trains\-\ of\-\ the\-\ Korteweg-de
$\\$\color{Green}$\%$\-\ Vries/Kuramoto-Sivashinsky\-\ equation\-\ in\-\ the\-\ Korteweg-de\-\ Vries\-\ limit
$\\$
$\\$
$\\$\color{Green}$\%$\-\ display\-\ type
$\\$\color{Green}$\%$\-\ intvalinit('DisplayMidRad');
$\\$\color{BrickRed}intvalinit('DisplayInfSup');\color{Green}
$\\$
$\\$
$\\$\color{BrickRed}total\_time\-\ =\-\ tic;\color{Green}
$\\$
$\\$
$\\$\color{Green}$\%$$\%$
$\\$
$\\$
$\\$\color{BrickRed}k\-\ =\-\ nm('0.99');\color{Green}
$\\$\color{BrickRed}p.k\-\ =\-\ k;\color{Green}
$\\$\color{BrickRed}pie\-\ =\-\ nm('pi');\color{Green}
$\\$\color{BrickRed}kappa\-\ =\-\ kappa\_of\_k(k);\color{Green}
$\\$\color{BrickRed}elipk\-\ =\-\ elliptic\_integral(k,1);\color{Green}
$\\$\color{BrickRed}elipk2\-\ =\-\ elliptic\_integral(sqrt(1-k\verb|^|2),1);\color{Green}
$\\$\color{BrickRed}omega\-\ =\-\ pie/kappa;\color{Green}
$\\$\color{BrickRed}omega\_prime\-\ =\-\ elipk2*pie/(elipk*kappa);\color{Green}
$\\$\color{BrickRed}X\-\ =\-\ 2*pie/kappa;\color{Green}
$\\$\color{BrickRed}w1\-\ =\-\ omega;\color{Green}
$\\$\color{BrickRed}w2\-\ =\-\ omega\_prime*1i;\color{Green}
$\\$\color{BrickRed}eta\_omega\-\ =\-\ weierstrass\_eta1(w1,w2);\color{Green}
$\\$\color{BrickRed}q\-\ =\-\ exp(-pie*elipk2/elipk);\color{Green}
$\\$
$\\$
$\\$\color{BrickRed}pie\-\ =\-\ nm('pi');\color{Green}
$\\$
$\\$
$\\$\color{Green}$\%$
$\\$\color{Green}$\%$\-\ Get\-\ the\-\ lower\-\ bound\-\ of\-\ theta
$\\$\color{Green}$\%$
$\\$
$\\$
$\\$\color{BrickRed}t1\-\ =\-\ tic;\color{Green}
$\\$\color{BrickRed}[dm,rho\_x]\-\ =\-\ lower\_bound(q);\-\ \color{Green}
$\\$\color{BrickRed}d.dm\_time\-\ =\-\ toc(t1);\color{Green}
$\\$\color{BrickRed}d.dm\-\ =\-\ dm;\color{Green}
$\\$\color{BrickRed}d.rho\_x\-\ =\-\ rho\_x;\color{Green}
$\\$
$\\$
$\\$
$\\$\-\ \color{Black}
Note that $\rho_{\psi}$ must be chosen so that $\xi(\omega + i\psi \omega')$ does not have a pole.
The poles of $\xi$ are $z = 2m \omega + 2n \omega'$. Setting
$2m\omega + 2n\omega' = \omega + i\psi \omega'$ with $\psi = 1/2 + \tilde \psi/2$, we
find that $|\Im(\tilde \psi)|< -\frac{\pi}{\log(q)}$ is necessary and sufficient to ensure analyticity.
\color{Green}
$\\$
$\\$\color{Green}$\%$$\%$
$\\$
$\\$
$\\$\color{Green}$\%$\-\ find\-\ rho\_psi
$\\$\color{BrickRed}c\_psi\-\ =\-\ nm('0.9')*pie/abs(log(q));\color{Green}
$\\$\color{BrickRed}rho\_psi\-\ =\-\ nm(inf(c\_psi\-\ +sqrt(c\_psi\verb|^|2+1)));\color{Green}
$\\$\color{BrickRed}d.rho\_psi\-\ =\-\ rho\_psi;\color{Green}
$\\$
$\\$
$\\$\color{Green}$\%$\-\ make\-\ sure\-\ rho\_psi\-\ is\-\ small\-\ enough\-\ that\-\ bounds
$\\$\color{Green}$\%$\-\ on\-\ xi(omega\-\ +\-\ 1i*psi*omega')\-\ are\-\ valid
$\\$\color{BrickRed}\color{NavyBlue}\-\ if\-\ \color{BrickRed}\-\ sup(rho\_psi)\-\ $>$=\-\ 3+sqrt(nm(2))\color{Green}
$\\$\color{BrickRed}\-\ \-\ \-\ \-\ rho\_psi\-\ =\-\ 3+sqrt(nm(2));\color{Green}
$\\$\color{BrickRed}\color{NavyBlue}\-\ end\-\ \color{BrickRed}\color{Green}
$\\$
$\\$
$\\$\color{BrickRed}\color{NavyBlue}\-\ if\-\ \color{BrickRed}\-\ sup(rho\_psi)\-\ $<$=\-\ 1\color{Green}
$\\$\color{BrickRed}\-\ \-\ \-\ \-\ error('problem');\color{Green}
$\\$\color{BrickRed}\color{NavyBlue}\-\ end\-\ \color{BrickRed}\color{Green}
$\\$\color{BrickRed}\-\ \color{Green}
$\\$\color{Green}$\%$
$\\$\color{Green}$\%$\-\ find\-\ rho\_q
$\\$\color{Green}$\%$
$\\$
$\\$
$\\$\color{BrickRed}q\_min\-\ =\-\ q;\color{Green}
$\\$\color{BrickRed}q\_max\-\ =\-\ q+1e-12;\color{Green}
$\\$\color{BrickRed}a\-\ =\-\ q\_min;\color{Green}
$\\$\color{BrickRed}b\-\ =\-\ q\_max;\color{Green}
$\\$\color{BrickRed}d.q\_min\-\ =\-\ q\_min;\color{Green}
$\\$\color{BrickRed}d.q\_max\-\ =\-\ q\_max;\color{Green}
$\\$
$\\$
$\\$\color{BrickRed}rho1\-\ =\-\ (b+a)/(b-a)-2*sqrt(a*b)/(b-a);\color{Green}
$\\$\color{BrickRed}rho2\-\ =\-\ (b+a)/(b-a)+2*sqrt(a*b)/(b-a);\color{Green}
$\\$\color{BrickRed}rho3\-\ =\-\ (2-a-b)/(b-a)-2*sqrt(1-a-b+a*b)/(b-a);\color{Green}
$\\$\color{BrickRed}rho4\-\ =\-\ (2-a-b)/(b-a)+2*sqrt(1-a-b+a*b)/(b-a);\color{Green}
$\\$
$\\$
$\\$\color{BrickRed}rho\_left\-\ =\-\ nm(rho1,rho3);\color{Green}
$\\$\color{BrickRed}rho\_right\-\ =\-\ nm(rho2,rho4);\color{Green}
$\\$\color{BrickRed}\color{NavyBlue}\-\ if\-\ \color{BrickRed}\-\ sup(rho\_left)\-\ $>$=\-\ inf(rho\_right)\color{Green}
$\\$\color{BrickRed}\-\ \-\ \-\ \-\ error('problem');\color{Green}
$\\$\color{BrickRed}\color{NavyBlue}\-\ end\-\ \color{BrickRed}\color{Green}
$\\$
$\\$
$\\$\color{BrickRed}rho\_left\-\ =\-\ nm(sup(rho\_left));\color{Green}
$\\$\color{BrickRed}rho\_right\-\ =\-\ nm(inf(rho\_right));\color{Green}
$\\$\color{BrickRed}rho\_q\-\ =\-\ rho\_left\-\ +\-\ nm('0.9')*(rho\_right-rho\_left);\color{Green}
$\\$\color{BrickRed}d.rho\_q\-\ =\-\ rho\_q;\color{Green}
$\\$
$\\$
$\\$\color{Green}$\%$
$\\$\color{Green}$\%$\-\ get\-\ bounds\-\ for\-\ alpha\-\ =\-\ omega\-\ +\-\ 1i*psi*omega'
$\\$\color{Green}$\%$
$\\$
$\\$
$\\$\color{Green}$\%$\-\ ntilde\-\ =\-\ 1
$\\$\color{BrickRed}ntilde\-\ =\-\ 1;\color{Green}
$\\$
$\\$
$\\$\color{BrickRed}time1\-\ =\-\ tic;\color{Green}
$\\$
$\\$
$\\$\color{BrickRed}[M\_psi,M\_x,M\_q]\-\ =\-\ bound\_numer(dm,rho\_psi,rho\_x,rho\_q,q\_min,q\_max,nm(0),nm(1),ntilde);\color{Green}
$\\$\color{BrickRed}toc(time1)\color{Green}
$\\$
$\\$
$\\$\color{BrickRed}abs\_tol\-\ =\-\ 1e-17;\color{Green}
$\\$\color{BrickRed}[N\_x,err\_x]\-\ =\-\ N\_nodes(rho\_x,M\_x,abs\_tol);\color{Green}
$\\$
$\\$
$\\$
$\\$
$\\$\color{BrickRed}abs\_tol\-\ =\-\ 1e-17;\color{Green}
$\\$\color{BrickRed}[N\_psi,err\_psi]\-\ =\-\ N\_nodes(rho\_psi,M\_psi,abs\_tol);\color{Green}
$\\$
$\\$
$\\$\color{BrickRed}[N\_q,err\_q]\-\ =\-\ N\_nodes(rho\_q,M\_q,abs\_tol);\color{Green}
$\\$
$\\$
$\\$\color{BrickRed}d.bd\_n1\_time\-\ =\-\ toc(time1);\color{Green}
$\\$
$\\$
$\\$
$\\$
$\\$\color{Green}$\%$------------------------------------------------------------
$\\$\color{Green}$\%$\-\ get\-\ interpolation\-\ coefficients\-\ when\-\ ntilde\-\ =\-\ 1
$\\$\color{Green}$\%$------------------------------------------------------------
$\\$
$\\$
$\\$\color{BrickRed}a\_q\-\ =\-\ a;\color{Green}
$\\$\color{BrickRed}b\_q\-\ =\-\ b;\color{Green}
$\\$\color{BrickRed}a\_psi\-\ =\-\ nm(0);\color{Green}
$\\$\color{BrickRed}b\_psi\-\ =\-\ nm(1);\color{Green}
$\\$
$\\$
$\\$\color{BrickRed}fun\-\ =\-\ \@(q,psi)(integrand\_numer(N\_x,err\_x,q*(b\_q-a\_q)/2+(a\_q+b\_q)/2,...\color{Green}
$\\$\color{BrickRed}\-\ \-\ \-\ \-\ psi*(b\_psi-a\_psi)/2+(a\_psi+b\_psi)/2,ntilde));\color{Green}
$\\$
$\\$
$\\$\color{Green}$\%$\-\ interpolation\-\ coefficients
$\\$\color{BrickRed}t1\-\ =\-\ tic;\color{Green}
$\\$\color{BrickRed}cfn1\-\ =\-\ cf\_biv\_cheby(N\_q,N\_psi,6,fun);\color{Green}
$\\$\color{BrickRed}d.cf1\_time\-\ =\-\ toc(t1);\color{Green}
$\\$\color{BrickRed}d.cfn1\-\ =\-\ cfn1;\color{Green}
$\\$
$\\$
$\\$\color{BrickRed}d.M\_psi\_n1\-\ \-\ =\-\ M\_psi;\-\ \-\ \-\ \-\ \-\ \-\ \-\ \-\ \-\ \-\ \-\ \-\ \-\ \-\ \color{Green}
$\\$\color{BrickRed}d.M\_q\_n1\-\ \-\ \-\ =\-\ M\_q;\-\ \-\ \-\ \-\ \-\ \-\ \-\ \-\ \-\ \-\ \-\ \-\ \-\ \-\ \-\ \-\ \-\ \-\ \-\ \-\ \-\ \-\ \-\ \-\ \-\ \-\ \-\ \color{Green}
$\\$\color{BrickRed}d.M\_x\_n1\-\ \-\ \-\ =\-\ M\_x;\-\ \-\ \-\ \-\ \-\ \-\ \-\ \-\ \-\ \-\ \-\ \-\ \-\ \-\ \-\ \-\ \-\ \-\ \-\ \-\ \-\ \color{Green}
$\\$\color{BrickRed}d.N\_psi\_n1\-\ =\-\ N\_psi;\-\ \-\ \-\ \-\ \-\ \-\ \-\ \-\ \-\ \-\ \-\ \-\ \-\ \-\ \-\ \-\ \-\ \-\ \-\ \-\ \-\ \color{Green}
$\\$\color{BrickRed}d.N\_q\_n1\-\ =\-\ N\_q;\-\ \-\ \-\ \-\ \-\ \-\ \-\ \-\ \-\ \-\ \-\ \-\ \-\ \-\ \-\ \-\ \-\ \-\ \-\ \-\ \-\ \-\ \-\ \-\ \-\ \-\ \-\ \-\ \-\ \-\ \-\ \-\ \color{Green}
$\\$\color{BrickRed}d.N\_x\_n1\-\ =\-\ N\_x;\-\ \-\ \color{Green}
$\\$\color{BrickRed}d.err\_psi\_n1\-\ \-\ =\-\ err\_psi;\-\ \-\ \-\ \-\ \-\ \-\ \-\ \-\ \-\ \-\ \-\ \-\ \-\ \-\ \-\ \color{Green}
$\\$\color{BrickRed}d.err\_q\_n1\-\ \-\ =\-\ err\_q;\-\ \-\ \-\ \-\ \-\ \-\ \-\ \-\ \-\ \-\ \-\ \-\ \-\ \-\ \-\ \-\ \-\ \-\ \-\ \-\ \-\ \-\ \-\ \color{Green}
$\\$\color{BrickRed}d.err\_x\_n1\-\ =\-\ err\_x;\-\ \-\ \color{Green}
$\\$\color{BrickRed}d.a\_q\-\ \-\ =\-\ a\_q;\-\ \-\ \-\ \-\ \-\ \-\ \-\ \-\ \-\ \-\ \-\ \-\ \-\ \-\ \-\ \-\ \-\ \-\ \-\ \-\ \-\ \-\ \-\ \-\ \-\ \-\ \-\ \-\ \-\ \-\ \-\ \-\ \-\ \-\ \-\ \-\ \-\ \-\ \-\ \-\ \-\ \-\ \-\ \-\ \-\ \-\ \-\ \-\ \-\ \-\ \-\ \-\ \-\ \-\ \-\ \-\ \-\ \-\ \-\ \-\ \-\ \-\ \-\ \-\ \-\ \color{Green}
$\\$\color{BrickRed}d.b\_q\-\ =\-\ b\_q;\-\ \-\ \-\ \-\ \-\ \-\ \-\ \-\ \-\ \-\ \-\ \-\ \-\ \-\ \-\ \-\ \-\ \-\ \-\ \-\ \-\ \-\ \-\ \color{Green}
$\\$\color{BrickRed}d.cfn1\-\ =\-\ cfn1;\-\ \-\ \-\ \-\ \-\ \-\ \-\ \-\ \-\ \-\ \-\ \-\ \-\ \-\ \-\ \-\ \-\ \-\ \-\ \-\ \-\ \-\ \-\ \-\ \-\ \-\ \-\ \-\ \-\ \-\ \-\ \-\ \-\ \-\ \-\ \-\ \-\ \-\ \color{Green}
$\\$\color{BrickRed}d.dm\-\ \-\ =\-\ dm;\-\ \-\ \-\ \-\ \-\ \-\ \-\ \-\ \-\ \-\ \-\ \-\ \-\ \-\ \-\ \-\ \-\ \-\ \-\ \-\ \-\ \-\ \-\ \-\ \-\ \color{Green}
$\\$\color{BrickRed}d.rho\_psi\_n1\-\ =\-\ rho\_psi;\-\ \-\ \-\ \-\ \-\ \-\ \-\ \-\ \-\ \-\ \-\ \-\ \-\ \-\ \-\ \-\ \-\ \-\ \-\ \-\ \-\ \-\ \-\ \-\ \-\ \-\ \-\ \-\ \-\ \-\ \-\ \-\ \-\ \-\ \-\ \color{Green}
$\\$\color{BrickRed}d.rho\_q\_n1\-\ =\-\ rho\_q;\-\ \-\ \-\ \-\ \-\ \-\ \-\ \-\ \-\ \-\ \-\ \-\ \-\ \-\ \-\ \-\ \-\ \-\ \-\ \-\ \-\ \-\ \-\ \-\ \-\ \-\ \-\ \-\ \-\ \-\ \-\ \-\ \-\ \color{Green}
$\\$\color{BrickRed}d.rho\_x\_n1\-\ \-\ =\-\ rho\_x;\-\ \-\ \-\ \-\ \color{Green}
$\\$
$\\$
$\\$\color{Green}$\%$
$\\$\color{Green}$\%$\-\ get\-\ bounds\-\ on\-\ 10\-\ integrands\-\ for\-\ alpha\-\ =\-\ 1i*psi*omega'
$\\$\color{Green}$\%$
$\\$
$\\$
$\\$\color{BrickRed}[M\_psi\_10,M\_x\_10,M\_q\_10]\-\ =\-\ bound\_sub\_integrals(dm,rho\_psi,rho\_x,rho\_q,a,b);\color{Green}
$\\$
$\\$
$\\$\color{BrickRed}d.M\_psi\_10\-\ =\-\ M\_psi\_10;\color{Green}
$\\$\color{BrickRed}d.M\_x\_10\-\ =\-\ M\_x\_10;\color{Green}
$\\$\color{BrickRed}d.M\_q\_10\-\ =\-\ M\_q\_10;\color{Green}
$\\$
$\\$
$\\$\color{BrickRed}abs\_tol\-\ =\-\ 1e-17;\color{Green}
$\\$\color{BrickRed}[N\_x\_10,err\_x\_10]\-\ =\-\ N\_nodes(rho\_x,M\_x\_10,abs\_tol);\color{Green}
$\\$\color{BrickRed}d.N\_x\_10\-\ =\-\ N\_x\_10;\color{Green}
$\\$\color{BrickRed}d.err\_x\_10\-\ =\-\ err\_x\_10;\color{Green}
$\\$
$\\$
$\\$\color{BrickRed}abs\_tol\-\ =\-\ 1e-4;\color{Green}
$\\$\color{BrickRed}[N\_psi\_10,err\_psi\_10]\-\ =\-\ N\_nodes(rho\_psi,M\_psi\_10,abs\_tol);\color{Green}
$\\$\color{BrickRed}d.N\_psi\_10\-\ =\-\ N\_psi\_10;\color{Green}
$\\$\color{BrickRed}d.err\_psi\_10\-\ =\-\ err\_psi\_10;\color{Green}
$\\$
$\\$
$\\$\color{BrickRed}[N\_q\_10,err\_q\_10]\-\ =\-\ N\_nodes(rho\_q,M\_q\_10,abs\_tol);\color{Green}
$\\$\color{BrickRed}d.N\_q\_10\-\ =\-\ N\_q\_10;\color{Green}
$\\$\color{BrickRed}d.err\_q\_10\-\ =\-\ err\_q\_10;\color{Green}
$\\$
$\\$
$\\$\color{Green}$\%$------------------------------------------------------------
$\\$\color{Green}$\%$\-\ get\-\ interpolation\-\ coefficients\-\ when\-\ ntilde\-\ =\-\ 0
$\\$\color{Green}$\%$\-\ for\-\ 10\-\ integrands
$\\$\color{Green}$\%$------------------------------------------------------------
$\\$
$\\$
$\\$\color{BrickRed}a\_q\-\ =\-\ a;\color{Green}
$\\$\color{BrickRed}b\_q\-\ =\-\ b;\color{Green}
$\\$\color{BrickRed}a\_psi\-\ =\-\ nm(0);\color{Green}
$\\$\color{BrickRed}b\_psi\-\ =\-\ nm(1);\color{Green}
$\\$
$\\$
$\\$\color{BrickRed}ntilde\-\ =\-\ 0;\color{Green}
$\\$\color{BrickRed}fun\-\ =\-\ \@(q,psi)(numer(N\_x,err\_x,q*(b\_q-a\_q)/2+(a\_q+b\_q)/2,...\color{Green}
$\\$\color{BrickRed}\-\ \-\ \-\ \-\ psi*(b\_psi-a\_psi)/2+(a\_psi+b\_psi)/2,ntilde));\color{Green}
$\\$
$\\$
$\\$\color{Green}$\%$\-\ interpolation\-\ coefficients
$\\$\color{BrickRed}t1\-\ =\-\ tic;\color{Green}
$\\$\color{BrickRed}cf10\-\ =\-\ cf\_biv\_cheby(N\_q\_10,N\_psi\_10,10,fun);\color{Green}
$\\$\color{BrickRed}d.cf10\_time\-\ =\-\ toc(t1);\color{Green}
$\\$\color{BrickRed}d.cf10\-\ =\-\ cf10;\color{Green}
$\\$
$\\$
$\\$\color{Green}$\%$------------------------------------------------------------
$\\$\color{Green}$\%$\-\ Find\-\ bound\-\ on\-\ numerator\-\ when\-\ ntilde\-\ =\-\ 0
$\\$\color{Green}$\%$------------------------------------------------------------
$\\$
$\\$
$\\$\color{BrickRed}ntilde\-\ =\-\ 0;\color{Green}
$\\$
$\\$
$\\$\color{BrickRed}a\_psi\-\ =\-\ inf(nm('0.5'));\color{Green}
$\\$\color{BrickRed}b\_psi\-\ =\-\ 1;\color{Green}
$\\$
$\\$
$\\$\color{BrickRed}time1\-\ =\-\ tic;\color{Green}
$\\$\color{BrickRed}[M\_psi,M\_x,M\_q]\-\ =\-\ bound\_numer...\color{Green}
$\\$\color{BrickRed}\-\ \-\ \-\ \-\ (dm,rho\_psi,rho\_x,rho\_q,q\_min,q\_max,a\_psi,b\_psi,ntilde);\color{Green}
$\\$\color{BrickRed}toc(time1)\color{Green}
$\\$
$\\$
$\\$\color{BrickRed}abs\_tol\-\ =\-\ 1e-17;\color{Green}
$\\$\color{BrickRed}[N\_x,err\_x]\-\ =\-\ N\_nodes(rho\_x,M\_x,abs\_tol);\color{Green}
$\\$
$\\$
$\\$\color{BrickRed}abs\_tol\-\ =\-\ 1e-17;\color{Green}
$\\$\color{BrickRed}[N\_psi,err\_psi]\-\ =\-\ N\_nodes(rho\_psi,M\_psi,abs\_tol);\color{Green}
$\\$
$\\$
$\\$\color{BrickRed}[N\_q,err\_q]\-\ =\-\ N\_nodes(rho\_q,M\_q,abs\_tol);\color{Green}
$\\$
$\\$
$\\$\color{BrickRed}d.bd\_n1\_time\-\ =\-\ toc(time1);\color{Green}
$\\$
$\\$
$\\$\color{BrickRed}d.M\_psi\_n0\-\ \-\ =\-\ M\_psi;\-\ \-\ \-\ \-\ \-\ \-\ \-\ \-\ \-\ \-\ \-\ \-\ \-\ \-\ \color{Green}
$\\$\color{BrickRed}d.M\_q\_n0\-\ \-\ \-\ =\-\ M\_q;\-\ \-\ \-\ \-\ \-\ \-\ \-\ \-\ \-\ \-\ \-\ \-\ \-\ \-\ \-\ \-\ \-\ \-\ \-\ \-\ \-\ \-\ \-\ \-\ \-\ \-\ \-\ \color{Green}
$\\$\color{BrickRed}d.M\_x\_n0\-\ \-\ \-\ =\-\ M\_x;\-\ \-\ \-\ \-\ \-\ \-\ \-\ \-\ \-\ \-\ \-\ \-\ \-\ \-\ \-\ \-\ \-\ \-\ \-\ \-\ \-\ \color{Green}
$\\$\color{BrickRed}d.N\_psi\_n0\-\ =\-\ N\_psi;\-\ \-\ \-\ \-\ \-\ \-\ \-\ \-\ \-\ \-\ \-\ \-\ \-\ \-\ \-\ \-\ \-\ \-\ \-\ \-\ \-\ \color{Green}
$\\$\color{BrickRed}d.N\_q\_n0\-\ =\-\ N\_q;\-\ \-\ \-\ \-\ \-\ \-\ \-\ \-\ \-\ \-\ \-\ \-\ \-\ \-\ \-\ \-\ \-\ \-\ \-\ \-\ \-\ \-\ \-\ \-\ \-\ \-\ \-\ \-\ \-\ \-\ \-\ \-\ \color{Green}
$\\$\color{BrickRed}d.N\_x\_n0\-\ =\-\ N\_x;\-\ \-\ \color{Green}
$\\$\color{BrickRed}d.err\_psi\_n0\-\ \-\ =\-\ err\_psi;\-\ \-\ \-\ \-\ \-\ \-\ \-\ \-\ \-\ \-\ \-\ \-\ \-\ \-\ \-\ \color{Green}
$\\$\color{BrickRed}d.err\_q\_n0\-\ \-\ =\-\ err\_q;\-\ \-\ \-\ \-\ \-\ \-\ \-\ \-\ \-\ \-\ \-\ \-\ \-\ \-\ \-\ \-\ \-\ \-\ \-\ \-\ \-\ \-\ \-\ \color{Green}
$\\$\color{BrickRed}d.err\_x\_n0\-\ =\-\ err\_x;\-\ \-\ \color{Green}
$\\$
$\\$
$\\$\color{Green}$\%$------------------------------------------------------------
$\\$\color{Green}$\%$\-\ get\-\ interpolation\-\ coefficients\-\ when\-\ ntilde\-\ =\-\ 0
$\\$\color{Green}$\%$------------------------------------------------------------
$\\$
$\\$
$\\$\color{BrickRed}fun\-\ =\-\ \@(q,psi)(integrand\_numer(N\_x,err\_x,q*(b\_q-a\_q)/2+(a\_q+b\_q)/2,...\color{Green}
$\\$\color{BrickRed}\-\ \-\ \-\ \-\ psi*(b\_psi-a\_psi)/2+(a\_psi+b\_psi)/2,ntilde));\color{Green}
$\\$
$\\$
$\\$\color{Green}$\%$\-\ interpolation\-\ coefficients
$\\$\color{BrickRed}t1\-\ =\-\ tic;\color{Green}
$\\$\color{BrickRed}cfn0\-\ =\-\ cf\_biv\_cheby(N\_q,N\_psi,6,fun);\color{Green}
$\\$\color{BrickRed}d.cf0\_time\-\ =\-\ toc(t1);\color{Green}
$\\$\color{BrickRed}d.cfn0\-\ =\-\ cfn0;\color{Green}
$\\$
$\\$
$\\$\color{Green}$\%$------------------------------------------------------------
$\\$\color{Green}$\%$\-\ save\-\ data
$\\$\color{Green}$\%$------------------------------------------------------------
$\\$
$\\$
$\\$\color{BrickRed}d.total\_time\-\ =\-\ toc(total\_time);\color{Green}
$\\$
$\\$
$\\$\color{BrickRed}file\_name\-\ =\-\ 'd\_11Nov2013';\color{Green}
$\\$\color{BrickRed}saveit(curr\_dir,'interval\_arithmetic',d,file\_name,'data');\color{Green}
$\\$
$\\$
$\\$\color{Black}\section{theta\_vec.m}

\color{Green}\color{BrickRed}\color{NavyBlue}\-\ function\-\ \color{BrickRed}\-\ out\-\ =\-\ theta\_vec(q,psi,x,m,ntilde)\color{Green}
$\\$\color{Green}$\%$\-\ out\-\ =\-\ theta\_vec(q,psi,x,m,ntilde)
$\\$\color{Green}$\%$
$\\$\color{Green}$\%$\-\ Returns\-\ an\-\ interval\-\ enclosure\-\ of\-\ f(x)\-\ and\-\ its\-\ first\-\ m\-\ derivatives\-\ where
$\\$\color{Green}$\%$\-\ f(x)\-\ =\-\ v(pie(x\-\ +\-\ n)/2+i*pi*omega\_prime*(1+psi)/(2*omega))
$\\$\color{Green}$\%$\-\ and\-\ v(z)\-\ is\-\ the\-\ first\-\ Jacobi\-\ Theta\-\ function\-\ with\-\ nome\-\ q.
$\\$\color{Green}$\%$
$\\$\color{Green}$\%$\-\ The\-\ input\-\ should\-\ satisify\-\ -1\-\ $<$=\-\ x\-\ $<$=\-\ 1,\-\ 0\-\ $<$=\-\ psi\-\ $<$=\-\ 1,\-\ 0$<$q$<$1,\-\ ntilde\-\ =\-\ 0\-\ or\-\ 1,\-\ m\-\ $>$=\-\ 0
$\\$
$\\$
$\\$
$\\$\color{Black}
The first Jacobi Theta function is given by the series,
\eqn{
\vartheta_1(z)&= 2\sum_{n=0}^{\infty} (-1)^n q^{(n+1/2)^2}\sin((2n+1)z).
}{}

$\\$
From 
\eq{
\vartheta_1(z) = 2\sum_{n=1}^{\infty} (-1)^{n+1}q^{(n-1/2)^2}\sin((2n-1)z),
}{\notag}

$\\$
we find that
\eq{
f(x)&:= \vartheta_1\left(\frac{\pi}{2\omega}(\omega x + i\omega' + \tilde n \omega + i\psi \omega')\right)
\\ &= -i\sum_{n=1}^{\infty}(-1)^{n+1}q^{(n-1/2)^2} \left(\hat v^{(2n-1)}-\hat v^{-(2n-1)}\right),
}{\notag}
where $\hat v := e^{i\pi(x+\tilde n)/2}q^{(1+\psi)/2}$. Hence

$\\$
\eq{
\pd{^m}{x^m} f(x)&=  -i(i\pi/2)^m\sum_{n=1}^{\infty}(-1)^{n+1}q^{(n-1/2)^2} (2n-1)^m\left(\hat v^{(2n-1)}-\hat v^{-(2n-1)}\right).
}{\notag}

$\\$
We find that
\eq{
Err &:= \left|-i(i\pi/2)^m\sum_{n=N+1}^{\infty}(-1)^{n+1}q^{(n-1/2)^2} (2n-1)^m\left(\hat v^{(2n-1)}-\hat v^{-(2n+1)}\right)\right| 
\\ &\leq 2q^{1/4}(\pi/2)^m\sum_{n = N+1}^{\infty}q^{n^2/2}\left((2n-1)^mq^{n^2/2-n-(2n-1)(1+\psi)/2}\right)\\
&\leq 2q^{1/4}(\pi/2)^m\sum_{n = N+1}^{\infty}q^{n^2/2}\\
&\leq2q^{1/4}(\pi/2)^mq^{(N+1)^2/2}\frac{1}{1-q}, 
}{\notag}
as long as we take $N$ large enough so that $g(x):= \left((2x-1)^mq^{n^2/2-x-(2x-1)(1+\psi)/2}\right)$ satisfies $g(N)< 1$, $g(x) < 0$ whenever $x> N$.
\color{Green}
$\\$
$\\$\color{Green}$\%$$\%$
$\\$
$\\$
$\\$\color{Green}$\%$$\%$\-\ Error\-\ checking
$\\$\color{Green}$\%$
$\\$\color{Green}$\%$\-\ I\-\ tested\-\ against\-\ theta1\_m\-\ for\-\ accuracy,\-\ which\-\ was\-\ tested\-\ against\-\ Maple\-\ and\-\ Mathematica.\-\ I
$\\$\color{Green}$\%$\-\ found\-\ that\-\ they\-\ agreed.\-\ In\-\ testing,\-\ I\-\ found\-\ that\-\ this\-\ function\-\ has\-\ error\-\ radius\-\ about\-\ twice\-\ that\-\ of\-\ 
$\\$\color{Green}$\%$\-\ theta1\_m\-\ for\-\ m\-\ =\-\ 4,\-\ but\-\ that\-\ it\-\ computed\-\ much\-\ faster.\-\ For\-\ example,\-\ it\-\ took\-\ nearly\-\ 2\-\ minutes\-\ to\-\ 
$\\$\color{Green}$\%$\-\ evaluate\-\ theta1\_m\-\ on\-\ 200\-\ x\-\ points\-\ but\-\ only\-\ took\-\ about\-\ 0.2\-\ seconds\-\ with\-\ this\-\ vectorized\-\ version.
$\\$\color{Green}$\%$
$\\$\color{Green}$\%$\-\ Below\-\ is\-\ code\-\ used\-\ for\-\ testing:
$\\$\color{Green}$\%$\-\ x\-\ =\-\ nm(linspace(-1,1,30));
$\\$\color{Green}$\%$\-\ q\-\ =\-\ nm('0.5');
$\\$\color{Green}$\%$\-\ psi\-\ =\-\ nm(linspace(0,1,20));
$\\$\color{Green}$\%$\-\ xicon\-\ =\-\ 0;
$\\$\color{Green}$\%$\-\ xicon\_der\-\ =\-\ 0;
$\\$\color{Green}$\%$\-\ 
$\\$\color{Green}$\%$\-\ tic
$\\$\color{Green}$\%$\-\ val\-\ =\-\ integrands(x,q,psi,xicon,xicon\_der);
$\\$\color{Green}$\%$\-\ toc
$\\$\color{Green}$\%$\-\ 
$\\$\color{Green}$\%$\-\ pie\-\ =\-\ nm('pi');
$\\$\color{Green}$\%$\-\ con\-\ =\-\ pie/2;
$\\$\color{Green}$\%$\-\ diff\-\ =\-\ 0;
$\\$\color{Green}$\%$\-\ for\-\ j\-\ =\-\ 1:length(x)
$\\$\color{Green}$\%$\-\ \-\ \-\ \-\ \-\ for\-\ k\-\ =\-\ 1:length(psi)
$\\$\color{Green}$\%$\-\ \-\ \-\ \-\ \-\ \-\ \-\ \-\ \-\ 
$\\$\color{Green}$\%$\-\ \-\ \-\ \-\ \-\ \-\ \-\ \-\ \-\ z\-\ =\-\ pie*(x(j)\-\ +\-\ 1)/2-log(q)*1i*(1+psi(k))/(2);
$\\$\color{Green}$\%$\-\ \-\ \-\ \-\ \-\ \-\ \-\ \-\ \-\ temp\-\ =\-\ theta1\_m(z,q,4,1e-17);
$\\$\color{Green}$\%$\-\ \-\ \-\ \-\ \-\ \-\ \-\ \-\ \-\ for\-\ ind\-\ =\-\ 1:5
$\\$\color{Green}$\%$\-\ \-\ \-\ \-\ \-\ \-\ \-\ \-\ \-\ \-\ \-\ \-\ \-\ diff\-\ =\-\ max(diff,sup(temp(ind)*con\verb|^|(ind-1)-val(j,k,ind)));
$\\$\color{Green}$\%$\-\ \-\ \-\ \-\ \-\ \-\ \-\ \-\ \-\ end
$\\$\color{Green}$\%$\-\ \-\ \-\ \-\ \-\ end
$\\$\color{Green}$\%$\-\ end
$\\$\color{Green}$\%$\-\ 
$\\$\color{Green}$\%$\-\ disp(diff);
$\\$\color{Green}$\%$\-\ 5.975892748039020e-11\-\ +\-\ 5.958839722380777e-11i
$\\$
$\\$
$\\$\color{Green}$\%$$\%$
$\\$
$\\$
$\\$\color{Green}$\%$\-\ error\-\ target
$\\$\color{BrickRed}tol\-\ =\-\ 1e-17;\color{Green}
$\\$
$\\$
$\\$\color{Green}$\%$
$\\$\color{Green}$\%$\-\ constants
$\\$\color{Green}$\%$
$\\$
$\\$
$\\$\color{BrickRed}psi0\-\ =\-\ max(sup(psi));\color{Green}
$\\$\color{BrickRed}pie\-\ =\-\ nm('pi');\color{Green}
$\\$\color{BrickRed}one\-\ =\-\ nm(1);\color{Green}
$\\$\color{BrickRed}req\_1\-\ =\-\ sup(real(-2*m/log(q)));\color{Green}
$\\$\color{BrickRed}c1\-\ =\-\ nm(2+psi0);\color{Green}
$\\$\color{BrickRed}c2\-\ =\-\ nm(2*q\verb|^|(one/4)*(pie/2)\verb|^|m/(1-q));\color{Green}
$\\$\color{BrickRed}qsqrt\-\ \-\ =\-\ q\verb|^|(one/2);\color{Green}
$\\$\color{BrickRed}half\-\ =\-\ one/2;\color{Green}
$\\$
$\\$
$\\$\color{Green}$\%$
$\\$\color{Green}$\%$\-\ find\-\ N\-\ large\-\ enough\-\ that\-\ truncation\-\ error\-\ is\-\ less\-\ than\-\ tol
$\\$\color{Green}$\%$
$\\$
$\\$
$\\$\color{BrickRed}err\-\ =\-\ tol\-\ +\-\ 1;\color{Green}
$\\$\color{BrickRed}N\-\ =\-\ 0;\color{Green}
$\\$\color{BrickRed}maxit\-\ =\-\ 1000;\color{Green}
$\\$\color{BrickRed}\color{NavyBlue}\-\ while\-\ \color{BrickRed}\-\ err\-\ $>$\-\ tol\color{Green}
$\\$\color{BrickRed}\-\ \-\ \-\ \-\ N\-\ =\-\ N\-\ +\-\ 1;\color{Green}
$\\$\color{BrickRed}\-\ \-\ \-\ \-\ \color{NavyBlue}\-\ if\-\ \color{BrickRed}\-\ N\-\ $>$\-\ maxit\-\ \color{Green}
$\\$\color{BrickRed}\-\ \-\ \-\ \-\ \-\ \-\ \-\ \-\ error('maximum\-\ iterations\-\ exceeded');\color{Green}
$\\$\color{BrickRed}\-\ \-\ \-\ \-\ \color{NavyBlue}\-\ end\-\ \color{BrickRed}\color{Green}
$\\$\color{BrickRed}\-\ \-\ \-\ \-\ \color{Green}
$\\$\color{BrickRed}\-\ \-\ \-\ \-\ \color{Green}$\%$\-\ check\-\ that\-\ derivative\-\ of\-\ \color{Black} $f(x):= q^{x^2/2-x-(2x-1)(1+\psi)/2}(2x-1)^m$ 
\color{BrickRed}\-\ \-\ \-\ \-\ \color{Black} is decreasing for $x \geq N$. \color{Green}
$\\$\color{BrickRed}\-\ \-\ \-\ \-\ \color{NavyBlue}\-\ if\-\ \color{BrickRed}\-\ inf(real((N-c1)*(2*N-1)))\-\ $<$=\-\ req\_1\color{Green}
$\\$\color{BrickRed}\-\ \-\ \-\ \-\ \-\ \-\ \-\ \-\ \color{NavyBlue}\-\ continue\-\ \color{BrickRed}\color{Green}
$\\$\color{BrickRed}\-\ \-\ \-\ \-\ \color{NavyBlue}\-\ end\-\ \color{BrickRed}\color{Green}
$\\$\color{BrickRed}\-\ \-\ \-\ \-\ \color{Green}
$\\$\color{BrickRed}\-\ \-\ \-\ \-\ \color{Green}$\%$\-\ check\-\ that\-\ \color{Black} $f(x):= q^{N^2/2-N-(2N-1)(1+\psi)/2}(2N-1)^m\leq 1$ \color{Green}
$\\$\color{BrickRed}\-\ \-\ \-\ \-\ \color{NavyBlue}\-\ if\-\ \color{BrickRed}\-\ sup(real(q\verb|^|(N\verb|^|2/2-N-(2*N-1)*(1+psi0)/2)*(2*N-1)\verb|^|m))\-\ $>$\-\ 1\color{Green}
$\\$\color{BrickRed}\-\ \-\ \-\ \-\ \-\ \-\ \-\ \-\ \color{NavyBlue}\-\ continue\-\ \color{BrickRed}\color{Green}
$\\$\color{BrickRed}\-\ \-\ \-\ \-\ \color{NavyBlue}\-\ end\-\ \color{BrickRed}\color{Green}
$\\$\color{BrickRed}\-\ \-\ \-\ \-\ \color{Green}
$\\$\color{BrickRed}\-\ \-\ \-\ \-\ \color{Green}$\%$\-\ turncation\-\ error
$\\$\color{BrickRed}\-\ \-\ \-\ \-\ err\-\ =\-\ sup(real(c2*qsqrt\verb|^|((N+1)\verb|^|2)));\color{Green}
$\\$\color{BrickRed}\-\ \-\ \-\ \-\ \color{Green}
$\\$\color{BrickRed}\color{NavyBlue}\-\ end\-\ \color{BrickRed}\color{Green}
$\\$
$\\$
$\\$\color{Green}$\%$
$\\$\color{Green}$\%$\-\ evaluate\-\ the\-\ partial\-\ sum\-\ 
$\\$\color{Green}$\%$
$\\$
$\\$
$\\$\color{Green}$\%$\-\ initialize\-\ vectors
$\\$\color{BrickRed}out\-\ =\-\ nm(zeros(length(x),length(psi),m+1));\color{Green}
$\\$\color{BrickRed}temp\-\ =\-\ out;\color{Green}
$\\$
$\\$
$\\$\color{Green}$\%$\-\ rearrange\-\ dimensions\-\ if\-\ needed
$\\$\color{BrickRed}sx\-\ =\-\ size(x,1);\color{Green}
$\\$\color{BrickRed}\color{NavyBlue}\-\ if\-\ \color{BrickRed}\-\ sx\-\ ==\-\ 1\color{Green}
$\\$\color{BrickRed}\-\ \-\ \-\ \-\ x\-\ =x.';\color{Green}
$\\$\color{BrickRed}\color{NavyBlue}\-\ end\-\ \color{BrickRed}\color{Green}
$\\$
$\\$
$\\$\color{BrickRed}sx\-\ =\-\ size(psi,1);\color{Green}
$\\$\color{BrickRed}\color{NavyBlue}\-\ if\-\ \color{BrickRed}\-\ sx\-\ $>$\-\ 1\color{Green}
$\\$\color{BrickRed}\-\ \-\ \-\ \-\ psi\-\ =psi.';\color{Green}
$\\$\color{BrickRed}\color{NavyBlue}\-\ end\-\ \color{BrickRed}\color{Green}
$\\$
$\\$
$\\$\color{Green}$\%$
$\\$\color{Green}$\%$\-\ add\-\ partial\-\ sum
$\\$\color{Green}$\%$
$\\$
$\\$
$\\$
$\\$
$\\$\color{BrickRed}vhat\-\ =\-\ exp(1i*pie*(x+ntilde)/2)*q.\verb|^|((1+psi)/2);\color{Green}
$\\$
$\\$
$\\$\color{BrickRed}\color{NavyBlue}\-\ for\-\ \color{BrickRed}\-\ n\-\ =\-\ 1:N\color{Green}
$\\$\color{BrickRed}\-\ \-\ \-\ \-\ \color{Green}
$\\$\color{BrickRed}\-\ \-\ \-\ \-\ vtemp\-\ =\-\ vhat.\verb|^|(2*n-1);\color{Green}
$\\$\color{BrickRed}\-\ \-\ \-\ \-\ \color{NavyBlue}\-\ for\-\ \color{BrickRed}\-\ k\-\ =\-\ 0:m\color{Green}
$\\$\color{BrickRed}\-\ \-\ \-\ \-\ \-\ \-\ \-\ \-\ temp(:,:,k+1)\-\ =\-\ (2*n-1)\verb|^|k*(vtemp\-\ -(-1)\verb|^|k*vtemp.\verb|^|(-1));\color{Green}
$\\$\color{BrickRed}\-\ \-\ \-\ \-\ \color{NavyBlue}\-\ end\-\ \color{BrickRed}\color{Green}
$\\$\color{BrickRed}\-\ \-\ \-\ \-\ \-\ \-\ \color{Green}
$\\$\color{BrickRed}\-\ \-\ \-\ \-\ out\-\ =\-\ out\-\ +(-1)\verb|^|(n+1)*q\verb|^|((n-half)\verb|^|2)*temp;\color{Green}
$\\$\color{BrickRed}\-\ \-\ \-\ \-\ \color{Green}
$\\$\color{BrickRed}\color{NavyBlue}\-\ end\-\ \color{BrickRed}\color{Green}
$\\$
$\\$
$\\$\color{Green}$\%$\-\ complete\-\ differentiation\-\ rule
$\\$\color{BrickRed}\color{NavyBlue}\-\ for\-\ \color{BrickRed}\-\ j\-\ =\-\ 0:m\color{Green}
$\\$\color{BrickRed}\-\ \-\ \-\ \-\ out(:,:,j+1)\-\ =\-\ out(:,:,j+1)*(1i*pie/2)\verb|^|j;\color{Green}
$\\$\color{BrickRed}\color{NavyBlue}\-\ end\-\ \color{BrickRed}\color{Green}
$\\$
$\\$
$\\$\color{Green}$\%$\-\ multiply\-\ by\-\ -1i\-\ coming\-\ from\-\ sin\-\ definition\-\ and\-\ add\-\ error\-\ interval
$\\$\color{BrickRed}out\-\ =\-\ -1i*out\-\ +\-\ nm(-err,err)+1i*nm(-err,err);\color{Green}
$\\$
$\\$
$\\$\color{Black}\section{theta\_vec\_z.m}

\color{Green}\color{BrickRed}\color{NavyBlue}\-\ function\-\ \color{BrickRed}\-\ out\-\ =\-\ theta\_vec\_z(q,z,m)\color{Green}
$\\$
$\\$
$\\$\color{Green}$\%$\-\ error\-\ target
$\\$\color{BrickRed}tol\-\ =\-\ 1e-17;\color{Green}
$\\$
$\\$
$\\$\color{Green}$\%$
$\\$\color{Green}$\%$\-\ constants
$\\$\color{Green}$\%$
$\\$
$\\$
$\\$\color{BrickRed}sig\-\ =\-\ -nm(max(sup(abs(imag(z)))))/nm(min(inf(abs(log(q)))));\color{Green}
$\\$
$\\$
$\\$\color{BrickRed}q0\-\ =\-\ nm(max(sup(q)));\color{Green}
$\\$
$\\$
$\\$\color{Green}$\%$
$\\$\color{Green}$\%$\-\ find\-\ N\-\ large\-\ enough\-\ that\-\ truncation\-\ error\-\ is\-\ less\-\ than\-\ tol
$\\$\color{Green}$\%$
$\\$
$\\$
$\\$\color{Green}$\%$\-\ Choose\-\ N\-\ so\-\ that\-\ derivative\-\ of\-\ \color{Black} $f(x) := q_0^{x^2+1/4+(2x+1)\sigma}(2x+1)^m $ 
\color{Black} is decreasing for $x \geq N$. \color{Green}
$\\$\color{BrickRed}c\-\ =\-\ m/log(q0);\color{Green}
$\\$\color{BrickRed}temp\-\ =\-\ (-(2*sig+1)+sqrt((2*sig+1)\verb|^|2-8*(sig+c)))/4;\color{Green}
$\\$\color{BrickRed}N\-\ =\-\ ceil(sup(temp));\color{Green}
$\\$\color{BrickRed}quarter\-\ =\-\ nm(1)/4;\color{Green}
$\\$\color{BrickRed}half\-\ =\-\ nm(1)/2;\color{Green}
$\\$
$\\$
$\\$\color{BrickRed}err\-\ =\-\ tol\-\ +\-\ 1;\color{Green}
$\\$\color{BrickRed}maxit\-\ =\-\ 1000;\color{Green}
$\\$\color{BrickRed}\color{NavyBlue}\-\ while\-\ \color{BrickRed}\-\ err\-\ $>$\-\ tol\color{Green}
$\\$\color{BrickRed}\-\ \-\ \-\ \-\ N\-\ =\-\ N\-\ +\-\ 1;\color{Green}
$\\$\color{BrickRed}\-\ \-\ \-\ \-\ \color{NavyBlue}\-\ if\-\ \color{BrickRed}\-\ N\-\ $>$\-\ maxit\-\ \color{Green}
$\\$\color{BrickRed}\-\ \-\ \-\ \-\ \-\ \-\ \-\ \-\ error('maximum\-\ iterations\-\ exceeded');\color{Green}
$\\$\color{BrickRed}\-\ \-\ \-\ \-\ \color{NavyBlue}\-\ end\-\ \color{BrickRed}\color{Green}
$\\$
$\\$
$\\$\color{BrickRed}\-\ \-\ \-\ \-\ \color{Green}$\%$\-\ turncation\-\ error
$\\$\color{BrickRed}\-\ \-\ \-\ \-\ err1\-\ =\-\ q0\verb|^|((N+1)\verb|^|2+quarter+(2*(N+1)+1)*sig)*(2*(N+1)+1)\verb|^|m;\color{Green}
$\\$\color{BrickRed}\-\ \-\ \-\ \-\ err2\-\ =\-\ q0\verb|^|(N+1)/(1-q0);\color{Green}
$\\$\color{BrickRed}\-\ \-\ \-\ \-\ err\-\ =\-\ sup(err1*err2);\color{Green}
$\\$\color{BrickRed}\-\ \-\ \-\ \-\ \color{Green}
$\\$\color{BrickRed}\color{NavyBlue}\-\ end\-\ \color{BrickRed}\color{Green}
$\\$
$\\$
$\\$
$\\$
$\\$\color{Green}$\%$
$\\$\color{Green}$\%$\-\ evaluate\-\ the\-\ partial\-\ sum\-\ 
$\\$\color{Green}$\%$
$\\$
$\\$
$\\$\color{Green}$\%$\-\ initialize\-\ vectors
$\\$\color{BrickRed}out\-\ =\-\ nm(zeros(length(z),length(q),m+1));\color{Green}
$\\$
$\\$
$\\$\color{Green}$\%$\-\ rearrange\-\ dimensions\-\ if\-\ needed
$\\$\color{BrickRed}sz\-\ =\-\ size(z,1);\color{Green}
$\\$\color{BrickRed}\color{NavyBlue}\-\ if\-\ \color{BrickRed}\-\ sz\-\ ==\-\ 1\color{Green}
$\\$\color{BrickRed}\-\ \-\ \-\ \-\ z\-\ =z.';\color{Green}
$\\$\color{BrickRed}\color{NavyBlue}\-\ end\-\ \color{BrickRed}\color{Green}
$\\$
$\\$
$\\$\color{BrickRed}sq\-\ =\-\ size(q,2);\color{Green}
$\\$\color{BrickRed}\color{NavyBlue}\-\ if\-\ \color{BrickRed}\-\ sq\-\ ==\-\ 1\color{Green}
$\\$\color{BrickRed}\-\ \-\ \-\ \-\ q\-\ =q.';\color{Green}
$\\$\color{BrickRed}\color{NavyBlue}\-\ end\-\ \color{BrickRed}\color{Green}
$\\$
$\\$
$\\$\color{Green}$\%$
$\\$\color{Green}$\%$\-\ add\-\ partial\-\ sum
$\\$\color{Green}$\%$
$\\$
$\\$
$\\$\color{BrickRed}V\-\ =\-\ exp(1i*z);\color{Green}
$\\$
$\\$
$\\$
$\\$
$\\$\color{BrickRed}temp\-\ =\-\ out;\color{Green}
$\\$
$\\$
$\\$\color{BrickRed}\color{NavyBlue}\-\ for\-\ \color{BrickRed}\-\ n\-\ =\-\ 0:N\color{Green}
$\\$\color{BrickRed}\-\ \-\ \-\ \-\ \color{Green}
$\\$\color{BrickRed}\-\ \-\ \-\ \-\ qtemp\-\ =\-\ (-1)\verb|^|n*q.\verb|^|((n+half)\verb|^|2);\color{Green}
$\\$
$\\$
$\\$\color{BrickRed}\-\ \-\ \-\ \-\ \color{NavyBlue}\-\ for\-\ \color{BrickRed}\-\ k\-\ =\-\ 0:m\color{Green}
$\\$\color{BrickRed}\-\ \-\ \-\ \-\ \-\ \-\ \-\ \-\ temp(:,:,k+1)\-\ =\-\ (2*n+1)\verb|^|k*(V.\verb|^|(2*n+1)\-\ -(-1)\verb|^|k*V.\verb|^|(-(2*n+1)))*qtemp;\color{Green}
$\\$\color{BrickRed}\-\ \-\ \-\ \-\ \color{NavyBlue}\-\ end\-\ \color{BrickRed}\color{Green}
$\\$\color{BrickRed}\-\ \-\ \-\ \-\ \-\ \-\ \color{Green}
$\\$\color{BrickRed}\-\ \-\ \-\ \-\ out\-\ =\-\ out\-\ +\-\ temp;\color{Green}
$\\$\color{BrickRed}\-\ \-\ \-\ \-\ \color{Green}
$\\$\color{BrickRed}\color{NavyBlue}\-\ end\-\ \color{BrickRed}\color{Green}
$\\$
$\\$
$\\$
$\\$
$\\$\color{Green}$\%$\-\ complete\-\ differentiation\-\ rule
$\\$\color{BrickRed}\color{NavyBlue}\-\ for\-\ \color{BrickRed}\-\ j\-\ =\-\ 0:m\color{Green}
$\\$\color{BrickRed}\-\ \-\ \-\ \-\ out(:,:,j+1)\-\ =\-\ out(:,:,j+1)*(1i)\verb|^|(j-1);\color{Green}
$\\$\color{BrickRed}\color{NavyBlue}\-\ end\-\ \color{BrickRed}\color{Green}
$\\$
$\\$
$\\$\color{Green}$\%$\-\ multiply\-\ by\-\ -1i\-\ coming\-\ from\-\ sin\-\ definition\-\ and\-\ add\-\ error\-\ interval
$\\$\color{BrickRed}out\-\ =\-\ out\-\ +\-\ nm(-err,err)+1i*nm(-err,err);\color{Green}
$\\$
$\\$
$\\$\color{Black}\section{verify\_data\_n0.m}

\color{Green}\color{BrickRed}\color{NavyBlue}\-\ function\-\ \color{BrickRed}\-\ verify\_data\_n0\color{Green}
$\\$
$\\$
$\\$\color{BrickRed}clear\-\ all;\-\ close\-\ all;\-\ beep\-\ off;\-\ clc;\-\ curr\_dir\-\ =\-\ cd;\color{Green}
$\\$\color{Green}$\%$$\%$\-\ startup\-\ commands
$\\$\color{BrickRed}cd('..');\color{Green}
$\\$\color{BrickRed}cd('..');\color{Green}
$\\$\color{BrickRed}startup('intlab','','start\-\ matlabpool','off');\color{Green}
$\\$\color{BrickRed}format\-\ long;\color{Green}
$\\$\color{BrickRed}clc;\color{Green}
$\\$\color{BrickRed}cd(curr\_dir);\color{Green}
$\\$\color{Green}$\%$\-\ Spectral\-\ stability\-\ of\-\ periodic\-\ wave\-\ trains\-\ of\-\ the\-\ Korteweg-de
$\\$\color{Green}$\%$\-\ Vries/Kuramoto-Sivashinsky\-\ equation\-\ in\-\ the\-\ Korteweg-de\-\ Vries\-\ limit
$\\$
$\\$
$\\$\color{Green}$\%$\-\ display\-\ type
$\\$\color{Green}$\%$\-\ intvalinit('DisplayMidRad');
$\\$\color{BrickRed}intvalinit('DisplayInfSup');\color{Green}
$\\$
$\\$
$\\$\color{BrickRed}file\_name\-\ =\-\ 'd\_31Oct2013';\-\ curr\_dir\-\ =\-\ cd;\color{Green}
$\\$
$\\$
$\\$\color{BrickRed}ld\-\ =\-\ retrieve\_it(curr\_dir,'interval\_arithmetic',file\_name,'data');\color{Green}
$\\$\color{BrickRed}d\-\ =\-\ ld.var;\color{Green}
$\\$
$\\$
$\\$\color{BrickRed}k\-\ =\-\ nm('0.99');\color{Green}
$\\$
$\\$
$\\$
$\\$
$\\$
$\\$
$\\$\color{BrickRed}j\-\ =\-\ 1;\-\ \color{Green}$\%$\-\ just\-\ one\-\ case\-\ since\-\ single\-\ wave
$\\$
$\\$
$\\$\color{BrickRed}left\_point\-\ =\-\ 1e-3;\color{Green}
$\\$\color{BrickRed}left\_verify(k,d,left\_point,j)\color{Green}
$\\$
$\\$
$\\$\color{BrickRed}a\-\ =\-\ left\_point;\color{Green}
$\\$\color{BrickRed}b\-\ =\-\ 0.1;\color{Green}
$\\$\color{BrickRed}[left1,right1]\-\ =\-\ verify(k,d,a,b,j)\color{Green}
$\\$
$\\$
$\\$\color{BrickRed}a\-\ =\-\ 0.09;\color{Green}
$\\$\color{BrickRed}b\-\ =\-\ 0.9;\color{Green}
$\\$\color{BrickRed}[left2,right2]\-\ =\-\ verify(k,d,a,b,j);\color{Green}
$\\$
$\\$
$\\$\color{BrickRed}a\-\ =\-\ 0.89;\color{Green}
$\\$\color{BrickRed}b\-\ =\-\ 0.96;\color{Green}
$\\$\color{BrickRed}[left3,right3]\-\ =\-\ verify(k,d,a,b,j);\color{Green}
$\\$
$\\$
$\\$\color{BrickRed}a\-\ =\-\ 0.95;\color{Green}
$\\$\color{BrickRed}b\-\ =\-\ 0.99;\color{Green}
$\\$\color{BrickRed}[left4,right4]\-\ =\-\ verify(k,d,a,b,j);\color{Green}
$\\$
$\\$
$\\$\color{BrickRed}a\-\ =\-\ 0.989;\color{Green}
$\\$\color{BrickRed}b\-\ =\-\ 1;\color{Green}
$\\$\color{BrickRed}[left5,right5]\-\ =\-\ verify(k,d,a,b,j)\color{Green}
$\\$
$\\$
$\\$\color{BrickRed}\color{NavyBlue}\-\ if\-\ \color{BrickRed}\-\ inf(right1-left2)\-\ $<$=0\color{Green}
$\\$\color{BrickRed}\-\ \-\ \-\ \-\ error('not\-\ verified');\color{Green}
$\\$\color{BrickRed}\color{NavyBlue}\-\ end\-\ \color{BrickRed}\color{Green}
$\\$
$\\$
$\\$\color{BrickRed}\color{NavyBlue}\-\ if\-\ \color{BrickRed}\-\ inf(right2-left3)\-\ $<$=0\color{Green}
$\\$\color{BrickRed}\-\ \-\ \-\ \-\ error('not\-\ verified');\color{Green}
$\\$\color{BrickRed}\color{NavyBlue}\-\ end\-\ \color{BrickRed}\color{Green}
$\\$
$\\$
$\\$\color{BrickRed}\color{NavyBlue}\-\ if\-\ \color{BrickRed}\-\ inf(right3-left4)\-\ $<$=0\color{Green}
$\\$\color{BrickRed}\-\ \-\ \-\ \-\ error('not\-\ verified');\color{Green}
$\\$\color{BrickRed}\color{NavyBlue}\-\ end\-\ \color{BrickRed}\color{Green}
$\\$\color{BrickRed}\-\ \-\ \-\ \-\ \color{Green}
$\\$\color{BrickRed}\color{NavyBlue}\-\ if\-\ \color{BrickRed}\-\ inf(right4-left5)\-\ $<$=0\color{Green}
$\\$\color{BrickRed}\-\ \-\ \-\ \-\ error('not\-\ verified');\color{Green}
$\\$\color{BrickRed}\color{NavyBlue}\-\ end\-\ \color{BrickRed}\color{Green}
$\\$
$\\$
$\\$\color{Green}$\%$\-\ 
$\\$\color{Green}$\%$\-\ verify\-\ on\-\ right
$\\$\color{Green}$\%$
$\\$
$\\$
$\\$\color{Green}$\%$TODO:\-\ Find\-\ exact\-\ bounds
$\\$\color{BrickRed}ntilde\-\ =\-\ 0;\color{Green}
$\\$\color{BrickRed}Nx\-\ =\-\ 250\color{Green}
$\\$\color{BrickRed}err\-\ =\-\ 1e-16\color{Green}
$\\$\color{BrickRed}pie\-\ =\-\ nm('pi');\color{Green}
$\\$\color{BrickRed}elipk\-\ =\-\ elliptic\_integral(k,1);\color{Green}
$\\$\color{BrickRed}elipk2\-\ =\-\ elliptic\_integral(sqrt(1-k.\verb|^|2),1);\color{Green}
$\\$\color{BrickRed}q\-\ =\-\ exp(-pie*elipk2./elipk);\color{Green}
$\\$\color{BrickRed}psi\_end\-\ =\-\ linspace(right5,1,1000);\color{Green}
$\\$\color{BrickRed}psi\_end\-\ =\-\ nm(psi\_end(1:end-1),psi\_end(2:end));\color{Green}
$\\$\color{BrickRed}vals\-\ =\-\ integrand\_numer(Nx,err,q,psi\_end,ntilde);\color{Green}
$\\$\color{BrickRed}kappa\-\ =\-\ kappa\_of\_k(k);\color{Green}
$\\$\color{BrickRed}omega\-\ =\-\ pie./kappa;\color{Green}
$\\$\color{BrickRed}numer\_der\-\ =\-\ vals(:,:,2)+vals(:,:,4)/omega\verb|^|2;\color{Green}
$\\$
$\\$
$\\$\color{BrickRed}\color{NavyBlue}\-\ if\-\ \color{BrickRed}\-\ any(sup(imag(numer\_der))$>$=0)\-\ ==\-\ 1\color{Green}
$\\$\color{BrickRed}\-\ \-\ \-\ \-\ error('not\-\ verified\-\ on\-\ right');\color{Green}
$\\$\color{BrickRed}\color{NavyBlue}\-\ end\-\ \color{BrickRed}\color{Green}
$\\$
$\\$
$\\$\color{Green}$\%$$\%$$\%$$\%$$\%$$\%$$\%$$\%$$\%$$\%$$\%$$\%$$\%$$\%$$\%$$\%$$\%$$\%$$\%$$\%$$\%$$\%$$\%$
$\\$\color{BrickRed}\color{NavyBlue}\-\ function\-\ \color{BrickRed}\-\ \-\ [left,\-\ right]\-\ =\-\ verify(k,d,a,b,j)\color{Green}
$\\$
$\\$
$\\$\color{BrickRed}pie\-\ =\-\ nm('pi');\color{Green}
$\\$
$\\$
$\\$\color{BrickRed}cf\-\ =\-\ d.cf10;\color{Green}
$\\$
$\\$
$\\$\color{Green}$\%$------------------------------------------------------------
$\\$\color{Green}$\%$\-\ plot
$\\$\color{Green}$\%$------------------------------------------------------------
$\\$\color{BrickRed}d.b\_psi\-\ =\-\ 1;\color{Green}
$\\$\color{BrickRed}d.a\_psi\-\ =\-\ 0;\color{Green}
$\\$
$\\$
$\\$\color{BrickRed}lam\_10\_q\-\ =\-\ (2/pie)*log(d.N\_q\_10)+(2/pie)*(nm('0.6')+log(8/pie)+pie/(72*d.N\_q\_10));\color{Green}
$\\$
$\\$
$\\$\color{BrickRed}err\-\ =\-\ d.err\_q\_10+lam\_10\_q*d.err\_psi\_10;\color{Green}
$\\$
$\\$
$\\$\color{BrickRed}psi\-\ =\-\ linspace(a,b,1000);\color{Green}
$\\$\color{BrickRed}psi\-\ =\-\ nm(psi);\color{Green}
$\\$\color{BrickRed}psi\-\ =\-\ nm(psi(1:end-1),psi(2:end));\color{Green}
$\\$
$\\$
$\\$\color{BrickRed}c1\_q\-\ =\-\ 2/(d.b\_q-d.a\_q);\color{Green}
$\\$\color{BrickRed}c2\_q\-\ =\-\ (d.a\_q+d.b\_q)/2;\color{Green}
$\\$\color{BrickRed}c1\_psi\-\ =\-\ 2/(d.b\_psi-d.a\_psi);\color{Green}
$\\$\color{BrickRed}c2\_psi\-\ =\-\ (d.a\_psi+d.b\_psi)/2;\color{Green}
$\\$
$\\$
$\\$\color{BrickRed}kappa\-\ =\-\ kappa\_of\_k(k);\color{Green}
$\\$\color{BrickRed}elipk\-\ =\-\ elliptic\_integral(k,1);\color{Green}
$\\$\color{BrickRed}elipk2\-\ =\-\ elliptic\_integral(sqrt(1-k.\verb|^|2),1);\color{Green}
$\\$\color{BrickRed}q\-\ =\-\ exp(-pie*elipk2./elipk);\color{Green}
$\\$\color{BrickRed}omega\-\ =\-\ pie./kappa;\color{Green}
$\\$\color{BrickRed}q\_tilde\-\ =\-\ c1\_q*(q-c2\_q);\color{Green}
$\\$\color{BrickRed}psi\_tilde\-\ =\-\ c1\_psi*(psi-c2\_psi);\color{Green}
$\\$
$\\$
$\\$\color{BrickRed}omega\-\ =\-\ omega.';\color{Green}
$\\$\color{BrickRed}omega\-\ =\-\ repmat(omega,[1\-\ length(psi)]);\color{Green}
$\\$\color{Green}$\%$$\%$$\%$$\%$$\%$$\%$$\%$$\%$$\%$
$\\$
$\\$
$\\$\color{BrickRed}c0\-\ =\-\ (cf\_eval(cf(:,:,1),q\_tilde,psi\_tilde))+nm(-err,err)+1i*nm(-err,err);\-\ \color{Green}
$\\$\color{BrickRed}c1\-\ =\-\ (cf\_eval(cf(:,:,2),q\_tilde,psi\_tilde))+nm(-err,err)+1i*nm(-err,err);\-\ \color{Green}
$\\$\color{BrickRed}c2\-\ =\-\ (cf\_eval(cf(:,:,3),q\_tilde,psi\_tilde))+nm(-err,err)+1i*nm(-err,err);\-\ \color{Green}
$\\$\color{BrickRed}c3\-\ =\-\ (cf\_eval(cf(:,:,4),q\_tilde,psi\_tilde))+nm(-err,err)+1i*nm(-err,err);\-\ \color{Green}
$\\$\color{BrickRed}h0\-\ =\-\ (cf\_eval(cf(:,:,5),q\_tilde,psi\_tilde))+nm(-err,err)+1i*nm(-err,err);\-\ \color{Green}
$\\$\color{BrickRed}h1\-\ =\-\ (cf\_eval(cf(:,:,6),q\_tilde,psi\_tilde))+nm(-err,err)+1i*nm(-err,err);\-\ \color{Green}
$\\$\color{BrickRed}h2\-\ =\-\ (cf\_eval(cf(:,:,7),q\_tilde,psi\_tilde))+nm(-err,err)+1i*nm(-err,err);\-\ \color{Green}
$\\$\color{BrickRed}h3\-\ =\-\ (cf\_eval(cf(:,:,8),q\_tilde,psi\_tilde))+nm(-err,err)+1i*nm(-err,err);\-\ \color{Green}
$\\$\color{BrickRed}h4\-\ =\-\ (cf\_eval(cf(:,:,9),q\_tilde,psi\_tilde))+nm(-err,err)+1i*nm(-err,err);\-\ \color{Green}
$\\$\color{BrickRed}h5\-\ =\-\ (cf\_eval(cf(:,:,10),q\_tilde,psi\_tilde))+nm(-err,err)+1i*nm(-err,err);\-\ \color{Green}
$\\$
$\\$
$\\$\color{Green}$\%$$\%$$\%$$\%$$\%$$\%$$\%$$\%$$\%$
$\\$
$\\$
$\\$\color{BrickRed}xicon\-\ =\-\ 1i*xi\_q\_psi(q(j),psi,0);\color{Green}
$\\$\color{BrickRed}numer1\-\ =\-\ c0(j,:)\-\ +\-\ c1(j,:).*xicon.\verb|^|1+\-\ c2(j,:).*xicon.\verb|^|2\-\ +\-\ c3(j,:).*xicon.\verb|^|3;\color{Green}
$\\$\color{BrickRed}numer2\-\ =\-\ h0(j,:)\-\ +\-\ h1(j,:).*xicon.\verb|^|1+\-\ h2(j,:).*xicon.\verb|^|2\-\ +\-\ h3(j,:).*xicon.\verb|^|3\-\ +\-\ ...\color{Green}
$\\$\color{BrickRed}\-\ \-\ \-\ \-\ \-\ \-\ \-\ \-\ \-\ \-\ \-\ \-\ \-\ \-\ \-\ \-\ h4(j,:).*xicon.\verb|^|4\-\ +\-\ h5(j,:).*xicon.\verb|^|5;\color{Green}
$\\$
$\\$
$\\$\color{BrickRed}numer\-\ =\-\ numer1\-\ +\-\ numer2./omega(j,:).\verb|^|2;\color{Green}
$\\$
$\\$
$\\$\color{BrickRed}numer\-\ =\-\ imag(numer);\color{Green}
$\\$
$\\$
$\\$\color{BrickRed}\color{NavyBlue}\-\ if\-\ \color{BrickRed}\-\ sum(isnan(numer))$>$0\-\ \color{Green}
$\\$\color{BrickRed}\-\ \-\ \-\ \-\ error('NaN\-\ present');\color{Green}
$\\$\color{BrickRed}\color{NavyBlue}\-\ end\-\ \color{BrickRed}\color{Green}
$\\$
$\\$
$\\$\color{BrickRed}ind1\-\ =\-\ find(inf(numer)$>$0);\color{Green}
$\\$
$\\$
$\\$\color{BrickRed}\color{NavyBlue}\-\ if\-\ \color{BrickRed}\-\ isempty(ind1)\color{Green}
$\\$\color{BrickRed}\-\ \-\ \-\ \-\ error('nothing\-\ verified');\color{Green}
$\\$\color{BrickRed}\color{NavyBlue}\-\ end\-\ \color{BrickRed}\color{Green}
$\\$
$\\$
$\\$\color{BrickRed}bg\-\ =\-\ ind1(1);\color{Green}
$\\$\color{BrickRed}\color{NavyBlue}\-\ for\-\ \color{BrickRed}\-\ jk\-\ =\-\ 0:length(ind1)-1\color{Green}
$\\$\color{BrickRed}\-\ \-\ \-\ \-\ \color{NavyBlue}\-\ if\-\ \color{BrickRed}\-\ norm(bg+jk-ind1(jk+1))\-\ $>$\-\ 0\color{Green}
$\\$\color{BrickRed}\-\ \-\ \-\ \-\ \-\ \-\ \-\ \-\ error('not\-\ consecutive');\color{Green}
$\\$\color{BrickRed}\-\ \-\ \-\ \-\ \color{NavyBlue}\-\ end\-\ \color{BrickRed}\color{Green}
$\\$\color{BrickRed}\color{NavyBlue}\-\ end\-\ \color{BrickRed}\color{Green}
$\\$
$\\$
$\\$\color{BrickRed}left\-\ =\-\ nm(sup(psi(ind1(1))));\color{Green}
$\\$\color{BrickRed}right\-\ =\-\ nm(inf(psi(ind1(end))));\color{Green}
$\\$
$\\$
$\\$\color{Green}$\%$$\%$$\%$$\%$$\%$$\%$$\%$$\%$$\%$$\%$$\%$$\%$$\%$$\%$$\%$$\%$$\%$$\%$$\%$$\%$$\%$$\%$$\%$$\%$$\%$$\%$$\%$$\%$$\%$$\%$$\%$$\%$$\%$$\%$$\%$$\%$$\%$$\%$$\%$$\%$$\%$$\%$$\%$$\%$$\%$$\%$
$\\$
$\\$
$\\$\color{Green}$\%$$\%$$\%$$\%$$\%$$\%$$\%$$\%$$\%$$\%$$\%$$\%$$\%$$\%$$\%$$\%$$\%$$\%$$\%$$\%$$\%$$\%$$\%$
$\\$\color{BrickRed}\color{NavyBlue}\-\ function\-\ \color{BrickRed}\-\ out\-\ =\-\ left\_verify(k,d,b,j)\color{Green}
$\\$
$\\$
$\\$\color{BrickRed}out\-\ =\-\ 0;\color{Green}
$\\$
$\\$
$\\$
$\\$
$\\$\color{BrickRed}pie\-\ =\-\ nm('pi');\color{Green}
$\\$\color{BrickRed}cf\-\ =\-\ d.cf10;\color{Green}
$\\$\color{BrickRed}d.b\_psi\-\ =\-\ 1;\color{Green}
$\\$\color{BrickRed}d.a\_psi\-\ =\-\ 0;\color{Green}
$\\$
$\\$
$\\$\color{BrickRed}lam\_10\_q\-\ =\-\ (2/pie)*log(d.N\_q\_10)+(2/pie)*(nm('0.6')+log(8/pie)+pie/(72*d.N\_q\_10));\color{Green}
$\\$
$\\$
$\\$\color{BrickRed}err\-\ =\-\ d.err\_q\_10+lam\_10\_q*d.err\_psi\_10;\color{Green}
$\\$
$\\$
$\\$\color{BrickRed}psi\-\ =\-\ linspace(0,b,1000);\color{Green}
$\\$\color{BrickRed}psi\-\ =\-\ nm(psi);\color{Green}
$\\$\color{BrickRed}psi\-\ =\-\ nm(psi(1:end-1),psi(2:end));\color{Green}
$\\$
$\\$
$\\$\color{BrickRed}c1\_q\-\ =\-\ 2/(d.b\_q-d.a\_q);\color{Green}
$\\$\color{BrickRed}c2\_q\-\ =\-\ (d.a\_q+d.b\_q)/2;\color{Green}
$\\$\color{BrickRed}c1\_psi\-\ =\-\ 2/(d.b\_psi-d.a\_psi);\color{Green}
$\\$\color{BrickRed}c2\_psi\-\ =\-\ (d.a\_psi+d.b\_psi)/2;\color{Green}
$\\$
$\\$
$\\$\color{BrickRed}kappa\-\ =\-\ kappa\_of\_k(k);\color{Green}
$\\$\color{BrickRed}elipk\-\ =\-\ elliptic\_integral(k,1);\color{Green}
$\\$\color{BrickRed}elipk2\-\ =\-\ elliptic\_integral(sqrt(1-k.\verb|^|2),1);\color{Green}
$\\$\color{BrickRed}q\-\ =\-\ exp(-pie*elipk2./elipk);\color{Green}
$\\$\color{BrickRed}omega\-\ =\-\ pie./kappa;\color{Green}
$\\$\color{BrickRed}q\_tilde\-\ =\-\ c1\_q*(q-c2\_q);\color{Green}
$\\$\color{BrickRed}psi\_tilde\-\ =\-\ c1\_psi*(psi-c2\_psi);\color{Green}
$\\$
$\\$
$\\$\color{BrickRed}omega\-\ =\-\ omega.';\color{Green}
$\\$\color{BrickRed}omega\-\ =\-\ repmat(omega,[1\-\ length(psi)]);\color{Green}
$\\$\color{Green}$\%$$\%$$\%$$\%$$\%$$\%$$\%$$\%$$\%$
$\\$
$\\$
$\\$\color{BrickRed}c0\-\ =\-\ (cf\_eval(cf(:,:,1),q\_tilde,psi\_tilde))+nm(-err,err)+1i*nm(-err,err);\-\ \color{Green}
$\\$\color{BrickRed}c1\-\ =\-\ (cf\_eval(cf(:,:,2),q\_tilde,psi\_tilde))+nm(-err,err)+1i*nm(-err,err);\-\ \color{Green}
$\\$\color{BrickRed}c2\-\ =\-\ (cf\_eval(cf(:,:,3),q\_tilde,psi\_tilde))+nm(-err,err)+1i*nm(-err,err);\-\ \color{Green}
$\\$\color{BrickRed}c3\-\ =\-\ (cf\_eval(cf(:,:,4),q\_tilde,psi\_tilde))+nm(-err,err)+1i*nm(-err,err);\-\ \color{Green}
$\\$\color{BrickRed}h0\-\ =\-\ (cf\_eval(cf(:,:,5),q\_tilde,psi\_tilde))+nm(-err,err)+1i*nm(-err,err);\-\ \color{Green}
$\\$\color{BrickRed}h1\-\ =\-\ (cf\_eval(cf(:,:,6),q\_tilde,psi\_tilde))+nm(-err,err)+1i*nm(-err,err);\-\ \color{Green}
$\\$\color{BrickRed}h2\-\ =\-\ (cf\_eval(cf(:,:,7),q\_tilde,psi\_tilde))+nm(-err,err)+1i*nm(-err,err);\-\ \color{Green}
$\\$\color{BrickRed}h3\-\ =\-\ (cf\_eval(cf(:,:,8),q\_tilde,psi\_tilde))+nm(-err,err)+1i*nm(-err,err);\-\ \color{Green}
$\\$\color{BrickRed}h4\-\ =\-\ (cf\_eval(cf(:,:,9),q\_tilde,psi\_tilde))+nm(-err,err)+1i*nm(-err,err);\-\ \color{Green}
$\\$\color{BrickRed}h5\-\ =\-\ (cf\_eval(cf(:,:,10),q\_tilde,psi\_tilde))+nm(-err,err)+1i*nm(-err,err);\-\ \color{Green}
$\\$
$\\$
$\\$\color{BrickRed}p0\-\ =\-\ max(sup(abs(c0+h0./omega(j,:).\verb|^|2)))\color{Green}
$\\$\color{BrickRed}p1\-\ =\-\ max(sup(abs(c1+h1./omega(j,:).\verb|^|2)))\color{Green}
$\\$\color{BrickRed}p2\-\ =\-\ max(sup(abs(c2+h2./omega(j,:).\verb|^|2)))\color{Green}
$\\$\color{BrickRed}p3\-\ =\-\ max(sup(abs(c3+h3./omega(j,:).\verb|^|2)))\color{Green}
$\\$\color{BrickRed}p4\-\ =\-\ max(sup(abs(h4./omega(j,:).\verb|^|2)))\color{Green}
$\\$\color{BrickRed}p5\-\ =\-\ min(inf(abs(h5./omega(j,:).\verb|^|2)))\color{Green}
$\\$
$\\$
$\\$\color{BrickRed}R\-\ =\-\ nm(1)+(1/nm(p5))*nm(max([p0,p1,p2,p3,p4]));\color{Green}
$\\$\color{BrickRed}R\-\ =\-\ sup(R)\color{Green}
$\\$\color{BrickRed}xicon\-\ =\-\ abs(xi\_q\_psi(q(j),b,0))\color{Green}
$\\$
$\\$
$\\$\color{BrickRed}\color{NavyBlue}\-\ if\-\ \color{BrickRed}\-\ inf(real(xicon))\-\ $>$\-\ R\color{Green}
$\\$\color{BrickRed}\-\ \-\ \-\ \-\ out\-\ =\-\ 1;\color{Green}
$\\$\color{BrickRed}\color{NavyBlue}\-\ end\-\ \color{BrickRed}\color{Green}
$\\$
$\\$
$\\$\color{Black}\section{verify\_data\_n1.m}

\color{Green}\color{BrickRed}clear\-\ all;\-\ close\-\ all;\-\ beep\-\ off;\-\ clc;\-\ curr\_dir\-\ =\-\ cd;\color{Green}
$\\$\color{Green}$\%$$\%$\-\ startup\-\ commands
$\\$\color{BrickRed}cd('..');\color{Green}
$\\$\color{BrickRed}cd('..');\color{Green}
$\\$\color{BrickRed}startup('intlab','','start\-\ matlabpool','off');\color{Green}
$\\$\color{BrickRed}format\-\ long;\color{Green}
$\\$\color{BrickRed}clc;\color{Green}
$\\$\color{BrickRed}cd(curr\_dir);\color{Green}
$\\$\color{Green}$\%$\-\ Spectral\-\ stability\-\ of\-\ periodic\-\ wave\-\ trains\-\ of\-\ the\-\ Korteweg-de
$\\$\color{Green}$\%$\-\ Vries/Kuramoto-Sivashinsky\-\ equation\-\ in\-\ the\-\ Korteweg-de\-\ Vries\-\ limit
$\\$
$\\$
$\\$\color{Green}$\%$\-\ display\-\ type
$\\$\color{BrickRed}intvalinit('DisplayMidRad');\color{Green}
$\\$\color{Green}$\%$\-\ intvalinit('DisplayInfSup');
$\\$\color{BrickRed}pie\-\ =\-\ nm('pi');\color{Green}
$\\$
$\\$
$\\$
$\\$
$\\$\color{BrickRed}file\_name\-\ =\-\ 'd\_31Oct2013';\-\ curr\_dir\-\ =\-\ cd;\color{Green}
$\\$
$\\$
$\\$\color{BrickRed}ld\-\ =\-\ retrieve\_it(curr\_dir,'interval\_arithmetic',file\_name,'data');\color{Green}
$\\$\color{BrickRed}d\-\ =\-\ ld.var;\color{Green}
$\\$
$\\$
$\\$\color{BrickRed}cf\-\ =\-\ d.cfn1;\color{Green}
$\\$
$\\$
$\\$\color{Green}$\%$------------------------------------------------------------
$\\$\color{Green}$\%$\-\ plot
$\\$\color{Green}$\%$------------------------------------------------------------
$\\$\color{BrickRed}d.b\_psi\-\ =\-\ 1;\color{Green}
$\\$\color{BrickRed}d.a\_psi\-\ =\-\ 0;\color{Green}
$\\$
$\\$
$\\$
$\\$
$\\$\color{BrickRed}lam\_n1\_q\-\ =\-\ (2/pie)*log(d.N\_q\_n1)+(2/pie)*(nm('0.6')+log(8/pie)+pie/(72*d.N\_q\_n1));\color{Green}
$\\$
$\\$
$\\$\color{BrickRed}err\-\ =\-\ d.err\_q\_n1+lam\_n1\_q*d.err\_psi\_n1;\color{Green}
$\\$
$\\$
$\\$\color{BrickRed}k\-\ =\-\ nm('0.99');\color{Green}
$\\$
$\\$
$\\$\color{BrickRed}pie\-\ =\-\ nm('pi');\color{Green}
$\\$
$\\$
$\\$\color{BrickRed}num\-\ =\-\ 300;\color{Green}
$\\$\color{BrickRed}psi\_vals\-\ =\-\ linspace(0,1,10*num);\color{Green}
$\\$\color{BrickRed}psi\_vals\-\ =\-\ nm(psi\_vals);\color{Green}
$\\$\color{BrickRed}psi\_vals\-\ =\-\ nm(psi\_vals(1:end-1),psi\_vals(2:end));\color{Green}
$\\$\color{BrickRed}\-\ \color{Green}
$\\$\color{BrickRed}c1\_q\-\ =\-\ 2/(d.b\_q-d.a\_q);\color{Green}
$\\$\color{BrickRed}c2\_q\-\ =\-\ (d.a\_q+d.b\_q)/2;\color{Green}
$\\$\color{BrickRed}c1\_psi\-\ =\-\ 2/(d.b\_psi-d.a\_psi);\color{Green}
$\\$\color{BrickRed}c2\_psi\-\ =\-\ (d.a\_psi+d.b\_psi)/2;\color{Green}
$\\$
$\\$
$\\$\color{BrickRed}kappa\-\ =\-\ kappa\_of\_k(k);\color{Green}
$\\$\color{BrickRed}elipk\-\ =\-\ elliptic\_integral(k,1);\color{Green}
$\\$\color{BrickRed}elipk2\-\ =\-\ elliptic\_integral(sqrt(1-k.\verb|^|2),1);\color{Green}
$\\$\color{BrickRed}q\-\ =\-\ exp(-pie*elipk2./elipk);\color{Green}
$\\$\color{BrickRed}omega\-\ =\-\ pie./kappa;\color{Green}
$\\$\color{BrickRed}q\_tilde\-\ =\-\ c1\_q*(q-c2\_q);\color{Green}
$\\$\color{BrickRed}psi\_tilde\-\ =\-\ c1\_psi*(psi\_vals-c2\_psi);\color{Green}
$\\$
$\\$
$\\$\color{Green}$\%$\-\ get\-\ f1\-\ interpolated\-\ value
$\\$\color{BrickRed}f1\-\ =\-\ imag(cf\_eval(cf(:,:,1),q\_tilde,psi\_tilde))+nm(-err,err)+1i*nm(-err,err);\-\ \color{Green}
$\\$\color{Green}$\%$\-\ get\-\ f1\_psi\-\ interpolated\-\ value
$\\$\color{BrickRed}f1\_psi\-\ =\-\ (cf\_eval(cf(:,:,2),q\_tilde,psi\_tilde))+nm(-err,err)+1i*nm(-err,err);\-\ \color{Green}
$\\$\color{Green}$\%$\-\ get\-\ f2\-\ interpolated\-\ value
$\\$\color{BrickRed}f2\-\ =\-\ imag(cf\_eval(cf(:,:,3),q\_tilde,psi\_tilde))+nm(-err,err)+1i*nm(-err,err);\-\ \color{Green}
$\\$\color{Green}$\%$\-\ get\-\ f2\_psi\-\ interpolated\-\ value
$\\$\color{BrickRed}f2\_psi\-\ =\-\ (cf\_eval(cf(:,:,4),q\_tilde,psi\_tilde))+nm(-err,err)+1i*nm(-err,err);\-\ \color{Green}
$\\$\color{Green}$\%$\-\ get\-\ g\-\ interpolated\-\ value
$\\$\color{BrickRed}g\-\ =\-\ imag(cf\_eval(cf(:,:,5),q\_tilde,psi\_tilde))+nm(-err,err)+1i*nm(-err,err);\-\ \color{Green}
$\\$\color{Green}$\%$\-\ get\-\ g\_psi\-\ interpolated\-\ value
$\\$\color{BrickRed}g\_psi\-\ =\-\ (cf\_eval(cf(:,:,6),q\_tilde,psi\_tilde))+nm(-err,err)+1i*nm(-err,err);\-\ \color{Green}
$\\$
$\\$
$\\$\color{BrickRed}g\-\ =\-\ real(g);\color{Green}
$\\$\color{BrickRed}g\_psi\-\ =\-\ imag(g\_psi);\color{Green}
$\\$
$\\$
$\\$\color{BrickRed}omega\-\ =\-\ omega.';\color{Green}
$\\$\color{BrickRed}omega\-\ =\-\ repmat(omega,[1\-\ length(psi\_vals)]);\color{Green}
$\\$
$\\$
$\\$\color{BrickRed}numer\-\ =\-\ real(f1+f2./omega.\verb|^|2);\color{Green}
$\\$\color{BrickRed}numer\_psi\-\ =\-\ imag(f1\_psi\-\ +\-\ f2\_psi./omega.\verb|^|2);\color{Green}
$\\$\color{BrickRed}\-\ \-\ \-\ \-\ \color{Green}
$\\$\color{Green}$\%$$\%$$\%$$\%$$\%$$\%$$\%$$\%$$\%$$\%$$\%$$\%$$\%$$\%$
$\\$
$\\$
$\\$\color{BrickRed}\color{NavyBlue}\-\ for\-\ \color{BrickRed}\-\ j\-\ =\-\ 1:length(q)\color{Green}
$\\$\color{BrickRed}\-\ \-\ \-\ \-\ \color{Green}
$\\$\color{BrickRed}\-\ \-\ \-\ \-\ numer\_psi\_left\_min\-\ =\-\ min(\-\ inf(numer\_psi(1:num)))\color{Green}
$\\$\color{BrickRed}\-\ \-\ \-\ \-\ numer\_mid\_min\-\ =\-\ min(inf(numer(j,num+1:9*num)))\color{Green}
$\\$\color{BrickRed}\-\ \-\ \-\ \-\ numer\_psi\_right\_max\-\ =\-\ max(sup(numer\_psi(9*num+1:10*num-1)))\color{Green}
$\\$\color{BrickRed}\-\ \-\ \-\ \-\ \color{Green}
$\\$\color{BrickRed}\-\ \-\ \-\ \-\ \color{NavyBlue}\-\ if\-\ \color{BrickRed}\-\ any(inf(numer\_psi(1:num))$<$=\-\ 0)\color{Green}
$\\$\color{BrickRed}\-\ \-\ \-\ \-\ \-\ \-\ \-\ \-\ error('failed\-\ to\-\ verify');\color{Green}
$\\$\color{BrickRed}\-\ \-\ \-\ \-\ \color{NavyBlue}\-\ end\-\ \color{BrickRed}\color{Green}
$\\$\color{BrickRed}\-\ \-\ \-\ \-\ \color{NavyBlue}\-\ if\-\ \color{BrickRed}\-\ any(inf(numer(j,num+1:9*num))$<$=\-\ 0)\color{Green}
$\\$\color{BrickRed}\-\ \-\ \-\ \-\ \-\ \-\ \-\ \-\ error('failed\-\ to\-\ verify');\color{Green}
$\\$\color{BrickRed}\-\ \-\ \-\ \-\ \color{NavyBlue}\-\ end\-\ \color{BrickRed}\color{Green}
$\\$\color{BrickRed}\-\ \-\ \-\ \-\ \color{NavyBlue}\-\ if\-\ \color{BrickRed}\-\ any(sup(numer\_psi(9*num+1:10*num-1))$>$=0)\color{Green}
$\\$\color{BrickRed}\-\ \-\ \-\ \-\ \-\ \-\ \-\ \-\ error('failed\-\ to\-\ verify');\color{Green}
$\\$\color{BrickRed}\-\ \-\ \-\ \-\ \color{NavyBlue}\-\ end\-\ \color{BrickRed}\color{Green}
$\\$\color{BrickRed}\color{NavyBlue}\-\ end\-\ \color{BrickRed}\color{Green}
$\\$
$\\$
$\\$\color{BrickRed}\color{NavyBlue}\-\ for\-\ \color{BrickRed}\-\ j\-\ =\-\ 1:length(q)\color{Green}
$\\$\color{BrickRed}\-\ \-\ \-\ \-\ \color{Green}
$\\$\color{BrickRed}\-\ \-\ \-\ \-\ g\_psi\_left\_max\-\ =\-\ max((sup(g\_psi(1:num))))\color{Green}
$\\$\color{BrickRed}\-\ \-\ \-\ \-\ g\_mid\_max\-\ =\-\ max(sup(g(j,num+1:9*num)))\color{Green}
$\\$\color{BrickRed}\-\ \-\ \-\ \-\ g\_psi\_right\_min\-\ =\-\ min(inf(g\_psi(9*num+1:10*num-1)))\color{Green}
$\\$\color{BrickRed}\-\ \-\ \-\ \-\ \color{Green}
$\\$\color{BrickRed}\-\ \-\ \-\ \-\ \color{NavyBlue}\-\ if\-\ \color{BrickRed}\-\ any(sup(g\_psi(1:num))$>$=\-\ 0)\color{Green}
$\\$\color{BrickRed}\-\ \-\ \-\ \-\ \-\ \-\ \-\ \-\ error('failed\-\ to\-\ verify');\color{Green}
$\\$\color{BrickRed}\-\ \-\ \-\ \-\ \color{NavyBlue}\-\ end\-\ \color{BrickRed}\color{Green}
$\\$\color{BrickRed}\-\ \-\ \-\ \-\ \color{NavyBlue}\-\ if\-\ \color{BrickRed}\-\ any(sup(g(j,num+1:9*num))$>$=\-\ 0)\color{Green}
$\\$\color{BrickRed}\-\ \-\ \-\ \-\ \-\ \-\ \-\ \-\ error('failed\-\ to\-\ verify');\color{Green}
$\\$\color{BrickRed}\-\ \-\ \-\ \-\ \color{NavyBlue}\-\ end\-\ \color{BrickRed}\color{Green}
$\\$\color{BrickRed}\-\ \-\ \-\ \-\ \color{NavyBlue}\-\ if\-\ \color{BrickRed}\-\ any(inf(g\_psi(9*num+1:10*num-1))$<$=0)\color{Green}
$\\$\color{BrickRed}\-\ \-\ \-\ \-\ \-\ \-\ \-\ \-\ error('failed\-\ to\-\ verify');\color{Green}
$\\$\color{BrickRed}\-\ \-\ \-\ \-\ \color{NavyBlue}\-\ end\-\ \color{BrickRed}\color{Green}
$\\$\color{BrickRed}\color{NavyBlue}\-\ end\-\ \color{BrickRed}\color{Green}
$\\$
$\\$
$\\$
$\\$
$\\$
$\\$
$\\$
$\\$
$\\$
$\\$
$\\$\color{Black}\section{xi\_der\_q\_psi.m}

\color{Green}\color{BrickRed}\color{NavyBlue}\-\ function\-\ \color{BrickRed}\-\ out\-\ =\-\ xi\_der\_q\_psi(q,psi,ntilde)\color{Green}
$\\$\color{Green}$\%$\-\ function\-\ out\-\ =\-\ xi\_der\_q\_psi(q,psi,ntilde)
$\\$\color{Green}$\%$
$\\$\color{Green}$\%$\-\ returns\-\ the\-\ derivative\-\ of\-\ \-\ omega*xi\-\ with\-\ respect\-\ to\-\ psi
$\\$
$\\$
$\\$
$\\$\color{Black}
Now 
\eq{
\pd{}{\psi}\omega \xi ( \omega + i\psi \omega')&= 2\omega \omega' \left(\left(\frac{\pi}{2\omega}\right)^2\sec^2\left(\frac{i\pi \psi \omega'}{2\omega}\right)-\frac{\pi^2}{\omega^2} \sum_{k=1}^{\infty}(-1)^k \frac{kq^{2k}}{1-q^{2k}}\left(q^{\psi k}+q^{-\psi k}\right) \right)\\
&= \frac{-\pi \log(q)}{2}\sec^2\left(\frac{\psi \log(q)i}{2}\right) + 2\pi \log(q)\sum_{k=1}^{\infty} (-1)^k \frac{kq^{2k}}{1-q^{2k}}\left(q^{\psi k}+q^{-\psi k}\right)
}{\notag}

$\\$
Note that 

$\\$
\eq{
\left|\sum_{k=N+1}^{\infty} (-1)^k \frac{kq^{2k}}{1-q^{2k}}\left(q^{\psi k}+q^{-\psi k}\right)\right|&\leq \sum_{k=N+1}^{\infty} \frac{kq^{2k}}{1-q^{2k}}\left(q^{\psi k}+q^{-\psi k}\right)\\
&\leq \frac{2}{1-q^{2(N+1)}}\sum_{k=0}^{\infty} (N+1+k)q^{(2-\psi)(N+1+k)}\\
&\leq \frac{2(N+1)q^{(2-\psi)(N+1)}}{1-q^{2(N+1)}}\sum_{k=0}^{\infty}q^{(2-\psi)k} + \frac{2q^{(2-\psi)N}}{1-q^{2(N+1)}}\sum_{k=0}^{\infty} k q^{(2-\psi)k}\\
&\leq  \frac{2(N+1)q^{(2-\psi)(N+1)}}{1-q^{2(N+1)}}\frac{1}{1-q^{(2-\psi)}} + \frac{2q^{(2-\psi)N}}{1-q^{2(N+1)}}\frac{q^{(2-\psi)}}{(1-q^{(2-\psi)})^2}
}{\notag}
\color{Green}
$\\$
$\\$\color{Green}$\%$$\%$
$\\$
$\\$
$\\$\color{Green}$\%$
$\\$\color{Green}$\%$\-\ constants
$\\$\color{Green}$\%$
$\\$
$\\$
$\\$\color{BrickRed}pie\-\ =\-\ nm('pi');\color{Green}
$\\$\color{BrickRed}psi0\-\ =\-\ max(sup(psi));\color{Green}
$\\$\color{BrickRed}q0\-\ =\-\ abs(q\verb|^|(2-psi0));\color{Green}
$\\$\color{BrickRed}qs0\-\ =\-\ q0\verb|^|(2-sup(abs(psi0)));\color{Green}
$\\$\color{BrickRed}q02\-\ =\-\ q0*q0;\color{Green}
$\\$
$\\$
$\\$\color{Green}$\%$\-\ 
$\\$\color{Green}$\%$\-\ find\-\ N\-\ for\-\ given\-\ error
$\\$\color{Green}$\%$
$\\$
$\\$
$\\$\color{BrickRed}N\-\ =\-\ 0;\color{Green}
$\\$\color{BrickRed}maxit\-\ =\-\ 1000;\color{Green}
$\\$\color{BrickRed}tol\-\ =\-\ 1e-17;\color{Green}
$\\$\color{BrickRed}err\-\ =\-\ tol\-\ +\-\ 1;\color{Green}
$\\$\color{BrickRed}\color{NavyBlue}\-\ while\-\ \color{BrickRed}\-\ err\-\ $>$\-\ tol\color{Green}
$\\$\color{BrickRed}\-\ \-\ \-\ \-\ N\-\ =\-\ N+1;\color{Green}
$\\$\color{BrickRed}\-\ \-\ \-\ \-\ \color{NavyBlue}\-\ if\-\ \color{BrickRed}\-\ N\-\ $>$\-\ maxit\color{Green}
$\\$\color{BrickRed}\-\ \-\ \-\ \-\ \-\ \-\ \-\ \-\ error('Maximum\-\ iterations\-\ exceeded');\color{Green}
$\\$\color{BrickRed}\-\ \-\ \-\ \-\ \color{NavyBlue}\-\ end\-\ \color{BrickRed}\color{Green}
$\\$\color{BrickRed}\-\ \-\ \-\ \-\ temp\-\ =\-\ 2*(N+1)*qs0\verb|^|(N+1)/((1-q02\verb|^|(N+1))*(1-qs0))\-\ +\-\ 2*qs0\verb|^|(N+1)/((1-q02\verb|^|(N+1))*(1-qs0)\verb|^|2);\color{Green}
$\\$\color{BrickRed}\-\ \-\ \-\ \-\ err\-\ =\-\ sup(temp);\color{Green}
$\\$\color{BrickRed}\color{NavyBlue}\-\ end\-\ \color{BrickRed}\color{Green}
$\\$
$\\$
$\\$\color{Green}$\%$\-\ initialize\-\ output
$\\$\color{BrickRed}out\-\ =\-\ nm(zeros(1,length(psi)));\color{Green}
$\\$
$\\$
$\\$\color{Green}$\%$\-\ find\-\ partial\-\ sum
$\\$\color{BrickRed}q2\-\ =\-\ q*q;\color{Green}
$\\$\color{BrickRed}\color{NavyBlue}\-\ for\-\ \color{BrickRed}\-\ k\-\ =\-\ 1:N\color{Green}
$\\$\color{BrickRed}\-\ \-\ \-\ \-\ q2k\-\ =\-\ q2\verb|^|k;\color{Green}
$\\$\color{BrickRed}\-\ \-\ \-\ \-\ out\-\ =\-\ out\-\ +\-\ ((-1)\verb|^|ntilde)\verb|^|k*(k*q2k/(1-q2k))*(q.\verb|^|(k*psi)+q.\verb|^|(-k*psi));\color{Green}
$\\$\color{BrickRed}\color{NavyBlue}\-\ end\-\ \color{BrickRed}\color{Green}
$\\$
$\\$
$\\$\color{Green}$\%$\-\ add\-\ error
$\\$\color{BrickRed}out\-\ =\-\ out\-\ +\-\ nm(-err,err)+1i*nm(-err,err);\color{Green}
$\\$
$\\$
$\\$\color{Green}$\%$\-\ add\-\ non\-\ infinite\-\ sum\-\ parts
$\\$\color{BrickRed}\color{NavyBlue}\-\ if\-\ \color{BrickRed}\-\ ntilde\-\ ==\-\ 0\color{Green}
$\\$\color{BrickRed}\-\ \-\ \-\ \-\ out\-\ =\-\ (-pie*log(q)/2)./(sin(1i*log(q)*psi/2).\verb|^|2)\-\ +\-\ 2*pie*log(q)*out;\color{Green}
$\\$\color{BrickRed}\color{NavyBlue}\-\ elseif\-\ \color{BrickRed}\-\ ntilde\-\ ==\-\ 1\color{Green}
$\\$\color{BrickRed}\-\ \-\ \-\ \-\ out\-\ =\-\ (-pie*log(q)/2)./(cos(1i*log(q)*psi/2).\verb|^|2)\-\ +\-\ 2*pie*log(q)*out;\color{Green}
$\\$\color{BrickRed}\color{NavyBlue}\-\ end\-\ \color{BrickRed}\color{Green}
$\\$
$\\$
$\\$
$\\$
$\\$
$\\$
$\\$
$\\$
$\\$
$\\$
$\\$
$\\$
$\\$
$\\$
$\\$
$\\$
$\\$
$\\$
$\\$
$\\$
$\\$
$\\$
$\\$
$\\$
$\\$
$\\$
$\\$
$\\$
$\\$
$\\$
$\\$
$\\$
$\\$
$\\$
$\\$
$\\$
$\\$
$\\$
$\\$
$\\$
$\\$
$\\$
$\\$
$\\$
$\\$\color{Black}\section{xi\_q\_psi.m}

\color{Green}\color{BrickRed}\color{NavyBlue}\-\ function\-\ \color{BrickRed}\-\ \-\ out\-\ =\-\ xi\_q\_psi(q,psi,ntilde)\color{Green}
$\\$\color{Green}$\%$\-\ computes\-\ omega*xi(ntilde*omega\-\ +\-\ 1i*psi*omega\_prime)
$\\$
$\\$
$\\$
$\\$\color{Black}
From 
\eq{
\vartheta_1(z) = 2\sum_{n=1}^{\infty} (-1)^{n+1}q^{(n-1/2)^2}\sin((2n-1)z),
}{\notag}

$\\$
we find that
\eq{
f(x)&:= \vartheta_1\left(\frac{\pi}{2\omega}(\omega x + i\omega' + \tilde n \omega + i\psi \omega')\right)
\\ &= -i\sum_{n=1}^{\infty}(-1)^{n+1}q^{(n-1/2)^2} \left(\hat v^{(2n-1)}-\hat v^{-(2n+1)}\right),
}{\notag}
where $\hat v := e^{i\pi(x+\tilde n)/2}q^{(1+\psi)/2}$. Hence

$\\$
\eq{
\pd{^m}{x^m} f(x)&=  -i(i\pi/2)^m\sum_{n=1}^{\infty}(-1)^{n+1}q^{(n-1/2)^2} (2n-1)\left(\hat v^{(2n-1)}-\hat v^{-(2n+1)}\right).
}{\notag}

$\\$
We find that
\eq{
Err &:= \left|-i(i\pi/2)^m\sum_{n=N+1}^{\infty}(-1)^{n+1}q^{(n-1/2)^2} (2n-1)\left(\hat v^{(2n-1)}-\hat v^{-(2n+1)}\right)\right| 
\\ &\leq 2q^{1/4}(\pi/2)^m\sum_{n = N+1}^{\infty}q^{n^2/2}\left((2n-1)^mq^{n^2/2-n-(2n-1)(1+\psi)/2}\right)\\
&\leq 2q^{1/4}(\pi/2)^m\sum_{n = N+1}^{\infty}q^{n^2/2}\\
&\leq2q^{1/4}(\pi/2)^mq^{(N+1)^2/2}\frac{1}{1-q}, 
}{\notag}
as long as we take $N$ large enough so that $g(x):= \left((2x-1)^mq^{n^2/2-x-(2x-1)(1+\psi)/2}\right)$ satisfies $g(N)< 1$, $g(x) < 0$ whenever $x> N$.

$\\$
\color{Green}
$\\$
$\\$\color{Green}$\%$$\%$
$\\$
$\\$
$\\$\color{Green}$\%$
$\\$\color{Green}$\%$\-\ constants
$\\$\color{Green}$\%$
$\\$
$\\$
$\\$\color{BrickRed}pie\-\ =\-\ nm('pi');\color{Green}
$\\$\color{BrickRed}psi0\-\ =\-\ max(sup(psi));\color{Green}
$\\$\color{BrickRed}q0\-\ =\-\ abs(q\verb|^|(2-psi0));\color{Green}
$\\$
$\\$
$\\$\color{Green}$\%$\-\ 
$\\$\color{Green}$\%$\-\ find\-\ N\-\ for\-\ given\-\ error
$\\$\color{Green}$\%$
$\\$
$\\$
$\\$\color{BrickRed}N\-\ =\-\ 0;\color{Green}
$\\$\color{BrickRed}maxit\-\ =\-\ 1000;\color{Green}
$\\$\color{BrickRed}tol\-\ =\-\ 1e-17;\color{Green}
$\\$\color{BrickRed}err\-\ =\-\ tol\-\ +\-\ 1;\color{Green}
$\\$\color{BrickRed}\color{NavyBlue}\-\ while\-\ \color{BrickRed}\-\ err\-\ $>$\-\ tol\color{Green}
$\\$\color{BrickRed}\-\ \-\ \-\ \-\ N\-\ =\-\ N+1;\color{Green}
$\\$\color{BrickRed}\-\ \-\ \-\ \-\ \color{NavyBlue}\-\ if\-\ \color{BrickRed}\-\ N\-\ $>$\-\ maxit\color{Green}
$\\$\color{BrickRed}\-\ \-\ \-\ \-\ \-\ \-\ \-\ \-\ error('Maximum\-\ iterations\-\ exceeded');\color{Green}
$\\$\color{BrickRed}\-\ \-\ \-\ \-\ \color{NavyBlue}\-\ end\-\ \color{BrickRed}\color{Green}
$\\$\color{BrickRed}\-\ \-\ \-\ \-\ temp\-\ =\-\ q0\verb|^|(N+1)/((1-q\verb|^|(2*(N+1)))*(1-q0));\color{Green}
$\\$\color{BrickRed}\-\ \-\ \-\ \-\ err\-\ =\-\ sup(temp);\color{Green}
$\\$\color{BrickRed}\color{NavyBlue}\-\ end\-\ \color{BrickRed}\color{Green}
$\\$
$\\$
$\\$\color{Green}$\%$\-\ initialize\-\ output
$\\$\color{BrickRed}out\-\ =\-\ nm(zeros(1,length(psi)));\color{Green}
$\\$
$\\$
$\\$\color{Green}$\%$\-\ add\-\ partial\-\ sum
$\\$\color{BrickRed}q2\-\ =\-\ q*q;\color{Green}
$\\$\color{BrickRed}logq\-\ =\-\ log(q);\color{Green}
$\\$\color{BrickRed}\color{NavyBlue}\-\ for\-\ \color{BrickRed}\-\ k\-\ =\-\ 1:N\color{Green}
$\\$\color{BrickRed}\-\ \-\ \-\ \-\ q2k\-\ =\-\ q2\verb|^|k;\color{Green}
$\\$\color{BrickRed}\-\ \-\ \-\ \-\ out\-\ =\-\ out\-\ +\-\ (q2k/(1-q2k))*sin(k*(pie*ntilde-1i*psi*logq));\color{Green}
$\\$\color{BrickRed}\color{NavyBlue}\-\ end\-\ \color{BrickRed}\color{Green}
$\\$
$\\$
$\\$\color{Green}$\%$\-\ add\-\ error
$\\$\color{BrickRed}out\-\ =\-\ out\-\ +\-\ nm(-err,err)\-\ +\-\ 1i*nm(-err,err);\color{Green}
$\\$
$\\$
$\\$\color{Green}$\%$\-\ multiply\-\ by\-\ constants
$\\$\color{BrickRed}out\-\ =\-\ 2*pi*out;\color{Green}
$\\$
$\\$
$\\$\color{Green}$\%$\-\ add\-\ non\-\ infinite\-\ sum\-\ parts
$\\$\color{BrickRed}out\-\ =\-\ out\-\ +\-\ (pie/2)*cot(pie*ntilde/2-1i*psi*logq/2);\color{Green}
$\\$
$\\$
$\\$\color{Green}$\%$\-\ multiply\-\ by\-\ constants\-\ to\-\ get\-\ omega*xi
$\\$\color{BrickRed}out\-\ =\-\ 2*1i*out;\color{Green}
$\\$
$\\$
$\\$
$\\$
$\\$
$\\$
$\\$
$\\$
$\\$
$\\$
$\\$
$\\$
$\\$
$\\$
$\\$
$\\$
$\\$
$\\$
$\\$
$\\$
$\\$
$\\$
$\\$

%% file: interval_arithmetic_all.tex
\color{Black}\chapter{Full study}

\color{Black}\section{N\_nodes.m}

\color{Green}\color{BrickRed}\color{NavyBlue}\-\ function\-\ \color{BrickRed}\-\ [N,err]\-\ =\-\ N\_nodes(rho,M,abs\_tol)\color{Green}
$\\$\color{Green}$\%$\-\ N\-\ =\-\ N\_nodes(rho,M,abs\_tol)
$\\$
$\\$\color{Black}
 Determine the number of Chebyshev nodes, N, needed to interpolate
 the analytic functon $f(z)$ on the interval $z \in [-1,1]$ with absolute
 interpolation error no more than abs tol, where $|f(z)|\leq M$ for 
 $z = \frac{1}{2}(\rho e^{i\theta}+e^{-i\theta}/\rho)$, with
 $\theta \in [0,2\pi]$.

$\\$
 $L_{\rho}$ is the length of the ellipse $E_{\rho}$ on which the bound M
 is computed. A bound is given by $L_{\rho} \leq \pi \sqrt{rho^2+1/rho^2}$.
 The distantce $D_{\rho}$ between the ellipse $E_{\rho}$ and the line $[-1,1]$
 is no smaller than $(rho+1/rho)/2-1$. An interpolation bound is given by,
 $|f(x)-p(x)| \leq M L_{\rho} /(2\pi D_{\rho} sinh(\eta) sinh(\eta N))$
 where $\eta = \log{\rho}$.
\color{Green}
$\\$
$\\$
$\\$
$\\$\color{BrickRed}eta\-\ =\-\ log(rho);\color{Green}
$\\$\color{BrickRed}denom\-\ =\-\ ((rho+1/rho)/2-1);\color{Green}
$\\$\color{BrickRed}bound\_constant\-\ =\-\ M*sqrt(rho\verb|^|2+1/rho\verb|^|2)/denom;\color{Green}
$\\$
$\\$
$\\$\color{BrickRed}N\-\ =\-\ 3;\color{Green}
$\\$\color{BrickRed}err\-\ =\-\ bound\_constant/sinh(eta*(N+1));\color{Green}
$\\$\color{BrickRed}\color{NavyBlue}\-\ while\-\ \color{BrickRed}\-\ err\-\ $>$\-\ 0.1*abs\_tol\color{Green}
$\\$\color{BrickRed}\-\ \-\ \-\ \-\ N\-\ =\-\ N+1;\color{Green}
$\\$\color{BrickRed}\-\ \-\ \-\ \-\ err\-\ =\-\ bound\_constant/sinh(eta*(N+1));\color{Green}
$\\$\color{BrickRed}\color{NavyBlue}\-\ end\-\ \color{BrickRed}\color{Green}
$\\$
$\\$
$\\$
$\\$
$\\$\color{Black}\section{agm.m}

\color{Green}\color{BrickRed}\color{NavyBlue}\-\ function\-\ \color{BrickRed}\-\ cnew\-\ =\-\ agm(a,b)\color{Green}
$\\$\color{Green}$\%$\-\ function\-\ out\-\ =\-\ agm(a,b)
$\\$\color{Green}$\%$
$\\$\color{Green}$\%$\-\ Evaluate\-\ the\-\ arithmetic\-\ geometric\-\ mean\-\ of\-\ a\-\ and\-\ b\-\ 
$\\$\color{Green}$\%$\-\ using\-\ interval\-\ arithmetic.
$\\$\color{Green}$\%$\-\ 
$\\$\color{Green}$\%$\-\ Input\-\ should\-\ be\-\ non-negative:\-\ a\-\ $>$=\-\ 0\-\ and\-\ \-\ b\-\ $>$=\-\ 0.
$\\$
$\\$
$\\$\color{Green}$\%$
$\\$\color{Green}$\%$\-\ check\-\ for\-\ user\-\ error
$\\$\color{Green}$\%$
$\\$
$\\$
$\\$\color{Green}$\%$\-\ check\-\ that\-\ input\-\ is\-\ real
$\\$\color{BrickRed}\color{NavyBlue}\-\ if\-\ \color{BrickRed}\-\ max(sup(abs(imag(a))))\-\ $>$\-\ 0\color{Green}
$\\$\color{BrickRed}\-\ \-\ \-\ \-\ error('\-\ a\-\ must\-\ be\-\ real');\color{Green}
$\\$\color{BrickRed}\color{NavyBlue}\-\ end\-\ \color{BrickRed}\color{Green}
$\\$
$\\$
$\\$\color{BrickRed}\color{NavyBlue}\-\ if\-\ \color{BrickRed}\-\ max(sup(abs(imag(b))))\-\ $>$\-\ 0\color{Green}
$\\$\color{BrickRed}\-\ \-\ \-\ \-\ error('\-\ b\-\ must\-\ be\-\ real');\color{Green}
$\\$\color{BrickRed}\color{NavyBlue}\-\ end\-\ \color{BrickRed}\color{Green}
$\\$
$\\$
$\\$\color{Green}$\%$\-\ check\-\ that\-\ input\-\ is\-\ non-negative
$\\$\color{BrickRed}\color{NavyBlue}\-\ if\-\ \color{BrickRed}\-\ min(inf(a))\-\ $<$\-\ 0\color{Green}
$\\$\color{BrickRed}\-\ \-\ \-\ \-\ error('\-\ a\-\ $>$=\-\ 0\-\ required');\color{Green}
$\\$\color{BrickRed}\color{NavyBlue}\-\ end\-\ \color{BrickRed}\color{Green}
$\\$
$\\$
$\\$\color{BrickRed}\color{NavyBlue}\-\ if\-\ \color{BrickRed}\-\ min(inf(b))\-\ $<$\-\ 0\color{Green}
$\\$\color{BrickRed}\-\ \-\ \-\ \-\ error('\-\ b\-\ $>$=\-\ 0\-\ required');\color{Green}
$\\$\color{BrickRed}\color{NavyBlue}\-\ end\-\ \color{BrickRed}\color{Green}
$\\$
$\\$
$\\$\color{Green}$\%$\-\ make\-\ initial\-\ interval\-\ containing\-\ AGM(a,b)
$\\$\color{BrickRed}cold\-\ =\-\ hull(nm(a/2,b/2),nm(2*a,2*b));\color{Green}
$\\$\color{BrickRed}cnew\-\ =\-\ hull(a,b);\color{Green}
$\\$
$\\$
$\\$\color{Green}$\%$\-\ iterate\-\ until\-\ agm\-\ has\-\ converged
$\\$\color{BrickRed}\color{NavyBlue}\-\ while\-\ \color{BrickRed}\-\ (max(sup(cold))-min(inf(cold)))-(max(sup(cnew))-min(inf(cnew)))\-\ $>$\-\ 0\color{Green}
$\\$\color{BrickRed}\-\ \-\ \-\ \-\ \color{Green}
$\\$\color{BrickRed}\-\ \-\ \-\ \-\ cold\-\ =\-\ cnew;\color{Green}
$\\$\color{BrickRed}\-\ \-\ \-\ \-\ atemp\-\ =\-\ (a+b)/2;\color{Green}
$\\$\color{BrickRed}\-\ \-\ \-\ \-\ btemp\-\ =\-\ sqrt(a.*b);\color{Green}
$\\$\color{BrickRed}\-\ \-\ \-\ \-\ a\-\ =\-\ atemp;\color{Green}
$\\$\color{BrickRed}\-\ \-\ \-\ \-\ b\-\ =\-\ btemp;\color{Green}
$\\$\color{BrickRed}\-\ \-\ \-\ \-\ cnew\-\ =\-\ hull(a,b);\color{Green}
$\\$\color{BrickRed}\-\ \-\ \-\ \-\ \color{Green}
$\\$\color{BrickRed}\color{NavyBlue}\-\ end\-\ \color{BrickRed}\color{Green}
$\\$
$\\$
$\\$
$\\$
$\\$
$\\$
$\\$
$\\$
$\\$\color{Black}\section{bound\_numer.m}

\color{Green}\color{BrickRed}\color{NavyBlue}\-\ function\-\ \color{BrickRed}\-\ [M\_psi,M\_x,M\_q]\-\ =\-\ bound\_numer(dm,rho\_psi,rho\_x,rho\_q,a\_q,b\_q,a\_psi,b\_psi,ntilde)\color{Green}
$\\$
$\\$
$\\$\color{Green}$\%$
$\\$\color{Green}$\%$\-\ constants\-\ 
$\\$\color{Green}$\%$
$\\$
$\\$
$\\$\color{BrickRed}min\_abs\_q\-\ =\-\ (a\_q+b\_q)/2\-\ -(b\_q-a\_q)*(rho\_q+1/rho\_q)/4;\color{Green}
$\\$
$\\$
$\\$\color{Green}$\%$
$\\$\color{Green}$\%$\-\ bound\-\ for\-\ interpolation\-\ in\-\ psi\-\ ------------------------------------------------------------
$\\$\color{Green}$\%$
$\\$
$\\$
$\\$\color{Green}$\%$\-\ top\-\ of\-\ ellipse\-\ \color{Black}$E_{\rho_{\psi}}$ \color{Green}
$\\$\color{BrickRed}max\_imag\_x\-\ =\-\ 0;\color{Green}
$\\$\color{BrickRed}max\_real\_psi\-\ =\-\ (1+(rho\_psi+1/rho\_psi)/2)/2;\color{Green}
$\\$\color{BrickRed}max\_abs\_q\-\ =\-\ sup(real(b\_q));\color{Green}
$\\$\color{BrickRed}prodq\-\ =\-\ abs(log(a\_q)/2);\color{Green}
$\\$
$\\$
$\\$\color{BrickRed}max\_xi\-\ =\-\ bound\_xi(a\_psi,b\_psi,rho\_psi,a\_q,b\_q,1,ntilde);\color{Green}
$\\$\color{BrickRed}max\_xi\_der\-\ =\-\ bound\_xi\_der...\color{Green}
$\\$\color{BrickRed}\-\ \-\ \-\ \-\ (max\_abs\_q,min\_abs\_q,a\_q,b\_q,max\_real\_psi,rho\_q,rho\_psi,a\_psi,b\_psi,ntilde);\color{Green}
$\\$
$\\$
$\\$\color{BrickRed}M\_psi\-\ =\-\ 2*local\_bounds(dm,max\_imag\_x,max\_real\_psi,max\_abs\_q,max\_xi,max\_xi\_der,prodq);\color{Green}
$\\$
$\\$
$\\$\color{Green}$\%$
$\\$\color{Green}$\%$\-\ bound\-\ for\-\ interpolation\-\ in\-\ x\-\ ------------------------------------------------------------
$\\$\color{Green}$\%$
$\\$
$\\$
$\\$\color{Green}$\%$\-\ top\-\ of\-\ ellipse\-\ \color{Black}$E_{\rho_{\beta}}$ \color{Green}
$\\$\color{BrickRed}max\_imag\_x\-\ =\-\ (rho\_x-1/rho\_x)/2;\color{Green}
$\\$\color{BrickRed}max\_real\_psi\-\ =\-\ 1;\color{Green}
$\\$\color{Green}$\%$\-\ max\_abs\_q\-\ the\-\ same
$\\$\color{Green}$\%$\-\ prodq\-\ the\-\ same
$\\$\color{Green}$\%$\-\ max\_xi\-\ the\-\ same
$\\$\color{Green}$\%$\-\ max\_xi\_der\-\ the\-\ same
$\\$
$\\$
$\\$\color{BrickRed}M\_x\-\ =\-\ local\_bounds(dm,max\_imag\_x,max\_real\_psi,max\_abs\_q,max\_xi,max\_xi\_der,prodq);\color{Green}
$\\$
$\\$
$\\$\color{Green}$\%$
$\\$\color{Green}$\%$\-\ bound\-\ for\-\ interpolation\-\ in\-\ q\-\ ------------------------------------------------------------
$\\$\color{Green}$\%$
$\\$
$\\$
$\\$\color{Green}$\%$\-\ top\-\ of\-\ ellipse\-\ \color{Black}$E_{\rho_{\beta}}$ \color{Green}
$\\$\color{BrickRed}max\_imag\_x\-\ =\-\ 0;\color{Green}
$\\$\color{BrickRed}max\_real\_psi\-\ =\-\ 1;\color{Green}
$\\$\color{BrickRed}max\_abs\_q\-\ =\-\ (a\_q+b\_q)/2\-\ +\-\ ((b\_q-a\_q)/4)*(rho\_q+1/rho\_q);\color{Green}
$\\$
$\\$
$\\$\color{BrickRed}max\_xi\-\ =\-\ bound\_xi(a\_psi,b\_psi,rho\_psi,a\_q,b\_q,1,ntilde);\color{Green}
$\\$\color{BrickRed}max\_xi\_der\-\ =\-\ bound\_xi\_der...\color{Green}
$\\$\color{BrickRed}\-\ \-\ \-\ \-\ (max\_abs\_q,min\_abs\_q,a\_q,b\_q,max\_real\_psi,rho\_q,rho\_psi,a\_psi,b\_psi,ntilde);\color{Green}
$\\$
$\\$
$\\$\color{BrickRed}min\_abs\_q\-\ =\-\ (a\_q+b\_q)/2\-\ -(b\_q-a\_q)*(rho\_q+1/rho\_q)/4;\color{Green}
$\\$\color{BrickRed}prodq\-\ =\-\ abs(log(min\_abs\_q)/2);\color{Green}
$\\$
$\\$
$\\$\color{BrickRed}M\_q\-\ =\-\ 2*local\_bounds(dm,max\_imag\_x,max\_real\_psi,max\_abs\_q,max\_xi,max\_xi\_der,prodq);\color{Green}
$\\$
$\\$
$\\$\color{Green}$\%$--------------------------------------------------------------------------------------------------
$\\$\color{Green}$\%$\-\ local\_bounds
$\\$\color{Green}$\%$--------------------------------------------------------------------------------------------------
$\\$
$\\$
$\\$\color{BrickRed}\color{NavyBlue}\-\ function\-\ \color{BrickRed}\-\ out\-\ =\-\ local\_bounds(dm,max\_imag\_x,max\_real\_psi,max\_abs\_q,xi,xi\_der,prodq)\color{Green}
$\\$
$\\$
$\\$\color{BrickRed}pie\-\ =\-\ nm('pi');\color{Green}
$\\$\color{BrickRed}prod\-\ =\-\ pie/2;\color{Green}
$\\$
$\\$
$\\$\color{Green}$\%$\-\ bound\-\ on\-\ \color{Black} $\vartheta_1^{(m)}(\frac{\pi}{2\omega}(\omega x \pm i\omega'))$  \color{Green}
$\\$\color{BrickRed}temp\-\ =\-\ bound\_theta1\_m(max\_imag\_x,max\_real\_psi,max\_abs\_q,4);\color{Green}
$\\$\color{BrickRed}n0\-\ =\-\ temp(1);\-\ \color{Green}
$\\$\color{BrickRed}n1\-\ =\-\ temp(2);\-\ \color{Green}$\%$\-\ first\-\ derivatve
$\\$\color{BrickRed}n2\-\ =\-\ temp(3);\-\ \color{Green}$\%$\-\ second\-\ derivative
$\\$\color{BrickRed}n3\-\ =\-\ temp(4);\-\ \color{Green}$\%$\-\ third\-\ derivative
$\\$\color{BrickRed}n4\-\ =\-\ temp(5);\color{Green}
$\\$
$\\$
$\\$\color{Green}$\%$\-\ bound\-\ on\-\ \color{Black} $\vartheta_1^{(m)}(\frac{\pi}{2\omega}
\color{Black} (\omega x \pm i\omega'))$  \color{Green}
$\\$\color{BrickRed}max\_real\_psi\-\ =\-\ 0;\color{Green}
$\\$\color{BrickRed}temp\-\ =\-\ bound\_theta1\_m(max\_imag\_x,max\_real\_psi,max\_abs\_q,4);\color{Green}
$\\$\color{BrickRed}d1\-\ =\-\ temp(2);\-\ \color{Green}$\%$\-\ first\-\ derivative
$\\$\color{BrickRed}d2\-\ =\-\ temp(3);\-\ \color{Green}$\%$\-\ second\-\ derivative
$\\$\color{BrickRed}d3\-\ =\-\ temp(4);\-\ \color{Green}$\%$\-\ third\-\ derivative
$\\$\color{BrickRed}d4\-\ =\-\ temp(5);\color{Green}
$\\$
$\\$
$\\$\color{Green}$\%$\-\ bound\-\ on\-\ w(x),\-\ the\-\ conjugate\-\ of\-\ w(x),\-\ and\-\ their\-\ derivatives
$\\$\color{Green}$\%$\-\ on\-\ the\-\ ellipse\-\ \color{Black}$E_{\rho_x}$ \color{Green}
$\\$
$\\$
$\\$\color{Green}$\%$\-\ w(x)
$\\$\color{BrickRed}w0\-\ =\-\ n0\verb|^|2/dm\verb|^|2;\color{Green}
$\\$
$\\$
$\\$\color{Green}$\%$\-\ w'(x)
$\\$\color{BrickRed}w1\-\ =\-\ 2*prod*(\-\ n0*n1/dm\verb|^|2\-\ +\-\ w0*d1/dm\-\ );\color{Green}
$\\$
$\\$
$\\$\color{Green}$\%$\-\ w''(x)
$\\$\color{BrickRed}w2\-\ =\-\ 4*prod\verb|^|2*(n1\verb|^|2/dm\verb|^|2+n0*n2/dm\verb|^|2+2*n0*n1*d1/dm\verb|^|3+w1*d1/dm+w0*d2/dm+w0*d1\verb|^|2/dm\verb|^|2);\color{Green}
$\\$
$\\$
$\\$\color{Green}$\%$\-\ w'''(x)
$\\$\color{BrickRed}w3\-\ =\-\ 8*prod\verb|^|3*(\-\ 2*n1*n2/dm\verb|^|2+2*n1\verb|^|2*d1/dm\verb|^|3+n1*n2/dm\verb|^|2+n0*n3/dm\verb|^|2+2*n0*n2*d1/dm\verb|^|3+...\color{Green}
$\\$\color{BrickRed}\-\ \-\ \-\ \-\ 2*n1\verb|^|2*d1/dm\verb|^|3+2*n0*n2*d1/dm\verb|^|3+2*n0*n1*d2/dm\verb|^|3+6*n0*n1*d1\verb|^|2/dm\verb|^|4+w2*d1/dm\-\ +\-\ w1*d2/dm+...\color{Green}
$\\$\color{BrickRed}\-\ \-\ \-\ \-\ w1*d1\verb|^|2/dm\verb|^|2+w1*d2/dm+\-\ w0*d3/dm+w0*d1*d2/dm\verb|^|2+w1*d1\verb|^|2/dm\verb|^|2+2*w0*d1*d2/dm\verb|^|2+...\color{Green}
$\\$\color{BrickRed}\-\ \-\ \-\ \-\ 2*w0*d1\verb|^|3/dm\verb|^|3);\color{Green}
$\\$
$\\$
$\\$\color{Green}$\%$\-\ bound\-\ on\-\ derivative\-\ of\-\ w(x)\-\ with\-\ respect\-\ to\-\ psi
$\\$\color{BrickRed}w0\_psi\-\ =\-\ 2*prodq*(n0*n1/dm\verb|^|2);\color{Green}
$\\$
$\\$
$\\$\color{Green}$\%$\-\ derivative\-\ of\-\ w'(x)\-\ with\-\ respect\-\ to\-\ psi
$\\$\color{BrickRed}w1\_psi\-\ =\-\ 2*prodq\verb|^|2*(n0*n2/dm\verb|^|2+n1\verb|^|2/dm\verb|^|2+w0\_psi*d1/dm);\color{Green}
$\\$
$\\$
$\\$\color{Green}$\%$\-\ bound\-\ on\-\ derivative\-\ of\-\ w''(x)\-\ with\-\ respect\-\ to\-\ psi
$\\$\color{Green}$\%$\-\ |w''(x)|\-\ $<$\-\ 4*prod\verb|^|2*(\-\ (2*n0*n1*d1)/dm\verb|^|3+\-\ (n1\verb|^|2+n0*n2+w0*d1\verb|^|2)/dm\verb|^|2\-\ +\-\ (w1*d1+\-\ w0*d2)/dm\-\ );
$\\$\color{BrickRed}w2\_psi\-\ =\-\ 4*prodq\verb|^|3*(\-\ (2*n1\verb|^|2*d1+2*n0*n2*d1+2*n0*n1*d2)/dm\verb|^|3\-\ +\-\ ...\color{Green}
$\\$\color{BrickRed}\-\ \-\ \-\ \-\ (2*n1*n2+n1*n2+n0*n3+w0\_psi*d1\verb|^|2+\-\ w0*2*d1*d2)/dm\verb|^|2\-\ +\-\ ...\color{Green}
$\\$\color{BrickRed}\-\ \-\ \-\ \-\ (w1\_psi*d1+w1*d2+w0\_psi*d2+w0*d3)/dm);\color{Green}
$\\$
$\\$
$\\$\color{Green}$\%$\-\ bound\-\ on\-\ derivative\-\ of\-\ w'''(x)\-\ with\-\ respect\-\ to\-\ psi
$\\$\color{BrickRed}w3\_psi\-\ =\-\ 8*prodq\verb|^|4*(...\color{Green}
$\\$\color{BrickRed}\-\ \-\ \-\ \-\ 2*n2*n2/dm\verb|^|2+2*n1*n3/dm\verb|^|2\-\ +\-\ ...\-\ \color{Green}$\%$
$\\$\color{BrickRed}\-\ \-\ \-\ \-\ 4*n1*n2*d1/dm\verb|^|3+2*n1\verb|^|2*d2/dm\verb|^|3+\-\ ...\color{Green}$\%$
$\\$\color{BrickRed}\-\ \-\ \-\ \-\ n2\verb|^|2/dm\verb|^|2+\-\ n1*n3/dm\verb|^|2\-\ +\-\ ...\-\ \color{Green}$\%$
$\\$\color{BrickRed}\-\ \-\ \-\ \-\ n1*n3/dm\verb|^|2+n0*n4/dm\verb|^|2\-\ +\-\ ...\-\ \color{Green}$\%$
$\\$\color{BrickRed}\-\ \-\ \-\ \-\ 2*n1*n2*d1/dm\verb|^|3\-\ +\-\ 2*n0*n3*d1/dm\verb|^|3\-\ +\-\ 2*n0*n2*d2/dm\verb|^|3\-\ +...\-\ \color{Green}$\%$
$\\$\color{BrickRed}\-\ \-\ \-\ \-\ 4*n1*n2*d1/dm\verb|^|3\-\ +\-\ 2*n1\verb|^|2*d2/dm\verb|^|3\-\ +\-\ ...\-\ \color{Green}$\%$
$\\$\color{BrickRed}\-\ \-\ \-\ \-\ 2*n1*n2*d1/dm\verb|^|3\-\ +\-\ 2*n0*n3*d1/dm\verb|^|3\-\ +\-\ 2*n0*n2*d2/dm\verb|^|3\-\ +\-\ ...\-\ \color{Green}$\%$
$\\$\color{BrickRed}\-\ \-\ \-\ \-\ 2*n1\verb|^|2*d2/dm\verb|^|3\-\ +\-\ 2*n0*n2*d2/dm\verb|^|3+2*n0*n1*d3/dm\verb|^|3+\-\ ...\-\ \color{Green}$\%$
$\\$\color{BrickRed}\-\ \-\ \-\ \-\ 6*n1\verb|^|2*d1\verb|^|2/dm\verb|^|4\-\ +\-\ 6*n0*n2*d1\verb|^|2/dm\verb|^|4+6*n0*n1*2*d1*d2/dm\verb|^|4\-\ +\-\ ...\-\ \color{Green}$\%$
$\\$\color{BrickRed}\-\ \-\ \-\ \-\ w2\_psi*d1/dm\-\ +\-\ w2*d2/dm\-\ +...\-\ \color{Green}$\%$
$\\$\color{BrickRed}\-\ \-\ \-\ \-\ w1\_psi*d2/dm\-\ +\-\ w1*d3/dm+\-\ ...\-\ \color{Green}$\%$
$\\$\color{BrickRed}\-\ \-\ \-\ \-\ w1\_psi*d1\verb|^|2/dm\verb|^|2\-\ +\-\ w1*2*d1*d2/dm\verb|^|2\-\ +\-\ ...\-\ \color{Green}$\%$
$\\$\color{BrickRed}\-\ \-\ \-\ \-\ w1\_psi*d2/dm\-\ +\-\ w1*d3/dm\-\ +\-\ ...\-\ \color{Green}$\%$
$\\$\color{BrickRed}\-\ \-\ \-\ \-\ w0\_psi*d3/dm\-\ +\-\ w0*d4/dm\-\ +\-\ ...\-\ \color{Green}$\%$
$\\$\color{BrickRed}\-\ \-\ \-\ \-\ w0\_psi*d1*d2/dm\verb|^|2\-\ +\-\ w0*d2*d2/dm\verb|^|2\-\ +\-\ w0*d1*d3/dm\verb|^|2\-\ +\-\ ...\-\ \color{Green}$\%$
$\\$\color{BrickRed}\-\ \-\ \-\ \-\ w1\_psi*d1\verb|^|2/dm\verb|^|2\-\ +\-\ w1*2*d1*d2/dm\verb|^|2\-\ +\-\ ...\-\ \color{Green}$\%$
$\\$\color{BrickRed}\-\ \-\ \-\ \-\ 2*w0\_psi*d1*d2/dm\verb|^|2\-\ +\-\ 2*w0*d2*d2/dm\verb|^|2\-\ +\-\ 2*w0*d1*d3/dm\verb|^|2\-\ +\-\ ...\color{Green}$\%$
$\\$\color{BrickRed}\-\ \-\ \-\ \-\ 2*w0\_psi*d1\verb|^|3/dm\verb|^|3\-\ +\-\ 2*w0*3*d1\verb|^|2*d2/dm\verb|^|3\-\ ...\-\ \color{Green}$\%$
$\\$\color{BrickRed}\-\ \-\ \-\ \-\ );\color{Green}
$\\$
$\\$
$\\$\color{BrickRed}xi2\-\ =\-\ xi*xi;\color{Green}
$\\$\color{BrickRed}xi3\-\ =\-\ xi2*xi;\color{Green}
$\\$
$\\$
$\\$\color{Green}$\%$\-\ derivatives\-\ of\-\ v\-\ with\-\ respect\-\ to\-\ x
$\\$\color{BrickRed}v1\-\ =\-\ w1+\-\ xi*w0;\color{Green}
$\\$\color{BrickRed}v2\-\ =\-\ w2+2*xi*w1+xi2*w0;\color{Green}
$\\$\color{BrickRed}v3\-\ =\-\ w3+3*xi*w2+3*xi2*w1+xi3*w0;\color{Green}
$\\$
$\\$
$\\$\color{Green}$\%$\-\ derivatives\-\ with\-\ respect\-\ to\-\ psi
$\\$\color{BrickRed}v1\_psi\-\ =\-\ w1\_psi+xi\_der*w0+xi*w0\_psi;\color{Green}
$\\$\color{BrickRed}v2\_psi\-\ =\-\ w2\_psi+2*xi\_der*w1+2*xi*w1\_psi+2*xi*xi\_der*w0+xi2*w0\_psi;\color{Green}
$\\$\color{BrickRed}v3\_psi\-\ =\-\ w3\_psi+3*xi\_der*w2+3*xi*w2\_psi+6*xi*xi\_der*w1...\color{Green}
$\\$\color{BrickRed}\-\ \-\ \-\ \-\ \-\ \-\ \-\ \-\ \-\ \-\ \-\ \-\ \-\ \-\ \-\ \-\ +3*xi2*w1\_psi\-\ +\-\ 3*xi2*xi\_der*w0+xi3*w0\_psi;\color{Green}
$\\$\color{BrickRed}\-\ \-\ \-\ \-\ \-\ \color{Green}
$\\$\color{Green}$\%$\-\ bound\-\ on\-\ functions\-\ to\-\ interpoate\-\ \-\ \-\ \-\ \-\ \-\ \-\ \-\ \-\ 
$\\$\color{BrickRed}f1\-\ =\-\ v1*v2;\color{Green}
$\\$\color{BrickRed}f2\-\ =\-\ v3*v2;\color{Green}
$\\$\color{BrickRed}g\-\ =\-\ w0*v1;\color{Green}
$\\$\color{BrickRed}\-\ \-\ \-\ \-\ \-\ \-\ \-\ \-\ \-\ \-\ \-\ \-\ \color{Green}
$\\$\color{Green}$\%$\-\ bound\-\ on\-\ derivative\-\ with\-\ respect\-\ to\-\ psi\-\ of\-\ 
$\\$\color{Green}$\%$\-\ functions\-\ to\-\ interpolate
$\\$\color{BrickRed}f1\_psi\-\ =\-\ v1\_psi*v2+v1*v2\_psi;\color{Green}
$\\$\color{BrickRed}f2\_psi\-\ =\-\ v3\_psi*v2+v3*v2\_psi;\color{Green}
$\\$\color{BrickRed}g\_psi\-\ =\-\ w0\_psi*v1+w0*v1\_psi;\color{Green}
$\\$
$\\$
$\\$\color{Green}$\%$\-\ here\-\ =\-\ f1(round(0.5*length(f1)))
$\\$
$\\$
$\\$\color{Green}$\%$\-\ bound\-\ on\-\ all\-\ sub\-\ bounds
$\\$\color{BrickRed}out\-\ =\-\ nm(max(sup([f1,f2,g,f1\_psi,f2\_psi,g\_psi])));\color{Green}
$\\$
$\\$
$\\$\color{Black}\section{bound\_numer\_unstable.m}

\color{Green}\color{BrickRed}\color{NavyBlue}\-\ function\-\ \color{BrickRed}\-\ [M\_psi,M\_x,M\_q]\-\ =\-\ bound\_numer\_unstable(dm,rho\_x,rho\_q,a\_q,b\_q)\color{Green}
$\\$
$\\$
$\\$\color{Green}$\%$
$\\$\color{Green}$\%$\-\ constants\-\ 
$\\$\color{Green}$\%$
$\\$
$\\$
$\\$\color{BrickRed}a\_psi\-\ =\-\ nm(1);\color{Green}
$\\$\color{BrickRed}b\_psi\-\ =\-\ nm(1);\color{Green}
$\\$\color{BrickRed}rho\_psi\-\ =\-\ 1;\color{Green}
$\\$\color{BrickRed}ntilde\-\ =\-\ 1;\color{Green}
$\\$
$\\$
$\\$\color{BrickRed}min\_abs\_q\-\ =\-\ (a\_q+b\_q)/2\-\ -(b\_q-a\_q)*(rho\_q+1/rho\_q)/4;\color{Green}
$\\$
$\\$
$\\$\color{Green}$\%$
$\\$\color{Green}$\%$\-\ bound\-\ for\-\ interpolation\-\ in\-\ psi\-\ ------------------------------------------------------------
$\\$\color{Green}$\%$
$\\$
$\\$
$\\$\color{Green}$\%$\-\ top\-\ of\-\ ellipse\-\ \color{Black}$E_{\rho_{\psi}}$ \color{Green}
$\\$\color{BrickRed}max\_imag\_x\-\ =\-\ 0;\color{Green}
$\\$\color{BrickRed}max\_real\_psi\-\ =\-\ 1;\color{Green}
$\\$\color{BrickRed}max\_abs\_q\-\ =\-\ sup(real(b\_q));\color{Green}
$\\$\color{BrickRed}prodq\-\ =\-\ abs(log(a\_q)/2);\color{Green}
$\\$
$\\$
$\\$\color{BrickRed}max\_xi\-\ =\-\ bound\_xi(a\_psi,b\_psi,rho\_psi,a\_q,b\_q,1,ntilde);\color{Green}
$\\$\color{BrickRed}max\_xi\_der\-\ =\-\ bound\_xi\_der...\color{Green}
$\\$\color{BrickRed}\-\ \-\ \-\ \-\ (max\_abs\_q,min\_abs\_q,a\_q,b\_q,max\_real\_psi,rho\_q,rho\_psi,a\_psi,b\_psi,ntilde);\color{Green}
$\\$
$\\$
$\\$\color{BrickRed}M\_psi\-\ =\-\ 2*local\_bounds(dm,max\_imag\_x,max\_real\_psi,max\_abs\_q,max\_xi,max\_xi\_der,prodq);\color{Green}
$\\$
$\\$
$\\$\color{Green}$\%$
$\\$\color{Green}$\%$\-\ bound\-\ for\-\ interpolation\-\ in\-\ x\-\ ------------------------------------------------------------
$\\$\color{Green}$\%$
$\\$
$\\$
$\\$\color{Green}$\%$\-\ top\-\ of\-\ ellipse\-\ \color{Black}$E_{\rho_{\beta}}$ \color{Green}
$\\$\color{BrickRed}max\_imag\_x\-\ =\-\ (rho\_x-1/rho\_x)/2;\color{Green}
$\\$\color{BrickRed}max\_real\_psi\-\ =\-\ 1;\color{Green}
$\\$\color{Green}$\%$\-\ max\_abs\_q\-\ the\-\ same
$\\$\color{Green}$\%$\-\ prodq\-\ the\-\ same
$\\$\color{Green}$\%$\-\ max\_xi\-\ the\-\ same
$\\$\color{Green}$\%$\-\ max\_xi\_der\-\ the\-\ same
$\\$
$\\$
$\\$\color{BrickRed}M\_x\-\ =\-\ local\_bounds(dm,max\_imag\_x,max\_real\_psi,max\_abs\_q,max\_xi,max\_xi\_der,prodq);\color{Green}
$\\$
$\\$
$\\$\color{Green}$\%$
$\\$\color{Green}$\%$\-\ bound\-\ for\-\ interpolation\-\ in\-\ q\-\ ------------------------------------------------------------
$\\$\color{Green}$\%$
$\\$
$\\$
$\\$\color{Green}$\%$\-\ top\-\ of\-\ ellipse\-\ \color{Black}$E_{\rho_{\beta}}$ \color{Green}
$\\$\color{BrickRed}max\_imag\_x\-\ =\-\ 0;\color{Green}
$\\$\color{BrickRed}max\_real\_psi\-\ =\-\ 1;\color{Green}
$\\$\color{BrickRed}max\_abs\_q\-\ =\-\ (a\_q+b\_q)/2\-\ +\-\ ((b\_q-a\_q)/4)*(rho\_q+1/rho\_q);\color{Green}
$\\$
$\\$
$\\$\color{BrickRed}max\_xi\-\ =\-\ bound\_xi(a\_psi,b\_psi,rho\_psi,a\_q,b\_q,1,ntilde);\color{Green}
$\\$\color{BrickRed}max\_xi\_der\-\ =\-\ bound\_xi\_der...\color{Green}
$\\$\color{BrickRed}\-\ \-\ \-\ \-\ (max\_abs\_q,min\_abs\_q,a\_q,b\_q,max\_real\_psi,rho\_q,rho\_psi,a\_psi,b\_psi,ntilde);\color{Green}
$\\$
$\\$
$\\$\color{BrickRed}min\_abs\_q\-\ =\-\ (a\_q+b\_q)/2\-\ -(b\_q-a\_q)*(rho\_q+1/rho\_q)/4;\color{Green}
$\\$\color{BrickRed}prodq\-\ =\-\ abs(log(min\_abs\_q)/2);\color{Green}
$\\$
$\\$
$\\$\color{BrickRed}M\_q\-\ =\-\ 2*local\_bounds(dm,max\_imag\_x,max\_real\_psi,max\_abs\_q,max\_xi,max\_xi\_der,prodq);\color{Green}
$\\$
$\\$
$\\$\color{Green}$\%$--------------------------------------------------------------------------------------------------
$\\$\color{Green}$\%$\-\ local\_bounds
$\\$\color{Green}$\%$--------------------------------------------------------------------------------------------------
$\\$
$\\$
$\\$\color{BrickRed}\color{NavyBlue}\-\ function\-\ \color{BrickRed}\-\ out\-\ =\-\ local\_bounds(dm,max\_imag\_x,max\_real\_psi,max\_abs\_q,xi,xi\_der,prodq)\color{Green}
$\\$
$\\$
$\\$\color{BrickRed}pie\-\ =\-\ nm('pi');\color{Green}
$\\$\color{BrickRed}prod\-\ =\-\ pie/2;\color{Green}
$\\$
$\\$
$\\$\color{Green}$\%$\-\ bound\-\ on\-\ \color{Black} $\vartheta_1^{(m)}(\frac{\pi}{2\omega}(\omega x \pm i\omega'))$  \color{Green}
$\\$\color{BrickRed}temp\-\ =\-\ bound\_theta1\_m(max\_imag\_x,max\_real\_psi,max\_abs\_q,4);\color{Green}
$\\$\color{BrickRed}n0\-\ =\-\ temp(1);\-\ \color{Green}
$\\$\color{BrickRed}n1\-\ =\-\ temp(2);\-\ \color{Green}$\%$\-\ first\-\ derivatve
$\\$\color{BrickRed}n2\-\ =\-\ temp(3);\-\ \color{Green}$\%$\-\ second\-\ derivative
$\\$\color{BrickRed}n3\-\ =\-\ temp(4);\-\ \color{Green}$\%$\-\ third\-\ derivative
$\\$\color{BrickRed}n4\-\ =\-\ temp(5);\color{Green}
$\\$
$\\$
$\\$\color{Green}$\%$\-\ bound\-\ on\-\ \color{Black} $\vartheta_1^{(m)}(\frac{\pi}{2\omega}
\color{Black} (\omega x \pm i\omega'))$  \color{Green}
$\\$\color{BrickRed}max\_real\_psi\-\ =\-\ 0;\color{Green}
$\\$\color{BrickRed}temp\-\ =\-\ bound\_theta1\_m(max\_imag\_x,max\_real\_psi,max\_abs\_q,4);\color{Green}
$\\$\color{BrickRed}d1\-\ =\-\ temp(2);\-\ \color{Green}$\%$\-\ first\-\ derivative
$\\$\color{BrickRed}d2\-\ =\-\ temp(3);\-\ \color{Green}$\%$\-\ second\-\ derivative
$\\$\color{BrickRed}d3\-\ =\-\ temp(4);\-\ \color{Green}$\%$\-\ third\-\ derivative
$\\$\color{BrickRed}d4\-\ =\-\ temp(5);\color{Green}
$\\$
$\\$
$\\$\color{Green}$\%$\-\ bound\-\ on\-\ w(x),\-\ the\-\ conjugate\-\ of\-\ w(x),\-\ and\-\ their\-\ derivatives
$\\$\color{Green}$\%$\-\ on\-\ the\-\ ellipse\-\ \color{Black}$E_{\rho_x}$ \color{Green}
$\\$
$\\$
$\\$\color{Green}$\%$\-\ w(x)
$\\$\color{BrickRed}w0\-\ =\-\ n0\verb|^|2/dm\verb|^|2;\color{Green}
$\\$
$\\$
$\\$\color{Green}$\%$\-\ w'(x)
$\\$\color{BrickRed}w1\-\ =\-\ 2*prod*(\-\ n0*n1/dm\verb|^|2\-\ +\-\ w0*d1/dm\-\ );\color{Green}
$\\$
$\\$
$\\$\color{Green}$\%$\-\ w''(x)
$\\$\color{BrickRed}w2\-\ =\-\ 4*prod\verb|^|2*(n1\verb|^|2/dm\verb|^|2+n0*n2/dm\verb|^|2+2*n0*n1*d1/dm\verb|^|3+w1*d1/dm+w0*d2/dm+w0*d1\verb|^|2/dm\verb|^|2);\color{Green}
$\\$
$\\$
$\\$\color{Green}$\%$\-\ w'''(x)
$\\$\color{BrickRed}w3\-\ =\-\ 8*prod\verb|^|3*(\-\ 2*n1*n2/dm\verb|^|2+2*n1\verb|^|2*d1/dm\verb|^|3+n1*n2/dm\verb|^|2+n0*n3/dm\verb|^|2+2*n0*n2*d1/dm\verb|^|3+...\color{Green}
$\\$\color{BrickRed}\-\ \-\ \-\ \-\ 2*n1\verb|^|2*d1/dm\verb|^|3+2*n0*n2*d1/dm\verb|^|3+2*n0*n1*d2/dm\verb|^|3+6*n0*n1*d1\verb|^|2/dm\verb|^|4+w2*d1/dm\-\ +\-\ w1*d2/dm+...\color{Green}
$\\$\color{BrickRed}\-\ \-\ \-\ \-\ w1*d1\verb|^|2/dm\verb|^|2+w1*d2/dm+\-\ w0*d3/dm+w0*d1*d2/dm\verb|^|2+w1*d1\verb|^|2/dm\verb|^|2+2*w0*d1*d2/dm\verb|^|2+...\color{Green}
$\\$\color{BrickRed}\-\ \-\ \-\ \-\ 2*w0*d1\verb|^|3/dm\verb|^|3);\color{Green}
$\\$
$\\$
$\\$\color{Green}$\%$\-\ bound\-\ on\-\ derivative\-\ of\-\ w(x)\-\ with\-\ respect\-\ to\-\ psi
$\\$\color{BrickRed}w0\_psi\-\ =\-\ 2*prodq*(n0*n1/dm\verb|^|2);\color{Green}
$\\$
$\\$
$\\$\color{Green}$\%$\-\ derivative\-\ of\-\ w'(x)\-\ with\-\ respect\-\ to\-\ psi
$\\$\color{BrickRed}w1\_psi\-\ =\-\ 2*prodq\verb|^|2*(n0*n2/dm\verb|^|2+n1\verb|^|2/dm\verb|^|2+w0\_psi*d1/dm);\color{Green}
$\\$
$\\$
$\\$\color{Green}$\%$\-\ bound\-\ on\-\ derivative\-\ of\-\ w''(x)\-\ with\-\ respect\-\ to\-\ psi
$\\$\color{Green}$\%$\-\ |w''(x)|\-\ $<$\-\ 4*prod\verb|^|2*(\-\ (2*n0*n1*d1)/dm\verb|^|3+\-\ (n1\verb|^|2+n0*n2+w0*d1\verb|^|2)/dm\verb|^|2\-\ +\-\ (w1*d1+\-\ w0*d2)/dm\-\ );
$\\$\color{BrickRed}w2\_psi\-\ =\-\ 4*prodq\verb|^|3*(\-\ (2*n1\verb|^|2*d1+2*n0*n2*d1+2*n0*n1*d2)/dm\verb|^|3\-\ +\-\ ...\color{Green}
$\\$\color{BrickRed}\-\ \-\ \-\ \-\ (2*n1*n2+n1*n2+n0*n3+w0\_psi*d1\verb|^|2+\-\ w0*2*d1*d2)/dm\verb|^|2\-\ +\-\ ...\color{Green}
$\\$\color{BrickRed}\-\ \-\ \-\ \-\ (w1\_psi*d1+w1*d2+w0\_psi*d2+w0*d3)/dm);\color{Green}
$\\$
$\\$
$\\$\color{Green}$\%$\-\ bound\-\ on\-\ derivative\-\ of\-\ w'''(x)\-\ with\-\ respect\-\ to\-\ psi
$\\$\color{BrickRed}w3\_psi\-\ =\-\ 8*prodq\verb|^|4*(...\color{Green}
$\\$\color{BrickRed}\-\ \-\ \-\ \-\ 2*n2*n2/dm\verb|^|2+2*n1*n3/dm\verb|^|2\-\ +\-\ ...\-\ \color{Green}$\%$
$\\$\color{BrickRed}\-\ \-\ \-\ \-\ 4*n1*n2*d1/dm\verb|^|3+2*n1\verb|^|2*d2/dm\verb|^|3+\-\ ...\color{Green}$\%$
$\\$\color{BrickRed}\-\ \-\ \-\ \-\ n2\verb|^|2/dm\verb|^|2+\-\ n1*n3/dm\verb|^|2\-\ +\-\ ...\-\ \color{Green}$\%$
$\\$\color{BrickRed}\-\ \-\ \-\ \-\ n1*n3/dm\verb|^|2+n0*n4/dm\verb|^|2\-\ +\-\ ...\-\ \color{Green}$\%$
$\\$\color{BrickRed}\-\ \-\ \-\ \-\ 2*n1*n2*d1/dm\verb|^|3\-\ +\-\ 2*n0*n3*d1/dm\verb|^|3\-\ +\-\ 2*n0*n2*d2/dm\verb|^|3\-\ +...\-\ \color{Green}$\%$
$\\$\color{BrickRed}\-\ \-\ \-\ \-\ 4*n1*n2*d1/dm\verb|^|3\-\ +\-\ 2*n1\verb|^|2*d2/dm\verb|^|3\-\ +\-\ ...\-\ \color{Green}$\%$
$\\$\color{BrickRed}\-\ \-\ \-\ \-\ 2*n1*n2*d1/dm\verb|^|3\-\ +\-\ 2*n0*n3*d1/dm\verb|^|3\-\ +\-\ 2*n0*n2*d2/dm\verb|^|3\-\ +\-\ ...\-\ \color{Green}$\%$
$\\$\color{BrickRed}\-\ \-\ \-\ \-\ 2*n1\verb|^|2*d2/dm\verb|^|3\-\ +\-\ 2*n0*n2*d2/dm\verb|^|3+2*n0*n1*d3/dm\verb|^|3+\-\ ...\-\ \color{Green}$\%$
$\\$\color{BrickRed}\-\ \-\ \-\ \-\ 6*n1\verb|^|2*d1\verb|^|2/dm\verb|^|4\-\ +\-\ 6*n0*n2*d1\verb|^|2/dm\verb|^|4+6*n0*n1*2*d1*d2/dm\verb|^|4\-\ +\-\ ...\-\ \color{Green}$\%$
$\\$\color{BrickRed}\-\ \-\ \-\ \-\ w2\_psi*d1/dm\-\ +\-\ w2*d2/dm\-\ +...\-\ \color{Green}$\%$
$\\$\color{BrickRed}\-\ \-\ \-\ \-\ w1\_psi*d2/dm\-\ +\-\ w1*d3/dm+\-\ ...\-\ \color{Green}$\%$
$\\$\color{BrickRed}\-\ \-\ \-\ \-\ w1\_psi*d1\verb|^|2/dm\verb|^|2\-\ +\-\ w1*2*d1*d2/dm\verb|^|2\-\ +\-\ ...\-\ \color{Green}$\%$
$\\$\color{BrickRed}\-\ \-\ \-\ \-\ w1\_psi*d2/dm\-\ +\-\ w1*d3/dm\-\ +\-\ ...\-\ \color{Green}$\%$
$\\$\color{BrickRed}\-\ \-\ \-\ \-\ w0\_psi*d3/dm\-\ +\-\ w0*d4/dm\-\ +\-\ ...\-\ \color{Green}$\%$
$\\$\color{BrickRed}\-\ \-\ \-\ \-\ w0\_psi*d1*d2/dm\verb|^|2\-\ +\-\ w0*d2*d2/dm\verb|^|2\-\ +\-\ w0*d1*d3/dm\verb|^|2\-\ +\-\ ...\-\ \color{Green}$\%$
$\\$\color{BrickRed}\-\ \-\ \-\ \-\ w1\_psi*d1\verb|^|2/dm\verb|^|2\-\ +\-\ w1*2*d1*d2/dm\verb|^|2\-\ +\-\ ...\-\ \color{Green}$\%$
$\\$\color{BrickRed}\-\ \-\ \-\ \-\ 2*w0\_psi*d1*d2/dm\verb|^|2\-\ +\-\ 2*w0*d2*d2/dm\verb|^|2\-\ +\-\ 2*w0*d1*d3/dm\verb|^|2\-\ +\-\ ...\color{Green}$\%$
$\\$\color{BrickRed}\-\ \-\ \-\ \-\ 2*w0\_psi*d1\verb|^|3/dm\verb|^|3\-\ +\-\ 2*w0*3*d1\verb|^|2*d2/dm\verb|^|3\-\ ...\-\ \color{Green}$\%$
$\\$\color{BrickRed}\-\ \-\ \-\ \-\ );\color{Green}
$\\$
$\\$
$\\$\color{BrickRed}xi2\-\ =\-\ xi*xi;\color{Green}
$\\$\color{BrickRed}xi3\-\ =\-\ xi2*xi;\color{Green}
$\\$
$\\$
$\\$\color{Green}$\%$\-\ derivatives\-\ of\-\ v\-\ with\-\ respect\-\ to\-\ x
$\\$\color{BrickRed}v1\-\ =\-\ w1+\-\ xi*w0;\color{Green}
$\\$\color{BrickRed}v2\-\ =\-\ w2+2*xi*w1+xi2*w0;\color{Green}
$\\$\color{BrickRed}v3\-\ =\-\ w3+3*xi*w2+3*xi2*w1+xi3*w0;\color{Green}
$\\$
$\\$
$\\$\color{Green}$\%$\-\ derivatives\-\ with\-\ respect\-\ to\-\ psi
$\\$\color{BrickRed}v1\_psi\-\ =\-\ w1\_psi+xi\_der*w0+xi*w0\_psi;\color{Green}
$\\$\color{BrickRed}v2\_psi\-\ =\-\ w2\_psi+2*xi\_der*w1+2*xi*w1\_psi+2*xi*xi\_der*w0+xi2*w0\_psi;\color{Green}
$\\$\color{BrickRed}v3\_psi\-\ =\-\ w3\_psi+3*xi\_der*w2+3*xi*w2\_psi+6*xi*xi\_der*w1...\color{Green}
$\\$\color{BrickRed}\-\ \-\ \-\ \-\ \-\ \-\ \-\ \-\ \-\ \-\ \-\ \-\ \-\ \-\ \-\ \-\ +3*xi2*w1\_psi\-\ +\-\ 3*xi2*xi\_der*w0+xi3*w0\_psi;\color{Green}
$\\$\color{BrickRed}\-\ \-\ \-\ \-\ \-\ \color{Green}
$\\$\color{Green}$\%$\-\ bound\-\ on\-\ functions\-\ to\-\ interpoate\-\ \-\ \-\ \-\ \-\ \-\ \-\ \-\ \-\ 
$\\$\color{BrickRed}f1\-\ =\-\ v1*v2;\color{Green}
$\\$\color{BrickRed}f2\-\ =\-\ v3*v2;\color{Green}
$\\$\color{BrickRed}g\-\ =\-\ w0*v1;\color{Green}
$\\$\color{BrickRed}\-\ \-\ \-\ \-\ \-\ \-\ \-\ \-\ \-\ \-\ \-\ \-\ \color{Green}
$\\$\color{Green}$\%$\-\ bound\-\ on\-\ derivative\-\ with\-\ respect\-\ to\-\ psi\-\ of\-\ 
$\\$\color{Green}$\%$\-\ functions\-\ to\-\ interpolate
$\\$\color{BrickRed}f1\_psi\-\ =\-\ v1\_psi*v2+v1*v2\_psi;\color{Green}
$\\$\color{BrickRed}f2\_psi\-\ =\-\ v3\_psi*v2+v3*v2\_psi;\color{Green}
$\\$\color{BrickRed}g\_psi\-\ =\-\ w0\_psi*v1+w0*v1\_psi;\color{Green}
$\\$
$\\$
$\\$
$\\$
$\\$\color{Green}$\%$\-\ bound\-\ on\-\ all\-\ sub\-\ bounds
$\\$\color{BrickRed}out\-\ =\-\ nm(max(sup([f1,f2,g,f1\_psi,f2\_psi,g\_psi])));\color{Green}
$\\$
$\\$
$\\$\color{Black}\section{bound\_sub\_integrals.m}

\color{Green}\color{BrickRed}\color{NavyBlue}\-\ function\-\ \color{BrickRed}\-\ [M\_psi,M\_x,M\_q]\-\ =\-\ bound\_sub\_integrals(dm,rho\_psi,rho\_x,rho\_q,a\_q,b\_q)\color{Green}
$\\$
$\\$
$\\$\color{Green}$\%$\-\ constants
$\\$\color{BrickRed}pie\-\ =\-\ nm('pi');\color{Green}
$\\$\color{BrickRed}prod\-\ =\-\ pie/2;\color{Green}
$\\$
$\\$
$\\$\color{Green}$\%$\-\ -----------------------------------------------------------
$\\$\color{Green}$\%$\-\ bound\-\ for\-\ interpolation\-\ in\-\ psi
$\\$\color{Green}$\%$\-\ -----------------------------------------------------------
$\\$
$\\$
$\\$\color{Green}$\%$\-\ top\-\ of\-\ ellipse\-\ \color{Black}$E_{\rho_x}$ \color{Green}
$\\$\color{BrickRed}max\_imag\_x\-\ =\-\ 0;\color{Green}
$\\$\color{BrickRed}max\_real\_psi\-\ =\-\ (1+(rho\_psi+1/rho\_psi)/2)/2;\color{Green}
$\\$\color{BrickRed}max\_abs\_q\-\ =\-\ sup(real(b\_q));\color{Green}
$\\$
$\\$
$\\$\color{BrickRed}vec\-\ =\-\ bound(prod,dm,max\_imag\_x,max\_real\_psi,max\_abs\_q);\color{Green}
$\\$
$\\$
$\\$\color{Green}$\%$\-\ largest\-\ bound\-\ on\-\ all\-\ the\-\ sub\-\ integrands\-\ when\-\ alpha\-\ =\-\ \-\ i\-\ beta
$\\$\color{BrickRed}M\_psi\-\ =\-\ sup(2*nm(max(vec)));\color{Green}
$\\$
$\\$
$\\$\color{Green}$\%$\-\ -----------------------------------------------------------
$\\$\color{Green}$\%$\-\ bound\-\ for\-\ interpolation\-\ in\-\ x
$\\$\color{Green}$\%$\-\ -----------------------------------------------------------
$\\$
$\\$
$\\$\color{Green}$\%$\-\ top\-\ of\-\ ellipse\-\ \color{Black}$E_{\rho_x}$ \color{Green}
$\\$\color{BrickRed}max\_imag\_x\-\ =\-\ (rho\_x-1/rho\_x)/2;\color{Green}
$\\$\color{BrickRed}max\_real\_psi\-\ =\-\ 1;\color{Green}
$\\$\color{BrickRed}max\_abs\_q\-\ =\-\ sup(real(b\_q));\color{Green}
$\\$
$\\$
$\\$\color{BrickRed}vec\-\ =\-\ bound(prod,dm,max\_imag\_x,max\_real\_psi,max\_abs\_q);\color{Green}
$\\$
$\\$
$\\$\color{Green}$\%$\-\ largest\-\ bound\-\ on\-\ all\-\ the\-\ sub\-\ integrands\-\ when\-\ alpha\-\ =\-\ \-\ i\-\ beta
$\\$\color{BrickRed}M\_x\-\ =\-\ max(vec);\color{Green}
$\\$
$\\$
$\\$\color{Green}$\%$\-\ -----------------------------------------------------------
$\\$\color{Green}$\%$\-\ bound\-\ for\-\ interpolation\-\ in\-\ q
$\\$\color{Green}$\%$\-\ -----------------------------------------------------------
$\\$
$\\$
$\\$\color{Green}$\%$\-\ top\-\ of\-\ ellipse\-\ \color{Black}$E_{\rho_x}$ \color{Green}
$\\$\color{BrickRed}max\_imag\_x\-\ =\-\ 0;\color{Green}
$\\$\color{BrickRed}max\_real\_psi\-\ =\-\ 1;\color{Green}
$\\$\color{BrickRed}max\_abs\_q\-\ =\-\ (a\_q+b\_q)/2\-\ +\-\ ((b\_q-a\_q)/4)*(rho\_q+1/rho\_q);\color{Green}
$\\$
$\\$
$\\$\color{BrickRed}vec\-\ =\-\ bound(prod,dm,max\_imag\_x,max\_real\_psi,max\_abs\_q);\color{Green}
$\\$
$\\$
$\\$\color{Green}$\%$\-\ largest\-\ bound\-\ on\-\ all\-\ the\-\ sub\-\ integrands\-\ when\-\ alpha\-\ =\-\ \-\ i\-\ beta
$\\$\color{BrickRed}M\_q\-\ =\-\ sup(2*nm(max(vec)));\color{Green}
$\\$
$\\$
$\\$\color{Green}$\%$\-\ -----------------------------------------------------------
$\\$\color{Green}$\%$\-\ function\-\ for\-\ bounds
$\\$\color{Green}$\%$\-\ -----------------------------------------------------------
$\\$\color{BrickRed}\color{NavyBlue}\-\ function\-\ \color{BrickRed}\-\ vec\-\ =\-\ bound(prod,dm,max\_imag\_x,max\_real\_psi,max\_abs\_q)\color{Green}
$\\$
$\\$
$\\$\color{Green}$\%$\-\ bound\-\ on\-\ \color{Black} $\vartheta_1^{(m)}(\frac{\pi}{2\omega}(\omega x \pm i\omega'))$  \color{Green}
$\\$\color{BrickRed}temp\-\ =\-\ bound\_theta1\_m(max\_imag\_x,max\_real\_psi,max\_abs\_q,4);\color{Green}
$\\$
$\\$
$\\$\color{BrickRed}n0\-\ =\-\ temp(1);\-\ \color{Green}
$\\$\color{BrickRed}n1\-\ =\-\ temp(2);\-\ \color{Green}$\%$\-\ first\-\ derivatve
$\\$\color{BrickRed}n2\-\ =\-\ temp(3);\-\ \color{Green}$\%$\-\ second\-\ derivative
$\\$\color{BrickRed}n3\-\ =\-\ temp(4);\-\ \color{Green}$\%$\-\ third\-\ derivative
$\\$
$\\$
$\\$\color{Green}$\%$\-\ bound\-\ on\-\ \color{Black} $\vartheta_1^{(m)}(\frac{\pi}{2\omega}
\color{Black} (\omega x \pm i\omega'+n\omega + i\beta))$  \color{Green}
$\\$\color{BrickRed}max\_real\_psi\-\ =\-\ 0;\color{Green}
$\\$\color{BrickRed}temp\-\ =\-\ bound\_theta1\_m(max\_imag\_x,max\_real\_psi,max\_abs\_q,4);\color{Green}
$\\$\color{BrickRed}d1\-\ =\-\ temp(2);\-\ \color{Green}$\%$\-\ first\-\ derivative
$\\$\color{BrickRed}d2\-\ =\-\ temp(3);\-\ \color{Green}$\%$\-\ second\-\ derivative
$\\$\color{BrickRed}d3\-\ =\-\ temp(4);\-\ \color{Green}$\%$\-\ third\-\ derivative
$\\$
$\\$
$\\$\color{Green}$\%$\-\ w(x)
$\\$\color{BrickRed}w0\-\ =\-\ n0\verb|^|2/dm\verb|^|2;\color{Green}
$\\$
$\\$
$\\$\color{Green}$\%$\-\ w'(x)
$\\$\color{BrickRed}w1\-\ =\-\ 2*prod*(\-\ n0*n1/dm\verb|^|2\-\ +\-\ w0*d1/dm\-\ );\color{Green}
$\\$
$\\$
$\\$\color{Green}$\%$\-\ w''(x)
$\\$\color{BrickRed}w2\-\ =\-\ 4*prod\verb|^|2*(n1\verb|^|2/dm\verb|^|2+n0*n2/dm\verb|^|2+2*n0*n1*d1/dm\verb|^|3+w1*d1/dm+w0*d2/dm+w0*d1\verb|^|2/dm\verb|^|2);\color{Green}
$\\$
$\\$
$\\$\color{Green}$\%$\-\ w'''(x)
$\\$\color{BrickRed}w3\-\ =\-\ 8*prod\verb|^|3*(\-\ 2*n1*n2/dm\verb|^|2+2*n1\verb|^|2*d1/dm\verb|^|3+n1*n2/dm\verb|^|2+n0*n3/dm\verb|^|2+2*n0*n2*d1/dm\verb|^|3+...\color{Green}
$\\$\color{BrickRed}\-\ \-\ \-\ \-\ 2*n1\verb|^|2*d1/dm\verb|^|3+2*n0*n2*d1/dm\verb|^|3+2*n0*n1*d2/dm\verb|^|3+6*n0*n1*d1\verb|^|2/dm\verb|^|4+w2*d1/dm\-\ +\-\ w1*d2/dm+...\color{Green}
$\\$\color{BrickRed}\-\ \-\ \-\ \-\ w1*d1\verb|^|2/dm\verb|^|2+w1*d2/dm+\-\ w0*d3/dm+w0*d1*d2/dm\verb|^|2+w1*d1\verb|^|2/dm\verb|^|2+2*w0*d1*d2/dm\verb|^|2+...\color{Green}
$\\$\color{BrickRed}\-\ \-\ \-\ \-\ 2*w0*d1\verb|^|3/dm\verb|^|3);\color{Green}
$\\$
$\\$
$\\$\color{Green}$\%$\-\ bounds\-\ on\-\ the\-\ 10\-\ sub\-\ integrands
$\\$
$\\$
$\\$\color{BrickRed}vec\-\ =\-\ [\-\ w1*w2;\-\ ...\color{Green}
$\\$\color{BrickRed}\-\ \-\ \-\ \-\ \-\ \-\ \-\ \-\ w0*w2+2*w1*w1;\-\ ...\color{Green}
$\\$\color{BrickRed}\-\ \-\ \-\ \-\ \-\ \-\ \-\ \-\ w1*w0+2*w0*w1;\-\ ...\color{Green}
$\\$\color{BrickRed}\-\ \-\ \-\ \-\ \-\ \-\ \-\ \-\ w0*w0;\-\ ...\color{Green}
$\\$\color{BrickRed}\-\ \-\ \-\ \-\ \-\ \-\ \-\ \-\ w3*w2;\-\ ...\color{Green}
$\\$\color{BrickRed}\-\ \-\ \-\ \-\ \-\ \-\ \-\ \-\ 3*w2*w2+2*w3*w1;\color{Green}
$\\$\color{BrickRed}\-\ \-\ \-\ \-\ \-\ \-\ \-\ \-\ w3*w0+6*w2*w1+3*w1*w2;\color{Green}
$\\$\color{BrickRed}\-\ \-\ \-\ \-\ \-\ \-\ \-\ \-\ 3*w2*w0+6*w1*w1+w0*w2;\color{Green}
$\\$\color{BrickRed}\-\ \-\ \-\ \-\ \-\ \-\ \-\ \-\ 3*w1*w0+2*w0*w1;\color{Green}
$\\$\color{BrickRed}\-\ \-\ \-\ \-\ \-\ \-\ \-\ \-\ w0*w0];\color{Green}
$\\$\color{BrickRed}\-\ \-\ \-\ \-\ \-\ \-\ \-\ \-\ \color{Green}
$\\$\color{Green}$\%$\-\ bound\-\ on\-\ 10\-\ sub\-\ integrands
$\\$\color{BrickRed}vec\-\ =\-\ sup(vec);\color{Green}
$\\$
$\\$
$\\$
$\\$
$\\$
$\\$
$\\$\color{Black}\section{bound\_theta1\_m.m}

\color{Green}\color{BrickRed}\color{NavyBlue}\-\ function\-\ \color{BrickRed}\-\ out\-\ =\-\ bound\_theta1\_m(max\_imag\_x,max\_real\_psi,max\_abs\_q,m)\color{Green}
$\\$
$\\$
$\\$
$\\$\color{Black}
The first Jacobi Theta function is given by the series,
\eqn{
\vartheta_1(z)&= 2\sum_{n=0}^{\infty} (-1)^n q^{(n+1/2)^2}\sin((2n+1)z),
}{}
and its $m$th derivative is given by,
\eqn{
\vartheta_1^{(m)}(z)&= 2\sum_{n=0}^{\infty}(-1)^n q^{(n+1/2)^2}(2n+1)^m f((2n+1)z),
}{}
where $f(\cdot) = \sin(\cdot)$ if $m\equiv 0 \mod 4$, $f(\cdot) = \cos(\cdot)$ if $m \equiv 1 \mod 4$,
$f(\cdot) = -\sin(\cdot)$ if $m\equiv 2 \mod 4$, and $f(\cdot) = -\cos(\cdot)$ if $m\equiv 3 \mod 4$.

$\\$
Now for $m \geq 0$, 
\eqn{
\left|(-1)^{n}q^{(n+1/2)^2}(2n+1)^m f((2n+1)z)\right| & \leq \left| q^{(n+1/2)^2}(2n+1)^m e^{(2n+1)|\Im(z)|}\right|\\
&\leq \left| q^{N^2+1/2}e^{(2N+1)|\Im(z)|}(2N+1)^m\right|q^n,
}{}
for $n\geq N$ with $N$ sufficiently large that
\eqn{
\left| q^{n^2+1/4}e^{(2n+1)|\Im(z)|}(2n+1)^m\right| & \leq \left| q^{N^2+1/4}e^{(2N+1)|\Im(z)|}(2N+1)^m\right|, 
}{}
whenever $n\geq N$. To determine how large $N$ must be, we define,
\eqn{
g(x):= q^{x^2+1/4}e^{(2x+1)|\Im(z)|}(2x+1)^m.
}{}
We will take $N$ large enough that $g'(x) < 0$ whenever $x\geq N$. If $m=0$, then
\eqn{
g'(x) = 2q^{x^2+1/4}e^{(2x+1)|\Im(z)|}\left( x\log(q)+|\Im(z)|\right),
}{}
and we see that 
\eqn{
N &> -\frac{|\Im(z)|}{\log(q)}
}{}
suffices. If $M > 0$,
\eqn{
g'(x) = 2q^{x^2+1/4}e^{(2x+1)|Im(z)|}(2x+1)^{m-1}\left((x\log(q)+|\Im(z)|)(2x+1)+m\right).
}{}
From 
\eqn{
(x\log(q)+|\Im(z)|)(2x+1)+m < 0,
}{}
we find that 
\eqn{
N > -\frac{2|\Im(z)|+\log(q) + \sqrt{(2|\Im(z)|+\log(q))^2-8\log(q)(|\Im(z)|+m)}}{4\log(q)}
}{}
suffices. For such an $N$, the error of the summation truncation is
\eqn{
q^{N^2+1/4}e^{(2N+1)|\Im(z)|}(2N+1)^m \frac{q^N}{1-q}.
}{}

$\\$
\color{Green}
$\\$
$\\$
$\\$
$\\$\color{Green}$\%$$\%$\-\ constants
$\\$\color{BrickRed}pie\-\ =\-\ nm('pi');\color{Green}
$\\$\color{BrickRed}q\-\ =\-\ nm(max\_abs\_q);\color{Green}
$\\$\color{Green}$\%$\-\ \color{Black}Find max of $\vartheta_1(\frac{\pi}{2\omega}(\omega x \pm i\omega' +n\omega +i\beta))$ on 
\color{Black} the ellipse $E_{\rho}$. \color{Green}
$\\$
$\\$
$\\$\color{Green}$\%$\-\ abz\-\ =\-\ pie*(omega*max\_imag\_x+omega\_prime+max\_real\_beta)/(2*omega);
$\\$\color{BrickRed}abz\-\ =\-\ sup(pie*max\_imag\_x\-\ +\-\ abs(log(max\_abs\_q))*(1+max\_real\_psi))/2;\color{Green}
$\\$
$\\$
$\\$\color{BrickRed}qlog\-\ =\-\ log(q);\color{Green}
$\\$\color{BrickRed}one\_fourth\-\ =\-\ nm(1)/4;\color{Green}
$\\$\color{BrickRed}one\_half\-\ =\-\ nm(1)/2;\color{Green}
$\\$\color{BrickRed}con\-\ =\-\ q\verb|^|one\_fourth*exp(abz);\color{Green}
$\\$
$\\$
$\\$\color{Green}$\%$$\%$
$\\$
$\\$
$\\$\color{Green}$\%$\-\ find\-\ N\-\ large\-\ enough\-\ that\-\ 
$\\$\color{Green}$\%$\-\ \color{Black}$f(x):= q^{x^2+1/4}e^{(2x+1)|Im(z)|}(2x+1)^m$ \color{Green}
$\\$\color{Green}$\%$\-\ is\-\ decreasing\-\ for\-\ \color{Black}$ x \geq N$. \color{Green}
$\\$\color{BrickRed}\color{NavyBlue}\-\ if\-\ \color{BrickRed}\-\ m\-\ ==\-\ 0\color{Green}
$\\$\color{BrickRed}\-\ \-\ \-\ \-\ N\-\ =\-\ ceil(sup(-abz/qlog));\color{Green}
$\\$\color{BrickRed}\color{NavyBlue}\-\ else\-\ \color{BrickRed}\color{Green}
$\\$\color{BrickRed}\-\ \-\ \-\ \-\ c1\-\ =\-\ 2*abz+qlog;\color{Green}
$\\$\color{BrickRed}\-\ \-\ \-\ \-\ disc\-\ =\-\ c1\verb|^|2-8*qlog*(abz+m);\color{Green}
$\\$\color{BrickRed}\-\ \-\ \-\ \-\ c2\-\ =\-\ -c1/(4*qlog);\color{Green}
$\\$\color{BrickRed}\-\ \-\ \-\ \-\ c3\-\ =\-\ disc/(4*qlog);\color{Green}
$\\$\color{BrickRed}\-\ \-\ \-\ \-\ N\-\ =\-\ ceil(sup(c2-c3));\color{Green}
$\\$\color{BrickRed}\color{NavyBlue}\-\ end\-\ \color{BrickRed}\color{Green}
$\\$
$\\$
$\\$\color{Green}$\%$\-\ compute\-\ theta(z,q)\-\ and\-\ its\-\ derivatives\-\ with\-\ partial\-\ sum
$\\$\color{Green}$\%$\-\ \color{Black}$\vartheta_1^{(m)}(z)= 2\sum_{n=0}^{N-1}(-1)^n q^{(n+1/2)^2}(2n+1)^m f((2n+1)z)$ \color{Green}
$\\$\color{BrickRed}out\-\ =\-\ nm(zeros(m+1,1));\color{Green}
$\\$\color{BrickRed}\color{NavyBlue}\-\ for\-\ \color{BrickRed}\-\ n\-\ =\-\ 0:N-1\color{Green}
$\\$\color{BrickRed}\-\ \-\ \-\ \-\ prod\-\ =\-\ 2*n+1;\color{Green}
$\\$\color{BrickRed}\-\ \-\ \-\ \-\ zprod\-\ =\-\ prod*abz;\color{Green}
$\\$\color{BrickRed}\-\ \-\ \-\ \-\ gen\-\ =\-\ exp(zprod\-\ +\-\ log(q)*(n+one\_half)\verb|^|2);\color{Green}
$\\$\color{BrickRed}\-\ \-\ \-\ \-\ \color{Green}
$\\$\color{BrickRed}\-\ \-\ \-\ \-\ \color{NavyBlue}\-\ for\-\ \color{BrickRed}\-\ ind\-\ =\-\ 1:m+1\color{Green}
$\\$\color{BrickRed}\-\ \-\ \-\ \-\ \-\ \-\ \-\ \-\ out(ind)\-\ =\-\ out(ind)\-\ +\-\ gen*prod\verb|^|(ind-1);\color{Green}
$\\$\color{BrickRed}\-\ \-\ \-\ \-\ \color{NavyBlue}\-\ end\-\ \color{BrickRed}\color{Green}
$\\$\color{BrickRed}\color{NavyBlue}\-\ end\-\ \color{BrickRed}\color{Green}
$\\$
$\\$
$\\$\color{Green}$\%$\-\ truncation\-\ error
$\\$\color{Green}$\%$\-\ \color{Black}$q^{N^2+1/4}e^{(2N+1)|\Im(z)|}(2N+1)^m \frac{q^N}{1-q}$ \color{Green}
$\\$\color{Green}$\%$\-\ note\-\ that\-\ con\-\ =\-\ \color{Black}$q^{1/4}e^{|\Im(z)|}$ \color{Green}
$\\$
$\\$
$\\$\color{BrickRed}\color{NavyBlue}\-\ for\-\ \color{BrickRed}\-\ ind\-\ =\-\ 0:m\color{Green}
$\\$\color{BrickRed}\-\ \-\ \-\ \-\ out(ind+1)\-\ =\-\ out(ind+1)\-\ +\-\ con*exp(N\verb|^|2*log(q)+2*abz*N)*(2*N+1)\verb|^|ind*q\verb|^|N/(1-q);\color{Green}
$\\$\color{BrickRed}\color{NavyBlue}\-\ end\-\ \color{BrickRed}\color{Green}
$\\$
$\\$
$\\$\color{BrickRed}out\-\ =\-\ 2*out;\color{Green}
$\\$\color{BrickRed}out\-\ =\-\ sup(out);\color{Green}
$\\$\color{Black}\section{bound\_xi.m}

\color{Green}\color{BrickRed}\color{NavyBlue}\-\ function\-\ \color{BrickRed}\-\ out\-\ =\-\ bound\_xi(a\_psi,b\_psi,rho\_psi,a\_q,b\_q,rho\_q,ntilde)\color{Green}
$\\$
$\\$
$\\$\color{Green}$\%$\-\ latex\-\ description
$\\$
$\\$\color{Black}

$\\$
Now 
\eqn{
\xi(\alpha) &= 2i\left(\zeta(\alpha)-\frac{\alpha}{\omega}\zeta(\omega)\right)\\
&= 2i\left(\frac{\pi}{2\omega}\cot\left(\frac{\pi \alpha}{2\omega}\right)+\frac{2\pi}{\omega}
\sum_{k=1}^{\infty} \frac{q^{2k}}{1-q^{2k}}\sin\left(\frac{k\pi \alpha}{\omega}\right)\right).
}{}

$\\$
When $\alpha = \omega + i\psi \omega'$ we have 
\eq{
\xi(\omega + i\psi \omega')&= 2i\left( \frac{\pi}{2\omega}\cot\left(\frac{\pi(\omega + i\psi \omega')}{2\omega}\right) 
+ \frac{2\pi}{\omega} \sum_{k=1}^{\infty} \frac{q^{2k}}{1-q^{2k}}
\sin\left(\frac{k\pi (\omega + i\psi \omega')}{\omega}\right)\right)\\
&= 2i\left(-\frac{\pi}{2\omega} \tan\left( \frac{i \pi \psi \omega'}{2\omega}\right) 
+ \frac{2\pi}{\omega} \sum_{k=1}^{\infty}
\frac{q^{2k}}{1-q^{2k}}(-1)^k \sin\left(\frac{ik\pi \psi \omega'}{\omega}\right)\right).
}{\label{efdef}}

$\\$
When $\alpha = i\psi \omega'$ we have,
\eqn{
\xi(i\psi \omega')&= 2i\left(\frac{\pi}{2\omega}\cot\left(\frac{\pi(i\psi \omega')}{2\omega}\right) 
+ \frac{2\pi}{\omega} \sum_{k=1}^{\infty} \frac{q^{2k}}{1-q^{2k}}
\sin\left(\frac{k\pi ( i\psi \omega')}{\omega}\right)\right).
}{\label{efdef2}}

$\\$

$\\$
Note that if either $\psi \in [0,1]$ and $0<|q|< 1$ with $q\in \C$, or if $q\in [q_a,q_b]\subset (0,1)$, 
$|\Re(\psi)| < 2$ and
 $|\Im(\psi)|< \frac{\pi}{|\log(q_a)|}$, then
\eqref{efdef} is analytic in that region. If either $\psi \in (0,1]$ and $0<|q|< 1$ with
 $q\in \C$, or if $q\in [q_a,q_b]\subset (0,1)$, $|\Re(\psi)| < 2$  and
 $|\psi|>0$, then
\eqref{efdef2} is analytic. 

$\\$
Note that $\sin\left(\frac{i\pi \psi \omega'k}{\omega}\right) = \frac{1}{2i}\left(q^{\psi k}-q^{-\psi k}\right)$
 so that in both cases the infinite sum is bounded by 
\eqn{
2\sum_{k=1}^{\infty} \frac{q_0^{\gamma k}}{1-q_0^{2k}}&\leq
\frac{2}{1-q_0^2}\sum_{k=1}^{\infty} q_0^{\gamma k}\\
&\leq \frac{2 q_0^{\gamma }}{(1-q_0^2)(1-q_0^{\gamma})}.
}{}
where $|q|\leq q_0 < 1$ and $\gamma := 2-|\Re(\psi)|$.

$\\$

$\\$
Let $M$ be a bound on 
 $\tan\left( \frac{-i\psi\log(q)}{2}\right)$ or $\cot\left( \frac{-i\psi\log(q)}{2}\right)$, depending on the
 choice of $\alpha$. Then

$\\$
\eqn{
\left|\omega \xi(\tilde n + i\psi \omega')\right| &
 \leq M\pi+\frac{8\pi q_0^{\gamma }}{(1-q_0^2)(1-q_0^{\gamma})}.
}{}

$\\$
\color{Green}
$\\$
$\\$\color{Green}$\%$$\%$
$\\$
$\\$
$\\$\color{BrickRed}pie\-\ =\-\ nm('pi');\color{Green}
$\\$
$\\$
$\\$\color{Green}$\%$
$\\$\color{Green}$\%$\-\ get\-\ bound\-\ on\-\ tan\-\ or\-\ cot
$\\$\color{Green}$\%$
$\\$
$\\$
$\\$\color{BrickRed}pts\-\ =\-\ 300;\color{Green}
$\\$\color{BrickRed}theta\-\ =\-\ nm(linspace(0,sup(2*pie),pts));\color{Green}
$\\$\color{BrickRed}\color{NavyBlue}\-\ if\-\ \color{BrickRed}\-\ rho\_psi\-\ ==\-\ 1\color{Green}
$\\$\color{BrickRed}\-\ \-\ \-\ \-\ psivec\-\ =\-\ nm(linspace(inf(a\_psi),sup(b\_psi),pts));\color{Green}
$\\$\color{BrickRed}\color{NavyBlue}\-\ else\-\ \color{BrickRed}\color{Green}
$\\$\color{BrickRed}\-\ \-\ \-\ \-\ psivec\-\ =\-\ nm(1)/2+\-\ (rho\_psi*exp(1i*theta)+exp(-1i*theta)/rho\_psi)/4;\color{Green}
$\\$\color{BrickRed}\color{NavyBlue}\-\ end\-\ \color{BrickRed}\color{Green}
$\\$
$\\$
$\\$\color{BrickRed}\color{NavyBlue}\-\ if\-\ \color{BrickRed}\-\ rho\_q\-\ ==\-\ 1\color{Green}
$\\$\color{BrickRed}\-\ \-\ \-\ \-\ qvec\-\ =\-\ nm(linspace(inf(a\_q),sup(b\_q),pts));\color{Green}
$\\$\color{BrickRed}\color{NavyBlue}\-\ else\-\ \color{BrickRed}\color{Green}
$\\$\color{BrickRed}\-\ \-\ \-\ \-\ qvec\-\ =\-\ (a\_q+b\_q)/2+(b\_q-a\_q)*(rho\_q*exp(1i*theta)+exp(-1i*theta)/rho\_q)/4;\color{Green}
$\\$\color{BrickRed}\color{NavyBlue}\-\ end\-\ \color{BrickRed}\color{Green}
$\\$
$\\$
$\\$\color{BrickRed}qv\-\ =\-\ nm(qvec(1:end-1),qvec(2:end));\color{Green}
$\\$\color{BrickRed}psiv\-\ =\-\ nm(psivec(1:end-1),psivec(2:end));\color{Green}
$\\$\color{BrickRed}\-\ \-\ \-\ \-\ \color{Green}
$\\$\color{Green}$\%$\-\ temp\-\ =\-\ sup(abs(tan(-1i*psiv.'*log(qv)/2)))
$\\$
$\\$
$\\$\color{BrickRed}\color{NavyBlue}\-\ if\-\ \color{BrickRed}\-\ ntilde\-\ ==\-\ 1\color{Green}
$\\$\color{BrickRed}\-\ \-\ \-\ \-\ \color{Green}$\%$\-\ get\-\ bound\-\ on\-\ tan(1i*pi*psi*omega\_prime/(2*omega))\-\ \-\ \-\ \-\ 
$\\$\color{BrickRed}\-\ \-\ \-\ \-\ temp\-\ =\-\ sup(abs(tan(-1i*psiv.'*log(qv)/2)));\color{Green}
$\\$\color{BrickRed}\color{NavyBlue}\-\ elseif\-\ \color{BrickRed}\-\ ntilde\-\ ==\-\ 0\color{Green}
$\\$\color{BrickRed}\-\ \-\ \-\ \-\ \color{Green}$\%$\-\ get\-\ bound\-\ on\-\ cot(1i*pi*psi*omega\_prime/(2*omega))\-\ \-\ \-\ \-\ 
$\\$\color{BrickRed}\-\ \-\ \-\ \-\ temp\-\ =\-\ sup(abs(cot(-1i*psiv.'*log(qv)/2)));\color{Green}
$\\$\color{BrickRed}\color{NavyBlue}\-\ end\-\ \color{BrickRed}\color{Green}
$\\$
$\\$
$\\$\color{BrickRed}\color{NavyBlue}\-\ if\-\ \color{BrickRed}\-\ sum(sum(isnan(temp)))\-\ $>$\-\ 0\color{Green}
$\\$\color{BrickRed}\-\ \-\ \-\ \-\ error('NaN\-\ present');\color{Green}
$\\$\color{BrickRed}\color{NavyBlue}\-\ end\-\ \color{BrickRed}\color{Green}
$\\$
$\\$
$\\$\color{BrickRed}M\-\ =\-\ max(max(temp));\color{Green}
$\\$
$\\$
$\\$\color{BrickRed}max\_abs\_real\_psi\-\ =\-\ (a\_psi\-\ +\-\ b\_psi)/2+(b\_psi-a\_psi)*(rho\_psi\-\ +1/rho\_psi)/4;\color{Green}
$\\$\color{BrickRed}max\_abs\_q\-\ =\-\ (a\_q\-\ +\-\ b\_q)/2+(b\_q-a\_q)*(rho\_q\-\ +1/rho\_q)/4;\color{Green}
$\\$
$\\$
$\\$\color{BrickRed}gamma\-\ =\-\ 2-max\_abs\_real\_psi;\color{Green}
$\\$\color{BrickRed}qg\-\ =\-\ max\_abs\_q\verb|^|gamma;\color{Green}
$\\$\color{BrickRed}out\-\ =\-\ pie*M+\-\ \-\ (8*pie*qg)/((1-max\_abs\_q\verb|^|2)*(1-qg));\color{Green}
$\\$
$\\$
$\\$
$\\$
$\\$
$\\$
$\\$
$\\$
$\\$
$\\$
$\\$
$\\$
$\\$
$\\$
$\\$\color{Black}\section{bound\_xi\_der.m}

\color{Green}\color{BrickRed}\color{NavyBlue}\-\ function\-\ \color{BrickRed}\-\ out\-\ =\-\ bound\_xi\_der(abs\_q,min\_abs\_q,a,b,max\_real\_psi,rho\_q,rho\_psi,a\_psi,b\_psi,ntilde)\color{Green}
$\\$
$\\$
$\\$
$\\$
$\\$
$\\$
$\\$\color{Green}$\%$\-\ latex\-\ comments
$\\$
$\\$
$\\$
$\\$\color{Black}
If $\tilde n = 1$, then from the q-series representation of the Weierstrass elliptic function, 
\eq{
\pd{}{\psi} \xi(\tilde n \omega + i\psi\omega') & = 2\omega'\left( \wp(\tilde\omega + i\psi\omega') + \frac{\zeta(\omega)}{\omega}\right)\\
&= 2\omega'\left( \left(\frac{\pi}{2\omega}\right)^2 \sec^2\left(\frac{i\pi \psi\omega'}{2\omega}\right) - \frac{2\pi^2}{\omega^2}
\sum_{k=1}^{\infty} (-1)^k\frac{kq^{2k}}{1-q^{2k}}\cos\left(\frac{ik \pi \psi\omega' }{\omega}\right)\right)\\
&= 2\omega'\left( \left(\frac{\pi}{2\omega}\right)^2 \sec^2\left(\frac{i\pi \psi\omega'}{2\omega}\right) - \frac{2\pi^2}{\omega^2}
\sum_{k=1}^{\infty} (-1)^k\frac{kq^{2k}}{1-q^{2k}}\left(\frac{q^{\psi k}+q^{-\psi k}}{2}\right)\right).
}{\label{efdef}}

$\\$
If $\tilde n = 0$, then from the q-series representation of the Weierstrass elliptic function, 
\eq{
\pd{}{\psi} \xi(\tilde n \omega + i\psi\omega') & = 2\omega'\left( \wp(\tilde\omega + i\psi\omega') + \frac{\zeta(\omega)}{\omega}\right)\\
&= 2\omega'\left( \left(\frac{\pi}{2\omega}\right)^2 \csc^2\left(\frac{i\pi \psi\omega'}{2\omega}\right) - \frac{2\pi^2}{\omega^2}
\sum_{k=1}^{\infty} \frac{kq^{2k}}{1-q^{2k}}\cos\left(\frac{ik \pi \psi\omega' }{\omega}\right)\right)\\
&= 2\omega'\left( \left(\frac{\pi}{2\omega}\right)^2 \csc^2\left(\frac{i\pi \psi\omega'}{2\omega}\right) - \frac{2\pi^2}{\omega^2}
\sum_{k=1}^{\infty} \frac{kq^{2k}}{1-q^{2k}}\left(\frac{q^{\psi k}-q^{-\psi k}}{2i}\right)\right).
}{\label{efdef2}}

$\\$

$\\$
Note that if either $\psi \in [0,1]$ is fixed and $|q|< 1$ with $q\in \C$, or if $q\in [q_a,q_b]\subset (0,1)$ is fixed and
 $|\Im(\psi)|< \frac{\pi}{|\log(q_a)|}$, then
\eqref{efdef} is analytic in that region. Note that if either $\psi \in [0,1]$ is fixed and $|q|< 1$ with $q\in \C$, or if $q\in [q_a,q_b]\subset (0,1)$ is fixed and
 $\Re(\psi)>0$, then
\eqref{efdef} is analytic in that region.

$\\$
Let $0\leq q_0 <1$, $\gamma \in \R$, 
 and define 
\eq{
f(x)&:= \sum_{k=0}^{\infty} \frac{1}{\gamma \log(q_0)} q_0^{\gamma k x}\\
&= \frac{1}{\gamma \log(q_0)} \frac{1}{1-q_0^{\gamma x}}.
}{}
Note that
\eq{
f'(x)& = \sum_{k=0}^{\infty}kq_0^{\gamma x}\\
&= \frac{q_0^{\gamma x}}{(1-q_0^{\gamma x})^2}.
}{}

$\\$
Then
\eq{
\left| \sum_{k=1}^{\infty}
(-1)^{\tilde nk} \frac{kq^{2k}}{1-q^{2k}}\left(q^{\psi k}+q^{-\psi k}\right)\right| &\leq
2\sum_{k=0}^{\infty} \frac{kq_0^{\gamma k}}{1-q_0^{2k}}\\
&\leq \frac{2}{1-q_0^2} \sum_{k=0}^{\infty} kq_0^{\gamma k}\\
&\leq \frac{2}{1-q_0^2}\frac{q_0^{\gamma}}{(1-q_0^{\gamma})^2},
}{}
where $|q|\leq q_0<1$ and $\gamma:= 2-\Re(\psi)$.

$\\$
\color{Green}
$\\$
$\\$\color{Green}$\%$$\%$
$\\$
$\\$
$\\$\color{Green}$\%$
$\\$\color{Green}$\%$\-\ constants
$\\$\color{Green}$\%$
$\\$
$\\$
$\\$\color{BrickRed}pie\-\ =\-\ nm('pi');\color{Green}
$\\$\color{BrickRed}gamma\-\ =\-\ 2-max\_real\_psi;\color{Green}
$\\$\color{BrickRed}qg\-\ =\-\ abs\_q\verb|^|gamma;\color{Green}
$\\$
$\\$
$\\$\color{Green}$\%$\-\ error\-\ check
$\\$\color{BrickRed}\color{NavyBlue}\-\ if\-\ \color{BrickRed}\-\ inf(gamma)\-\ $<$=\-\ 0\color{Green}
$\\$\color{BrickRed}\-\ \-\ \-\ \-\ error('max\_psi\-\ too\-\ big');\color{Green}
$\\$\color{BrickRed}\color{NavyBlue}\-\ end\-\ \color{BrickRed}\color{Green}
$\\$
$\\$
$\\$\color{Green}$\%$\-\ 
$\\$\color{Green}$\%$\-\ get\-\ bound\-\ on\-\ (sec/csc)(1i*pi*psi*omega\_prime/(2*omega))
$\\$\color{Green}$\%$
$\\$
$\\$
$\\$\color{BrickRed}pnts\-\ =\-\ 500;\color{Green}
$\\$\color{Green}$\%$\-\ bound\-\ on\-\ stadium\-\ in\-\ variable\-\ psi
$\\$\color{BrickRed}qvec\-\ =\-\ nm(linspace(inf(a),sup(b),pnts));\color{Green}
$\\$\color{BrickRed}theta\-\ =\-\ nm(linspace(0,sup(2*pie),pnts));\color{Green}
$\\$\color{BrickRed}psivec\-\ =\-\ (a\_psi+b\_psi)/2+\-\ (b\_psi-a\_psi)*(rho\_psi*exp(1i*theta)+exp(-1i*theta)/rho\_psi)/4;\color{Green}
$\\$\color{BrickRed}qv\-\ =\-\ nm(qvec(1:end-1),qvec(2:end));\color{Green}
$\\$\color{BrickRed}psiv\-\ =\-\ nm(psivec(1:end-1),psivec(2:end));\color{Green}
$\\$\color{BrickRed}\color{NavyBlue}\-\ if\-\ \color{BrickRed}\-\ ntilde\-\ ==\-\ 1\color{Green}
$\\$\color{BrickRed}\-\ \-\ \-\ \-\ temp\-\ =\-\ sup(abs(sec(-1i*psiv.'*log(qv)/2)));\color{Green}
$\\$\color{BrickRed}\color{NavyBlue}\-\ elseif\-\ \color{BrickRed}\-\ ntilde\-\ ==\-\ 0\color{Green}
$\\$\color{BrickRed}\-\ \-\ \-\ \-\ temp\-\ =\-\ sup(abs(csc(-1i*psiv.'*log(qv)/2)));\color{Green}
$\\$\color{BrickRed}\color{NavyBlue}\-\ end\-\ \color{BrickRed}\color{Green}
$\\$\color{BrickRed}\color{NavyBlue}\-\ if\-\ \color{BrickRed}\-\ sum(sum(isnan(temp)))\-\ $>$\-\ 0\color{Green}
$\\$\color{BrickRed}\-\ \-\ \-\ \-\ error('NaN\-\ present');\color{Green}
$\\$\color{BrickRed}\color{NavyBlue}\-\ end\-\ \color{BrickRed}\color{Green}
$\\$\color{BrickRed}max\_temp1\-\ =\-\ max(max(temp));\color{Green}
$\\$
$\\$
$\\$\color{Green}$\%$\-\ bound\-\ on\-\ stadium\-\ in\-\ variable\-\ q
$\\$\color{BrickRed}psivec\-\ =\-\ nm(linspace(a\_psi,b\_psi,pnts));\color{Green}
$\\$\color{BrickRed}theta\-\ =\-\ nm(linspace(0,sup(2*pie),pnts));\color{Green}
$\\$\color{BrickRed}qvec\-\ =\-\ (a+b)/2+(b-a)*(rho\_q*exp(1i*theta)+exp(-1i*theta)/rho\_q)/4;\color{Green}
$\\$\color{BrickRed}qv\-\ =\-\ nm(qvec(1:end-1),qvec(2:end));\color{Green}
$\\$\color{BrickRed}psiv\-\ =\-\ nm(psivec(1:end-1),psivec(2:end));\color{Green}
$\\$
$\\$
$\\$\color{BrickRed}\color{NavyBlue}\-\ if\-\ \color{BrickRed}\-\ ntilde\-\ ==\-\ 1\color{Green}
$\\$\color{BrickRed}\-\ \-\ \-\ \-\ temp\-\ =\-\ sup(abs(sec(-1i*psiv.'*log(qv)/2)));\color{Green}
$\\$\color{BrickRed}\color{NavyBlue}\-\ elseif\-\ \color{BrickRed}\-\ ntilde\-\ ==\-\ 0\color{Green}
$\\$\color{BrickRed}\-\ \-\ \-\ \-\ temp\-\ =\-\ sup(abs(csc(-1i*psiv.'*log(qv)/2)));\color{Green}
$\\$\color{BrickRed}\color{NavyBlue}\-\ end\-\ \color{BrickRed}\color{Green}
$\\$\color{BrickRed}\color{NavyBlue}\-\ if\-\ \color{BrickRed}\-\ sum(sum(isnan(temp)))\-\ $>$\-\ 0\color{Green}
$\\$\color{BrickRed}\-\ \-\ \-\ \-\ error('NaN\-\ present');\color{Green}
$\\$\color{BrickRed}\color{NavyBlue}\-\ end\-\ \color{BrickRed}\color{Green}
$\\$\color{BrickRed}max\_temp2\-\ =\-\ max(max(temp));\color{Green}
$\\$
$\\$
$\\$\color{Green}$\%$\-\ take\-\ maximum\-\ of\-\ two\-\ bounds
$\\$\color{BrickRed}temp\_bd\-\ =\-\ nm(max(max\_temp1,max\_temp2));\color{Green}
$\\$
$\\$
$\\$\color{Green}$\%$\-\ 
$\\$\color{Green}$\%$\-\ bound\-\ on\-\ infinite\-\ sum\-\ part
$\\$\color{Green}$\%$
$\\$
$\\$
$\\$\color{BrickRed}out\-\ =\-\ (2/(1-abs\_q\verb|^|2))*(qg/(1-qg)\verb|^|2);\color{Green}
$\\$
$\\$
$\\$\color{Green}$\%$
$\\$\color{Green}$\%$\-\ combine\-\ all\-\ parts
$\\$\color{Green}$\%$
$\\$
$\\$
$\\$\color{BrickRed}out\-\ =\-\ 2*abs(log(min\_abs\_q))*pie*(out\-\ +\-\ temp\_bd\verb|^|2/4);\color{Green}
$\\$
$\\$
$\\$
$\\$
$\\$
$\\$
$\\$
$\\$
$\\$
$\\$
$\\$
$\\$
$\\$
$\\$
$\\$
$\\$
$\\$
$\\$
$\\$\color{Black}\section{cf\_biv\_cheby.m}

\color{Green}\color{BrickRed}\color{NavyBlue}\-\ function\-\ \color{BrickRed}\-\ cf\-\ =\-\ cf\_biv\_cheby(m,n,num\_funs,fun)\color{Green}
$\\$\color{Green}$\%$\-\ function\-\ cf\_padau(m,n,fun)
$\\$\color{Green}$\%$
$\\$\color{Green}$\%$\-\ Returns\-\ the\-\ coefficients\-\ for\-\ bivariate\-\ Chebyshev\-\ interpolation
$\\$
$\\$
$\\$
$\\$\color{Black}
Let $f(x,y):[-1,1]\times [-1,1] \to \mathbb{C}$. Let $T_i(x)$, $T_j(y)$ be
Chebyshev polynomials of the first kind. This function returns a matrix 
cf of coefficients to the bivariate interpolation polynomial, 
$p(x,y) = \sum_{i=0}^m\sum_{j=0}^n c_{i,j} T_i(x)T_j(y)$,
where $c_{i,j}$ is the $ith+1$, $jth+1$ row and column of the matrix cf.
\color{Green}
$\\$
$\\$
$\\$
$\\$\color{Green}$\%$\-\ constants
$\\$\color{BrickRed}pie\-\ =\-\ nm('pi');\color{Green}
$\\$\color{BrickRed}c\_1\-\ =\-\ pie/(2*(m+1));\color{Green}
$\\$\color{BrickRed}c\_2\-\ =\-\ pie/(2*(n+1));\color{Green}
$\\$
$\\$
$\\$\color{Green}$\%$\-\ theta\-\ for\-\ x\-\ and\-\ y
$\\$\color{BrickRed}theta\_xr\-\ =\-\ c\_1*(2*(0:1:m)+1);\color{Green}
$\\$\color{BrickRed}theta\_ys\-\ =\-\ c\_2*(2*(0:1:n)+1);\color{Green}
$\\$
$\\$
$\\$\color{Green}$\%$\-\ x\-\ and\-\ y\-\ interpolation\-\ points
$\\$\color{BrickRed}xr\-\ =\-\ cos(theta\_xr);\color{Green}
$\\$\color{BrickRed}ys\-\ =\-\ cos(theta\_ys);\-\ \color{Green}
$\\$
$\\$
$\\$\color{Green}$\%$\-\ evaluate\-\ function\-\ at\-\ grid\-\ points
$\\$\color{BrickRed}f\_xy\-\ =\-\ nm(zeros(m+1,n+1,num\_funs));\color{Green}
$\\$
$\\$
$\\$\color{BrickRed}fun2\-\ =\-\ \@(x)(fun(x,ys));\color{Green}
$\\$
$\\$
$\\$\color{BrickRed}\color{NavyBlue}\-\ for\-\ \color{BrickRed}\-\ r\-\ =\-\ 1:m+1\color{Green}
$\\$\color{BrickRed}\-\ \-\ \-\ \-\ \-\ \-\ \-\ \color{Green}
$\\$\color{BrickRed}\-\ \-\ \-\ \-\ \color{Green}$\%$
$\\$\color{BrickRed}\-\ \-\ \-\ \-\ \color{Green}$\%$\-\ get\-\ function\-\ values
$\\$\color{BrickRed}\-\ \-\ \-\ \-\ \color{Green}$\%$
$\\$\color{BrickRed}\-\ \-\ \-\ \-\ \color{Green}
$\\$\color{BrickRed}\-\ \-\ \-\ \-\ f\_xy(r,:,:)\-\ =\-\ fun2(xr(r));\color{Green}
$\\$\color{BrickRed}\-\ \-\ \-\ \-\ \-\ \-\ \-\ \-\ \color{Green}
$\\$\color{BrickRed}\color{NavyBlue}\-\ end\-\ \color{BrickRed}\color{Green}
$\\$
$\\$
$\\$\color{Green}$\%$\-\ get\-\ Chebyshev\-\ polynomials\-\ evaluated\-\ at\-\ points;
$\\$\color{BrickRed}Tx\-\ =\-\ cos((0:1:m).'*theta\_xr);\color{Green}
$\\$\color{BrickRed}Ty\-\ =\-\ cos(theta\_ys.'*(0:1:n));\color{Green}
$\\$
$\\$
$\\$\color{BrickRed}cf\-\ =\-\ nm(zeros(m+1,n+1,num\_funs));\color{Green}
$\\$
$\\$
$\\$\color{BrickRed}\color{NavyBlue}\-\ for\-\ \color{BrickRed}\-\ j\-\ =\-\ 1:num\_funs\color{Green}
$\\$\color{BrickRed}\-\ \-\ \-\ \-\ cf(:,:,j)\-\ =\-\ (4/((m+1)*(n+1)))*Tx*f\_xy(:,:,j)*Ty;\color{Green}
$\\$\color{BrickRed}\-\ \-\ \-\ \-\ cf(:,1,j)\-\ =\-\ cf(:,1,j)/2;\color{Green}
$\\$\color{BrickRed}\-\ \-\ \-\ \-\ cf(1,:,j)\-\ =\-\ cf(1,:,j)/2;\color{Green}
$\\$\color{BrickRed}\color{NavyBlue}\-\ end\-\ \color{BrickRed}\color{Green}
$\\$
$\\$
$\\$\color{Black}\section{cf\_eval.m}

\color{Green}\color{BrickRed}\color{NavyBlue}\-\ function\-\ \color{BrickRed}\-\ out\-\ =\-\ cf\_eval(cf,x,y)\color{Green}
$\\$\color{Green}$\%$\-\ function\-\ out\-\ =\-\ cf\_eval(cf,x,y)
$\\$\color{Green}$\%$
$\\$\color{Green}$\%$\-\ Evaluates\-\ the\-\ two\-\ dimensional\-\ Chebyshev\-\ polynomial\-\ at\-\ (x,y)
$\\$
$\\$
$\\$
$\\$\color{Black}
Let $f(x,y):[-1,1]\times [-1,1] \to \mathbb{C}$. Let $T_i(x)$, $T_j(y)$ be
Chebyshev polynomials of the first kind. This function evaluates  
$p(x,y) = \sum_{i=0}^m\sum_{j=0}^n c_{i,j} T_i(x)T_j(y)$,
where $c_{i,j}$ is the $ith+1$, $jth+1$ row and column of the matrix cf.
\color{Green}
$\\$
$\\$
$\\$
$\\$
$\\$
$\\$
$\\$
$\\$\color{BrickRed}m\-\ =\-\ size(cf,1);\color{Green}
$\\$\color{BrickRed}n\-\ =\-\ size(cf,2);\color{Green}
$\\$\color{BrickRed}theta\_x\-\ =\-\ acos(x);\color{Green}
$\\$\color{BrickRed}theta\_y\-\ =\-\ acos(y);\color{Green}
$\\$\color{BrickRed}ind\_x\-\ =\-\ 0:1:(m-1);\color{Green}
$\\$\color{BrickRed}ind\_y\-\ =\-\ 0:1:(n-1);\color{Green}
$\\$
$\\$
$\\$
$\\$
$\\$\color{BrickRed}Tx\-\ =\-\ cos(ind\_x.'*theta\_x);\color{Green}
$\\$\color{BrickRed}Ty\-\ =\-\ cos(theta\_y.'*ind\_y);\color{Green}
$\\$
$\\$
$\\$\color{BrickRed}out\-\ =\-\ nm(zeros(length(x),length(y)));\color{Green}
$\\$\color{BrickRed}\color{NavyBlue}\-\ for\-\ \color{BrickRed}\-\ j\-\ =\-\ 1:length(x)\color{Green}
$\\$\color{BrickRed}\-\ \-\ \-\ \-\ \color{NavyBlue}\-\ for\-\ \color{BrickRed}\-\ k\-\ =\-\ 1:length(y)\color{Green}
$\\$\color{BrickRed}\-\ \-\ \-\ \-\ \-\ \-\ \-\ \-\ out(j,k)\-\ =sum(sum(cf.*(Tx(:,j)*Ty(k,:))));\color{Green}
$\\$\color{BrickRed}\-\ \-\ \-\ \-\ \color{NavyBlue}\-\ end\-\ \color{BrickRed}\color{Green}
$\\$\color{BrickRed}\color{NavyBlue}\-\ end\-\ \color{BrickRed}\color{Green}
$\\$
$\\$
$\\$\color{Black}\section{cf\_eval\_theta.m}

\color{Green}\color{BrickRed}\color{NavyBlue}\-\ function\-\ \color{BrickRed}\-\ out\-\ =\-\ cf\_eval\_theta(cf,theta\_x,theta\_y)\color{Green}
$\\$\color{Green}$\%$\-\ function\-\ out\-\ =\-\ cf\_eval(cf,x,y)
$\\$\color{Green}$\%$
$\\$\color{Green}$\%$\-\ Evaluates\-\ the\-\ two\-\ dimensional\-\ Chebyshev\-\ polynomial\-\ at\-\ (x,y)
$\\$
$\\$
$\\$
$\\$\color{Black}
Let $f(x,y):[-1,1]\times [-1,1] \to \mathbb{C}$. Let $T_i(x)$, $T_j(y)$ be
Chebyshev polynomials of the first kind. This function evaluates  
$p(x,y) = \sum_{i=0}^m\sum_{j=0}^n c_{i,j} T_i(x)T_j(y)$,
where $c_{i,j}$ is the $ith+1$, $jth+1$ row and column of the matrix cf.
\color{Green}
$\\$
$\\$
$\\$
$\\$\color{BrickRed}m\-\ =\-\ size(cf,1);\color{Green}
$\\$\color{BrickRed}n\-\ =\-\ size(cf,2);\color{Green}
$\\$\color{Green}$\%$\-\ theta\_x\-\ =\-\ acos(x);
$\\$\color{Green}$\%$\-\ theta\_y\-\ =\-\ acos(y);
$\\$\color{BrickRed}ind\_x\-\ =\-\ 0:1:(m-1);\color{Green}
$\\$\color{BrickRed}ind\_y\-\ =\-\ 0:1:(n-1);\color{Green}
$\\$
$\\$
$\\$
$\\$
$\\$\color{BrickRed}Tx\-\ =\-\ cos(ind\_x.'*theta\_x);\color{Green}
$\\$\color{BrickRed}Ty\-\ =\-\ cos(theta\_y.'*ind\_y);\color{Green}
$\\$
$\\$
$\\$\color{BrickRed}out\-\ =\-\ nm(zeros(length(theta\_x),length(theta\_y)));\color{Green}
$\\$\color{BrickRed}\color{NavyBlue}\-\ for\-\ \color{BrickRed}\-\ j\-\ =\-\ 1:length(theta\_x)\color{Green}
$\\$\color{BrickRed}\-\ \-\ \-\ \-\ \color{NavyBlue}\-\ for\-\ \color{BrickRed}\-\ k\-\ =\-\ 1:length(theta\_y)\color{Green}
$\\$\color{BrickRed}\-\ \-\ \-\ \-\ \-\ \-\ \-\ \-\ out(j,k)\-\ =sum(sum(cf.*(Tx(:,j)*Ty(k,:))));\color{Green}
$\\$\color{BrickRed}\-\ \-\ \-\ \-\ \color{NavyBlue}\-\ end\-\ \color{BrickRed}\color{Green}
$\\$\color{BrickRed}\color{NavyBlue}\-\ end\-\ \color{BrickRed}\color{Green}
$\\$
$\\$
$\\$\color{Black}\section{cheby\_eval.m}

\color{Green}\color{BrickRed}\color{NavyBlue}\-\ function\-\ \color{BrickRed}\-\ out\-\ =\-\ cheby\_eval(a,ltx,rtx,lty,rty,ints\_x,ints\_y)\color{Green}
$\\$\color{Green}$\%$\-\ cheby\_taylor(a,ltx,rtx,lty,rty,ints\_x,ints\_y)
$\\$\color{Green}$\%$
$\\$\color{Green}$\%$\-\ a\-\ =\-\ Chebyshev\-\ coefficients
$\\$\color{Green}$\%$\-\ ltx\-\ =\-\ left\-\ theta\-\ value\-\ for\-\ variable\-\ x
$\\$\color{Green}$\%$\-\ rtx\-\ =\-\ right\-\ theta\-\ value\-\ for\-\ variable\-\ x
$\\$\color{Green}$\%$\-\ lty\-\ =\-\ left\-\ theta\-\ value\-\ for\-\ variable\-\ y
$\\$\color{Green}$\%$\-\ rty\-\ =\-\ right\-\ theta\-\ value\-\ for\-\ variable\-\ y
$\\$\color{Green}$\%$\-\ ints\_x\-\ =\-\ number\-\ of\-\ intervals\-\ in\-\ variable\-\ x
$\\$\color{Green}$\%$\-\ ints\_y\-\ =\-\ number\-\ of\-\ intervals\-\ in\-\ variable\-\ y
$\\$
$\\$
$\\$\color{BrickRed}Nx\-\ =\-\ size(a,1)-1;\-\ \color{Green}$\%$\-\ degree\-\ of\-\ polynomial\-\ in\-\ x
$\\$\color{BrickRed}Ny\-\ =\-\ size(a,2)-1;\-\ \color{Green}$\%$\-\ degeree\-\ of\-\ polynomial\-\ in\-\ y
$\\$\color{BrickRed}tx\-\ =\-\ linspace(ltx,rtx,ints\_x+1).';\-\ \color{Green}$\%$\-\ theta\-\ points\-\ in\-\ x
$\\$\color{BrickRed}ty\-\ =\-\ linspace(lty,rty,ints\_y+1);\-\ \color{Green}$\%$\-\ theta\-\ points\-\ in\-\ y
$\\$\color{BrickRed}indx\-\ =\-\ 0:1:Nx;\-\ \color{Green}
$\\$\color{BrickRed}indy\-\ =\-\ (0:1:Ny).';\color{Green}
$\\$
$\\$
$\\$\color{BrickRed}zx\-\ =\-\ \-\ nm(tx)*indx;\-\ \color{Green}$\%$\-\ function\-\ argument\-\ for\-\ x
$\\$\color{BrickRed}zy\-\ =\-\ indy*nm(ty);\-\ \color{Green}$\%$\-\ function\-\ argument\-\ for\-\ y
$\\$
$\\$
$\\$\color{Green}$\%$\-\ cos\-\ and\-\ sin
$\\$\color{BrickRed}cosx\-\ =\-\ cos(zx);\color{Green}
$\\$\color{BrickRed}cosy\-\ =\-\ cos(zy);\color{Green}
$\\$
$\\$
$\\$\color{BrickRed}out\-\ =\-\ cosx*a*cosy;\color{Green}
$\\$
$\\$
$\\$
$\\$
$\\$
$\\$
$\\$
$\\$
$\\$
$\\$
$\\$
$\\$
$\\$\color{BrickRed}\-\ \-\ \-\ \color{Green}
$\\$\color{Black}\section{cheby\_taylor.m}

\color{Green}\color{BrickRed}\color{NavyBlue}\-\ function\-\ \color{BrickRed}\-\ out\-\ =\-\ cheby\_taylor(a,ltx,rtx,lty,rty,ints\_x,ints\_y)\color{Green}
$\\$\color{Green}$\%$\-\ cheby\_taylor(a,ltx,rtx,lty,rty,ints\_x,ints\_y)
$\\$\color{Green}$\%$
$\\$\color{Green}$\%$\-\ a\-\ =\-\ Chebyshev\-\ coefficients
$\\$\color{Green}$\%$\-\ ltx\-\ =\-\ left\-\ theta\-\ value\-\ for\-\ variable\-\ x
$\\$\color{Green}$\%$\-\ rtx\-\ =\-\ right\-\ theta\-\ value\-\ for\-\ variable\-\ x
$\\$\color{Green}$\%$\-\ lty\-\ =\-\ left\-\ theta\-\ value\-\ for\-\ variable\-\ y
$\\$\color{Green}$\%$\-\ rty\-\ =\-\ right\-\ theta\-\ value\-\ for\-\ variable\-\ y
$\\$\color{Green}$\%$\-\ ints\_x\-\ =\-\ number\-\ of\-\ intervals\-\ in\-\ variable\-\ x
$\\$\color{Green}$\%$\-\ ints\_y\-\ =\-\ number\-\ of\-\ intervals\-\ in\-\ variable\-\ y
$\\$
$\\$
$\\$\color{BrickRed}Nx\-\ =\-\ size(a,1)-1;\-\ \color{Green}$\%$\-\ degree\-\ of\-\ polynomial\-\ in\-\ x
$\\$\color{BrickRed}Ny\-\ =\-\ size(a,2)-1;\-\ \color{Green}$\%$\-\ degeree\-\ of\-\ polynomial\-\ in\-\ y
$\\$\color{BrickRed}tx\-\ =\-\ linspace(ltx,rtx,ints\_x+1).';\-\ \color{Green}$\%$\-\ theta\-\ points\-\ in\-\ x
$\\$\color{BrickRed}ty\-\ =\-\ linspace(lty,rty,ints\_y+1);\-\ \color{Green}$\%$\-\ theta\-\ points\-\ in\-\ y
$\\$\color{BrickRed}indx\-\ =\-\ 0:1:Nx;\-\ \color{Green}
$\\$\color{BrickRed}indy\-\ =\-\ (0:1:Ny).';\color{Green}
$\\$
$\\$
$\\$\color{BrickRed}zx\-\ =\-\ \-\ nm(tx(1:end-1))*indx;\-\ \color{Green}$\%$\-\ function\-\ argument\-\ for\-\ x
$\\$\color{BrickRed}zy\-\ =\-\ indy*nm(ty(1:end-1));\-\ \color{Green}$\%$\-\ function\-\ argument\-\ for\-\ y
$\\$
$\\$
$\\$\color{Green}$\%$\-\ function\-\ argument\-\ for\-\ error
$\\$\color{BrickRed}ztx\-\ =\-\ \-\ nm(tx(1:end-1),tx(2:end))*indx;\-\ \color{Green}
$\\$\color{BrickRed}zty\-\ =\-\ indy*nm(ty(1:end-1),ty(2:end));\color{Green}
$\\$
$\\$
$\\$\color{Green}$\%$\-\ cos\-\ and\-\ sin
$\\$\color{BrickRed}cosx\-\ =\-\ cos(zx);\color{Green}
$\\$\color{BrickRed}cosy\-\ =\-\ cos(zy);\color{Green}
$\\$\color{BrickRed}sinx\-\ =\-\ sin(zx);\color{Green}
$\\$\color{BrickRed}siny\-\ =\-\ sin(zy);\color{Green}
$\\$
$\\$
$\\$\color{Green}$\%$\-\ find\-\ delta\-\ theta\_x\-\ and\-\ delta\-\ theta\_y
$\\$\color{BrickRed}txint\-\ =\-\ nm(tx);\color{Green}
$\\$\color{BrickRed}tyint\-\ =\-\ nm(ty);\color{Green}
$\\$\color{BrickRed}hx\-\ =\-\ nm(0,max(sup(txint(2:end)-txint(1:end-1))));\color{Green}
$\\$\color{BrickRed}hy\-\ =\-\ nm(0,max(sup(tyint(2:end)-tyint(1:end-1))));\color{Green}
$\\$\color{BrickRed}\-\ \color{Green}
$\\$\color{BrickRed}indx\-\ =\-\ repmat(indx,length(tx)-1,1);\color{Green}
$\\$\color{BrickRed}indy\-\ =\-\ repmat(indy,1,length(ty)-1);\color{Green}
$\\$
$\\$
$\\$\color{BrickRed}dx1\-\ =\-\ -(indx.*sinx);\color{Green}
$\\$\color{BrickRed}dx2\-\ =\-\ -(indx.\verb|^|2.*cosx);\color{Green}
$\\$\color{BrickRed}dx3\-\ =\-\ (indx.\verb|^|3.*sinx);\color{Green}
$\\$\color{BrickRed}dx4\-\ =\-\ (indx.\verb|^|4.*cosx);\color{Green}
$\\$
$\\$
$\\$\color{BrickRed}dy1\-\ =\-\ -(indy.*siny);\color{Green}
$\\$\color{BrickRed}dy2\-\ =\-\ -(indy.\verb|^|2.*cosy);\color{Green}
$\\$\color{BrickRed}dy3\-\ =\-\ (indy.\verb|^|3.*siny);\color{Green}
$\\$\color{BrickRed}dy4\-\ =\-\ (indy.\verb|^|4.*cosy);\color{Green}
$\\$
$\\$
$\\$\color{Green}$\%$\-\ Taylor\-\ expansion\-\ in\-\ two\-\ variables
$\\$\color{BrickRed}out\-\ =\-\ cosx*a*cosy\-\ ...\color{Green}
$\\$\color{BrickRed}\-\ \-\ \-\ \-\ +\-\ dx1*a*cosy*hx+cosx*a*dy1*hy\-\ ...\color{Green}
$\\$\color{BrickRed}\-\ \-\ \-\ \-\ +\-\ (\-\ dx2*a*cosy*hx\verb|^|2+2*dx1*a*dy1*hx*hy+cosx*a*dy2*hy\verb|^|2\-\ \-\ )/2\-\ ...\color{Green}
$\\$\color{BrickRed}\-\ \-\ \-\ \-\ +\-\ (\-\ \-\ dx3*a*cosy*hx\verb|^|3\-\ +\-\ 3*dx2*a*dy1*hx\verb|^|2*hy\-\ +\-\ 3*dx1*a*dy2*hx*hy\verb|^|2+cosx*a*dy3*hy\verb|^|3\-\ \-\ )/6\-\ ...\color{Green}
$\\$\color{BrickRed}\-\ \-\ \-\ \-\ +\-\ (\-\ \-\ dx4*a*cosy*hx\verb|^|4+4*dx3*a*dy1*hx\verb|^|3*hy+6*dx2*a*dy2*hx\verb|^|2*hy\verb|^|2+\-\ ...\-\ \color{Green}
$\\$\color{BrickRed}\-\ \-\ \-\ \-\ \-\ \-\ \-\ \-\ \-\ \-\ \-\ 4*dx1*a*dy3*hx*hy\verb|^|3\-\ +\-\ cosx*a*dy4*hy\verb|^|4\-\ \-\ )/24;\color{Green}
$\\$\color{BrickRed}\-\ \-\ \-\ \-\ \-\ \-\ \-\ \color{Green}
$\\$\color{Green}$\%$\-\ for\-\ error\-\ term
$\\$\color{BrickRed}costx\-\ =\-\ cos(ztx);\color{Green}
$\\$\color{BrickRed}costy\-\ =\-\ cos(zty);\color{Green}
$\\$\color{BrickRed}sintx\-\ =\-\ sin(ztx);\color{Green}
$\\$\color{BrickRed}sinty\-\ =\-\ sin(zty);\color{Green}
$\\$
$\\$
$\\$\color{Green}$\%$\-\ for\-\ error\-\ term
$\\$\color{BrickRed}ft\_1\-\ =\-\ -hx\verb|^|5*(indx.\verb|^|5.*sintx)*a*costy/120;\color{Green}
$\\$\color{BrickRed}ft\_2\-\ =\-\ -hx\verb|^|4*hy*(indx.\verb|^|4.*costx)*a*(indy.*sinty)/24;\color{Green}
$\\$\color{BrickRed}ft\_3\-\ =\-\ -hx\verb|^|3*hy\verb|^|2*(indx.\verb|^|3.*sintx)*a*(indy.\verb|^|2.*costy)/12;\color{Green}
$\\$\color{BrickRed}ft\_4\-\ =\-\ -hx\verb|^|2*hy\verb|^|3*(indx.\verb|^|2.*costx)*a*(indy.\verb|^|3.*sinty)/12;\color{Green}
$\\$\color{BrickRed}ft\_5\-\ =\-\ -hx*hy\verb|^|4*(indx.*sintx)*a*(indy.\verb|^|4.*costy)/24;\color{Green}
$\\$\color{BrickRed}ft\_6\-\ =\-\ -hy\verb|^|5*costx*a*(indy.\verb|^|5.*sinty)/120;\color{Green}
$\\$
$\\$
$\\$\color{Green}$\%$\-\ find\-\ err\-\ interval
$\\$\color{BrickRed}err\-\ =\-\ ft\_1+ft\_2+ft\_3+ft\_4+ft\_5+ft\_6;\color{Green}
$\\$
$\\$
$\\$\color{Green}$\%$\-\ combine\-\ partial\-\ sum\-\ and\-\ error\-\ interval
$\\$\color{BrickRed}out\-\ =\-\ out\-\ +\-\ err;\color{Green}
$\\$
$\\$
$\\$
$\\$
$\\$\color{Black}\section{check\_alpha\_distinct.m}

\color{Green}\color{BrickRed}\color{NavyBlue}\-\ function\-\ \color{BrickRed}\-\ success\-\ =\-\ check\_alpha\_distinct(k)\color{Green}
$\\$
$\\$
$\\$\color{BrickRed}success\-\ =\-\ 1;\color{Green}
$\\$
$\\$
$\\$\color{BrickRed}K\-\ =\-\ elliptic\_integral(k,1);\color{Green}
$\\$\color{BrickRed}E\-\ =\-\ elliptic\_integral(k,2);\color{Green}
$\\$\color{BrickRed}k2\-\ =\-\ k*k;\color{Green}
$\\$\color{BrickRed}c1\-\ =\-\ 1-k2;\color{Green}
$\\$\color{BrickRed}b1\-\ =\-\ k2*K/(E-K);\color{Green}
$\\$\color{BrickRed}b2\-\ =\-\ k2*c1*K/(c1*K-E);\color{Green}
$\\$\color{BrickRed}b3\-\ =\-\ c1*K/E;\color{Green}
$\\$
$\\$
$\\$\color{BrickRed}\color{NavyBlue}\-\ if\-\ \color{BrickRed}\-\ sup(b1)\-\ $>$=\-\ inf(b2)\color{Green}
$\\$\color{BrickRed}\-\ \-\ \-\ \-\ success\-\ =\-\ 0;\color{Green}
$\\$\color{BrickRed}\-\ \-\ \-\ \-\ \color{NavyBlue}\-\ return\-\ \color{BrickRed}\color{Green}
$\\$\color{BrickRed}\color{NavyBlue}\-\ end\-\ \color{BrickRed}\color{Green}
$\\$
$\\$
$\\$\color{BrickRed}\color{NavyBlue}\-\ if\-\ \color{BrickRed}\-\ sup(b2)\-\ $>$=\-\ inf(b3)\color{Green}
$\\$\color{BrickRed}\-\ \-\ \-\ \-\ success\-\ =\-\ 0;\color{Green}
$\\$\color{BrickRed}\-\ \-\ \-\ \-\ \color{NavyBlue}\-\ return\-\ \color{BrickRed}\color{Green}
$\\$\color{BrickRed}\color{NavyBlue}\-\ end\-\ \color{BrickRed}\color{Green}
$\\$
$\\$
$\\$
$\\$
$\\$
$\\$
$\\$
$\\$
$\\$
$\\$
$\\$
$\\$
$\\$
$\\$
$\\$
$\\$
$\\$
$\\$
$\\$
$\\$
$\\$
$\\$
$\\$
$\\$
$\\$\color{Black}\section{create\_tableA.m}

\color{Green}\color{Green}$\%$\-\ create\-\ Table\-\ of\-\ lower\-\ instability\-\ for\-\ the\-\ paper
$\\$\color{BrickRed}clear\-\ all;\-\ clc;\-\ curr\_dir\-\ =\-\ local\_startup;\color{Green}
$\\$
$\\$
$\\$
$\\$
$\\$\color{BrickRed}file\_name\-\ =\-\ 'instability\_result\_lower.mat';\color{Green}
$\\$\color{BrickRed}ld\-\ =\-\ retrieve\_it(curr\_dir,'interval\_arithmetic',file\_name,'data\_final');\color{Green}
$\\$\color{BrickRed}d\-\ =\-\ ld.var;\color{Green}
$\\$
$\\$
$\\$\color{BrickRed}fprintf('\textbackslash n\textbackslash n\textbackslash n\textbackslash \textbackslash begin{table}[!b]\textbackslash n\textbackslash \textbackslash begin{tabular}{|c|c|c|c|c|c|c|c|c|c|}\textbackslash n\textbackslash \textbackslash hline');\color{Green}
$\\$\color{BrickRed}fprintf('\textbackslash n$k\_L$\&$k\_R$\&$q\_L$\&$q\_R$\&$M\_x$\&$M\_q$\&$\textbackslash \textbackslash rho\_q$\&$N\_x$\&$N\_q$\&$M\_{\textbackslash \textbackslash lambda}$\textbackslash \textbackslash \textbackslash \textbackslash ');\color{Green}
$\\$
$\\$
$\\$\color{BrickRed}\color{NavyBlue}\-\ for\-\ \color{BrickRed}\-\ j\-\ =\-\ 1:length(d.RK)\color{Green}
$\\$
$\\$
$\\$\color{BrickRed}\-\ \-\ \-\ \-\ fprintf('\textbackslash n\textbackslash \textbackslash hline\textbackslash n');\color{Green}
$\\$\color{BrickRed}\-\ \-\ \-\ \-\ fprintf('$\%$5.5g',d.KL(j));\color{Green}
$\\$\color{BrickRed}\-\ \-\ \-\ \-\ fprintf('\-\ \&\-\ ');\color{Green}
$\\$\color{BrickRed}\-\ \-\ \-\ \-\ fprintf('$\%$5.5g',d.RK(j));\color{Green}
$\\$\color{BrickRed}\-\ \-\ \-\ \-\ fprintf('\-\ \&\-\ ');\color{Green}
$\\$\color{BrickRed}\-\ \-\ \-\ \-\ \color{Green}
$\\$\color{BrickRed}\-\ \-\ \-\ \-\ file\_name\-\ =\-\ ['d\_lower\_unstable\_',num2str(j),'.mat'];\color{Green}
$\\$\color{BrickRed}\-\ \-\ \-\ \-\ ld\-\ =\-\ retrieve\_it(curr\_dir,'interval\_arithmetic',file\_name,'data\_final');\color{Green}
$\\$\color{BrickRed}\-\ \-\ \-\ \-\ r\-\ =\-\ ld.var;\color{Green}
$\\$\color{BrickRed}\-\ \-\ \-\ \-\ \color{Green}
$\\$\color{BrickRed}\-\ \-\ \-\ \-\ fprintf('$\%$5.5g',sup(r.q\_min));\color{Green}
$\\$\color{BrickRed}\-\ \-\ \-\ \-\ fprintf('\-\ \&\-\ ');\color{Green}
$\\$\color{BrickRed}\-\ \-\ \-\ \-\ fprintf('$\%$5.5g',inf(r.q\_max));\color{Green}
$\\$\color{BrickRed}\-\ \-\ \-\ \-\ fprintf('\-\ \&\-\ ');\color{Green}
$\\$\color{BrickRed}\-\ \-\ \-\ \-\ fprintf('$\%$3.3g',sup(r.M\_x\_n1));\color{Green}
$\\$\color{BrickRed}\-\ \-\ \-\ \-\ fprintf('\-\ \&\-\ ');\color{Green}
$\\$\color{BrickRed}\-\ \-\ \-\ \-\ fprintf('$\%$3.3g',sup(r.M\_q\_n1));\color{Green}
$\\$\color{BrickRed}\-\ \-\ \-\ \-\ fprintf('\-\ \&\-\ ');\color{Green}
$\\$\color{BrickRed}\-\ \-\ \-\ \-\ fprintf('$\%$3.3g',sup(r.rho\_q));\color{Green}
$\\$\color{BrickRed}\-\ \-\ \-\ \-\ fprintf('\-\ \&\-\ ');\color{Green}
$\\$\color{BrickRed}\-\ \-\ \-\ \-\ fprintf('$\%$g',r.N\_x\_n1);\color{Green}
$\\$\color{BrickRed}\-\ \-\ \-\ \-\ fprintf('\-\ \&\-\ ');\color{Green}
$\\$\color{BrickRed}\-\ \-\ \-\ \-\ fprintf('$\%$g',r.N\_q\_n1);\color{Green}
$\\$\color{BrickRed}\-\ \-\ \-\ \-\ fprintf('\-\ \&\-\ ');\color{Green}
$\\$\color{BrickRed}\-\ \-\ \-\ \-\ fprintf('$\%$3.3g',d.MN(j));\color{Green}
$\\$\color{BrickRed}\-\ \-\ \-\ \-\ fprintf('\-\ \textbackslash \textbackslash \textbackslash \textbackslash \textbackslash ');\color{Green}
$\\$\color{BrickRed}\-\ \-\ \-\ \-\ \color{Green}
$\\$\color{BrickRed}\-\ \-\ \-\ \-\ \color{NavyBlue}\-\ if\-\ \color{BrickRed}\-\ sup(r.err\_x\_n1)\-\ $>$\-\ 1e-17\color{Green}
$\\$\color{BrickRed}\-\ \-\ \-\ \-\ \-\ \-\ \-\ \-\ error('interpolation\-\ error\-\ greater\-\ than\-\ claimed')\color{Green}
$\\$\color{BrickRed}\-\ \-\ \-\ \-\ \color{NavyBlue}\-\ end\-\ \color{BrickRed}\color{Green}
$\\$\color{BrickRed}\-\ \-\ \-\ \-\ \color{Green}
$\\$\color{BrickRed}\-\ \-\ \-\ \-\ \color{NavyBlue}\-\ if\-\ \color{BrickRed}\-\ sup(r.err\_q\_n1)\-\ $>$\-\ 1e-17\color{Green}
$\\$\color{BrickRed}\-\ \-\ \-\ \-\ \-\ \-\ \-\ \-\ error('interpolation\-\ error\-\ greater\-\ than\-\ claimed')\color{Green}
$\\$\color{BrickRed}\-\ \-\ \-\ \-\ \color{NavyBlue}\-\ end\-\ \color{BrickRed}\color{Green}
$\\$\color{BrickRed}\-\ \-\ \-\ \-\ \color{Green}
$\\$\color{BrickRed}\color{NavyBlue}\-\ end\-\ \color{BrickRed}\color{Green}
$\\$
$\\$
$\\$\color{BrickRed}fprintf('\textbackslash n\textbackslash \textbackslash hline\textbackslash n\textbackslash \textbackslash end{tabular}\textbackslash n\textbackslash \textbackslash caption{TODO\-\ }\-\ $\%$$\%$Table\-\ created\-\ by\-\ create\_tableA.m\textbackslash n\textbackslash \textbackslash label{table:lower-instability}\textbackslash n\textbackslash \textbackslash end{table}');\color{Green}
$\\$
$\\$
$\\$
$\\$
$\\$\color{BrickRed}fprintf('\textbackslash n\textbackslash n\textbackslash n');\color{Green}
$\\$\color{Black}\section{distinct.m}

\color{Green}\color{BrickRed}clear\-\ all;\-\ curr\_dir\-\ =\-\ local\_startup;\color{Green}
$\\$
$\\$
$\\$\color{BrickRed}pie\-\ =\-\ nm('pi');\color{Green}
$\\$
$\\$
$\\$\color{BrickRed}Kvals\-\ =\-\ [\color{Green}
$\\$\color{BrickRed}\-\ \-\ \-\ \-\ nm('0.999999'),\-\ nm('0.9999995');\color{Green}
$\\$\color{BrickRed}\-\ \-\ \-\ \-\ nm('0.99999'),\-\ nm('0.999999');\color{Green}
$\\$\color{BrickRed}\-\ \-\ \-\ \-\ nm('0.9999'),\-\ nm('0.99999');\color{Green}
$\\$\color{BrickRed}\-\ \-\ \-\ \-\ nm('0.999'),\-\ nm('0.9999');\color{Green}
$\\$\color{BrickRed}\-\ \-\ \-\ \-\ nm('0.99'),\-\ nm('0.999');\color{Green}
$\\$\color{BrickRed}\-\ \-\ \-\ \-\ nm('0.9'),\-\ nm('0.99');\color{Green}
$\\$\color{BrickRed}];\color{Green}
$\\$
$\\$
$\\$\color{BrickRed}N\-\ =\-\ size(Kvals,1);\color{Green}
$\\$\color{BrickRed}n\-\ =\-\ 10;\color{Green}
$\\$
$\\$
$\\$\color{BrickRed}\color{NavyBlue}\-\ for\-\ \color{BrickRed}\-\ ind\-\ =\-\ 1:N\color{Green}
$\\$\color{BrickRed}\-\ \-\ \-\ \-\ \color{Green}
$\\$\color{BrickRed}\-\ \-\ \-\ \-\ left\-\ \-\ =\-\ Kvals(ind,1);\-\ \color{Green}$\%$\-\ left\-\ side\-\ of\-\ interval\-\ in\-\ k
$\\$\color{BrickRed}\-\ \-\ \-\ \-\ right\-\ =\-\ Kvals(ind,2);\-\ \color{Green}$\%$\-\ right\-\ side\-\ of\-\ interval\-\ in\-\ k
$\\$\color{BrickRed}\-\ \-\ \-\ \-\ del\-\ =\-\ (right-left)/n;\-\ \color{Green}$\%$\-\ width\-\ of\-\ sub\-\ interval\-\ in\-\ k
$\\$
$\\$
$\\$\color{BrickRed}\-\ \-\ \-\ \-\ \color{NavyBlue}\-\ for\-\ \color{BrickRed}\-\ j\-\ =\-\ 1:n\color{Green}
$\\$
$\\$
$\\$\color{BrickRed}\-\ \-\ \-\ \-\ \-\ \-\ \-\ \-\ clc;\color{Green}
$\\$\color{BrickRed}\-\ \-\ \-\ \-\ \-\ \-\ \-\ \-\ fprintf('Verifying\-\ alpha\_j\-\ distinct:\-\ Percent\-\ done\-\ =\-\ $\%$4.4g\textbackslash n',100*((ind-1)*n+j)/(N*n));\color{Green}
$\\$\color{BrickRed}\-\ \-\ \-\ \-\ \-\ \-\ \-\ \-\ \color{Green}$\%$\-\ interval\-\ in\-\ k
$\\$\color{BrickRed}\-\ \-\ \-\ \-\ \-\ \-\ \-\ \-\ k\-\ =\-\ left\-\ +\-\ \-\ nm((j-1)*del,j*del);\color{Green}
$\\$
$\\$
$\\$\color{BrickRed}\-\ \-\ \-\ \-\ \-\ \-\ \-\ \-\ success\-\ =\-\ check\_alpha\_distinct(k);\color{Green}
$\\$\color{BrickRed}\-\ \-\ \-\ \-\ \-\ \-\ \-\ \-\ \color{NavyBlue}\-\ if\-\ \color{BrickRed}\-\ success\-\ ==\-\ 0\color{Green}
$\\$\color{BrickRed}\-\ \-\ \-\ \-\ \-\ \-\ \-\ \-\ \-\ \-\ \-\ \-\ error('Failed\-\ to\-\ verify\-\ the\-\ alpha\_j\-\ are\-\ distinct.');\color{Green}
$\\$\color{BrickRed}\-\ \-\ \-\ \-\ \-\ \-\ \-\ \-\ \color{NavyBlue}\-\ end\-\ \color{BrickRed}\color{Green}
$\\$
$\\$
$\\$\color{BrickRed}\-\ \-\ \-\ \-\ \color{NavyBlue}\-\ end\-\ \color{BrickRed}\color{Green}
$\\$\color{BrickRed}\-\ \-\ \-\ \-\ \color{Green}
$\\$\color{BrickRed}\color{NavyBlue}\-\ end\-\ \color{BrickRed}\color{Green}
$\\$
$\\$
$\\$
$\\$
$\\$
$\\$
$\\$\color{Black}\section{divide\_interpolation.m}

\color{Green}\color{BrickRed}clear\-\ all;\-\ curr\_dir\-\ =\-\ local\_startup;\color{Green}
$\\$
$\\$
$\\$
$\\$\color{Black}
    Finds the 2d Chebyshev polynomials with $q\in[q_{\min},q_{\max}]$. 
\color{Green}
$\\$
$\\$
$\\$
$\\$\color{BrickRed}cnt\-\ =\-\ 0;\color{Green}
$\\$
$\\$
$\\$\color{BrickRed}cnt\-\ =\-\ cnt\-\ +\-\ 1;\color{Green}
$\\$\color{BrickRed}q\_min{cnt}\-\ =\-\ nm('0.65');\-\ \color{Green}$\%$\-\ q\_min
$\\$\color{BrickRed}q\_max{cnt}\-\ =\-\ nm('0.71');\-\ \color{Green}$\%$\-\ q\_max
$\\$\color{Green}$\%$\-\ these\-\ next\-\ two\-\ numbers\-\ break\-\ up\-\ the\-\ domain\-\ for\-\ 
$\\$\color{Green}$\%$\-\ finding\-\ a\-\ lower\-\ bound\-\ on\-\ the\-\ theta\-\ function\-\ that\-\ 
$\\$\color{Green}$\%$\-\ shows\-\ up\-\ in\-\ the\-\ definition\-\ of\-\ v(x)
$\\$\color{BrickRed}num\_steps{cnt}\-\ =\-\ 5000\-\ ;\-\ \color{Green}
$\\$\color{BrickRed}steps{cnt}\-\ =\-\ 20000\-\ ;\color{Green}
$\\$\color{Green}$\%$\-\ name\-\ by\-\ which\-\ this\-\ will\-\ be\-\ saved
$\\$\color{BrickRed}file\_name{cnt}\-\ =\-\ 'interp2d\_650\_to\_710';\color{Green}
$\\$
$\\$
$\\$\color{BrickRed}cnt\-\ =\-\ cnt\-\ +\-\ 1;\color{Green}
$\\$\color{BrickRed}q\_min{cnt}\-\ =\-\ nm('0.59');\color{Green}
$\\$\color{BrickRed}q\_max{cnt}\-\ =\-\ nm('0.66');\color{Green}
$\\$\color{BrickRed}num\_steps{cnt}\-\ =\-\ 12000\-\ ;\color{Green}
$\\$\color{BrickRed}steps{cnt}\-\ =\-\ 10000\-\ ;\color{Green}
$\\$\color{BrickRed}file\_name{cnt}\-\ =\-\ 'interp2d\_590\_to\_660';\color{Green}
$\\$
$\\$
$\\$\color{BrickRed}cnt\-\ =\-\ cnt\-\ +\-\ 1;\color{Green}
$\\$\color{BrickRed}q\_min{cnt}\-\ =\-\ nm('0.53');\color{Green}
$\\$\color{BrickRed}q\_max{cnt}\-\ =\-\ nm('0.6');\color{Green}
$\\$\color{BrickRed}num\_steps{cnt}\-\ =\-\ 12000\-\ ;\color{Green}
$\\$\color{BrickRed}steps{cnt}\-\ =\-\ 10000\-\ ;\color{Green}
$\\$\color{BrickRed}file\_name{cnt}\-\ =\-\ 'interp2d\_530\_to\_600';\color{Green}
$\\$
$\\$
$\\$\color{BrickRed}cnt\-\ =\-\ cnt\-\ +\-\ 1;\color{Green}
$\\$\color{BrickRed}q\_min{cnt}\-\ =\-\ nm('0.49');\color{Green}
$\\$\color{BrickRed}q\_max{cnt}\-\ =\-\ nm('0.538');\color{Green}
$\\$\color{BrickRed}num\_steps{cnt}\-\ =\-\ 600\-\ ;\color{Green}
$\\$\color{BrickRed}steps{cnt}\-\ =\-\ 8000\-\ ;\color{Green}
$\\$\color{BrickRed}file\_name{cnt}\-\ =\-\ 'interp2d\_490\_to\_538';\color{Green}
$\\$
$\\$
$\\$\color{BrickRed}cnt\-\ =\-\ cnt\-\ +\-\ 1;\color{Green}
$\\$\color{BrickRed}q\_min{cnt}\-\ =\-\ nm('0.35');\color{Green}
$\\$\color{BrickRed}q\_max{cnt}\-\ =\-\ nm('0.5');\color{Green}
$\\$\color{BrickRed}num\_steps{cnt}\-\ =\-\ 600\-\ ;\color{Green}
$\\$\color{BrickRed}steps{cnt}\-\ =\-\ 8000\-\ ;\color{Green}
$\\$\color{BrickRed}file\_name{cnt}\-\ =\-\ 'interp2d\_350\_to\_500';\color{Green}
$\\$
$\\$
$\\$\color{BrickRed}cnt\-\ =\-\ cnt\-\ +\-\ 1;\color{Green}
$\\$\color{BrickRed}q\_min{cnt}\-\ =\-\ nm('0.1');\color{Green}
$\\$\color{BrickRed}q\_max{cnt}\-\ =\-\ nm('0.4');\color{Green}
$\\$\color{BrickRed}num\_steps{cnt}\-\ =\-\ 100\-\ ;\color{Green}
$\\$\color{BrickRed}steps{cnt}\-\ =\-\ 8000\-\ ;\color{Green}
$\\$\color{BrickRed}file\_name{cnt}\-\ =\-\ 'interp2d\_100\_to\_400';\color{Green}
$\\$
$\\$
$\\$\color{BrickRed}\color{NavyBlue}\-\ parfor\-\ \color{BrickRed}\-\ j\-\ =\-\ 1:length(q\_min)\color{Green}
$\\$\color{BrickRed}\-\ \-\ \-\ \-\ \color{Green}
$\\$\color{BrickRed}\-\ \-\ \-\ \-\ interpolation\_2d(q\_min{j},q\_max{j},num\_steps{j},steps{j},file\_name{j});\color{Green}
$\\$\color{BrickRed}\-\ \-\ \-\ \-\ \color{Green}
$\\$\color{BrickRed}\color{NavyBlue}\-\ end\-\ \color{BrickRed}\color{Green}
$\\$
$\\$
$\\$
$\\$
$\\$
$\\$
$\\$
$\\$
$\\$\color{Black}\section{driver.m}

\color{Green}
$\\$
$\\$
$\\$
$\\$\color{Green}$\%$------------------------------------------------------------
$\\$\color{Green}$\%$\-\ alpha\_j\-\ distinct
$\\$\color{Green}$\%$------------------------------------------------------------
$\\$
$\\$
$\\$
$\\$\color{Black}
Verify that the $\alpha_j$ are distinct for $k\in[0.9,0.9999995]$. 
Distinctness of the $\alpha_j$ is required for showing stability of
the periodic waves.
\color{Green}
$\\$
$\\$
$\\$
$\\$\color{BrickRed}distinct;\color{Green}
$\\$
$\\$
$\\$\color{Green}$\%$------------------------------------------------------------
$\\$\color{Green}$\%$\-\ simplicity\-\ of\-\ eigenvalues
$\\$\color{Green}$\%$------------------------------------------------------------
$\\$
$\\$
$\\$\color{BrickRed}simplicity;\color{Green}
$\\$
$\\$
$\\$\color{Green}$\%$------------------------------------------------------------
$\\$\color{Green}$\%$\-\ lower\-\ instability
$\\$\color{Green}$\%$------------------------------------------------------------
$\\$
$\\$
$\\$
$\\$Verify\-\ spectral\-\ instability\-\ for\-\ a\-\ region\-\ below\-\ the\-\ stability\-\ region\-\ by\-\ 
$\\$interpolating\-\ in\-\ the\-\ variable\-\ $q$\-\ with\-\ the\-\ variable\-\ $\textbackslash psi$\-\ fixed\-\ at\-\ 1.
$\\$
$\\$\color{BrickRed}\-\ \color{Green}
$\\$\color{BrickRed}lower\_instability\_interpolation;\color{Green}
$\\$
$\\$
$\\$\color{Green}$\%$------------------------------------------------------------
$\\$\color{Green}$\%$\-\ 2d\-\ interpolation
$\\$\color{Green}$\%$------------------------------------------------------------
$\\$
$\\$
$\\$\color{BrickRed}divide\_interpolation;\color{Green}
$\\$
$\\$
$\\$\color{Green}$\%$------------------------------------------------------------
$\\$\color{Green}$\%$\-\ midle\-\ stability\-\ region
$\\$\color{Green}$\%$------------------------------------------------------------
$\\$
$\\$
$\\$\color{Green}$\%$\-\ verify\-\ for\-\ k\-\ \textbackslash in\-\ [0.93,\-\ 0.9999983]\-\ that\-\ the\-\ case\-\ 
$\\$\color{Green}$\%$\-\ alpha\-\ =\-\ i\textbackslash psi\textbackslash omega'\-\ is\-\ consistent\-\ with\-\ stability
$\\$
$\\$
$\\$\color{Green}$\%$\-\ alpha\-\ =\-\ i\textbackslash psi\textbackslash omega'
$\\$\color{BrickRed}driver\_stability\_n0;\-\ \color{Green}
$\\$
$\\$
$\\$\color{Green}$\%$\-\ alpha\-\ =\-\ omega+i\textbackslash psi\textbackslash omega'
$\\$\color{BrickRed}driver\_stability\_n1;\color{Green}
$\\$
$\\$
$\\$\color{Green}$\%$------------------------------------------------------------
$\\$\color{Green}$\%$\-\ midle\-\ stability\-\ region
$\\$\color{Green}$\%$------------------------------------------------------------
$\\$
$\\$
$\\$\color{BrickRed}driver\_instability\_upper;\color{Green}
$\\$
$\\$
$\\$\color{Green}$\%$------------------------------------------------------------
$\\$\color{Green}$\%$\-\ strict\-\ upper
$\\$\color{Green}$\%$------------------------------------------------------------
$\\$
$\\$
$\\$\color{BrickRed}driver\_strict\_upper\_taylor;\color{Green}
$\\$
$\\$
$\\$\color{BrickRed}driver\_pinpoint;\color{Green}
$\\$
$\\$
$\\$\color{BrickRed}driver\_strict\_lower\color{Green}
$\\$
$\\$
$\\$\color{Green}$\%$------------------------------------------------------------
$\\$\color{Green}$\%$\-\ aux\-\ lemma
$\\$\color{Green}$\%$------------------------------------------------------------
$\\$
$\\$
$\\$\color{BrickRed}kappa\_lemma;\color{Green}
$\\$
$\\$
$\\$\color{BrickRed}lemma\_qk;\color{Green}
$\\$
$\\$
$\\$
$\\$
$\\$
$\\$
$\\$
$\\$
$\\$
$\\$
$\\$
$\\$
$\\$
$\\$
$\\$\color{Black}\section{driver\_instability\_upper.m}

\color{Green}\color{BrickRed}clear\-\ all;\-\ curr\_dir\-\ =\-\ local\_startup;\color{Green}
$\\$
$\\$
$\\$\color{BrickRed}psi\_left\-\ =\-\ 0.6;\-\ \color{Green}
$\\$\color{BrickRed}psi\_right\-\ =\-\ 0.8;\color{Green}
$\\$
$\\$
$\\$\color{BrickRed}ints\_x\-\ =\-\ 1;\color{Green}
$\\$\color{BrickRed}ints\_y\-\ =\-\ 1000;\color{Green}
$\\$
$\\$
$\\$
$\\$
$\\$\color{Green}$\%$$\%$$\%$$\%$$\%$$\%$$\%$$\%$$\%$$\%$$\%$$\%$$\%$$\%$$\%$$\%$$\%$$\%$$\%$$\%$$\%$$\%$$\%$$\%$$\%$$\%$$\%$$\%$$\%$$\%$$\%$$\%$$\%$$\%$$\%$$\%$
$\\$
$\\$
$\\$\color{BrickRed}file\_name\-\ =\-\ 'interp2d\_490\_to\_538';\color{Green}
$\\$\color{BrickRed}ld\-\ =\-\ retrieve\_it(curr\_dir,'interval\_arithmetic',file\_name,'data\_final');\color{Green}
$\\$\color{BrickRed}d\-\ =\-\ ld.var;\color{Green}
$\\$\color{BrickRed}cnt\-\ =\-\ 0;\color{Green}
$\\$
$\\$
$\\$\color{BrickRed}cnt\-\ =\-\ cnt\-\ +\-\ 1;\color{Green}
$\\$\color{BrickRed}st{cnt}{1}\-\ =\-\ nm('0.99999839');\color{Green}
$\\$\color{BrickRed}st{cnt}{2}\-\ =\-\ nm('0.9999984');\color{Green}
$\\$\color{BrickRed}st{cnt}{3}\-\ =\-\ 200;\-\ \color{Green}$\%$\-\ number\-\ of\-\ k\-\ intervals
$\\$
$\\$
$\\$\color{BrickRed}cnt\-\ =\-\ cnt\-\ +\-\ 1;\color{Green}
$\\$\color{BrickRed}st{cnt}{1}\-\ =\-\ nm('0.9999984');\color{Green}
$\\$\color{BrickRed}st{cnt}{2}\-\ =\-\ nm('0.9999985');\color{Green}
$\\$\color{BrickRed}st{cnt}{3}\-\ =\-\ 500;\-\ \color{Green}$\%$\-\ number\-\ of\-\ k\-\ intervals
$\\$
$\\$
$\\$\color{BrickRed}cnt\-\ =\-\ cnt\-\ +\-\ 1;\color{Green}
$\\$\color{BrickRed}st{cnt}{1}\-\ =\-\ nm('0.9999985');\color{Green}
$\\$\color{BrickRed}st{cnt}{2}\-\ =\-\ nm('0.9999986');\color{Green}
$\\$\color{BrickRed}st{cnt}{3}\-\ =\-\ 100;\-\ \color{Green}$\%$\-\ number\-\ of\-\ k\-\ intervals
$\\$
$\\$
$\\$\color{BrickRed}cnt\-\ =\-\ cnt\-\ +\-\ 1;\color{Green}
$\\$\color{BrickRed}st{cnt}{1}\-\ =\-\ nm('0.9999986');\color{Green}
$\\$\color{BrickRed}st{cnt}{2}\-\ =\-\ nm('0.9999987');\color{Green}
$\\$\color{BrickRed}st{cnt}{3}\-\ =\-\ 50;\-\ \color{Green}$\%$\-\ number\-\ of\-\ k\-\ intervals
$\\$
$\\$
$\\$\color{BrickRed}cnt\-\ =\-\ cnt\-\ +\-\ 1;\color{Green}
$\\$\color{BrickRed}st{cnt}{1}\-\ =\-\ nm('0.9999987');\color{Green}
$\\$\color{BrickRed}st{cnt}{2}\-\ =\-\ nm('0.9999988');\color{Green}
$\\$\color{BrickRed}st{cnt}{3}\-\ =\-\ 50;\-\ \color{Green}$\%$\-\ number\-\ of\-\ k\-\ intervals
$\\$
$\\$
$\\$\color{BrickRed}cnt\-\ =\-\ cnt\-\ +\-\ 1;\color{Green}
$\\$\color{BrickRed}st{cnt}{1}\-\ =\-\ nm('0.9999988');\color{Green}
$\\$\color{BrickRed}st{cnt}{2}\-\ =\-\ nm('0.9999989');\color{Green}
$\\$\color{BrickRed}st{cnt}{3}\-\ =\-\ 50;\-\ \color{Green}$\%$\-\ number\-\ of\-\ k\-\ intervals
$\\$
$\\$
$\\$\color{BrickRed}cnt\-\ =\-\ cnt\-\ +\-\ 1;\color{Green}
$\\$\color{BrickRed}st{cnt}{1}\-\ =\-\ nm('0.9999989');\color{Green}
$\\$\color{BrickRed}st{cnt}{2}\-\ =\-\ nm('0.999999');\color{Green}
$\\$\color{BrickRed}st{cnt}{3}\-\ =\-\ 50;\-\ \color{Green}$\%$\-\ number\-\ of\-\ k\-\ intervals
$\\$
$\\$
$\\$\color{BrickRed}\color{NavyBlue}\-\ for\-\ \color{BrickRed}\-\ ind\-\ =\-\ 1:cnt\color{Green}
$\\$\color{BrickRed}\-\ \-\ \-\ \-\ kvals\-\ =\-\ linspace(inf(st{ind}{1}),sup(st{ind}{2}),st{ind}{3}+1);\color{Green}
$\\$\color{BrickRed}\-\ \-\ \-\ \-\ \color{NavyBlue}\-\ for\-\ \color{BrickRed}\-\ j\-\ =\-\ length(kvals)-1\color{Green}
$\\$\color{BrickRed}\-\ \-\ \-\ \-\ \-\ \-\ \-\ \-\ clc;\color{Green}
$\\$\color{BrickRed}\-\ \-\ \-\ \-\ \-\ \-\ \-\ \-\ fprintf('upper\-\ instability,\-\ k:\-\ $\%$16.16g\textbackslash n',mid(kvals(j)));\color{Green}
$\\$\color{BrickRed}\-\ \-\ \-\ \-\ \-\ \-\ \-\ \-\ verify\_instability\_upper(d,\-\ kvals(j),\-\ kvals(j+1),psi\_left,psi\_right,ints\_y);\-\ \color{Green}
$\\$\color{BrickRed}\-\ \-\ \-\ \-\ \color{NavyBlue}\-\ end\-\ \color{BrickRed}\color{Green}
$\\$\color{BrickRed}\color{NavyBlue}\-\ end\-\ \color{BrickRed}\color{Green}
$\\$
$\\$
$\\$\color{Green}$\%$$\%$$\%$$\%$$\%$$\%$$\%$$\%$$\%$$\%$$\%$$\%$$\%$$\%$$\%$$\%$$\%$$\%$$\%$$\%$$\%$$\%$$\%$
$\\$
$\\$
$\\$\color{BrickRed}file\_name\-\ =\-\ 'interp2d\_530\_to\_600';\color{Green}
$\\$\color{BrickRed}ld\-\ =\-\ retrieve\_it(curr\_dir,'interval\_arithmetic',file\_name,'data\_final');\color{Green}
$\\$\color{BrickRed}d\-\ =\-\ ld.var;\color{Green}
$\\$\color{BrickRed}cnt\-\ =\-\ 0;\color{Green}
$\\$
$\\$
$\\$\color{BrickRed}cnt\-\ =\-\ cnt\-\ +\-\ 1;\color{Green}
$\\$\color{BrickRed}st{cnt}{1}\-\ =\-\ nm('0.999999');\color{Green}
$\\$\color{BrickRed}st{cnt}{2}\-\ =\-\ nm('0.9999999');\color{Green}
$\\$\color{BrickRed}st{cnt}{3}\-\ =\-\ 500;\-\ \color{Green}$\%$\-\ number\-\ of\-\ k\-\ intervals
$\\$
$\\$
$\\$\color{BrickRed}cnt\-\ =\-\ cnt\-\ +\-\ 1;\color{Green}
$\\$\color{BrickRed}st{cnt}{1}\-\ =\-\ nm('0.9999999');\color{Green}
$\\$\color{BrickRed}st{cnt}{2}\-\ =\-\ nm('0.99999995');\color{Green}
$\\$\color{BrickRed}st{cnt}{3}\-\ =\-\ 500;\-\ \color{Green}$\%$\-\ number\-\ of\-\ k\-\ intervals
$\\$
$\\$
$\\$\color{BrickRed}\color{NavyBlue}\-\ for\-\ \color{BrickRed}\-\ ind\-\ =\-\ 1:cnt\color{Green}
$\\$\color{BrickRed}\-\ \-\ \-\ \-\ kvals\-\ =\-\ linspace(inf(st{ind}{1}),sup(st{ind}{2}),st{ind}{3}+1);\color{Green}
$\\$\color{BrickRed}\-\ \-\ \-\ \-\ \color{NavyBlue}\-\ for\-\ \color{BrickRed}\-\ j\-\ =\-\ length(kvals)-1\color{Green}
$\\$\color{BrickRed}\-\ \-\ \-\ \-\ \-\ \-\ \-\ \-\ clc;\color{Green}
$\\$\color{BrickRed}\-\ \-\ \-\ \-\ \-\ \-\ \-\ \-\ fprintf('upper\-\ instability,\-\ k:\-\ $\%$16.16g\textbackslash n',mid(kvals(j)));\color{Green}
$\\$\color{BrickRed}\-\ \-\ \-\ \-\ \-\ \-\ \-\ \-\ verify\_instability\_upper(d,\-\ kvals(j),\-\ kvals(j+1),psi\_left,psi\_right,ints\_y);\-\ \color{Green}
$\\$\color{BrickRed}\-\ \-\ \-\ \-\ \color{NavyBlue}\-\ end\-\ \color{BrickRed}\color{Green}
$\\$\color{BrickRed}\color{NavyBlue}\-\ end\-\ \color{BrickRed}\color{Green}
$\\$
$\\$
$\\$\color{Green}$\%$$\%$$\%$$\%$$\%$$\%$$\%$$\%$$\%$$\%$$\%$$\%$$\%$$\%$$\%$$\%$$\%$$\%$$\%$$\%$$\%$$\%$$\%$
$\\$
$\\$
$\\$\color{BrickRed}file\_name\-\ =\-\ 'interp2d\_590\_to\_660';\color{Green}
$\\$\color{BrickRed}ld\-\ =\-\ retrieve\_it(curr\_dir,'interval\_arithmetic',file\_name,'data\_final');\color{Green}
$\\$\color{BrickRed}d\-\ =\-\ ld.var;\color{Green}
$\\$\color{BrickRed}cnt\-\ =\-\ 0;\color{Green}
$\\$
$\\$
$\\$\color{BrickRed}cnt\-\ =\-\ cnt\-\ +\-\ 1;\color{Green}
$\\$\color{BrickRed}st{cnt}{1}\-\ =\-\ nm('0.99999995');\color{Green}
$\\$\color{BrickRed}st{cnt}{2}\-\ =\-\ nm('0.99999999');\color{Green}
$\\$\color{BrickRed}st{cnt}{3}\-\ =\-\ 500;\-\ \color{Green}$\%$\-\ number\-\ of\-\ k\-\ intervals
$\\$
$\\$
$\\$\color{BrickRed}cnt\-\ =\-\ cnt\-\ +\-\ 1;\color{Green}
$\\$\color{BrickRed}st{cnt}{1}\-\ =\-\ nm('0.99999999');\color{Green}
$\\$\color{BrickRed}st{cnt}{2}\-\ =\-\ nm('0.9999999993');\color{Green}
$\\$\color{BrickRed}st{cnt}{3}\-\ =\-\ 500;\-\ \color{Green}$\%$\-\ number\-\ of\-\ k\-\ intervals
$\\$
$\\$
$\\$\color{BrickRed}\color{NavyBlue}\-\ for\-\ \color{BrickRed}\-\ ind\-\ =\-\ 1:cnt\color{Green}
$\\$\color{BrickRed}\-\ \-\ \-\ \-\ kvals\-\ =\-\ linspace(inf(st{ind}{1}),sup(st{ind}{2}),st{ind}{3}+1);\color{Green}
$\\$\color{BrickRed}\-\ \-\ \-\ \-\ \color{NavyBlue}\-\ for\-\ \color{BrickRed}\-\ j\-\ =\-\ length(kvals)-1\color{Green}
$\\$\color{BrickRed}\-\ \-\ \-\ \-\ \-\ \-\ \-\ \-\ clc;\color{Green}
$\\$\color{BrickRed}\-\ \-\ \-\ \-\ \-\ \-\ \-\ \-\ fprintf('upper\-\ instability,\-\ k:\-\ $\%$16.16g\textbackslash n',mid(kvals(j)));\color{Green}
$\\$\color{BrickRed}\-\ \-\ \-\ \-\ \-\ \-\ \-\ \-\ verify\_instability\_upper(d,\-\ kvals(j),\-\ kvals(j+1),psi\_left,psi\_right,ints\_y);\-\ \color{Green}
$\\$\color{BrickRed}\-\ \-\ \-\ \-\ \color{NavyBlue}\-\ end\-\ \color{BrickRed}\color{Green}
$\\$\color{BrickRed}\color{NavyBlue}\-\ end\-\ \color{BrickRed}\color{Green}
$\\$
$\\$
$\\$\color{Green}$\%$$\%$$\%$$\%$$\%$$\%$$\%$$\%$$\%$$\%$$\%$$\%$$\%$$\%$$\%$$\%$$\%$$\%$$\%$$\%$$\%$$\%$$\%$$\%$
$\\$
$\\$
$\\$\color{BrickRed}file\_name\-\ =\-\ 'interp2d\_650\_to\_710';\color{Green}
$\\$\color{BrickRed}ld\-\ =\-\ retrieve\_it(curr\_dir,'interval\_arithmetic',file\_name,'data\_final');\color{Green}
$\\$\color{BrickRed}d\-\ =\-\ ld.var;\color{Green}
$\\$\color{BrickRed}cnt\-\ =\-\ 0;\color{Green}
$\\$
$\\$
$\\$\color{BrickRed}cnt\-\ =\-\ cnt\-\ +\-\ 1;\color{Green}
$\\$\color{BrickRed}st{cnt}{1}\-\ =\-\ nm('0.9999999993');\color{Green}
$\\$\color{BrickRed}st{cnt}{2}\-\ =\-\ nm('0.9999999999');\color{Green}
$\\$\color{BrickRed}st{cnt}{3}\-\ =\-\ 500;\-\ \color{Green}$\%$\-\ number\-\ of\-\ k\-\ intervals
$\\$
$\\$
$\\$\color{BrickRed}cnt\-\ =\-\ cnt\-\ +\-\ 1;\color{Green}
$\\$\color{BrickRed}st{cnt}{1}\-\ =\-\ nm('0.9999999999');\color{Green}
$\\$\color{BrickRed}st{cnt}{2}\-\ =\-\ nm('0.99999999999');\color{Green}
$\\$\color{BrickRed}st{cnt}{3}\-\ =\-\ 500;\-\ \color{Green}$\%$\-\ number\-\ of\-\ k\-\ intervals
$\\$
$\\$
$\\$\color{BrickRed}cnt\-\ =\-\ cnt\-\ +\-\ 1;\color{Green}
$\\$\color{BrickRed}st{cnt}{1}\-\ =\-\ nm('0.99999999999');\color{Green}
$\\$\color{BrickRed}st{cnt}{2}\-\ =\-\ nm('0.999999999997');\color{Green}
$\\$\color{BrickRed}st{cnt}{3}\-\ =\-\ 500;\-\ \color{Green}$\%$\-\ number\-\ of\-\ k\-\ intervals
$\\$
$\\$
$\\$
$\\$
$\\$\color{BrickRed}\color{NavyBlue}\-\ for\-\ \color{BrickRed}\-\ ind\-\ =\-\ 1:cnt\color{Green}
$\\$\color{BrickRed}\-\ \-\ \-\ \-\ kvals\-\ =\-\ linspace(inf(st{ind}{1}),sup(st{ind}{2}),st{ind}{3}+1);\color{Green}
$\\$\color{BrickRed}\-\ \-\ \-\ \-\ \color{NavyBlue}\-\ for\-\ \color{BrickRed}\-\ j\-\ =\-\ length(kvals)-1\color{Green}
$\\$\color{BrickRed}\-\ \-\ \-\ \-\ \-\ \-\ \-\ \-\ clc;\color{Green}
$\\$\color{BrickRed}\-\ \-\ \-\ \-\ \-\ \-\ \-\ \-\ fprintf('upper\-\ instability,\-\ k:\-\ $\%$16.16g\textbackslash n',mid(kvals(j)));\color{Green}
$\\$\color{BrickRed}\-\ \-\ \-\ \-\ \-\ \-\ \-\ \-\ verify\_instability\_upper(d,\-\ kvals(j),\-\ kvals(j+1),psi\_left,psi\_right,ints\_y);\-\ \color{Green}
$\\$\color{BrickRed}\-\ \-\ \-\ \-\ \color{NavyBlue}\-\ end\-\ \color{BrickRed}\color{Green}
$\\$\color{BrickRed}\color{NavyBlue}\-\ end\-\ \color{BrickRed}\color{Green}
$\\$
$\\$
$\\$\color{Black}\section{driver\_stability\_n0.m}

\color{Green}\color{Green}$\%$\-\ clear\-\ all;\-\ curr\_dir\-\ =\-\ local\_startup;
$\\$\color{Green}$\%$\-\ 
$\\$\color{Green}$\%$\-\ $\%$\-\ -----------------------------------------------------------
$\\$\color{Green}$\%$\-\ $\%$\-\ k\-\ =\-\ 0.93\-\ -\-\ 0.9997
$\\$\color{Green}$\%$\-\ $\%$\-\ -----------------------------------------------------------
$\\$
$\\$
$\\$\color{BrickRed}file\_name\-\ =\-\ 'interp2d\_100\_to\_400';\color{Green}
$\\$\color{BrickRed}ld\-\ =\-\ retrieve\_it(curr\_dir,'interval\_arithmetic',file\_name,'data\_final');\color{Green}
$\\$\color{BrickRed}d\-\ =\-\ ld.var;\color{Green}
$\\$
$\\$
$\\$\color{BrickRed}ints\_yL\-\ =\-\ 1200;\color{Green}
$\\$\color{BrickRed}ints\_yR\-\ =\-\ 1000;\color{Green}
$\\$\color{BrickRed}psi\_L\-\ =\-\ nm('0.001');\color{Green}
$\\$\color{BrickRed}psi\_M\-\ =\-\ nm('0.5');\color{Green}
$\\$\color{BrickRed}psi\_R\-\ =\-\ nm('0.7');\color{Green}
$\\$\color{BrickRed}psi\_R2\-\ =\-\ nm('0.8');\color{Green}
$\\$\color{BrickRed}psi\_R3\-\ =\-\ nm('0.9');\color{Green}
$\\$\color{BrickRed}stats\-\ =\-\ 'off';\color{Green}
$\\$\color{BrickRed}pnts\-\ =\-\ 1500;\color{Green}
$\\$\color{BrickRed}cnt\-\ =\-\ 0;\color{Green}
$\\$
$\\$
$\\$\color{BrickRed}cnt\-\ =\-\ cnt\-\ +\-\ 1;\color{Green}
$\\$\color{BrickRed}st{cnt}{1}\-\ =\-\ nm('0.93');\-\ \color{Green}$\%$\-\ left\-\ k\-\ value
$\\$\color{BrickRed}st{cnt}{2}\-\ =\-\ nm('0.99');\-\ \color{Green}$\%$\-\ right\-\ k\-\ value
$\\$\color{BrickRed}st{cnt}{3}\-\ =\-\ 800;\-\ \color{Green}$\%$\-\ kvals
$\\$
$\\$
$\\$\color{BrickRed}cnt\-\ =\-\ cnt\-\ +\-\ 1;\color{Green}
$\\$\color{BrickRed}st{cnt}{1}\-\ =\-\ nm('0.99');\color{Green}
$\\$\color{BrickRed}st{cnt}{2}\-\ =\-\ nm('0.999');\color{Green}
$\\$\color{BrickRed}st{cnt}{3}\-\ =\-\ 100;\color{Green}
$\\$
$\\$
$\\$\color{BrickRed}cnt\-\ =\-\ cnt\-\ +\-\ 1;\color{Green}
$\\$\color{BrickRed}st{cnt}{1}\-\ =\-\ nm('0.999');\color{Green}
$\\$\color{BrickRed}st{cnt}{2}\-\ =\-\ nm('0.9997');\color{Green}
$\\$\color{BrickRed}st{cnt}{3}\-\ =\-\ 100;\color{Green}
$\\$
$\\$
$\\$\color{BrickRed}\color{NavyBlue}\-\ for\-\ \color{BrickRed}\-\ ind\-\ =\-\ 1:cnt\color{Green}
$\\$\color{BrickRed}\-\ \-\ \-\ \-\ kvals\-\ =\-\ (linspace(inf(st{cnt}{1}),sup(st{cnt}{2}),st{cnt}{3}));\color{Green}
$\\$\color{BrickRed}\-\ \-\ \-\ \-\ \color{NavyBlue}\-\ parfor\-\ \color{BrickRed}\-\ j\-\ =\-\ 1:length(kvals)-1\color{Green}
$\\$
$\\$
$\\$\color{BrickRed}\-\ \-\ \-\ \-\ \-\ \-\ \-\ \-\ local\_startup\_batch(curr\_dir);\color{Green}
$\\$\color{BrickRed}\-\ \-\ \-\ \-\ \-\ \-\ \-\ \-\ clc;\color{Green}
$\\$\color{BrickRed}\-\ \-\ \-\ \-\ \-\ \-\ \-\ \-\ j\color{Green}
$\\$\color{BrickRed}\-\ \-\ \-\ \-\ \-\ \-\ \-\ \-\ \color{Green}
$\\$\color{BrickRed}\-\ \-\ \-\ \-\ \-\ \-\ \-\ \-\ verify\_stability\_n0\_strict(d,\-\ kvals(j),\-\ kvals(j+1),ints\_yL,...\color{Green}
$\\$\color{BrickRed}\-\ \-\ \-\ \-\ \-\ \-\ \-\ \-\ \-\ \-\ \-\ \-\ \-\ ints\_yR,psi\_L,psi\_M,psi\_R,psi\_R2,psi\_R3,stats);\color{Green}
$\\$
$\\$
$\\$\color{BrickRed}\-\ \-\ \-\ \-\ \-\ \-\ \-\ \-\ \-\ verify\_stability\_single(d,nm(kvals(j),kvals(j+1)),pnts,psi\_R,psi\_R2)\color{Green}
$\\$
$\\$
$\\$\color{BrickRed}\-\ \-\ \-\ \-\ \color{NavyBlue}\-\ end\-\ \color{BrickRed}\color{Green}
$\\$\color{BrickRed}\color{NavyBlue}\-\ end\-\ \color{BrickRed}\color{Green}
$\\$
$\\$
$\\$\color{Green}$\%$\-\ -----------------------------------------------------------
$\\$\color{Green}$\%$\-\ k\-\ =\-\ 0.9997\-\ -\-\ 0.99999
$\\$\color{Green}$\%$\-\ -----------------------------------------------------------
$\\$
$\\$
$\\$\color{BrickRed}clc;\color{Green}
$\\$\color{BrickRed}file\_name\-\ =\-\ 'interp2d\_350\_to\_500';\color{Green}
$\\$\color{BrickRed}ld\-\ =\-\ retrieve\_it(curr\_dir,'interval\_arithmetic',file\_name,'data\_final');\color{Green}
$\\$\color{BrickRed}d\-\ =\-\ ld.var;\color{Green}
$\\$\color{BrickRed}cnt\-\ =\-\ 0;\color{Green}
$\\$
$\\$
$\\$\color{BrickRed}cnt\-\ =\-\ cnt\-\ +\-\ 1;\color{Green}
$\\$\color{BrickRed}st{cnt}{1}\-\ =\-\ nm('0.9997');\color{Green}
$\\$\color{BrickRed}st{cnt}{2}\-\ =\-\ nm('0.9999');\color{Green}
$\\$\color{BrickRed}st{cnt}{3}\-\ =\-\ 600;\color{Green}
$\\$
$\\$
$\\$\color{BrickRed}cnt\-\ =\-\ cnt\-\ +\-\ 1;\color{Green}
$\\$\color{BrickRed}st{cnt}{1}\-\ =\-\ nm('0.9999');\color{Green}
$\\$\color{BrickRed}st{cnt}{2}\-\ =\-\ nm('0.999993');\color{Green}
$\\$\color{BrickRed}st{cnt}{3}\-\ =\-\ 600;\color{Green}
$\\$
$\\$
$\\$\color{BrickRed}\color{NavyBlue}\-\ for\-\ \color{BrickRed}\-\ ind\-\ =\-\ 1:cnt\color{Green}
$\\$\color{BrickRed}\-\ \-\ \-\ \-\ kvals\-\ =\-\ fliplr(linspace(inf(st{cnt}{1}),sup(st{cnt}{2}),st{cnt}{3}));\color{Green}
$\\$\color{BrickRed}\-\ \-\ \-\ \-\ \color{NavyBlue}\-\ parfor\-\ \color{BrickRed}\-\ j\-\ =\-\ 1:length(kvals)-1\color{Green}
$\\$
$\\$
$\\$\color{BrickRed}\-\ \-\ \-\ \-\ \-\ \-\ \-\ \-\ local\_startup\_batch(curr\_dir);\color{Green}
$\\$\color{BrickRed}\-\ \-\ \-\ \-\ \-\ \-\ \-\ \-\ clc;\color{Green}
$\\$\color{BrickRed}\-\ \-\ \-\ \-\ \-\ \-\ \-\ \-\ \color{Green}
$\\$\color{BrickRed}\-\ \-\ \-\ \-\ \-\ \-\ \-\ \-\ verify\_stability\_n0\_strict(d,\-\ kvals(j),\-\ kvals(j+1),ints\_yL,...\color{Green}
$\\$\color{BrickRed}\-\ \-\ \-\ \-\ \-\ \-\ \-\ \-\ \-\ \-\ \-\ \-\ \-\ ints\_yR,psi\_L,psi\_M,psi\_R,psi\_R2,psi\_R3,stats);\color{Green}
$\\$
$\\$
$\\$\color{BrickRed}\-\ \-\ \-\ \-\ \-\ \-\ \-\ \-\ \-\ verify\_stability\_single(d,nm(kvals(j),kvals(j+1)),pnts,psi\_R,psi\_R2)\color{Green}
$\\$
$\\$
$\\$\color{BrickRed}\-\ \-\ \-\ \-\ \color{NavyBlue}\-\ end\-\ \color{BrickRed}\color{Green}
$\\$\color{BrickRed}\color{NavyBlue}\-\ end\-\ \color{BrickRed}\color{Green}
$\\$
$\\$
$\\$
$\\$
$\\$\color{Green}$\%$\-\ -----------------------------------------------------------
$\\$\color{Green}$\%$\-\ k\-\ =\-\ 0.99999\-\ -\-\ 0.9999983
$\\$\color{Green}$\%$\-\ -----------------------------------------------------------
$\\$
$\\$
$\\$\color{BrickRed}clc;\color{Green}
$\\$\color{BrickRed}file\_name\-\ =\-\ 'interp2d\_490\_to\_538';\color{Green}
$\\$\color{BrickRed}ld\-\ =\-\ retrieve\_it(curr\_dir,'interval\_arithmetic',file\_name,'data\_final');\color{Green}
$\\$\color{BrickRed}d\-\ =\-\ ld.var;\color{Green}
$\\$\color{BrickRed}cnt\-\ =\-\ 0;\color{Green}
$\\$
$\\$
$\\$\color{BrickRed}cnt\-\ =\-\ cnt\-\ +\-\ 1;\color{Green}
$\\$\color{BrickRed}st{cnt}{1}\-\ =\-\ nm('0.999993');\color{Green}
$\\$\color{BrickRed}st{cnt}{2}\-\ =\-\ nm('0.999997');\color{Green}
$\\$\color{BrickRed}st{cnt}{3}\-\ =\-\ 200;\color{Green}
$\\$
$\\$
$\\$\color{BrickRed}cnt\-\ =\-\ cnt\-\ +\-\ 1;\color{Green}
$\\$\color{BrickRed}st{cnt}{1}\-\ =\-\ nm('0.999997');\color{Green}
$\\$\color{BrickRed}st{cnt}{2}\-\ =\-\ nm('0.9999978');\color{Green}
$\\$\color{BrickRed}st{cnt}{3}\-\ =\-\ 100;\color{Green}
$\\$
$\\$
$\\$\color{BrickRed}cnt\-\ =\-\ cnt\-\ +\-\ 1;\color{Green}
$\\$\color{BrickRed}st{cnt}{1}\-\ =\-\ nm('0.9999977');\color{Green}
$\\$\color{BrickRed}st{cnt}{2}\-\ =\-\ nm('0.9999978');\color{Green}
$\\$\color{BrickRed}st{cnt}{3}\-\ =\-\ 10;\color{Green}
$\\$
$\\$
$\\$\color{BrickRed}cnt\-\ =\-\ cnt\-\ +\-\ 1;\color{Green}
$\\$\color{BrickRed}st{cnt}{1}\-\ =\-\ nm('0.9999978');\color{Green}
$\\$\color{BrickRed}st{cnt}{2}\-\ =\-\ nm('0.9999979');\color{Green}
$\\$\color{BrickRed}st{cnt}{3}\-\ =\-\ 10;\color{Green}
$\\$
$\\$
$\\$\color{BrickRed}cnt\-\ =\-\ cnt\-\ +\-\ 1;\color{Green}
$\\$\color{BrickRed}st{cnt}{1}\-\ =\-\ nm('0.9999979');\color{Green}
$\\$\color{BrickRed}st{cnt}{2}\-\ =\-\ nm('0.999998');\color{Green}
$\\$\color{BrickRed}st{cnt}{3}\-\ =\-\ 20;\color{Green}
$\\$
$\\$
$\\$\color{BrickRed}cnt\-\ =\-\ cnt\-\ +\-\ 1;\color{Green}
$\\$\color{BrickRed}st{cnt}{1}\-\ =\-\ nm('0.999998');\color{Green}
$\\$\color{BrickRed}st{cnt}{2}\-\ =\-\ nm('0.9999981');\color{Green}
$\\$\color{BrickRed}st{cnt}{3}\-\ =\-\ 20;\color{Green}
$\\$
$\\$
$\\$\color{BrickRed}cnt\-\ =\-\ cnt\-\ +\-\ 1;\color{Green}
$\\$\color{BrickRed}st{cnt}{1}\-\ =\-\ nm('0.9999981');\color{Green}
$\\$\color{BrickRed}st{cnt}{2}\-\ =\-\ nm('0.9999982');\color{Green}
$\\$\color{BrickRed}st{cnt}{3}\-\ =\-\ 30;\color{Green}
$\\$
$\\$
$\\$\color{BrickRed}cnt\-\ =\-\ cnt\-\ +\-\ 1;\color{Green}
$\\$\color{BrickRed}st{cnt}{1}\-\ =\-\ nm('0.9999982');\color{Green}
$\\$\color{BrickRed}st{cnt}{2}\-\ =\-\ nm('0.9999983');\color{Green}
$\\$\color{BrickRed}st{cnt}{3}\-\ =\-\ 100;\color{Green}
$\\$
$\\$
$\\$\color{BrickRed}\color{NavyBlue}\-\ for\-\ \color{BrickRed}\-\ ind\-\ =\-\ 1:cnt\color{Green}
$\\$\color{BrickRed}\-\ \-\ \-\ \-\ kvals\-\ =\-\ fliplr(linspace(inf(st{cnt}{1}),sup(st{cnt}{2}),st{cnt}{3}));\color{Green}
$\\$\color{BrickRed}\-\ \-\ \-\ \-\ \color{NavyBlue}\-\ parfor\-\ \color{BrickRed}\-\ j\-\ =\-\ 1:length(kvals)-1\color{Green}
$\\$\color{BrickRed}\-\ \-\ \-\ \-\ \-\ \-\ \-\ \-\ \color{Green}
$\\$\color{BrickRed}\-\ \-\ \-\ \-\ \-\ \-\ \-\ \-\ local\_startup\_batch(curr\_dir);\color{Green}
$\\$\color{BrickRed}\-\ \-\ \-\ \-\ \-\ \-\ \-\ \-\ clc;\color{Green}
$\\$\color{BrickRed}\-\ \-\ \-\ \-\ \-\ \-\ \-\ \-\ \color{Green}
$\\$\color{BrickRed}\-\ \-\ \-\ \-\ \-\ \-\ \-\ \-\ verify\_stability\_n0\_strict(d,\-\ kvals(j),\-\ kvals(j+1),ints\_yL,...\color{Green}
$\\$\color{BrickRed}\-\ \-\ \-\ \-\ \-\ \-\ \-\ \-\ \-\ \-\ \-\ \-\ \-\ ints\_yR,psi\_L,psi\_M,psi\_R,psi\_R2,psi\_R3,stats);\color{Green}
$\\$
$\\$
$\\$\color{BrickRed}\-\ \-\ \-\ \-\ \-\ \-\ \-\ \-\ \-\ verify\_stability\_single(d,nm(kvals(j),kvals(j+1)),pnts,psi\_R,psi\_R2)\color{Green}
$\\$
$\\$
$\\$\color{BrickRed}\-\ \-\ \-\ \-\ \color{NavyBlue}\-\ end\-\ \color{BrickRed}\color{Green}
$\\$\color{BrickRed}\color{NavyBlue}\-\ end\-\ \color{BrickRed}\color{Green}
$\\$
$\\$
$\\$
$\\$
$\\$
$\\$
$\\$
$\\$
$\\$\color{Black}\section{driver\_stability\_n1.m}

\color{Green}\color{BrickRed}clear\-\ all;\-\ curr\_dir\-\ =\-\ local\_startup;\color{Green}
$\\$
$\\$
$\\$\color{Green}$\%$\-\ keepf\-\ fixed
$\\$\color{BrickRed}ints\_x\-\ =\-\ 1;\color{Green}
$\\$
$\\$
$\\$\color{Green}$\%$\-\ -----------------------------------------------------------
$\\$\color{Green}$\%$\-\ k\-\ =\-\ 0.93\-\ -\-\ 0.9997
$\\$\color{Green}$\%$\-\ -----------------------------------------------------------
$\\$
$\\$
$\\$\color{BrickRed}clc;\color{Green}
$\\$\color{BrickRed}file\_name\-\ =\-\ 'interp2d\_100\_to\_400';\color{Green}
$\\$\color{BrickRed}ld\-\ =\-\ retrieve\_it(curr\_dir,'interval\_arithmetic',file\_name,'data\_final');\color{Green}
$\\$\color{BrickRed}d\-\ =\-\ ld.var;\color{Green}
$\\$\color{BrickRed}cnt\-\ =\-\ 0;\color{Green}
$\\$
$\\$
$\\$\color{BrickRed}cnt\-\ =\-\ cnt\-\ +\-\ 1;\color{Green}
$\\$\color{BrickRed}st{cnt}{1}\-\ =\-\ nm('0.9426');\color{Green}
$\\$\color{BrickRed}st{cnt}{2}\-\ =\-\ nm('0.943');\color{Green}
$\\$\color{BrickRed}st{cnt}{3}\-\ =\-\ 11;\-\ \color{Green}
$\\$\color{BrickRed}st{cnt}{4}\-\ =\-\ 3000;\-\ \color{Green}
$\\$
$\\$
$\\$\color{BrickRed}cnt\-\ =\-\ cnt\-\ +\-\ 1;\color{Green}
$\\$\color{BrickRed}st{cnt}{1}\-\ =\-\ nm('0.943');\color{Green}
$\\$\color{BrickRed}st{cnt}{2}\-\ =\-\ nm('0.944');\color{Green}
$\\$\color{BrickRed}st{cnt}{3}\-\ =\-\ 11;\-\ \color{Green}
$\\$\color{BrickRed}st{cnt}{4}\-\ =\-\ 3000;\color{Green}
$\\$
$\\$
$\\$\color{BrickRed}cnt\-\ =\-\ cnt\-\ +\-\ 1;\color{Green}
$\\$\color{BrickRed}st{cnt}{1}\-\ =\-\ nm('0.944');\color{Green}
$\\$\color{BrickRed}st{cnt}{2}\-\ =\-\ nm('0.945');\color{Green}
$\\$\color{BrickRed}st{cnt}{3}\-\ =\-\ 11;\-\ \color{Green}
$\\$\color{BrickRed}st{cnt}{4}\-\ =\-\ 3000;\color{Green}
$\\$\color{BrickRed}\-\ \color{Green}
$\\$\color{BrickRed}cnt\-\ =\-\ cnt\-\ +\-\ 1;\color{Green}
$\\$\color{BrickRed}st{cnt}{1}\-\ =\-\ nm('0.945');\color{Green}
$\\$\color{BrickRed}st{cnt}{2}\-\ =\-\ nm('0.946');\color{Green}
$\\$\color{BrickRed}st{cnt}{3}\-\ =\-\ 11;\-\ \color{Green}
$\\$\color{BrickRed}st{cnt}{4}\-\ =\-\ 3000;\color{Green}
$\\$
$\\$
$\\$\color{BrickRed}cnt\-\ =\-\ cnt\-\ +\-\ 1;\color{Green}
$\\$\color{BrickRed}st{cnt}{1}\-\ =\-\ nm('0.946');\color{Green}
$\\$\color{BrickRed}st{cnt}{2}\-\ =\-\ nm('0.947');\color{Green}
$\\$\color{BrickRed}st{cnt}{3}\-\ =\-\ 11;\-\ \color{Green}
$\\$\color{BrickRed}st{cnt}{4}\-\ =\-\ 3000;\color{Green}
$\\$
$\\$
$\\$\color{BrickRed}cnt\-\ =\-\ cnt\-\ +\-\ 1;\color{Green}
$\\$\color{BrickRed}st{cnt}{1}\-\ =\-\ nm('0.947');\color{Green}
$\\$\color{BrickRed}st{cnt}{2}\-\ =\-\ nm('0.948');\color{Green}
$\\$\color{BrickRed}st{cnt}{3}\-\ =\-\ 11;\-\ \color{Green}
$\\$\color{BrickRed}st{cnt}{4}\-\ =\-\ 3000;\color{Green}
$\\$
$\\$
$\\$\color{BrickRed}cnt\-\ =\-\ cnt\-\ +\-\ 1;\color{Green}
$\\$\color{BrickRed}st{cnt}{1}\-\ =\-\ nm('0.948');\color{Green}
$\\$\color{BrickRed}st{cnt}{2}\-\ =\-\ nm('0.949');\color{Green}
$\\$\color{BrickRed}st{cnt}{3}\-\ =\-\ 11;\-\ \color{Green}
$\\$\color{BrickRed}st{cnt}{4}\-\ =\-\ 3000;\color{Green}
$\\$
$\\$
$\\$\color{BrickRed}cnt\-\ =\-\ cnt\-\ +\-\ 1;\color{Green}
$\\$\color{BrickRed}st{cnt}{1}\-\ =\-\ nm('0.949');\color{Green}
$\\$\color{BrickRed}st{cnt}{2}\-\ =\-\ nm('0.95');\color{Green}
$\\$\color{BrickRed}st{cnt}{3}\-\ =\-\ 20;\-\ \color{Green}
$\\$\color{BrickRed}st{cnt}{4}\-\ =\-\ 3000;\color{Green}
$\\$
$\\$
$\\$\color{BrickRed}cnt\-\ =\-\ cnt\-\ +\-\ 1;\color{Green}
$\\$\color{BrickRed}st{cnt}{1}\-\ =\-\ nm('0.95');\color{Green}
$\\$\color{BrickRed}st{cnt}{2}\-\ =\-\ nm('0.96');\color{Green}
$\\$\color{BrickRed}st{cnt}{3}\-\ =\-\ 11;\-\ \color{Green}
$\\$\color{BrickRed}st{cnt}{4}\-\ =\-\ 3000;\color{Green}
$\\$
$\\$
$\\$\color{BrickRed}cnt\-\ =\-\ cnt\-\ +\-\ 1;\color{Green}
$\\$\color{BrickRed}st{cnt}{1}\-\ =\-\ nm('0.96');\color{Green}
$\\$\color{BrickRed}st{cnt}{2}\-\ =\-\ nm('0.97');\color{Green}
$\\$\color{BrickRed}st{cnt}{3}\-\ =\-\ 11;\-\ \color{Green}
$\\$\color{BrickRed}st{cnt}{4}\-\ =\-\ 1000;\color{Green}
$\\$
$\\$
$\\$\color{BrickRed}cnt\-\ =\-\ cnt\-\ +\-\ 1;\color{Green}
$\\$\color{BrickRed}st{cnt}{1}\-\ =\-\ nm('0.97');\color{Green}
$\\$\color{BrickRed}st{cnt}{2}\-\ =\-\ nm('0.999');\color{Green}
$\\$\color{BrickRed}st{cnt}{3}\-\ =\-\ 200;\-\ \color{Green}
$\\$\color{BrickRed}st{cnt}{4}\-\ =\-\ 200;\color{Green}
$\\$
$\\$
$\\$\color{BrickRed}cnt\-\ =\-\ cnt\-\ +\-\ 1;\color{Green}
$\\$\color{BrickRed}st{cnt}{1}\-\ =\-\ nm('0.999');\color{Green}
$\\$\color{BrickRed}st{cnt}{2}\-\ =\-\ nm('0.9997');\color{Green}
$\\$\color{BrickRed}st{cnt}{3}\-\ =\-\ 20;\-\ \color{Green}
$\\$\color{BrickRed}st{cnt}{4}\-\ =\-\ 200;\color{Green}
$\\$
$\\$
$\\$
$\\$
$\\$\color{BrickRed}\color{NavyBlue}\-\ for\-\ \color{BrickRed}\-\ ind\-\ =\-\ 1:cnt\color{Green}
$\\$\color{BrickRed}\-\ \-\ \-\ \-\ \color{Green}
$\\$\color{BrickRed}\-\ \-\ \-\ \-\ kvals\-\ =\-\ linspace(inf(st{ind}{1}),sup(st{ind}{2}),st{ind}{3}+1);\color{Green}
$\\$\color{BrickRed}\-\ \-\ \-\ \-\ ints\_y\-\ =\-\ st{ind}{4};\color{Green}
$\\$\color{BrickRed}\-\ \-\ \-\ \-\ \color{NavyBlue}\-\ for\-\ \color{BrickRed}\-\ j\-\ =\-\ 1:length(kvals)-1\color{Green}
$\\$\color{BrickRed}\-\ \-\ \-\ \-\ \-\ \-\ \-\ \-\ \color{Green}
$\\$\color{BrickRed}\-\ \-\ \-\ \-\ \-\ \-\ \-\ \-\ clc;\color{Green}
$\\$\color{BrickRed}\-\ \-\ \-\ \-\ \-\ \-\ \-\ \-\ fprintf('k\-\ =\-\ $\%$16.16g\textbackslash n',mid(kvals(j)));\color{Green}
$\\$\color{BrickRed}\-\ \-\ \-\ \-\ \-\ \-\ \-\ \-\ \color{Green}
$\\$\color{BrickRed}\-\ \-\ \-\ \-\ \-\ \-\ \-\ \-\ verify\_stability\_n1(d,\-\ kvals(j),\-\ kvals(j+1),ints\_x,ints\_y);\color{Green}
$\\$
$\\$
$\\$\color{BrickRed}\-\ \-\ \-\ \-\ \color{NavyBlue}\-\ end\-\ \color{BrickRed}\color{Green}
$\\$\color{BrickRed}\color{NavyBlue}\-\ end\-\ \color{BrickRed}\color{Green}
$\\$
$\\$
$\\$
$\\$
$\\$
$\\$
$\\$\color{Green}$\%$\-\ -----------------------------------------------------------
$\\$\color{Green}$\%$\-\ k\-\ =\-\ 0.9997\-\ -\-\ 0.99999
$\\$\color{Green}$\%$\-\ -----------------------------------------------------------
$\\$
$\\$
$\\$\color{BrickRed}clc;\color{Green}
$\\$\color{BrickRed}file\_name\-\ =\-\ 'interp2d\_350\_to\_500';\color{Green}
$\\$\color{BrickRed}ld\-\ =\-\ retrieve\_it(curr\_dir,'interval\_arithmetic',file\_name,'data\_final');\color{Green}
$\\$\color{BrickRed}d\-\ =\-\ ld.var;\color{Green}
$\\$\color{BrickRed}cnt\-\ =\-\ 0;\color{Green}
$\\$
$\\$
$\\$\color{BrickRed}cnt\-\ =\-\ cnt\-\ +\-\ 1;\color{Green}
$\\$\color{BrickRed}st{cnt}{1}\-\ =\-\ nm('0.9997');\color{Green}
$\\$\color{BrickRed}st{cnt}{2}\-\ =\-\ nm('0.9999');\color{Green}
$\\$\color{BrickRed}st{cnt}{3}\-\ =\-\ 11;\-\ \color{Green}
$\\$\color{BrickRed}st{cnt}{4}\-\ =\-\ 3000;\color{Green}
$\\$
$\\$
$\\$\color{BrickRed}cnt\-\ =\-\ cnt\-\ +\-\ 1;\color{Green}
$\\$\color{BrickRed}st{cnt}{1}\-\ =\-\ nm('0.9999');\color{Green}
$\\$\color{BrickRed}st{cnt}{2}\-\ =\-\ nm('0.999993');\color{Green}
$\\$\color{BrickRed}st{cnt}{3}\-\ =\-\ 100;\-\ \color{Green}
$\\$\color{BrickRed}st{cnt}{4}\-\ =\-\ 3000;\color{Green}
$\\$
$\\$
$\\$\color{BrickRed}\color{NavyBlue}\-\ for\-\ \color{BrickRed}\-\ ind\-\ =\-\ 1:cnt\color{Green}
$\\$\color{BrickRed}\-\ \-\ \-\ \-\ \color{Green}
$\\$\color{BrickRed}\-\ \-\ \-\ \-\ kvals\-\ =\-\ linspace(inf(st{ind}{1}),sup(st{ind}{2}),st{ind}{3}+1);\color{Green}
$\\$\color{BrickRed}\-\ \-\ \-\ \-\ ints\_y\-\ =\-\ st{ind}{4};\color{Green}
$\\$\color{BrickRed}\-\ \-\ \-\ \-\ \color{NavyBlue}\-\ for\-\ \color{BrickRed}\-\ j\-\ =\-\ 1:length(kvals)-1\color{Green}
$\\$\color{BrickRed}\-\ \-\ \-\ \-\ \-\ \-\ \-\ \-\ \color{Green}
$\\$\color{BrickRed}\-\ \-\ \-\ \-\ \-\ \-\ \-\ \-\ clc;\color{Green}
$\\$\color{BrickRed}\-\ \-\ \-\ \-\ \-\ \-\ \-\ \-\ fprintf('k\-\ =\-\ $\%$16.16g\textbackslash n',mid(kvals(j)));\color{Green}
$\\$\color{BrickRed}\-\ \-\ \-\ \-\ \-\ \-\ \-\ \-\ \color{Green}
$\\$\color{BrickRed}\-\ \-\ \-\ \-\ \-\ \-\ \-\ \-\ verify\_stability\_n1(d,\-\ kvals(j),\-\ kvals(j+1),ints\_x,ints\_y);\color{Green}
$\\$
$\\$
$\\$\color{BrickRed}\-\ \-\ \-\ \-\ \color{NavyBlue}\-\ end\-\ \color{BrickRed}\color{Green}
$\\$\color{BrickRed}\color{NavyBlue}\-\ end\-\ \color{BrickRed}\color{Green}
$\\$
$\\$
$\\$
$\\$
$\\$
$\\$
$\\$\color{Green}$\%$\-\ -----------------------------------------------------------
$\\$\color{Green}$\%$\-\ k\-\ =\-\ 0.99999\-\ -\-\ 0.9999983
$\\$\color{Green}$\%$\-\ -----------------------------------------------------------
$\\$
$\\$
$\\$\color{BrickRed}clc;\color{Green}
$\\$\color{BrickRed}file\_name\-\ =\-\ 'interp2d\_490\_to\_538';\color{Green}
$\\$\color{BrickRed}ld\-\ =\-\ retrieve\_it(curr\_dir,'interval\_arithmetic',file\_name,'data\_final');\color{Green}
$\\$\color{BrickRed}d\-\ =\-\ ld.var;\color{Green}
$\\$\color{BrickRed}cnt\-\ =\-\ 0;\color{Green}
$\\$
$\\$
$\\$\color{BrickRed}cnt\-\ =\-\ cnt\-\ +\-\ 1;\color{Green}
$\\$\color{BrickRed}st{cnt}{1}\-\ =\-\ nm('0.999993');\color{Green}
$\\$\color{BrickRed}st{cnt}{2}\-\ =\-\ nm('0.999994');\color{Green}
$\\$\color{BrickRed}st{cnt}{3}\-\ =\-\ 20;\-\ \color{Green}
$\\$\color{BrickRed}st{cnt}{4}\-\ =\-\ 3000;\color{Green}
$\\$
$\\$
$\\$\color{BrickRed}cnt\-\ =\-\ cnt\-\ +\-\ 1;\color{Green}
$\\$\color{BrickRed}st{cnt}{1}\-\ =\-\ nm('0.999994');\color{Green}
$\\$\color{BrickRed}st{cnt}{2}\-\ =\-\ nm('0.999995');\color{Green}
$\\$\color{BrickRed}st{cnt}{3}\-\ =\-\ 20;\-\ \color{Green}
$\\$\color{BrickRed}st{cnt}{4}\-\ =\-\ 3000;\color{Green}
$\\$
$\\$
$\\$\color{BrickRed}cnt\-\ =\-\ cnt\-\ +\-\ 1;\color{Green}
$\\$\color{BrickRed}st{cnt}{1}\-\ =\-\ nm('0.999995');\color{Green}
$\\$\color{BrickRed}st{cnt}{2}\-\ =\-\ nm('0.999996');\color{Green}
$\\$\color{BrickRed}st{cnt}{3}\-\ =\-\ 20;\-\ \color{Green}
$\\$\color{BrickRed}st{cnt}{4}\-\ =\-\ 3000;\color{Green}
$\\$
$\\$
$\\$\color{BrickRed}cnt\-\ =\-\ cnt\-\ +\-\ 1;\color{Green}
$\\$\color{BrickRed}st{cnt}{1}\-\ =\-\ nm('0.999996');\color{Green}
$\\$\color{BrickRed}st{cnt}{2}\-\ =\-\ nm('0.999997');\color{Green}
$\\$\color{BrickRed}st{cnt}{3}\-\ =\-\ 20;\-\ \color{Green}
$\\$\color{BrickRed}st{cnt}{4}\-\ =\-\ 3000;\color{Green}
$\\$
$\\$
$\\$\color{BrickRed}cnt\-\ =\-\ cnt\-\ +\-\ 1;\color{Green}
$\\$\color{BrickRed}st{cnt}{1}\-\ =\-\ nm('0.999997');\color{Green}
$\\$\color{BrickRed}st{cnt}{2}\-\ =\-\ nm('0.999998');\color{Green}
$\\$\color{BrickRed}st{cnt}{3}\-\ =\-\ 20;\-\ \color{Green}
$\\$\color{BrickRed}st{cnt}{4}\-\ =\-\ 3000;\color{Green}
$\\$
$\\$
$\\$\color{BrickRed}cnt\-\ =\-\ cnt\-\ +\-\ 1;\color{Green}
$\\$\color{BrickRed}st{cnt}{1}\-\ =\-\ nm('0.999998');\color{Green}
$\\$\color{BrickRed}st{cnt}{2}\-\ =\-\ nm('0.999999');\color{Green}
$\\$\color{BrickRed}st{cnt}{3}\-\ =\-\ 100;\-\ \color{Green}
$\\$\color{BrickRed}st{cnt}{4}\-\ =\-\ 3000;\color{Green}
$\\$
$\\$
$\\$\color{BrickRed}\color{NavyBlue}\-\ for\-\ \color{BrickRed}\-\ ind\-\ =\-\ 1:cnt\color{Green}
$\\$\color{BrickRed}\-\ \-\ \-\ \-\ \color{Green}
$\\$\color{BrickRed}\-\ \-\ \-\ \-\ kvals\-\ =\-\ linspace(inf(st{ind}{1}),sup(st{ind}{2}),st{ind}{3}+1);\color{Green}
$\\$\color{BrickRed}\-\ \-\ \-\ \-\ ints\_y\-\ =\-\ st{ind}{4};\color{Green}
$\\$\color{BrickRed}\-\ \-\ \-\ \-\ \color{NavyBlue}\-\ for\-\ \color{BrickRed}\-\ j\-\ =\-\ 1:length(kvals)-1\color{Green}
$\\$\color{BrickRed}\-\ \-\ \-\ \-\ \-\ \-\ \-\ \-\ \color{Green}
$\\$\color{BrickRed}\-\ \-\ \-\ \-\ \-\ \-\ \-\ \-\ clc;\color{Green}
$\\$\color{BrickRed}\-\ \-\ \-\ \-\ \-\ \-\ \-\ \-\ fprintf('k\-\ =\-\ $\%$16.16g\textbackslash n',mid(kvals(j)));\color{Green}
$\\$\color{BrickRed}\-\ \-\ \-\ \-\ \-\ \-\ \-\ \-\ \color{Green}
$\\$\color{BrickRed}\-\ \-\ \-\ \-\ \-\ \-\ \-\ \-\ verify\_stability\_n1(d,\-\ kvals(j),\-\ kvals(j+1),ints\_x,ints\_y);\color{Green}
$\\$
$\\$
$\\$\color{BrickRed}\-\ \-\ \-\ \-\ \color{NavyBlue}\-\ end\-\ \color{BrickRed}\color{Green}
$\\$\color{BrickRed}\color{NavyBlue}\-\ end\-\ \color{BrickRed}\color{Green}
$\\$
$\\$
$\\$\color{Green}$\%$$\%$$\%$$\%$$\%$$\%$$\%$$\%$$\%$$\%$$\%$$\%$$\%$$\%$$\%$$\%$$\%$$\%$$\%$$\%$$\%$$\%$$\%$$\%$$\%$$\%$$\%$$\%$$\%$$\%$$\%$$\%$$\%$
$\\$
$\\$
$\\$
$\\$
$\\$
$\\$
$\\$
$\\$
$\\$
$\\$
$\\$
$\\$
$\\$\color{Black}\section{driver\_strict\_lower.m}

\color{Green}\color{BrickRed}clear\-\ all;\-\ curr\_dir\-\ =\-\ local\_startup;\color{Green}
$\\$\color{BrickRed}intvalinit('DisplayMidRad');\color{Green}
$\\$\color{Green}$\%$\-\ return
$\\$
$\\$
$\\$\color{BrickRed}file\_name\-\ =\-\ 'd\_stability10';\color{Green}
$\\$\color{BrickRed}ld\-\ =\-\ retrieve\_it(curr\_dir,'interval\_arithmetic',file\_name,'data');\color{Green}
$\\$\color{BrickRed}d\-\ =\-\ ld.var;\color{Green}
$\\$
$\\$
$\\$\color{BrickRed}pie\-\ =\-\ nm('pi');\color{Green}
$\\$
$\\$
$\\$\color{Green}$\%$
$\\$\color{Green}$\%$\-\ \-\ minimum\-\ distance\-\ between\-\ the\-\ two\-\ stadiums
$\\$\color{Green}$\%$
$\\$
$\\$
$\\$\color{BrickRed}rho\_q\_sm\-\ =\-\ ((d.rho\_q\-\ -\-\ 1/d.rho\_q)+sqrt((d.rho\_q-1/d.rho\_q)\verb|^|2+16))/4;\color{Green}
$\\$\color{BrickRed}rho\_psi\_sm\-\ =\-\ ((d.rho\_psi\-\ -\-\ 1/d.rho\_psi)+sqrt((d.rho\_psi-1/d.rho\_psi)\verb|^|2+16))/4;\color{Green}
$\\$\color{BrickRed}rho\_q\-\ =\-\ d.rho\_q;\color{Green}
$\\$\color{BrickRed}rho\_psi\-\ =\-\ d.rho\_psi;\color{Green}
$\\$
$\\$
$\\$\color{BrickRed}nm\_pts\-\ =\-\ 100;\color{Green}
$\\$\color{BrickRed}min\_diff\_q\-\ =\-\ inf((rho\_q-1/rho\_q)/2);\color{Green}
$\\$\color{BrickRed}\color{NavyBlue}\-\ for\-\ \color{BrickRed}\-\ j\-\ =\-\ 0:nm\_pts-1\color{Green}
$\\$\color{BrickRed}\-\ \-\ \-\ \-\ theta1\-\ =\-\ nm(2*(j*pie)/nm\_pts,2*((j+1)*pie)/nm\_pts);\color{Green}
$\\$\color{BrickRed}\-\ \-\ \-\ \-\ \color{NavyBlue}\-\ for\-\ \color{BrickRed}\-\ q\-\ =\-\ 0:nm\_pts-1\color{Green}
$\\$\color{BrickRed}\-\ \-\ \-\ \-\ \-\ \-\ \-\ \-\ \-\ \-\ \-\ \-\ theta2\-\ =\-\ nm(2*(q*pie)/nm\_pts,2*((q+1)*pie)/nm\_pts);\color{Green}
$\\$\color{BrickRed}\-\ \-\ \-\ \-\ \-\ \-\ \-\ \-\ \-\ \-\ \-\ \-\ del\-\ =\-\ (exp(1i*theta1)*rho\_q+exp(-1i*theta1)/rho\_q\-\ ...\color{Green}
$\\$\color{BrickRed}\-\ \-\ \-\ \-\ \-\ \-\ \-\ \-\ \-\ \-\ \-\ \-\ \-\ \-\ \-\ \-\ -(exp(1i*theta2)*rho\_q\_sm+exp(-1i*theta2)/rho\_q\_sm))/2;\color{Green}
$\\$\color{BrickRed}\-\ \-\ \-\ \-\ \-\ \-\ \-\ \-\ \-\ \-\ \-\ \-\ min\_diff\_q\-\ =\-\ min(min\_diff\_q,inf(abs(del)));\color{Green}
$\\$\color{BrickRed}\-\ \-\ \-\ \-\ \-\ \-\ \-\ \-\ \-\ \-\ \-\ \-\ \color{NavyBlue}\-\ if\-\ \color{BrickRed}\-\ min\_diff\_q\-\ ==\-\ 0\color{Green}
$\\$\color{BrickRed}\-\ \-\ \-\ \-\ \-\ \-\ \-\ \-\ \-\ \-\ \-\ \-\ \-\ \-\ \-\ \-\ error('min\-\ is\-\ 0')\color{Green}
$\\$\color{BrickRed}\-\ \-\ \-\ \-\ \-\ \-\ \-\ \-\ \-\ \-\ \-\ \-\ \color{NavyBlue}\-\ end\-\ \color{BrickRed}\color{Green}
$\\$\color{BrickRed}\-\ \-\ \-\ \-\ \color{NavyBlue}\-\ end\-\ \color{BrickRed}\color{Green}
$\\$\color{BrickRed}\color{NavyBlue}\-\ end\-\ \color{BrickRed}\color{Green}
$\\$\color{BrickRed}min\_diff\_q\-\ =\-\ min\_diff\_q/2;\color{Green}
$\\$\color{Green}$\%$\-\ min\_diff\_q\-\ =\-\ 0.150639673896165;
$\\$
$\\$
$\\$\color{BrickRed}kvals\-\ =\-\ [nm('0.942197747747748','0.9422'),\-\ ...\color{Green}
$\\$\color{BrickRed}\-\ \-\ \-\ \-\ nm('0.9422','0.9423'),\-\ nm('0.9423','0.9424'),\-\ ...\color{Green}
$\\$\color{BrickRed}\-\ \-\ \-\ \-\ nm('0.9424','0.9425'),\-\ nm('0.9425','0.9426')];\color{Green}
$\\$
$\\$
$\\$
$\\$
$\\$\color{BrickRed}psi0\-\ =\-\ nm('1');\color{Green}
$\\$\color{BrickRed}psiL\-\ =\-\ nm('0.99');\color{Green}
$\\$\color{BrickRed}\-\ \color{Green}
$\\$\color{BrickRed}tic\color{Green}
$\\$\color{BrickRed}\color{NavyBlue}\-\ for\-\ \color{BrickRed}\-\ j\-\ =\-\ 1:length(kvals)\color{Green}
$\\$\color{BrickRed}\-\ \-\ \-\ \-\ \color{Green}
$\\$\color{BrickRed}\-\ \-\ \-\ \-\ fprintf('\textbackslash n\textbackslash n\textbackslash n');\color{Green}
$\\$\color{BrickRed}\-\ \-\ \-\ \-\ strict\_transition\_lower(d,kvals(j),psi0,psiL,min\_diff\_q,rho\_q\_sm)\color{Green}
$\\$
$\\$
$\\$\color{BrickRed}\color{NavyBlue}\-\ end\-\ \color{BrickRed}\color{Green}
$\\$\color{BrickRed}toc\color{Green}
$\\$
$\\$
$\\$\color{BrickRed}disp('verify\-\ stability')\color{Green}
$\\$
$\\$
$\\$\color{BrickRed}ints\_x\-\ =\-\ 1;\color{Green}
$\\$\color{BrickRed}ints\_y\-\ =\-\ 8000;\color{Green}
$\\$\color{BrickRed}divisions\-\ =\-\ 2;\color{Green}
$\\$
$\\$
$\\$\color{BrickRed}\color{NavyBlue}\-\ for\-\ \color{BrickRed}\-\ k\-\ =\-\ 1:length(kvals)\color{Green}
$\\$\color{BrickRed}\-\ \-\ \-\ \-\ kv\-\ =\-\ linspace(inf(kvals(k)),sup(kvals(k)),divisions);\color{Green}
$\\$\color{BrickRed}\-\ \-\ \-\ \-\ \color{NavyBlue}\-\ for\-\ \color{BrickRed}\-\ j\-\ =\-\ 1:length(kv)-1\color{Green}
$\\$\color{BrickRed}\-\ \-\ \-\ \-\ \-\ \-\ \-\ \-\ \color{Green}
$\\$\color{BrickRed}\-\ \-\ \-\ \-\ \-\ \-\ \-\ \-\ kleft\-\ =\-\ kv(j);\color{Green}
$\\$\color{BrickRed}\-\ \-\ \-\ \-\ \-\ \-\ \-\ \-\ kright\-\ =\-\ kv(j+1);\color{Green}
$\\$\color{BrickRed}\-\ \-\ \-\ \-\ \-\ \-\ \-\ \-\ \color{Green}
$\\$\color{BrickRed}\-\ \-\ \-\ \-\ \-\ \-\ \-\ \-\ verify\_stability\_n1\_strict(d,\-\ kleft,\-\ kright,\-\ ints\_x,ints\_y,psiL)\color{Green}
$\\$
$\\$
$\\$\color{BrickRed}\-\ \-\ \-\ \-\ \color{NavyBlue}\-\ end\-\ \color{BrickRed}\color{Green}
$\\$\color{BrickRed}\color{NavyBlue}\-\ end\-\ \color{BrickRed}\color{Green}
$\\$
$\\$
$\\$
$\\$
$\\$
$\\$
$\\$
$\\$
$\\$\color{Black}\section{driver\_strict\_upper\_taylor.m}

\color{Green}\color{BrickRed}clear\-\ all;\-\ curr\_dir\-\ =\-\ local\_startup;\color{Green}
$\\$\color{Green}$\%$$\%$$\%$$\%$$\%$$\%$$\%$$\%$$\%$$\%$$\%$$\%$$\%$$\%$$\%$$\%$$\%$$\%$$\%$$\%$$\%$$\%$$\%$
$\\$
$\\$
$\\$\color{BrickRed}file\_name\-\ =\-\ 'interp2d\_490\_to\_538';\color{Green}
$\\$
$\\$
$\\$\color{Green}$\%$$\%$$\%$$\%$$\%$$\%$$\%$$\%$$\%$$\%$$\%$$\%$$\%$$\%$$\%$$\%$$\%$$\%$$\%$$\%$$\%$$\%$$\%$
$\\$
$\\$
$\\$\color{Green}$\%$-----------------------------------------------------------
$\\$\color{Green}$\%$\-\ \-\ k\-\ and\-\ psi\-\ points/intervals
$\\$\color{Green}$\%$-----------------------------------------------------------
$\\$
$\\$
$\\$\color{BrickRed}num\_k\-\ =\-\ 1001;\color{Green}
$\\$\color{BrickRed}kv\-\ =\-\ linspace(inf(nm('0.9999983')),sup(nm('0.99999839')),num\_k);\color{Green}
$\\$\color{BrickRed}kv\-\ =\-\ nm(kv(1:end-1),kv(2:end));\color{Green}
$\\$
$\\$
$\\$\color{BrickRed}psiv\-\ =\-\ nm('0.7','0.8');\color{Green}
$\\$\color{BrickRed}num\-\ =\-\ 100;\color{Green}
$\\$
$\\$
$\\$\color{BrickRed}psi0\-\ =\-\ linspace(inf(psiv(1)),sup(psiv(end)),num);\color{Green}
$\\$\color{BrickRed}psi0\-\ =\-\ nm(psi0(1:end-1),psi0(2:end));\color{Green}
$\\$
$\\$
$\\$\color{Green}$\%$-----------------------------------------------------------
$\\$\color{Green}$\%$\-\ \-\ retrieve\-\ file
$\\$\color{Green}$\%$-----------------------------------------------------------
$\\$
$\\$
$\\$\color{BrickRed}ld\-\ =\-\ retrieve\_it(curr\_dir,'interval\_arithmetic',file\_name,'data\_final');\color{Green}
$\\$\color{BrickRed}d\-\ =\-\ ld.var;\color{Green}
$\\$\color{BrickRed}pie\-\ =\-\ nm('pi');\color{Green}
$\\$
$\\$
$\\$\color{Green}$\%$-----------------------------------------------------------
$\\$\color{Green}$\%$\-\ \-\ diff\-\ 
$\\$\color{Green}$\%$-----------------------------------------------------------
$\\$
$\\$
$\\$\color{BrickRed}rho\_q\_sm\-\ =\-\ ((d.rho\_q\-\ -\-\ 1/d.rho\_q)+sqrt((d.rho\_q-1/d.rho\_q)\verb|^|2+16))/4;\color{Green}
$\\$\color{BrickRed}rho\_psi\_sm\-\ =\-\ ((d.rho\_psi\-\ -\-\ 1/d.rho\_psi)+sqrt((d.rho\_psi-1/d.rho\_psi)\verb|^|2+16))/4;\color{Green}
$\\$\color{BrickRed}rho\_q\-\ =\-\ d.rho\_q;\color{Green}
$\\$\color{BrickRed}rho\_psi\-\ =\-\ d.rho\_psi;\color{Green}
$\\$
$\\$
$\\$\color{BrickRed}nm\_pts\-\ =\-\ 100;\color{Green}
$\\$\color{BrickRed}min\_diff\_q\-\ =\-\ inf((rho\_q-1/rho\_q)/2);\color{Green}
$\\$\color{BrickRed}\color{NavyBlue}\-\ for\-\ \color{BrickRed}\-\ j\-\ =\-\ 0:nm\_pts-1\color{Green}
$\\$\color{BrickRed}\-\ \-\ \-\ \-\ theta1\-\ =\-\ nm(2*(j*pie)/nm\_pts,2*((j+1)*pie)/nm\_pts);\color{Green}
$\\$\color{BrickRed}\-\ \-\ \-\ \-\ \color{NavyBlue}\-\ for\-\ \color{BrickRed}\-\ q\-\ =\-\ 0:nm\_pts-1\color{Green}
$\\$\color{BrickRed}\-\ \-\ \-\ \-\ \-\ \-\ \-\ \-\ \-\ \-\ \-\ \-\ theta2\-\ =\-\ nm(2*(q*pie)/nm\_pts,2*((q+1)*pie)/nm\_pts);\color{Green}
$\\$\color{BrickRed}\-\ \-\ \-\ \-\ \-\ \-\ \-\ \-\ \-\ \-\ \-\ \-\ del\-\ =\-\ (exp(1i*theta1)*rho\_q+exp(-1i*theta1)/rho\_q\-\ ...\color{Green}
$\\$\color{BrickRed}\-\ \-\ \-\ \-\ \-\ \-\ \-\ \-\ \-\ \-\ \-\ \-\ \-\ \-\ \-\ \-\ -(exp(1i*theta2)*rho\_q\_sm+exp(-1i*theta2)/rho\_q\_sm))/2;\color{Green}
$\\$\color{BrickRed}\-\ \-\ \-\ \-\ \-\ \-\ \-\ \-\ \-\ \-\ \-\ \-\ min\_diff\_q\-\ =\-\ min(min\_diff\_q,inf(abs(del)));\color{Green}
$\\$\color{BrickRed}\-\ \-\ \-\ \-\ \-\ \-\ \-\ \-\ \-\ \-\ \-\ \-\ \color{NavyBlue}\-\ if\-\ \color{BrickRed}\-\ min\_diff\_q\-\ ==\-\ 0\color{Green}
$\\$\color{BrickRed}\-\ \-\ \-\ \-\ \-\ \-\ \-\ \-\ \-\ \-\ \-\ \-\ \-\ \-\ \-\ \-\ error('min\-\ is\-\ 0')\color{Green}
$\\$\color{BrickRed}\-\ \-\ \-\ \-\ \-\ \-\ \-\ \-\ \-\ \-\ \-\ \-\ \color{NavyBlue}\-\ end\-\ \color{BrickRed}\color{Green}
$\\$\color{BrickRed}\-\ \-\ \-\ \-\ \color{NavyBlue}\-\ end\-\ \color{BrickRed}\color{Green}
$\\$\color{BrickRed}\color{NavyBlue}\-\ end\-\ \color{BrickRed}\color{Green}
$\\$\color{BrickRed}min\_diff\_q\-\ =\-\ min\_diff\_q/2;\color{Green}
$\\$
$\\$
$\\$\color{BrickRed}nm\_pts\-\ =\-\ 100;\color{Green}
$\\$\color{BrickRed}min\_diff\_psi\-\ =\-\ inf((rho\_psi-1/rho\_psi)/2);\color{Green}
$\\$\color{BrickRed}\color{NavyBlue}\-\ for\-\ \color{BrickRed}\-\ j\-\ =\-\ 0:nm\_pts-1\color{Green}
$\\$\color{BrickRed}\-\ \-\ \-\ \-\ theta1\-\ =\-\ nm(2*(j*pie)/nm\_pts,2*((j+1)*pie)/nm\_pts);\color{Green}
$\\$\color{BrickRed}\-\ \-\ \-\ \-\ \color{NavyBlue}\-\ for\-\ \color{BrickRed}\-\ q\-\ =\-\ 0:nm\_pts-1\color{Green}
$\\$\color{BrickRed}\-\ \-\ \-\ \-\ \-\ \-\ \-\ \-\ \-\ \-\ \-\ \-\ theta2\-\ =\-\ nm(2*(q*pie)/nm\_pts,2*((q+1)*pie)/nm\_pts);\color{Green}
$\\$\color{BrickRed}\-\ \-\ \-\ \-\ \-\ \-\ \-\ \-\ \-\ \-\ \-\ \-\ del\-\ =\-\ (exp(1i*theta1)*rho\_psi+exp(-1i*theta1)/rho\_psi\-\ ...\color{Green}
$\\$\color{BrickRed}\-\ \-\ \-\ \-\ \-\ \-\ \-\ \-\ \-\ \-\ \-\ \-\ \-\ \-\ \-\ \-\ -(exp(1i*theta2)*rho\_psi\_sm+exp(-1i*theta2)/rho\_psi\_sm))/2;\color{Green}
$\\$\color{BrickRed}\-\ \-\ \-\ \-\ \-\ \-\ \-\ \-\ \-\ \-\ \-\ \-\ min\_diff\_psi\-\ =\-\ min(min\_diff\_psi,inf(abs(del)));\color{Green}
$\\$\color{BrickRed}\-\ \-\ \-\ \-\ \-\ \-\ \-\ \-\ \-\ \-\ \-\ \-\ \color{NavyBlue}\-\ if\-\ \color{BrickRed}\-\ min\_diff\_psi\-\ ==\-\ 0\color{Green}
$\\$\color{BrickRed}\-\ \-\ \-\ \-\ \-\ \-\ \-\ \-\ \-\ \-\ \-\ \-\ \-\ \-\ \-\ \-\ error('min\-\ is\-\ 0')\color{Green}
$\\$\color{BrickRed}\-\ \-\ \-\ \-\ \-\ \-\ \-\ \-\ \-\ \-\ \-\ \-\ \color{NavyBlue}\-\ end\-\ \color{BrickRed}\color{Green}
$\\$\color{BrickRed}\-\ \-\ \-\ \-\ \color{NavyBlue}\-\ end\-\ \color{BrickRed}\color{Green}
$\\$\color{BrickRed}\color{NavyBlue}\-\ end\-\ \color{BrickRed}\color{Green}
$\\$\color{BrickRed}min\_diff\_psi\-\ =\-\ min\_diff\_psi/2;\color{Green}
$\\$
$\\$
$\\$\color{BrickRed}\color{NavyBlue}\-\ for\-\ \color{BrickRed}\-\ j\-\ =\-\ 1:length(kv)\color{Green}
$\\$\color{BrickRed}\-\ \-\ \-\ \-\ \color{Green}
$\\$\color{BrickRed}\-\ \-\ \-\ \-\ fprintf('\textbackslash npercent\-\ done:\-\ $\%$4.4g',100*j/num\_k);\color{Green}
$\\$\color{BrickRed}\-\ \-\ \-\ \-\ k\-\ =\-\ kv(j);\color{Green}
$\\$
$\\$
$\\$\color{BrickRed}\-\ \-\ \-\ \-\ [mx\_fk,mn\_fpsipsi]\-\ =\-\ strict\_transition(k,psi0,min\_diff\_q,min\_diff\_psi,d);\color{Green}
$\\$
$\\$
$\\$\color{BrickRed}\-\ \-\ \-\ \-\ \color{NavyBlue}\-\ if\-\ \color{BrickRed}\-\ mx\_fk\-\ $>$=\-\ 0\color{Green}
$\\$\color{BrickRed}\-\ \-\ \-\ \-\ \-\ \-\ \-\ \-\ error('failed\-\ to\-\ verify\-\ strict\-\ transition\-\ on\-\ the\-\ region')\color{Green}
$\\$\color{BrickRed}\-\ \-\ \-\ \-\ \color{NavyBlue}\-\ end\-\ \color{BrickRed}\color{Green}
$\\$\color{BrickRed}\-\ \-\ \-\ \-\ \color{NavyBlue}\-\ if\-\ \color{BrickRed}\-\ mn\_fpsipsi\-\ $<$=\-\ 0\color{Green}
$\\$\color{BrickRed}\-\ \-\ \-\ \-\ \-\ \-\ \-\ \-\ error('failed\-\ to\-\ verify\-\ strict\-\ transition\-\ on\-\ the\-\ region')\color{Green}
$\\$\color{BrickRed}\-\ \-\ \-\ \-\ \color{NavyBlue}\-\ end\-\ \color{BrickRed}\color{Green}
$\\$\color{BrickRed}\-\ \-\ \-\ \-\ \color{Green}
$\\$\color{BrickRed}\color{NavyBlue}\-\ end\-\ \color{BrickRed}\color{Green}
$\\$
$\\$
$\\$
$\\$
$\\$
$\\$
$\\$
$\\$
$\\$
$\\$
$\\$
$\\$
$\\$
$\\$
$\\$
$\\$
$\\$
$\\$
$\\$
$\\$
$\\$\color{Black}\section{elliptic\_integral.m}

\color{Green}\color{BrickRed}\color{NavyBlue}\-\ function\-\ \color{BrickRed}\-\ out\-\ =\-\ elliptic\_integral(k,id)\color{Green}
$\\$\color{Green}$\%$\-\ function\-\ out\-\ =\-\ elliptic\_integral(k,id)
$\\$\color{Green}$\%$
$\\$\color{Green}$\%$\-\ Compute\-\ elliptic\-\ integrals\-\ using\-\ interval\-\ arithmetic.
$\\$\color{Green}$\%$
$\\$\color{Green}$\%$\-\ input\-\ k\-\ should\-\ be\-\ an\-\ intlab\-\ interval.\-\ The\-\ input\-\ id
$\\$\color{Green}$\%$\-\ denotes\-\ the\-\ type\-\ of\-\ elliptic\-\ interval.
$\\$\color{Green}$\%$
$\\$\color{Green}$\%$\-\ id\-\ =\-\ 1,\-\ complete\-\ elliptic\-\ integral\-\ of\-\ the\-\ first\-\ kind
$\\$\color{Green}$\%$\-\ id\-\ =\-\ 2,\-\ complete\-\ elliptic\-\ integral\-\ of\-\ the\-\ second\-\ kind
$\\$\color{Green}$\%$
$\\$\color{Green}$\%$\-\ The\-\ complete\-\ elliptic\-\ interval\-\ of\-\ the\-\ first\-\ kind\-\ is\-\ solved\-\ using
$\\$\color{Green}$\%$\-\ the\-\ arithmetic\-\ geometric\-\ mean.\-\ The\-\ complete\-\ elliptic\-\ integral
$\\$\color{Green}$\%$\-\ of\-\ the\-\ second\-\ kind\-\ is\-\ solved\-\ using\-\ Carlson's\-\ algorithm.
$\\$
$\\$
$\\$\color{Green}$\%$\-\ complete\-\ elliptic\-\ integral\-\ of\-\ the\-\ first\-\ kind
$\\$\color{Green}$\%$
$\\$\color{Green}$\%$\-\ $\textbackslash int\_0\verb|^|1\-\ \textbackslash frac{1}{\textbackslash sqrt{(1-x\verb|^|2)(1-k\verb|^|2x\verb|^|2)}}dx$
$\\$\color{BrickRed}\-\ \-\ \-\ \color{Green}
$\\$\color{BrickRed}M\-\ =\-\ agm(ones(size(k)),sqrt(1-k.\verb|^|2));\color{Green}
$\\$\color{BrickRed}out\-\ =\-\ nm('pi')./(2*M);\color{Green}
$\\$
$\\$
$\\$\color{BrickRed}\color{NavyBlue}\-\ if\-\ \color{BrickRed}\-\ id\-\ ==\-\ 1\color{Green}
$\\$\color{BrickRed}\-\ \-\ \-\ \-\ return;\color{Green}
$\\$\color{BrickRed}\color{NavyBlue}\-\ end\-\ \color{BrickRed}\color{Green}
$\\$
$\\$
$\\$\color{Green}$\%$
$\\$\color{Green}$\%$\-\ complete\-\ elliptic\-\ integral\-\ of\-\ the\-\ second\-\ kind
$\\$\color{Green}$\%$\-\ 
$\\$\color{Green}$\%$\-\ \-\ $\textbackslash int\_0\verb|^|{\textbackslash pi/2}\-\ \textbackslash sqrt{1-k\verb|^|2\-\ sin\verb|^|2(\textbackslash theta)}d\textbackslash theta$
$\\$
$\\$
$\\$\color{Green}$\%$\-\ compute\-\ m\-\ =\-\ 0\-\ case
$\\$\color{BrickRed}x\-\ =\-\ nm(zeros(size(k)));\color{Green}
$\\$\color{BrickRed}y\-\ =\-\ 1-k.\verb|^|2;\color{Green}
$\\$\color{BrickRed}z\-\ =\-\ nm(ones(size(k)));\color{Green}
$\\$\color{BrickRed}lambda\-\ =\-\ sqrt(x.*y)+sqrt(x.*z)+sqrt(y.*z);\color{Green}
$\\$\color{Green}$\%$\-\ mu\-\ =\-\ (x+y+3*z)/5;
$\\$\color{BrickRed}sum\-\ =\-\ 1./(sqrt(z).*(z+lambda));\color{Green}
$\\$\color{Green}$\%$\-\ X\-\ =\-\ 1-x./mu;
$\\$\color{Green}$\%$\-\ Y\-\ =\-\ 1-y./mu;
$\\$\color{Green}$\%$\-\ Z\-\ =\-\ 1-z./mu;
$\\$
$\\$
$\\$\color{Green}$\%$\-\ t1\-\ =\-\ hull(nm(zeros(size(k))),abs(X));
$\\$\color{Green}$\%$\-\ t2\-\ =\-\ hull(abs(Y),abs(Z));
$\\$\color{Green}$\%$\-\ eps\-\ =\-\ hull(t1,t2);
$\\$
$\\$
$\\$\color{Green}$\%$\-\ r\-\ =\-\ abs(3*eps.\verb|^|6./((1-eps).\verb|^|(3/nm(2))));
$\\$\color{Green}$\%$\-\ 
$\\$\color{Green}$\%$\-\ S2\-\ =\-\ (X.\verb|^|2+Y.\verb|^|2+3*Z.\verb|^|2)/4;
$\\$\color{Green}$\%$\-\ S3\-\ =\-\ (X.\verb|^|3+Y.\verb|^|3+3*Z.\verb|^|3)/6;
$\\$\color{Green}$\%$\-\ S4\-\ =\-\ (X.\verb|^|4+Y.\verb|^|4+3*Z.\verb|^|4)/8;
$\\$\color{Green}$\%$\-\ S5\-\ =\-\ (X.\verb|^|5+Y.\verb|^|5+3*Z.\verb|^|5)/10;
$\\$
$\\$
$\\$\color{Green}$\%$\-\ compute\-\ m\-\ =\-\ 1\-\ case;
$\\$\color{BrickRed}m\-\ =\-\ 1;\color{Green}
$\\$
$\\$
$\\$\color{BrickRed}x\-\ =\-\ (x+lambda)/4;\color{Green}
$\\$\color{BrickRed}y\-\ =\-\ (y+lambda)/4;\color{Green}
$\\$\color{BrickRed}z\-\ =\-\ (z+lambda)/4;\color{Green}
$\\$\color{BrickRed}lambda\-\ =\-\ sqrt(x.*y)+sqrt(x.*z)+sqrt(y.*z);\color{Green}
$\\$\color{BrickRed}mu\-\ =\-\ (x+y+3*z)/5;\color{Green}
$\\$\color{BrickRed}X\-\ =\-\ 1-x./mu;\color{Green}
$\\$\color{BrickRed}Y\-\ =\-\ 1-y./mu;\color{Green}
$\\$\color{BrickRed}Z\-\ =\-\ 1-z./mu;\color{Green}
$\\$
$\\$
$\\$\color{BrickRed}t1\-\ =\-\ hull(nm(zeros(size(k))),abs(X));\color{Green}
$\\$\color{BrickRed}t2\-\ =\-\ hull(abs(Y),abs(Z));\color{Green}
$\\$\color{BrickRed}eps\-\ =\-\ hull(t1,t2);\color{Green}
$\\$\color{BrickRed}r\-\ =\-\ abs(3*eps.\verb|^|6./((1-eps).\verb|^|(3/nm(2))));\color{Green}
$\\$
$\\$
$\\$\color{BrickRed}S2\-\ =\-\ (X.\verb|^|2+Y.\verb|^|2+3*Z.\verb|^|2)/4;\color{Green}
$\\$\color{BrickRed}S3\-\ =\-\ (X.\verb|^|3+Y.\verb|^|3+3*Z.\verb|^|3)/6;\color{Green}
$\\$\color{BrickRed}S4\-\ =\-\ (X.\verb|^|4+Y.\verb|^|4+3*Z.\verb|^|4)/8;\color{Green}
$\\$\color{BrickRed}S5\-\ =\-\ (X.\verb|^|5+Y.\verb|^|5+3*Z.\verb|^|5)/10;\color{Green}
$\\$\color{BrickRed}\-\ \-\ \-\ \-\ \-\ \-\ \-\ \-\ \-\ \color{Green}
$\\$\color{BrickRed}factor\-\ =\-\ 4\verb|^|nm(-m)*mu.\verb|^|(-nm(3)/2).*(1+...\color{Green}
$\\$\color{BrickRed}\-\ \-\ \-\ \-\ (nm(3)/7)*S2+(nm(1)/3)*S3+...\color{Green}
$\\$\color{BrickRed}\-\ \-\ \-\ \-\ \-\ \-\ \-\ \-\ \-\ (nm(3)/22)*S2.\verb|^|2\-\ +\-\ (nm(3)/11)*S4+...\color{Green}
$\\$\color{BrickRed}\-\ \-\ \-\ \-\ \-\ \-\ \-\ \-\ \-\ (nm(3)/13)*S2.*S3+(nm(3)/13)*S5\-\ +\-\ r);\-\ \-\ \-\ \color{Green}
$\\$
$\\$
$\\$\color{BrickRed}Rd\_new\-\ =\-\ 3*sum\-\ +\-\ factor;\color{Green}
$\\$
$\\$
$\\$\color{BrickRed}sum\-\ =\-\ sum\-\ +\-\ 4\verb|^|(-nm(m))./(sqrt(z).*(z+lambda));\color{Green}
$\\$
$\\$
$\\$\color{BrickRed}Rd\_old\-\ =\-\ infsup(-1+inf(Rd\_new),1+sup(Rd\_new));\color{Green}
$\\$
$\\$
$\\$\color{BrickRed}m\-\ =\-\ 2;\color{Green}
$\\$\color{BrickRed}\color{NavyBlue}\-\ while\-\ \color{BrickRed}\-\ (max(sup(Rd\_old))-min(inf(Rd\_old)))-\-\ (max(sup(Rd\_new))-min(inf(Rd\_new)))\-\ $>$\-\ 0\color{Green}
$\\$\color{BrickRed}\-\ \-\ \-\ \-\ \color{Green}
$\\$\color{BrickRed}\-\ \-\ \-\ \-\ \color{Green}
$\\$\color{BrickRed}\-\ \-\ \-\ \-\ Rd\_old\-\ =\-\ Rd\_new;\color{Green}
$\\$\color{BrickRed}\-\ \-\ \-\ \-\ \color{Green}
$\\$\color{BrickRed}\-\ \-\ \-\ \-\ x\-\ =\-\ (x+lambda)/4;\color{Green}
$\\$\color{BrickRed}\-\ \-\ \-\ \-\ y\-\ =\-\ (y+lambda)/4;\color{Green}
$\\$\color{BrickRed}\-\ \-\ \-\ \-\ z\-\ =\-\ (z+lambda)/4;\color{Green}
$\\$\color{BrickRed}\-\ \-\ \-\ \-\ lambda\-\ =\-\ sqrt(x.*y)+sqrt(x.*z)+sqrt(y.*z);\color{Green}
$\\$\color{BrickRed}\-\ \-\ \-\ \-\ mu\-\ =\-\ (x+y+3*z)./5;\color{Green}
$\\$\color{BrickRed}\-\ \-\ \-\ \-\ X\-\ =\-\ 1-x./mu;\color{Green}
$\\$\color{BrickRed}\-\ \-\ \-\ \-\ Y\-\ =\-\ 1-y./mu;\color{Green}
$\\$\color{BrickRed}\-\ \-\ \-\ \-\ Z\-\ =\-\ 1-z./mu;\color{Green}
$\\$
$\\$
$\\$\color{BrickRed}\-\ \-\ \-\ \-\ t1\-\ =\-\ hull(nm(zeros(size(k))),abs(X));\color{Green}
$\\$\color{BrickRed}\-\ \-\ \-\ \-\ t2\-\ =\-\ hull(abs(Y),abs(Z));\color{Green}
$\\$\color{BrickRed}\-\ \-\ \-\ \-\ eps\-\ =\-\ hull(t1,t2);\color{Green}
$\\$\color{BrickRed}\-\ \-\ \-\ \-\ r\-\ =\-\ abs(3*eps.\verb|^|6./(1-eps).\verb|^|(nm(3)/2));\color{Green}
$\\$
$\\$
$\\$\color{BrickRed}\-\ \-\ \-\ \-\ S2\-\ =\-\ (X.\verb|^|2+Y.\verb|^|2+3*Z.\verb|^|2)/4;\color{Green}
$\\$\color{BrickRed}\-\ \-\ \-\ \-\ S3\-\ =\-\ (X.\verb|^|3+Y.\verb|^|3+3*Z.\verb|^|3)/6;\color{Green}
$\\$\color{BrickRed}\-\ \-\ \-\ \-\ S4\-\ =\-\ (X.\verb|^|4+Y.\verb|^|4+3*Z.\verb|^|4)/8;\color{Green}
$\\$\color{BrickRed}\-\ \-\ \-\ \-\ S5\-\ =\-\ (X.\verb|^|5+Y.\verb|^|5+3*Z.\verb|^|5)/10;\color{Green}
$\\$
$\\$
$\\$\color{BrickRed}\-\ \-\ \-\ \-\ factor\-\ =\-\ 4\verb|^|nm(-m)*mu.\verb|^|(-nm(3)/2).*(1+...\color{Green}
$\\$\color{BrickRed}\-\ \-\ \-\ \-\ \-\ \-\ \-\ \-\ (nm(3)/7)*S2+(nm(1)/3)*S3+...\color{Green}
$\\$\color{BrickRed}\-\ \-\ \-\ \-\ \-\ \-\ \-\ \-\ \-\ \-\ \-\ \-\ \-\ (nm(3)/22)*S2.\verb|^|2\-\ +\-\ (nm(3)/11)*S4+...\color{Green}
$\\$\color{BrickRed}\-\ \-\ \-\ \-\ \-\ \-\ \-\ \-\ \-\ \-\ \-\ \-\ \-\ (nm(3)/13)*S2.*S3+(nm(3)/13)*S5\-\ +\-\ r);\-\ \-\ \-\ \-\ \-\ \color{Green}
$\\$
$\\$
$\\$\color{BrickRed}\-\ \-\ \-\ \-\ Rd\_new\-\ =\-\ 3*sum\-\ +\-\ factor;\color{Green}
$\\$
$\\$
$\\$\color{BrickRed}\-\ \-\ \-\ \-\ sum\-\ =\-\ sum\-\ +\-\ 4\verb|^|(-nm(m))./(sqrt(z).*(z+lambda));\color{Green}
$\\$
$\\$
$\\$\color{BrickRed}\-\ \-\ \-\ \-\ m\-\ =\-\ m\-\ +\-\ 1;\color{Green}
$\\$\color{BrickRed}\-\ \-\ \-\ \-\ \color{Green}
$\\$\color{BrickRed}\color{NavyBlue}\-\ end\-\ \color{BrickRed}\color{Green}
$\\$
$\\$
$\\$\color{BrickRed}\color{NavyBlue}\-\ if\-\ \color{BrickRed}\-\ id\-\ ==\-\ 2\color{Green}
$\\$\color{BrickRed}\-\ \-\ \-\ \-\ \color{Green}$\%$\-\ complete\-\ elliptic\-\ integral\-\ of\-\ the\-\ second\-\ kind.
$\\$\color{BrickRed}\-\ \-\ \-\ \-\ out\-\ =\-\ out\-\ -(k.\verb|^|2/3).*Rd\_new;\color{Green}
$\\$\color{BrickRed}\color{NavyBlue}\-\ elseif\-\ \color{BrickRed}\-\ id\-\ ==\-\ 3\color{Green}
$\\$\color{BrickRed}\-\ \-\ \-\ \-\ \color{Green}$\%$\-\ The\-\ difference\-\ of\-\ elliptic\-\ integrals,\-\ K-E
$\\$\color{BrickRed}\-\ \-\ \-\ \-\ out\-\ \-\ =\-\ (k.\verb|^|2/3).*Rd\_new;\color{Green}
$\\$\color{BrickRed}\color{NavyBlue}\-\ end\-\ \color{BrickRed}\color{Green}
$\\$
$\\$
$\\$\color{Black}\section{instability\_interpolation.m}

\color{Green}\color{BrickRed}\color{NavyBlue}\-\ function\-\ \color{BrickRed}\-\ d\-\ =\-\ instability\_interpolation(d,q\_min,q\_max)\color{Green}
$\\$
$\\$
$\\$\color{BrickRed}pie\-\ =\-\ nm('pi');\color{Green}
$\\$\color{Green}$\%$$\%$$\%$$\%$$\%$$\%$$\%$$\%$$\%$$\%$$\%$$\%$$\%$$\%$$\%$$\%$
$\\$\color{BrickRed}d.a\_q\-\ =\-\ q\_min;\color{Green}
$\\$\color{BrickRed}d.b\_q\-\ =\-\ q\_max;\color{Green}
$\\$
$\\$
$\\$
$\\$\-\ \color{Black}
Note that $\rho_{\psi}$ must be chosen so that $\xi(\omega + i\psi \omega')$ does not have a pole.
The poles of $\xi$ are $z = 2m \omega + 2n \omega'$. Setting
$2m\omega + 2n\omega' = \omega + i\psi \omega'$ with $\psi = 1/2 + \tilde \psi/2$, we
find that $|\Im(\tilde \psi)|< -\frac{\pi}{\log(q)}$ is necessary and sufficient to ensure analyticity.
\color{Green}
$\\$
$\\$\color{Green}$\%$$\%$
$\\$
$\\$
$\\$\color{Green}$\%$\-\ find\-\ rho\_psi
$\\$\color{BrickRed}c\_psi\-\ =\-\ nm('0.9')*pie/abs(log(q\_min));\color{Green}
$\\$\color{BrickRed}rho\_psi\-\ =\-\ nm(inf(c\_psi\-\ +sqrt(c\_psi\verb|^|2+1)));\color{Green}
$\\$\color{BrickRed}d.rho\_psi\-\ =\-\ rho\_psi;\color{Green}
$\\$
$\\$
$\\$\color{Green}$\%$\-\ make\-\ sure\-\ rho\_psi\-\ is\-\ small\-\ enough\-\ that\-\ bounds
$\\$\color{Green}$\%$\-\ on\-\ xi(omega\-\ +\-\ 1i*psi*omega')\-\ are\-\ valid
$\\$\color{BrickRed}\color{NavyBlue}\-\ if\-\ \color{BrickRed}\-\ sup(rho\_psi)\-\ $>$=\-\ 3+sqrt(nm(2))\color{Green}
$\\$\color{BrickRed}\-\ \-\ \-\ \-\ rho\_psi\-\ =\-\ 3+sqrt(nm(2));\color{Green}
$\\$\color{BrickRed}\color{NavyBlue}\-\ end\-\ \color{BrickRed}\color{Green}
$\\$
$\\$
$\\$\color{BrickRed}\color{NavyBlue}\-\ if\-\ \color{BrickRed}\-\ sup(rho\_psi)\-\ $<$=\-\ 1\color{Green}
$\\$\color{BrickRed}\-\ \-\ \-\ \-\ error('problem');\color{Green}
$\\$\color{BrickRed}\color{NavyBlue}\-\ end\-\ \color{BrickRed}\color{Green}
$\\$\color{BrickRed}\-\ \color{Green}
$\\$\color{Green}$\%$
$\\$\color{Green}$\%$\-\ find\-\ rho\_q
$\\$\color{Green}$\%$
$\\$
$\\$
$\\$\color{BrickRed}a\-\ =\-\ q\_min;\color{Green}
$\\$\color{BrickRed}b\-\ =\-\ q\_max;\color{Green}
$\\$\color{BrickRed}d.q\_min\-\ =\-\ q\_min;\color{Green}
$\\$\color{BrickRed}d.q\_max\-\ =\-\ q\_max;\color{Green}
$\\$
$\\$
$\\$\color{BrickRed}rho1\-\ =\-\ (b+a)/(b-a)-2*sqrt(a*b)/(b-a);\color{Green}
$\\$\color{BrickRed}rho2\-\ =\-\ (b+a)/(b-a)+2*sqrt(a*b)/(b-a);\color{Green}
$\\$\color{BrickRed}rho3\-\ =\-\ (2-a-b)/(b-a)-2*sqrt(1-a-b+a*b)/(b-a);\color{Green}
$\\$\color{BrickRed}rho4\-\ =\-\ (2-a-b)/(b-a)+2*sqrt(1-a-b+a*b)/(b-a);\color{Green}
$\\$
$\\$
$\\$\color{BrickRed}rho\_left\-\ =\-\ nm(rho1,rho3);\color{Green}
$\\$\color{BrickRed}rho\_right\-\ =\-\ nm(rho2,rho4);\color{Green}
$\\$\color{BrickRed}\color{NavyBlue}\-\ if\-\ \color{BrickRed}\-\ sup(rho\_left)\-\ $>$=\-\ inf(rho\_right)\color{Green}
$\\$\color{BrickRed}\-\ \-\ \-\ \-\ error('problem');\color{Green}
$\\$\color{BrickRed}\color{NavyBlue}\-\ end\-\ \color{BrickRed}\color{Green}
$\\$
$\\$
$\\$\color{BrickRed}rho\_left\-\ =\-\ nm(sup(rho\_left));\color{Green}
$\\$\color{BrickRed}rho\_right\-\ =\-\ nm(inf(rho\_right));\color{Green}
$\\$\color{BrickRed}rho\_q\-\ =\-\ rho\_left\-\ +\-\ nm('0.9')*(rho\_right-rho\_left);\color{Green}
$\\$\color{BrickRed}d.rho\_q\-\ =\-\ rho\_q;\color{Green}
$\\$
$\\$
$\\$\color{Green}$\%$
$\\$\color{Green}$\%$\-\ get\-\ bounds\-\ for\-\ alpha\-\ =\-\ omega\-\ +\-\ 1i*psi*omega'
$\\$\color{Green}$\%$
$\\$
$\\$
$\\$\color{BrickRed}ntilde\-\ =\-\ 1;\color{Green}
$\\$
$\\$
$\\$\color{Green}$\%$\-\ \-\ bound\_numer(dm,rho\_psi,rho\_x,rho\_q,q\_min,q\_max,nm(0),nm(1),ntilde);
$\\$\color{BrickRed}[M\_psi,M\_x,M\_q]\-\ =\-\ bound\_numer\_unstable(d.dm,d.rho\_x,d.rho\_q,q\_min,q\_max);\color{Green}
$\\$
$\\$
$\\$\color{BrickRed}\color{NavyBlue}\-\ if\-\ \color{BrickRed}\-\ isnan(sup(M\_psi))\color{Green}
$\\$\color{BrickRed}\-\ \-\ \-\ \-\ error('Nan\-\ \color{NavyBlue}\-\ for\-\ \color{BrickRed}\-\ bound')\color{Green}
$\\$\color{BrickRed}\color{NavyBlue}\-\ end\-\ \color{BrickRed}\color{Green}
$\\$\color{BrickRed}\color{NavyBlue}\-\ if\-\ \color{BrickRed}\-\ isnan(sup(M\_x))\color{Green}
$\\$\color{BrickRed}\-\ \-\ \-\ \-\ error('Nan\-\ \color{NavyBlue}\-\ for\-\ \color{BrickRed}\-\ bound')\color{Green}
$\\$\color{BrickRed}\color{NavyBlue}\-\ end\-\ \color{BrickRed}\color{Green}
$\\$\color{BrickRed}\color{NavyBlue}\-\ if\-\ \color{BrickRed}\-\ isnan(sup(M\_q))\color{Green}
$\\$\color{BrickRed}\-\ \-\ \-\ \-\ error('Nan\-\ \color{NavyBlue}\-\ for\-\ \color{BrickRed}\-\ bound')\color{Green}
$\\$\color{BrickRed}\color{NavyBlue}\-\ end\-\ \color{BrickRed}\color{Green}
$\\$\color{BrickRed}\color{NavyBlue}\-\ if\-\ \color{BrickRed}\-\ isinf(sup(M\_psi))\color{Green}
$\\$\color{BrickRed}\-\ \-\ \-\ \-\ error('Infinity\-\ \color{NavyBlue}\-\ for\-\ \color{BrickRed}\-\ bound')\color{Green}
$\\$\color{BrickRed}\color{NavyBlue}\-\ end\-\ \color{BrickRed}\color{Green}
$\\$\color{BrickRed}\color{NavyBlue}\-\ if\-\ \color{BrickRed}\-\ isinf(sup(M\_x))\color{Green}
$\\$\color{BrickRed}\-\ \-\ \-\ \-\ error('Infinity\-\ \color{NavyBlue}\-\ for\-\ \color{BrickRed}\-\ bound')\color{Green}
$\\$\color{BrickRed}\color{NavyBlue}\-\ end\-\ \color{BrickRed}\color{Green}
$\\$\color{BrickRed}\color{NavyBlue}\-\ if\-\ \color{BrickRed}\-\ isinf(sup(M\_q))\color{Green}
$\\$\color{BrickRed}\-\ \-\ \-\ \-\ error('Infinity\-\ \color{NavyBlue}\-\ for\-\ \color{BrickRed}\-\ bound')\color{Green}
$\\$\color{BrickRed}\color{NavyBlue}\-\ end\-\ \color{BrickRed}\color{Green}
$\\$
$\\$
$\\$\color{BrickRed}abs\_tol\-\ =\-\ 1e-17;\color{Green}
$\\$\color{BrickRed}[N\_x,err\_x]\-\ =\-\ N\_nodes(d.rho\_x,M\_x,abs\_tol);\color{Green}
$\\$\color{BrickRed}d.N\_x\-\ =\-\ N\_x;\color{Green}
$\\$\color{BrickRed}d.err\_x\-\ =\-\ err\_x;\color{Green}
$\\$
$\\$
$\\$\color{BrickRed}[N\_q,err\_q]\-\ =\-\ N\_nodes(rho\_q,M\_q,abs\_tol);\color{Green}
$\\$
$\\$
$\\$\color{Green}$\%$------------------------------------------------------------
$\\$\color{Green}$\%$\-\ get\-\ interpolation\-\ coefficients\-\ when\-\ ntilde\-\ =\-\ 1
$\\$\color{Green}$\%$------------------------------------------------------------
$\\$
$\\$
$\\$\color{BrickRed}a\_q\-\ =\-\ a;\color{Green}
$\\$\color{BrickRed}b\_q\-\ =\-\ b;\color{Green}
$\\$
$\\$
$\\$\color{BrickRed}fun\-\ =\-\ \@(q,psi)(integrand\_numer(N\_x,err\_x,q*(b\_q-a\_q)/2+(a\_q+b\_q)/2,1,ntilde));\color{Green}
$\\$
$\\$
$\\$\color{Green}$\%$\-\ interpolation\-\ coefficients
$\\$\color{BrickRed}t1\-\ =\-\ tic;\color{Green}
$\\$
$\\$
$\\$\color{BrickRed}N\_psi\-\ =\-\ 0;\-\ \color{Green}$\%$\-\ KEEP!\-\ this\-\ is\-\ sufficient\-\ for\-\ instability\-\ computation
$\\$
$\\$
$\\$\color{BrickRed}cfn1\-\ =\-\ cf\_biv\_cheby(N\_q,N\_psi,6,fun);\color{Green}
$\\$\color{BrickRed}d.cf1\_time\-\ =\-\ toc(t1);\color{Green}
$\\$\color{BrickRed}d.cfn1\-\ =\-\ cfn1;\color{Green}
$\\$
$\\$
$\\$\color{BrickRed}d.M\_psi\_n1\-\ \-\ =\-\ M\_psi;\-\ \-\ \-\ \-\ \-\ \-\ \-\ \-\ \-\ \-\ \-\ \-\ \-\ \-\ \color{Green}
$\\$\color{BrickRed}d.M\_q\_n1\-\ \-\ \-\ =\-\ M\_q;\-\ \-\ \-\ \-\ \-\ \-\ \-\ \-\ \-\ \-\ \-\ \-\ \-\ \-\ \-\ \-\ \-\ \-\ \-\ \-\ \-\ \-\ \-\ \-\ \-\ \-\ \-\ \color{Green}
$\\$\color{BrickRed}d.M\_x\_n1\-\ \-\ \-\ =\-\ M\_x;\-\ \-\ \-\ \-\ \-\ \-\ \-\ \-\ \-\ \-\ \-\ \-\ \-\ \-\ \-\ \-\ \-\ \-\ \-\ \-\ \-\ \color{Green}
$\\$\color{BrickRed}d.N\_psi\_n1\-\ =\-\ N\_psi;\-\ \-\ \-\ \-\ \-\ \-\ \-\ \-\ \-\ \-\ \-\ \-\ \-\ \-\ \-\ \-\ \-\ \-\ \-\ \-\ \-\ \color{Green}
$\\$\color{BrickRed}d.N\_q\_n1\-\ =\-\ N\_q;\-\ \-\ \-\ \-\ \-\ \-\ \-\ \-\ \-\ \-\ \-\ \-\ \-\ \-\ \-\ \-\ \-\ \-\ \-\ \-\ \-\ \-\ \-\ \-\ \-\ \-\ \-\ \-\ \-\ \-\ \-\ \-\ \color{Green}
$\\$\color{BrickRed}d.N\_x\_n1\-\ =\-\ N\_x;\-\ \-\ \-\ \-\ \-\ \-\ \-\ \-\ \-\ \-\ \-\ \-\ \color{Green}
$\\$\color{BrickRed}d.err\_q\_n1\-\ \-\ =\-\ err\_q;\-\ \-\ \-\ \-\ \-\ \-\ \-\ \-\ \-\ \-\ \-\ \-\ \-\ \-\ \-\ \-\ \-\ \-\ \-\ \-\ \-\ \-\ \-\ \color{Green}
$\\$\color{BrickRed}d.err\_x\_n1\-\ =\-\ err\_x;\-\ \-\ \color{Green}
$\\$\color{BrickRed}d.a\_q\-\ \-\ =\-\ a\_q;\-\ \-\ \-\ \-\ \-\ \-\ \-\ \-\ \-\ \-\ \-\ \-\ \-\ \-\ \-\ \-\ \-\ \-\ \-\ \-\ \-\ \-\ \-\ \-\ \-\ \-\ \-\ \-\ \-\ \-\ \-\ \-\ \-\ \-\ \-\ \-\ \-\ \-\ \-\ \-\ \-\ \-\ \-\ \-\ \-\ \-\ \-\ \-\ \-\ \-\ \-\ \-\ \-\ \-\ \-\ \-\ \-\ \-\ \-\ \-\ \-\ \-\ \-\ \-\ \-\ \color{Green}
$\\$\color{BrickRed}d.b\_q\-\ =\-\ b\_q;\-\ \-\ \-\ \-\ \-\ \-\ \-\ \-\ \-\ \-\ \-\ \-\ \-\ \-\ \-\ \-\ \-\ \-\ \-\ \-\ \-\ \-\ \-\ \color{Green}
$\\$\color{BrickRed}d.cfn1\-\ =\-\ cfn1;\-\ \-\ \-\ \-\ \-\ \-\ \-\ \-\ \-\ \-\ \-\ \-\ \-\ \-\ \-\ \-\ \-\ \-\ \-\ \-\ \-\ \-\ \-\ \-\ \-\ \-\ \-\ \-\ \-\ \-\ \-\ \-\ \-\ \-\ \-\ \-\ \-\ \-\ \-\ \-\ \-\ \-\ \-\ \-\ \-\ \-\ \-\ \-\ \-\ \-\ \-\ \-\ \-\ \-\ \-\ \-\ \-\ \-\ \-\ \-\ \-\ \-\ \-\ \-\ \-\ \-\ \-\ \-\ \-\ \-\ \-\ \-\ \-\ \-\ \-\ \-\ \-\ \-\ \-\ \-\ \-\ \-\ \-\ \-\ \-\ \-\ \-\ \-\ \-\ \-\ \color{Green}
$\\$\color{BrickRed}d.rho\_q\_n1\-\ =\-\ rho\_q;\-\ \-\ \-\ \-\ \-\ \-\ \-\ \-\ \-\ \-\ \-\ \-\ \-\ \-\ \-\ \-\ \-\ \-\ \-\ \-\ \-\ \-\ \-\ \-\ \-\ \-\ \-\ \-\ \-\ \-\ \-\ \-\ \-\ \color{Green}
$\\$\color{BrickRed}\-\ \-\ \color{Green}
$\\$
$\\$
$\\$\color{Black}\section{integrand\_numer.m}

\color{Green}\color{BrickRed}\color{NavyBlue}\-\ function\-\ \color{BrickRed}\-\ out\-\ =\-\ integrand\_numer(Nx,err,q,psi,ntilde)\color{Green}
$\\$
$\\$
$\\$\color{BrickRed}xi\-\ =\-\ xi\_q\_psi(q,psi,ntilde);\color{Green}
$\\$\color{BrickRed}xi\_der\-\ =\-\ xi\_der\_q\_psi(q,psi,ntilde);\color{Green}
$\\$
$\\$
$\\$\color{BrickRed}pie\-\ =\-\ nm('pi');\color{Green}
$\\$\color{BrickRed}half\-\ =\-\ nm(1)/2;\color{Green}
$\\$\color{BrickRed}theta\-\ =\-\ ((0:1:Nx)+half)*pie/(Nx+1);\-\ \color{Green}
$\\$\color{BrickRed}x\-\ =\-\ cos(theta);\color{Green}
$\\$
$\\$
$\\$\color{Green}$\%$\-\ alpha\-\ =\-\ 0
$\\$\color{BrickRed}J\-\ =\-\ theta\_vec(q,0,x,3,0);\color{Green}
$\\$\color{BrickRed}J\-\ =\-\ repmat(J,[1\-\ length(psi)]);\color{Green}
$\\$
$\\$
$\\$\color{BrickRed}E0\-\ =\-\ 1./J(:,:,1);\color{Green}
$\\$\color{BrickRed}E1\-\ =\-\ -J(:,:,2)./J(:,:,1).\verb|^|2;\color{Green}
$\\$\color{BrickRed}E2\-\ =\-\ 2*J(:,:,2).\verb|^|2./J(:,:,1).\verb|^|3-J(:,:,3)./J(:,:,1).\verb|^|2;\color{Green}
$\\$\color{BrickRed}E3\-\ =\-\ -6*J(:,:,2).\verb|^|3./J(:,:,1).\verb|^|4+6*J(:,:,2).*J(:,:,3)./J(:,:,1).\verb|^|3-J(:,:,4)./J(:,:,1).\verb|^|2;\color{Green}
$\\$
$\\$
$\\$\color{BrickRed}B0\-\ =\-\ E0.*E0;\color{Green}
$\\$\color{BrickRed}B1\-\ =\-\ 2*E0.*E1;\color{Green}
$\\$\color{BrickRed}B2\-\ =\-\ 2*(E0.*E2+E1.*E1);\color{Green}
$\\$\color{BrickRed}B3\-\ =\-\ 2*(E3.*E0+3*E1.*E2);\color{Green}
$\\$
$\\$
$\\$\color{Green}$\%$\-\ alpha\-\ \textbackslash neq\-\ 0
$\\$\color{BrickRed}L\-\ =\-\ theta\_vec(q,psi,x,4,ntilde);\color{Green}
$\\$
$\\$
$\\$\color{BrickRed}A0\-\ =\-\ L(:,:,1).*L(:,:,1);\color{Green}
$\\$\color{BrickRed}A1\-\ =\-\ 2*L(:,:,1).*L(:,:,2.);\color{Green}
$\\$\color{BrickRed}A2\-\ =\-\ 2*(L(:,:,1).*L(:,:,3)+L(:,:,2).*L(:,:,2));\color{Green}
$\\$\color{BrickRed}A3\-\ =\-\ 2*(L(:,:,4).*L(:,:,1)+3*L(:,:,2).*L(:,:,3));\color{Green}
$\\$\color{BrickRed}A4\-\ =\-\ 2*(L(:,:,1).*L(:,:,5)+4*L(:,:,2).*L(:,:,4)+3*L(:,:,3).*L(:,:,3));\color{Green}
$\\$
$\\$
$\\$\color{BrickRed}w0\-\ =\-\ A0.*B0;\color{Green}
$\\$\color{BrickRed}w1\-\ =\-\ A0.*B1+A1.*B0;\color{Green}
$\\$\color{BrickRed}w2\-\ =\-\ A0.*B2+2*A1.*B1+A2.*B0;\color{Green}
$\\$\color{BrickRed}w3\-\ =\-\ A0.*B3+3*A1.*B2+3*A2.*B1+\-\ A3.*B0;\color{Green}
$\\$
$\\$
$\\$\color{BrickRed}con\-\ =\-\ log(q)/(1i*pie);\color{Green}
$\\$\color{BrickRed}D0\-\ =\-\ con*(A1.*B0);\color{Green}
$\\$\color{BrickRed}D1\-\ =\-\ con*(A1.*B1+A2.*B0);\color{Green}
$\\$\color{BrickRed}D2\-\ =\-\ con*(A1.*B2+2*A2.*B1+A3.*B0);\color{Green}
$\\$\color{BrickRed}D3\-\ =\-\ con*(A1.*B3+3*A2.*B2+3*A3.*B1+\-\ A4.*B0);\color{Green}
$\\$
$\\$
$\\$\color{BrickRed}xicon\-\ =\-\ 1i*repmat(xi,length(x),1);\color{Green}
$\\$\color{BrickRed}xicon\_der\-\ =\-\ 1i*repmat(xi\_der,length(x),1);\-\ \color{Green}
$\\$
$\\$
$\\$\color{BrickRed}xicon2\-\ =\-\ xicon.*xicon;\color{Green}
$\\$\color{BrickRed}xicon3\-\ =\-\ xicon.*xicon2;\color{Green}
$\\$
$\\$
$\\$\color{BrickRed}v1\-\ =\-\ w1+xicon.*w0;\color{Green}
$\\$\color{BrickRed}v2\-\ =\-\ w2+2*xicon.*w1+xicon2.*w0;\color{Green}
$\\$\color{BrickRed}v3\-\ =\-\ w3+3*xicon.*w2+3*xicon2.*w1+xicon3.*w0;\color{Green}
$\\$
$\\$
$\\$\color{BrickRed}v1\_psi\-\ =\-\ D1\-\ +\-\ xicon\_der.*w0+xicon.*D0;\color{Green}
$\\$
$\\$
$\\$\color{BrickRed}v2\_psi\-\ =\-\ D2+2*xicon\_der.*w1+2*xicon.*D1+...\color{Green}
$\\$\color{BrickRed}\-\ \-\ \-\ \-\ 2*xicon.*xicon\_der.*w0+xicon2.*D0;\color{Green}
$\\$
$\\$
$\\$\color{BrickRed}v3\_psi\-\ =\-\ D3\-\ +\-\ 3*xicon\_der.*w2+3*xicon.*D2+6*xicon.*xicon\_der.*w1...\color{Green}
$\\$\color{BrickRed}\-\ \-\ \-\ \-\ +3*xicon2.*D1+3*xicon2.*xicon\_der.*w0+xicon3.*D0;\color{Green}
$\\$
$\\$
$\\$
$\\$
$\\$\color{BrickRed}f1\-\ =\-\ v1.*conj(v2);\color{Green}
$\\$\color{BrickRed}f1\_psi\-\ =\-\ v1\_psi.*conj(v2)+v1.*conj(v2\_psi);\color{Green}
$\\$
$\\$
$\\$\color{BrickRed}f2\-\ =\-\ v3.*conj(v2);\color{Green}
$\\$\color{BrickRed}f2\_psi\-\ =\-\ v3\_psi.*conj(v2)+v3.*conj(v2\_psi);\color{Green}
$\\$
$\\$
$\\$\color{BrickRed}g\-\ =\-\ w0.*conj(v1);\color{Green}
$\\$\color{BrickRed}g\_psi\-\ =\-\ D0.*conj(v1)+w0.*conj(v1\_psi);\color{Green}
$\\$
$\\$
$\\$\color{Green}$\%$\-\ Chebyshev\-\ polynomials\-\ evaluated\-\ at\-\ the\-\ points
$\\$\color{BrickRed}Tx\-\ =\-\ cos(theta.'*(0:2:Nx));\color{Green}
$\\$
$\\$
$\\$\color{BrickRed}cf1\-\ =\-\ 2*f1.'*Tx/(Nx+1);\color{Green}
$\\$\color{BrickRed}cf1(:,1)\-\ =\-\ cf1(:,1)/2;\color{Green}
$\\$\color{BrickRed}cf2\-\ =\-\ 2*f2.'*Tx/(Nx+1);\color{Green}
$\\$\color{BrickRed}cf2(:,1)\-\ =\-\ cf2(:,1)/2;\color{Green}
$\\$\color{BrickRed}cg\-\ =\-\ 2*g.'*Tx/(Nx+1);\color{Green}
$\\$\color{BrickRed}cg(:,1)\-\ =\-\ cg(:,1)/2;\color{Green}
$\\$
$\\$
$\\$\color{BrickRed}cf1\_psi\-\ =\-\ 2*f1\_psi.'*Tx/(Nx+1);\color{Green}
$\\$\color{BrickRed}cf1\_psi(:,1)\-\ =\-\ cf1\_psi(:,1)/2;\color{Green}
$\\$\color{BrickRed}cf2\_psi\-\ =\-\ 2*f2\_psi.'*Tx/(Nx+1);\color{Green}
$\\$\color{BrickRed}cf2\_psi(:,1)\-\ =\-\ cf2\_psi(:,1)/2;\color{Green}
$\\$\color{BrickRed}cg\_psi\-\ =\-\ 2*g\_psi.'*Tx/(Nx+1);\color{Green}
$\\$\color{BrickRed}cg\_psi(:,1)\-\ =\-\ cg\_psi(:,1)/2;\color{Green}
$\\$
$\\$
$\\$
$\\$
$\\$\color{BrickRed}out\-\ =\-\ nm(zeros(length(psi),1,6));\color{Green}
$\\$
$\\$
$\\$\color{BrickRed}out(:,1)\-\ =\-\ 2*cf1*(1./(1-(0:2:Nx).\verb|^|2)).';\color{Green}
$\\$\color{BrickRed}out(:,3)\-\ =\-\ 2*cf2*(1./(1-(0:2:Nx).\verb|^|2)).';\color{Green}
$\\$\color{BrickRed}out(:,5)\-\ =\-\ 2*cg*(1./(1-(0:2:Nx).\verb|^|2)).';\color{Green}
$\\$
$\\$
$\\$\color{BrickRed}out(:,2)\-\ =\-\ 2*cf1\_psi*(1./(1-(0:2:Nx).\verb|^|2)).';\color{Green}
$\\$\color{BrickRed}out(:,4)\-\ =\-\ 2*cf2\_psi*(1./(1-(0:2:Nx).\verb|^|2)).';\color{Green}
$\\$\color{BrickRed}out(:,6)\-\ =\-\ 2*cg\_psi*(1./(1-(0:2:Nx).\verb|^|2)).';\color{Green}
$\\$
$\\$
$\\$\color{Green}$\%$\-\ add\-\ integration\-\ (\-\ in\-\ x)\-\ error
$\\$\color{BrickRed}out\-\ =\-\ out\-\ +\-\ 2*(nm(-err,err)+1i*nm(-err,err));\color{Green}
$\\$
$\\$
$\\$
$\\$
$\\$
$\\$
$\\$
$\\$
$\\$
$\\$
$\\$
$\\$
$\\$
$\\$
$\\$
$\\$
$\\$\color{Black}\section{interpolation\_2d.m}

\color{Green}\color{BrickRed}\color{NavyBlue}\-\ function\-\ \color{BrickRed}\-\ interpolation\_2d(q\_min,q\_max,num\_steps,steps,file\_name)\color{Green}
$\\$
$\\$
$\\$\color{BrickRed}curr\_dir\-\ =\-\ cd;\color{Green}
$\\$\color{BrickRed}local\_startup\_batch(cd);\color{Green}
$\\$
$\\$
$\\$\color{BrickRed}total\_time\-\ =\-\ tic;\color{Green}
$\\$
$\\$
$\\$\color{Green}$\%$
$\\$\color{Green}$\%$\-\ Get\-\ the\-\ lower\-\ bound\-\ of\-\ theta
$\\$\color{Green}$\%$
$\\$
$\\$
$\\$\color{BrickRed}pie\-\ =\-\ nm('pi');\color{Green}
$\\$\color{BrickRed}frac\-\ =\-\ nm('0.9');\color{Green}
$\\$
$\\$
$\\$\color{BrickRed}t1\-\ =\-\ tic;\color{Green}
$\\$\color{BrickRed}[dm,rho\_x]\-\ =\-\ lower\_bound(frac,q\_min,q\_max,num\_steps,steps);\color{Green}
$\\$\color{BrickRed}d.dm\_time\-\ =\-\ toc(t1);\-\ \color{Green}$\%$\-\ 
$\\$
$\\$
$\\$\color{BrickRed}\-\ \color{Green}
$\\$\color{BrickRed}d.dm\-\ =\-\ dm;\color{Green}
$\\$\color{BrickRed}d.rho\_x\-\ =\-\ rho\_x;\color{Green}
$\\$
$\\$
$\\$
$\\$\-\ \color{Black}
Note that $\rho_{\psi}$ must be chosen so that $\xi(\omega + i\psi \omega')$ does not have a pole.
The poles of $\xi$ are $z = 2m \omega + 2n \omega'$. Setting
$2m\omega + 2n\omega' = \omega + i\psi \omega'$ with $\psi = 1/2 + \tilde \psi/2$, we
find that $|\Im(\tilde \psi)|< -\frac{\pi}{\log(q)}$ is necessary and sufficient to ensure analyticity.
\color{Green}
$\\$
$\\$\color{Green}$\%$$\%$
$\\$
$\\$
$\\$\color{Green}$\%$\-\ find\-\ rho\_psi
$\\$\color{BrickRed}c\_psi\-\ =\-\ nm('0.9')*pie/abs(log(q\_min));\color{Green}
$\\$\color{BrickRed}rho\_psi\-\ =\-\ nm(inf(c\_psi\-\ +sqrt(c\_psi\verb|^|2+1)));\color{Green}
$\\$\color{BrickRed}d.rho\_psi\-\ =\-\ rho\_psi;\color{Green}
$\\$
$\\$
$\\$\color{Green}$\%$\-\ make\-\ sure\-\ rho\_psi\-\ is\-\ small\-\ enough\-\ that\-\ bounds
$\\$\color{Green}$\%$\-\ on\-\ xi(omega\-\ +\-\ 1i*psi*omega')\-\ are\-\ valid
$\\$\color{BrickRed}\color{NavyBlue}\-\ if\-\ \color{BrickRed}\-\ sup(rho\_psi)\-\ $>$=\-\ 3+sqrt(nm(2))\color{Green}
$\\$\color{BrickRed}\-\ \-\ \-\ \-\ rho\_psi\-\ =\-\ 3+sqrt(nm(2));\color{Green}
$\\$\color{BrickRed}\color{NavyBlue}\-\ end\-\ \color{BrickRed}\color{Green}
$\\$
$\\$
$\\$\color{BrickRed}\color{NavyBlue}\-\ if\-\ \color{BrickRed}\-\ sup(rho\_psi)\-\ $<$=\-\ 1\color{Green}
$\\$\color{BrickRed}\-\ \-\ \-\ \-\ error('problem');\color{Green}
$\\$\color{BrickRed}\color{NavyBlue}\-\ end\-\ \color{BrickRed}\color{Green}
$\\$\color{BrickRed}\-\ \color{Green}
$\\$\color{Green}$\%$
$\\$\color{Green}$\%$\-\ find\-\ rho\_q
$\\$\color{Green}$\%$
$\\$
$\\$
$\\$\color{BrickRed}a\-\ =\-\ q\_min;\color{Green}
$\\$\color{BrickRed}b\-\ =\-\ q\_max;\color{Green}
$\\$\color{BrickRed}d.q\_min\-\ =\-\ q\_min;\color{Green}
$\\$\color{BrickRed}d.q\_max\-\ =\-\ q\_max;\color{Green}
$\\$
$\\$
$\\$\color{BrickRed}rho1\-\ =\-\ (b+a)/(b-a)-2*sqrt(a*b)/(b-a);\color{Green}
$\\$\color{BrickRed}rho2\-\ =\-\ (b+a)/(b-a)+2*sqrt(a*b)/(b-a);\color{Green}
$\\$\color{BrickRed}rho3\-\ =\-\ (2-a-b)/(b-a)-2*sqrt(1-a-b+a*b)/(b-a);\color{Green}
$\\$\color{BrickRed}rho4\-\ =\-\ (2-a-b)/(b-a)+2*sqrt(1-a-b+a*b)/(b-a);\color{Green}
$\\$
$\\$
$\\$\color{BrickRed}rho\_left\-\ =\-\ nm(rho1,rho3);\color{Green}
$\\$\color{BrickRed}rho\_right\-\ =\-\ nm(rho2,rho4);\color{Green}
$\\$\color{BrickRed}\color{NavyBlue}\-\ if\-\ \color{BrickRed}\-\ sup(rho\_left)\-\ $>$=\-\ inf(rho\_right)\color{Green}
$\\$\color{BrickRed}\-\ \-\ \-\ \-\ error('problem');\color{Green}
$\\$\color{BrickRed}\color{NavyBlue}\-\ end\-\ \color{BrickRed}\color{Green}
$\\$
$\\$
$\\$\color{BrickRed}rho\_left\-\ =\-\ nm(sup(rho\_left));\color{Green}
$\\$\color{BrickRed}rho\_right\-\ =\-\ nm(inf(rho\_right));\color{Green}
$\\$\color{BrickRed}rho\_q\-\ =\-\ rho\_left\-\ +\-\ nm('0.9')*(rho\_right-rho\_left);\color{Green}
$\\$\color{BrickRed}d.rho\_q\-\ =\-\ rho\_q;\color{Green}
$\\$
$\\$
$\\$\color{Green}$\%$
$\\$\color{Green}$\%$\-\ get\-\ bounds\-\ for\-\ alpha\-\ =\-\ omega\-\ +\-\ 1i*psi*omega'
$\\$\color{Green}$\%$
$\\$
$\\$
$\\$\color{Green}$\%$\-\ ntilde\-\ =\-\ 1
$\\$\color{BrickRed}ntilde\-\ =\-\ 1;\color{Green}
$\\$
$\\$
$\\$\color{BrickRed}time1\-\ =\-\ tic;\color{Green}
$\\$
$\\$
$\\$\color{BrickRed}[M\_psi,M\_x,M\_q]\-\ =\-\ bound\_numer(dm,rho\_psi,rho\_x,rho\_q,q\_min,q\_max,nm(0),nm(1),ntilde);\color{Green}
$\\$\color{BrickRed}toc(time1)\color{Green}
$\\$
$\\$
$\\$\color{BrickRed}abs\_tol\-\ =\-\ 1e-17;\color{Green}
$\\$\color{BrickRed}[N\_x,err\_x]\-\ =\-\ N\_nodes(rho\_x,M\_x,abs\_tol);\color{Green}
$\\$
$\\$
$\\$
$\\$
$\\$\color{BrickRed}abs\_tol\-\ =\-\ 1e-17;\color{Green}
$\\$\color{BrickRed}[N\_psi,err\_psi]\-\ =\-\ N\_nodes(rho\_psi,M\_psi,abs\_tol);\color{Green}
$\\$
$\\$
$\\$\color{BrickRed}[N\_q,err\_q]\-\ =\-\ N\_nodes(rho\_q,M\_q,abs\_tol);\color{Green}
$\\$
$\\$
$\\$\color{BrickRed}d.bd\_n1\_time\-\ =\-\ toc(time1);\color{Green}
$\\$
$\\$
$\\$\color{Green}$\%$------------------------------------------------------------
$\\$\color{Green}$\%$\-\ get\-\ interpolation\-\ coefficients\-\ when\-\ ntilde\-\ =\-\ 1
$\\$\color{Green}$\%$------------------------------------------------------------
$\\$
$\\$
$\\$\color{BrickRed}a\_q\-\ =\-\ a;\color{Green}
$\\$\color{BrickRed}b\_q\-\ =\-\ b;\color{Green}
$\\$\color{BrickRed}a\_psi\-\ =\-\ nm(0);\color{Green}
$\\$\color{BrickRed}b\_psi\-\ =\-\ nm(1);\color{Green}
$\\$
$\\$
$\\$\color{BrickRed}fun\-\ =\-\ \@(q,psi)(integrand\_numer(N\_x,err\_x,q*(b\_q-a\_q)/2+(a\_q+b\_q)/2,...\color{Green}
$\\$\color{BrickRed}\-\ \-\ \-\ \-\ psi*(b\_psi-a\_psi)/2+(a\_psi+b\_psi)/2,ntilde));\color{Green}
$\\$
$\\$
$\\$
$\\$
$\\$\color{Green}$\%$\-\ interpolation\-\ coefficients
$\\$\color{BrickRed}t1\-\ =\-\ tic;\color{Green}
$\\$\color{BrickRed}cfn1\-\ =\-\ cf\_biv\_cheby(N\_q,N\_psi,6,fun);\color{Green}
$\\$\color{BrickRed}d.cf1\_time\-\ =\-\ toc(t1);\color{Green}
$\\$\color{BrickRed}d.cfn1\-\ =\-\ cfn1;\color{Green}
$\\$
$\\$
$\\$\color{BrickRed}d.M\_psi\_n1\-\ \-\ =\-\ M\_psi;\-\ \-\ \-\ \-\ \-\ \-\ \-\ \-\ \-\ \-\ \-\ \-\ \-\ \-\ \color{Green}
$\\$\color{BrickRed}d.M\_q\_n1\-\ \-\ \-\ =\-\ M\_q;\-\ \-\ \-\ \-\ \-\ \-\ \-\ \-\ \-\ \-\ \-\ \-\ \-\ \-\ \-\ \-\ \-\ \-\ \-\ \-\ \-\ \-\ \-\ \-\ \-\ \-\ \-\ \color{Green}
$\\$\color{BrickRed}d.M\_x\_n1\-\ \-\ \-\ =\-\ M\_x;\-\ \-\ \-\ \-\ \-\ \-\ \-\ \-\ \-\ \-\ \-\ \-\ \-\ \-\ \-\ \-\ \-\ \-\ \-\ \-\ \-\ \color{Green}
$\\$\color{BrickRed}d.N\_psi\_n1\-\ =\-\ N\_psi;\-\ \-\ \-\ \-\ \-\ \-\ \-\ \-\ \-\ \-\ \-\ \-\ \-\ \-\ \-\ \-\ \-\ \-\ \-\ \-\ \-\ \color{Green}
$\\$\color{BrickRed}d.N\_q\_n1\-\ =\-\ N\_q;\-\ \-\ \-\ \-\ \-\ \-\ \-\ \-\ \-\ \-\ \-\ \-\ \-\ \-\ \-\ \-\ \-\ \-\ \-\ \-\ \-\ \-\ \-\ \-\ \-\ \-\ \-\ \-\ \-\ \-\ \-\ \-\ \color{Green}
$\\$\color{BrickRed}d.N\_x\_n1\-\ =\-\ N\_x;\-\ \-\ \color{Green}
$\\$\color{BrickRed}d.err\_psi\_n1\-\ \-\ =\-\ err\_psi;\-\ \-\ \-\ \-\ \-\ \-\ \-\ \-\ \-\ \-\ \-\ \-\ \-\ \-\ \-\ \color{Green}
$\\$\color{BrickRed}d.err\_q\_n1\-\ \-\ =\-\ err\_q;\-\ \-\ \-\ \-\ \-\ \-\ \-\ \-\ \-\ \-\ \-\ \-\ \-\ \-\ \-\ \-\ \-\ \-\ \-\ \-\ \-\ \-\ \-\ \color{Green}
$\\$\color{BrickRed}d.err\_x\_n1\-\ =\-\ err\_x;\-\ \-\ \color{Green}
$\\$\color{BrickRed}d.a\_q\-\ \-\ =\-\ a\_q;\-\ \-\ \-\ \-\ \-\ \-\ \-\ \-\ \-\ \-\ \-\ \-\ \-\ \-\ \-\ \-\ \-\ \-\ \-\ \-\ \-\ \-\ \-\ \-\ \-\ \-\ \-\ \-\ \-\ \-\ \-\ \-\ \-\ \-\ \-\ \-\ \-\ \-\ \-\ \-\ \-\ \-\ \-\ \-\ \-\ \-\ \-\ \-\ \-\ \-\ \-\ \-\ \-\ \-\ \-\ \-\ \-\ \-\ \-\ \-\ \-\ \-\ \-\ \-\ \-\ \color{Green}
$\\$\color{BrickRed}d.b\_q\-\ =\-\ b\_q;\-\ \-\ \-\ \-\ \-\ \-\ \-\ \-\ \-\ \-\ \-\ \-\ \-\ \-\ \-\ \-\ \-\ \-\ \-\ \-\ \-\ \-\ \-\ \color{Green}
$\\$\color{BrickRed}d.cfn1\-\ =\-\ cfn1;\-\ \-\ \-\ \-\ \-\ \-\ \-\ \-\ \-\ \-\ \-\ \-\ \-\ \-\ \-\ \-\ \-\ \-\ \-\ \-\ \-\ \-\ \-\ \-\ \-\ \-\ \-\ \-\ \-\ \-\ \-\ \-\ \-\ \-\ \-\ \-\ \-\ \-\ \color{Green}
$\\$\color{BrickRed}d.dm\-\ \-\ =\-\ dm;\-\ \-\ \-\ \-\ \-\ \-\ \-\ \-\ \-\ \-\ \-\ \-\ \-\ \-\ \-\ \-\ \-\ \-\ \-\ \-\ \-\ \-\ \-\ \-\ \-\ \color{Green}
$\\$\color{BrickRed}d.rho\_psi\_n1\-\ =\-\ rho\_psi;\-\ \-\ \-\ \-\ \-\ \-\ \-\ \-\ \-\ \-\ \-\ \-\ \-\ \-\ \-\ \-\ \-\ \-\ \-\ \-\ \-\ \-\ \-\ \-\ \-\ \-\ \-\ \-\ \-\ \-\ \-\ \-\ \-\ \-\ \-\ \color{Green}
$\\$\color{BrickRed}d.rho\_q\_n1\-\ =\-\ rho\_q;\-\ \-\ \-\ \-\ \-\ \-\ \-\ \-\ \-\ \-\ \-\ \-\ \-\ \-\ \-\ \-\ \-\ \-\ \-\ \-\ \-\ \-\ \-\ \-\ \-\ \-\ \-\ \-\ \-\ \-\ \-\ \-\ \-\ \color{Green}
$\\$\color{BrickRed}d.rho\_x\_n1\-\ \-\ =\-\ rho\_x;\-\ \-\ \-\ \-\ \color{Green}
$\\$
$\\$
$\\$\color{Green}$\%$
$\\$\color{Green}$\%$\-\ get\-\ bounds\-\ on\-\ 10\-\ integrands\-\ for\-\ alpha\-\ =\-\ 1i*psi*omega'
$\\$\color{Green}$\%$
$\\$
$\\$
$\\$\color{BrickRed}[M\_psi\_10,M\_x\_10,M\_q\_10]\-\ =\-\ bound\_sub\_integrals(dm,rho\_psi,rho\_x,rho\_q,a,b);\color{Green}
$\\$
$\\$
$\\$\color{BrickRed}d.M\_psi\_10\-\ =\-\ M\_psi\_10;\color{Green}
$\\$\color{BrickRed}d.M\_x\_10\-\ =\-\ M\_x\_10;\color{Green}
$\\$\color{BrickRed}d.M\_q\_10\-\ =\-\ M\_q\_10;\color{Green}
$\\$
$\\$
$\\$\color{BrickRed}abs\_tol\-\ =\-\ 1e-17;\color{Green}
$\\$\color{BrickRed}[N\_x\_10,err\_x\_10]\-\ =\-\ N\_nodes(rho\_x,M\_x\_10,abs\_tol);\color{Green}
$\\$\color{BrickRed}d.N\_x\_10\-\ =\-\ N\_x\_10;\color{Green}
$\\$\color{BrickRed}d.err\_x\_10\-\ =\-\ err\_x\_10;\color{Green}
$\\$
$\\$
$\\$\color{BrickRed}abs\_tol\-\ =\-\ 1e-17;\color{Green}
$\\$\color{BrickRed}[N\_psi\_10,err\_psi\_10]\-\ =\-\ N\_nodes(rho\_psi,M\_psi\_10,abs\_tol);\color{Green}
$\\$\color{BrickRed}d.N\_psi\_10\-\ =\-\ N\_psi\_10;\color{Green}
$\\$\color{BrickRed}d.err\_psi\_10\-\ =\-\ err\_psi\_10;\color{Green}
$\\$
$\\$
$\\$\color{BrickRed}[N\_q\_10,err\_q\_10]\-\ =\-\ N\_nodes(rho\_q,M\_q\_10,abs\_tol);\color{Green}
$\\$\color{BrickRed}d.N\_q\_10\-\ =\-\ N\_q\_10;\color{Green}
$\\$\color{BrickRed}d.err\_q\_10\-\ =\-\ err\_q\_10;\color{Green}
$\\$
$\\$
$\\$\color{Green}$\%$------------------------------------------------------------
$\\$\color{Green}$\%$\-\ get\-\ interpolation\-\ coefficients\-\ when\-\ ntilde\-\ =\-\ 0
$\\$\color{Green}$\%$\-\ for\-\ 10\-\ integrands
$\\$\color{Green}$\%$------------------------------------------------------------
$\\$
$\\$
$\\$\color{BrickRed}a\_q\-\ =\-\ a;\color{Green}
$\\$\color{BrickRed}b\_q\-\ =\-\ b;\color{Green}
$\\$\color{BrickRed}a\_psi\-\ =\-\ nm(0);\color{Green}
$\\$\color{BrickRed}b\_psi\-\ =\-\ nm(1);\color{Green}
$\\$
$\\$
$\\$\color{BrickRed}ntilde\-\ =\-\ 0;\color{Green}
$\\$\color{BrickRed}fun\-\ =\-\ \@(q,psi)(numer(N\_x,err\_x,q*(b\_q-a\_q)/2+(a\_q+b\_q)/2,...\color{Green}
$\\$\color{BrickRed}\-\ \-\ \-\ \-\ psi*(b\_psi-a\_psi)/2+(a\_psi+b\_psi)/2,ntilde));\color{Green}
$\\$
$\\$
$\\$\color{Green}$\%$\-\ interpolation\-\ coefficients
$\\$\color{BrickRed}t1\-\ =\-\ tic;\color{Green}
$\\$\color{BrickRed}cf10\-\ =\-\ cf\_biv\_cheby(N\_q\_10,N\_psi\_10,10,fun);\color{Green}
$\\$\color{BrickRed}d.cf10\_time\-\ =\-\ toc(t1);\color{Green}
$\\$\color{BrickRed}d.cf10\-\ =\-\ cf10;\color{Green}
$\\$
$\\$
$\\$\color{Green}$\%$------------------------------------------------------------
$\\$\color{Green}$\%$\-\ Find\-\ bound\-\ on\-\ numerator\-\ when\-\ ntilde\-\ =\-\ 0
$\\$\color{Green}$\%$------------------------------------------------------------
$\\$
$\\$
$\\$\color{BrickRed}ntilde\-\ =\-\ 0;\color{Green}
$\\$
$\\$
$\\$\color{BrickRed}a\_psi\-\ =\-\ inf(nm('0.5'));\color{Green}
$\\$\color{BrickRed}b\_psi\-\ =\-\ 1;\color{Green}
$\\$
$\\$
$\\$\color{BrickRed}time1\-\ =\-\ tic;\color{Green}
$\\$\color{BrickRed}[M\_psi,M\_x,M\_q]\-\ =\-\ bound\_numer...\color{Green}
$\\$\color{BrickRed}\-\ \-\ \-\ \-\ (dm,rho\_psi,rho\_x,rho\_q,q\_min,q\_max,a\_psi,b\_psi,ntilde);\color{Green}
$\\$\color{BrickRed}toc(time1)\color{Green}
$\\$
$\\$
$\\$\color{BrickRed}abs\_tol\-\ =\-\ 1e-17;\color{Green}
$\\$\color{BrickRed}[N\_x,err\_x]\-\ =\-\ N\_nodes(rho\_x,M\_x,abs\_tol);\color{Green}
$\\$
$\\$
$\\$\color{BrickRed}abs\_tol\-\ =\-\ 1e-17;\color{Green}
$\\$\color{BrickRed}[N\_psi,err\_psi]\-\ =\-\ N\_nodes(rho\_psi,M\_psi,abs\_tol);\color{Green}
$\\$
$\\$
$\\$\color{BrickRed}[N\_q,err\_q]\-\ =\-\ N\_nodes(rho\_q,M\_q,abs\_tol);\color{Green}
$\\$
$\\$
$\\$\color{BrickRed}d.bd\_n1\_time\-\ =\-\ toc(time1);\color{Green}
$\\$
$\\$
$\\$\color{BrickRed}d.M\_psi\_n0\-\ \-\ =\-\ M\_psi;\-\ \-\ \-\ \-\ \-\ \-\ \-\ \-\ \-\ \-\ \-\ \-\ \-\ \-\ \color{Green}
$\\$\color{BrickRed}d.M\_q\_n0\-\ \-\ \-\ =\-\ M\_q;\-\ \-\ \-\ \-\ \-\ \-\ \-\ \-\ \-\ \-\ \-\ \-\ \-\ \-\ \-\ \-\ \-\ \-\ \-\ \-\ \-\ \-\ \-\ \-\ \-\ \-\ \-\ \color{Green}
$\\$\color{BrickRed}d.M\_x\_n0\-\ \-\ \-\ =\-\ M\_x;\-\ \-\ \-\ \-\ \-\ \-\ \-\ \-\ \-\ \-\ \-\ \-\ \-\ \-\ \-\ \-\ \-\ \-\ \-\ \-\ \-\ \color{Green}
$\\$\color{BrickRed}d.N\_psi\_n0\-\ =\-\ N\_psi;\-\ \-\ \-\ \-\ \-\ \-\ \-\ \-\ \-\ \-\ \-\ \-\ \-\ \-\ \-\ \-\ \-\ \-\ \-\ \-\ \-\ \color{Green}
$\\$\color{BrickRed}d.N\_q\_n0\-\ =\-\ N\_q;\-\ \-\ \-\ \-\ \-\ \-\ \-\ \-\ \-\ \-\ \-\ \-\ \-\ \-\ \-\ \-\ \-\ \-\ \-\ \-\ \-\ \-\ \-\ \-\ \-\ \-\ \-\ \-\ \-\ \-\ \-\ \-\ \color{Green}
$\\$\color{BrickRed}d.N\_x\_n0\-\ =\-\ N\_x;\-\ \-\ \color{Green}
$\\$\color{BrickRed}d.err\_psi\_n0\-\ \-\ =\-\ err\_psi;\-\ \-\ \-\ \-\ \-\ \-\ \-\ \-\ \-\ \-\ \-\ \-\ \-\ \-\ \-\ \color{Green}
$\\$\color{BrickRed}d.err\_q\_n0\-\ \-\ =\-\ err\_q;\-\ \-\ \-\ \-\ \-\ \-\ \-\ \-\ \-\ \-\ \-\ \-\ \-\ \-\ \-\ \-\ \-\ \-\ \-\ \-\ \-\ \-\ \-\ \color{Green}
$\\$\color{BrickRed}d.err\_x\_n0\-\ =\-\ err\_x;\-\ \-\ \color{Green}
$\\$
$\\$
$\\$\color{Green}$\%$------------------------------------------------------------
$\\$\color{Green}$\%$\-\ get\-\ interpolation\-\ coefficients\-\ when\-\ ntilde\-\ =\-\ 0
$\\$\color{Green}$\%$------------------------------------------------------------
$\\$
$\\$
$\\$\color{BrickRed}fun\-\ =\-\ \@(q,psi)(integrand\_numer(N\_x,err\_x,q*(b\_q-a\_q)/2+(a\_q+b\_q)/2,...\color{Green}
$\\$\color{BrickRed}\-\ \-\ \-\ \-\ psi*(b\_psi-a\_psi)/2+(a\_psi+b\_psi)/2,ntilde));\color{Green}
$\\$
$\\$
$\\$\color{Green}$\%$\-\ interpolation\-\ coefficients
$\\$\color{BrickRed}tstart\-\ =\-\ tic;\color{Green}
$\\$\color{BrickRed}cfn0\-\ =\-\ cf\_biv\_cheby(N\_q,N\_psi,6,fun);\color{Green}
$\\$\color{BrickRed}d.cf0\_time\-\ =\-\ toc(t1);\color{Green}
$\\$\color{BrickRed}d.cfn0\-\ =\-\ cfn0;\color{Green}
$\\$\color{BrickRed}d.cfn0\_time\-\ =\-\ toc(tstart);\color{Green}
$\\$
$\\$
$\\$\color{Green}$\%$------------------------------------------------------------
$\\$\color{Green}$\%$\-\ save\-\ data
$\\$\color{Green}$\%$------------------------------------------------------------
$\\$
$\\$
$\\$\color{BrickRed}d.total\_time\-\ =\-\ toc(total\_time);\color{Green}
$\\$\color{BrickRed}d.date\-\ =\-\ date;\color{Green}
$\\$
$\\$
$\\$\color{BrickRed}saveit(curr\_dir,'interval\_arithmetic',d,file\_name,'data\_final');\color{Green}
$\\$
$\\$
$\\$\color{Black}\section{kappa\_lemma.m}

\color{Green}\color{Green}$\%$\-\ TODO
$\\$\color{BrickRed}curr\_dir\-\ =\-\ local\_startup;\color{Green}
$\\$
$\\$
$\\$
$\\$\color{Black}
Now 
\eqn{
\kappa^2(k) = \left(\frac{\pi^2}{K^2(k)}\right)\left(\frac{7}{20}\right)\left(\frac{A(k)-B(k)}{C(k)+D(k)}\right),
}{}
where 
\eqn{
A(k)&:= 2(k^4-k^2+1)E(k)\\
B(k)&:= (1-k^2)(2-k^2)K(k)\\
C(k)&:= (-2+3k^2+3k^4-2k^6)E(k)\\
D(k)&:= (k^6+k^4-4k^2+2)K(k),
}{}
where $K(k)$ and $E(k)$ are respectively the complete elliptic integrals of the  first and second kind.

$\\$
The derivative of $\kappa^2(k)$ is given by,
\eqn{
\pd{}{k}\kappa^2(k)&= \left(\frac{-2\pi K'(k)}{K^3(k)}\right)\left(\frac{7}{20}\right)\left(\frac{A(k)-B(k)}{C(k)+D(k)}\right)\\
&+ \left(\frac{\pi^2}{K^2(k)}\right)\left(\frac{7}{20}\right)\left(\frac{(C(k)+D(k))(A'(k)-B'(k))-(A(k)-B(k))(C'(k)+D'(k))}{(C(k)+D(k))^2} \right),
}{}
where,
\eqn{
A'(k)&= 2(4k^3-2k)E(k)+2(k^4-k^2+1)E'(k),\\
B'(k)&= -2k(2-k^2)K(k)-2k(1-k^2)K(k)+(1-k^2)(2-k^2)K'(k),\\
C'(k)&= (6k+12k^3-12k^5)E(k) + (-2+3k^2+3k^4-2k^6)E'(k),\\
D'(k)&= (6k^5+4k^3-8k)K(k)+(k^6+k^4-4k^2+2)K'(k).
}{}
The elliptic integrals have derivatives,
\eqn{
K'(k) &= \frac{E(k)}{k(1-k^2)}-\frac{K(k)}{k}\\
E'(k)&= \frac{E(k)-K(k)}{k}.
}{}
\color{Green}
$\\$
$\\$\color{Green}$\%$$\%$
$\\$
$\\$
$\\$\color{BrickRed}pie\-\ =\-\ nm('pi');\color{Green}
$\\$
$\\$
$\\$\color{Green}$\%$\-\ k\-\ interval\-\ ends
$\\$\color{BrickRed}s0\-\ =\-\ \-\ linspace(0.9,0.93,100);\color{Green}
$\\$\color{BrickRed}s1\-\ =\-\ \-\ linspace(0.93,0.99,100);\color{Green}
$\\$\color{BrickRed}s2\-\ =\-\ \-\ linspace(0.99,0.999,100);\color{Green}
$\\$\color{BrickRed}s3\-\ =\-\ \-\ linspace(0.999,0.9999,100);\color{Green}
$\\$\color{BrickRed}s4\-\ =\-\ linspace(0.9999,0.99999,100);\color{Green}
$\\$\color{BrickRed}s5\-\ =\-\ linspace(0.99999,0.999999,100);\color{Green}
$\\$\color{BrickRed}s6\-\ =\-\ linspace(0.999999,0.9999999,100);\color{Green}
$\\$\color{BrickRed}k\-\ =\-\ [s0,\-\ s1,\-\ s2,\-\ s3,\-\ s4,\-\ s5,\-\ s6];\color{Green}
$\\$\color{Green}$\%$\-\ make\-\ k\-\ intervals\-\ from\-\ end\-\ poitns
$\\$\color{BrickRed}k\-\ =\-\ nm(k(1:end-1),k(2:end));\color{Green}
$\\$
$\\$
$\\$\color{Green}$\%$\-\ complete\-\ elliptic\-\ integral\-\ of\-\ the\-\ first\-\ kind
$\\$\color{BrickRed}K\-\ =\-\ elliptic\_integral(k,1);\color{Green}
$\\$\color{Green}$\%$\-\ complete\-\ elliptic\-\ integral\-\ of\-\ the\-\ second\-\ kind
$\\$\color{BrickRed}E\-\ =\-\ elliptic\_integral(k,2);\color{Green}
$\\$
$\\$
$\\$\color{Green}$\%$\-\ derivatives\-\ of\-\ K\-\ and\-\ E\-\ with\-\ resepct\-\ to\-\ k
$\\$\color{BrickRed}Kp\-\ =\-\ E./(k.*(1-k.\verb|^|2))-K./k;\color{Green}
$\\$\color{BrickRed}Ep\-\ =\-\ (E-K)./k;\color{Green}
$\\$
$\\$
$\\$\color{Green}$\%$\-\ auxiliary\-\ functions
$\\$\color{BrickRed}A\-\ =\-\ 2*(k.\verb|^|4-k.\verb|^|2+1).*E;\color{Green}
$\\$
$\\$
$\\$\color{BrickRed}B\-\ =\-\ (1-k.\verb|^|2).*(2-k.\verb|^|2).*K;\color{Green}
$\\$
$\\$
$\\$\color{BrickRed}C\-\ =\-\ (-2+3*k.\verb|^|2+3*k.\verb|^|4-2*k.\verb|^|6).*E;\color{Green}
$\\$
$\\$
$\\$\color{BrickRed}D\-\ =\-\ (k.\verb|^|6+k.\verb|^|4-4*k.\verb|^|2+2).*K;\color{Green}
$\\$
$\\$
$\\$\color{Green}$\%$\-\ derivatives\-\ of\-\ auxiliary\-\ functions\-\ with\-\ respect\-\ to\-\ k
$\\$\color{BrickRed}Ap\-\ =\-\ 2*(4*k.\verb|^|3-2*k).*E+2*(k.\verb|^|4-k.\verb|^|2+1).*Ep;\color{Green}
$\\$
$\\$
$\\$\color{BrickRed}Bp\-\ =\-\ -2*k.*(2-k.\verb|^|2).*K-2*k.*(1-k.\verb|^|2).*K+(1-k.\verb|^|2).*(2-k.\verb|^|2).*Kp;\color{Green}
$\\$
$\\$
$\\$\color{BrickRed}Cp\-\ =\-\ (6*k+12*k.\verb|^|3-12*k.\verb|^|5).*E+(-2+3*k.\verb|^|2+3*k.\verb|^|4-2*k.\verb|^|6).*Ep;\color{Green}
$\\$
$\\$
$\\$\color{BrickRed}Dp\-\ =\-\ (6*k.\verb|^|5+4*k.\verb|^|3-8*k).*K+(k.\verb|^|6+k.\verb|^|4-4*k.\verb|^|2+2).*Kp;\color{Green}
$\\$
$\\$
$\\$\color{Green}$\%$\-\ derivative\-\ of\-\ kappa\verb|^|2\-\ with\-\ repsect\-\ to\-\ k
$\\$\color{BrickRed}kappa2p\-\ =\-\ ((-2*pie\verb|^|2*Kp)./K.\verb|^|3)*(nm(7)/20).*((A-B)./(C+D))\-\ \-\ ...\color{Green}
$\\$\color{BrickRed}\-\ \-\ \-\ \-\ \-\ \-\ \-\ \-\ \-\ \-\ \-\ \-\ \-\ \-\ \-\ \-\ \-\ \-\ \-\ +\-\ (pie\verb|^|2./K.\verb|^|2)*(nm(7)/20).*(((C+D).*(Ap-Bp)-(A-B).*(Cp+Dp))./(C+D).\verb|^|2);\color{Green}
$\\$\color{BrickRed}\-\ \-\ \-\ \-\ \-\ \-\ \-\ \-\ \-\ \-\ \-\ \-\ \-\ \-\ \-\ \color{Green}
$\\$\color{Green}$\%$\-\ check\-\ that\-\ NaN\-\ is\-\ not\-\ present\-\ \-\ \-\ \-\ \-\ \-\ \-\ \-\ 
$\\$\color{BrickRed}\color{NavyBlue}\-\ if\-\ \color{BrickRed}\-\ sum(isnan(kappa2p))\-\ $>$\-\ 0\color{Green}
$\\$\color{BrickRed}\-\ \-\ \-\ \-\ error('NaN\-\ present');\color{Green}
$\\$\color{BrickRed}\color{NavyBlue}\-\ end\-\ \color{BrickRed}\color{Green}
$\\$\color{BrickRed}\-\ \-\ \-\ \-\ \-\ \-\ \-\ \-\ \-\ \-\ \-\ \-\ \color{Green}
$\\$\color{Green}$\%$\-\ find\-\ maximum\-\ value\-\ of\-\ the\-\ derivative\-\ of\-\ kappa\verb|^|2\-\ on\-\ the\-\ k\-\ intervals
$\\$\color{BrickRed}mx\-\ =\-\ max(sup(kappa2p));\color{Green}
$\\$\color{BrickRed}\-\ \-\ \-\ \-\ \-\ \-\ \-\ \-\ \-\ \-\ \-\ \-\ \-\ \-\ \-\ \color{Green}
$\\$\color{Green}$\%$\-\ check\-\ that\-\ mx\-\ $<$\-\ 0\-\ implying\-\ that\-\ kappa\verb|^|2\-\ is\-\ monotone\-\ decreasing
$\\$\color{BrickRed}\color{NavyBlue}\-\ if\-\ \color{BrickRed}\-\ mx\-\ $>$=\-\ 0\color{Green}
$\\$\color{BrickRed}\-\ \-\ \-\ \-\ error('failed\-\ to\-\ verify\-\ monotonicity');\color{Green}
$\\$\color{BrickRed}\color{NavyBlue}\-\ end\-\ \color{BrickRed}\color{Green}
$\\$\color{BrickRed}\-\ \-\ \-\ \-\ \-\ \color{Green}
$\\$\color{Green}$\%$\-\ find\-\ the\-\ left\-\ end\-\ of\-\ the\-\ k\-\ intervals
$\\$\color{BrickRed}kleft\-\ =\-\ min(inf(k));\color{Green}
$\\$\color{Green}$\%$\-\ find\-\ the\-\ right\-\ end\-\ of\-\ the\-\ k\-\ intervals
$\\$\color{BrickRed}kright\-\ =\-\ max(sup(k));\color{Green}
$\\$
$\\$
$\\$\color{BrickRed}fprintf('\textbackslash nThe\-\ derivative\-\ of\-\ kappa\verb|^|2\-\ with\-\ respect\-\ to\-\ k\-\ is\-\ $<$=\-\ ');\color{Green}
$\\$\color{BrickRed}disp(mx);\color{Green}
$\\$\color{BrickRed}fprintf('\-\ \color{NavyBlue}\-\ for\-\ \color{BrickRed}\-\ k\-\ in\-\ the\-\ interval\-\ with\-\ left\-\ \color{NavyBlue}\-\ end\-\ \color{BrickRed}\-\ point,');\color{Green}
$\\$\color{BrickRed}disp(kleft);\color{Green}
$\\$\color{BrickRed}fprintf('and\-\ right\-\ \color{NavyBlue}\-\ end\-\ \color{BrickRed}\-\ point,\-\ ');\color{Green}
$\\$\color{BrickRed}disp(kright);\color{Green}
$\\$\color{BrickRed}fprintf('\textbackslash n\textbackslash n');\color{Green}
$\\$
$\\$
$\\$\color{BrickRed}\-\ \-\ \-\ \-\ \-\ \-\ \-\ \-\ \-\ \-\ \-\ \-\ \-\ \-\ \-\ \color{Green}
$\\$\color{BrickRed}\-\ \-\ \-\ \-\ \-\ \-\ \-\ \-\ \-\ \-\ \-\ \-\ \-\ \-\ \-\ \color{Green}
$\\$\color{BrickRed}\-\ \-\ \-\ \-\ \-\ \-\ \-\ \-\ \-\ \-\ \-\ \-\ \-\ \-\ \-\ \color{Green}
$\\$\color{BrickRed}\-\ \-\ \-\ \-\ \-\ \-\ \-\ \-\ \-\ \-\ \-\ \-\ \-\ \-\ \-\ \color{Green}
$\\$\color{BrickRed}\-\ \-\ \-\ \-\ \-\ \-\ \-\ \-\ \-\ \-\ \-\ \-\ \-\ \-\ \-\ \color{Green}
$\\$
$\\$
$\\$\color{Black}\section{kappa\_of\_k.m}

\color{Green}\color{BrickRed}\color{NavyBlue}\-\ function\-\ \color{BrickRed}\-\ kappa\-\ =\-\ kappa\_of\_k(k)\color{Green}
$\\$\color{Green}$\%$\-\ function\-\ kappa\-\ =\-\ kappa\_of\_k(k)
$\\$\color{Green}$\%$
$\\$\color{Green}$\%$\-\ Returns\-\ kappa(k)\-\ 
$\\$
$\\$
$\\$
$\\$\color{Black}
From
\eqn{
\left(\frac{K(k)\mathcal{G}(k)}{\pi}\right)^2=
   \frac{7}{20}\frac{2(k^4-k^2+1)E(k)-(1-k^2)(2-k^2)K(k)}{(-2+3k^2+3k^4-2k^6)E(k)+(k^6+k^4-4k^2+2)K(k)}
}{}
we may determine $\kappa = \mathcal{G}(k)$.
\color{Green}
$\\$
$\\$
$\\$
$\\$\color{BrickRed}pie\-\ =\-\ nm('pi');\color{Green}
$\\$\color{Green}$\%$\-\ elliptic\-\ integrals
$\\$\color{BrickRed}K\-\ =\-\ elliptic\_integral(k,1);\color{Green}
$\\$\color{BrickRed}E\-\ =\-\ elliptic\_integral(k,2);\color{Green}
$\\$\color{Green}$\%$\-\ KmE\-\ =\-\ elliptic\_integral(k,3)
$\\$
$\\$
$\\$\color{Green}$\%$\-\ kappa(k)
$\\$\color{BrickRed}k2\-\ =\-\ k.*k;\color{Green}
$\\$\color{BrickRed}k4\-\ =\-\ k2.*k2;\color{Green}
$\\$\color{BrickRed}k6\-\ =\-\ k2.*k4;\color{Green}
$\\$\color{BrickRed}c1\-\ =\-\ 2*(k4-k2+1).*E-(1-k2).*(2-k2).*K;\color{Green}
$\\$\color{BrickRed}c2\-\ =\-\ (-2+3*k2+3*k4-2*k6).*E+(k6+k4-4*k2+2).*K;\color{Green}
$\\$\color{BrickRed}kappa\-\ =\-\ pie*sqrt(7*c1./(20*c2))./K;\color{Green}
$\\$
$\\$
$\\$
$\\$
$\\$
$\\$
$\\$\color{Black}\section{lambda\_xi.m}

\color{Green}\color{BrickRed}\color{NavyBlue}\-\ function\-\ \color{BrickRed}\-\ [f,fd,fdd,g,gd,gdd]\-\ =\-\ lambda\_xi(q,omega,omega\_prime,psi,ntilde)\color{Green}
$\\$\color{Green}$\%$
$\\$\color{Green}$\%$\-\ out\-\ =\-\ lambda\_xi(q,omega,ntilde,psi\_tilde)
$\\$\color{Green}$\%$
$\\$\color{Green}$\%$\-\ Returns\-\ in\-\ the\-\ first\-\ three\-\ components\-\ 
$\\$\color{Green}$\%$\-\ xi(ntilde*omega\_prime+1i*psi*omega\_prime)\-\ 
$\\$\color{Green}$\%$\-\ and\-\ its\-\ first\-\ two\-\ derviatives\-\ with\-\ respect\-\ to\-\ psi.\-\ Returns\-\ in\-\ the\-\ next\-\ three\-\ 
$\\$\color{Green}$\%$\-\ components\-\ 1i*c*lambda\_0(ntilde*omega+1i*psi*omega\_prime)\-\ where\-\ c\-\ is\-\ a\-\ real,\-\ nonzero\-\ constant.
$\\$
$\\$
$\\$
$\\$\color{Black}
Now
\eqn{
\wp'(z)&= -\frac{\sigma(2z)}{\sigma^4(z)}\\
\sigma(z) &= \frac{2\omega}{\pi}e^{\eta_1 z^2/2\omega}\vartheta_1(\pi z/2\omega)/\vartheta_1'(0)\\
\implies \wp'(z)&= -\frac{(\pi\vartheta_1'(0))^3}{8\omega^3}
\frac{\vartheta_1(\pi z/\omega)}{\vartheta_1^4(\pi z/2\omega)} }{}
\color{Green}
$\\$
$\\$\color{Green}$\%$$\%$
$\\$
$\\$
$\\$\color{Green}$\%$\-\ pi
$\\$\color{BrickRed}pie\-\ =\-\ nm('pi');\color{Green}
$\\$
$\\$
$\\$\color{Green}$\%$\-\ \color{Black}$\vartheta_1'(0)$ \color{Green}
$\\$\color{BrickRed}v0\-\ =\-\ theta\_vec\_z(q,0,1);\color{Green}
$\\$
$\\$
$\\$\color{Green}$\%$\-\ number\-\ of\-\ derivatives\-\ to\-\ take
$\\$\color{BrickRed}m\-\ =\-\ 2;\-\ \color{Green}
$\\$
$\\$
$\\$\color{Green}$\%$\-\ \color{Black}$\vartheta_1(\pi z/\omega)$ \color{Green}
$\\$\color{BrickRed}con\-\ =\-\ pie./(omega);\color{Green}
$\\$\color{BrickRed}z\-\ =\-\ con.*(ntilde*omega+1i*psi.*omega\_prime);\color{Green}
$\\$\color{BrickRed}vn\-\ =\-\ theta\_vec\_z(q,z,m);\color{Green}
$\\$
$\\$
$\\$\color{Green}$\%$\-\ \color{Black}$\vartheta_1(\pi z/(2\omega))$ \color{Green}
$\\$\color{BrickRed}con\-\ =\-\ pie./(2*omega);\color{Green}
$\\$\color{BrickRed}z\-\ =\-\ con.*(ntilde*omega+1i*psi*omega\_prime);\color{Green}
$\\$\color{BrickRed}vd\-\ =\-\ theta\_vec\_z(q,z,m);\color{Green}
$\\$
$\\$
$\\$\color{Green}$\%$\-\ \color{Black}$\wp'(z)$ \color{Green}
$\\$\color{BrickRed}gdd\-\ =\-\ -v0(:,:,2).\verb|^|3.*pie\verb|^|3.*vn(:,:,1)./(8*omega.\verb|^|3.*vd(:,:,1).\verb|^|4);\color{Green}
$\\$\color{Green}$\%$\-\ \color{Black}$\pd{^2}{\psi^2} \omega\zeta(\tilde n \omega + i\psi\omega')$ \color{Green}
$\\$\color{BrickRed}gdd\-\ =\-\ 2*1i*omega\_prime.\verb|^|2.*omega.*gdd;\color{Green}
$\\$
$\\$
$\\$\color{Green}$\%$\-\ \color{Black}$ \xi(\tilde n + i\psi \omega')$ \color{Green}
$\\$\color{BrickRed}g\-\ =\-\ xi\_q\_psi(q,psi,ntilde);\color{Green}
$\\$\color{BrickRed}g\-\ =\-\ g.';\color{Green}
$\\$
$\\$
$\\$\color{Green}$\%$\-\ \color{Black}$\pd{}{\psi} \xi(\tilde n + i\psi \omega')$ \color{Green}
$\\$\color{BrickRed}gd\-\ =\-\ \-\ xi\_der\_q\_psi(q,psi,ntilde);\color{Green}
$\\$\color{BrickRed}gd\-\ =\-\ gd.';\color{Green}
$\\$
$\\$
$\\$\color{Green}$\%$\-\ auxiliary\-\ quantities
$\\$
$\\$
$\\$\color{BrickRed}A\-\ =\-\ ((pie./omega).*vd(:,:,1).\verb|^|4.*vn(:,:,2)-(2*pie./omega).*vn(:,:,1).*vd(:,:,1).\verb|^|3.*vd(:,:,2));\color{Green}
$\\$
$\\$
$\\$\color{BrickRed}Ad\-\ =\-\ (\-\ (4*pie./omega).*vd(:,:,1).\verb|^|3.*vd(:,:,2).*vn(:,:,2)*(pie./(2*omega))\-\ ...\color{Green}
$\\$\color{BrickRed}\-\ \-\ \-\ \-\ +\-\ (pie./omega).*vd(:,:,1).\verb|^|4.*vn(:,:,3).*(pie./omega)\-\ ...\color{Green}
$\\$\color{BrickRed}\-\ \-\ \-\ \-\ -\-\ (2*pie./omega).*vn(:,:,2).*(pie./omega).*vd(:,:,1).\verb|^|3.*vd(:,:,2)\-\ ...\color{Green}
$\\$\color{BrickRed}\-\ \-\ \-\ \-\ -(2*pie./omega).*vn(:,:,1).*3.*vd(:,:,1).\verb|^|2.*vd(:,:,2).*(pie./(2*omega)).*vd(:,:,2)\-\ ...\color{Green}
$\\$\color{BrickRed}\-\ \-\ \-\ \-\ -(2*pie./omega).*vn(:,:,1).*vd(:,:,1).\verb|^|3.*vd(:,:,3).*(pie./(2*omega)));\color{Green}
$\\$
$\\$
$\\$\color{BrickRed}B\-\ =\-\ vd(:,:,1).\verb|^|8;\color{Green}
$\\$
$\\$
$\\$\color{BrickRed}Bd\-\ =\-\ 8*vd(:,:,1).\verb|^|7.*vd(:,:,2).*(pie./(2*omega));\color{Green}
$\\$
$\\$
$\\$\color{Green}$\%$\-\ \color{Black} $ ic\lambda_0(\tilde n\omega+i\psi\omega') $  \color{Green}
$\\$\color{BrickRed}f\-\ =\-\ 1i*vn(:,:,1)./vd(:,:,1).\verb|^|4;\color{Green}
$\\$
$\\$
$\\$\color{Green}$\%$\-\ \color{Black} $\pd{}{\psi} ic\lambda_0(\tilde n\omega+i\psi\omega') $  \color{Green}
$\\$\color{BrickRed}fd\-\ =\-\ A./B;\color{Green}
$\\$\color{BrickRed}fd\-\ =\-\ 1i*fd.*(1i*omega\_prime);\color{Green}
$\\$
$\\$
$\\$\color{Green}$\%$\-\ \color{Black} $\pd{^2}{\psi^2} ic\lambda_0(\tilde n\omega+i\psi\omega') $  \color{Green}
$\\$\color{BrickRed}fdd\-\ =\-\ (B.*Ad-A.*Bd)./B.\verb|^|2;\color{Green}
$\\$\color{BrickRed}fdd\-\ =\-\ 1i*fdd.*(1i*omega\_prime).\verb|^|2;\color{Green}
$\\$
$\\$
$\\$
$\\$
$\\$
$\\$
$\\$\color{Black}\section{lemma\_qk.m}

\color{Green}
$\\$
$\\$\color{BrickRed}curr\_dir\-\ =\-\ local\_startup;\color{Green}
$\\$
$\\$
$\\$\color{BrickRed}pie\-\ =\-\ nm('pi');\color{Green}
$\\$
$\\$
$\\$\color{Green}$\%$\-\ k\-\ interval\-\ ends
$\\$\color{BrickRed}s0\-\ =\-\ \-\ linspace(0.9,0.93,400);\color{Green}
$\\$\color{BrickRed}s1\-\ =\-\ \-\ linspace(0.93,0.99,400);\color{Green}
$\\$\color{BrickRed}s2\-\ =\-\ \-\ linspace(0.99,0.999,400);\color{Green}
$\\$\color{BrickRed}s3\-\ =\-\ \-\ linspace(0.999,0.9999,400);\color{Green}
$\\$\color{BrickRed}s4\-\ =\-\ linspace(0.9999,0.99999,400);\color{Green}
$\\$\color{BrickRed}s5\-\ =\-\ linspace(0.99999,0.999999,400);\color{Green}
$\\$\color{BrickRed}s6\-\ =\-\ linspace(0.999999,0.9999999,400);\color{Green}
$\\$\color{BrickRed}k\-\ =\-\ [s0,\-\ s1,\-\ s2,\-\ s3,\-\ s4,\-\ s5,\-\ s6];\color{Green}
$\\$\color{Green}$\%$\-\ make\-\ k\-\ intervals\-\ from\-\ end\-\ poitns
$\\$\color{BrickRed}k\-\ =\-\ nm(k(1:end-1),k(2:end));\color{Green}
$\\$
$\\$
$\\$\color{Green}$\%$\-\ complete\-\ elliptic\-\ integral\-\ of\-\ the\-\ first\-\ kind
$\\$\color{BrickRed}K\-\ =\-\ elliptic\_integral(k,1);\color{Green}
$\\$\color{Green}$\%$\-\ complete\-\ elliptic\-\ integral\-\ of\-\ the\-\ second\-\ kind
$\\$\color{BrickRed}E\-\ =\-\ elliptic\_integral(k,2);\color{Green}
$\\$
$\\$
$\\$\color{Green}$\%$\-\ complete\-\ elliptic\-\ integral\-\ of\-\ the\-\ first\-\ kind
$\\$\color{BrickRed}K2\-\ =\-\ elliptic\_integral(1-k.\verb|^|2,1);\color{Green}
$\\$\color{Green}$\%$\-\ complete\-\ elliptic\-\ integral\-\ of\-\ the\-\ second\-\ kind
$\\$\color{BrickRed}E2\-\ =\-\ elliptic\_integral(1-k.\verb|^|2,2);\color{Green}
$\\$
$\\$
$\\$\color{Green}$\%$\-\ derivatives\-\ of\-\ K\-\ and\-\ E\-\ with\-\ resepct\-\ to\-\ k
$\\$\color{BrickRed}Kp\-\ =\-\ E./(k.*(1-k.\verb|^|2))-K./k;\color{Green}
$\\$\color{BrickRed}Ep\-\ =\-\ (E-K)./k;\color{Green}
$\\$
$\\$
$\\$\color{Green}$\%$\-\ derivatives\-\ of\-\ K2\-\ and\-\ E2\-\ with\-\ resepct\-\ to\-\ k
$\\$\color{BrickRed}K2p\-\ =\-\ -2*k.*(E2./((1-k.\verb|^|2).*(1-(1-k.\verb|^|2).\verb|^|2))-K2./(1-k.\verb|^|2));\color{Green}
$\\$\color{BrickRed}E2p\-\ =\-\ -2*k.*(E2-K2)./(1-k.\verb|^|2);\color{Green}
$\\$
$\\$
$\\$\color{Green}$\%$\-\ part\-\ of\-\ derivative\-\ of\-\ q(k)\-\ with\-\ respect\-\ to\-\ k
$\\$\color{BrickRed}T\-\ =\-\ -2*k.*K.*K2p-K2.*Kp;\color{Green}
$\\$
$\\$
$\\$\color{BrickRed}mx\-\ =\-\ max(sup(T));\color{Green}
$\\$\color{Green}$\%$\-\ find\-\ the\-\ left\-\ end\-\ of\-\ the\-\ k\-\ intervals
$\\$\color{BrickRed}kleft\-\ =\-\ min(inf(k));\color{Green}
$\\$\color{Green}$\%$\-\ find\-\ the\-\ right\-\ end\-\ of\-\ the\-\ k\-\ intervals
$\\$\color{BrickRed}kright\-\ =\-\ max(sup(k));\color{Green}
$\\$
$\\$
$\\$\color{Green}$\%$\-\ mx\-\ negative\-\ indicates\-\ that\-\ the\-\ derivative\-\ of\-\ q(k)\-\ with\-\ respect\-\ 
$\\$\color{Green}$\%$\-\ to\-\ k\-\ is\-\ strictly\-\ increasing\-\ on\-\ interval\-\ [kleft,kright]
$\\$
$\\$
$\\$\color{Black}\section{local\_startup.m}

\color{Green}\color{BrickRed}\color{NavyBlue}\-\ function\-\ \color{BrickRed}\-\ curr\_dir\-\ =\-\ local\_startup\color{Green}
$\\$
$\\$
$\\$\color{BrickRed}clear\-\ all;\-\ close\-\ all;\-\ beep\-\ off;\-\ clc;\-\ curr\_dir\-\ =\-\ cd;\color{Green}
$\\$\color{Green}$\%$$\%$\-\ startup\-\ commands
$\\$\color{BrickRed}cd('..');\color{Green}
$\\$\color{BrickRed}cd('..');\color{Green}
$\\$\color{BrickRed}startup('intlab','','start\-\ matlabpool','off');\color{Green}
$\\$\color{BrickRed}format\-\ long;\color{Green}
$\\$\color{BrickRed}clc;\color{Green}
$\\$\color{BrickRed}cd(curr\_dir);\color{Green}
$\\$
$\\$
$\\$\color{Green}$\%$\-\ display\-\ type
$\\$\color{BrickRed}intvalinit('DisplayMidRad');\color{Green}
$\\$\color{Green}$\%$\-\ intvalinit('DisplayInfSup');
$\\$\color{BrickRed}curr\_dir\-\ =\-\ cd;\color{Green}
$\\$\color{Black}\section{local\_startup\_batch.m}

\color{Green}\color{BrickRed}\color{NavyBlue}\-\ function\-\ \color{BrickRed}\-\ local\_startup\_batch(curr\_dir)\color{Green}
$\\$
$\\$
$\\$\color{Green}$\%$$\%$\-\ startup\-\ commands
$\\$\color{BrickRed}cd('..');\color{Green}
$\\$\color{BrickRed}cd('..');\color{Green}
$\\$\color{BrickRed}startup('intlab','','start\-\ matlabpool','off');\color{Green}
$\\$\color{BrickRed}format\-\ long;\color{Green}
$\\$
$\\$
$\\$\color{BrickRed}cd(curr\_dir);\color{Green}
$\\$
$\\$
$\\$\color{Green}$\%$\-\ display\-\ type
$\\$\color{Green}$\%$\-\ intvalinit('DisplayMidRad');
$\\$\color{BrickRed}intvalinit('DisplayInfSup');\color{Green}
$\\$\color{BrickRed}clc;\color{Green}
$\\$
$\\$
$\\$\color{Black}\section{lower\_bound.m}

\color{Green}\color{BrickRed}\color{NavyBlue}\-\ function\-\ \color{BrickRed}\-\ [dm,rho\_x]\-\ =\-\ lower\_bound(frac,q\_min,q\_max,num\_steps,steps)\color{Green}
$\\$\color{Green}$\%$\-\ frac\-\ \-\ must\-\ be\-\ strictly\-\ between\-\ 0\-\ and\-\ 1\-\ (for\-\ choosing\-\ rho\_x)
$\\$
$\\$
$\\$\color{Green}$\%$\-\ Get\-\ a\-\ lower\-\ bound\-\ on\-\ theta\-\ function\-\ that\-\ shows\-\ up\-\ in\-\ the\-\ denominator
$\\$\color{Green}$\%$\-\ in\-\ the\-\ definition\-\ of\-\ v(x)
$\\$
$\\$
$\\$\color{BrickRed}temp\-\ =\-\ lower\_bound\_local(q\_min,q\_max,num\_steps,frac,steps);\color{Green}
$\\$
$\\$
$\\$\color{BrickRed}dm\-\ =\-\ min(temp);\color{Green}
$\\$
$\\$
$\\$\color{BrickRed}pie\-\ =\-\ nm('pi');\color{Green}
$\\$\color{BrickRed}rho\_con\-\ =\-\ -frac*log(q\_max)/pie;\color{Green}
$\\$\color{BrickRed}rho\_x\-\ =\-\ rho\_con+sqrt(rho\_con\verb|^|2+1);\color{Green}
$\\$
$\\$
$\\$\color{Green}$\%$------------------------------------------------------------
$\\$\color{Green}$\%$\-\ lower\_bound\_local
$\\$\color{Green}$\%$------------------------------------------------------------
$\\$\color{BrickRed}\color{NavyBlue}\-\ function\-\ \color{BrickRed}\-\ out\-\ =\-\ lower\_bound\_local(q\_min,q\_max,num\_steps,frac,steps)\color{Green}
$\\$
$\\$
$\\$\color{BrickRed}pie\-\ =\-\ nm('pi');\color{Green}
$\\$\color{Green}$\%$\-\ parity\-\ and\-\ mirror\-\ symmetry\-\ of\-\ \color{Black}$\vartheta_1$ \color{Green}only\-\ requries\-\ we\-\ go
$\\$\color{Green}$\%$\-\ to\-\ pi/2\-\ instead\-\ of\-\ 2\-\ pi.
$\\$\color{BrickRed}half\-\ =\-\ nm('0.5');\color{Green}
$\\$\color{BrickRed}con\-\ =\-\ (half*pie/steps);\-\ \color{Green}
$\\$
$\\$
$\\$\color{BrickRed}ind\-\ =\-\ 0:1:steps;\color{Green}
$\\$\color{BrickRed}theta\-\ =\-\ nm(ind(1:end-1)*con,ind(2:end)*con);\color{Green}
$\\$
$\\$
$\\$\color{BrickRed}q\_del\-\ =\-\ (q\_max-q\_min)/num\_steps;\color{Green}
$\\$\color{BrickRed}psi\-\ =\-\ 0;\-\ ntilde\-\ =\-\ 0;\color{Green}
$\\$\color{BrickRed}fun\-\ =\-\ \@(q,x)(theta\_vec(q,psi,x,0,ntilde));\color{Green}
$\\$
$\\$
$\\$\color{BrickRed}out\-\ =\-\ 1000;\color{Green}
$\\$\color{BrickRed}\color{NavyBlue}\-\ for\-\ \color{BrickRed}\-\ j\-\ =\-\ 1:num\_steps-1\color{Green}
$\\$\color{BrickRed}\-\ \-\ \-\ \-\ \color{Green}
$\\$\color{Green}$\%$\-\ \-\ \-\ \-\ \-\ if\-\ mod(j,10)\-\ ==\-\ 0
$\\$\color{Green}$\%$\-\ \-\ \-\ \-\ \-\ \-\ \-\ \-\ \-\ fprintf('percent\-\ done\-\ =\-\ $\%$$\%$\-\ $\%$4.4g\textbackslash n',100*j/num\_steps);
$\\$\color{Green}$\%$\-\ \-\ \-\ \-\ \-\ end
$\\$\color{BrickRed}\-\ \-\ \-\ \-\ \color{Green}
$\\$\color{BrickRed}\-\ \-\ \-\ \-\ q\-\ =\-\ q\_min\-\ +\-\ nm((j-1)*q\_del,j*q\_del);\color{Green}
$\\$\color{BrickRed}\-\ \-\ \-\ \-\ rho\_x\-\ =\-\ get\_rho(q,frac);\color{Green}
$\\$\color{BrickRed}\-\ \-\ \-\ \-\ \-\ \-\ \-\ \-\ \color{Green}
$\\$\color{BrickRed}\-\ \-\ \-\ \-\ x\-\ =\-\ half*(rho\_x*exp(1i*theta)+exp(-1i*theta)/rho\_x);\-\ \color{Green}
$\\$\color{BrickRed}\-\ \-\ \-\ \-\ out\-\ =\-\ min(out,\-\ min(inf(abs(fun(q,x)))));\color{Green}
$\\$\color{BrickRed}\-\ \-\ \-\ \-\ \color{NavyBlue}\-\ if\-\ \color{BrickRed}\-\ out\-\ ==\-\ 0\color{Green}
$\\$\color{BrickRed}\-\ \-\ \-\ \-\ \-\ \-\ \-\ error('problem')\color{Green}
$\\$\color{BrickRed}\-\ \-\ \-\ \-\ \color{NavyBlue}\-\ end\-\ \color{BrickRed}\color{Green}
$\\$
$\\$
$\\$\color{BrickRed}\color{NavyBlue}\-\ end\-\ \color{BrickRed}\color{Green}
$\\$
$\\$
$\\$\color{BrickRed}\color{NavyBlue}\-\ if\-\ \color{BrickRed}\-\ isnan(out)\color{Green}
$\\$\color{BrickRed}\-\ \-\ \-\ \-\ error('NaN\-\ given\-\ \color{NavyBlue}\-\ for\-\ \color{BrickRed}\-\ error\-\ bound');\color{Green}
$\\$\color{BrickRed}\color{NavyBlue}\-\ end\-\ \color{BrickRed}\color{Green}
$\\$
$\\$
$\\$\color{Green}$\%$------------------------------------------------------------
$\\$\color{Green}$\%$\-\ get\-\ rho
$\\$\color{Green}$\%$------------------------------------------------------------
$\\$\color{BrickRed}\color{NavyBlue}\-\ function\-\ \color{BrickRed}\-\ rho\_x\-\ =\-\ get\_rho(q,frac)\color{Green}
$\\$
$\\$
$\\$\color{Green}$\%$
$\\$\color{Green}$\%$$\%$\-\ compute\-\ rho\-\ for\-\ ellipse\-\ used\-\ in\-\ getting\-\ Chebyshev\-\ bounds\-\ for\-\ x\-\ variable
$\\$\color{Green}$\%$
$\\$
$\\$
$\\$
$\\$\color{Black}
To interpolate in $x\in[-1,1]$ we need a bound on the integrands. The integrands 
are analytic in and on an ellipise that does not intersect zeros of
$\vartheta_1(\pi(x\pm i\omega')/2)$. Now the zeros of $\vartheta_1$ are the set
$\{m\pi +n\pi i\omega'/\omega | m,n\in \N \}$ .
Then from
\eq{
\frac{\pi}{2\omega} \left(\omega x\pm i\omega'\right) &= m\pi + n\pi i \omega'/\omega
}{}
we find
\eq{
\Im(x)< \omega'/\omega.
}{}
is required.

$\\$
Now if $0<c<\omega'/\omega$ is the height of the top of the ellipse $E_{\rho}$, then
\eq{
\frac{1}{2} ( \rho - 1/\rho) = c,
}{}
then
\eq{
\rho = c + \sqrt{c^2+1}.
}{}

$\\$
\color{Green}
$\\$
$\\$
$\\$
$\\$\color{BrickRed}pie\-\ =\-\ nm('pi');\color{Green}
$\\$\color{BrickRed}rho\_con\-\ =\-\ -frac*log(q)/pie;\color{Green}
$\\$\color{BrickRed}rho\_x\-\ =\-\ rho\_con+sqrt(rho\_con\verb|^|2+1);\color{Green}
$\\$
$\\$
$\\$
$\\$
$\\$
$\\$
$\\$
$\\$
$\\$\color{Black}\section{lower\_bound\_unstable.m}

\color{Green}\color{BrickRed}\color{NavyBlue}\-\ function\-\ \color{BrickRed}\-\ [dm,rho\_x,q\_min\_out,q\_max\_out]\-\ =\-\ lower\_bound\_unstable\color{Green}
$\\$
$\\$
$\\$
$\\$
$\\$\color{BrickRed}temp\-\ =\-\ zeros(5,1);\color{Green}
$\\$\color{BrickRed}frac\-\ =\-\ nm('0.9');\-\ \color{Green}$\%$\-\ strictly\-\ between\-\ 0\-\ and\-\ 1\-\ (for\-\ choosing\-\ rho\_x)
$\\$
$\\$
$\\$\color{BrickRed}q\_min\-\ =\-\ nm('0.1');\color{Green}
$\\$\color{BrickRed}q\_max\-\ =\-\ nm('0.15');\color{Green}
$\\$\color{BrickRed}q\_max\_out\-\ =\-\ q\_max;\color{Green}
$\\$\color{BrickRed}num\_steps\-\ =\-\ 100;\color{Green}
$\\$\color{BrickRed}temp(1)\-\ =\-\ lower\_bound\_local(q\_min,q\_max,num\_steps,frac);\color{Green}
$\\$
$\\$
$\\$\color{BrickRed}q\_min\-\ =\-\ nm('0.01');\color{Green}
$\\$\color{BrickRed}q\_max\-\ =\-\ nm('0.1');\color{Green}
$\\$\color{BrickRed}num\_steps\-\ =\-\ 100;\color{Green}
$\\$\color{BrickRed}temp(2)\-\ =\-\ lower\_bound\_local(q\_min,q\_max,num\_steps,frac);\color{Green}
$\\$
$\\$
$\\$\color{BrickRed}q\_min\-\ =\-\ nm('0.001');\color{Green}
$\\$\color{BrickRed}q\_max\-\ =\-\ nm('0.01');\color{Green}
$\\$\color{BrickRed}num\_steps\-\ =\-\ 100;\color{Green}
$\\$\color{BrickRed}temp(3)\-\ =\-\ lower\_bound\_local(q\_min,q\_max,num\_steps,frac);\color{Green}
$\\$
$\\$
$\\$\color{BrickRed}q\_min\-\ =\-\ nm('0.0001');\color{Green}
$\\$\color{BrickRed}q\_max\-\ =\-\ nm('0.001');\color{Green}
$\\$\color{BrickRed}num\_steps\-\ =\-\ 100;\color{Green}
$\\$\color{BrickRed}temp(4)\-\ =\-\ lower\_bound\_local(q\_min,q\_max,num\_steps,frac);\color{Green}
$\\$
$\\$
$\\$\color{BrickRed}q\_min\-\ =\-\ nm('1e-7');\color{Green}
$\\$\color{BrickRed}q\_max\-\ =\-\ nm('0.0001');\color{Green}
$\\$\color{BrickRed}q\_min\_out\-\ =\-\ q\_min;\color{Green}
$\\$\color{BrickRed}num\_steps\-\ =\-\ 100;\color{Green}
$\\$\color{BrickRed}temp(5)\-\ =\-\ lower\_bound\_local(q\_min,q\_max,num\_steps,frac);\color{Green}
$\\$
$\\$
$\\$\color{BrickRed}dm\-\ =\-\ min(temp);\color{Green}
$\\$
$\\$
$\\$\color{BrickRed}pie\-\ =\-\ nm('pi');\color{Green}
$\\$\color{BrickRed}rho\_con\-\ =\-\ -frac*log(q\_max)/pie;\color{Green}
$\\$\color{BrickRed}rho\_x\-\ =\-\ rho\_con+sqrt(rho\_con\verb|^|2+1);\color{Green}
$\\$
$\\$
$\\$\color{Green}$\%$------------------------------------------------------------
$\\$\color{Green}$\%$\-\ lower\_bound\_local
$\\$\color{Green}$\%$------------------------------------------------------------
$\\$\color{BrickRed}\color{NavyBlue}\-\ function\-\ \color{BrickRed}\-\ out\-\ =\-\ lower\_bound\_local(q\_min,q\_max,num\_steps,frac)\color{Green}
$\\$
$\\$
$\\$
$\\$
$\\$\color{BrickRed}steps\-\ =\-\ 8000;\color{Green}
$\\$\color{BrickRed}pie\-\ =\-\ nm('pi');\color{Green}
$\\$\color{Green}$\%$\-\ parity\-\ and\-\ mirror\-\ symmetry\-\ of\-\ \color{Black}$\vartheta_1$ \color{Green}only\-\ requries\-\ we\-\ go
$\\$\color{Green}$\%$\-\ to\-\ pi/2\-\ instead\-\ of\-\ 2\-\ pi.
$\\$\color{BrickRed}half\-\ =\-\ nm('0.5');\color{Green}
$\\$\color{BrickRed}con\-\ =\-\ (half*pie/steps);\-\ \color{Green}
$\\$
$\\$
$\\$\color{BrickRed}ind\-\ =\-\ 0:1:steps;\color{Green}
$\\$\color{BrickRed}theta\-\ =\-\ nm(ind(1:end-1)*con,ind(2:end)*con);\color{Green}
$\\$
$\\$
$\\$\color{BrickRed}q\_del\-\ =\-\ (q\_max-q\_min)/num\_steps;\color{Green}
$\\$\color{BrickRed}psi\-\ =\-\ 0;\-\ ntilde\-\ =\-\ 0;\color{Green}
$\\$\color{BrickRed}fun\-\ =\-\ \@(q,x)(theta\_vec(q,psi,x,0,ntilde));\color{Green}
$\\$
$\\$
$\\$\color{BrickRed}out\-\ =\-\ 1000;\color{Green}
$\\$\color{BrickRed}\color{NavyBlue}\-\ for\-\ \color{BrickRed}\-\ j\-\ =\-\ 1:num\_steps-1\color{Green}
$\\$\color{BrickRed}\-\ \-\ \-\ \-\ \color{Green}
$\\$\color{BrickRed}\-\ \-\ \-\ \-\ q\-\ =\-\ q\_min\-\ +\-\ nm((j-1)*q\_del,j*q\_del);\color{Green}
$\\$\color{BrickRed}\-\ \-\ \-\ \-\ rho\_x\-\ =\-\ get\_rho(q,frac);\color{Green}
$\\$\color{BrickRed}\-\ \-\ \-\ \-\ \-\ \-\ \-\ \-\ \color{Green}
$\\$\color{BrickRed}\-\ \-\ \-\ \-\ x\-\ =\-\ half*(rho\_x*exp(1i*theta)+exp(-1i*theta)/rho\_x);\-\ \color{Green}
$\\$\color{BrickRed}\-\ \-\ \-\ \-\ out\-\ =\-\ min(out,\-\ min(inf(abs(fun(q,x)))));\color{Green}
$\\$
$\\$
$\\$\color{BrickRed}\color{NavyBlue}\-\ end\-\ \color{BrickRed}\color{Green}
$\\$
$\\$
$\\$\color{Green}$\%$\-\ check\-\ if\-\ lower\-\ bound\-\ is\-\ not\-\ helpful
$\\$\color{BrickRed}\color{NavyBlue}\-\ if\-\ \color{BrickRed}\-\ out\-\ ==\-\ 0\color{Green}
$\\$\color{BrickRed}\-\ \-\ \-\ \-\ error('lower\-\ bound\-\ is\-\ 0');\color{Green}
$\\$\color{BrickRed}\color{NavyBlue}\-\ end\-\ \color{BrickRed}\color{Green}
$\\$
$\\$
$\\$\color{Green}$\%$\-\ check\-\ that\-\ out\-\ is\-\ not\-\ NaN
$\\$\color{BrickRed}\color{NavyBlue}\-\ if\-\ \color{BrickRed}\-\ ~(out\-\ $<$\-\ 10)\color{Green}
$\\$\color{BrickRed}\-\ \-\ \-\ \-\ error('error\-\ apparent')\color{Green}
$\\$\color{BrickRed}\color{NavyBlue}\-\ end\-\ \color{BrickRed}\color{Green}
$\\$
$\\$
$\\$
$\\$
$\\$\color{Green}$\%$------------------------------------------------------------
$\\$\color{Green}$\%$\-\ get\-\ rho
$\\$\color{Green}$\%$------------------------------------------------------------
$\\$\color{BrickRed}\color{NavyBlue}\-\ function\-\ \color{BrickRed}\-\ rho\_x\-\ =\-\ get\_rho(q,frac)\color{Green}
$\\$
$\\$
$\\$\color{Green}$\%$
$\\$\color{Green}$\%$$\%$\-\ compute\-\ rho\-\ for\-\ ellipse\-\ used\-\ in\-\ getting\-\ Chebyshev\-\ bounds\-\ for\-\ x\-\ variable
$\\$\color{Green}$\%$
$\\$
$\\$
$\\$
$\\$\color{Black}
To interpolate in $x\in[-1,1]$ we need a bound on the integrands. The integrands 
are analytic in and on an ellipise that does not intersect zeros of
$\vartheta_1(\pi(x\pm i\omega')/2)$. Now the zeros of $\vartheta_1$ are the set
$\{m\pi +n\pi i\omega'/\omega | m,n\in \N \}$ .
Then from
\eq{
\frac{\pi}{2\omega} \left(\omega x\pm i\omega'\right) &= m\pi + n\pi i \omega'/\omega
}{}
we find
\eq{
\Im(x)< \omega'/\omega.
}{}
is required.

$\\$
Now if $0<c<\omega'/\omega$ is the height of the top of the ellipse $E_{\rho}$, then
\eq{
\frac{1}{2} ( \rho - 1/\rho) = c,
}{}
then
\eq{
\rho = c + \sqrt{c^2+1}.
}{}

$\\$
\color{Green}
$\\$
$\\$
$\\$
$\\$\color{BrickRed}pie\-\ =\-\ nm('pi');\color{Green}
$\\$\color{BrickRed}rho\_con\-\ =\-\ -frac*log(q)/pie;\color{Green}
$\\$\color{BrickRed}rho\_x\-\ =\-\ rho\_con+sqrt(rho\_con\verb|^|2+1);\color{Green}
$\\$
$\\$
$\\$
$\\$
$\\$
$\\$
$\\$
$\\$
$\\$\color{Black}\section{lower\_instability\_interpolation.m}

\color{Green}\color{Green}$\%$\-\ verify\-\ instability\-\ lower
$\\$\color{BrickRed}clear\-\ all;\-\ curr\_dir\-\ =\-\ local\_startup;\color{Green}
$\\$
$\\$
$\\$\color{BrickRed}pie\-\ =\-\ nm('pi');\color{Green}
$\\$
$\\$
$\\$\color{BrickRed}tstart\-\ =\-\ tic;\color{Green}
$\\$\color{BrickRed}[dm,rho\_x,q\_min\_out,q\_max\_out]\-\ =\-\ lower\_bound\_unstable;\-\ \color{Green}
$\\$\color{BrickRed}d.dm\_time\-\ =\-\ toc(tstart);\color{Green}
$\\$\color{BrickRed}d.dm\-\ =\-\ dm;\color{Green}
$\\$\color{BrickRed}d.rho\_x\-\ =\-\ rho\_x;\color{Green}
$\\$
$\\$
$\\$\color{Green}$\%$
$\\$\color{Green}$\%$\-\ choices\-\ of\-\ q
$\\$\color{Green}$\%$
$\\$
$\\$
$\\$\color{BrickRed}num\-\ =\-\ 7;\color{Green}
$\\$\color{BrickRed}KL\-\ =\-\ zeros(num,1);\color{Green}
$\\$\color{BrickRed}RK\-\ =\-\ zeros(num,1);\color{Green}
$\\$\color{BrickRed}MN\-\ =\-\ zeros(num,1);\color{Green}
$\\$\color{BrickRed}time\-\ =\-\ zeros(num,1);\color{Green}
$\\$
$\\$
$\\$\color{BrickRed}cnt\-\ =\-\ 0;\color{Green}
$\\$
$\\$
$\\$\color{BrickRed}cnt\-\ =\-\ cnt\-\ +\-\ 1;\color{Green}
$\\$\color{BrickRed}st{cnt}{1}\-\ =\-\ nm('0.122');\-\ \color{Green}$\%$\-\ qmin
$\\$\color{BrickRed}st{cnt}{2}\-\ =\-\ nm('0.139');\-\ \color{Green}$\%$\-\ qmax
$\\$\color{BrickRed}st{cnt}{3}\-\ =\-\ [linspace(inf(nm('0.93')),sup(nm('0.9422')),100),sup(nm('0.9423'))];\-\ \color{Green}$\%$\-\ k\-\ points
$\\$\color{BrickRed}st{cnt}{4}\-\ =\-\ 1e-4;\color{Green}
$\\$
$\\$
$\\$\color{BrickRed}cnt\-\ =\-\ cnt\-\ +\-\ 1;\color{Green}
$\\$\color{BrickRed}st{cnt}{1}\-\ =\-\ nm('0.1');\color{Green}
$\\$\color{BrickRed}st{cnt}{2}\-\ =\-\ nm('0.2');\color{Green}
$\\$\color{BrickRed}st{cnt}{3}\-\ =\-\ [\-\ 0.899,\-\ 0.93];\color{Green}
$\\$\color{BrickRed}st{cnt}{4}\-\ =\-\ 1e-4;\color{Green}
$\\$
$\\$
$\\$\color{BrickRed}cnt\-\ =\-\ cnt\-\ +\-\ 1;\color{Green}
$\\$\color{BrickRed}st{cnt}{1}\-\ =\-\ nm('0.05');\color{Green}
$\\$\color{BrickRed}st{cnt}{2}\-\ =\-\ nm('0.11');\color{Green}
$\\$\color{BrickRed}st{cnt}{3}\-\ =\-\ [\-\ 0.75,\-\ 0.8\-\ 0.9];\color{Green}
$\\$\color{BrickRed}st{cnt}{4}\-\ =\-\ 1e-4;\color{Green}
$\\$
$\\$
$\\$\color{BrickRed}cnt\-\ =\-\ cnt\-\ +\-\ 1;\color{Green}
$\\$\color{BrickRed}st{cnt}{1}\-\ =\-\ nm('0.01');\color{Green}
$\\$\color{BrickRed}st{cnt}{2}\-\ =\-\ nm('0.06');\color{Green}
$\\$\color{BrickRed}st{cnt}{3}\-\ =linspace(0.4,0.751,10);\-\ \color{Green}
$\\$\color{BrickRed}st{cnt}{4}\-\ =\-\ 1e-4;\color{Green}
$\\$
$\\$
$\\$\color{BrickRed}cnt\-\ =\-\ cnt\-\ +\-\ 1;\color{Green}
$\\$\color{BrickRed}st{cnt}{1}\-\ =\-\ nm('0.005');\color{Green}
$\\$\color{BrickRed}st{cnt}{2}\-\ =\-\ nm('0.011');\color{Green}
$\\$\color{BrickRed}st{cnt}{3}\-\ =\-\ [0.3\-\ 0.35\-\ 0.4\-\ ];\color{Green}
$\\$\color{BrickRed}st{cnt}{4}\-\ =\-\ 1e-4;\color{Green}
$\\$
$\\$
$\\$\color{BrickRed}cnt\-\ =\-\ cnt\-\ +\-\ 1;\color{Green}
$\\$\color{BrickRed}st{cnt}{1}\-\ =\-\ nm('0.003');\color{Green}
$\\$\color{BrickRed}st{cnt}{2}\-\ =\-\ nm('0.006');\color{Green}
$\\$\color{BrickRed}st{cnt}{3}\-\ =\-\ [0.24\-\ 0.25\-\ 0.26\-\ 0.27\-\ 0.28\-\ 0.29\-\ 0.3\-\ ];\color{Green}
$\\$\color{BrickRed}st{cnt}{4}\-\ =\-\ 1e-4;\color{Green}
$\\$
$\\$
$\\$\color{BrickRed}cnt\-\ =\-\ cnt\-\ +\-\ 1;\color{Green}
$\\$\color{BrickRed}st{cnt}{1}\-\ =\-\ nm('0.002');\color{Green}
$\\$\color{BrickRed}st{cnt}{2}\-\ =\-\ nm('0.0038');\color{Green}
$\\$\color{BrickRed}st{cnt}{3}\-\ =\-\ [\-\ \-\ 0.2\-\ 0.21\-\ 0.22\-\ 0.23\-\ 0.24\-\ ];\color{Green}
$\\$\color{BrickRed}st{cnt}{4}\-\ =\-\ 1e-4;\color{Green}
$\\$
$\\$
$\\$\color{BrickRed}total\_time\_start\-\ =\-\ tic;\color{Green}
$\\$\color{BrickRed}\color{NavyBlue}\-\ for\-\ \color{BrickRed}\-\ ind\-\ =\-\ 1:num\color{Green}
$\\$\color{BrickRed}\-\ \-\ \-\ \-\ \color{Green}
$\\$\color{BrickRed}\-\ \-\ \-\ \-\ fprintf('Lower\-\ instability\-\ interpolation\-\ -\-\ percent\-\ done:\-\ $\%$3.3g\textbackslash n',100*(ind-1)/7);\color{Green}
$\\$\color{BrickRed}\-\ \-\ \-\ \-\ \color{Green}
$\\$\color{BrickRed}\-\ \-\ \-\ \-\ tstart\-\ =\-\ tic;\color{Green}
$\\$\color{BrickRed}\-\ \-\ \-\ \-\ \color{Green}$\%$\-\ interpolate
$\\$\color{BrickRed}\-\ \-\ \-\ \-\ d.q\_min\-\ =\-\ st{ind}{1};\color{Green}
$\\$\color{BrickRed}\-\ \-\ \-\ \-\ d.q\_max\-\ =\-\ st{ind}{2};\color{Green}
$\\$\color{BrickRed}\-\ \-\ \-\ \-\ d\-\ =\-\ instability\_interpolation(d,st{ind}{1},st{ind}{2});\color{Green}
$\\$\color{BrickRed}\-\ \-\ \-\ \-\ \color{Green}
$\\$\color{BrickRed}\-\ \-\ \-\ \-\ file\_name\-\ =\-\ ['d\_lower\_unstable\_',num2str(ind)];\color{Green}
$\\$\color{BrickRed}\-\ \-\ \-\ \-\ saveit(curr\_dir,'interval\_arithmetic',d,file\_name,'data\_final');\color{Green}
$\\$\color{BrickRed}\-\ \-\ \-\ \-\ \color{Green}
$\\$\color{BrickRed}\-\ \-\ \-\ \-\ \color{Green}$\%$\-\ divide\-\ up\-\ into\-\ k\-\ intervals
$\\$\color{BrickRed}\-\ \-\ \-\ \-\ kpts\-\ =\-\ st{ind}{3};\color{Green}
$\\$\color{BrickRed}\-\ \-\ \-\ \-\ left\_k\-\ =\-\ zeros(length(kpts)-1,1);\color{Green}
$\\$\color{BrickRed}\-\ \-\ \-\ \-\ right\_k\-\ =\-\ zeros(length(kpts)-1,1);\color{Green}
$\\$\color{BrickRed}\-\ \-\ \-\ \-\ min\_quot\-\ =\-\ zeros(length(kpts)-1,1);\color{Green}
$\\$
$\\$
$\\$\color{BrickRed}\-\ \-\ \-\ \-\ \color{Green}$\%$\-\ verify\-\ stability\-\ 
$\\$\color{BrickRed}\-\ \-\ \-\ \-\ eps\-\ =\-\ st{ind}{4};\color{Green}
$\\$\color{BrickRed}\-\ \-\ \-\ \-\ \color{NavyBlue}\-\ for\-\ \color{BrickRed}\-\ j\-\ =\-\ 1:length(kpts)-1\color{Green}
$\\$\color{BrickRed}\-\ \-\ \-\ \-\ \-\ \-\ \-\ \-\ k\-\ =\-\ linspace(kpts(j)-eps,kpts(j+1)+eps,1000);\color{Green}
$\\$\color{BrickRed}\-\ \-\ \-\ \-\ \-\ \-\ \-\ \-\ [left\_k(j),right\_k(j),min\_quot(j)]\-\ =\-\ verify\_instability\_lower(d,k);\color{Green}
$\\$\color{BrickRed}\-\ \-\ \-\ \-\ \color{NavyBlue}\-\ end\-\ \color{BrickRed}\color{Green}
$\\$
$\\$
$\\$\color{BrickRed}\-\ \-\ \-\ \-\ \color{Green}$\%$\-\ check\-\ for\-\ gap\-\ in\-\ inner\-\ k\-\ intervals
$\\$\color{BrickRed}\-\ \-\ \-\ \-\ \color{NavyBlue}\-\ for\-\ \color{BrickRed}\-\ j\-\ =\-\ 1:length(left\_k)-1\color{Green}
$\\$\color{BrickRed}\-\ \-\ \-\ \-\ \-\ \-\ \-\ \-\ \color{NavyBlue}\-\ if\-\ \color{BrickRed}\-\ right\_k(j)\-\ $<$\-\ left\_k(j+1)\color{Green}
$\\$\color{BrickRed}\-\ \-\ \-\ \-\ \-\ \-\ \-\ \-\ \-\ \-\ \-\ \-\ error('gap\-\ present');\color{Green}
$\\$\color{BrickRed}\-\ \-\ \-\ \-\ \-\ \-\ \-\ \-\ \color{NavyBlue}\-\ end\-\ \color{BrickRed}\color{Green}
$\\$\color{BrickRed}\-\ \-\ \-\ \-\ \color{NavyBlue}\-\ end\-\ \color{BrickRed}\color{Green}
$\\$
$\\$
$\\$\color{BrickRed}\-\ \-\ \-\ \-\ \color{Green}$\%$\-\ record\-\ data
$\\$\color{BrickRed}\-\ \-\ \-\ \-\ KL(ind)\-\ =\-\ left\_k(1);\color{Green}
$\\$\color{BrickRed}\-\ \-\ \-\ \-\ RK(ind)\-\ =\-\ right\_k(end);\color{Green}
$\\$\color{BrickRed}\-\ \-\ \-\ \-\ MN(ind)\-\ =\-\ min(min\_quot);\color{Green}
$\\$
$\\$
$\\$\color{BrickRed}\-\ \-\ \-\ \-\ \color{Green}$\%$\-\ error\-\ check\-\ for\-\ gap\-\ between\-\ outer\-\ k\-\ intervals
$\\$\color{BrickRed}\-\ \-\ \-\ \-\ \color{NavyBlue}\-\ for\-\ \color{BrickRed}\-\ j\-\ =\-\ 1:ind-1\color{Green}
$\\$\color{BrickRed}\-\ \-\ \-\ \-\ \-\ \-\ \-\ \-\ \color{NavyBlue}\-\ if\-\ \color{BrickRed}\-\ RK(j+1)\-\ $<$\-\ KL(j)\color{Green}
$\\$\color{BrickRed}\-\ \-\ \-\ \-\ \-\ \-\ \-\ \-\ \-\ \-\ \-\ \-\ error('gap\-\ present');\color{Green}
$\\$\color{BrickRed}\-\ \-\ \-\ \-\ \-\ \-\ \-\ \-\ \color{NavyBlue}\-\ end\-\ \color{BrickRed}\color{Green}
$\\$\color{BrickRed}\-\ \-\ \-\ \-\ \color{NavyBlue}\-\ end\-\ \color{BrickRed}\color{Green}
$\\$\color{BrickRed}\-\ \-\ \-\ \-\ \color{Green}
$\\$\color{BrickRed}\-\ \-\ \-\ \-\ \color{Green}$\%$\-\ record\-\ time
$\\$\color{BrickRed}\-\ \-\ \-\ \-\ time(ind)\-\ =\-\ toc(tstart);\color{Green}
$\\$\color{BrickRed}\-\ \-\ \-\ \-\ \color{Green}
$\\$\color{BrickRed}\color{NavyBlue}\-\ end\-\ \color{BrickRed}\color{Green}
$\\$
$\\$
$\\$\color{BrickRed}data.total\_time\-\ =\-\ toc(total\_time\_start);\color{Green}
$\\$
$\\$
$\\$\color{Green}$\%$$\%$$\%$$\%$$\%$$\%$$\%$$\%$$\%$$\%$$\%$$\%$$\%$$\%$$\%$$\%$$\%$$\%$$\%$$\%$$\%$$\%$$\%$$\%$
$\\$
$\\$
$\\$\color{BrickRed}Left\_K\-\ =\-\ KL(end);\color{Green}
$\\$\color{BrickRed}Right\_K\-\ =\-\ RK(1);\color{Green}
$\\$\color{BrickRed}MNtotal\-\ =\-\ min(MN);\color{Green}
$\\$
$\\$
$\\$\color{BrickRed}file\_name\-\ =\-\ 'instability\_result\_lower';\color{Green}
$\\$\color{BrickRed}data.Left\_K\-\ =\-\ Left\_K;\color{Green}
$\\$\color{BrickRed}data.Right\_K\-\ =\-\ Right\_K;\color{Green}
$\\$\color{BrickRed}data.KL\-\ =\-\ KL;\color{Green}
$\\$\color{BrickRed}data.RK\-\ =\-\ RK;\color{Green}
$\\$\color{BrickRed}data.MNtotal\-\ =\-\ MNtotal;\color{Green}
$\\$\color{BrickRed}data.MN\-\ =\-\ MN;\color{Green}
$\\$\color{BrickRed}data.st\-\ =\-\ st;\color{Green}
$\\$\color{BrickRed}data.time\-\ =\-\ time;\color{Green}
$\\$\color{BrickRed}saveit(curr\_dir,'interval\_arithmetic',data,file\_name,'data\_final');\color{Green}
$\\$
$\\$
$\\$
$\\$
$\\$
$\\$
$\\$
$\\$
$\\$
$\\$
$\\$
$\\$
$\\$
$\\$
$\\$\color{Black}\section{numer.m}

\color{Green}\color{BrickRed}\color{NavyBlue}\-\ function\-\ \color{BrickRed}\-\ out\-\ =\-\ numer(Nx,err,q,psi,ntilde)\color{Green}
$\\$\color{Green}$\%$\-\ out\-\ =\-\ numer(Nx,err,q,psi,ntilde)
$\\$\color{Green}$\%$
$\\$\color{Green}$\%$\-\ 
$\\$
$\\$
$\\$
$\\$\color{Black}

 Definition of $\lambda_1$

 \eq{
\lambda_1 
&= \frac{\int_{-1}^{1} \left[\frac{\partial}{\partial x}v(\omega x)+\frac{1}{\omega^2}\frac{\partial^3}{\partial x^3}v(\omega x)\right]\frac{\partial^2}{\partial x^2}\bar v(\omega x)dx}{\omega^2\int_{-1}^{1} v(\omega x)\frac{\partial}{\partial x}\bar v(\omega x)dx}.
}{\label{innerprod2}}

Now
\eq{
v(x) &= w(x)e^{\gamma x}\\
v'(x) & = w'(x) e^{\gamma x} + \gamma w(x)e^{\gamma x}\\
v''(x) & = w''(x) e^{\gamma x} + 2\gamma w'(x) e^{\gamma x} + \gamma^2 w(x) e^{\gamma x}\\
v'''(x) & = w'''(x) e^{\gamma x} +3\gamma w''(x) e^{\gamma x} + 3\gamma^2 w'(x) e^{\gamma x} + \gamma^3 w(x) e^{\gamma x}.
}{\notag}

Then
\eq{
v(\omega x) \bar v'(\omega x) &= w(\omega x) \bar w'(\omega x) + (i\xi) w(\omega x) \bar w(\omega x),\\
v(\omega x) \bar v''(\omega x) &= \sum_{n=0}^{3} \gamma^n c_n(x),\\
v'''(\omega x) \bar v''(\omega x) &= \sum_{n=0}^{5} \gamma^n h_n(x),
}{}
where

\eq{
c_0(x)&:= w'(\omega x) \bar w''(\omega x)\\
c_1(x)&:= w(\omega x) \bar w''(\omega x)-2w'(\omega x)\bar w'(\omega x)\\
c_2(x)&:= w'(\omega x)\bar w(\omega x) -2w(\omega x)\bar w'(\omega x)\\
c_3(x)&:= w(\omega x)\bar w(\omega x)\\
h_0(x)&:= w'''(\omega x)\bar w''(\omega x)\\
h_1(x)&:= 3w''(\omega x)\bar w''(\omega x)-2w'''(\omega x)\bar w'(\omega x)\\
h_2(x)&:= w'''(\omega x)\bar w(\omega x)-6w''(\omega x)\bar w'(\omega x)+3 w'(\omega x)\bar w''(\omega x)\\
h_3(x)&:= 3w''(\omega x)\bar w(\omega x)-6w'(\omega x)\bar w'(\omega x)+w(\omega x)\bar w''(\omega x)\\
h_4(x)&:= 3w'(\omega x)\bar w(\omega x)-2w(\omega x)\bar w'(\omega x)\\
h_5(x)&:= w(\omega x)\bar w(\omega x).
}{}

$\\$
\color{Green}
$\\$
$\\$\color{Green}$\%$$\%$
$\\$
$\\$
$\\$\color{Green}$\%$\-\ -----------------------------------------------------------
$\\$\color{Green}$\%$\-\ Get\-\ theta\-\ and\-\ derivatives
$\\$\color{Green}$\%$\-\ -----------------------------------------------------------
$\\$
$\\$
$\\$\color{Green}$\%$\-\ constants
$\\$\color{BrickRed}pie\-\ =\-\ nm('pi');\color{Green}
$\\$\color{BrickRed}half\-\ =\-\ nm(1)/2;\color{Green}
$\\$
$\\$
$\\$\color{Green}$\%$\-\ interpolation\-\ points\-\ in\-\ x\-\ for\-\ integration
$\\$\color{BrickRed}theta\-\ =\-\ ((0:1:Nx)+half)*pie/(Nx+1);\-\ \color{Green}
$\\$\color{BrickRed}x\-\ =\-\ cos(theta);\color{Green}
$\\$
$\\$
$\\$\color{Green}$\%$\-\ denominator\-\ term\-\ (alpha\-\ =\-\ 0)
$\\$\color{BrickRed}J\-\ =\-\ theta\_vec(q,0,x,3,0);\color{Green}
$\\$\color{BrickRed}J\-\ =\-\ repmat(J,[1\-\ length(psi)]);\color{Green}
$\\$\color{BrickRed}\-\ \-\ \color{Green}
$\\$\color{BrickRed}E0\-\ =\-\ 1./J(:,:,1);\color{Green}
$\\$\color{BrickRed}E1\-\ =\-\ -J(:,:,2)./J(:,:,1).\verb|^|2;\color{Green}
$\\$\color{BrickRed}E2\-\ =\-\ 2*J(:,:,2).\verb|^|2./J(:,:,1).\verb|^|3-J(:,:,3)./J(:,:,1).\verb|^|2;\color{Green}
$\\$\color{BrickRed}E3\-\ =\-\ -6*J(:,:,2).\verb|^|3./J(:,:,1).\verb|^|4+6*J(:,:,2).*J(:,:,3)./J(:,:,1).\verb|^|3-J(:,:,4)./J(:,:,1).\verb|^|2;\color{Green}
$\\$
$\\$
$\\$\color{BrickRed}B0\-\ =\-\ E0.*E0;\color{Green}
$\\$\color{BrickRed}B1\-\ =\-\ 2*E0.*E1;\color{Green}
$\\$\color{BrickRed}B2\-\ =\-\ 2*(E0.*E2+E1.*E1);\color{Green}
$\\$\color{BrickRed}B3\-\ =\-\ 2*(E3.*E0+3*E1.*E2);\color{Green}
$\\$
$\\$
$\\$\color{Green}$\%$\-\ numerator\-\ term\-\ (alpha\-\ =\-\ ntilde*omega\-\ +\-\ 1i*psi*omega')
$\\$\color{BrickRed}L\-\ =\-\ theta\_vec(q,psi,x,4,ntilde);\color{Green}
$\\$
$\\$
$\\$\color{BrickRed}A0\-\ =\-\ L(:,:,1).*L(:,:,1);\color{Green}
$\\$\color{BrickRed}A1\-\ =\-\ 2*L(:,:,1).*L(:,:,2.);\color{Green}
$\\$\color{BrickRed}A2\-\ =\-\ 2*(L(:,:,1).*L(:,:,3)+L(:,:,2).*L(:,:,2));\color{Green}
$\\$\color{BrickRed}A3\-\ =\-\ 2*(L(:,:,4).*L(:,:,1)+3*L(:,:,2).*L(:,:,3));\color{Green}
$\\$
$\\$
$\\$\color{BrickRed}w0\-\ =\-\ A0.*B0;\color{Green}
$\\$\color{BrickRed}w1\-\ =\-\ A0.*B1+A1.*B0;\color{Green}
$\\$\color{BrickRed}w2\-\ =\-\ A0.*B2+2*A1.*B1+A2.*B0;\color{Green}
$\\$\color{BrickRed}w3\-\ =\-\ A0.*B3+3*A1.*B2+3*A2.*B1+\-\ A3.*B0;\color{Green}
$\\$
$\\$
$\\$\color{Green}$\%$\-\ -----------------------------------------------------------
$\\$\color{Green}$\%$\-\ functions\-\ in\-\ expansions
$\\$\color{Green}$\%$\-\ -----------------------------------------------------------
$\\$\color{BrickRed}\-\ \color{Green}
$\\$\color{Green}$\%$\-\ \color{Black}$v(\omega x) \bar v''(\omega x) = \sum_{n=0}^{3} \gamma^n c_n(x)$ \color{Green}
$\\$\color{BrickRed}c0\-\ =\-\ w1.*conj(w2);\color{Green}
$\\$\color{BrickRed}c1\-\ =\-\ w0.*conj(w2)-2*w1.*conj(w1);\color{Green}
$\\$\color{BrickRed}c2\-\ =\-\ w1.*conj(w0)-2*w0.*conj(w1);\color{Green}
$\\$\color{BrickRed}c3\-\ =\-\ w0.*conj(w0);\color{Green}
$\\$
$\\$
$\\$\color{Green}$\%$\-\ \color{Black}$v'''(\omega x) \bar v''(\omega x) = \sum_{n=0}^{5} \gamma^n h_n(x)$ \color{Green}
$\\$\color{BrickRed}h0\-\ =\-\ w3.*conj(w2);\color{Green}
$\\$\color{BrickRed}h1\-\ =\-\ 3*w2.*conj(w2)-2*w3.*conj(w1);\color{Green}
$\\$\color{BrickRed}h2\-\ =\-\ w3.*conj(w0)-6*w2.*conj(w1)+3*w1.*conj(w2);\color{Green}
$\\$\color{BrickRed}h3\-\ =\-\ 3*w2.*conj(w0)-6*w1.*conj(w1)+w0.*conj(w2);\color{Green}
$\\$\color{BrickRed}h4\-\ =\-\ 3*w1.*conj(w0)-2*w0.*conj(w1);\color{Green}
$\\$\color{BrickRed}h5\-\ =\-\ w0.*conj(w0);\color{Green}
$\\$
$\\$
$\\$\color{Green}$\%$\-\ Chebyshev\-\ polynomials\-\ evaluated\-\ at\-\ the\-\ interpolation\-\ points
$\\$\color{BrickRed}Tx\-\ =\-\ cos(theta.'*(0:2:Nx));\color{Green}
$\\$
$\\$
$\\$\color{Green}$\%$\-\ Chebyshev\-\ coefficients
$\\$\color{BrickRed}cf\_c0\-\ =\-\ 2*c0.'*Tx/(Nx+1);\color{Green}
$\\$\color{BrickRed}cf\_c0(:,1)\-\ =\-\ cf\_c0(:,1)/2;\color{Green}
$\\$
$\\$
$\\$\color{BrickRed}cf\_c1\-\ =\-\ 2*c1.'*Tx/(Nx+1);\color{Green}
$\\$\color{BrickRed}cf\_c1(:,1)\-\ =\-\ cf\_c1(:,1)/2;\color{Green}
$\\$
$\\$
$\\$\color{BrickRed}cf\_c2\-\ =\-\ 2*c2.'*Tx/(Nx+1);\color{Green}
$\\$\color{BrickRed}cf\_c2(:,1)\-\ =\-\ cf\_c2(:,1)/2;\color{Green}
$\\$
$\\$
$\\$\color{BrickRed}cf\_c3\-\ =\-\ 2*c3.'*Tx/(Nx+1);\color{Green}
$\\$\color{BrickRed}cf\_c3(:,1)\-\ =\-\ cf\_c3(:,1)/2;\color{Green}
$\\$
$\\$
$\\$\color{BrickRed}cf\_h0\-\ =\-\ 2*h0.'*Tx/(Nx+1);\color{Green}
$\\$\color{BrickRed}cf\_h0(:,1)\-\ =\-\ cf\_h0(:,1)/2;\color{Green}
$\\$
$\\$
$\\$\color{BrickRed}cf\_h1\-\ =\-\ 2*h1.'*Tx/(Nx+1);\color{Green}
$\\$\color{BrickRed}cf\_h1(:,1)\-\ =\-\ cf\_h1(:,1)/2;\color{Green}
$\\$
$\\$
$\\$\color{BrickRed}cf\_h2\-\ =\-\ 2*h2.'*Tx/(Nx+1);\color{Green}
$\\$\color{BrickRed}cf\_h2(:,1)\-\ =\-\ cf\_h2(:,1)/2;\color{Green}
$\\$
$\\$
$\\$\color{BrickRed}cf\_h3\-\ =\-\ 2*h3.'*Tx/(Nx+1);\color{Green}
$\\$\color{BrickRed}cf\_h3(:,1)\-\ =\-\ cf\_h3(:,1)/2;\color{Green}
$\\$
$\\$
$\\$\color{BrickRed}cf\_h4\-\ =\-\ 2*h4.'*Tx/(Nx+1);\color{Green}
$\\$\color{BrickRed}cf\_h4(:,1)\-\ =\-\ cf\_h4(:,1)/2;\color{Green}
$\\$
$\\$
$\\$\color{BrickRed}cf\_h5\-\ =\-\ 2*h5.'*Tx/(Nx+1);\color{Green}
$\\$\color{BrickRed}cf\_h5(:,1)\-\ =\-\ cf\_h5(:,1)/2;\color{Green}
$\\$
$\\$
$\\$\color{BrickRed}out\-\ =\-\ nm(zeros(length(psi),1,10));\color{Green}
$\\$\color{BrickRed}out(:,1)\-\ =\-\ 2*cf\_c0*(1./(1-(0:2:Nx).\verb|^|2)).';\color{Green}
$\\$\color{BrickRed}out(:,2)\-\ =\-\ 2*cf\_c1*(1./(1-(0:2:Nx).\verb|^|2)).';\color{Green}
$\\$\color{BrickRed}out(:,3)\-\ =\-\ 2*cf\_c2*(1./(1-(0:2:Nx).\verb|^|2)).';\color{Green}
$\\$\color{BrickRed}out(:,4)\-\ =\-\ 2*cf\_c3*(1./(1-(0:2:Nx).\verb|^|2)).';\color{Green}
$\\$\color{BrickRed}out(:,5)\-\ =\-\ 2*cf\_h0*(1./(1-(0:2:Nx).\verb|^|2)).';\color{Green}
$\\$\color{BrickRed}out(:,6)\-\ =\-\ 2*cf\_h1*(1./(1-(0:2:Nx).\verb|^|2)).';\color{Green}
$\\$\color{BrickRed}out(:,7)\-\ =\-\ 2*cf\_h2*(1./(1-(0:2:Nx).\verb|^|2)).';\color{Green}
$\\$\color{BrickRed}out(:,8)\-\ =\-\ 2*cf\_h3*(1./(1-(0:2:Nx).\verb|^|2)).';\color{Green}
$\\$\color{BrickRed}out(:,9)\-\ =\-\ 2*cf\_h4*(1./(1-(0:2:Nx).\verb|^|2)).';\color{Green}
$\\$\color{BrickRed}out(:,10)\-\ =\-\ 2*cf\_h5*(1./(1-(0:2:Nx).\verb|^|2)).';\color{Green}
$\\$
$\\$
$\\$\color{Green}$\%$\-\ add\-\ integration\-\ (\-\ in\-\ x)\-\ error
$\\$\color{BrickRed}out\-\ =\-\ out\-\ +\-\ 2*(nm(-err,err)+1i*nm(-err,err));\color{Green}
$\\$
$\\$
$\\$
$\\$
$\\$\color{Black}\section{save\_figure.m}

\color{Green}\color{Green}$\%$\-\ save\-\ figure\-\ in\-\ BFZstudy\-\ folder
$\\$
$\\$
$\\$
$\\$
$\\$\color{BrickRed}sf.name\-\ =\-\ 'intervalarithmetic';\color{Green}
$\\$\color{BrickRed}\color{NavyBlue}\-\ try\-\ \color{BrickRed}\color{Green}
$\\$\color{BrickRed}\-\ \-\ \-\ \-\ savefig(sf,cd,'interval\_arithmetic',gcf,p)\color{Green}
$\\$\color{BrickRed}\color{NavyBlue}\-\ catch\-\ \color{BrickRed}\color{Green}
$\\$\color{BrickRed}\-\ \-\ \-\ \-\ savefig(sf,cd,'interval\_arithmetic',gcf)\color{Green}
$\\$\color{BrickRed}\color{NavyBlue}\-\ end\-\ \color{BrickRed}\color{Green}
$\\$
$\\$
$\\$\color{Black}\section{simplicity.m}

\color{Green}\color{BrickRed}clear\-\ all;\-\ close\-\ all;\-\ beep\-\ off;\-\ clc;\-\ curr\_dir\-\ =\-\ cd;\color{Green}
$\\$\color{Green}$\%$$\%$\-\ startup\-\ commands
$\\$\color{BrickRed}cd('..');\color{Green}
$\\$\color{BrickRed}cd('..');\color{Green}
$\\$\color{BrickRed}startup('intlab','','start\-\ matlabpool','off');\color{Green}
$\\$\color{BrickRed}format\-\ long;\color{Green}
$\\$\color{BrickRed}clc;\color{Green}
$\\$\color{BrickRed}cd(curr\_dir);\color{Green}
$\\$
$\\$
$\\$\color{Green}$\%$\-\ display\-\ type
$\\$\color{BrickRed}intvalinit('DisplayMidRad');\color{Green}
$\\$\color{Green}$\%$\-\ intvalinit('DisplayInfSup');
$\\$
$\\$
$\\$\color{Green}$\%$
$\\$\color{Green}$\%$$\%$\-\ parameters
$\\$\color{Green}$\%$
$\\$
$\\$
$\\$\color{BrickRed}st{1}\-\ =\-\ linspace(inf(nm('0.942')),sup(nm('0.97')),28001);\color{Green}
$\\$\color{BrickRed}st{2}\-\ =\-\ linspace(inf(nm('0.97')),sup(nm('0.99')),20001);\color{Green}
$\\$
$\\$
$\\$\color{BrickRed}st{3}\-\ =\-\ linspace(inf(nm('0.99')),sup(nm('0.995')),10000);\color{Green}
$\\$\color{BrickRed}st{4}\-\ =\-\ linspace(inf(nm('0.995')),sup(nm('0.999')),10000);\color{Green}
$\\$\color{BrickRed}st{5}\-\ =\-\ linspace(inf(nm('0.999')),sup(nm('0.9995')),10000);\color{Green}
$\\$\color{BrickRed}st{6}\-\ =\-\ linspace(inf(nm('0.9995')),sup(nm('0.9999')),10000);\color{Green}
$\\$\color{BrickRed}st{7}\-\ =\-\ linspace(inf(nm('0.9999')),sup(nm('0.99992')),10000);\color{Green}
$\\$\color{BrickRed}st{8}\-\ =\-\ linspace(inf(nm('0.99992')),sup(nm('0.99993')),10000);\color{Green}
$\\$\color{BrickRed}st{9}\-\ =\-\ linspace(inf(nm('0.99993')),sup(nm('0.999934')),10000);\color{Green}
$\\$\color{BrickRed}st{10}\-\ =\-\ linspace(inf(nm('0.999934')),sup(nm('0.999938')),10000);\color{Green}
$\\$
$\\$
$\\$\color{BrickRed}st{11}\-\ =\-\ linspace(inf(nm('0.999938')),sup(nm('0.99999')),20000);\color{Green}
$\\$\color{BrickRed}st{12}\-\ =\-\ linspace(inf(nm('0.99999')),sup(nm('0.999998')),20000);\color{Green}
$\\$\color{BrickRed}st{13}\-\ =\-\ linspace(inf(nm('0.999998')),sup(nm('0.999998134')),2500);\color{Green}
$\\$\color{BrickRed}st{14}\-\ =\-\ linspace(inf(nm('0.999998134')),sup(nm('0.999998267')),2500);\color{Green}
$\\$\color{BrickRed}st{15}\-\ =\-\ linspace(inf(nm('0.999998267')),sup(nm('0.9999984')),2500);\color{Green}
$\\$
$\\$
$\\$\color{BrickRed}out\-\ =\-\ zeros(length(st),1);\color{Green}
$\\$\color{BrickRed}\color{NavyBlue}\-\ parfor\-\ \color{BrickRed}\-\ j\-\ =\-\ 1:length(out)\color{Green}
$\\$\color{BrickRed}\-\ \-\ \-\ \-\ out(j)\-\ =\-\ simplicity\_aux(st{j},curr\_dir);\color{Green}
$\\$\color{BrickRed}\color{NavyBlue}\-\ end\-\ \color{BrickRed}\color{Green}
$\\$
$\\$
$\\$\color{Green}$\%$\-\ 1\-\ indicates\-\ success\-\ and\-\ 0\-\ indicates\-\ failure\-\ in\-\ verifying\-\ simplicity
$\\$\color{BrickRed}\color{NavyBlue}\-\ for\-\ \color{BrickRed}\-\ j\-\ =\-\ 1:length(out)\color{Green}
$\\$\color{BrickRed}\-\ \-\ \-\ \-\ \color{NavyBlue}\-\ if\-\ \color{BrickRed}\-\ out(j)\-\ ==\-\ 0\color{Green}
$\\$\color{BrickRed}\-\ \-\ \-\ \-\ \-\ \-\ \-\ \-\ error('Failed\-\ to\-\ verify\-\ simplicity')\color{Green}
$\\$\color{BrickRed}\-\ \-\ \-\ \-\ \color{NavyBlue}\-\ end\-\ \color{BrickRed}\color{Green}
$\\$\color{BrickRed}\color{NavyBlue}\-\ end\-\ \color{BrickRed}\color{Green}
$\\$
$\\$
$\\$
$\\$
$\\$
$\\$
$\\$
$\\$
$\\$
$\\$
$\\$
$\\$
$\\$
$\\$
$\\$
$\\$
$\\$
$\\$
$\\$
$\\$
$\\$
$\\$
$\\$
$\\$
$\\$\color{Black}\section{simplicity\_aux.m}

\color{Green}\color{BrickRed}\color{NavyBlue}\-\ function\-\ \color{BrickRed}\-\ out\-\ =\-\ simplicity\_aux(k\_vals,curr\_dir)\color{Green}
$\\$
$\\$
$\\$\color{BrickRed}out\-\ =\-\ 0;\color{Green}
$\\$
$\\$
$\\$\color{Green}$\%$$\%$\-\ startup\-\ commands
$\\$
$\\$
$\\$\color{BrickRed}local\_startup\_batch(curr\_dir);\color{Green}
$\\$
$\\$
$\\$\color{BrickRed}\color{NavyBlue}\-\ for\-\ \color{BrickRed}\-\ j\-\ =\-\ 1:length(k\_vals)-1\color{Green}
$\\$
$\\$
$\\$\color{BrickRed}\-\ \-\ \-\ \-\ k\-\ =\-\ nm(k\_vals(j),k\_vals(j+1));\color{Green}
$\\$
$\\$
$\\$\color{BrickRed}\-\ \-\ \-\ \-\ \color{Green}$\%$
$\\$\color{BrickRed}\-\ \-\ \-\ \-\ \color{Green}$\%$$\%$\-\ derived\-\ constants
$\\$\color{BrickRed}\-\ \-\ \-\ \-\ \color{Green}$\%$
$\\$
$\\$
$\\$\color{BrickRed}\-\ \-\ \-\ \-\ p.k\-\ =\-\ k;\color{Green}
$\\$\color{BrickRed}\-\ \-\ \-\ \-\ pie\-\ =\-\ nm('pi');\color{Green}
$\\$\color{BrickRed}\-\ \-\ \-\ \-\ kappa\-\ =\-\ kappa\_of\_k(k);\color{Green}
$\\$\color{BrickRed}\-\ \-\ \-\ \-\ elipk\-\ =\-\ elliptic\_integral(k,1);\color{Green}
$\\$\color{BrickRed}\-\ \-\ \-\ \-\ elipk2\-\ =\-\ elliptic\_integral(sqrt(1-k.\verb|^|2),1);\color{Green}
$\\$\color{BrickRed}\-\ \-\ \-\ \-\ omega\-\ =\-\ pie./kappa;\color{Green}
$\\$\color{BrickRed}\-\ \-\ \-\ \-\ omega\_prime\-\ =\-\ elipk2*pie./(elipk.*kappa);\color{Green}
$\\$\color{BrickRed}\-\ \-\ \-\ \-\ q\-\ =\-\ exp(-pie*elipk2./elipk);\color{Green}
$\\$
$\\$
$\\$\color{BrickRed}\-\ \-\ \-\ \-\ \color{Green}$\%$
$\\$\color{BrickRed}\-\ \-\ \-\ \-\ \color{Green}$\%$\-\ ----------------
$\\$\color{BrickRed}\-\ \-\ \-\ \-\ \color{Green}$\%$
$\\$
$\\$
$\\$\color{BrickRed}\-\ \-\ \-\ \-\ ntilde\-\ =\-\ 1;\color{Green}
$\\$
$\\$
$\\$\color{BrickRed}\-\ \-\ \-\ \-\ num\-\ =\-\ 100;\color{Green}
$\\$\color{BrickRed}\-\ \-\ \-\ \-\ L\-\ =\-\ 20;\-\ \-\ \-\ \color{Green}$\%$\-\ \color{Black}$\beta_0 = 0.95$ \color{Green}\-\ 
$\\$\color{BrickRed}\-\ \-\ \-\ \-\ L2\-\ =\-\ 1;\-\ \-\ \-\ \color{Green}$\%$\-\ \color{Black}$\beta \in [0,\omega']$ \color{Green}
$\\$
$\\$
$\\$\color{BrickRed}\-\ \-\ \-\ \-\ psi\_x\-\ =\-\ nm(0:1:num)/(L2*num)+nm(L2-1)/L2;\color{Green}
$\\$\color{BrickRed}\-\ \-\ \-\ \-\ psi\_x\-\ =\-\ nm(psi\_x(1:end-1),psi\_x(2:end));\-\ \color{Green}$\%$\-\ make\-\ intervals
$\\$\color{BrickRed}\-\ \-\ \-\ \-\ \-\ \-\ \-\ \color{Green}
$\\$\color{BrickRed}\-\ \-\ \-\ \-\ psi\_y\-\ =\-\ nm(0:1:num)/(L*num)+nm(L-1)/L;\color{Green}
$\\$\color{BrickRed}\-\ \-\ \-\ \-\ psi\_y\-\ =\-\ nm(psi\_y(1:end-1),psi\_y(2:end));\-\ \color{Green}$\%$\-\ make\-\ intervals
$\\$\color{BrickRed}\-\ \-\ \-\ \-\ \color{Green}
$\\$\color{BrickRed}\-\ \-\ \-\ \-\ num\-\ =\-\ num\-\ -1;\color{Green}
$\\$
$\\$
$\\$\color{BrickRed}\-\ \-\ \-\ \-\ \color{Green}$\%$\-\ get\-\ components\-\ of\-\ h(x,y)
$\\$\color{BrickRed}\-\ \-\ \-\ \-\ [f1,fd1,fdd1,g1,gd1,gdd1]\-\ =\-\ lambda\_xi(q,omega,omega\_prime,psi\_x,ntilde);\color{Green}
$\\$
$\\$
$\\$\color{BrickRed}\-\ \-\ \-\ \-\ F1\-\ =\-\ repmat(f1,1,num+1);\color{Green}
$\\$\color{BrickRed}\-\ \-\ \-\ \-\ Fd1\-\ =\-\ repmat(fd1,1,num+1);\color{Green}
$\\$\color{BrickRed}\-\ \-\ \-\ \-\ Fdd1\-\ =\-\ repmat(fdd1,1,num+1);\color{Green}
$\\$\color{BrickRed}\-\ \-\ \-\ \-\ G1\-\ =\-\ repmat(g1,1,num+1);\color{Green}
$\\$\color{BrickRed}\-\ \-\ \-\ \-\ Gd1\-\ =\-\ repmat(gd1,1,num+1);\color{Green}
$\\$\color{BrickRed}\-\ \-\ \-\ \-\ Gdd1\-\ =\-\ repmat(gdd1,1,num+1);\color{Green}
$\\$
$\\$
$\\$\color{BrickRed}\-\ \-\ \-\ \-\ \color{Green}$\%$
$\\$\color{BrickRed}\-\ \-\ \-\ \-\ \color{Green}$\%$\-\ ----------------
$\\$\color{BrickRed}\-\ \-\ \-\ \-\ \color{Green}$\%$
$\\$
$\\$
$\\$\color{BrickRed}\-\ \-\ \-\ \-\ ntilde\-\ =\-\ 0;\color{Green}
$\\$
$\\$
$\\$\color{BrickRed}\-\ \-\ \-\ \-\ \color{Green}$\%$\-\ get\-\ components\-\ of\-\ h(x,y)
$\\$\color{BrickRed}\-\ \-\ \-\ \-\ [f2,fd2,fdd2,g2,gd2,gdd2]\-\ =\-\ lambda\_xi(q,omega,omega\_prime,psi\_y,ntilde);\color{Green}
$\\$\color{BrickRed}\-\ \-\ \-\ \-\ f2\-\ =\-\ f2.';\-\ fd2\-\ =\-\ fd2.';\-\ fdd2\-\ =\-\ fdd2.';\-\ g2\-\ =\-\ g2.';\-\ gd2\-\ =\-\ gd2.';\-\ gdd2\-\ =\-\ gdd2.';\color{Green}
$\\$
$\\$
$\\$\color{BrickRed}\-\ \-\ \-\ \-\ F2\-\ =\-\ repmat(f2,num+1,1);\color{Green}
$\\$\color{BrickRed}\-\ \-\ \-\ \-\ Fd2\-\ =\-\ repmat(fd2,num+1,1);\color{Green}
$\\$\color{BrickRed}\-\ \-\ \-\ \-\ Fdd2\-\ =\-\ repmat(fdd2,num+1,1);\color{Green}
$\\$\color{BrickRed}\-\ \-\ \-\ \-\ G2\-\ =\-\ repmat(g2,num+1,1);\color{Green}
$\\$\color{BrickRed}\-\ \-\ \-\ \-\ Gd2\-\ =\-\ repmat(gd2,num+1,1);\color{Green}
$\\$\color{BrickRed}\-\ \-\ \-\ \-\ Gdd2\-\ =\-\ repmat(gdd2,num+1,1);\color{Green}
$\\$
$\\$
$\\$\color{BrickRed}\-\ \-\ \-\ \-\ \color{Green}$\%$
$\\$\color{BrickRed}\-\ \-\ \-\ \-\ \color{Green}$\%$\-\ ----------------
$\\$\color{BrickRed}\-\ \-\ \-\ \-\ \color{Green}$\%$
$\\$
$\\$
$\\$\color{BrickRed}\-\ \-\ \-\ \-\ \color{Green}$\%$\-\ \color{Black}Show that $\pd{}{\psi} \xi(\omega+i\psi\omega') > 0$ for $\psi \in [0,1]$. \color{Green}
$\\$
$\\$
$\\$\color{BrickRed}\-\ \-\ \-\ \-\ con\-\ =\-\ 1;\color{Green}
$\\$
$\\$
$\\$\color{BrickRed}\-\ \-\ \-\ \-\ T\-\ =\-\ con*(F1+F2);\color{Green}
$\\$\color{BrickRed}\-\ \-\ \-\ \-\ Tx\-\ =\-\ con*(Fd1);\color{Green}
$\\$\color{BrickRed}\-\ \-\ \-\ \-\ Txx\-\ =\-\ con*(Fdd1);\color{Green}
$\\$\color{BrickRed}\-\ \-\ \-\ \-\ \color{Green}$\%$\-\ Txy\-\ =\-\ 0
$\\$\color{BrickRed}\-\ \-\ \-\ \-\ Ty\-\ =\-\ con*(Fd2);\color{Green}
$\\$\color{BrickRed}\-\ \-\ \-\ \-\ Tyy\-\ =\-\ con*(Fdd2);\color{Green}
$\\$
$\\$
$\\$\color{BrickRed}\-\ \-\ \-\ \-\ S\-\ =\-\ G1+G2-2*pie;\color{Green}
$\\$\color{BrickRed}\-\ \-\ \-\ \-\ Sx\-\ =\-\ Gd1;\color{Green}
$\\$\color{BrickRed}\-\ \-\ \-\ \-\ \color{Green}$\%$\-\ Sxy\-\ =\-\ 0
$\\$\color{BrickRed}\-\ \-\ \-\ \-\ Sxx\-\ =\-\ Gdd1;\color{Green}
$\\$\color{BrickRed}\-\ \-\ \-\ \-\ Sy\-\ =\-\ Gd2;\color{Green}
$\\$\color{BrickRed}\-\ \-\ \-\ \-\ Syy\-\ =\-\ Gdd2;\color{Green}
$\\$
$\\$
$\\$\color{Green}$\%$\-\ \-\ \-\ \-\ \-\ h\-\ =\-\ T.\verb|^|2+S.\verb|^|2;
$\\$\color{Green}$\%$\-\ \-\ \-\ \-\ \-\ hx\-\ =\-\ 2*T.*Tx+2*S.*Sx;
$\\$\color{BrickRed}\-\ \-\ \-\ \-\ hxx\-\ =\-\ 2*Tx.\verb|^|2+2*T.*Txx+2*Sx.\verb|^|2+2*S.*Sxx;\color{Green}
$\\$\color{Green}$\%$\-\ \-\ \-\ \-\ \-\ hy\-\ =\-\ 2*T.*Ty+2*S.*Sy;
$\\$\color{BrickRed}\-\ \-\ \-\ \-\ hyy\-\ =\-\ 2*Ty.\verb|^|2+2*T.*Tyy+2*Sy.\verb|^|2+2*S.*Syy;\color{Green}
$\\$\color{BrickRed}\-\ \-\ \-\ \-\ hxy\-\ =\-\ 2*Ty.*Tx+2*Sy.*Sx;\color{Green}
$\\$
$\\$
$\\$\color{BrickRed}\-\ \-\ \-\ \-\ \color{Green}$\%$Del
$\\$
$\\$
$\\$\color{BrickRed}\-\ \-\ \-\ \-\ Del\-\ =\-\ (hxx.*hyy-hxy.\verb|^|2);\color{Green}
$\\$\color{BrickRed}\-\ \-\ \-\ \-\ Del\-\ =\-\ inf(real(Del));\color{Green}
$\\$
$\\$
$\\$\color{BrickRed}\-\ \-\ \-\ \-\ \color{NavyBlue}\-\ if\-\ \color{BrickRed}\-\ \-\ abs(sum(sum(isfinite(Del)-1)))\-\ $>$\-\ 0\color{Green}
$\\$\color{BrickRed}\-\ \-\ \-\ \-\ \-\ \-\ \-\ \-\ error('Del\-\ is\-\ not\-\ finite');\color{Green}
$\\$\color{BrickRed}\-\ \-\ \-\ \-\ \color{NavyBlue}\-\ end\-\ \color{BrickRed}\color{Green}
$\\$
$\\$
$\\$\color{BrickRed}\-\ \-\ \-\ \-\ min\_Del\-\ =\-\ min(min(Del));\color{Green}
$\\$
$\\$
$\\$\color{BrickRed}\-\ \-\ \-\ \-\ \color{Green}$\%$\-\ hxx
$\\$\color{BrickRed}\-\ \-\ \-\ \-\ hxx\-\ =\-\ inf(real(hxx));\color{Green}
$\\$
$\\$
$\\$\color{BrickRed}\-\ \-\ \-\ \-\ \color{NavyBlue}\-\ if\-\ \color{BrickRed}\-\ \-\ abs(sum(sum(isfinite(hxx)-1)))\-\ $>$\-\ 0\color{Green}
$\\$\color{BrickRed}\-\ \-\ \-\ \-\ \-\ \-\ \-\ \-\ error('hxx\-\ is\-\ not\-\ finite');\color{Green}
$\\$\color{BrickRed}\-\ \-\ \-\ \-\ \color{NavyBlue}\-\ end\-\ \color{BrickRed}\color{Green}
$\\$
$\\$
$\\$\color{BrickRed}\-\ \-\ \-\ \-\ min\_hxx\-\ =\-\ min(min(hxx));\color{Green}
$\\$
$\\$
$\\$\color{BrickRed}\-\ \-\ \-\ \-\ temp1\-\ =\-\ -4*gdd1/(2*1i*omega\_prime\verb|^|2*omega);\color{Green}
$\\$\color{BrickRed}\-\ \-\ \-\ \-\ max\_gdd1\-\ =\-\ max(sup(abs(imag(temp1))));\color{Green}
$\\$
$\\$
$\\$\color{BrickRed}\-\ \-\ \-\ \-\ temp2\-\ =\-\ -4*gdd2(1)/(2*1i*omega\_prime\verb|^|2*omega);\color{Green}
$\\$\color{BrickRed}\-\ \-\ \-\ \-\ gdd1\_one\-\ =\-\ inf(imag(temp2));\color{Green}
$\\$
$\\$
$\\$\color{BrickRed}\-\ \-\ \-\ \-\ \color{Green}$\%$\-\ throw\-\ errors\-\ if\-\ we\-\ do\-\ not\-\ verify\-\ the\-\ necessary\-\ properties
$\\$
$\\$
$\\$\color{BrickRed}\-\ \-\ \-\ \-\ \color{NavyBlue}\-\ if\-\ \color{BrickRed}\-\ min\_Del\-\ $<$=0\color{Green}
$\\$\color{BrickRed}\-\ \-\ \-\ \-\ \-\ \-\ \-\ \-\ \color{NavyBlue}\-\ return\-\ \color{BrickRed}\color{Green}
$\\$\color{BrickRed}\-\ \-\ \-\ \-\ \color{NavyBlue}\-\ end\-\ \color{BrickRed}\color{Green}
$\\$
$\\$
$\\$\color{BrickRed}\-\ \-\ \-\ \-\ \color{NavyBlue}\-\ if\-\ \color{BrickRed}\-\ min\_hxx\-\ $<$=\-\ 0\color{Green}
$\\$\color{BrickRed}\-\ \-\ \-\ \-\ \-\ \-\ \-\ \-\ \color{NavyBlue}\-\ return\-\ \color{BrickRed}\color{Green}
$\\$\color{BrickRed}\-\ \-\ \-\ \-\ \color{NavyBlue}\-\ end\-\ \color{BrickRed}\color{Green}
$\\$
$\\$
$\\$\color{BrickRed}\-\ \-\ \-\ \-\ \color{NavyBlue}\-\ if\-\ \color{BrickRed}\-\ inf(nm(gdd1\_one)-nm(max\_gdd1))\-\ $<$=\-\ 0\color{Green}
$\\$\color{BrickRed}\-\ \-\ \-\ \-\ \-\ \-\ \-\ \-\ \color{NavyBlue}\-\ return\-\ \color{BrickRed}\color{Green}
$\\$\color{BrickRed}\-\ \-\ \-\ \-\ \color{NavyBlue}\-\ end\-\ \color{BrickRed}\color{Green}
$\\$
$\\$
$\\$\color{BrickRed}\-\ \-\ \-\ \-\ \color{NavyBlue}\-\ if\-\ \color{BrickRed}\-\ -inf(g2(1)-3*pie/2)\-\ $<$=\-\ 0\color{Green}
$\\$\color{BrickRed}\-\ \-\ \-\ \-\ \-\ \-\ \-\ \-\ \color{NavyBlue}\-\ return\-\ \color{BrickRed}\color{Green}
$\\$\color{BrickRed}\-\ \-\ \-\ \-\ \color{NavyBlue}\-\ end\-\ \color{BrickRed}\color{Green}
$\\$
$\\$
$\\$\color{BrickRed}\color{NavyBlue}\-\ end\-\ \color{BrickRed}\color{Green}
$\\$
$\\$
$\\$\color{BrickRed}out\-\ =\-\ 1;\color{Green}
$\\$
$\\$
$\\$
$\\$
$\\$\color{Black}\section{strict\_taylor.m}

\color{Green}\color{BrickRed}\color{NavyBlue}\-\ function\-\ \color{BrickRed}\-\ out\-\ =\-\ strict\_taylor(cf,q\_tilde,psi\_tilde,id)\color{Green}
$\\$
$\\$
$\\$\color{Green}$\%$\-\ points
$\\$\color{BrickRed}q\_tilde\-\ =\-\ nm(q\_tilde);\color{Green}
$\\$\color{BrickRed}psi\_tilde\-\ =\-\ nm(psi\_tilde);\color{Green}
$\\$\color{BrickRed}tx\-\ =\-\ acos(q\_tilde).';\color{Green}
$\\$\color{BrickRed}ty\-\ =\-\ acos(psi\_tilde);\color{Green}
$\\$
$\\$
$\\$\color{Green}$\%$\-\ 
$\\$\color{BrickRed}Nx\-\ =\-\ size(cf,1)-1;\-\ \color{Green}$\%$\-\ degree\-\ of\-\ polynomial\-\ in\-\ x
$\\$\color{BrickRed}Ny\-\ =\-\ size(cf,2)-1;\-\ \color{Green}$\%$\-\ degeree\-\ of\-\ polynomial\-\ in\-\ y
$\\$\color{BrickRed}indx\-\ =\-\ 0:1:Nx;\-\ \color{Green}
$\\$\color{BrickRed}indy\-\ =\-\ (0:1:Ny).';\color{Green}
$\\$
$\\$
$\\$\color{BrickRed}zx\-\ =\-\ \-\ nm(tx)*indx;\-\ \color{Green}$\%$\-\ function\-\ argument\-\ for\-\ x
$\\$\color{BrickRed}zy\-\ =\-\ indy*nm(ty);\-\ \color{Green}$\%$\-\ function\-\ argument\-\ for\-\ y
$\\$
$\\$
$\\$\color{BrickRed}\color{NavyBlue}\-\ if\-\ \color{BrickRed}\-\ id\-\ ==\-\ 1\-\ \color{Green}$\%$\-\ f
$\\$\color{BrickRed}\-\ \-\ \-\ \-\ \color{Green}
$\\$\color{BrickRed}\-\ \-\ \-\ \-\ cosx\-\ =\-\ cos(zx);\color{Green}
$\\$\color{BrickRed}\-\ \-\ \-\ \-\ cosy\-\ =\-\ cos(zy);\color{Green}
$\\$\color{BrickRed}\-\ \-\ \-\ \-\ out\-\ =\-\ cosx*cf*cosy;\color{Green}
$\\$\color{BrickRed}\-\ \-\ \-\ \-\ \color{Green}
$\\$\color{BrickRed}\color{NavyBlue}\-\ elseif\-\ \color{BrickRed}\-\ id\-\ ==\-\ 2\-\ \color{Green}$\%$\-\ f\_psi
$\\$\color{BrickRed}\-\ \-\ \-\ \-\ \color{Green}
$\\$\color{BrickRed}\-\ \-\ \-\ \-\ cosx\-\ =\-\ cos(zx);\color{Green}
$\\$\color{BrickRed}\-\ \-\ \-\ \-\ nsiny\-\ =\-\ repmat(indy,1,length(ty)).*sin(zy);\color{Green}
$\\$\color{BrickRed}\-\ \-\ \-\ \-\ out\-\ =\-\ cosx*cf*(nsiny.*repmat(1./sin(ty),length(indy),1));\color{Green}
$\\$\color{BrickRed}\-\ \-\ \-\ \-\ \color{Green}
$\\$\color{BrickRed}\color{NavyBlue}\-\ elseif\-\ \color{BrickRed}\-\ id\-\ ==\-\ 3\-\ \color{Green}$\%$\-\ f\_q
$\\$\color{BrickRed}\-\ \-\ \-\ \-\ \color{Green}
$\\$\color{BrickRed}\-\ \-\ \-\ \-\ nsinx\-\ =\-\ repmat(indx,length(tx),1).*sin(zx);\color{Green}
$\\$\color{BrickRed}\-\ \-\ \-\ \-\ cosy\-\ =\-\ cos(zy);\color{Green}
$\\$\color{BrickRed}\-\ \-\ \-\ \-\ out\-\ =\-\ (repmat(1./sin(tx),1,length(indx)).*nsinx)*cf*cosy;\color{Green}
$\\$\color{BrickRed}\-\ \-\ \-\ \-\ \color{Green}
$\\$\color{BrickRed}\color{NavyBlue}\-\ elseif\-\ \color{BrickRed}\-\ id\-\ ==\-\ 4\-\ \color{Green}$\%$\-\ f\_q\_q
$\\$\color{BrickRed}\-\ \-\ \-\ \-\ \color{Green}
$\\$\color{BrickRed}\-\ \-\ \-\ \-\ n2cosx\-\ =\-\ repmat(indx.\verb|^|2,length(tx),1).*cos(zx);\color{Green}
$\\$\color{BrickRed}\-\ \-\ \-\ \-\ cosy\-\ =\-\ cos(zy);\color{Green}
$\\$\color{BrickRed}\-\ \-\ \-\ \-\ out\-\ =\-\ (repmat(-1./sin(tx).\verb|^|2,1,length(indx)).*n2cosx)*cf*cosy;\color{Green}
$\\$\color{BrickRed}\-\ \-\ \-\ \-\ \color{Green}
$\\$\color{BrickRed}\color{NavyBlue}\-\ end\-\ \color{BrickRed}\color{Green}
$\\$
$\\$
$\\$
$\\$
$\\$
$\\$
$\\$
$\\$
$\\$\color{Black}\section{strict\_transition.m}

\color{Green}\color{BrickRed}\color{NavyBlue}\-\ function\-\ \color{BrickRed}\-\ [mx\_fk,mn\_fpsipsi]\-\ =\-\ strict\_transition(k,psi0,min\_diff\_q,min\_diff\_psi,d)\color{Green}
$\\$
$\\$
$\\$\color{Green}$\%$-----------------------------------------------------------
$\\$\color{Green}$\%$\-\ \-\ background
$\\$\color{Green}$\%$-----------------------------------------------------------
$\\$
$\\$The\-\ Taylor\-\ expansion\-\ is\-\ give\-\ by,
$\\$
$\\$
$\\$\textbackslash eq{
$\\$f(k,\textbackslash psi)\-\ \&=\-\ f(k\_*,\textbackslash psi\_*)\-\ +\-\ f\_k(k\_*,\textbackslash psi\_*)(k-k\_*)\-\ +\-\ f\_{\textbackslash psi}(k\_*,\textbackslash psi\_*)(\textbackslash psi-\textbackslash psi\_*)\-\ +\-\ \textbackslash \textbackslash 
$\\$\&\textbackslash frac{1}{2}\textbackslash left(f\_{kk}(\textbackslash hat\-\ k,\textbackslash hat\-\ \textbackslash psi)(k-k\_*)\verb|^|2+2f\_{\textbackslash psi\-\ k}(\textbackslash hat\-\ k,\textbackslash hat\-\ \textbackslash psi)(k-k\_*)(\textbackslash psi-\textbackslash psi\_*)+f\_{\textbackslash psi\textbackslash psi}(\textbackslash hat\-\ k,\-\ \textbackslash hat\-\ \textbackslash psi)(\textbackslash psi-\textbackslash psi\_*)\verb|^|2\-\ \textbackslash right).
$\\$}{}
$\\$
$\\$
$\\$
$\\$
$\\$
$\\$
$\\$
$\\$\color{Green}$\%$-----------------------------------------------------------
$\\$\color{Green}$\%$\-\ \-\ constants
$\\$\color{Green}$\%$-----------------------------------------------------------
$\\$
$\\$
$\\$\color{BrickRed}k0\-\ =\-\ k;\color{Green}
$\\$\color{BrickRed}pie\-\ =\-\ nm('pi');\color{Green}
$\\$
$\\$
$\\$\color{BrickRed}psi\_tilde0\-\ =\-\ 4*psi0-3;\color{Green}
$\\$
$\\$
$\\$\color{Green}$\%$\-\ constants\-\ for\-\ transformation\-\ in\-\ q
$\\$\color{BrickRed}c1\_q\-\ =\-\ 2/(d.b\_q-d.a\_q);\color{Green}
$\\$\color{BrickRed}c2\_q\-\ =\-\ (d.a\_q+d.b\_q)/2;\color{Green}
$\\$
$\\$
$\\$\color{Green}$\%$\-\ elliptic\-\ integrals
$\\$\color{BrickRed}elipk\-\ =\-\ elliptic\_integral(k,1);\color{Green}
$\\$\color{BrickRed}elipk2\-\ =\-\ elliptic\_integral(sqrt(1-k.\verb|^|2),1);\color{Green}
$\\$\color{Green}$\%$\-\ q
$\\$\color{BrickRed}q0\-\ =\-\ exp(-pie*elipk2./elipk);\color{Green}
$\\$\color{Green}$\%$\-\ omega0\-\ =\-\ pie/kappa;
$\\$
$\\$
$\\$\color{BrickRed}\color{NavyBlue}\-\ if\-\ \color{BrickRed}\-\ inf(q0)\-\ $<$\-\ d.a\_q\color{Green}
$\\$\color{Green}$\%$\-\ \-\ \-\ \-\ \-\ q0
$\\$\color{BrickRed}\-\ \-\ \-\ \-\ error('q\-\ out\-\ of\-\ range')\color{Green}
$\\$\color{BrickRed}\color{NavyBlue}\-\ end\-\ \color{BrickRed}\color{Green}
$\\$\color{BrickRed}\color{NavyBlue}\-\ if\-\ \color{BrickRed}\-\ sup(q0)\-\ $>$\-\ d.b\_q\color{Green}
$\\$\color{BrickRed}\-\ \-\ \-\ \-\ error('q\-\ out\-\ of\-\ range')\color{Green}
$\\$\color{BrickRed}\color{NavyBlue}\-\ end\-\ \color{BrickRed}\color{Green}
$\\$
$\\$
$\\$\color{Green}$\%$\-\ get\-\ theta\-\ values\-\ for\-\ q
$\\$\color{BrickRed}q\_tilde0\-\ =\-\ c1\_q*(q0-c2\_q);\color{Green}
$\\$
$\\$
$\\$\color{Green}$\%$\-\ interpolation\-\ coefficients
$\\$\color{BrickRed}cf\-\ =\-\ d.cfn0;\color{Green}
$\\$
$\\$
$\\$\color{Green}$\%$\-\ interpolation\-\ error
$\\$\color{BrickRed}lam\_n0\_q\-\ =\-\ (2/pie)*log(d.N\_q\_n0)+(2/pie)*(nm('0.6')+log(8/pie)+pie/(72*d.N\_q\_n0));\color{Green}
$\\$\color{BrickRed}err\-\ =\-\ d.err\_q\_n0+lam\_n0\_q*d.err\_psi\_n0;\color{Green}
$\\$\color{BrickRed}err\-\ =\-\ nm(-err,err)+1i*nm(-err,err);\color{Green}
$\\$
$\\$
$\\$\color{Green}$\%$-----------------------------------------------------------
$\\$\color{Green}$\%$\-\ \-\ interpolation\-\ error\-\ of\-\ interpolant\-\ derivative
$\\$\color{Green}$\%$-----------------------------------------------------------
$\\$
$\\$
$\\$\color{Green}$\%$\-\ To\-\ bound\-\ the\-\ interpoaltion\-\ error\-\ coming\-\ from\-\ approximating\-\ the\-\ derivative\-\ of\-\ f
$\\$\color{Green}$\%$\-\ with\-\ the\-\ derivative\-\ of\-\ the\-\ interpolating\-\ polynomial\-\ p,\-\ we\-\ need\-\ a\-\ bound\-\ on\-\ 
$\\$\color{Green}$\%$\-\ the\-\ derivative\-\ of\-\ f\-\ on\-\ a\-\ stadium.\-\ We\-\ use\-\ Cauchy's\-\ integral\-\ formula\-\ to\-\ get\-\ that\-\ bound.\-\ 
$\\$\color{Green}$\%$\-\ We\-\ already\-\ have\-\ a\-\ bound\-\ on\-\ f\-\ on\-\ a\-\ stadium\-\ of\-\ radius\-\ rho\_q.\-\ We\-\ choose\-\ rho\_q\_sm\-\ and
$\\$\color{Green}$\%$\-\ rho\_psi\_sm\-\ so\-\ that\-\ the\-\ smaller\-\ ellipse\-\ has\-\ a\-\ minor\-\ axis\-\ of\-\ half\-\ the\-\ length\-\ of\-\ the\-\ larger\-\ one.
$\\$\color{Green}$\%$\-\ Then\-\ we\-\ use\-\ Cauchy's\-\ integral\-\ formula\-\ with\-\ f\-\ on\-\ the\-\ larger\-\ stadium\-\ to\-\ get\-\ a\-\ bound\-\ on\-\ the
$\\$\color{Green}$\%$\-\ derivative\-\ of\-\ f\-\ on\-\ the\-\ smaller\-\ stadium.\-\ We\-\ then\-\ use\-\ the\-\ radius\-\ of\-\ the\-\ smaller\-\ stadium\-\ to\-\ get
$\\$\color{Green}$\%$\-\ the\-\ needed\-\ interpolation\-\ error\-\ bounds\-\ for\-\ the\-\ derivative\-\ of\-\ f.
$\\$\color{BrickRed}rho\_q\_sm\-\ =\-\ ((d.rho\_q\-\ -\-\ 1/d.rho\_q)+sqrt((d.rho\_q-1/d.rho\_q)\verb|^|2+16))/4;\color{Green}
$\\$\color{BrickRed}rho\_psi\_sm\-\ =\-\ ((d.rho\_psi\-\ -\-\ 1/d.rho\_psi)+sqrt((d.rho\_psi-1/d.rho\_psi)\verb|^|2+16))/4;\color{Green}
$\\$\color{BrickRed}rho\_q\-\ =\-\ d.rho\_q;\color{Green}
$\\$\color{BrickRed}rho\_psi\-\ =\-\ d.rho\_psi;\color{Green}
$\\$
$\\$
$\\$\color{Green}$\%$\-\ Next\-\ we\-\ find\-\ the\-\ minimum\-\ distance\-\ between\-\ the\-\ larger\-\ and\-\ smaller\-\ stadiums\-\ to\-\ 
$\\$\color{Green}$\%$\-\ provide\-\ a\-\ bound\-\ on\-\ the\-\ integrand\-\ in\-\ Taylor's\-\ integral\-\ formula.
$\\$
$\\$
$\\$\color{Green}$\%$\-\ This\-\ is\-\ a\-\ bound\-\ on\-\ the\-\ derivative\-\ of\-\ f\-\ on\-\ the\-\ smaller\-\ satdium
$\\$\color{Green}$\%$\-\ derived\-\ using\-\ Cauchy's\-\ integral\-\ formual
$\\$\color{BrickRed}M\_Dq\-\ =\-\ (1/nm(2))*sqrt(rho\_q\verb|^|2+1/rho\_q\verb|^|2)*d.M\_q\_n0/min\_diff\_q\verb|^|2;\color{Green}
$\\$\color{BrickRed}M\_Dpsi\-\ =\-\ (1/nm(2))*sqrt(rho\_psi\verb|^|2+1/rho\_psi\verb|^|2)*d.M\_psi\_n0/min\_diff\_psi\verb|^|2;\color{Green}
$\\$
$\\$
$\\$\color{Green}$\%$\-\ This\-\ is\-\ a\-\ bound\-\ on\-\ the\-\ length\-\ of\-\ the\-\ smaller\-\ stadium
$\\$\color{BrickRed}Lq\-\ =\-\ pie*sqrt(rho\_q\_sm\verb|^|2+1/rho\_q\_sm\verb|^|2);\color{Green}
$\\$\color{Green}$\%$\-\ This\-\ is\-\ a\-\ lower\-\ bound\-\ on\-\ the\-\ distance\-\ between\-\ the
$\\$\color{Green}$\%$\-\ stadium\-\ and\-\ the\-\ line\-\ on\-\ which\-\ we\-\ interpolate
$\\$\color{BrickRed}Ddq\-\ =\-\ (rho\_q\_sm+1/rho\_q\_sm)/2-1;\color{Green}
$\\$\color{BrickRed}eta\_q\-\ =\-\ log(rho\_q\_sm);\color{Green}
$\\$\color{BrickRed}err\_Dq\-\ =\-\ M\_Dq*Lq/(pie*Ddq*sinh(eta\_q*(d.N\_q\_n0+1)));\color{Green}
$\\$
$\\$
$\\$\color{Green}$\%$\-\ This\-\ is\-\ a\-\ bound\-\ on\-\ the\-\ length\-\ of\-\ the\-\ smaller\-\ stadium
$\\$\color{BrickRed}Lpsi\-\ =\-\ pie*sqrt(rho\_psi\_sm\verb|^|2+1/rho\_psi\_sm\verb|^|2);\color{Green}
$\\$\color{Green}$\%$\-\ This\-\ is\-\ a\-\ lower\-\ bound\-\ on\-\ the\-\ distance\-\ between\-\ the
$\\$\color{Green}$\%$\-\ stadium\-\ and\-\ the\-\ line\-\ on\-\ which\-\ we\-\ interpolate
$\\$\color{BrickRed}Ddpsi\-\ =\-\ (rho\_psi\_sm+1/rho\_psi\_sm)/2-1;\color{Green}
$\\$\color{BrickRed}eta\_psi\-\ =\-\ log(rho\_psi\_sm);\color{Green}
$\\$\color{BrickRed}err\_Dpsi\-\ =\-\ M\_Dpsi*Lpsi/(pie*Ddpsi*sinh(eta\_psi*(d.N\_psi\_n0+1)));\color{Green}
$\\$
$\\$
$\\$\color{Green}$\%$\-\ we\-\ take\-\ the\-\ error\-\ to\-\ be\-\ the\-\ larger\-\ so\-\ we\-\ can\-\ use\-\ either\-\ variable
$\\$\color{BrickRed}err\_interp\_der\-\ =\-\ nm(err\_Dq,err\_Dpsi);\color{Green}
$\\$
$\\$
$\\$\color{Green}$\%$\-\ Next\-\ we\-\ find\-\ the\-\ 2-d\-\ interpolation\-\ error\-\ for\-\ the\-\ derivative\-\ of\-\ f\-\ with
$\\$\color{Green}$\%$\-\ respect\-\ to\-\ q
$\\$
$\\$
$\\$\color{BrickRed}Drhoq\-\ =\-\ (d.rho\_q+1/d.rho\_q)/2-1;\color{Green}
$\\$\color{BrickRed}fc\-\ =\-\ (d.N\_q\_n0+1)*(d.N\_q\_n0+3)/Drhoq+1/Drhoq\verb|^|2;\color{Green}
$\\$\color{BrickRed}err\_Dqf\-\ =\-\ fc*((pie*sqrt(rho\_q\verb|^|2+1/rho\_q\verb|^|2))*d.M\_q\_n0/(2*pie*sinh(log(d.rho\_q)*(d.N\_q\_n0+1))));\color{Green}
$\\$
$\\$
$\\$\color{Green}$\%$\-\ interpolation\-\ error
$\\$\color{BrickRed}lam\_n0\_psi\-\ =\-\ (2/pie)*log(d.N\_psi\_n0)+(2/pie)*(nm('0.6')+log(8/pie)+pie/(72*d.N\_psi\_n0));\color{Green}
$\\$\color{BrickRed}err\_k\_q\-\ =\-\ err\_interp\_der+lam\_n0\_psi*err\_Dqf;\color{Green}
$\\$\color{BrickRed}err\_k\_q\-\ =\-\ nm(-err\_k\_q,err\_k\_q)+1i*nm(-err\_k\_q,err\_k\_q);\color{Green}
$\\$
$\\$
$\\$\color{Green}$\%$-----------------------------------------------------------
$\\$\color{Green}$\%$\-\ \-\ $f0\_k$
$\\$\color{Green}$\%$-----------------------------------------------------------
$\\$
$\\$
$\\$\color{BrickRed}k\-\ =\-\ k0;\color{Green}
$\\$
$\\$
$\\$\color{BrickRed}kappa\-\ =\-\ kappa\_of\_k(k);\color{Green}
$\\$\color{BrickRed}K\-\ =\-\ elliptic\_integral(k,1);\color{Green}
$\\$\color{BrickRed}E\-\ =\-\ elliptic\_integral(k,2);\color{Green}
$\\$
$\\$
$\\$\color{BrickRed}R\-\ =\-\ elliptic\_integral(sqrt(1-k\verb|^|2),1);\color{Green}
$\\$\color{BrickRed}T\-\ =\-\ elliptic\_integral(sqrt(1-k\verb|^|2),2);\color{Green}
$\\$
$\\$
$\\$\color{Green}$\%$\-\ Derivative\-\ of\-\ complete\-\ elliptic\-\ integral\-\ of\-\ the\-\ first\-\ kind
$\\$\color{Green}$\%$\-\ \color{Black}$\pd{}{k}K(k)$
\color{BrickRed}K\_k\-\ =\-\ -K/k+E/(k*(1-k\verb|^|2));\color{Black}

$\\$
\color{BrickRed}sqk\-\ =\-\ sqrt(1-k\verb|^|2);\color{Black}
\color{BrickRed}R\_k\-\ =\-\ -k*(-R/sqk+T/(k\verb|^|2*sqk))/sqk;\color{Black}

$\\$
\color{BrickRed}q\_k\-\ =\-\ (-pie*R\_k/K\-\ +\-\ pie*K\_k*R/K\verb|^|2)*exp(-pie*R/K);\color{Black}

$\\$
\color{BrickRed}qtilde\_q\-\ =\-\ 2/(d.b\_q-d.a\_q);\color{Black}

$\\$
\color{BrickRed}omega\-\ =\-\ pie/kappa;\color{Black}

$\\$
\color{BrickRed}kap\_k\-\ =\-\ \-\ pie*sqrt((E*(2*k\verb|^|4\-\ -\-\ 2*k\verb|^|2\-\ +\-\ 2)\-\ +\-\ K*(-k\verb|^|2\-\ +\-\ 2)*(k\verb|^|2\-\ -\-\ 1))/(E*(-2*k\verb|^|6\-\ +\-\ 3*k\verb|^|4\-\ +\-\ 3*k\verb|^|2\-\ -\-\ 2)\-\ +\-\ K*(k\verb|^|6\-\ +\-\ k\verb|^|4\-\ -\-\ 4*k\verb|^|2\-\ +\-\ 2)))*(E*(-2*k\verb|^|6\-\ +\-\ 3*k\verb|^|4\-\ +\-\ 3*k\verb|^|2\-\ -\-\ 2)\-\ +\-\ K*(k\verb|^|6\-\ +\-\ k\verb|^|4\-\ -\-\ 4*k\verb|^|2\-\ +\-\ 2))*((E*(2*k\verb|^|4\-\ -\-\ 2*k\verb|^|2\-\ +\-\ 2)\-\ +\-\ K*(-k\verb|^|2\-\ +\-\ 2)*(k\verb|^|2\-\ -\-\ 1))*(-E*(-12*k\verb|^|5\-\ +\-\ 12*k\verb|^|3\-\ +\-\ 6*k)\-\ -\-\ K*(6*k\verb|^|5\-\ +\-\ 4*k\verb|^|3\-\ -\-\ 8*k)\-\ -\-\ (E/(k*(-k\verb|^|2\-\ +\-\ 1))\-\ -\-\ K/k)*(k\verb|^|6\-\ +\-\ k\verb|^|4\-\ -\-\ 4*k\verb|^|2\-\ +\-\ 2)\-\ -\-\ (E\-\ -\-\ K)*(-2*k\verb|^|6\-\ +\-\ 3*k\verb|^|4\-\ +\-\ 3*k\verb|^|2\-\ -\-\ 2)/k)/(2*(E*(-2*k\verb|^|6\-\ +\-\ 3*k\verb|^|4\-\ +\-\ 3*k\verb|^|2\-\ -\-\ 2)\-\ +\-\ K*(k\verb|^|6\-\ +\-\ k\verb|^|4\-\ -\-\ 4*k\verb|^|2\-\ +\-\ 2))\verb|^|2)\-\ +\-\ (E*(8*k\verb|^|3\-\ -\-\ 4*k)\-\ +\-\ 2*K*k*(-k\verb|^|2\-\ +\-\ 2)\-\ -\-\ 2*K*k*(k\verb|^|2\-\ -\-\ 1)\-\ +\-\ (-k\verb|^|2\-\ +\-\ 2)*(k\verb|^|2\-\ -\-\ 1)*(E/(k*(-k\verb|^|2\-\ +\-\ 1))\-\ -\-\ K/k)\-\ +\-\ (E\-\ -\-\ K)*(2*k\verb|^|4\-\ -\-\ 2*k\verb|^|2\-\ +\-\ 2)/k)/(2*(E*(-2*k\verb|^|6\-\ +\-\ 3*k\verb|^|4\-\ +\-\ 3*k\verb|^|2\-\ -\-\ 2)\-\ +\-\ K*(k\verb|^|6\-\ +\-\ k\verb|^|4\-\ -\-\ 4*k\verb|^|2\-\ +\-\ 2))))/(K*(E*(2*k\verb|^|4\-\ -\-\ 2*k\verb|^|2\-\ +\-\ 2)\-\ +\-\ K*(-k\verb|^|2\-\ +\-\ 2)*(k\verb|^|2\-\ -\-\ 1)))\-\ -\-\ pie*sqrt((E*(2*k\verb|^|4\-\ -\-\ 2*k\verb|^|2\-\ +\-\ 2)\-\ +\-\ K*(-k\verb|^|2\-\ +\-\ 2)*(k\verb|^|2\-\ -\-\ 1))/(E*(-2*k\verb|^|6\-\ +\-\ 3*k\verb|^|4\-\ +\-\ 3*k\verb|^|2\-\ -\-\ 2)\-\ +\-\ K*(k\verb|^|6\-\ +\-\ k\verb|^|4\-\ -\-\ 4*k\verb|^|2\-\ +\-\ 2)))*(E/(k*(-k\verb|^|2\-\ +\-\ 1))\-\ -\-\ K/k)/K\verb|^|2\-\ ;\color{Black}
\color{BrickRed}kappa\_k\-\ =\-\ sqrt(nm(7)/20)*kap\_k;\color{Black}
\color{BrickRed}omega\_k\-\ =\-\ -pi*kappa\_k/kappa\verb|^|2;\color{Black}

$\\$

$\\$
\color{BrickRed}f1\_tilde\_q\-\ =\-\ strict\_taylor(cf(:,:,1),q\_tilde0,psi\_tilde0,3)+err\_k\_q;\color{Black}
\color{BrickRed}f2\_tilde\_q\-\ =\-\ strict\_taylor(cf(:,:,3),q\_tilde0,psi\_tilde0,3)+err\_k\_q;\color{Black}
\color{BrickRed}f2\-\ =\-\ strict\_taylor(cf(:,:,3),q\_tilde0,psi\_tilde0,1)+err;\color{Black}

$\\$
\color{BrickRed}f1\_k\-\ =\-\ q\_k*qtilde\_q*f1\_tilde\_q;\color{Black}
\color{BrickRed}f2\_k\-\ =\-\ q\_k*qtilde\_q*f2\_tilde\_q;\color{Black}

$\\$
\color{BrickRed}f\_k\-\ =\-\ f1\_k+f2\_k/omega\verb|^|2-2*f2*omega\_k/omega\verb|^|3;\color{Black}

$\\$
\color{Black}-----------------------------------------------------------
\color{Black}  $f0_{\psi\psi}$
\color{Black}-----------------------------------------------------------

$\\$
\color{Black} Next we find the 2-d interpolation error for the derivative of f with
\color{Black} respect to q

$\\$
\color{BrickRed}Drhopsi\-\ =\-\ (d.rho\_psi+1/d.rho\_psi)/2-1;\color{Black}
\color{BrickRed}fc\-\ =\-\ (d.N\_psi\_n0+1)*(d.N\_psi\_n0+3)/Drhopsi+1/Drhopsi\verb|^|2;\color{Black}
\color{BrickRed}err\_Dpsif\-\ =\-\ fc*((pie*sqrt(rho\_psi\verb|^|2+1/rho\_psi\verb|^|2))*d.M\_psi\_n0/(2*pie*sinh(log(d.rho\_psi)*(d.N\_psi\_n0+1))));\color{Black}

$\\$
\color{Black} interpolation error
\color{BrickRed}lam\_n0\_q\-\ =\-\ (2/pie)*log(d.N\_q\_n0)+(2/pie)*(nm('0.6')+log(8/pie)+pie/(72*d.N\_q\_n0));\color{Black}
\color{BrickRed}err\_k\_psi\-\ =\-\ err\_interp\_der+lam\_n0\_q*err\_Dpsif;\color{Black}
\color{BrickRed}err\_k\_psi\-\ =\-\ nm(-err\_k\_psi,err\_k\_psi)+1i*nm(-err\_k\_psi,err\_k\_psi);\color{Black}

$\\$
\color{BrickRed}psitilde\_psi\-\ =\-\ nm(4);\color{Black}

$\\$
\color{BrickRed}f1\_psi\_psitilde\-\ =\-\ strict\_taylor(cf(:,:,2),q\_tilde0,psi\_tilde0,2)+err\_k\_psi;\color{Black}
\color{BrickRed}f2\_psi\_psitilde\-\ =\-\ strict\_taylor(cf(:,:,4),q\_tilde0,psi\_tilde0,2)+err\_k\_psi;\color{Black}

$\\$
\color{BrickRed}f1\_psi\_psi\-\ =\-\ psitilde\_psi*f1\_psi\_psitilde;\color{Black}
\color{BrickRed}f2\_psi\_psi\-\ =\-\ psitilde\_psi*f2\_psi\_psitilde;\color{Black}

$\\$
\color{BrickRed}f\_psi\_psi\-\ =\-\ f1\_psi\_psi\-\ +\-\ f2\_psi\_psi/omega\verb|^|2;\color{Black}

$\\$
\color{BrickRed}mx\_fk\-\ =\-\ max(sup(imag(f\_k)));\color{Black}
\color{BrickRed}mn\_fpsipsi\-\ =\-\ min(inf(imag(f\_psi\_psi)));\color{Black}

$\\$
\color{Black}\section{strict\_transition\_lower.m}

\color{Green}\color{BrickRed}\color{NavyBlue}\-\ function\-\ \color{BrickRed}\-\ strict\_transition\_lower(d,k,psi0,psiL,min\_diff\_q,rho\_q\_sm)\color{Green}
$\\$
$\\$
$\\$\color{Green}$\%$-----------------------------------------------------------
$\\$\color{Green}$\%$\-\ \-\ constants
$\\$\color{Green}$\%$-----------------------------------------------------------
$\\$
$\\$
$\\$\color{BrickRed}pie\-\ =\-\ nm('pi');\color{Green}
$\\$
$\\$
$\\$\color{BrickRed}psi\_tilde0\-\ =\-\ 2*psi0-1;\color{Green}
$\\$
$\\$
$\\$\color{Green}$\%$\-\ constants\-\ for\-\ transformation\-\ in\-\ q
$\\$\color{BrickRed}c1\_q\-\ =\-\ 2/(d.b\_q-d.a\_q);\color{Green}
$\\$\color{BrickRed}c2\_q\-\ =\-\ (d.a\_q+d.b\_q)/2;\color{Green}
$\\$\color{Green}$\%$\-\ elliptic\-\ integrals
$\\$\color{BrickRed}kappa\-\ =\-\ kappa\_of\_k(k);\color{Green}
$\\$\color{BrickRed}elipk\-\ =\-\ elliptic\_integral(k,1);\color{Green}
$\\$\color{BrickRed}elipk2\-\ =\-\ elliptic\_integral(sqrt(1-k.\verb|^|2),1);\color{Green}
$\\$\color{BrickRed}q0\-\ =\-\ exp(-pie*elipk2./elipk);\color{Green}
$\\$\color{BrickRed}omega0\-\ =\-\ pie/kappa;\color{Green}
$\\$
$\\$
$\\$\color{BrickRed}\color{NavyBlue}\-\ if\-\ \color{BrickRed}\-\ inf(q0)\-\ $<$\-\ d.a\_q\color{Green}
$\\$\color{BrickRed}\-\ \-\ \-\ \-\ error('q\-\ out\-\ of\-\ range')\color{Green}
$\\$\color{BrickRed}\color{NavyBlue}\-\ end\-\ \color{BrickRed}\color{Green}
$\\$\color{BrickRed}\color{NavyBlue}\-\ if\-\ \color{BrickRed}\-\ sup(q0)\-\ $>$\-\ d.b\_q\color{Green}
$\\$\color{BrickRed}\-\ \-\ \-\ \-\ error('q\-\ out\-\ of\-\ range')\color{Green}
$\\$\color{BrickRed}\color{NavyBlue}\-\ end\-\ \color{BrickRed}\color{Green}
$\\$
$\\$
$\\$\color{Green}$\%$\-\ get\-\ theta\-\ values\-\ for\-\ q
$\\$\color{BrickRed}q\_tilde0\-\ =\-\ c1\_q*(q0-c2\_q);\color{Green}
$\\$
$\\$
$\\$\color{Green}$\%$\-\ interpolation\-\ coefficients
$\\$\color{BrickRed}cf\-\ =\-\ d.cfn1;\color{Green}
$\\$
$\\$
$\\$\color{Green}$\%$\-\ interpolation\-\ error
$\\$\color{BrickRed}lam\_n1\_q\-\ =\-\ (2/pie)*log(d.N\_q\_n1)+(2/pie)*(nm('0.6')+log(8/pie)+pie/(72*d.N\_q\_n1));\color{Green}
$\\$\color{BrickRed}err\-\ =\-\ d.err\_q\_n1+lam\_n1\_q*d.err\_psi\_n1;\color{Green}
$\\$\color{BrickRed}err\-\ =\-\ nm(-err,err)+1i*nm(-err,err);\color{Green}
$\\$
$\\$
$\\$\color{Green}$\%$\-\ $\%$-----------------------------------------------------------
$\\$\color{Green}$\%$\-\ $\%$\-\ \-\ $f0\_{\textbackslash psi}$\-\ and\-\ $g0\_{\textbackslash psi}$
$\\$\color{Green}$\%$\-\ $\%$-----------------------------------------------------------
$\\$
$\\$
$\\$\color{BrickRed}f1\_psi\-\ =\-\ strict\_taylor(cf(:,:,2),q\_tilde0,psi\_tilde0,1)+err;\color{Green}
$\\$\color{BrickRed}f2\_psi\-\ =\-\ strict\_taylor(cf(:,:,4),q\_tilde0,psi\_tilde0,1)+err;\color{Green}
$\\$
$\\$
$\\$\color{Green}$\%$\-\ derivative\-\ with\-\ respect\-\ to\-\ psi\-\ (not\-\ tilde\-\ psi)
$\\$\color{BrickRed}f0\_psi\-\ =\-\ f1\_psi\-\ +\-\ f2\_psi/omega0\verb|^|2;\color{Green}
$\\$
$\\$
$\\$\color{BrickRed}g0\_psi\-\ =\-\ strict\_taylor(cf(:,:,6),q\_tilde0,psi\_tilde0,1)+err;\color{Green}
$\\$
$\\$
$\\$\color{Green}$\%$-----------------------------------------------------------
$\\$\color{Green}$\%$\-\ \-\ interpolation\-\ error\-\ of\-\ interpolant\-\ derivative
$\\$\color{Green}$\%$-----------------------------------------------------------
$\\$
$\\$
$\\$\color{Green}$\%$\-\ To\-\ bound\-\ the\-\ interpoaltion\-\ error\-\ coming\-\ from\-\ approximating\-\ the\-\ derivative\-\ of\-\ f
$\\$\color{Green}$\%$\-\ with\-\ the\-\ derivative\-\ of\-\ the\-\ interpolating\-\ polynomial\-\ p,\-\ we\-\ need\-\ a\-\ bound\-\ on\-\ 
$\\$\color{Green}$\%$\-\ the\-\ derivative\-\ of\-\ f\-\ on\-\ a\-\ stadium.\-\ We\-\ use\-\ Cauchy's\-\ integral\-\ formula\-\ to\-\ get\-\ that\-\ bound.\-\ 
$\\$\color{Green}$\%$\-\ We\-\ already\-\ have\-\ a\-\ bound\-\ on\-\ f\-\ on\-\ a\-\ stadium\-\ of\-\ radius\-\ rho\_q.\-\ We\-\ choose\-\ rho\_q\_sm\-\ and
$\\$\color{Green}$\%$\-\ rho\_psi\_sm\-\ so\-\ that\-\ the\-\ smaller\-\ ellipse\-\ has\-\ a\-\ minor\-\ axis\-\ of\-\ half\-\ the\-\ length\-\ of\-\ the\-\ larger\-\ one.
$\\$\color{Green}$\%$\-\ Then\-\ we\-\ use\-\ Cauchy's\-\ integral\-\ formula\-\ with\-\ f\-\ on\-\ the\-\ larger\-\ stadium\-\ to\-\ get\-\ a\-\ bound\-\ on\-\ the
$\\$\color{Green}$\%$\-\ derivative\-\ of\-\ f\-\ on\-\ the\-\ smaller\-\ stadium.\-\ We\-\ then\-\ use\-\ the\-\ radius\-\ of\-\ the\-\ smaller\-\ stadium\-\ to\-\ get
$\\$\color{Green}$\%$\-\ the\-\ needed\-\ interpolation\-\ error\-\ bounds\-\ for\-\ the\-\ derivative\-\ of\-\ f.
$\\$
$\\$
$\\$\color{Green}$\%$\-\ This\-\ is\-\ a\-\ bound\-\ on\-\ the\-\ derivative\-\ of\-\ f\-\ on\-\ the\-\ smaller\-\ stadium
$\\$\color{Green}$\%$\-\ derived\-\ using\-\ Cauchy's\-\ integral\-\ formula
$\\$\color{BrickRed}M\_Dq\-\ =\-\ (1/nm(2))*sqrt(d.rho\_q\verb|^|2+1/d.rho\_q\verb|^|2)*d.M\_q\_n1/min\_diff\_q\verb|^|2;\color{Green}
$\\$\color{Green}$\%$\-\ This\-\ is\-\ a\-\ bound\-\ on\-\ the\-\ length\-\ of\-\ the\-\ smaller\-\ stadium
$\\$\color{BrickRed}Lq\-\ =\-\ pie*sqrt(rho\_q\_sm\verb|^|2+1/rho\_q\_sm\verb|^|2);\color{Green}
$\\$\color{Green}$\%$\-\ This\-\ is\-\ a\-\ lower\-\ bound\-\ on\-\ the\-\ distance\-\ between\-\ the
$\\$\color{Green}$\%$\-\ stadium\-\ and\-\ the\-\ line\-\ on\-\ which\-\ we\-\ interpolate
$\\$\color{BrickRed}Ddq\-\ =\-\ (rho\_q\_sm+1/rho\_q\_sm)/2-1;\color{Green}
$\\$\color{BrickRed}eta\_q\-\ =\-\ log(rho\_q\_sm);\color{Green}
$\\$\color{BrickRed}err\_Dq\-\ =\-\ M\_Dq*Lq/(pie*Ddq*sinh(eta\_q*(d.N\_q\_n1+1)));\color{Green}
$\\$
$\\$
$\\$\color{Green}$\%$\-\ Next\-\ we\-\ find\-\ the\-\ 2-d\-\ interpolation\-\ error\-\ for\-\ the\-\ derivative\-\ of\-\ f\-\ with
$\\$\color{Green}$\%$\-\ respect\-\ to\-\ q
$\\$
$\\$
$\\$\color{BrickRed}Drhoq\-\ =\-\ (d.rho\_q+1/d.rho\_q)/2-1;\color{Green}
$\\$\color{BrickRed}fc\-\ =\-\ (d.N\_q\_n1+1)*(d.N\_q\_n1+3)/Drhoq+1/Drhoq\verb|^|2;\color{Green}
$\\$\color{BrickRed}err\_Dqf\-\ =\-\ fc*((pie*sqrt(d.rho\_q\verb|^|2+1/d.rho\_q\verb|^|2))*d.M\_q\_n1/(2*pie*sinh(log(d.rho\_q)*(d.N\_q\_n1+1))));\color{Green}
$\\$
$\\$
$\\$\color{Green}$\%$\-\ interpolation\-\ error
$\\$\color{BrickRed}lam\_n1\_psi\-\ =\-\ (2/pie)*log(d.N\_psi\_n1)+(2/pie)*(nm('0.6')+log(8/pie)+pie/(72*d.N\_psi\_n1));\color{Green}
$\\$\color{BrickRed}err\_k\_q\-\ =\-\ err\_Dq+lam\_n1\_psi*err\_Dqf;\color{Green}
$\\$\color{BrickRed}err\_k\_q\-\ =\-\ nm(-err\_k\_q,err\_k\_q)+1i*nm(-err\_k\_q,err\_k\_q);\color{Green}
$\\$
$\\$
$\\$\color{Green}$\%$-----------------------------------------------------------
$\\$\color{Green}$\%$\-\ \-\ auxiliary\-\ for\-\ taking\-\ derivative\-\ in\-\ k
$\\$\color{Green}$\%$-----------------------------------------------------------
$\\$
$\\$
$\\$\color{BrickRed}kappa\-\ =\-\ kappa\_of\_k(k);\color{Green}
$\\$\color{BrickRed}K\-\ =\-\ elliptic\_integral(k,1);\color{Green}
$\\$\color{BrickRed}E\-\ =\-\ elliptic\_integral(k,2);\color{Green}
$\\$
$\\$
$\\$\color{BrickRed}R\-\ =\-\ elliptic\_integral(sqrt(1-k\verb|^|2),1);\color{Green}
$\\$\color{BrickRed}T\-\ =\-\ elliptic\_integral(sqrt(1-k\verb|^|2),2);\color{Green}
$\\$
$\\$
$\\$\color{Green}$\%$\-\ Derivative\-\ of\-\ complete\-\ elliptic\-\ integral\-\ of\-\ the\-\ first\-\ kind
$\\$\color{Green}$\%$\-\ \color{Black}$\pd{}{k}K(k)$
\color{BrickRed}K\_k\-\ =\-\ -K/k+E/(k*(1-k\verb|^|2));\color{Black}

$\\$
\color{BrickRed}sqk\-\ =\-\ sqrt(1-k\verb|^|2);\color{Black}
\color{BrickRed}R\_k\-\ =\-\ -k*(-R/sqk+T/(k\verb|^|2*sqk))/sqk;\color{Black}

$\\$
\color{BrickRed}q\_k\-\ =\-\ (-pie*R\_k/K\-\ +\-\ pie*K\_k*R/K\verb|^|2)*exp(-pie*R/K);\color{Black}

$\\$
\color{BrickRed}qtilde\_q\-\ =\-\ 2/(d.b\_q-d.a\_q);\color{Black}

$\\$
\color{BrickRed}omega\-\ =\-\ pie/kappa;\color{Black}

$\\$
\color{BrickRed}kap\_k\-\ =\-\ \-\ pie*sqrt((E*(2*k\verb|^|4\-\ -\-\ 2*k\verb|^|2\-\ +\-\ 2)\-\ +\-\ K*(-k\verb|^|2\-\ +\-\ 2)*(k\verb|^|2\-\ -\-\ 1))/(E*(-2*k\verb|^|6\-\ +\-\ 3*k\verb|^|4\-\ +\-\ 3*k\verb|^|2\-\ -\-\ 2)\-\ +\-\ K*(k\verb|^|6\-\ +\-\ k\verb|^|4\-\ -\-\ 4*k\verb|^|2\-\ +\-\ 2)))*(E*(-2*k\verb|^|6\-\ +\-\ 3*k\verb|^|4\-\ +\-\ 3*k\verb|^|2\-\ -\-\ 2)\-\ +\-\ K*(k\verb|^|6\-\ +\-\ k\verb|^|4\-\ -\-\ 4*k\verb|^|2\-\ +\-\ 2))*((E*(2*k\verb|^|4\-\ -\-\ 2*k\verb|^|2\-\ +\-\ 2)\-\ +\-\ K*(-k\verb|^|2\-\ +\-\ 2)*(k\verb|^|2\-\ -\-\ 1))*(-E*(-12*k\verb|^|5\-\ +\-\ 12*k\verb|^|3\-\ +\-\ 6*k)\-\ -\-\ K*(6*k\verb|^|5\-\ +\-\ 4*k\verb|^|3\-\ -\-\ 8*k)\-\ -\-\ (E/(k*(-k\verb|^|2\-\ +\-\ 1))\-\ -\-\ K/k)*(k\verb|^|6\-\ +\-\ k\verb|^|4\-\ -\-\ 4*k\verb|^|2\-\ +\-\ 2)\-\ -\-\ (E\-\ -\-\ K)*(-2*k\verb|^|6\-\ +\-\ 3*k\verb|^|4\-\ +\-\ 3*k\verb|^|2\-\ -\-\ 2)/k)/(2*(E*(-2*k\verb|^|6\-\ +\-\ 3*k\verb|^|4\-\ +\-\ 3*k\verb|^|2\-\ -\-\ 2)\-\ +\-\ K*(k\verb|^|6\-\ +\-\ k\verb|^|4\-\ -\-\ 4*k\verb|^|2\-\ +\-\ 2))\verb|^|2)\-\ +\-\ (E*(8*k\verb|^|3\-\ -\-\ 4*k)\-\ +\-\ 2*K*k*(-k\verb|^|2\-\ +\-\ 2)\-\ -\-\ 2*K*k*(k\verb|^|2\-\ -\-\ 1)\-\ +\-\ (-k\verb|^|2\-\ +\-\ 2)*(k\verb|^|2\-\ -\-\ 1)*(E/(k*(-k\verb|^|2\-\ +\-\ 1))\-\ -\-\ K/k)\-\ +\-\ (E\-\ -\-\ K)*(2*k\verb|^|4\-\ -\-\ 2*k\verb|^|2\-\ +\-\ 2)/k)/(2*(E*(-2*k\verb|^|6\-\ +\-\ 3*k\verb|^|4\-\ +\-\ 3*k\verb|^|2\-\ -\-\ 2)\-\ +\-\ K*(k\verb|^|6\-\ +\-\ k\verb|^|4\-\ -\-\ 4*k\verb|^|2\-\ +\-\ 2))))/(K*(E*(2*k\verb|^|4\-\ -\-\ 2*k\verb|^|2\-\ +\-\ 2)\-\ +\-\ K*(-k\verb|^|2\-\ +\-\ 2)*(k\verb|^|2\-\ -\-\ 1)))\-\ -\-\ pie*sqrt((E*(2*k\verb|^|4\-\ -\-\ 2*k\verb|^|2\-\ +\-\ 2)\-\ +\-\ K*(-k\verb|^|2\-\ +\-\ 2)*(k\verb|^|2\-\ -\-\ 1))/(E*(-2*k\verb|^|6\-\ +\-\ 3*k\verb|^|4\-\ +\-\ 3*k\verb|^|2\-\ -\-\ 2)\-\ +\-\ K*(k\verb|^|6\-\ +\-\ k\verb|^|4\-\ -\-\ 4*k\verb|^|2\-\ +\-\ 2)))*(E/(k*(-k\verb|^|2\-\ +\-\ 1))\-\ -\-\ K/k)/K\verb|^|2\-\ ;\color{Black}
\color{BrickRed}kappa\_k\-\ =\-\ sqrt(nm(7)/20)*kap\_k;\color{Black}
\color{BrickRed}omega\_k\-\ =\-\ -pi*kappa\_k/kappa\verb|^|2;\color{Black}

$\\$
\color{Black}-----------------------------------------------------------
\color{Black}  $f0_{\psi k}$ and $g_{\psi k}$
\color{Black}-----------------------------------------------------------

$\\$
\color{BrickRed}f1\_psi\_tilde\_q\-\ =\-\ strict\_taylor(cf(:,:,2),q\_tilde0,psi\_tilde0,3)+err\_k\_q;\color{Black}
\color{BrickRed}f2\_psi\_tilde\_q\-\ =\-\ strict\_taylor(cf(:,:,4),q\_tilde0,psi\_tilde0,3)+err\_k\_q;\color{Black}

$\\$
\color{BrickRed}f1\_psi\_k\-\ =\-\ q\_k*qtilde\_q*f1\_psi\_tilde\_q;\color{Black}
\color{BrickRed}f2\_psi\_k\-\ =\-\ q\_k*qtilde\_q*f2\_psi\_tilde\_q;\color{Black}

$\\$
\color{BrickRed}f0\_psi\_k\-\ =\-\ f1\_psi\_k+f2\_psi\_k/omega\verb|^|2-2*f2\_psi*omega\_k/omega\verb|^|3;\color{Black}

$\\$
\color{BrickRed}g\_psi\_tilde\_q\-\ =\-\ strict\_taylor(cf(:,:,6),q\_tilde0,psi\_tilde0,3)+err\_k\_q;\color{Black}
\color{BrickRed}g0\_psi\_k\-\ =\-\ q\_k*qtilde\_q*g\_psi\_tilde\_q;\color{Black}

$\\$
\color{Black}-----------------------------------------------------------
\color{Black}  bound on $f_{\psi\psi k}$ and $g_{\psi\psi k}$
\color{Black}-----------------------------------------------------------

$\\$
\color{BrickRed}psi\_tildeL\-\ =\-\ 2*psiL-1;\color{Black}
\color{BrickRed}temp\-\ =\-\ nm(psi\_tildeL,psi\_tilde0);\color{Black}
\color{BrickRed}tilde\_psi\-\ =\-\ linspace(inf(temp),sup(temp),1000);\color{Black}

$\\$
\color{BrickRed}f1\_psi\_psi\_tilde\_q\-\ =\-\ strict\_taylor(cf(:,:,7),q\_tilde0,tilde\_psi,3)+err\_k\_q;\color{Black}
\color{BrickRed}f2\_psi\_psi\_tilde\_q\-\ =\-\ strict\_taylor(cf(:,:,8),q\_tilde0,tilde\_psi,3)+err\_k\_q;\color{Black}
\color{BrickRed}f2\_psi\_psi\-\ =\-\ strict\_taylor(cf(:,:,8),q\_tilde0,tilde\_psi,1)+err;\color{Black}

$\\$
\color{BrickRed}f1\_psi\_psi\_k\-\ =\-\ q\_k*qtilde\_q*f1\_psi\_psi\_tilde\_q;\color{Black}
\color{BrickRed}f2\_psi\_psi\_k\-\ =\-\ q\_k*qtilde\_q*f2\_psi\_psi\_tilde\_q;\color{Black}

$\\$
\color{BrickRed}f\_psi\_psi\_k\-\ =\-\ f1\_psi\_psi\_k+f2\_psi\_psi\_k/omega\verb|^|2-2*f2\_psi\_psi*omega\_k/omega\verb|^|3;\color{Black}

$\\$
\color{BrickRed}g\_psi\_psi\_tilde\_q\-\ =\-\ strict\_taylor(cf(:,:,9),q\_tilde0,tilde\_psi,3)+err\_k\_q;\color{Black}
\color{BrickRed}g\_psi\_psi\_k\-\ =\-\ q\_k*qtilde\_q*g\_psi\_psi\_tilde\_q;\color{Black}

$\\$
\color{BrickRed}M\_f\_psi\_psi\_k\-\ =\-\ max(sup(abs(f\_psi\_psi\_k)));\color{Black}

$\\$
\color{BrickRed}M\_g\_psi\_psi\_k\-\ =\-\ max(sup(abs(g\_psi\_psi\_k)));\color{Black}

$\\$
\color{Black}-----------------------------------------------------------
\color{Black}  bound on $f_{\psi\psi}$ and $g_{\psi\psi}$
\color{Black}-----------------------------------------------------------

$\\$
\color{BrickRed}f1\_psi\_psi\-\ =\-\ strict\_taylor(cf(:,:,7),q\_tilde0,tilde\_psi,1)+err;\color{Black}
\color{BrickRed}f2\_psi\_psi\-\ =\-\ strict\_taylor(cf(:,:,8),q\_tilde0,tilde\_psi,1)+err;\color{Black}
\color{BrickRed}f\_psi\_psi\-\ =\-\ f1\_psi\_psi+f2\_psi\_psi/omega\verb|^|2;\color{Black}

$\\$
\color{BrickRed}g\_psi\_psi\-\ =\-\ strict\_taylor(cf(:,:,9),q\_tilde0,tilde\_psi,1)+err;\color{Black}

$\\$
\color{BrickRed}M\_f\_psi\_psi\-\ =\-\ max(sup(abs(f\_psi\_psi)));\color{Black}

$\\$
\color{BrickRed}M\_g\_psi\_psi\-\ =\-\ max(sup(abs(g\_psi\_psi)));\color{Black}

$\\$

$\\$
\color{Black}-----------------------------------------------------------
\color{Black}  combine
\color{Black}-----------------------------------------------------------

$\\$
\color{BrickRed}wid\-\ =\-\ psi0-psiL;\color{Black}
\color{BrickRed}wid\-\ =\-\ nm(-wid,wid);\color{Black}

$\\$
\color{BrickRed}t1\-\ =\-\ f0\_psi\_k+M\_f\_psi\_psi\_k*wid/2;\color{Black}
\color{BrickRed}t2\-\ =\-\ g0\_psi+\-\ M\_g\_psi\_psi*wid/2;\color{Black}
\color{BrickRed}t3\-\ =\-\ f0\_psi\-\ +\-\ M\_f\_psi\_psi*wid/2;\color{Black}
\color{BrickRed}t4\-\ =\-\ g0\_psi\-\ +\-\ M\_g\_psi\_psi*wid/2;\color{Black}
\color{BrickRed}t5\-\ =\-\ g0\_psi\_k\-\ +\-\ M\_g\_psi\_psi\_k*wid/2;\color{Black}

$\\$
\color{BrickRed}out\-\ =\-\ t1/t2\-\ -\-\ (t3/t4)*(t5/t4);\color{Black}

$\\$

$\\$
\color{BrickRed}\color{NavyBlue}\-\ if\-\ \color{BrickRed}\-\ sup(real(out))\-\ $>$=\-\ 0\color{Black}
\color{BrickRed}\-\ \-\ \-\ \-\ error('failed\-\ to\-\ verify\-\ the\-\ derivative\-\ is\-\ negative')\color{Black}
\color{BrickRed}\color{NavyBlue}\-\ end\-\ \color{BrickRed}\color{Black}

$\\$

$\\$

$\\$

$\\$

$\\$
\color{Black}\section{theta\_vec.m}

\color{Green}\color{BrickRed}\color{NavyBlue}\-\ function\-\ \color{BrickRed}\-\ out\-\ =\-\ theta\_vec(q,psi,x,m,ntilde)\color{Green}
$\\$\color{Green}$\%$\-\ out\-\ =\-\ theta\_vec(q,psi,x,m,ntilde)
$\\$\color{Green}$\%$
$\\$\color{Green}$\%$\-\ Returns\-\ an\-\ interval\-\ enclosure\-\ of\-\ f(x)\-\ and\-\ its\-\ first\-\ m\-\ derivatives\-\ where
$\\$\color{Green}$\%$\-\ f(x)\-\ =\-\ v(pie(x\-\ +\-\ n)/2+i*pi*omega\_prime*(1+psi)/(2*omega))
$\\$\color{Green}$\%$\-\ and\-\ v(z)\-\ is\-\ the\-\ first\-\ Jacobi\-\ Theta\-\ function\-\ with\-\ nome\-\ q.
$\\$\color{Green}$\%$
$\\$\color{Green}$\%$\-\ The\-\ input\-\ should\-\ satisify\-\ -1\-\ $<$=\-\ x\-\ $<$=\-\ 1,\-\ 0\-\ $<$=\-\ psi\-\ $<$=\-\ 1,\-\ 0$<$q$<$1,\-\ ntilde\-\ =\-\ 0\-\ or\-\ 1,\-\ m\-\ $>$=\-\ 0
$\\$
$\\$
$\\$
$\\$\color{Black}
The first Jacobi Theta function is given by the series,
\eqn{
\vartheta_1(z)&= 2\sum_{n=0}^{\infty} (-1)^n q^{(n+1/2)^2}\sin((2n+1)z).
}{}

$\\$
From 
\eq{
\vartheta_1(z) = 2\sum_{n=1}^{\infty} (-1)^{n+1}q^{(n-1/2)^2}\sin((2n-1)z),
}{\notag}

$\\$
we find that
\eq{
f(x)&:= \vartheta_1\left(\frac{\pi}{2\omega}(\omega x + i\omega' + \tilde n \omega + i\psi \omega')\right)
\\ &= -i\sum_{n=1}^{\infty}(-1)^{n+1}q^{(n-1/2)^2} \left(\hat v^{(2n-1)}-\hat v^{-(2n-1)}\right),
}{\notag}
where $\hat v := e^{i\pi(x+\tilde n)/2}q^{(1+\psi)/2}$. Hence

$\\$
\eq{
\pd{^m}{x^m} f(x)&=  -i(i\pi/2)^m\sum_{n=1}^{\infty}(-1)^{n+1}q^{(n-1/2)^2} (2n-1)^m\left(\hat v^{(2n-1)}-\hat v^{-(2n-1)}\right).
}{\notag}

$\\$
We find that
\eq{
Err &:= \left|-i(i\pi/2)^m\sum_{n=N+1}^{\infty}(-1)^{n+1}q^{(n-1/2)^2} (2n-1)^m\left(\hat v^{(2n-1)}-\hat v^{-(2n+1)}\right)\right| 
\\ &\leq 2q^{1/4}(\pi/2)^m\sum_{n = N+1}^{\infty}q^{n^2/2}\left((2n-1)^mq^{n^2/2-n-(2n-1)(1+\psi)/2}\right)\\
&\leq 2q^{1/4}(\pi/2)^m\sum_{n = N+1}^{\infty}q^{n^2/2}\\
&\leq2q^{1/4}(\pi/2)^mq^{(N+1)^2/2}\frac{1}{1-q}, 
}{\notag}
as long as we take $N$ large enough so that $g(x):= \left((2x-1)^mq^{n^2/2-x-(2x-1)(1+\psi)/2}\right)$ satisfies $g(N)< 1$, $g(x) < 0$ whenever $x> N$.
\color{Green}
$\\$
$\\$\color{Green}$\%$$\%$
$\\$
$\\$
$\\$\color{Green}$\%$$\%$\-\ Error\-\ checking
$\\$\color{Green}$\%$
$\\$\color{Green}$\%$\-\ I\-\ tested\-\ against\-\ theta1\_m\-\ for\-\ accuracy,\-\ which\-\ was\-\ tested\-\ against\-\ Maple\-\ and\-\ Mathematica.\-\ I
$\\$\color{Green}$\%$\-\ found\-\ that\-\ they\-\ agreed.\-\ In\-\ testing,\-\ I\-\ found\-\ that\-\ this\-\ function\-\ has\-\ error\-\ radius\-\ about\-\ twice\-\ that\-\ of\-\ 
$\\$\color{Green}$\%$\-\ theta1\_m\-\ for\-\ m\-\ =\-\ 4,\-\ but\-\ that\-\ it\-\ computed\-\ much\-\ faster.\-\ For\-\ example,\-\ it\-\ took\-\ nearly\-\ 2\-\ minutes\-\ to\-\ 
$\\$\color{Green}$\%$\-\ evaluate\-\ theta1\_m\-\ on\-\ 200\-\ x\-\ points\-\ but\-\ only\-\ took\-\ about\-\ 0.2\-\ seconds\-\ with\-\ this\-\ vectorized\-\ version.
$\\$\color{Green}$\%$
$\\$\color{Green}$\%$\-\ Below\-\ is\-\ code\-\ used\-\ for\-\ testing:
$\\$\color{Green}$\%$\-\ x\-\ =\-\ nm(linspace(-1,1,30));
$\\$\color{Green}$\%$\-\ q\-\ =\-\ nm('0.5');
$\\$\color{Green}$\%$\-\ psi\-\ =\-\ nm(linspace(0,1,20));
$\\$\color{Green}$\%$\-\ xicon\-\ =\-\ 0;
$\\$\color{Green}$\%$\-\ xicon\_der\-\ =\-\ 0;
$\\$\color{Green}$\%$\-\ 
$\\$\color{Green}$\%$\-\ tic
$\\$\color{Green}$\%$\-\ val\-\ =\-\ integrands(x,q,psi,xicon,xicon\_der);
$\\$\color{Green}$\%$\-\ toc
$\\$\color{Green}$\%$\-\ 
$\\$\color{Green}$\%$\-\ pie\-\ =\-\ nm('pi');
$\\$\color{Green}$\%$\-\ con\-\ =\-\ pie/2;
$\\$\color{Green}$\%$\-\ diff\-\ =\-\ 0;
$\\$\color{Green}$\%$\-\ for\-\ j\-\ =\-\ 1:length(x)
$\\$\color{Green}$\%$\-\ \-\ \-\ \-\ \-\ for\-\ k\-\ =\-\ 1:length(psi)
$\\$\color{Green}$\%$\-\ \-\ \-\ \-\ \-\ \-\ \-\ \-\ \-\ 
$\\$\color{Green}$\%$\-\ \-\ \-\ \-\ \-\ \-\ \-\ \-\ \-\ z\-\ =\-\ pie*(x(j)\-\ +\-\ 1)/2-log(q)*1i*(1+psi(k))/(2);
$\\$\color{Green}$\%$\-\ \-\ \-\ \-\ \-\ \-\ \-\ \-\ \-\ temp\-\ =\-\ theta1\_m(z,q,4,1e-17);
$\\$\color{Green}$\%$\-\ \-\ \-\ \-\ \-\ \-\ \-\ \-\ \-\ for\-\ ind\-\ =\-\ 1:5
$\\$\color{Green}$\%$\-\ \-\ \-\ \-\ \-\ \-\ \-\ \-\ \-\ \-\ \-\ \-\ \-\ diff\-\ =\-\ max(diff,sup(temp(ind)*con\verb|^|(ind-1)-val(j,k,ind)));
$\\$\color{Green}$\%$\-\ \-\ \-\ \-\ \-\ \-\ \-\ \-\ \-\ end
$\\$\color{Green}$\%$\-\ \-\ \-\ \-\ \-\ end
$\\$\color{Green}$\%$\-\ end
$\\$\color{Green}$\%$\-\ 
$\\$\color{Green}$\%$\-\ disp(diff);
$\\$\color{Green}$\%$\-\ 5.975892748039020e-11\-\ +\-\ 5.958839722380777e-11i
$\\$
$\\$
$\\$\color{Green}$\%$$\%$
$\\$
$\\$
$\\$\color{Green}$\%$\-\ error\-\ target
$\\$\color{BrickRed}tol\-\ =\-\ 1e-17;\color{Green}
$\\$
$\\$
$\\$\color{Green}$\%$
$\\$\color{Green}$\%$\-\ constants
$\\$\color{Green}$\%$
$\\$
$\\$
$\\$\color{BrickRed}psi0\-\ =\-\ max(sup(psi));\color{Green}
$\\$\color{BrickRed}pie\-\ =\-\ nm('pi');\color{Green}
$\\$\color{BrickRed}one\-\ =\-\ nm(1);\color{Green}
$\\$\color{BrickRed}req\_1\-\ =\-\ sup(real(-2*m/log(q)));\color{Green}
$\\$\color{BrickRed}c1\-\ =\-\ nm(2+psi0);\color{Green}
$\\$\color{BrickRed}c2\-\ =\-\ nm(2*q\verb|^|(one/4)*(pie/2)\verb|^|m/(1-q));\color{Green}
$\\$\color{BrickRed}qsqrt\-\ \-\ =\-\ q\verb|^|(one/2);\color{Green}
$\\$\color{BrickRed}half\-\ =\-\ one/2;\color{Green}
$\\$
$\\$
$\\$\color{Green}$\%$
$\\$\color{Green}$\%$\-\ find\-\ N\-\ large\-\ enough\-\ that\-\ truncation\-\ error\-\ is\-\ less\-\ than\-\ tol
$\\$\color{Green}$\%$
$\\$
$\\$
$\\$\color{BrickRed}err\-\ =\-\ tol\-\ +\-\ 1;\color{Green}
$\\$\color{BrickRed}N\-\ =\-\ 0;\color{Green}
$\\$\color{BrickRed}maxit\-\ =\-\ 1000;\color{Green}
$\\$\color{BrickRed}\color{NavyBlue}\-\ while\-\ \color{BrickRed}\-\ err\-\ $>$\-\ tol\color{Green}
$\\$\color{BrickRed}\-\ \-\ \-\ \-\ N\-\ =\-\ N\-\ +\-\ 1;\color{Green}
$\\$\color{BrickRed}\-\ \-\ \-\ \-\ \color{NavyBlue}\-\ if\-\ \color{BrickRed}\-\ N\-\ $>$\-\ maxit\-\ \color{Green}
$\\$\color{BrickRed}\-\ \-\ \-\ \-\ \-\ \-\ \-\ \-\ error('maximum\-\ iterations\-\ exceeded');\color{Green}
$\\$\color{BrickRed}\-\ \-\ \-\ \-\ \color{NavyBlue}\-\ end\-\ \color{BrickRed}\color{Green}
$\\$\color{BrickRed}\-\ \-\ \-\ \-\ \color{Green}
$\\$\color{BrickRed}\-\ \-\ \-\ \-\ \color{Green}$\%$\-\ check\-\ that\-\ derivative\-\ of\-\ \color{Black} $f(x):= q^{x^2/2-x-(2x-1)(1+\psi)/2}(2x-1)^m$ 
\color{BrickRed}\-\ \-\ \-\ \-\ \color{Black} is decreasing for $x \geq N$. \color{Green}
$\\$\color{BrickRed}\-\ \-\ \-\ \-\ \color{NavyBlue}\-\ if\-\ \color{BrickRed}\-\ inf(real((N-c1)*(2*N-1)))\-\ $<$=\-\ req\_1\color{Green}
$\\$\color{BrickRed}\-\ \-\ \-\ \-\ \-\ \-\ \-\ \-\ \color{NavyBlue}\-\ continue\-\ \color{BrickRed}\color{Green}
$\\$\color{BrickRed}\-\ \-\ \-\ \-\ \color{NavyBlue}\-\ end\-\ \color{BrickRed}\color{Green}
$\\$\color{BrickRed}\-\ \-\ \-\ \-\ \color{Green}
$\\$\color{BrickRed}\-\ \-\ \-\ \-\ \color{Green}$\%$\-\ check\-\ that\-\ \color{Black} $f(x):= q^{N^2/2-N-(2N-1)(1+\psi)/2}(2N-1)^m\leq 1$ \color{Green}
$\\$\color{BrickRed}\-\ \-\ \-\ \-\ \color{NavyBlue}\-\ if\-\ \color{BrickRed}\-\ sup(real(q\verb|^|(N\verb|^|2/2-N-(2*N-1)*(1+psi0)/2)*(2*N-1)\verb|^|m))\-\ $>$\-\ 1\color{Green}
$\\$\color{BrickRed}\-\ \-\ \-\ \-\ \-\ \-\ \-\ \-\ \color{NavyBlue}\-\ continue\-\ \color{BrickRed}\color{Green}
$\\$\color{BrickRed}\-\ \-\ \-\ \-\ \color{NavyBlue}\-\ end\-\ \color{BrickRed}\color{Green}
$\\$\color{BrickRed}\-\ \-\ \-\ \-\ \color{Green}
$\\$\color{BrickRed}\-\ \-\ \-\ \-\ \color{Green}$\%$\-\ turncation\-\ error
$\\$\color{BrickRed}\-\ \-\ \-\ \-\ err\-\ =\-\ sup(real(c2*qsqrt\verb|^|((N+1)\verb|^|2)));\color{Green}
$\\$\color{BrickRed}\-\ \-\ \-\ \-\ \color{Green}
$\\$\color{BrickRed}\color{NavyBlue}\-\ end\-\ \color{BrickRed}\color{Green}
$\\$
$\\$
$\\$\color{Green}$\%$
$\\$\color{Green}$\%$\-\ evaluate\-\ the\-\ partial\-\ sum\-\ 
$\\$\color{Green}$\%$
$\\$
$\\$
$\\$\color{Green}$\%$\-\ initialize\-\ vectors
$\\$\color{BrickRed}out\-\ =\-\ nm(zeros(length(x),length(psi),m+1));\color{Green}
$\\$\color{BrickRed}temp\-\ =\-\ out;\color{Green}
$\\$
$\\$
$\\$\color{Green}$\%$\-\ rearrange\-\ dimensions\-\ if\-\ needed
$\\$\color{BrickRed}sx\-\ =\-\ size(x,1);\color{Green}
$\\$\color{BrickRed}\color{NavyBlue}\-\ if\-\ \color{BrickRed}\-\ sx\-\ ==\-\ 1\color{Green}
$\\$\color{BrickRed}\-\ \-\ \-\ \-\ x\-\ =x.';\color{Green}
$\\$\color{BrickRed}\color{NavyBlue}\-\ end\-\ \color{BrickRed}\color{Green}
$\\$
$\\$
$\\$\color{BrickRed}sx\-\ =\-\ size(psi,1);\color{Green}
$\\$\color{BrickRed}\color{NavyBlue}\-\ if\-\ \color{BrickRed}\-\ sx\-\ $>$\-\ 1\color{Green}
$\\$\color{BrickRed}\-\ \-\ \-\ \-\ psi\-\ =psi.';\color{Green}
$\\$\color{BrickRed}\color{NavyBlue}\-\ end\-\ \color{BrickRed}\color{Green}
$\\$
$\\$
$\\$\color{Green}$\%$
$\\$\color{Green}$\%$\-\ add\-\ partial\-\ sum
$\\$\color{Green}$\%$
$\\$
$\\$
$\\$
$\\$
$\\$\color{BrickRed}vhat\-\ =\-\ exp(1i*pie*(x+ntilde)/2)*q.\verb|^|((1+psi)/2);\color{Green}
$\\$
$\\$
$\\$\color{BrickRed}\color{NavyBlue}\-\ for\-\ \color{BrickRed}\-\ n\-\ =\-\ 1:N\color{Green}
$\\$\color{BrickRed}\-\ \-\ \-\ \-\ \color{Green}
$\\$\color{BrickRed}\-\ \-\ \-\ \-\ vtemp\-\ =\-\ vhat.\verb|^|(2*n-1);\color{Green}
$\\$\color{BrickRed}\-\ \-\ \-\ \-\ \color{NavyBlue}\-\ for\-\ \color{BrickRed}\-\ k\-\ =\-\ 0:m\color{Green}
$\\$\color{BrickRed}\-\ \-\ \-\ \-\ \-\ \-\ \-\ \-\ temp(:,:,k+1)\-\ =\-\ (2*n-1)\verb|^|k*(vtemp\-\ -(-1)\verb|^|k*vtemp.\verb|^|(-1));\color{Green}
$\\$\color{BrickRed}\-\ \-\ \-\ \-\ \color{NavyBlue}\-\ end\-\ \color{BrickRed}\color{Green}
$\\$\color{BrickRed}\-\ \-\ \-\ \-\ \-\ \-\ \color{Green}
$\\$\color{BrickRed}\-\ \-\ \-\ \-\ out\-\ =\-\ out\-\ +(-1)\verb|^|(n+1)*q\verb|^|((n-half)\verb|^|2)*temp;\color{Green}
$\\$\color{BrickRed}\-\ \-\ \-\ \-\ \color{Green}
$\\$\color{BrickRed}\color{NavyBlue}\-\ end\-\ \color{BrickRed}\color{Green}
$\\$
$\\$
$\\$\color{Green}$\%$\-\ complete\-\ differentiation\-\ rule
$\\$\color{BrickRed}\color{NavyBlue}\-\ for\-\ \color{BrickRed}\-\ j\-\ =\-\ 0:m\color{Green}
$\\$\color{BrickRed}\-\ \-\ \-\ \-\ out(:,:,j+1)\-\ =\-\ out(:,:,j+1)*(1i*pie/2)\verb|^|j;\color{Green}
$\\$\color{BrickRed}\color{NavyBlue}\-\ end\-\ \color{BrickRed}\color{Green}
$\\$
$\\$
$\\$\color{Green}$\%$\-\ multiply\-\ by\-\ -1i\-\ coming\-\ from\-\ sin\-\ definition\-\ and\-\ add\-\ error\-\ interval
$\\$\color{BrickRed}out\-\ =\-\ -1i*out\-\ +\-\ nm(-err,err)+1i*nm(-err,err);\color{Green}
$\\$
$\\$
$\\$\color{Black}\section{upper\_pinpoint.m}

\color{Green}\color{BrickRed}curr\_dir\-\ =\-\ local\_startup;\-\ clc;\color{Green}
$\\$\color{BrickRed}intvalinit('DisplayMidRad');\color{Green}
$\\$
$\\$
$\\$\color{Green}$\%$\-\ retrieve\-\ bounds\-\ on\-\ function
$\\$
$\\$
$\\$\color{BrickRed}file\_name\-\ =\-\ 'interp2d\_490\_to\_538';\color{Green}
$\\$
$\\$
$\\$\color{BrickRed}ld\-\ =\-\ retrieve\_it(curr\_dir,'interval\_arithmetic',file\_name,'data\_final');\color{Green}
$\\$\color{BrickRed}d\-\ =\-\ ld.var;\color{Green}
$\\$
$\\$
$\\$\color{Green}$\%$\-\ constants
$\\$\color{BrickRed}pie\-\ =\-\ nm('pi');\color{Green}
$\\$
$\\$
$\\$\color{BrickRed}ntilde\-\ =\-\ 0;\color{Green}
$\\$\color{Green}$\%$\-\ initialize\-\ function
$\\$\color{BrickRed}fun\-\ =\-\ \@(q,psi)(integrand\_numer(d.N\_x\_n0,d.err\_x\_n0,q,psi,ntilde));\color{Green}
$\\$
$\\$
$\\$\color{BrickRed}min\_wid\-\ =\-\ 1e-8;\color{Green}
$\\$
$\\$
$\\$\color{BrickRed}kstart\-\ =\-\ nm('0.9999989');\-\ \color{Green}
$\\$\color{BrickRed}psiL\-\ =\-\ inf(nm('0.7'));\color{Green}
$\\$\color{BrickRed}psiR\-\ =\-\ sup(nm('0.8'));\color{Green}
$\\$
$\\$
$\\$\color{Green}$\%$\-\ confirm\-\ stability\-\ above
$\\$\color{BrickRed}\color{NavyBlue}\-\ for\-\ \color{BrickRed}\-\ j\-\ =\-\ 18:53\color{Green}
$\\$\color{BrickRed}\-\ \-\ \-\ \-\ \color{Green}
$\\$\color{BrickRed}\-\ \-\ \-\ \-\ k\-\ =\-\ nm(mid(kstart-nm(1)/2\verb|^|j));\color{Green}
$\\$\color{BrickRed}\-\ \color{Green}
$\\$\color{BrickRed}\-\ \-\ \-\ \-\ kappa\-\ =\-\ kappa\_of\_k(k);\color{Green}
$\\$\color{BrickRed}\-\ \-\ \-\ \-\ elipk\-\ =\-\ elliptic\_integral(k,1);\color{Green}
$\\$\color{BrickRed}\-\ \-\ \-\ \-\ elipk2\-\ =\-\ elliptic\_integral(sqrt(1-k\verb|^|2),1);\color{Green}
$\\$\color{BrickRed}\-\ \-\ \-\ \-\ omega\-\ =\-\ pie/kappa;\color{Green}
$\\$\color{BrickRed}\-\ \-\ \-\ \-\ X\-\ =\-\ 2*pie/kappa;\color{Green}
$\\$\color{BrickRed}\-\ \-\ \-\ \-\ q\-\ =\-\ exp(-pie*elipk2/elipk);\color{Green}
$\\$
$\\$
$\\$\color{BrickRed}\-\ \-\ \-\ \-\ \color{NavyBlue}\-\ if\-\ \color{BrickRed}\-\ (sup(q)\-\ $>$\-\ inf(d.b\_q))||(inf(q)$<$\-\ sup(d.a\_q))\color{Green}
$\\$\color{BrickRed}\-\ \-\ \-\ \-\ \-\ \-\ \-\ \-\ error('q\-\ out\-\ of\-\ range');\color{Green}
$\\$\color{BrickRed}\-\ \-\ \-\ \-\ \color{NavyBlue}\-\ end\-\ \color{BrickRed}\color{Green}
$\\$\color{BrickRed}\-\ \-\ \-\ \-\ \color{Green}
$\\$\color{BrickRed}\-\ \-\ \-\ \-\ psiv\-\ =\-\ linspace(psiL,psiR,200);\color{Green}
$\\$\color{BrickRed}\-\ \-\ \-\ \-\ out\-\ =\-\ fun(q,psiv);\color{Green}
$\\$\color{BrickRed}\-\ \-\ \-\ \-\ f\-\ =\-\ out(:,:,1)+out(:,:,3)/omega\verb|^|2;\color{Green}
$\\$\color{BrickRed}\-\ \-\ \-\ \-\ f\-\ =\-\ sup(imag(f));\color{Green}
$\\$\color{BrickRed}\-\ \-\ \-\ \-\ \color{Green}
$\\$\color{BrickRed}\-\ \-\ \-\ \-\ \color{NavyBlue}\-\ if\-\ \color{BrickRed}\-\ sum(any(f$<$0))$>$0\color{Green}
$\\$\color{BrickRed}\-\ \-\ \-\ \-\ \-\ \-\ \-\ \-\ kstart\-\ =\-\ k;\color{Green}
$\\$\color{BrickRed}\-\ \-\ \-\ \-\ \-\ \-\ \-\ \-\ ind\-\ =\-\ find(f\-\ $<$\-\ 0);\color{Green}
$\\$\color{BrickRed}\-\ \-\ \-\ \-\ \-\ \-\ \-\ \-\ psiL\-\ =\-\ psiv(max(ind(1)-1,1));\color{Green}
$\\$\color{BrickRed}\-\ \-\ \-\ \-\ \-\ \-\ \-\ \-\ psiR\-\ =\-\ psiv(min(ind(end)+1,length(psiv)));\color{Green}
$\\$\color{BrickRed}\-\ \-\ \-\ \-\ \color{NavyBlue}\-\ end\-\ \color{BrickRed}\color{Green}
$\\$\color{BrickRed}\-\ \-\ \-\ \-\ \color{Green}
$\\$\color{BrickRed}\color{NavyBlue}\-\ end\-\ \color{BrickRed}\color{Green}
$\\$
$\\$
$\\$\color{BrickRed}kstart\-\ =\-\ sup(kstart);\color{Green}
$\\$
$\\$
$\\$\color{Green}$\%$\-\ confirm\-\ stability\-\ below
$\\$\color{BrickRed}\color{NavyBlue}\-\ for\-\ \color{BrickRed}\-\ j\-\ =\-\ 30:1:53\color{Green}
$\\$\color{BrickRed}\-\ \-\ \-\ \-\ \color{Green}
$\\$\color{BrickRed}\-\ \-\ \-\ \-\ k\-\ =\-\ nm(mid(kstart-nm(1)/2\verb|^|j));\color{Green}
$\\$\color{BrickRed}\-\ \-\ \-\ \-\ \-\ \color{Green}
$\\$\color{BrickRed}\-\ \-\ \-\ \-\ kappa\-\ =\-\ kappa\_of\_k(k);\color{Green}
$\\$\color{BrickRed}\-\ \-\ \-\ \-\ elipk\-\ =\-\ elliptic\_integral(k,1);\color{Green}
$\\$\color{BrickRed}\-\ \-\ \-\ \-\ elipk2\-\ =\-\ elliptic\_integral(sqrt(1-k\verb|^|2),1);\color{Green}
$\\$\color{BrickRed}\-\ \-\ \-\ \-\ omega\-\ =\-\ pie/kappa;\color{Green}
$\\$\color{BrickRed}\-\ \-\ \-\ \-\ X\-\ =\-\ 2*pie/kappa;\color{Green}
$\\$\color{BrickRed}\-\ \-\ \-\ \-\ q\-\ =\-\ exp(-pie*elipk2/elipk);\color{Green}
$\\$
$\\$
$\\$\color{BrickRed}\-\ \-\ \-\ \-\ \color{NavyBlue}\-\ if\-\ \color{BrickRed}\-\ (sup(q)\-\ $>$\-\ inf(d.b\_q))||(inf(q)$<$\-\ sup(d.a\_q))\color{Green}
$\\$\color{BrickRed}\-\ \-\ \-\ \-\ \-\ \-\ \-\ \-\ error('q\-\ out\-\ of\-\ range');\color{Green}
$\\$\color{BrickRed}\-\ \-\ \-\ \-\ \color{NavyBlue}\-\ end\-\ \color{BrickRed}\color{Green}
$\\$\color{BrickRed}\-\ \-\ \-\ \-\ \color{Green}
$\\$\color{BrickRed}\-\ \-\ \-\ \-\ psiL\-\ =\-\ inf(nm('0.7'));\color{Green}
$\\$\color{BrickRed}\-\ \-\ \-\ \-\ psiR\-\ =\-\ sup(nm('0.8'));\color{Green}
$\\$\color{BrickRed}\-\ \-\ \-\ \-\ \color{Green}
$\\$\color{BrickRed}\-\ \-\ \-\ \-\ psi\_keep\_going\-\ =\-\ 1;\color{Green}
$\\$\color{BrickRed}\-\ \-\ \-\ \-\ \color{NavyBlue}\-\ while\-\ \color{BrickRed}\-\ psi\_keep\_going\-\ ==\-\ 1\color{Green}
$\\$\color{BrickRed}\-\ \-\ \-\ \-\ \-\ \-\ \-\ \-\ \color{Green}
$\\$\color{BrickRed}\-\ \-\ \-\ \-\ \-\ \-\ \-\ \-\ psiv\-\ =\-\ linspace(psiL,psiR,200);\color{Green}
$\\$
$\\$
$\\$\color{BrickRed}\-\ \-\ \-\ \-\ \-\ \-\ \-\ \-\ out\-\ =\-\ fun(q,psiv);\color{Green}
$\\$\color{BrickRed}\-\ \-\ \-\ \-\ \-\ \-\ \-\ \-\ f\_psi\-\ =\-\ out(:,:,2)+out(:,:,4)/omega\verb|^|2;\color{Green}
$\\$
$\\$
$\\$\color{BrickRed}\-\ \-\ \-\ \-\ \-\ \-\ \-\ \-\ indL\-\ =\-\ find(sup(imag(f\_psi))\-\ $<$\-\ 0);\color{Green}
$\\$\color{BrickRed}\-\ \-\ \-\ \-\ \-\ \-\ \-\ \-\ indR\-\ =\-\ find(inf(imag(f\_psi))\-\ $>$\-\ 0);\color{Green}
$\\$
$\\$
$\\$\color{BrickRed}\-\ \-\ \-\ \-\ \-\ \-\ \-\ \-\ disp('\-\ ')\color{Green}
$\\$\color{BrickRed}\-\ \-\ \-\ \-\ \-\ \-\ \-\ \-\ psiL\-\ =\-\ psiv(indL(end));\color{Green}
$\\$\color{BrickRed}\-\ \-\ \-\ \-\ \-\ \-\ \-\ \-\ psiR\-\ =\-\ psiv(indR(1));\color{Green}
$\\$
$\\$
$\\$\color{BrickRed}\-\ \-\ \-\ \-\ \-\ \-\ \-\ \-\ temp\-\ =\-\ fun(q,nm(psiL,psiR));\color{Green}
$\\$
$\\$
$\\$\color{BrickRed}\-\ \-\ \-\ \-\ \-\ \-\ \-\ \-\ test\-\ =\-\ temp(:,:,1)+temp(:,:,3)/omega\verb|^|2;\color{Green}
$\\$
$\\$
$\\$\color{BrickRed}\-\ \-\ \-\ \-\ \-\ \-\ \-\ \-\ \color{NavyBlue}\-\ if\-\ \color{BrickRed}\-\ inf(imag(test))\-\ $>$\-\ 0\color{Green}
$\\$\color{BrickRed}\-\ \-\ \-\ \-\ \-\ \-\ \-\ \-\ \-\ \-\ \-\ \-\ psi\_keep\_going\-\ =\-\ 0;\color{Green}
$\\$\color{BrickRed}\-\ \-\ \-\ \-\ \-\ \-\ \-\ \-\ \color{NavyBlue}\-\ elseif\-\ \color{BrickRed}\-\ sup(imag(test))\-\ $<$\-\ 0\color{Green}
$\\$\color{BrickRed}\-\ \-\ \-\ \-\ \-\ \-\ \-\ \-\ \-\ \-\ \-\ \-\ k\_stab\-\ =\-\ nm(mid(kstart-nm(1)/2\verb|^|(j-1)));\color{Green}
$\\$\color{BrickRed}\-\ \-\ \-\ \-\ \-\ \-\ \-\ \-\ \-\ \-\ \-\ \-\ \color{NavyBlue}\-\ return\-\ \color{BrickRed}\color{Green}
$\\$\color{BrickRed}\-\ \-\ \-\ \-\ \-\ \-\ \-\ \-\ \color{NavyBlue}\-\ elseif\-\ \color{BrickRed}\-\ psiR-psiL\-\ $<$\-\ min\_wid\color{Green}
$\\$\color{BrickRed}\-\ \-\ \-\ \-\ \-\ \-\ \-\ \-\ \-\ \-\ \-\ \-\ k\_stab\-\ =\-\ nm(mid(kstart-nm(1)/2\verb|^|(j-1)));\color{Green}
$\\$\color{BrickRed}\-\ \-\ \-\ \-\ \-\ \-\ \-\ \-\ \-\ \-\ \-\ \-\ \color{NavyBlue}\-\ return\-\ \color{BrickRed}\color{Green}
$\\$\color{BrickRed}\-\ \-\ \-\ \-\ \-\ \-\ \-\ \-\ \color{NavyBlue}\-\ end\-\ \color{BrickRed}\color{Green}
$\\$\color{BrickRed}\-\ \-\ \-\ \-\ \-\ \-\ \-\ \-\ \color{Green}
$\\$\color{BrickRed}\-\ \-\ \-\ \-\ \color{NavyBlue}\-\ end\-\ \color{BrickRed}\color{Green}
$\\$\color{BrickRed}\color{NavyBlue}\-\ end\-\ \color{BrickRed}\color{Green}
$\\$\color{BrickRed}\-\ \-\ \-\ \-\ \-\ \color{Green}
$\\$\color{BrickRed}kstab\-\ =\-\ inf(k\_stab);\color{Green}
$\\$
$\\$
$\\$\color{Green}$\%$\-\ \-\ \-\ \-\ \-\ psiL\-\ =\-\ \-\ \-\ \-\ 0.759188696240224
$\\$\color{Green}$\%$\-\ \-\ \-\ \-\ \-\ psiR\-\ =\-\ \-\ \-\ \-\ 0.759188699683576
$\\$
$\\$
$\\$\color{Green}$\%$\-\ \-\ \-\ \-\ verified\-\ stable\-\ at:\-\ \-\ \-\ \-\ \-\ \-\ \-\ \-\ \-\ k\-\ =\-\ 0.999998385205026,\-\ X\-\ =\-\ 26.05736207014433
$\\$\color{Green}$\%$\-\ \-\ \-\ \-\ verified\-\ unstable\-\ at\-\ :\-\ \-\ \-\ \-\ k\-\ =\-\ 0.999998385263233,\-\ \-\ X\-\ =\-\ 26.05742300506267
$\\$
$\\$
$\\$\color{Black}\section{verify\_instability\_lower.m}

\color{Green}\color{BrickRed}\color{NavyBlue}\-\ function\-\ \color{BrickRed}\-\ [k\_left,k\_right,min\_quot]\-\ =\-\ verify\_instability\_lower(d,k)\color{Green}
$\\$\color{Green}$\%$------------------------------------------------------------
$\\$\color{Green}$\%$\-\ verify\-\ instability
$\\$\color{Green}$\%$------------------------------------------------------------
$\\$
$\\$
$\\$\color{Green}$\%$\-\ interval\-\ pi
$\\$\color{BrickRed}pie\-\ =\-\ nm('pi');\color{Green}
$\\$\color{Green}$\%$\-\ interpolation\-\ error
$\\$\color{BrickRed}err\-\ =\-\ d.err\_q\_n1;\color{Green}
$\\$
$\\$
$\\$\color{BrickRed}kappa\-\ =\-\ kappa\_of\_k(k);\color{Green}
$\\$\color{BrickRed}omega\-\ =\-\ pie./nm(kappa(1:end-1),kappa(2:end));\color{Green}
$\\$
$\\$
$\\$\color{Green}$\%$\-\ make\-\ intervals\-\ between\-\ k\-\ values
$\\$\color{BrickRed}k\-\ =\-\ nm(k(1:end-1),k(2:end));\color{Green}
$\\$\color{Green}$\%$\-\ constants\-\ for\-\ transforming\-\ coordinats
$\\$\color{BrickRed}c1\_q\-\ =\-\ 2/(d.b\_q-d.a\_q);\color{Green}
$\\$\color{BrickRed}c2\_q\-\ =\-\ (d.a\_q+d.b\_q)/2;\color{Green}
$\\$\color{Green}$\%$\-\ get\-\ q\-\ values\-\ and\-\ omega\-\ values
$\\$\color{BrickRed}elipk\-\ =\-\ elliptic\_integral(k,1);\color{Green}
$\\$\color{BrickRed}elipk2\-\ =\-\ elliptic\_integral(sqrt(1-k.\verb|^|2),1);\color{Green}
$\\$\color{BrickRed}q\-\ =\-\ exp(-pie*elipk2./elipk);\color{Green}
$\\$\color{Green}$\%$\-\ make\-\ transformation\-\ in\-\ q\-\ to\-\ qtilde\-\ in\-\ [-1,1]
$\\$\color{BrickRed}q\_tilde\-\ =\-\ c1\_q*(q-c2\_q);\color{Green}
$\\$\color{BrickRed}psi\_tilde\-\ =\-\ nm(1);\color{Green}
$\\$
$\\$
$\\$\color{Green}$\%$\-\ make\-\ sure\-\ q\-\ is\-\ within\-\ range\-\ of\-\ data\-\ in\-\ structure\-\ d
$\\$\color{BrickRed}\color{NavyBlue}\-\ if\-\ \color{BrickRed}\-\ min(inf(q))\-\ $<$\-\ sup(d.q\_min)\color{Green}
$\\$\color{BrickRed}\-\ \-\ \-\ \-\ error('q\-\ out\-\ of\-\ range');\color{Green}
$\\$\color{BrickRed}\color{NavyBlue}\-\ end\-\ \color{BrickRed}\color{Green}
$\\$
$\\$
$\\$\color{Green}$\%$\-\ make\-\ sure\-\ q\-\ is\-\ within\-\ range\-\ of\-\ data\-\ in\-\ structure\-\ d
$\\$\color{BrickRed}\color{NavyBlue}\-\ if\-\ \color{BrickRed}\-\ max(sup(q))\-\ $>$\-\ d.q\_max\color{Green}
$\\$\color{BrickRed}\-\ \-\ \-\ \-\ error('q\-\ out\-\ of\-\ range');\color{Green}
$\\$\color{BrickRed}\color{NavyBlue}\-\ end\-\ \color{BrickRed}\color{Green}
$\\$
$\\$
$\\$\color{Green}$\%$\-\ get\-\ interpolated\-\ values\-\ and\-\ add\-\ error
$\\$\color{BrickRed}f1\_psi\-\ =\-\ (cf\_eval(d.cfn1(:,:,2),q\_tilde,psi\_tilde))+nm(-err,err)+1i*nm(-err,err);\-\ \color{Green}
$\\$\color{BrickRed}f2\_psi\-\ =\-\ (cf\_eval(d.cfn1(:,:,4),q\_tilde,psi\_tilde))+nm(-err,err)+1i*nm(-err,err);\-\ \color{Green}
$\\$\color{BrickRed}g\_psi\-\ =\-\ (cf\_eval(d.cfn1(:,:,6),q\_tilde,psi\_tilde))+nm(-err,err)+1i*nm(-err,err);\-\ \color{Green}
$\\$
$\\$
$\\$\color{Green}$\%$\-\ $\%$\-\ process\-\ data\-\ to\-\ get\-\ function\-\ values
$\\$
$\\$
$\\$\color{BrickRed}omega\-\ =\-\ omega.';\color{Green}
$\\$\color{BrickRed}quot\-\ =real((f1\_psi./g\_psi)\-\ +\-\ (f2\_psi./g\_psi)./omega.\verb|^|2);\color{Green}
$\\$
$\\$
$\\$\color{Green}$\%$\-\ find\-\ where\-\ instability\-\ is\-\ verified
$\\$\color{BrickRed}ind\-\ =\-\ find(inf(quot)$>$0);\color{Green}
$\\$
$\\$
$\\$\color{Green}$\%$\-\ throw\-\ error\-\ if\-\ no\-\ instability\-\ verified
$\\$\color{BrickRed}\color{NavyBlue}\-\ if\-\ \color{BrickRed}\-\ isempty(ind)\color{Green}
$\\$\color{BrickRed}\-\ \-\ \-\ \-\ error('failed\-\ to\-\ verify');\color{Green}
$\\$\color{BrickRed}\color{NavyBlue}\-\ end\-\ \color{BrickRed}\color{Green}
$\\$
$\\$
$\\$\color{Green}$\%$\-\ make\-\ sure\-\ there\-\ are\-\ no\-\ gaps\-\ on\-\ what\-\ we\-\ report
$\\$\color{Green}$\%$\-\ as\-\ having\-\ been\-\ verified\-\ as\-\ unstable
$\\$\color{BrickRed}left\-\ =\-\ ind(1);\color{Green}
$\\$\color{BrickRed}right\-\ =\-\ ind(1);\color{Green}
$\\$\color{BrickRed}\color{NavyBlue}\-\ for\-\ \color{BrickRed}\-\ j\-\ =\-\ 1:length(ind)-1\color{Green}
$\\$\color{BrickRed}\-\ \-\ \-\ \-\ \color{NavyBlue}\-\ if\-\ \color{BrickRed}\-\ ind(j+1)\-\ ==\-\ ind(j)+1\color{Green}
$\\$\color{BrickRed}\-\ \-\ \-\ \-\ \-\ \-\ \-\ \-\ right\-\ =\-\ ind(j+1);\color{Green}
$\\$\color{BrickRed}\-\ \-\ \-\ \-\ \color{NavyBlue}\-\ else\-\ \color{BrickRed}\color{Green}
$\\$\color{BrickRed}\-\ \-\ \-\ \-\ \-\ \-\ \-\ \-\ break\color{Green}
$\\$\color{BrickRed}\-\ \-\ \-\ \-\ \color{NavyBlue}\-\ end\-\ \color{BrickRed}\color{Green}
$\\$\color{BrickRed}\color{NavyBlue}\-\ end\-\ \color{BrickRed}\color{Green}
$\\$
$\\$
$\\$\color{Green}$\%$\-\ get\-\ left\-\ and\-\ right\-\ k\-\ values\-\ that\-\ were\-\ verified.
$\\$\color{BrickRed}k\_left\-\ =\-\ mid(k(left));\color{Green}
$\\$\color{BrickRed}k\_right\-\ =\-\ mid(k(right));\color{Green}
$\\$
$\\$
$\\$\color{Green}$\%$\-\ find\-\ the\-\ minimum\-\ of\-\ the\-\ quotient\-\ for\-\ verified\-\ values
$\\$\color{BrickRed}min\_quot\-\ =\-\ min(inf(quot(left:right)));\color{Green}
$\\$
$\\$
$\\$
$\\$
$\\$\color{Black}\section{verify\_instability\_upper.m}

\color{Green}\color{BrickRed}\color{NavyBlue}\-\ function\-\ \color{BrickRed}\-\ \-\ verify\_instability\_upper(d,\-\ kleft,\-\ kright,psi\_left,psi\_right,ints\_y)\color{Green}
$\\$
$\\$
$\\$\color{Green}$\%$$\%$$\%$$\%$$\%$$\%$$\%$$\%$$\%$$\%$$\%$$\%$$\%$$\%$$\%$$\%$$\%$$\%$$\%$$\%$$\%$$\%$$\%$$\%$
$\\$\color{Green}$\%$\-\ form\-\ k\-\ interval
$\\$\color{BrickRed}k\-\ =\-\ [nm(kleft),\-\ nm(kright)];\color{Green}
$\\$\color{Green}$\%$\-\ coefficients
$\\$\color{BrickRed}cf\-\ =\-\ d.cf10;\color{Green}
$\\$\color{Green}$\%$\-\ constants
$\\$\color{BrickRed}pie\-\ =\-\ nm('pi');\color{Green}
$\\$\color{Green}$\%$\-\ interpolation\-\ error
$\\$\color{BrickRed}lam\_10\_q\-\ =\-\ (2/pie)*log(d.N\_q\_10)+(2/pie)*(nm('0.6')+log(8/pie)+pie/(72*d.N\_q\_10));\color{Green}
$\\$\color{BrickRed}err\-\ =\-\ d.err\_q\_10+lam\_10\_q*d.err\_psi\_10;\color{Green}
$\\$\color{Green}$\%$$\%$$\%$$\%$$\%$$\%$$\%$$\%$$\%$$\%$$\%$$\%$$\%$$\%$$\%$$\%$$\%$$\%$$\%$$\%$$\%$$\%$$\%$$\%$
$\\$
$\\$
$\\$\color{Green}$\%$\-\ constants\-\ for\-\ transformation\-\ in\-\ q
$\\$\color{BrickRed}c1\_q\-\ =\-\ 2/(d.b\_q-d.a\_q);\color{Green}
$\\$\color{BrickRed}c2\_q\-\ =\-\ (d.a\_q+d.b\_q)/2;\color{Green}
$\\$\color{Green}$\%$\-\ kappa
$\\$\color{BrickRed}kappa\-\ =\-\ kappa\_of\_k(k);\color{Green}
$\\$\color{Green}$\%$\-\ period
$\\$\color{BrickRed}X\-\ =\-\ 2*pie./kappa;\color{Green}
$\\$\color{Green}$\%$\-\ fprintf('\textbackslash nX:\-\ ')
$\\$\color{Green}$\%$\-\ sup(X)
$\\$\color{Green}$\%$\-\ elliptic\-\ integrals
$\\$\color{BrickRed}elipk\-\ =\-\ elliptic\_integral(k,1);\color{Green}
$\\$\color{BrickRed}elipk2\-\ =\-\ elliptic\_integral(sqrt(1-k.\verb|^|2),1);\color{Green}
$\\$\color{Green}$\%$\-\ q
$\\$\color{BrickRed}q\-\ =\-\ exp(-pie*elipk2./elipk);\color{Green}
$\\$\color{Green}$\%$\-\ omega
$\\$\color{BrickRed}omega\-\ =\-\ pie./kappa;\color{Green}
$\\$\color{Green}$\%$\-\ get\-\ theta\-\ values\-\ for\-\ q
$\\$\color{BrickRed}q\_tilde\-\ =\-\ c1\_q*(q-c2\_q);\color{Green}
$\\$\color{BrickRed}theta\_q\-\ =\-\ fliplr(acos(q\_tilde));\color{Green}
$\\$\color{BrickRed}c1\_psi\-\ =\-\ nm(2);\color{Green}
$\\$\color{BrickRed}c2\_psi\-\ =\-\ nm(1)/2;\color{Green}
$\\$
$\\$
$\\$\color{Green}$\%$\-\ check\-\ user\-\ input\-\ is\-\ correct
$\\$\color{BrickRed}\color{NavyBlue}\-\ if\-\ \color{BrickRed}\-\ inf(q(1))\-\ $<$\-\ inf(d.a\_q)\color{Green}
$\\$\color{BrickRed}\-\ \-\ \-\ \-\ error('k\-\ out\-\ of\-\ range');\color{Green}
$\\$\color{BrickRed}\color{NavyBlue}\-\ end\-\ \color{BrickRed}\color{Green}
$\\$
$\\$
$\\$\color{BrickRed}\color{NavyBlue}\-\ if\-\ \color{BrickRed}\-\ sup(q(2))\-\ $>$\-\ sup(d.b\_q)\color{Green}
$\\$\color{BrickRed}\-\ \-\ \-\ \-\ error('k\-\ out\-\ of\-\ range');\color{Green}
$\\$\color{BrickRed}\color{NavyBlue}\-\ end\-\ \color{BrickRed}\color{Green}
$\\$
$\\$
$\\$\color{BrickRed}lty\-\ =\-\ acos((2*psi\_left-1));\color{Green}
$\\$\color{BrickRed}rty\-\ =\-\ acos((2*psi\_right-1));\color{Green}
$\\$
$\\$
$\\$\color{BrickRed}ltx\-\ =\-\ inf(theta\_q(1));\color{Green}
$\\$\color{BrickRed}rtx\-\ =\-\ sup(theta\_q(2));\color{Green}
$\\$
$\\$
$\\$\color{Green}$\%$\-\ interpolate\-\ 
$\\$\color{BrickRed}c0\-\ =\-\ cheby\_eval(cf(:,:,1),ltx,rtx,lty,rty,0,ints\_y)+nm(-err,err)+1i*nm(-err,err);\color{Green}
$\\$\color{BrickRed}c1\-\ =\-\ cheby\_eval(cf(:,:,2),ltx,rtx,lty,rty,0,ints\_y)+nm(-err,err)+1i*nm(-err,err);\color{Green}
$\\$\color{BrickRed}c2\-\ =\-\ cheby\_eval(cf(:,:,3),ltx,rtx,lty,rty,0,ints\_y)+nm(-err,err)+1i*nm(-err,err);\color{Green}
$\\$\color{BrickRed}c3\-\ =\-\ cheby\_eval(cf(:,:,4),ltx,rtx,lty,rty,0,ints\_y)+nm(-err,err)+1i*nm(-err,err);\color{Green}
$\\$\color{BrickRed}h0\-\ =\-\ cheby\_eval(cf(:,:,5),ltx,rtx,lty,rty,0,ints\_y)+nm(-err,err)+1i*nm(-err,err);\color{Green}
$\\$\color{BrickRed}h1\-\ =\-\ cheby\_eval(cf(:,:,6),ltx,rtx,lty,rty,0,ints\_y)+nm(-err,err)+1i*nm(-err,err);\color{Green}
$\\$\color{BrickRed}h2\-\ =\-\ cheby\_eval(cf(:,:,7),ltx,rtx,lty,rty,0,ints\_y)+nm(-err,err)+1i*nm(-err,err);\color{Green}
$\\$\color{BrickRed}h3\-\ =\-\ cheby\_eval(cf(:,:,8),ltx,rtx,lty,rty,0,ints\_y)+nm(-err,err)+1i*nm(-err,err);\color{Green}
$\\$\color{BrickRed}h4\-\ =\-\ cheby\_eval(cf(:,:,9),ltx,rtx,lty,rty,0,ints\_y)+nm(-err,err)+1i*nm(-err,err);\color{Green}
$\\$\color{BrickRed}h5\-\ =\-\ cheby\_eval(cf(:,:,10),ltx,rtx,lty,rty,0,ints\_y)+nm(-err,err)+1i*nm(-err,err);\color{Green}
$\\$
$\\$
$\\$\color{BrickRed}ty\-\ =\-\ linspace(lty,rty,ints\_y+1);\color{Green}
$\\$\color{BrickRed}psi\_tilde\-\ =\-\ cos(nm(ty));\color{Green}
$\\$\color{BrickRed}psi\-\ =\-\ psi\_tilde/c1\_psi+c2\_psi;\color{Green}
$\\$
$\\$
$\\$\color{Green}$\%$\-\ 1i*xi\-\ -\-\ 1i*(Floquet\-\ parameter)
$\\$\color{BrickRed}xicon\-\ =\-\ 1i*xi\_q\_psi(nm(q(1),q(2)),psi,0);\color{Green}
$\\$
$\\$
$\\$\color{Green}$\%$\-\ sub\-\ functions
$\\$\color{BrickRed}f1\-\ =\-\ c0\-\ +\-\ c1.*xicon.\verb|^|1+\-\ c2.*xicon.\verb|^|2\-\ +\-\ c3.*xicon.\verb|^|3;\color{Green}
$\\$\color{BrickRed}f2\-\ =\-\ h0\-\ +\-\ h1.*xicon.\verb|^|1+\-\ h2.*xicon.\verb|^|2\-\ +\-\ h3.*xicon.\verb|^|3\-\ +\-\ ...\color{Green}
$\\$\color{BrickRed}\-\ \-\ \-\ \-\ \-\ \-\ \-\ \-\ \-\ \-\ \-\ \-\ \-\ \-\ \-\ \-\ h4.*xicon.\verb|^|4\-\ +\-\ h5.*xicon.\verb|^|5;\color{Green}
$\\$\color{BrickRed}\-\ \-\ \-\ \-\ \-\ \-\ \-\ \-\ \-\ \-\ \-\ \-\ \color{Green}
$\\$\color{Green}$\%$-------------------------------------------------------------
$\\$\color{Green}$\%$\-\ check\-\ that\-\ the\-\ middle\-\ of\-\ numerator\-\ is\-\ positive
$\\$\color{Green}$\%$-------------------------------------------------------------
$\\$
$\\$
$\\$\color{Green}$\%$\-\ numerator\-\ of\-\ lambda\_1
$\\$\color{BrickRed}numer\-\ =\-\ f1+f2/nm(omega(1),omega(2))\verb|^|2;\color{Green}
$\\$
$\\$
$\\$\color{BrickRed}numer\-\ =\-\ sup(imag(numer));\color{Green}
$\\$
$\\$
$\\$
$\\$
$\\$\color{Green}$\%$\-\ check\-\ for\-\ NaNs
$\\$\color{BrickRed}\color{NavyBlue}\-\ if\-\ \color{BrickRed}\-\ sum(any(isnan(numer)))\-\ $>$\-\ 0\color{Green}
$\\$\color{BrickRed}\-\ \-\ \-\ \-\ error('NaN\-\ present');\color{Green}
$\\$\color{BrickRed}\color{NavyBlue}\-\ end\-\ \color{BrickRed}\color{Green}
$\\$
$\\$
$\\$\color{Green}$\%$\-\ instability\-\ test:\-\ WTS\-\ \-\ f1\-\ +\-\ f2/omega\verb|^|2\-\ $<$\-\ 0\-\ 
$\\$\color{BrickRed}\color{NavyBlue}\-\ if\-\ \color{BrickRed}\-\ sum(any(numer\-\ $<$\-\ 0))\-\ ==\-\ 0\color{Green}
$\\$\color{BrickRed}\-\ \-\ \-\ \-\ error('failed\-\ to\-\ verify\-\ instability');\color{Green}
$\\$\color{BrickRed}\color{NavyBlue}\-\ end\-\ \color{BrickRed}\-\ \-\ \-\ \-\ \-\ \-\ \-\ \-\ \-\ \-\ \-\ \-\ \color{Green}
$\\$\color{Black}\section{verify\_stability\_n0\_strict.m}

\color{Green}\color{BrickRed}\color{NavyBlue}\-\ function\-\ \color{BrickRed}\-\ verify\_stability\_n0\_strict(d,\-\ kleft,\-\ kright,\-\ ints\_yL,ints\_yR,\-\ ...\color{Green}
$\\$\color{BrickRed}\-\ \-\ \-\ \-\ psi\_L,psi\_M,psi\_R,psi\_R2,psi\_R3,stats)\color{Green}
$\\$
$\\$
$\\$
$\\$
$\\$\color{Green}$\%$\-\ number\-\ of\-\ intervals\-\ in\-\ q\-\ (keep\-\ fixed\-\ at\-\ 1)
$\\$\color{BrickRed}ints\_x\-\ =\-\ 1;\color{Green}
$\\$\color{Green}$\%$\-\ ensure\-\ that\-\ psi\_L,\-\ psi\_M,\-\ and\-\ psi\_R\-\ are\-\ intervals
$\\$\color{BrickRed}psi\_L\-\ =\-\ nm(psi\_L);\color{Green}
$\\$\color{BrickRed}psi\_M\-\ =\-\ nm(psi\_M);\color{Green}
$\\$\color{BrickRed}psi\_R\-\ =\-\ nm(psi\_R);\color{Green}
$\\$\color{Green}$\%$$\%$$\%$$\%$$\%$$\%$$\%$$\%$$\%$$\%$$\%$$\%$$\%$$\%$$\%$$\%$$\%$$\%$$\%$$\%$$\%$$\%$$\%$$\%$
$\\$\color{Green}$\%$\-\ form\-\ k\-\ interval
$\\$\color{BrickRed}k\-\ =\-\ [nm(kleft),\-\ nm(kright)];\color{Green}
$\\$\color{Green}$\%$\-\ coefficients
$\\$\color{BrickRed}cf\-\ =\-\ d.cf10;\color{Green}
$\\$\color{Green}$\%$\-\ constants
$\\$\color{BrickRed}pie\-\ =\-\ nm('pi');\color{Green}
$\\$\color{Green}$\%$\-\ interpolation\-\ error
$\\$\color{BrickRed}lam\_10\_q\-\ =\-\ (2/pie)*log(d.N\_q\_10)+(2/pie)*(nm('0.6')+log(8/pie)+pie/(72*d.N\_q\_10));\color{Green}
$\\$\color{BrickRed}err\-\ =\-\ d.err\_q\_10+lam\_10\_q*d.err\_psi\_10;\color{Green}
$\\$\color{Green}$\%$$\%$$\%$$\%$$\%$$\%$$\%$$\%$$\%$$\%$$\%$$\%$$\%$$\%$$\%$$\%$$\%$$\%$$\%$$\%$$\%$$\%$$\%$$\%$
$\\$
$\\$
$\\$\color{Green}$\%$\-\ constants\-\ for\-\ transformation\-\ in\-\ q
$\\$\color{BrickRed}c1\_q\-\ =\-\ 2/(d.b\_q-d.a\_q);\color{Green}
$\\$\color{BrickRed}c2\_q\-\ =\-\ (d.a\_q+d.b\_q)/2;\color{Green}
$\\$\color{Green}$\%$\-\ kappa
$\\$\color{BrickRed}kappa\-\ =\-\ kappa\_of\_k(k);\color{Green}
$\\$\color{Green}$\%$\-\ period
$\\$\color{BrickRed}X\-\ =\-\ 2*pie./kappa;\color{Green}
$\\$\color{Green}$\%$\-\ elliptic\-\ integrals
$\\$\color{BrickRed}elipk\-\ =\-\ elliptic\_integral(k,1);\color{Green}
$\\$\color{BrickRed}elipk2\-\ =\-\ elliptic\_integral(sqrt(1-k.\verb|^|2),1);\color{Green}
$\\$\color{Green}$\%$\-\ q
$\\$\color{BrickRed}q\-\ =\-\ exp(-pie*elipk2./elipk);\color{Green}
$\\$\color{Green}$\%$\-\ omega
$\\$\color{BrickRed}omega\-\ =\-\ pie./kappa;\color{Green}
$\\$\color{Green}$\%$\-\ get\-\ theta\-\ values\-\ for\-\ q
$\\$\color{BrickRed}q\_tilde\-\ =\-\ c1\_q*(q-c2\_q);\color{Green}
$\\$\color{BrickRed}theta\_q\-\ =\-\ fliplr(acos(q\_tilde));\color{Green}
$\\$\color{BrickRed}c1\_psi\-\ =\-\ nm(2);\color{Green}
$\\$\color{BrickRed}c2\_psi\-\ =\-\ nm(1)/2;\color{Green}
$\\$\color{BrickRed}ltx\-\ =\-\ inf(theta\_q(1));\color{Green}
$\\$\color{BrickRed}rtx\-\ =\-\ sup(theta\_q(2));\color{Green}
$\\$
$\\$
$\\$\color{Green}$\%$\-\ check\-\ user\-\ input\-\ is\-\ correct
$\\$\color{BrickRed}\color{NavyBlue}\-\ if\-\ \color{BrickRed}\-\ inf(q(1))\-\ $<$\-\ inf(d.a\_q)\color{Green}
$\\$\color{BrickRed}\-\ \-\ \-\ \-\ error('k\-\ out\-\ of\-\ range');\color{Green}
$\\$\color{BrickRed}\color{NavyBlue}\-\ end\-\ \color{BrickRed}\color{Green}
$\\$
$\\$
$\\$\color{BrickRed}\color{NavyBlue}\-\ if\-\ \color{BrickRed}\-\ sup(q(2))\-\ $>$\-\ sup(d.b\_q)\color{Green}
$\\$\color{BrickRed}\-\ \-\ \-\ \-\ error('k\-\ out\-\ of\-\ range');\color{Green}
$\\$\color{BrickRed}\color{NavyBlue}\-\ end\-\ \color{BrickRed}\color{Green}
$\\$
$\\$
$\\$\color{BrickRed}psi\_tilde\_L\-\ =\-\ 2*psi\_L-1;\color{Green}
$\\$\color{BrickRed}psi\_tilde\_M\-\ =\-\ 2*psi\_M-1;\color{Green}
$\\$\color{BrickRed}rty\-\ =\-\ inf(acos(psi\_tilde\_L));\color{Green}
$\\$\color{BrickRed}lty\-\ =\-\ sup(acos(psi\_tilde\_M));\color{Green}
$\\$
$\\$
$\\$\color{Green}$\%$\-\ interpolate\-\ 
$\\$\color{BrickRed}c0\-\ =\-\ cheby\_taylor(cf(:,:,1),ltx,rtx,lty,rty,ints\_x,ints\_yL)+nm(-err,err)+1i*nm(-err,err);\color{Green}
$\\$\color{BrickRed}c1\-\ =\-\ cheby\_taylor(cf(:,:,2),ltx,rtx,lty,rty,ints\_x,ints\_yL)+nm(-err,err)+1i*nm(-err,err);\color{Green}
$\\$\color{BrickRed}c2\-\ =\-\ cheby\_taylor(cf(:,:,3),ltx,rtx,lty,rty,ints\_x,ints\_yL)+nm(-err,err)+1i*nm(-err,err);\color{Green}
$\\$\color{BrickRed}c3\-\ =\-\ cheby\_taylor(cf(:,:,4),ltx,rtx,lty,rty,ints\_x,ints\_yL)+nm(-err,err)+1i*nm(-err,err);\color{Green}
$\\$\color{BrickRed}h0\-\ =\-\ cheby\_taylor(cf(:,:,5),ltx,rtx,lty,rty,ints\_x,ints\_yL)+nm(-err,err)+1i*nm(-err,err);\color{Green}
$\\$\color{BrickRed}h1\-\ =\-\ cheby\_taylor(cf(:,:,6),ltx,rtx,lty,rty,ints\_x,ints\_yL)+nm(-err,err)+1i*nm(-err,err);\color{Green}
$\\$\color{BrickRed}h2\-\ =\-\ cheby\_taylor(cf(:,:,7),ltx,rtx,lty,rty,ints\_x,ints\_yL)+nm(-err,err)+1i*nm(-err,err);\color{Green}
$\\$\color{BrickRed}h3\-\ =\-\ cheby\_taylor(cf(:,:,8),ltx,rtx,lty,rty,ints\_x,ints\_yL)+nm(-err,err)+1i*nm(-err,err);\color{Green}
$\\$\color{BrickRed}h4\-\ =\-\ cheby\_taylor(cf(:,:,9),ltx,rtx,lty,rty,ints\_x,ints\_yL)+nm(-err,err)+1i*nm(-err,err);\color{Green}
$\\$\color{BrickRed}h5\-\ =\-\ cheby\_taylor(cf(:,:,10),ltx,rtx,lty,rty,ints\_x,ints\_yL)+nm(-err,err)+1i*nm(-err,err);\color{Green}
$\\$
$\\$
$\\$\color{BrickRed}ty\-\ =\-\ linspace(lty,rty,ints\_yL+1);\color{Green}
$\\$\color{BrickRed}psi\_tilde\-\ =\-\ cos(nm(ty));\color{Green}
$\\$\color{BrickRed}psi\-\ =\-\ psi\_tilde/c1\_psi+c2\_psi;\color{Green}
$\\$
$\\$
$\\$\color{Green}$\%$\-\ 1i*xi\-\ -\-\ 1i*(Floquet\-\ parameter)
$\\$\color{BrickRed}xicon\-\ =\-\ 1i*xi\_q\_psi(nm(q(1),q(2)),psi,0);\color{Green}
$\\$\color{BrickRed}xicon\-\ =\-\ nm(xicon(1:end-1),xicon(2:end));\color{Green}
$\\$\color{BrickRed}xicon\-\ =\-\ repmat(xicon,ints\_x,1);\color{Green}
$\\$
$\\$
$\\$\color{Green}$\%$\-\ sub\-\ functions
$\\$\color{BrickRed}f1\-\ =\-\ c0\-\ +\-\ c1.*xicon.\verb|^|1+\-\ c2.*xicon.\verb|^|2\-\ +\-\ c3.*xicon.\verb|^|3;\color{Green}
$\\$\color{BrickRed}f2\-\ =\-\ h0\-\ +\-\ h1.*xicon.\verb|^|1+\-\ h2.*xicon.\verb|^|2\-\ +\-\ h3.*xicon.\verb|^|3\-\ +\-\ ...\color{Green}
$\\$\color{BrickRed}\-\ \-\ \-\ \-\ \-\ \-\ \-\ \-\ \-\ \-\ \-\ \-\ \-\ \-\ \-\ \-\ h4.*xicon.\verb|^|4\-\ +\-\ h5.*xicon.\verb|^|5;\color{Green}
$\\$
$\\$
$\\$\color{Green}$\%$-------------------------------------------------------------
$\\$\color{Green}$\%$\-\ check\-\ that\-\ the\-\ middle\-\ of\-\ numerator\-\ is\-\ positive
$\\$\color{Green}$\%$-------------------------------------------------------------
$\\$
$\\$
$\\$\color{Green}$\%$\-\ numerator\-\ of\-\ lambda\_1
$\\$\color{BrickRed}numer\-\ =\-\ f1+f2/nm(omega(1),omega(2))\verb|^|2;\color{Green}
$\\$
$\\$
$\\$\color{BrickRed}\color{NavyBlue}\-\ if\-\ \color{BrickRed}\-\ strcmp(stats,'on')\color{Green}
$\\$\color{BrickRed}\-\ \-\ \-\ \-\ figure\color{Green}
$\\$\color{BrickRed}\-\ \-\ \-\ \-\ hold\-\ on\color{Green}
$\\$\color{BrickRed}\-\ \-\ \-\ \-\ psi\_tildep\-\ =\-\ cos(ty(1:end-1));\color{Green}
$\\$\color{BrickRed}\-\ \-\ \-\ \-\ plot((psi\_tildep+1)/2,inf(imag(numer)),'-r','LineWidth',2);\color{Green}
$\\$\color{BrickRed}\-\ \-\ \-\ \-\ plot((psi\_tildep+1)/2,sup(imag(numer)),'-b','LineWidth',2);\color{Green}
$\\$\color{BrickRed}\-\ \-\ \-\ \-\ h\-\ =\-\ xlabel('psi');\color{Green}
$\\$\color{BrickRed}\-\ \-\ \-\ \-\ set(h,'FontSize',22);\color{Green}
$\\$\color{BrickRed}\-\ \-\ \-\ \-\ h\-\ =\-\ ylabel('f');\color{Green}
$\\$\color{BrickRed}\-\ \-\ \-\ \-\ set(h,'FontSize',22);\color{Green}
$\\$\color{BrickRed}\-\ \-\ \-\ \-\ h\-\ =\-\ gca;\color{Green}
$\\$\color{BrickRed}\-\ \-\ \-\ \-\ set(h,'FontSize',22);\color{Green}
$\\$\color{BrickRed}\color{NavyBlue}\-\ end\-\ \color{BrickRed}\color{Green}
$\\$
$\\$
$\\$\color{BrickRed}numer\-\ =\-\ inf(imag(numer));\color{Green}
$\\$
$\\$
$\\$\color{Green}$\%$\-\ check\-\ for\-\ NaNs
$\\$\color{BrickRed}\color{NavyBlue}\-\ if\-\ \color{BrickRed}\-\ sum(any(isnan(numer)))\-\ $>$\-\ 0\color{Green}
$\\$\color{BrickRed}\-\ \-\ \-\ \-\ error('NaN\-\ present');\color{Green}
$\\$\color{BrickRed}\color{NavyBlue}\-\ end\-\ \color{BrickRed}\color{Green}
$\\$
$\\$
$\\$\color{Green}$\%$\-\ stability\-\ test:\-\ WTS\-\ \-\ f1\-\ +\-\ f2/omega\verb|^|2\-\ $>$\-\ 0\-\ 
$\\$\color{BrickRed}\color{NavyBlue}\-\ if\-\ \color{BrickRed}\-\ sum(any(numer\-\ $<$=0))\-\ $>$\-\ 0\color{Green}
$\\$\color{BrickRed}\-\ \-\ \-\ \-\ \-\ error('failed\-\ to\-\ verify\-\ stability\-\ on\-\ middle\-\ part\-\ of\-\ numerator');\color{Green}
$\\$\color{BrickRed}\color{NavyBlue}\-\ end\-\ \color{BrickRed}\-\ \-\ \-\ \-\ \-\ \-\ \-\ \-\ \-\ \-\ \-\ \-\ \color{Green}
$\\$
$\\$
$\\$\color{Green}$\%$-------------------------------------------------------------
$\\$\color{Green}$\%$\-\ check\-\ numerator\-\ on\-\ left
$\\$\color{Green}$\%$-------------------------------------------------------------
$\\$
$\\$
$\\$\color{BrickRed}psi\_tilde\_L\-\ =\-\ 2*psi\_L-1;\color{Green}
$\\$\color{BrickRed}rty\-\ =\-\ sup(pie);\color{Green}
$\\$\color{BrickRed}lty\-\ =\-\ sup(acos(psi\_tilde\_L));\color{Green}
$\\$
$\\$
$\\$\color{Green}$\%$\-\ cut-off\-\ left\-\ psi\-\ value
$\\$\color{BrickRed}left\_psi\-\ =\-\ cos(nm(lty))/c1\_psi+c2\_psi;\color{Green}
$\\$
$\\$
$\\$\color{Green}$\%$\-\ interpolate
$\\$\color{BrickRed}c0\-\ =\-\ cheby\_taylor(cf(:,:,1),ltx,rtx,lty,rty,ints\_x,ints\_yL)+nm(-err,err)+1i*nm(-err,err);\color{Green}
$\\$\color{BrickRed}c1\-\ =\-\ cheby\_taylor(cf(:,:,2),ltx,rtx,lty,rty,ints\_x,ints\_yL)+nm(-err,err)+1i*nm(-err,err);\color{Green}
$\\$\color{BrickRed}c2\-\ =\-\ cheby\_taylor(cf(:,:,3),ltx,rtx,lty,rty,ints\_x,ints\_yL)+nm(-err,err)+1i*nm(-err,err);\color{Green}
$\\$\color{BrickRed}c3\-\ =\-\ cheby\_taylor(cf(:,:,4),ltx,rtx,lty,rty,ints\_x,ints\_yL)+nm(-err,err)+1i*nm(-err,err);\color{Green}
$\\$\color{BrickRed}h0\-\ =\-\ cheby\_taylor(cf(:,:,5),ltx,rtx,lty,rty,ints\_x,ints\_yL)+nm(-err,err)+1i*nm(-err,err);\color{Green}
$\\$\color{BrickRed}h1\-\ =\-\ cheby\_taylor(cf(:,:,6),ltx,rtx,lty,rty,ints\_x,ints\_yL)+nm(-err,err)+1i*nm(-err,err);\color{Green}
$\\$\color{BrickRed}h2\-\ =\-\ cheby\_taylor(cf(:,:,7),ltx,rtx,lty,rty,ints\_x,ints\_yL)+nm(-err,err)+1i*nm(-err,err);\color{Green}
$\\$\color{BrickRed}h3\-\ =\-\ cheby\_taylor(cf(:,:,8),ltx,rtx,lty,rty,ints\_x,ints\_yL)+nm(-err,err)+1i*nm(-err,err);\color{Green}
$\\$\color{BrickRed}h4\-\ =\-\ cheby\_taylor(cf(:,:,9),ltx,rtx,lty,rty,ints\_x,ints\_yL)+nm(-err,err)+1i*nm(-err,err);\color{Green}
$\\$\color{BrickRed}h5\-\ =\-\ cheby\_taylor(cf(:,:,10),ltx,rtx,lty,rty,ints\_x,ints\_yL)+nm(-err,err)+1i*nm(-err,err);\color{Green}
$\\$
$\\$
$\\$
$\\$
$\\$\color{BrickRed}\color{NavyBlue}\-\ if\-\ \color{BrickRed}\-\ sum(any(isnan(c0)))\-\ $>$\-\ 0\color{Green}
$\\$\color{BrickRed}\-\ \-\ \-\ \-\ error('NaN\-\ present');\color{Green}
$\\$\color{BrickRed}\color{NavyBlue}\-\ end\-\ \color{BrickRed}\color{Green}
$\\$\color{BrickRed}\color{NavyBlue}\-\ if\-\ \color{BrickRed}\-\ sum(any(isnan(c1)))\-\ $>$\-\ 0\color{Green}
$\\$\color{BrickRed}\-\ \-\ \-\ \-\ error('NaN\-\ present');\color{Green}
$\\$\color{BrickRed}\color{NavyBlue}\-\ end\-\ \color{BrickRed}\color{Green}
$\\$\color{BrickRed}\color{NavyBlue}\-\ if\-\ \color{BrickRed}\-\ sum(any(isnan(c2)))\-\ $>$\-\ 0\color{Green}
$\\$\color{BrickRed}\-\ \-\ \-\ \-\ error('NaN\-\ present');\color{Green}
$\\$\color{BrickRed}\color{NavyBlue}\-\ end\-\ \color{BrickRed}\color{Green}
$\\$\color{BrickRed}\color{NavyBlue}\-\ if\-\ \color{BrickRed}\-\ sum(any(isnan(c3)))\-\ $>$\-\ 0\color{Green}
$\\$\color{BrickRed}\-\ \-\ \-\ \-\ error('NaN\-\ present');\color{Green}
$\\$\color{BrickRed}\color{NavyBlue}\-\ end\-\ \color{BrickRed}\color{Green}
$\\$\color{BrickRed}\color{NavyBlue}\-\ if\-\ \color{BrickRed}\-\ sum(any(isnan(h0)))\-\ $>$\-\ 0\color{Green}
$\\$\color{BrickRed}\-\ \-\ \-\ \-\ error('NaN\-\ present');\color{Green}
$\\$\color{BrickRed}\color{NavyBlue}\-\ end\-\ \color{BrickRed}\color{Green}
$\\$\color{BrickRed}\color{NavyBlue}\-\ if\-\ \color{BrickRed}\-\ sum(any(isnan(h1)))\-\ $>$\-\ 0\color{Green}
$\\$\color{BrickRed}\-\ \-\ \-\ \-\ error('NaN\-\ present');\color{Green}
$\\$\color{BrickRed}\color{NavyBlue}\-\ end\-\ \color{BrickRed}\color{Green}
$\\$\color{BrickRed}\color{NavyBlue}\-\ if\-\ \color{BrickRed}\-\ sum(any(isnan(h2)))\-\ $>$\-\ 0\color{Green}
$\\$\color{BrickRed}\-\ \-\ \-\ \-\ error('NaN\-\ present');\color{Green}
$\\$\color{BrickRed}\color{NavyBlue}\-\ end\-\ \color{BrickRed}\color{Green}
$\\$\color{BrickRed}\color{NavyBlue}\-\ if\-\ \color{BrickRed}\-\ sum(any(isnan(h3)))\-\ $>$\-\ 0\color{Green}
$\\$\color{BrickRed}\-\ \-\ \-\ \-\ error('NaN\-\ present');\color{Green}
$\\$\color{BrickRed}\color{NavyBlue}\-\ end\-\ \color{BrickRed}\color{Green}
$\\$\color{BrickRed}\color{NavyBlue}\-\ if\-\ \color{BrickRed}\-\ sum(any(isnan(h4)))\-\ $>$\-\ 0\color{Green}
$\\$\color{BrickRed}\-\ \-\ \-\ \-\ error('NaN\-\ present');\color{Green}
$\\$\color{BrickRed}\color{NavyBlue}\-\ end\-\ \color{BrickRed}\color{Green}
$\\$\color{BrickRed}\color{NavyBlue}\-\ if\-\ \color{BrickRed}\-\ sum(any(isnan(h5)))\-\ $>$\-\ 0\color{Green}
$\\$\color{BrickRed}\-\ \-\ \-\ \-\ error('NaN\-\ present');\color{Green}
$\\$\color{BrickRed}\color{NavyBlue}\-\ end\-\ \color{BrickRed}\color{Green}
$\\$
$\\$
$\\$\color{Green}$\%$\-\ get\-\ polynomial\-\ coefficient\-\ extremes
$\\$\color{BrickRed}omeg\-\ =\-\ nm(omega(1),omega(2));\color{Green}
$\\$\color{BrickRed}p0\-\ =\-\ max(max(sup(abs(c0+h0./omeg.\verb|^|2))));\color{Green}
$\\$\color{BrickRed}p1\-\ =\-\ max(max(sup(abs(c1+h1./omeg.\verb|^|2))));\color{Green}
$\\$\color{BrickRed}p2\-\ =\-\ max(max(sup(abs(c2+h2./omeg.\verb|^|2))));\color{Green}
$\\$\color{BrickRed}p3\-\ =\-\ max(max(sup(abs(c3+h3./omeg.\verb|^|2))));\color{Green}
$\\$\color{BrickRed}p4\-\ =\-\ max(max(sup(abs(h4./omeg.\verb|^|2))));\color{Green}
$\\$\color{BrickRed}p5\-\ =\-\ min(min(inf(abs(h5./omeg.\verb|^|2))));\color{Green}
$\\$
$\\$
$\\$\color{Green}$\%$\-\ form\-\ bound\-\ on\-\ roots
$\\$\color{BrickRed}R\-\ =\-\ nm(1)+(1/nm(p5))*nm(max([p0,p1,p2,p3,p4]));\color{Green}
$\\$\color{BrickRed}R\-\ =\-\ sup(R);\color{Green}
$\\$\color{Green}$\%$\-\ determine\-\ |xi|\-\ at\-\ left\_psi
$\\$\color{BrickRed}xicon\-\ =\-\ abs(xi\_q\_psi(nm(q(1),q(2)),left\_psi,0));\color{Green}
$\\$
$\\$
$\\$\color{Green}$\%$\-\ check\-\ if\-\ |xi|\-\ $<$\-\ R
$\\$\color{BrickRed}\color{NavyBlue}\-\ if\-\ \color{BrickRed}\-\ inf(real(xicon))\-\ $<$=\-\ R\color{Green}
$\\$\color{BrickRed}\-\ \-\ \-\ \-\ error('failed\-\ to\-\ verify\-\ on\-\ left')\color{Green}
$\\$\color{BrickRed}\color{NavyBlue}\-\ end\-\ \color{BrickRed}\color{Green}
$\\$
$\\$
$\\$\color{Green}$\%$-------------------------------------------------------------
$\\$\color{Green}$\%$\-\ middle\-\ region
$\\$\color{Green}$\%$-------------------------------------------------------------
$\\$
$\\$
$\\$\color{BrickRed}lam\_q\-\ =\-\ (2/pie)*log(d.N\_q\_n0)+(2/pie)*(nm('0.6')+log(8/pie)+pie/(72*d.N\_q\_n0));\color{Green}
$\\$\color{BrickRed}err\-\ =\-\ d.err\_q\_n0+lam\_q*d.err\_psi\_n0;\color{Green}
$\\$
$\\$
$\\$\color{Green}$\%$\-\ left\-\ theta\-\ region\-\ corresponding\-\ to\-\ right\-\ x\-\ region
$\\$\color{BrickRed}psi\_tilde\_R\-\ =\-\ 4*psi\_R-3;\color{Green}
$\\$\color{BrickRed}psi\_tilde\_M\-\ =\-\ 4*psi\_M-3;\color{Green}
$\\$\color{BrickRed}rty\-\ =\-\ inf(acos(psi\_tilde\_M));\color{Green}
$\\$\color{BrickRed}lty\-\ =\-\ sup(acos(psi\_tilde\_R));\color{Green}
$\\$
$\\$
$\\$\color{BrickRed}f1\-\ =\-\ cheby\_taylor(d.cfn0(:,:,1),ltx,rtx,lty,rty,ints\_x,ints\_yR)+nm(-err,err)+1i*nm(-err,err);\color{Green}
$\\$\color{BrickRed}f2\-\ =\-\ cheby\_taylor(d.cfn0(:,:,3),ltx,rtx,lty,rty,ints\_x,ints\_yR)+nm(-err,err)+1i*nm(-err,err);\color{Green}
$\\$\color{BrickRed}numer\-\ =\-\ f1+\-\ f2/nm(omega(1),omega(2))\verb|^|2;\color{Green}
$\\$
$\\$
$\\$\color{BrickRed}\color{NavyBlue}\-\ if\-\ \color{BrickRed}\-\ strcmp(stats,'on')\color{Green}
$\\$\color{BrickRed}\-\ \-\ \-\ \-\ hold\-\ on\color{Green}
$\\$\color{BrickRed}\-\ \-\ \-\ \-\ typ\-\ =\-\ linspace(lty,rty,ints\_yR+1);\color{Green}
$\\$\color{BrickRed}\-\ \-\ \-\ \-\ psi\_tildep\-\ =\-\ cos(typ(1:end-1));\color{Green}
$\\$\color{BrickRed}\-\ \-\ \-\ \-\ plot((psi\_tildep+3)/4,inf(imag(numer)),'-m','LineWidth',2);\color{Green}
$\\$\color{BrickRed}\-\ \-\ \-\ \-\ plot((psi\_tildep+3)/4,sup(imag(numer)),'-c','LineWidth',2);\color{Green}
$\\$\color{BrickRed}\-\ \-\ \-\ \-\ h\-\ =\-\ xlabel('psi');\color{Green}
$\\$\color{BrickRed}\-\ \-\ \-\ \-\ set(h,'FontSize',22);\color{Green}
$\\$\color{BrickRed}\-\ \-\ \-\ \-\ h\-\ =\-\ ylabel('f');\color{Green}
$\\$\color{BrickRed}\-\ \-\ \-\ \-\ set(h,'FontSize',22);\color{Green}
$\\$\color{BrickRed}\-\ \-\ \-\ \-\ h\-\ =\-\ gca;\color{Green}
$\\$\color{BrickRed}\-\ \-\ \-\ \-\ set(h,'FontSize',22);\color{Green}
$\\$\color{BrickRed}\color{NavyBlue}\-\ end\-\ \color{BrickRed}\color{Green}
$\\$
$\\$
$\\$\color{BrickRed}numer\-\ =\-\ inf(imag(numer));\color{Green}
$\\$
$\\$
$\\$\color{Green}$\%$\-\ check\-\ for\-\ NaNs
$\\$\color{BrickRed}\color{NavyBlue}\-\ if\-\ \color{BrickRed}\-\ sum(any(isnan(numer)))\-\ $>$\-\ 0\color{Green}
$\\$\color{BrickRed}\-\ \-\ \-\ \-\ error('NaN\-\ present');\color{Green}
$\\$\color{BrickRed}\color{NavyBlue}\-\ end\-\ \color{BrickRed}\color{Green}
$\\$
$\\$
$\\$\color{Green}$\%$\-\ stability\-\ test:\-\ WTS\-\ \-\ f1\-\ +\-\ f2/omega\verb|^|2\-\ $>$\-\ 0\-\ 
$\\$\color{BrickRed}\color{NavyBlue}\-\ if\-\ \color{BrickRed}\-\ sum(any(numer\-\ $<$=0))\-\ $>$\-\ 0\color{Green}
$\\$\color{BrickRed}\-\ \-\ \-\ \-\ \-\ error('failed\-\ to\-\ verify\-\ stability\-\ on\-\ middle\-\ part\-\ of\-\ numerator');\color{Green}
$\\$\color{BrickRed}\color{NavyBlue}\-\ end\-\ \color{BrickRed}\color{Green}
$\\$
$\\$
$\\$\color{Green}$\%$-------------------------------------------------------------
$\\$\color{Green}$\%$\-\ derivatives\-\ on\-\ right
$\\$\color{Green}$\%$-------------------------------------------------------------
$\\$
$\\$
$\\$\color{BrickRed}lam\_q\-\ =\-\ (2/pie)*log(d.N\_q\_n0)+(2/pie)*(nm('0.6')+log(8/pie)+pie/(72*d.N\_q\_n0));\color{Green}
$\\$\color{BrickRed}err\-\ =\-\ d.err\_q\_n0+lam\_q*d.err\_psi\_n0;\color{Green}
$\\$
$\\$
$\\$\color{BrickRed}psi\_tilde\_R\-\ =\-\ 4*psi\_R3-3;\color{Green}
$\\$\color{BrickRed}lty\-\ =\-\ 0;\color{Green}
$\\$\color{BrickRed}rty\-\ =\-\ sup(acos(psi\_tilde\_R));\color{Green}
$\\$
$\\$
$\\$\color{Green}$\%$\-\ get\-\ interpolated\-\ polynomials
$\\$\color{BrickRed}f1\_psi\-\ =\-\ cheby\_taylor(d.cfn0(:,:,2),ltx,ltx,lty,rty,ints\_x,ints\_yL)+nm(-err,err)+1i*nm(-err,err);\color{Green}
$\\$\color{BrickRed}f2\_psi\-\ =\-\ cheby\_taylor(d.cfn0(:,:,4),ltx,rtx,lty,rty,ints\_x,ints\_yL)+nm(-err,err)+1i*nm(-err,err);\color{Green}
$\\$
$\\$
$\\$\color{Green}$\%$-------------------------------------------------------------
$\\$\color{Green}$\%$\-\ derivative\-\ of\-\ numerator\-\ on\-\ right
$\\$\color{Green}$\%$-------------------------------------------------------------
$\\$
$\\$
$\\$\color{BrickRed}numer\_psi\-\ =\-\ f1\_psi+f2\_psi/nm(omega(1),omega(2))\verb|^|2;\color{Green}
$\\$
$\\$
$\\$\color{BrickRed}\color{NavyBlue}\-\ if\-\ \color{BrickRed}\-\ strcmp(stats,'on')\color{Green}
$\\$\color{BrickRed}\-\ \-\ \-\ \-\ hold\-\ on\color{Green}
$\\$\color{BrickRed}\-\ \-\ \-\ \-\ typ\-\ =\-\ linspace(lty,rty,ints\_yL+1);\color{Green}
$\\$\color{BrickRed}\-\ \-\ \-\ \-\ psi\_tildep\-\ =\-\ cos(typ(1:end-1));\color{Green}
$\\$\color{BrickRed}\-\ \-\ \-\ \-\ plot((psi\_tildep+3)/4,sup(imag(numer\_psi)),'-r')\color{Green}
$\\$\color{BrickRed}\-\ \-\ \-\ \-\ plot((psi\_tildep+3)/4,inf(imag(numer\_psi)),'-g')\color{Green}
$\\$\color{BrickRed}\-\ \-\ \-\ \-\ h\-\ =\-\ xlabel('psi');\color{Green}
$\\$\color{BrickRed}\-\ \-\ \-\ \-\ set(h,'FontSize',22);\color{Green}
$\\$\color{BrickRed}\-\ \-\ \-\ \-\ h\-\ =\-\ ylabel('f');\color{Green}
$\\$\color{BrickRed}\-\ \-\ \-\ \-\ set(h,'FontSize',22);\color{Green}
$\\$\color{BrickRed}\-\ \-\ \-\ \-\ h\-\ =\-\ gca;\color{Green}
$\\$\color{BrickRed}\-\ \-\ \-\ \-\ set(h,'FontSize',22);\color{Green}
$\\$\color{BrickRed}\color{NavyBlue}\-\ end\-\ \color{BrickRed}\color{Green}
$\\$
$\\$
$\\$\color{BrickRed}numer\_psi\-\ =\-\ sup(imag(numer\_psi));\color{Green}
$\\$
$\\$
$\\$\color{BrickRed}\color{NavyBlue}\-\ if\-\ \color{BrickRed}\-\ sum(any(isnan(numer\_psi)))\-\ $>$\-\ 0\color{Green}
$\\$\color{BrickRed}\-\ \-\ \-\ \-\ error('NaN\-\ present');\color{Green}
$\\$\color{BrickRed}\color{NavyBlue}\-\ end\-\ \color{BrickRed}\color{Green}
$\\$
$\\$
$\\$\color{Green}$\%$\-\ stability\-\ test:\-\ equivlanet\-\ to\-\ showing\-\ f1+f2/omega\verb|^|2\-\ $<$\-\ 0
$\\$\color{BrickRed}\color{NavyBlue}\-\ if\-\ \color{BrickRed}\-\ sum(any(numer\_psi\-\ $>$=0))\color{Green}
$\\$\color{BrickRed}\-\ \-\ \-\ \-\ error('failed\-\ to\-\ verify\-\ stability\-\ on\-\ right\-\ part\-\ of\-\ numerator');\color{Green}
$\\$\color{BrickRed}\color{NavyBlue}\-\ end\-\ \color{BrickRed}\color{Green}
$\\$
$\\$
$\\$\color{Green}$\%$-------------------------------------------------------------
$\\$\color{Green}$\%$\-\ right\-\ middle\-\ part
$\\$\color{Green}$\%$-------------------------------------------------------------
$\\$
$\\$
$\\$\color{BrickRed}lam\_q\-\ =\-\ (2/pie)*log(d.N\_q\_n0)+(2/pie)*(nm('0.6')+log(8/pie)+pie/(72*d.N\_q\_n0));\color{Green}
$\\$\color{BrickRed}err\-\ =\-\ d.err\_q\_n0+lam\_q*d.err\_psi\_n0;\color{Green}
$\\$
$\\$
$\\$\color{BrickRed}psi\_tilde\_R\-\ =\-\ 4*psi\_R3-3;\color{Green}
$\\$\color{BrickRed}psi\_tilde\_L\-\ =\-\ 4*psi\_R2-3;\color{Green}
$\\$\color{BrickRed}lty\-\ =\-\ sup(acos(psi\_tilde\_L));\color{Green}
$\\$\color{BrickRed}rty\-\ =\-\ sup(acos(psi\_tilde\_R));\color{Green}
$\\$
$\\$
$\\$\color{Green}$\%$\-\ get\-\ interpolated\-\ polynomials
$\\$\color{BrickRed}f1\-\ =\-\ cheby\_taylor(d.cfn0(:,:,1),ltx,ltx,lty,rty,ints\_x,ints\_yL)+nm(-err,err)+1i*nm(-err,err);\color{Green}
$\\$\color{BrickRed}f2\-\ =\-\ cheby\_taylor(d.cfn0(:,:,3),ltx,rtx,lty,rty,ints\_x,ints\_yL)+nm(-err,err)+1i*nm(-err,err);\color{Green}
$\\$
$\\$
$\\$\color{BrickRed}numer\-\ =\-\ f1+f2/nm(omega(1),omega(2))\verb|^|2;\color{Green}
$\\$
$\\$
$\\$\color{BrickRed}numer\-\ =\-\ inf(imag(numer));\color{Green}
$\\$
$\\$
$\\$\color{BrickRed}\color{NavyBlue}\-\ if\-\ \color{BrickRed}\-\ sum(any(isnan(numer)))\-\ $>$\-\ 0\color{Green}
$\\$\color{BrickRed}\-\ \-\ \-\ \-\ error('NaN\-\ present');\color{Green}
$\\$\color{BrickRed}\color{NavyBlue}\-\ end\-\ \color{BrickRed}\color{Green}
$\\$
$\\$
$\\$\color{Green}$\%$\-\ stability\-\ test:\-\ equivlanet\-\ to\-\ showing\-\ f1+f2/omega\verb|^|2\-\ $<$\-\ 0
$\\$\color{BrickRed}\color{NavyBlue}\-\ if\-\ \color{BrickRed}\-\ sum(any(numer\-\ $<$=0))\color{Green}
$\\$\color{BrickRed}\-\ \-\ \-\ \-\ error('failed\-\ to\-\ verify\-\ stability\-\ on\-\ right\-\ part\-\ of\-\ numerator');\color{Green}
$\\$\color{BrickRed}\color{NavyBlue}\-\ end\-\ \color{BrickRed}\color{Green}
$\\$\color{Black}\section{verify\_stability\_n1.m}

\color{Green}\color{BrickRed}\color{NavyBlue}\-\ function\-\ \color{BrickRed}\-\ verify\_stability\_n1(d,\-\ kleft,\-\ kright,\-\ ints\_x,ints\_y)\color{Green}
$\\$
$\\$
$\\$\color{Green}$\%$$\%$$\%$$\%$$\%$$\%$$\%$$\%$$\%$$\%$$\%$$\%$$\%$$\%$$\%$$\%$$\%$$\%$$\%$$\%$$\%$$\%$$\%$$\%$
$\\$\color{Green}$\%$\-\ form\-\ k\-\ interval
$\\$\color{BrickRed}k\-\ =\-\ [nm(kleft),\-\ nm(kright)];\color{Green}
$\\$\color{Green}$\%$\-\ coefficients
$\\$\color{BrickRed}cf\-\ =\-\ d.cfn1;\color{Green}
$\\$\color{BrickRed}pie\-\ =\-\ nm('pi');\color{Green}
$\\$\color{Green}$\%$\-\ interpolation\-\ error
$\\$\color{BrickRed}lam\_n1\_q\-\ =\-\ (2/pie)*log(d.N\_q\_n1)+(2/pie)*(nm('0.6')+log(8/pie)+pie/(72*d.N\_q\_n1));\color{Green}
$\\$\color{BrickRed}err\-\ =\-\ d.err\_q\_n1+lam\_n1\_q*d.err\_psi\_n1;\color{Green}
$\\$\color{Green}$\%$$\%$$\%$$\%$$\%$$\%$$\%$$\%$$\%$$\%$$\%$$\%$$\%$$\%$$\%$$\%$$\%$$\%$$\%$$\%$$\%$$\%$$\%$$\%$
$\\$
$\\$
$\\$\color{Green}$\%$\-\ constants\-\ for\-\ transformation\-\ in\-\ q
$\\$\color{BrickRed}c1\_q\-\ =\-\ 2/(d.b\_q-d.a\_q);\color{Green}
$\\$\color{BrickRed}c2\_q\-\ =\-\ (d.a\_q+d.b\_q)/2;\color{Green}
$\\$\color{Green}$\%$\-\ kappa
$\\$\color{BrickRed}kappa\-\ =\-\ kappa\_of\_k(k);\color{Green}
$\\$\color{Green}$\%$\-\ elliptic\-\ integrals
$\\$\color{Green}$\%$\-\ X\-\ =\-\ 2*pie./kappa;
$\\$\color{BrickRed}elipk\-\ =\-\ elliptic\_integral(k,1);\color{Green}
$\\$\color{BrickRed}elipk2\-\ =\-\ elliptic\_integral(sqrt(1-k.\verb|^|2),1);\color{Green}
$\\$\color{Green}$\%$\-\ q
$\\$\color{BrickRed}q\-\ =\-\ exp(-pie*elipk2./elipk);\color{Green}
$\\$\color{Green}$\%$\-\ omega
$\\$\color{BrickRed}omega\-\ =\-\ pie./kappa;\color{Green}
$\\$\color{Green}$\%$\-\ get\-\ theta\-\ values\-\ for\-\ q
$\\$\color{BrickRed}q\_tilde\-\ =\-\ c1\_q*(q-c2\_q);\color{Green}
$\\$\color{BrickRed}theta\_q\-\ =\-\ fliplr(acos(q\_tilde));\color{Green}
$\\$
$\\$
$\\$\color{Green}$\%$\-\ check\-\ user\-\ input\-\ is\-\ correct
$\\$\color{BrickRed}\color{NavyBlue}\-\ if\-\ \color{BrickRed}\-\ inf(q(1))\-\ $<$\-\ inf(d.a\_q)\color{Green}
$\\$\color{BrickRed}\-\ \-\ \-\ \-\ error('k\-\ out\-\ of\-\ range');\color{Green}
$\\$\color{BrickRed}\color{NavyBlue}\-\ end\-\ \color{BrickRed}\color{Green}
$\\$
$\\$
$\\$\color{BrickRed}\color{NavyBlue}\-\ if\-\ \color{BrickRed}\-\ sup(q(2))\-\ $>$\-\ sup(d.b\_q)\color{Green}
$\\$\color{BrickRed}\-\ \-\ \-\ \-\ error('k\-\ out\-\ of\-\ range');\color{Green}
$\\$\color{BrickRed}\color{NavyBlue}\-\ end\-\ \color{BrickRed}\color{Green}
$\\$
$\\$
$\\$\color{Green}$\%$\-\ specify\-\ middle\-\ of\-\ three\-\ theta\-\ regions
$\\$\color{Green}$\%$\-\ lty\-\ =\-\ inf(acos(psi\_tilde\_L))
$\\$\color{BrickRed}lty\-\ =\-\ inf(2*pie/10);\color{Green}
$\\$\color{BrickRed}rty\-\ =\-\ sup((9*pie)/10);\color{Green}
$\\$\color{BrickRed}ltx\-\ =\-\ inf(theta\_q(1));\color{Green}
$\\$\color{BrickRed}rtx\-\ =\-\ sup(theta\_q(2));\color{Green}
$\\$
$\\$
$\\$\color{Green}$\%$\-\ evalute\-\ interpolating\-\ polynomials
$\\$\color{BrickRed}f1\-\ =\-\ cheby\_taylor(cf(:,:,1),ltx,rtx,lty,rty,ints\_x,ints\_y)+nm(-err,err)+1i*nm(-err,err);\color{Green}
$\\$\color{BrickRed}f2\-\ =\-\ cheby\_taylor(cf(:,:,3),ltx,rtx,lty,rty,ints\_x,ints\_y)+nm(-err,err)+1i*nm(-err,err);\color{Green}
$\\$\color{BrickRed}g\-\ =\-\ cheby\_taylor(cf(:,:,5),ltx,rtx,lty,rty,ints\_x,ints\_y)+nm(-err,err)+1i*nm(-err,err);\color{Green}
$\\$
$\\$
$\\$\color{Green}$\%$-------------------------------------------------------------
$\\$\color{Green}$\%$\-\ check\-\ that\-\ the\-\ middle\-\ of\-\ numerator\-\ is\-\ positive
$\\$\color{Green}$\%$-------------------------------------------------------------
$\\$
$\\$
$\\$\color{BrickRed}f1\-\ =\-\ imag(f1);\color{Green}
$\\$\color{BrickRed}f2\-\ =\-\ imag(f2);\color{Green}
$\\$
$\\$
$\\$\color{BrickRed}numer\-\ =\-\ f1+f2/nm(omega(1),omega(2))\verb|^|2;\color{Green}
$\\$
$\\$
$\\$\color{BrickRed}\color{NavyBlue}\-\ if\-\ \color{BrickRed}\-\ sum(any(isnan(numer)))\-\ $>$\-\ 0\color{Green}
$\\$\color{BrickRed}\-\ \-\ \-\ \-\ error('NaN\-\ present');\color{Green}
$\\$\color{BrickRed}\color{NavyBlue}\-\ end\-\ \color{BrickRed}\color{Green}
$\\$
$\\$
$\\$\color{Green}$\%$\-\ stability\-\ test:\-\ equivalent\-\ to\-\ showing\-\ f1\-\ +\-\ f2/omega\verb|^|2\-\ $>$\-\ 0
$\\$\color{BrickRed}\color{NavyBlue}\-\ if\-\ \color{BrickRed}\-\ sum(any(inf(numer)$<$=0))\-\ $>$\-\ 0\color{Green}
$\\$\color{BrickRed}\-\ \-\ \-\ \-\ error('failed\-\ to\-\ verify\-\ stability\-\ on\-\ middle\-\ part\-\ of\-\ numerator');\color{Green}
$\\$\color{BrickRed}\color{NavyBlue}\-\ end\-\ \color{BrickRed}\color{Green}
$\\$
$\\$
$\\$\color{Green}$\%$-------------------------------------------------------------
$\\$\color{Green}$\%$\-\ check\-\ that\-\ the\-\ middle\-\ of\-\ denominator\-\ is\-\ positive
$\\$\color{Green}$\%$-------------------------------------------------------------
$\\$
$\\$
$\\$\color{BrickRed}g\-\ =\-\ imag(g);\color{Green}
$\\$
$\\$
$\\$\color{BrickRed}\color{NavyBlue}\-\ if\-\ \color{BrickRed}\-\ sum(any(isnan(g)))\-\ $>$\-\ 0\color{Green}
$\\$\color{BrickRed}\-\ \-\ \-\ \-\ error('NaN\-\ present');\color{Green}
$\\$\color{BrickRed}\color{NavyBlue}\-\ end\-\ \color{BrickRed}\color{Green}
$\\$
$\\$
$\\$\color{Green}$\%$\-\ stability\-\ test
$\\$\color{BrickRed}\color{NavyBlue}\-\ if\-\ \color{BrickRed}\-\ sum(any(sup(g)\-\ $>$=0))\-\ $>$\-\ 0\color{Green}
$\\$\color{BrickRed}\-\ \-\ \-\ \-\ error('failed\-\ to\-\ verify\-\ stability\-\ on\-\ middle\-\ part\-\ of\-\ numerator');\color{Green}
$\\$\color{BrickRed}\color{NavyBlue}\-\ end\-\ \color{BrickRed}\color{Green}
$\\$
$\\$
$\\$\color{Green}$\%$-------------------------------------------------------------
$\\$\color{Green}$\%$\-\ derivatives\-\ on\-\ left
$\\$\color{Green}$\%$-------------------------------------------------------------
$\\$
$\\$
$\\$\color{Green}$\%$\-\ specify\-\ theta\-\ region\-\ on\-\ right,\-\ corresponding\-\ to\-\ region\-\ in\-\ x\-\ on\-\ left
$\\$\color{BrickRed}rty\-\ =pie;\color{Green}
$\\$\color{BrickRed}lty\-\ =\-\ inf(9*pie/10);\color{Green}
$\\$
$\\$
$\\$\color{Green}$\%$\-\ evaluate\-\ interpolation\-\ polynomials
$\\$\color{BrickRed}f1\_psi\-\ =\-\ cheby\_taylor(cf(:,:,2),ltx,ltx,lty,rty,ints\_x,ints\_y)+nm(-err,err)+1i*nm(-err,err);\color{Green}
$\\$\color{BrickRed}f2\_psi\-\ =\-\ cheby\_taylor(cf(:,:,4),ltx,rtx,lty,rty,ints\_x,ints\_y)+nm(-err,err)+1i*nm(-err,err);\color{Green}
$\\$\color{BrickRed}g\_psi\-\ =\-\ cheby\_taylor(cf(:,:,6),ltx,rtx,lty,rty,ints\_x,ints\_y)+nm(-err,err)+1i*nm(-err,err);\color{Green}
$\\$
$\\$
$\\$\color{Green}$\%$-------------------------------------------------------------
$\\$\color{Green}$\%$\-\ derivative\-\ of\-\ numerator\-\ on\-\ left
$\\$\color{Green}$\%$-------------------------------------------------------------
$\\$
$\\$
$\\$\color{BrickRed}f1\_psi\-\ =\-\ imag(f1\_psi);\color{Green}
$\\$\color{BrickRed}f2\_psi\-\ =\-\ imag(f2\_psi);\color{Green}
$\\$
$\\$
$\\$\color{Green}$\%$\-\ error\-\ checking
$\\$\color{BrickRed}\color{NavyBlue}\-\ if\-\ \color{BrickRed}\-\ sum(any(inf(f2\_psi)\-\ $<$=\-\ 0))\-\ $>$\-\ 0\color{Green}
$\\$\color{BrickRed}\-\ \-\ \-\ \-\ error('f2\_psi\-\ has\-\ non-positive\-\ part')\color{Green}
$\\$\color{BrickRed}\color{NavyBlue}\-\ end\-\ \color{BrickRed}\color{Green}
$\\$
$\\$
$\\$\color{BrickRed}quot\-\ =\-\ inf(f1\_psi./f2\_psi+1/omega(2)\verb|^|2);\color{Green}
$\\$
$\\$
$\\$\color{BrickRed}\color{NavyBlue}\-\ if\-\ \color{BrickRed}\-\ sum(any(isnan(quot)))\-\ $>$\-\ 0\color{Green}
$\\$\color{BrickRed}\-\ \-\ \-\ \-\ error('NaN\-\ present');\color{Green}
$\\$\color{BrickRed}\color{NavyBlue}\-\ end\-\ \color{BrickRed}\color{Green}
$\\$
$\\$
$\\$\color{Green}$\%$\-\ stability\-\ test:\-\ equivalent\-\ to\-\ showing\-\ f1+f2/omega\verb|^|2\-\ $>$\-\ 0
$\\$\color{BrickRed}\color{NavyBlue}\-\ if\-\ \color{BrickRed}\-\ any(quot\-\ $<$=0)\color{Green}
$\\$\color{BrickRed}\-\ \-\ \-\ \-\ error('failed\-\ to\-\ verify\-\ stability\-\ on\-\ left\-\ part\-\ of\-\ numerator');\color{Green}
$\\$\color{BrickRed}\color{NavyBlue}\-\ end\-\ \color{BrickRed}\color{Green}
$\\$
$\\$
$\\$\color{Green}$\%$-------------------------------------------------------------
$\\$\color{Green}$\%$\-\ derivative\-\ of\-\ denominator\-\ on\-\ left
$\\$\color{Green}$\%$-------------------------------------------------------------
$\\$
$\\$
$\\$\color{BrickRed}g\_psi\-\ =\-\ imag(g\_psi);\color{Green}
$\\$
$\\$
$\\$\color{Green}$\%$\-\ error\-\ checking
$\\$\color{BrickRed}\color{NavyBlue}\-\ if\-\ \color{BrickRed}\-\ sum(any(isnan(g\_psi)))\-\ $>$\-\ 0\color{Green}
$\\$\color{BrickRed}\-\ \-\ \-\ \-\ error('NaN\-\ present');\color{Green}
$\\$\color{BrickRed}\color{NavyBlue}\-\ end\-\ \color{BrickRed}\color{Green}
$\\$
$\\$
$\\$\color{Green}$\%$\-\ stability\-\ test
$\\$\color{BrickRed}\color{NavyBlue}\-\ if\-\ \color{BrickRed}\-\ any(\-\ sup(g\_psi)\-\ $>$=0)\color{Green}
$\\$\color{BrickRed}\-\ \-\ \-\ \-\ error('failed\-\ to\-\ verify\-\ stability\-\ on\-\ left\-\ part\-\ of\-\ denominator');\color{Green}
$\\$\color{BrickRed}\color{NavyBlue}\-\ end\-\ \color{BrickRed}\color{Green}
$\\$
$\\$
$\\$\color{Green}$\%$-------------------------------------------------------------
$\\$\color{Green}$\%$\-\ derivatives\-\ on\-\ right
$\\$\color{Green}$\%$-------------------------------------------------------------
$\\$
$\\$
$\\$\color{Green}$\%$\-\ left\-\ theta\-\ regon\-\ corresponding\-\ to\-\ right\-\ x\-\ region
$\\$\color{BrickRed}rty\-\ =sup(2*pie/10);\color{Green}
$\\$\color{BrickRed}lty\-\ =\-\ 0;\color{Green}
$\\$
$\\$
$\\$\color{Green}$\%$\-\ get\-\ interpolated\-\ polynomials
$\\$\color{BrickRed}f1\_psi\-\ =\-\ cheby\_taylor(cf(:,:,2),ltx,ltx,lty,rty,ints\_x,ints\_y)+nm(-err,err)+1i*nm(-err,err);\color{Green}
$\\$\color{BrickRed}f2\_psi\-\ =\-\ cheby\_taylor(cf(:,:,4),ltx,rtx,lty,rty,ints\_x,ints\_y)+nm(-err,err)+1i*nm(-err,err);\color{Green}
$\\$
$\\$
$\\$\color{Green}$\%$-------------------------------------------------------------
$\\$\color{Green}$\%$\-\ derivative\-\ of\-\ numerator\-\ on\-\ right
$\\$\color{Green}$\%$-------------------------------------------------------------
$\\$
$\\$
$\\$\color{BrickRed}f1\_psi\-\ =\-\ imag(f1\_psi);\color{Green}
$\\$\color{BrickRed}f2\_psi\-\ =\-\ imag(f2\_psi);\color{Green}
$\\$
$\\$
$\\$\color{Green}$\%$\-\ error\-\ checking
$\\$\color{BrickRed}\color{NavyBlue}\-\ if\-\ \color{BrickRed}\-\ sum(any(sup(f2\_psi)\-\ $>$=\-\ 0))\-\ $>$\-\ 0\color{Green}
$\\$\color{BrickRed}\-\ \-\ \-\ \-\ error('f2\_psi\-\ has\-\ non-negative\-\ part')\color{Green}
$\\$\color{BrickRed}\color{NavyBlue}\-\ end\-\ \color{BrickRed}\color{Green}
$\\$
$\\$
$\\$\color{BrickRed}quot\-\ =\-\ inf(f1\_psi./f2\_psi+1/omega(2)\verb|^|2);\color{Green}
$\\$
$\\$
$\\$\color{BrickRed}\color{NavyBlue}\-\ if\-\ \color{BrickRed}\-\ sum(any(isnan(quot)))\-\ $>$\-\ 0\color{Green}
$\\$\color{BrickRed}\-\ \-\ \-\ \-\ error('NaN\-\ present');\color{Green}
$\\$\color{BrickRed}\color{NavyBlue}\-\ end\-\ \color{BrickRed}\color{Green}
$\\$
$\\$
$\\$\color{Green}$\%$\-\ stability\-\ test:\-\ equivlanet\-\ to\-\ showing\-\ f1+f2/omega\verb|^|2\-\ $<$\-\ 0
$\\$\color{BrickRed}\color{NavyBlue}\-\ if\-\ \color{BrickRed}\-\ any(quot\-\ $<$=0)\color{Green}
$\\$\color{BrickRed}\-\ \-\ \-\ \-\ this\-\ =\-\ quot(find(quot$<$=0));\color{Green}
$\\$\color{BrickRed}\-\ \-\ \-\ \-\ this(1:min(10,length(this)))\color{Green}
$\\$\color{BrickRed}\-\ \-\ \-\ \-\ error('failed\-\ to\-\ verify\-\ stability\-\ on\-\ right\-\ part\-\ of\-\ numerator');\color{Green}
$\\$\color{BrickRed}\color{NavyBlue}\-\ end\-\ \color{BrickRed}\color{Green}
$\\$
$\\$
$\\$\color{Green}$\%$-------------------------------------------------------------
$\\$\color{Green}$\%$\-\ derivative\-\ of\-\ denominator\-\ on\-\ right
$\\$\color{Green}$\%$-------------------------------------------------------------
$\\$
$\\$
$\\$\color{Green}$\%$\-\ left\-\ theta\-\ regon\-\ corresponding\-\ to\-\ right\-\ x\-\ region
$\\$\color{BrickRed}rty\-\ =sup(2*pie/10);\color{Green}
$\\$\color{BrickRed}lty\-\ =\-\ 0;\color{Green}
$\\$
$\\$
$\\$\color{BrickRed}g\_psi\-\ =\-\ cheby\_taylor(cf(:,:,6),ltx,rtx,lty,rty,ints\_x,ints\_y)+nm(-err,err)+1i*nm(-err,err);\color{Green}
$\\$\color{BrickRed}g\_psi\-\ =\-\ imag(g\_psi);\color{Green}
$\\$
$\\$
$\\$\color{Green}$\%$\-\ error\-\ checking
$\\$\color{BrickRed}\color{NavyBlue}\-\ if\-\ \color{BrickRed}\-\ sum(any(isnan(g\_psi)))\-\ $>$\-\ 0\color{Green}
$\\$\color{BrickRed}\-\ \-\ \-\ \-\ error('NaN\-\ present');\color{Green}
$\\$\color{BrickRed}\color{NavyBlue}\-\ end\-\ \color{BrickRed}\color{Green}
$\\$
$\\$
$\\$\color{Green}$\%$\-\ stability\-\ test
$\\$\color{BrickRed}\color{NavyBlue}\-\ if\-\ \color{BrickRed}\-\ sum(any(\-\ sup(g\_psi))\-\ $<$=0)\color{Green}
$\\$\color{BrickRed}\-\ \-\ \-\ \-\ error('failed\-\ to\-\ verify\-\ stability\-\ on\-\ right\-\ part\-\ of\-\ denominator');\color{Green}
$\\$\color{BrickRed}\color{NavyBlue}\-\ end\-\ \color{BrickRed}\color{Green}
$\\$
$\\$
$\\$\color{Green}$\%$$\%$$\%$$\%$$\%$$\%$$\%$$\%$$\%$$\%$$\%$$\%$$\%$$\%$$\%$$\%$$\%$$\%$$\%$$\%$$\%$$\%$$\%$$\%$$\%$$\%$$\%$$\%$$\%$$\%$$\%$
$\\$
$\\$
$\\$
$\\$
$\\$\color{Black}\section{verify\_stability\_n1\_strict.m}

\color{Green}\color{BrickRed}\color{NavyBlue}\-\ function\-\ \color{BrickRed}\-\ verify\_stability\_n1\_strict(d,\-\ kleft,\-\ kright,\-\ ints\_x,ints\_y,psiL)\color{Green}
$\\$
$\\$
$\\$\color{Green}$\%$$\%$$\%$$\%$$\%$$\%$$\%$$\%$$\%$$\%$$\%$$\%$$\%$$\%$$\%$$\%$$\%$$\%$$\%$$\%$$\%$$\%$$\%$$\%$
$\\$\color{Green}$\%$\-\ form\-\ k\-\ interval
$\\$\color{BrickRed}k\-\ =\-\ [nm(kleft),\-\ nm(kright)];\color{Green}
$\\$\color{Green}$\%$\-\ coefficients
$\\$\color{BrickRed}cf\-\ =\-\ d.cfn1;\color{Green}
$\\$\color{BrickRed}pie\-\ =\-\ nm('pi');\color{Green}
$\\$\color{Green}$\%$\-\ interpolation\-\ error
$\\$\color{BrickRed}lam\_n1\_q\-\ =\-\ (2/pie)*log(d.N\_q\_n1)+(2/pie)*(nm('0.6')+log(8/pie)+pie/(72*d.N\_q\_n1));\color{Green}
$\\$\color{BrickRed}err\-\ =\-\ d.err\_q\_n1+lam\_n1\_q*d.err\_psi\_n1;\color{Green}
$\\$\color{Green}$\%$$\%$$\%$$\%$$\%$$\%$$\%$$\%$$\%$$\%$$\%$$\%$$\%$$\%$$\%$$\%$$\%$$\%$$\%$$\%$$\%$$\%$$\%$$\%$
$\\$
$\\$
$\\$\color{Green}$\%$\-\ constants\-\ for\-\ transformation\-\ in\-\ q
$\\$\color{BrickRed}c1\_q\-\ =\-\ 2/(d.b\_q-d.a\_q);\color{Green}
$\\$\color{BrickRed}c2\_q\-\ =\-\ (d.a\_q+d.b\_q)/2;\color{Green}
$\\$\color{Green}$\%$\-\ kappa
$\\$\color{BrickRed}kappa\-\ =\-\ kappa\_of\_k(k);\color{Green}
$\\$\color{Green}$\%$\-\ elliptic\-\ integrals
$\\$\color{Green}$\%$\-\ X\-\ =\-\ 2*pie./kappa
$\\$\color{BrickRed}elipk\-\ =\-\ elliptic\_integral(k,1);\color{Green}
$\\$\color{BrickRed}elipk2\-\ =\-\ elliptic\_integral(sqrt(1-k.\verb|^|2),1);\color{Green}
$\\$\color{Green}$\%$\-\ q
$\\$\color{Green}$\%$\-\ q\-\ =\-\ exp(-pie*elipk2./elipk)
$\\$\color{Green}$\%$\-\ omega
$\\$\color{BrickRed}omega\-\ =\-\ pie./kappa;\color{Green}
$\\$\color{Green}$\%$\-\ get\-\ theta\-\ values\-\ for\-\ q
$\\$\color{BrickRed}q\_tilde\-\ =\-\ c1\_q*(q-c2\_q);\color{Green}
$\\$\color{BrickRed}theta\_q\-\ =\-\ fliplr(acos(q\_tilde));\color{Green}
$\\$
$\\$
$\\$\color{Green}$\%$\-\ check\-\ user\-\ input\-\ is\-\ correct
$\\$\color{BrickRed}\color{NavyBlue}\-\ if\-\ \color{BrickRed}\-\ inf(q(1))\-\ $<$\-\ inf(d.a\_q)\color{Green}
$\\$\color{BrickRed}\-\ \-\ \-\ \-\ error('k\-\ out\-\ of\-\ range');\color{Green}
$\\$\color{BrickRed}\color{NavyBlue}\-\ end\-\ \color{BrickRed}\color{Green}
$\\$
$\\$
$\\$\color{BrickRed}\color{NavyBlue}\-\ if\-\ \color{BrickRed}\-\ sup(q(2))\-\ $>$\-\ sup(d.b\_q)\color{Green}
$\\$\color{BrickRed}\-\ \-\ \-\ \-\ error('k\-\ out\-\ of\-\ range');\color{Green}
$\\$\color{BrickRed}\color{NavyBlue}\-\ end\-\ \color{BrickRed}\color{Green}
$\\$
$\\$
$\\$\color{Green}$\%$\-\ specify\-\ middle\-\ of\-\ three\-\ theta\-\ regions
$\\$\color{Green}$\%$\-\ lty\-\ =\-\ inf(acos(psi\_tilde\_L))
$\\$\color{BrickRed}lty\-\ =\-\ inf(2*pie/10);\color{Green}
$\\$\color{BrickRed}rty\-\ =\-\ sup((9*pie)/10);\color{Green}
$\\$\color{BrickRed}ltx\-\ =\-\ inf(theta\_q(1));\color{Green}
$\\$\color{BrickRed}rtx\-\ =\-\ sup(theta\_q(2));\color{Green}
$\\$
$\\$
$\\$\color{Green}$\%$\-\ evalute\-\ interpolating\-\ polynomials
$\\$\color{BrickRed}f1\-\ =\-\ cheby\_taylor(cf(:,:,1),ltx,rtx,lty,rty,ints\_x,ints\_y)+nm(-err,err)+1i*nm(-err,err);\color{Green}
$\\$\color{BrickRed}f2\-\ =\-\ cheby\_taylor(cf(:,:,3),ltx,rtx,lty,rty,ints\_x,ints\_y)+nm(-err,err)+1i*nm(-err,err);\color{Green}
$\\$\color{BrickRed}g\-\ =\-\ cheby\_taylor(cf(:,:,5),ltx,rtx,lty,rty,ints\_x,ints\_y)+nm(-err,err)+1i*nm(-err,err);\color{Green}
$\\$
$\\$
$\\$\color{Green}$\%$-------------------------------------------------------------
$\\$\color{Green}$\%$\-\ check\-\ that\-\ the\-\ middle\-\ of\-\ numerator\-\ is\-\ positive
$\\$\color{Green}$\%$-------------------------------------------------------------
$\\$
$\\$
$\\$\color{BrickRed}f1\-\ =\-\ imag(f1);\color{Green}
$\\$\color{BrickRed}f2\-\ =\-\ imag(f2);\color{Green}
$\\$
$\\$
$\\$\color{Green}$\%$\-\ error\-\ checking
$\\$\color{BrickRed}\color{NavyBlue}\-\ if\-\ \color{BrickRed}\-\ sum(any(inf(f2)\-\ $<$=\-\ 0))\-\ $>$\-\ 0\color{Green}
$\\$\color{BrickRed}\-\ \-\ \-\ \-\ ind=\-\ find(inf(f2)$<$=0);\color{Green}
$\\$\color{BrickRed}\-\ \-\ \-\ \-\ length(ind)\color{Green}
$\\$\color{BrickRed}\-\ \-\ \-\ \-\ error('f2\-\ has\-\ non-positive\-\ part')\color{Green}
$\\$\color{BrickRed}\color{NavyBlue}\-\ end\-\ \color{BrickRed}\color{Green}
$\\$
$\\$
$\\$\color{BrickRed}quot\-\ =\-\ inf(f1./f2+1/omega(2)\verb|^|2);\color{Green}
$\\$
$\\$
$\\$\color{BrickRed}\color{NavyBlue}\-\ if\-\ \color{BrickRed}\-\ sum(any(isnan(quot)))\-\ $>$\-\ 0\color{Green}
$\\$\color{BrickRed}\-\ \-\ \-\ \-\ error('NaN\-\ present');\color{Green}
$\\$\color{BrickRed}\color{NavyBlue}\-\ end\-\ \color{BrickRed}\color{Green}
$\\$
$\\$
$\\$\color{Green}$\%$\-\ stability\-\ test:\-\ equivalent\-\ to\-\ showing\-\ f1\-\ +\-\ f2/omega\verb|^|2\-\ $>$\-\ 0
$\\$\color{BrickRed}\color{NavyBlue}\-\ if\-\ \color{BrickRed}\-\ sum(any(quot\-\ $<$=0))\-\ $>$\-\ 0\color{Green}
$\\$\color{BrickRed}\-\ \-\ \-\ \-\ error('failed\-\ to\-\ verify\-\ stability\-\ on\-\ middle\-\ part\-\ of\-\ numerator');\color{Green}
$\\$\color{BrickRed}\color{NavyBlue}\-\ end\-\ \color{BrickRed}\color{Green}
$\\$
$\\$
$\\$\color{Green}$\%$-------------------------------------------------------------
$\\$\color{Green}$\%$\-\ check\-\ that\-\ the\-\ middle\-\ of\-\ denominator\-\ is\-\ positive
$\\$\color{Green}$\%$-------------------------------------------------------------
$\\$
$\\$
$\\$\color{BrickRed}g\-\ =\-\ imag(g);\color{Green}
$\\$
$\\$
$\\$\color{BrickRed}\color{NavyBlue}\-\ if\-\ \color{BrickRed}\-\ sum(any(isnan(g)))\-\ $>$\-\ 0\color{Green}
$\\$\color{BrickRed}\-\ \-\ \-\ \-\ error('NaN\-\ present');\color{Green}
$\\$\color{BrickRed}\color{NavyBlue}\-\ end\-\ \color{BrickRed}\color{Green}
$\\$
$\\$
$\\$\color{Green}$\%$\-\ stability\-\ test
$\\$\color{BrickRed}\color{NavyBlue}\-\ if\-\ \color{BrickRed}\-\ sum(any(sup(g)\-\ $>$=0))\-\ $>$\-\ 0\color{Green}
$\\$\color{BrickRed}\-\ \-\ \-\ \-\ error('failed\-\ to\-\ verify\-\ stability\-\ on\-\ middle\-\ part\-\ of\-\ numerator');\color{Green}
$\\$\color{BrickRed}\color{NavyBlue}\-\ end\-\ \color{BrickRed}\color{Green}
$\\$
$\\$
$\\$\color{Green}$\%$-------------------------------------------------------------
$\\$\color{Green}$\%$\-\ derivatives\-\ on\-\ left
$\\$\color{Green}$\%$-------------------------------------------------------------
$\\$
$\\$
$\\$\color{Green}$\%$\-\ specify\-\ theta\-\ region\-\ on\-\ right,\-\ corresponding\-\ to\-\ region\-\ in\-\ x\-\ on\-\ left
$\\$\color{BrickRed}rty\-\ =pie;\color{Green}
$\\$\color{BrickRed}lty\-\ =\-\ inf(9*pie/10);\color{Green}
$\\$
$\\$
$\\$\color{Green}$\%$\-\ evaluate\-\ interpolation\-\ polynomials
$\\$\color{BrickRed}f1\_psi\-\ =\-\ cheby\_taylor(cf(:,:,2),ltx,ltx,lty,rty,ints\_x,ints\_y)+nm(-err,err)+1i*nm(-err,err);\color{Green}
$\\$\color{BrickRed}f2\_psi\-\ =\-\ cheby\_taylor(cf(:,:,4),ltx,rtx,lty,rty,ints\_x,ints\_y)+nm(-err,err)+1i*nm(-err,err);\color{Green}
$\\$\color{BrickRed}g\_psi\-\ =\-\ cheby\_taylor(cf(:,:,6),ltx,rtx,lty,rty,ints\_x,ints\_y)+nm(-err,err)+1i*nm(-err,err);\color{Green}
$\\$
$\\$
$\\$\color{Green}$\%$-------------------------------------------------------------
$\\$\color{Green}$\%$\-\ derivative\-\ of\-\ numerator\-\ on\-\ left
$\\$\color{Green}$\%$-------------------------------------------------------------
$\\$
$\\$
$\\$\color{BrickRed}f1\_psi\-\ =\-\ imag(f1\_psi);\color{Green}
$\\$\color{BrickRed}f2\_psi\-\ =\-\ imag(f2\_psi);\color{Green}
$\\$
$\\$
$\\$\color{Green}$\%$\-\ error\-\ checking
$\\$\color{BrickRed}\color{NavyBlue}\-\ if\-\ \color{BrickRed}\-\ sum(any(inf(f2\_psi)\-\ $<$=\-\ 0))\-\ $>$\-\ 0\color{Green}
$\\$\color{BrickRed}\-\ \-\ \-\ \-\ error('f2\_psi\-\ has\-\ non-positive\-\ part')\color{Green}
$\\$\color{BrickRed}\color{NavyBlue}\-\ end\-\ \color{BrickRed}\color{Green}
$\\$
$\\$
$\\$\color{BrickRed}quot\-\ =\-\ inf(f1\_psi./f2\_psi+1/omega(2)\verb|^|2);\color{Green}
$\\$
$\\$
$\\$\color{BrickRed}\color{NavyBlue}\-\ if\-\ \color{BrickRed}\-\ sum(any(isnan(quot)))\-\ $>$\-\ 0\color{Green}
$\\$\color{BrickRed}\-\ \-\ \-\ \-\ error('NaN\-\ present');\color{Green}
$\\$\color{BrickRed}\color{NavyBlue}\-\ end\-\ \color{BrickRed}\color{Green}
$\\$
$\\$
$\\$\color{Green}$\%$\-\ stability\-\ test:\-\ equivalent\-\ to\-\ showing\-\ f1+f2/omega\verb|^|2\-\ $>$\-\ 0
$\\$\color{BrickRed}\color{NavyBlue}\-\ if\-\ \color{BrickRed}\-\ any(quot\-\ $<$=0)\color{Green}
$\\$\color{BrickRed}\-\ \-\ \-\ \-\ error('failed\-\ to\-\ verify\-\ stability\-\ on\-\ left\-\ part\-\ of\-\ numerator');\color{Green}
$\\$\color{BrickRed}\color{NavyBlue}\-\ end\-\ \color{BrickRed}\color{Green}
$\\$
$\\$
$\\$\color{Green}$\%$-------------------------------------------------------------
$\\$\color{Green}$\%$\-\ derivative\-\ of\-\ denominator\-\ on\-\ left
$\\$\color{Green}$\%$-------------------------------------------------------------
$\\$
$\\$
$\\$\color{BrickRed}g\_psi\-\ =\-\ imag(g\_psi);\color{Green}
$\\$
$\\$
$\\$\color{Green}$\%$\-\ error\-\ checking
$\\$\color{BrickRed}\color{NavyBlue}\-\ if\-\ \color{BrickRed}\-\ sum(any(isnan(g\_psi)))\-\ $>$\-\ 0\color{Green}
$\\$\color{BrickRed}\-\ \-\ \-\ \-\ error('NaN\-\ present');\color{Green}
$\\$\color{BrickRed}\color{NavyBlue}\-\ end\-\ \color{BrickRed}\color{Green}
$\\$
$\\$
$\\$\color{Green}$\%$\-\ stability\-\ test
$\\$\color{BrickRed}\color{NavyBlue}\-\ if\-\ \color{BrickRed}\-\ any(\-\ sup(g\_psi)\-\ $>$=0)\color{Green}
$\\$\color{BrickRed}\-\ \-\ \-\ \-\ error('failed\-\ to\-\ verify\-\ stability\-\ on\-\ left\-\ part\-\ of\-\ denominator');\color{Green}
$\\$\color{BrickRed}\color{NavyBlue}\-\ end\-\ \color{BrickRed}\color{Green}
$\\$
$\\$
$\\$\color{Green}$\%$-------------------------------------------------------------
$\\$\color{Green}$\%$\-\ derivatives\-\ on\-\ right
$\\$\color{Green}$\%$-------------------------------------------------------------
$\\$
$\\$
$\\$\color{Green}$\%$\-\ left\-\ theta\-\ regon\-\ corresponding\-\ to\-\ right\-\ x\-\ region
$\\$\color{BrickRed}rty\-\ =sup(2*pie/10);\color{Green}
$\\$\color{BrickRed}lty\-\ =\-\ inf(acos(2*psiL-1));\color{Green}
$\\$
$\\$
$\\$\color{Green}$\%$\-\ get\-\ interpolated\-\ polynomials
$\\$\color{BrickRed}f1\_psi\-\ =\-\ cheby\_taylor(cf(:,:,2),ltx,ltx,lty,rty,ints\_x,ints\_y)+nm(-err,err)+1i*nm(-err,err);\color{Green}
$\\$\color{BrickRed}f2\_psi\-\ =\-\ cheby\_taylor(cf(:,:,4),ltx,rtx,lty,rty,ints\_x,ints\_y)+nm(-err,err)+1i*nm(-err,err);\color{Green}
$\\$
$\\$
$\\$\color{Green}$\%$-------------------------------------------------------------
$\\$\color{Green}$\%$\-\ derivative\-\ of\-\ numerator\-\ on\-\ right
$\\$\color{Green}$\%$-------------------------------------------------------------
$\\$
$\\$
$\\$\color{BrickRed}f1\_psi\-\ =\-\ imag(f1\_psi);\color{Green}
$\\$\color{BrickRed}f2\_psi\-\ =\-\ imag(f2\_psi);\color{Green}
$\\$
$\\$
$\\$\color{Green}$\%$\-\ error\-\ checking
$\\$\color{BrickRed}\color{NavyBlue}\-\ if\-\ \color{BrickRed}\-\ sum(any(sup(f2\_psi)\-\ $>$=\-\ 0))\-\ $>$\-\ 0\color{Green}
$\\$\color{BrickRed}\-\ \-\ \-\ \-\ error('f2\_psi\-\ has\-\ non-negative\-\ part')\color{Green}
$\\$\color{BrickRed}\color{NavyBlue}\-\ end\-\ \color{BrickRed}\color{Green}
$\\$
$\\$
$\\$\color{BrickRed}quot\-\ =\-\ inf(f1\_psi./f2\_psi+1/omega(2)\verb|^|2);\color{Green}
$\\$
$\\$
$\\$\color{BrickRed}\color{NavyBlue}\-\ if\-\ \color{BrickRed}\-\ sum(any(isnan(quot)))\-\ $>$\-\ 0\color{Green}
$\\$\color{BrickRed}\-\ \-\ \-\ \-\ error('NaN\-\ present');\color{Green}
$\\$\color{BrickRed}\color{NavyBlue}\-\ end\-\ \color{BrickRed}\color{Green}
$\\$
$\\$
$\\$\color{Green}$\%$\-\ stability\-\ test:\-\ equivlanet\-\ to\-\ showing\-\ f1+f2/omega\verb|^|2\-\ $<$\-\ 0
$\\$\color{BrickRed}\color{NavyBlue}\-\ if\-\ \color{BrickRed}\-\ any(quot\-\ $<$=0)\color{Green}
$\\$\color{BrickRed}\-\ \-\ \-\ \-\ error('failed\-\ to\-\ verify\-\ stability\-\ on\-\ right\-\ part\-\ of\-\ numerator');\color{Green}
$\\$\color{BrickRed}\color{NavyBlue}\-\ end\-\ \color{BrickRed}\color{Green}
$\\$
$\\$
$\\$\color{Green}$\%$-------------------------------------------------------------
$\\$\color{Green}$\%$\-\ derivative\-\ of\-\ denominator\-\ on\-\ right
$\\$\color{Green}$\%$-------------------------------------------------------------
$\\$
$\\$
$\\$\color{Green}$\%$\-\ left\-\ theta\-\ regon\-\ corresponding\-\ to\-\ right\-\ x\-\ region
$\\$\color{BrickRed}rty\-\ =sup(2*pie/10);\color{Green}
$\\$\color{BrickRed}lty\-\ =\-\ 0;\color{Green}
$\\$
$\\$
$\\$\color{BrickRed}g\_psi\-\ =\-\ cheby\_taylor(cf(:,:,6),ltx,rtx,lty,rty,ints\_x,ints\_y)+nm(-err,err)+1i*nm(-err,err);\color{Green}
$\\$\color{BrickRed}g\_psi\-\ =\-\ imag(g\_psi);\color{Green}
$\\$
$\\$
$\\$\color{Green}$\%$\-\ error\-\ checking
$\\$\color{BrickRed}\color{NavyBlue}\-\ if\-\ \color{BrickRed}\-\ sum(any(isnan(g\_psi)))\-\ $>$\-\ 0\color{Green}
$\\$\color{BrickRed}\-\ \-\ \-\ \-\ error('NaN\-\ present');\color{Green}
$\\$\color{BrickRed}\color{NavyBlue}\-\ end\-\ \color{BrickRed}\color{Green}
$\\$
$\\$
$\\$\color{Green}$\%$\-\ stability\-\ test
$\\$\color{BrickRed}\color{NavyBlue}\-\ if\-\ \color{BrickRed}\-\ sum(any(\-\ sup(g\_psi))\-\ $<$=0)\color{Green}
$\\$\color{BrickRed}\-\ \-\ \-\ \-\ error('failed\-\ to\-\ verify\-\ stability\-\ on\-\ right\-\ part\-\ of\-\ denominator');\color{Green}
$\\$\color{BrickRed}\color{NavyBlue}\-\ end\-\ \color{BrickRed}\color{Green}
$\\$
$\\$
$\\$\color{Green}$\%$$\%$$\%$$\%$$\%$$\%$$\%$$\%$$\%$$\%$$\%$$\%$$\%$$\%$$\%$$\%$$\%$$\%$$\%$$\%$$\%$$\%$$\%$$\%$$\%$$\%$$\%$$\%$$\%$$\%$$\%$
$\\$
$\\$
$\\$
$\\$
$\\$\color{Black}\section{verify\_stability\_single.m}

\color{Green}\color{BrickRed}\color{NavyBlue}\-\ function\-\ \color{BrickRed}\-\ verify\_stability\_single(d,k,pnts,psi\_L,psi\_R)\color{Green}
$\\$
$\\$
$\\$\color{Green}$\%$\-\ number\-\ of\-\ intervals\-\ in\-\ q\-\ (keep\-\ fixed\-\ at\-\ 1)
$\\$\color{BrickRed}ints\_x\-\ =\-\ 1;\color{Green}
$\\$
$\\$
$\\$\color{Green}$\%$\-\ ensure\-\ that\-\ psi\_L,\-\ psi\_M,\-\ and\-\ psi\_R\-\ are\-\ intervals
$\\$\color{BrickRed}psi\_L\-\ =\-\ nm(psi\_L);\color{Green}
$\\$\color{BrickRed}psi\_R\-\ =\-\ nm(psi\_R);\color{Green}
$\\$
$\\$
$\\$\color{Green}$\%$\-\ constants
$\\$\color{BrickRed}pie\-\ =\-\ nm('pi');\color{Green}
$\\$\color{Green}$\%$\-\ interpolation\-\ error
$\\$
$\\$
$\\$\color{Green}$\%$$\%$$\%$$\%$$\%$$\%$$\%$$\%$$\%$$\%$$\%$$\%$$\%$$\%$$\%$$\%$$\%$$\%$$\%$$\%$$\%$$\%$$\%$$\%$
$\\$
$\\$
$\\$\color{Green}$\%$\-\ constants\-\ for\-\ transformation\-\ in\-\ q
$\\$\color{BrickRed}c1\_q\-\ =\-\ 2/(d.b\_q-d.a\_q);\color{Green}
$\\$\color{BrickRed}c2\_q\-\ =\-\ (d.a\_q+d.b\_q)/2;\color{Green}
$\\$\color{Green}$\%$\-\ kappa
$\\$\color{BrickRed}kappa\-\ =\-\ kappa\_of\_k(k);\color{Green}
$\\$\color{Green}$\%$\-\ period
$\\$\color{Green}$\%$\-\ X\-\ =\-\ 2*pie./kappa;
$\\$\color{Green}$\%$\-\ elliptic\-\ integrals
$\\$\color{BrickRed}elipk\-\ =\-\ elliptic\_integral(k,1);\color{Green}
$\\$\color{BrickRed}elipk2\-\ =\-\ elliptic\_integral(sqrt(1-k.\verb|^|2),1);\color{Green}
$\\$\color{Green}$\%$\-\ q
$\\$\color{BrickRed}q\-\ =\-\ exp(-pie*elipk2./elipk);\color{Green}
$\\$\color{Green}$\%$\-\ omega
$\\$\color{BrickRed}omega\-\ =\-\ pie./kappa;\color{Green}
$\\$\color{Green}$\%$\-\ get\-\ theta\-\ values\-\ for\-\ q
$\\$\color{BrickRed}q\_tilde\-\ =\-\ c1\_q*(q-c2\_q);\color{Green}
$\\$\color{BrickRed}theta\_q\-\ =\-\ fliplr(acos(q\_tilde));\color{Green}
$\\$\color{BrickRed}ltx\-\ =\-\ inf(theta\_q(1));\color{Green}
$\\$\color{BrickRed}rtx\-\ =\-\ sup(theta\_q(1));\color{Green}
$\\$
$\\$
$\\$\color{Green}$\%$\-\ check\-\ user\-\ input\-\ is\-\ correct
$\\$\color{BrickRed}\color{NavyBlue}\-\ if\-\ \color{BrickRed}\-\ inf(q)\-\ $<$\-\ inf(d.a\_q)\color{Green}
$\\$\color{BrickRed}\-\ \-\ \-\ \-\ error('k\-\ out\-\ of\-\ range');\color{Green}
$\\$\color{BrickRed}\color{NavyBlue}\-\ end\-\ \color{BrickRed}\color{Green}
$\\$
$\\$
$\\$\color{Green}$\%$-------------------------------------------------------------
$\\$\color{Green}$\%$\-\ midde\-\ middle\-\ part
$\\$\color{Green}$\%$-------------------------------------------------------------
$\\$
$\\$
$\\$\color{BrickRed}lam\_q\-\ =\-\ (2/pie)*log(d.N\_q\_n0)+(2/pie)*(nm('0.6')+log(8/pie)+pie/(72*d.N\_q\_n0));\color{Green}
$\\$\color{BrickRed}err\-\ =\-\ d.err\_q\_n0+lam\_q*d.err\_psi\_n0;\color{Green}
$\\$
$\\$
$\\$\color{BrickRed}psi\_tilde\_R\-\ =\-\ 4*psi\_R-3;\color{Green}
$\\$\color{BrickRed}psi\_tilde\_L\-\ =\-\ 4*psi\_L-3;\color{Green}
$\\$\color{BrickRed}lty\-\ =\-\ sup(acos(psi\_tilde\_L));\color{Green}
$\\$\color{BrickRed}rty\-\ =\-\ sup(acos(psi\_tilde\_R));\color{Green}
$\\$
$\\$
$\\$\color{Green}$\%$\-\ get\-\ interpolated\-\ polynomials
$\\$\color{BrickRed}f1\-\ =\-\ cheby\_taylor(d.cfn0(:,:,1),ltx,ltx,lty,rty,ints\_x,pnts)+nm(-err,err)+1i*nm(-err,err);\color{Green}
$\\$\color{BrickRed}f2\-\ =\-\ cheby\_taylor(d.cfn0(:,:,3),ltx,rtx,lty,rty,ints\_x,pnts)+nm(-err,err)+1i*nm(-err,err);\color{Green}
$\\$
$\\$
$\\$\color{BrickRed}numer\-\ =\-\ f1+f2/omega\verb|^|2;\color{Green}
$\\$
$\\$
$\\$\color{BrickRed}numer\-\ =\-\ inf(imag(numer));\color{Green}
$\\$
$\\$
$\\$\color{BrickRed}\color{NavyBlue}\-\ if\-\ \color{BrickRed}\-\ sum(any(isnan(numer)))\-\ $>$\-\ 0\color{Green}
$\\$\color{BrickRed}\-\ \-\ \-\ \-\ error('NaN\-\ present');\color{Green}
$\\$\color{BrickRed}\color{NavyBlue}\-\ end\-\ \color{BrickRed}\color{Green}
$\\$
$\\$
$\\$\color{Green}$\%$\-\ stability\-\ test:\-\ equivlanet\-\ to\-\ showing\-\ f1+f2/omega\verb|^|2\-\ $<$\-\ 0
$\\$\color{BrickRed}\color{NavyBlue}\-\ if\-\ \color{BrickRed}\-\ sum(any(numer\-\ $<$=0))\-\ $>$\-\ 0\color{Green}
$\\$\color{BrickRed}\-\ \-\ \-\ \-\ min(numer)\color{Green}
$\\$\color{BrickRed}\-\ \-\ \-\ \-\ error('failed\-\ to\-\ verify\-\ stability');\color{Green}
$\\$\color{BrickRed}\color{NavyBlue}\-\ end\-\ \color{BrickRed}\color{Green}
$\\$
$\\$
$\\$
$\\$
$\\$
$\\$
$\\$
$\\$
$\\$
$\\$
$\\$\color{Black}\section{weierstrass\_eta1.m}

\color{Green}\color{BrickRed}\color{NavyBlue}\-\ function\-\ \color{BrickRed}\-\ out\-\ =\-\ weierstrass\_eta1(omega\_1,omega\_2)\color{Green}
$\\$\color{Green}$\%$\-\ out\-\ =\-\ weierstrass\_eta1(omega\_1,omega\_2)
$\\$\color{Green}$\%$\-\ 
$\\$\color{Green}$\%$\-\ Returns,\-\ using\-\ interval\-\ arithmetic,\-\ $z(omega\_1;g\_2,g\_3)$\-\ where\-\ $z$\-\ is\-\ 
$\\$\color{Green}$\%$\-\ the\-\ Weierstrass\-\ zeta\-\ function.
$\\$
$\\$
$\\$\color{Green}$\%$\-\ nome
$\\$\color{BrickRed}q\-\ =\-\ exp(nm('pi')*nm('1i')*omega\_2/omega\_1);\color{Green}
$\\$
$\\$
$\\$\color{Green}$\%$\-\ check\-\ for\-\ input\-\ error
$\\$\color{BrickRed}\color{NavyBlue}\-\ if\-\ \color{BrickRed}\-\ real(q)\-\ $<$=\-\ nm('0')\color{Green}
$\\$\color{BrickRed}\-\ \-\ \-\ \-\ error('q\-\ not\-\ postive')\color{Green}
$\\$\color{BrickRed}\color{NavyBlue}\-\ end\-\ \color{BrickRed}\color{Green}
$\\$\color{BrickRed}\color{NavyBlue}\-\ if\-\ \color{BrickRed}\-\ real(q)\-\ $>$=\-\ nm('1')\color{Green}
$\\$\color{BrickRed}\-\ \-\ \-\ \-\ error('q\-\ not\-\ less\-\ than\-\ 1');\color{Green}
$\\$\color{BrickRed}\color{NavyBlue}\-\ end\-\ \color{BrickRed}\color{Green}
$\\$\color{BrickRed}\color{NavyBlue}\-\ if\-\ \color{BrickRed}\-\ sup(abs(imag(q))$>$0)\color{Green}
$\\$\color{BrickRed}\-\ \-\ \-\ \-\ error('q\-\ not\-\ real');\color{Green}
$\\$\color{BrickRed}\color{NavyBlue}\-\ end\-\ \color{BrickRed}\color{Green}
$\\$\color{BrickRed}\-\ \-\ \-\ \-\ \color{Green}
$\\$\color{Green}$\%$\-\ find\-\ k\-\ so\-\ that\-\ $kq\verb|^|k$<$1$\-\ and\-\ $q(1+q\verb|^|k)$<$1$
$\\$\color{BrickRed}k\-\ =\-\ 1;\color{Green}
$\\$\color{BrickRed}proceed\-\ =\-\ 1;\color{Green}
$\\$\color{BrickRed}\color{NavyBlue}\-\ while\-\ \color{BrickRed}\-\ proceed\-\ ==\-\ 1\color{Green}
$\\$\color{BrickRed}\-\ \-\ \-\ \-\ k\-\ =\-\ k+1;\color{Green}
$\\$\color{BrickRed}\-\ \-\ \-\ \-\ ks\-\ =\-\ nm(num2str(k));\-\ \-\ \-\ \-\ \color{Green}
$\\$\color{BrickRed}\-\ \-\ \-\ \-\ \color{NavyBlue}\-\ if\-\ \color{BrickRed}\-\ sup(ks*q\verb|^|ks-nm('1'))$<$0\color{Green}
$\\$\color{BrickRed}\-\ \-\ \-\ \-\ \-\ \-\ \-\ \-\ \color{NavyBlue}\-\ if\-\ \color{BrickRed}\-\ sup(q*(1+q\verb|^|ks)-nm('1'))$<$0\color{Green}
$\\$\color{BrickRed}\-\ \-\ \-\ \-\ \-\ \-\ \-\ \-\ \-\ \-\ \-\ \-\ proceed\-\ =0;\color{Green}
$\\$\color{BrickRed}\-\ \-\ \-\ \-\ \-\ \-\ \-\ \-\ \color{NavyBlue}\-\ end\-\ \color{BrickRed}\color{Green}
$\\$\color{BrickRed}\-\ \-\ \-\ \-\ \color{NavyBlue}\-\ end\-\ \color{BrickRed}\color{Green}
$\\$\color{BrickRed}\color{NavyBlue}\-\ end\-\ \color{BrickRed}\color{Green}
$\\$
$\\$
$\\$\color{Green}$\%$\-\ find\-\ k\-\ so\-\ that\-\ the\-\ error\-\ bound\-\ is\-\ less\-\ than\-\ tol\-\ =\-\ 1e-16
$\\$\color{BrickRed}error\_bound\-\ =\-\ abs((q\verb|^|ks/(1-q\verb|^|(nm('2')*ks)))*nm('1')/(nm('1')-q));\color{Green}
$\\$\color{BrickRed}\color{NavyBlue}\-\ while\-\ \color{BrickRed}\-\ sup(error\_bound)\-\ $>$\-\ 1e-16\color{Green}
$\\$\color{BrickRed}\-\ \-\ \-\ \-\ k\-\ =\-\ k+1;\color{Green}
$\\$\color{BrickRed}\-\ \-\ \-\ \-\ ks\-\ =\-\ nm(num2str(k));\color{Green}
$\\$\color{BrickRed}\-\ \-\ \-\ \-\ error\_bound\-\ =\-\ abs((q\verb|^|ks/(1-q\verb|^|(nm('2')*ks)))/(nm('1')-q));\color{Green}
$\\$\color{BrickRed}\color{NavyBlue}\-\ end\-\ \color{BrickRed}\color{Green}
$\\$
$\\$
$\\$\color{Green}$\%$\-\ add\-\ the\-\ first\-\ k-1\-\ terms\-\ of\-\ the\-\ sum
$\\$\color{BrickRed}sum\-\ =\-\ nm('0');\color{Green}
$\\$\color{BrickRed}\color{NavyBlue}\-\ for\-\ \color{BrickRed}\-\ j\-\ =\-\ 1:k-1\color{Green}
$\\$\color{BrickRed}\-\ \-\ \-\ \-\ js\-\ =\-\ nm(num2str(j));\color{Green}
$\\$\color{BrickRed}\-\ \-\ \-\ \-\ sum\-\ =\-\ sum\-\ +\-\ js*q\verb|^|(nm('2')*js)/(nm('1')-q\verb|^|(nm('2')*js));\color{Green}
$\\$\color{BrickRed}\color{NavyBlue}\-\ end\-\ \color{BrickRed}\color{Green}
$\\$
$\\$
$\\$\color{Green}$\%$\-\ add\-\ error\-\ to\-\ sum
$\\$\color{BrickRed}sum\-\ =\-\ sum\-\ +\-\ hull(-error\_bound,error\_bound);\color{Green}
$\\$
$\\$
$\\$\color{Green}$\%$\-\ compute\-\ eta\_1\-\ with\-\ interval\-\ sum
$\\$\color{BrickRed}pie\-\ =\-\ nm('pi');\color{Green}
$\\$\color{BrickRed}out\-\ =\-\ pie\verb|^|nm('2')/(nm('12')*omega\_1)-nm('2')*pie\verb|^|nm('2')*sum/omega\_1;\color{Green}
$\\$
$\\$
$\\$
$\\$
$\\$
$\\$
$\\$
$\\$
$\\$
$\\$
$\\$
$\\$
$\\$
$\\$
$\\$
$\\$
$\\$
$\\$
$\\$
$\\$
$\\$
$\\$
$\\$
$\\$
$\\$\color{Black}\section{weierstrass\_invariants.m}

\color{Green}\color{BrickRed}\color{NavyBlue}\-\ function\-\ \color{BrickRed}\-\ [g2,g3]\-\ =\-\ weierstrass\_invariants(omega\_1,omega\_2)\color{Green}
$\\$\color{Green}$\%$\-\ [g2,g3]\-\ =\-\ weierstrass\_invariants(omega\_1,omega\_2)
$\\$\color{Green}$\%$
$\\$\color{Green}$\%$\-\ Returns,\-\ using\-\ interval\-\ arithmetic,\-\ the\-\ Weierstrass\-\ elliptic\-\ function\-\ invariants\-\ corresponding\-\ to
$\\$\color{Green}$\%$\-\ half\-\ periods\-\ omega\_1\-\ and\-\ omega\_2.\-\ The\-\ half\-\ period\-\ omega\_1\-\ should\-\ be
$\\$\color{Green}$\%$\-\ a\-\ positive\-\ real\-\ and\-\ the\-\ half\-\ period\-\ omega\_2\-\ should\-\ be\-\ purely\-\ imaginary
$\\$\color{Green}$\%$\-\ and\-\ lie\-\ in\-\ the\-\ upper\-\ half\-\ of\-\ the\-\ complex\-\ plane.\-\ 
$\\$
$\\$
$\\$\color{Green}$\%$\-\ period\-\ ratio
$\\$\color{BrickRed}tau\-\ =\-\ omega\_2/omega\_1;\color{Green}
$\\$
$\\$
$\\$\color{Green}$\%$\-\ Jacobi\-\ theta\-\ functions
$\\$\color{BrickRed}z\-\ =\-\ nm('0');\color{Green}
$\\$\color{BrickRed}theta\_3\-\ =\-\ theta\_func(z,tau);\color{Green}
$\\$\color{BrickRed}theta\_2\-\ =\-\ exp(nm('0.25')*nm('pi')*nm('1i')*tau+nm('pi')*nm('1i')*z)*theta\_func(z+nm('0.5')*tau,tau);\color{Green}
$\\$
$\\$
$\\$\color{Green}$\%$\-\ for\-\ convenience
$\\$\color{BrickRed}pie\-\ =\-\ nm('pi');\color{Green}
$\\$
$\\$
$\\$\color{Green}$\%$\-\ Weierstrass\-\ elliptic\-\ function\-\ invariants
$\\$\color{BrickRed}g2\-\ =\-\ pie\verb|^|nm('4')/(nm('12')*omega\_1\verb|^|nm('4'))*...\color{Green}
$\\$\color{BrickRed}\-\ \-\ \-\ \-\ (theta\_2\verb|^|nm('8')-theta\_3\verb|^|nm('4')*theta\_2\verb|^|nm('4')+theta\_3\verb|^|nm('8'));\color{Green}
$\\$
$\\$
$\\$\color{BrickRed}g3\-\ =\-\ (pie\verb|^|nm('6')/(nm('2')*omega\_1)\verb|^|nm('6'))*((nm('8')/nm('27'))*(theta\_2\verb|^|nm('12')+theta\_3\verb|^|nm('12'))...\color{Green}
$\\$\color{BrickRed}\-\ \-\ \-\ \-\ -(nm('4')/nm('9'))*(theta\_2\verb|^|nm('4')+theta\_3\verb|^|nm('4'))*theta\_2\verb|^|nm('4')*theta\_3\verb|^|nm('4'));\color{Green}
$\\$
$\\$
$\\$
$\\$
$\\$
$\\$
$\\$
$\\$
$\\$
$\\$
$\\$
$\\$
$\\$\color{Black}\section{weierstrass\_p.m}

\color{Green}\color{BrickRed}\color{NavyBlue}\-\ function\-\ \color{BrickRed}\-\ out\-\ =\-\ weierstrass\_p(zin,w1,w2)\color{Green}
$\\$\color{Green}$\%$\-\ out\-\ =\-\ weierstrass\_p(z,w1,w2)
$\\$\color{Green}$\%$
$\\$\color{Green}$\%$\-\ Returns,\-\ computed\-\ with\-\ ineterval\-\ arithemtic,\-\ the\-\ Weierstrass\-\ elliptic\-\ function
$\\$\color{Green}$\%$\-\ with\-\ half\-\ periods\-\ w1\-\ and\-\ w2\-\ evaluated\-\ at\-\ z.\-\ Here\-\ w1\-\ is\-\ a\-\ postive\-\ real\-\ and\-\ 
$\\$\color{Green}$\%$\-\ w2\-\ is\-\ purely\-\ imaginary\-\ lying\-\ above\-\ the\-\ real\-\ axis.\-\ 
$\\$
$\\$
$\\$\color{Green}$\%$\-\ use\-\ periodicity\-\ of\-\ the\-\ Weierstrass\-\ elliptic\-\ function\-\ to\-\ bring\-\ argument
$\\$\color{Green}$\%$\-\ close\-\ to\-\ origin\-\ where\-\ the\-\ theta\_func\-\ program\-\ can\-\ succesffuly\-\ compute
$\\$\color{BrickRed}\color{NavyBlue}\-\ while\-\ \color{BrickRed}\-\ sup(real(zin))\-\ $>$\-\ sup(real(w1))\color{Green}
$\\$\color{BrickRed}\-\ \-\ \-\ \-\ zin\-\ =\-\ zin\-\ -\-\ nm('2')*w1;\color{Green}
$\\$\color{BrickRed}\color{NavyBlue}\-\ end\-\ \color{BrickRed}\color{Green}
$\\$\color{BrickRed}\color{NavyBlue}\-\ while\-\ \color{BrickRed}\-\ inf(real(zin))\-\ $<$\-\ -inf(real(w1))\color{Green}
$\\$\color{BrickRed}\-\ \-\ \-\ \-\ zin\-\ =\-\ zin\-\ +\-\ nm('2')*w1;\color{Green}
$\\$\color{BrickRed}\color{NavyBlue}\-\ end\-\ \color{BrickRed}\color{Green}
$\\$\color{BrickRed}\color{NavyBlue}\-\ while\-\ \color{BrickRed}\-\ sup(imag(zin))\-\ $>$\-\ sup(imag(w2))\color{Green}
$\\$\color{BrickRed}\-\ \-\ \-\ \-\ zin\-\ =\-\ zin\-\ -\-\ nm('2')*w2;\color{Green}
$\\$\color{BrickRed}\color{NavyBlue}\-\ end\-\ \color{BrickRed}\color{Green}
$\\$\color{BrickRed}\color{NavyBlue}\-\ while\-\ \color{BrickRed}\-\ inf(imag(zin))\-\ $<$\-\ inf(-imag(w2))\color{Green}
$\\$\color{BrickRed}\-\ \-\ \-\ \-\ zin\-\ =\-\ zin\-\ +nm('2')*w2;\color{Green}
$\\$\color{BrickRed}\color{NavyBlue}\-\ end\-\ \color{BrickRed}\color{Green}
$\\$
$\\$
$\\$
$\\$
$\\$\color{BrickRed}tau\-\ =\-\ w2/w1;\color{Green}
$\\$\color{BrickRed}z\-\ =\-\ zin/(nm('2')*w1);\color{Green}
$\\$
$\\$
$\\$\color{BrickRed}v01z\-\ =\-\ theta\_func(z+nm('0.5'),tau);\color{Green}
$\\$\color{BrickRed}v11z\-\ =\-\ exp(nm('0.25')*nm('pi')*nm('1i')*tau+nm('pi')*nm('1i')*(z+nm('0.5')))...\color{Green}
$\\$\color{BrickRed}\-\ \-\ \-\ \-\ *theta\_func(z+nm('0.5')*tau+nm('0.5'),tau);\color{Green}
$\\$
$\\$
$\\$\color{BrickRed}v00zzero\-\ =\-\ theta\_func(nm('0'),tau);\color{Green}
$\\$\color{BrickRed}v10zzero\-\ =\-\ exp(nm('0.25')*nm('pi')*nm('1i')*tau)*theta\_func(nm('0.5')*tau,tau);\color{Green}
$\\$
$\\$
$\\$\color{BrickRed}p\-\ =\-\ nm('pi')\verb|^|nm('2')*v00zzero\verb|^|nm('2')*v10zzero\verb|^|nm('2')*(v01z\verb|^|nm('2')/v11z\verb|^|nm('2'))...\color{Green}
$\\$\color{BrickRed}\-\ \-\ \-\ \-\ -(nm('pi')\verb|^|nm('2')/nm('3'))*(v00zzero\verb|^|nm('4')+v10zzero\verb|^|nm('4'));\color{Green}
$\\$
$\\$
$\\$\color{BrickRed}out\-\ =\-\ p/(nm('2')*w1)\verb|^|nm('2');\color{Green}
$\\$\color{Black}\section{weierstrass\_p\_der.m}

\color{Green}\color{BrickRed}\color{NavyBlue}\-\ function\-\ \color{BrickRed}\-\ out\-\ =\-\ weierstrass\_p\_der(z,w1,w2,varargin)\color{Green}
$\\$\color{Green}$\%$\-\ function\-\ out\-\ =\-\ weierstrass\_p\_der(z,w1,w2,varargin)
$\\$\color{Green}$\%$
$\\$\color{Green}$\%$\-\ Returns,\-\ computed\-\ with\-\ ineterval\-\ arithemtic,\-\ the\-\ derivative\-\ of\-\ the\-\ Weierstrass\-\ elliptic\-\ function
$\\$\color{Green}$\%$\-\ with\-\ half\-\ periods\-\ w1\-\ and\-\ w2\-\ evaluated\-\ at\-\ z.\-\ Here\-\ w1\-\ is\-\ a\-\ postive\-\ real\-\ number\-\ and\-\ 
$\\$\color{Green}$\%$\-\ w2\-\ is\-\ purely\-\ imaginary\-\ number\-\ lying\-\ above\-\ the\-\ real\-\ axis.\-\ 
$\\$
$\\$
$\\$\color{Green}$\%$\-\ use\-\ periodicity\-\ of\-\ the\-\ Weierstrass\-\ elliptic\-\ prime\-\ function\-\ to\-\ bring\-\ argument
$\\$\color{Green}$\%$\-\ close\-\ to\-\ origin\-\ where\-\ the\-\ program\-\ can\-\ compute\-\ more\-\ easily
$\\$\color{BrickRed}\color{NavyBlue}\-\ while\-\ \color{BrickRed}\-\ sup(real(z))\-\ $>$\-\ sup(real(w1))\color{Green}
$\\$\color{BrickRed}\-\ \-\ \-\ \-\ z\-\ =\-\ z\-\ -\-\ nm('2')*w1;\color{Green}
$\\$\color{BrickRed}\color{NavyBlue}\-\ end\-\ \color{BrickRed}\color{Green}
$\\$\color{BrickRed}\color{NavyBlue}\-\ while\-\ \color{BrickRed}\-\ inf(real(z))\-\ $<$\-\ -inf(real(w1))\color{Green}
$\\$\color{BrickRed}\-\ \-\ \-\ \-\ z\-\ =\-\ z\-\ +\-\ nm('2')*w1;\color{Green}
$\\$\color{BrickRed}\color{NavyBlue}\-\ end\-\ \color{BrickRed}\color{Green}
$\\$\color{BrickRed}\color{NavyBlue}\-\ while\-\ \color{BrickRed}\-\ sup(imag(z))\-\ $>$\-\ sup(imag(w2))\color{Green}
$\\$\color{BrickRed}\-\ \-\ \-\ \-\ z\-\ =\-\ z\-\ -\-\ nm('2')*w2;\color{Green}
$\\$\color{BrickRed}\color{NavyBlue}\-\ end\-\ \color{BrickRed}\color{Green}
$\\$\color{BrickRed}\color{NavyBlue}\-\ while\-\ \color{BrickRed}\-\ inf(imag(z))\-\ $<$\-\ inf(-imag(w2))\color{Green}
$\\$\color{BrickRed}\-\ \-\ \-\ \-\ z\-\ =\-\ z\-\ +nm('2')*w2;\color{Green}
$\\$\color{BrickRed}\color{NavyBlue}\-\ end\-\ \color{BrickRed}\color{Green}
$\\$
$\\$
$\\$\color{Green}$\%$
$\\$\color{Green}$\%$\-\ default\-\ values
$\\$\color{Green}$\%$
$\\$
$\\$
$\\$\color{BrickRed}abstol\-\ =\-\ 1e-16;\color{Green}
$\\$
$\\$
$\\$\color{Green}$\%$
$\\$\color{Green}$\%$\-\ constants
$\\$\color{Green}$\%$
$\\$
$\\$
$\\$\color{BrickRed}tau\-\ =\-\ w2/w1;\color{Green}
$\\$\color{BrickRed}z\-\ =\-\ z/(nm('2')*w1);\color{Green}
$\\$
$\\$
$\\$\color{Green}$\%$\-\ 
$\\$\color{Green}$\%$\-\ process\-\ optional\-\ user\-\ input
$\\$\color{Green}$\%$
$\\$
$\\$
$\\$\color{BrickRed}alen\-\ =\-\ length(varargin);\color{Green}
$\\$\color{BrickRed}pos\-\ =\-\ 0;\color{Green}
$\\$\color{BrickRed}\color{NavyBlue}\-\ while\-\ \color{BrickRed}\-\ pos\-\ $<$\-\ alen\color{Green}
$\\$\color{BrickRed}\-\ \-\ \-\ \-\ pos\-\ =\-\ pos+1;\color{Green}
$\\$\color{BrickRed}\-\ \-\ \-\ \-\ com\-\ =\-\ varargin{pos};\color{Green}
$\\$\color{BrickRed}\-\ \-\ \-\ \-\ \color{NavyBlue}\-\ if\-\ \color{BrickRed}\-\ isa(com,'char')\color{Green}
$\\$\color{BrickRed}\-\ \-\ \-\ \-\ \-\ \-\ \-\ \-\ \color{Green}
$\\$\color{BrickRed}\-\ \-\ \-\ \-\ \-\ \-\ \-\ \-\ switch\-\ \-\ lower(com)\color{Green}
$\\$\color{BrickRed}\-\ \-\ \-\ \-\ \-\ \-\ \-\ \-\ \-\ \-\ \-\ \-\ \color{Green}$\%$\-\ the\-\ smallest\-\ requirement\-\ on\-\ truncation\-\ error
$\\$\color{BrickRed}\-\ \-\ \-\ \-\ \-\ \-\ \-\ \-\ \-\ \-\ \-\ \-\ case\-\ 'abstol'\color{Green}
$\\$\color{BrickRed}\-\ \-\ \-\ \-\ \-\ \-\ \-\ \-\ \-\ \-\ \-\ \-\ \-\ \-\ \-\ \-\ pos\-\ =\-\ pos\-\ +\-\ 1;\color{Green}
$\\$\color{BrickRed}\-\ \-\ \-\ \-\ \-\ \-\ \-\ \-\ \-\ \-\ \-\ \-\ \-\ \-\ \-\ \-\ abstol\-\ =\-\ varargin{pos};\color{Green}
$\\$\color{BrickRed}\-\ \-\ \-\ \-\ \-\ \-\ \-\ \-\ \-\ \-\ \-\ \-\ otherwise\color{Green}
$\\$\color{BrickRed}\-\ \-\ \-\ \-\ \-\ \-\ \-\ \-\ \-\ \-\ \-\ \-\ \-\ \-\ \-\ \-\ error(['user\-\ input,\-\ ',com,'\-\ is\-\ not\-\ an\-\ option']);\color{Green}
$\\$\color{BrickRed}\-\ \-\ \-\ \-\ \-\ \-\ \-\ \-\ \color{NavyBlue}\-\ end\-\ \color{BrickRed}\color{Green}
$\\$\color{BrickRed}\-\ \-\ \-\ \-\ \-\ \-\ \-\ \-\ \color{Green}
$\\$\color{BrickRed}\-\ \-\ \-\ \-\ \color{NavyBlue}\-\ else\-\ \color{BrickRed}\color{Green}
$\\$\color{BrickRed}\-\ \-\ \-\ \-\ \-\ \-\ \-\ \-\ error('User\-\ error\-\ in\-\ optional\-\ arguments');\color{Green}
$\\$\color{BrickRed}\-\ \-\ \-\ \-\ \color{NavyBlue}\-\ end\-\ \color{BrickRed}\color{Green}
$\\$\color{BrickRed}\color{NavyBlue}\-\ end\-\ \color{BrickRed}\color{Green}
$\\$
$\\$
$\\$\color{Green}$\%$\-\ auxiliary\-\ Jacobi\-\ theta\-\ functions
$\\$\color{BrickRed}v01z\-\ =\-\ theta\_func(z+nm('0.5'),tau);\color{Green}
$\\$\color{BrickRed}v11z\-\ =\-\ exp(nm('0.25')*nm('pi')*nm('1i')*tau+nm('pi')*nm('1i')*(z+nm('0.5')))...\color{Green}
$\\$\color{BrickRed}\-\ \-\ \-\ \-\ *theta\_func(z+nm('0.5')*tau+nm('0.5'),tau);\color{Green}
$\\$
$\\$
$\\$\color{BrickRed}v00zzero\-\ =\-\ theta\_func(nm('0'),tau);\color{Green}
$\\$\color{BrickRed}v10zzero\-\ =\-\ exp(nm('0.25')*nm('pi')*nm('1i')*tau)*theta\_func(nm('0.5')*tau,tau);\color{Green}
$\\$
$\\$
$\\$\color{BrickRed}v01z\_der\-\ =\-\ theta\_func\_der(z+nm('0.5'),tau,'abstol',abstol);\color{Green}
$\\$\color{BrickRed}v11z\_der\-\ =\-\ nm('1i')*nm('pi')*v11z+...\color{Green}
$\\$\color{BrickRed}\-\ \-\ \-\ \-\ exp(nm('0.25')*nm('pi')*nm('1i')*tau+nm('pi')*nm('1i')*(z+nm('0.5')))...\color{Green}
$\\$\color{BrickRed}\-\ \-\ \-\ \-\ *theta\_func\_der(z+nm('0.5')*tau+nm('0.5'),tau,'AbsTol',abstol);\color{Green}
$\\$
$\\$
$\\$\color{Green}$\%$\-\ constant
$\\$\color{BrickRed}A\-\ =\-\ nm('pi')\verb|^|nm('2')*v00zzero\verb|^|nm('2')*v10zzero\verb|^|nm('2');\color{Green}
$\\$
$\\$
$\\$\color{Green}$\%$\-\ derivative\-\ 
$\\$\color{BrickRed}pz\-\ =\-\ nm('2')*A*(v01z*v01z\_der/v11z\verb|^|nm('2')-v01z\verb|^|nm('2')*v11z\_der/v11z\verb|^|nm('3'));\color{Green}
$\\$
$\\$
$\\$\color{Green}$\%$\-\ change\-\ of\-\ variables
$\\$\color{BrickRed}out\-\ =\-\ pz/(nm('2')*w1)\verb|^|nm('3');\color{Green}
$\\$
$\\$
$\\$
$\\$
$\\$\color{Black}\section{weierstrass\_sigma.m}

\color{Green}\color{BrickRed}\color{NavyBlue}\-\ function\-\ \color{BrickRed}\-\ out\-\ =\-\ weierstrass\_sigma(z,omega\_1,omega\_2,varargin)\color{Green}
$\\$
$\\$
$\\$\color{Green}$\%$\-\ use\-\ periodicity\-\ of\-\ the\-\ Weierstrass\-\ sigma\-\ function\-\ to\-\ bring\-\ argument
$\\$\color{Green}$\%$\-\ close\-\ to\-\ origin\-\ where\-\ the\-\ theta\_func\-\ program\-\ can\-\ compute\-\ more\-\ easily
$\\$\color{BrickRed}m1\-\ =\-\ 0;\-\ \color{Green}
$\\$\color{BrickRed}m2\-\ =\-\ 0;\color{Green}
$\\$\color{BrickRed}\color{NavyBlue}\-\ while\-\ \color{BrickRed}\-\ sup(real(z))\-\ $>$\-\ sup(real(omega\_1))\color{Green}
$\\$\color{BrickRed}\-\ \-\ \-\ \-\ z\-\ =\-\ z\-\ -\-\ nm('2')*omega\_1;\color{Green}
$\\$\color{BrickRed}\-\ \-\ \-\ \-\ m1\-\ =\-\ m1\-\ +\-\ 1;\color{Green}
$\\$\color{BrickRed}\color{NavyBlue}\-\ end\-\ \color{BrickRed}\color{Green}
$\\$\color{BrickRed}\color{NavyBlue}\-\ while\-\ \color{BrickRed}\-\ inf(real(z))\-\ $<$\-\ -inf(real(omega\_1))\color{Green}
$\\$\color{BrickRed}\-\ \-\ \-\ \-\ z\-\ =\-\ z\-\ +\-\ nm('2')*omega\_1;\color{Green}
$\\$\color{BrickRed}\-\ \-\ \-\ \-\ m1\-\ =\-\ m1\-\ -\-\ 1;\color{Green}
$\\$\color{BrickRed}\color{NavyBlue}\-\ end\-\ \color{BrickRed}\color{Green}
$\\$\color{BrickRed}\color{NavyBlue}\-\ while\-\ \color{BrickRed}\-\ sup(imag(z))\-\ $>$\-\ sup(imag(omega\_2))\color{Green}
$\\$\color{BrickRed}\-\ \-\ \-\ \-\ z\-\ =\-\ z\-\ -\-\ nm('2')*omega\_2;\color{Green}
$\\$\color{BrickRed}\-\ \-\ \-\ \-\ m2\-\ =\-\ m2\-\ +\-\ 1;\color{Green}
$\\$\color{BrickRed}\color{NavyBlue}\-\ end\-\ \color{BrickRed}\color{Green}
$\\$\color{BrickRed}\color{NavyBlue}\-\ while\-\ \color{BrickRed}\-\ inf(imag(z))\-\ $<$\-\ inf(-imag(omega\_2))\color{Green}
$\\$\color{BrickRed}\-\ \-\ \-\ \-\ z\-\ =\-\ z\-\ +nm('2')*omega\_2;\color{Green}
$\\$\color{BrickRed}\-\ \-\ \-\ \-\ m2\-\ =\-\ m2\-\ -\-\ 1;\color{Green}
$\\$\color{BrickRed}\color{NavyBlue}\-\ end\-\ \color{BrickRed}\color{Green}
$\\$
$\\$
$\\$\color{Green}$\%$
$\\$\color{Green}$\%$\-\ default\-\ values
$\\$\color{Green}$\%$
$\\$
$\\$
$\\$\color{BrickRed}AbsTol\-\ =\-\ 1e-16;\color{Green}
$\\$
$\\$
$\\$\color{Green}$\%$
$\\$\color{Green}$\%$\-\ constants
$\\$\color{Green}$\%$
$\\$
$\\$
$\\$\color{BrickRed}omega\_1\-\ =\-\ real(omega\_1);\color{Green}
$\\$\color{BrickRed}tau\-\ =\-\ omega\_2/omega\_1;\color{Green}
$\\$\color{BrickRed}q\-\ =\-\ exp(-nm('pi')*imag(tau));\color{Green}
$\\$
$\\$
$\\$\color{Green}$\%$\-\ 
$\\$\color{Green}$\%$\-\ process\-\ optional\-\ user\-\ input
$\\$\color{Green}$\%$
$\\$
$\\$
$\\$\color{BrickRed}alen\-\ =\-\ length(varargin);\color{Green}
$\\$\color{BrickRed}pos\-\ =\-\ 0;\color{Green}
$\\$\color{BrickRed}\color{NavyBlue}\-\ while\-\ \color{BrickRed}\-\ pos\-\ $<$\-\ alen\color{Green}
$\\$\color{BrickRed}\-\ \-\ \-\ \-\ pos\-\ =\-\ pos+1;\color{Green}
$\\$\color{BrickRed}\-\ \-\ \-\ \-\ com\-\ =\-\ varargin{pos};\color{Green}
$\\$\color{BrickRed}\-\ \-\ \-\ \-\ \color{NavyBlue}\-\ if\-\ \color{BrickRed}\-\ isa(com,'char')\color{Green}
$\\$\color{BrickRed}\-\ \-\ \-\ \-\ \-\ \-\ \-\ \-\ \color{Green}
$\\$\color{BrickRed}\-\ \-\ \-\ \-\ \-\ \-\ \-\ \-\ switch\-\ \-\ lower(com)\color{Green}
$\\$\color{BrickRed}\-\ \-\ \-\ \-\ \-\ \-\ \-\ \-\ \-\ \-\ \-\ \-\ \color{Green}$\%$\-\ the\-\ smallest\-\ requirement\-\ on\-\ truncation\-\ error
$\\$\color{BrickRed}\-\ \-\ \-\ \-\ \-\ \-\ \-\ \-\ \-\ \-\ \-\ \-\ case\-\ 'abstol'\color{Green}
$\\$\color{BrickRed}\-\ \-\ \-\ \-\ \-\ \-\ \-\ \-\ \-\ \-\ \-\ \-\ \-\ \-\ \-\ \-\ pos\-\ =\-\ pos\-\ +\-\ 1;\color{Green}
$\\$\color{BrickRed}\-\ \-\ \-\ \-\ \-\ \-\ \-\ \-\ \-\ \-\ \-\ \-\ \-\ \-\ \-\ \-\ abstol\-\ =\-\ varargin{pos};\color{Green}
$\\$\color{BrickRed}\-\ \-\ \-\ \-\ \-\ \-\ \-\ \-\ \-\ \-\ \-\ \-\ \color{Green}$\%$\-\ Weierstrass\-\ eta\_1
$\\$\color{BrickRed}\-\ \-\ \-\ \-\ \-\ \-\ \-\ \-\ \-\ \-\ \-\ \-\ case\-\ 'eta1'\color{Green}
$\\$\color{BrickRed}\-\ \-\ \-\ \-\ \-\ \-\ \-\ \-\ \-\ \-\ \-\ \-\ \-\ \-\ \-\ \-\ pos\-\ =\-\ pos\-\ +\-\ 1;\color{Green}
$\\$\color{BrickRed}\-\ \-\ \-\ \-\ \-\ \-\ \-\ \-\ \-\ \-\ \-\ \-\ \-\ \-\ \-\ \-\ eta\_1\-\ =\-\ varargin{pos};\color{Green}
$\\$\color{BrickRed}\-\ \-\ \-\ \-\ \-\ \-\ \-\ \-\ \-\ \-\ \-\ \-\ otherwise\color{Green}
$\\$\color{BrickRed}\-\ \-\ \-\ \-\ \-\ \-\ \-\ \-\ \-\ \-\ \-\ \-\ \-\ \-\ \-\ \-\ error(['user\-\ input,\-\ ',com,'\-\ is\-\ not\-\ an\-\ option']);\color{Green}
$\\$\color{BrickRed}\-\ \-\ \-\ \-\ \-\ \-\ \-\ \-\ \color{NavyBlue}\-\ end\-\ \color{BrickRed}\color{Green}
$\\$\color{BrickRed}\-\ \-\ \-\ \-\ \-\ \-\ \-\ \-\ \color{Green}
$\\$\color{BrickRed}\-\ \-\ \-\ \-\ \color{NavyBlue}\-\ else\-\ \color{BrickRed}\color{Green}
$\\$\color{BrickRed}\-\ \-\ \-\ \-\ \-\ \-\ \-\ \-\ error('User\-\ error\-\ in\-\ optional\-\ arguments');\color{Green}
$\\$\color{BrickRed}\-\ \-\ \-\ \-\ \color{NavyBlue}\-\ end\-\ \color{BrickRed}\color{Green}
$\\$\color{BrickRed}\color{NavyBlue}\-\ end\-\ \color{BrickRed}\color{Green}
$\\$
$\\$
$\\$\color{Green}$\%$
$\\$\color{Green}$\%$\-\ choose\-\ k\-\ large\-\ enough\-\ that\-\ error\-\ is\-\ less\-\ than\-\ AbsTol
$\\$\color{Green}$\%$
$\\$
$\\$
$\\$\color{BrickRed}k\-\ =1;\color{Green}
$\\$\color{BrickRed}ks\-\ =\-\ nm(num2str(k));\color{Green}
$\\$\color{Green}$\%$\-\ enusre\-\ k\-\ is\-\ large\-\ enough\-\ that\-\ error\_bound\-\ $<$\-\ AbsTol
$\\$\color{BrickRed}c1\-\ =\-\ nm('1')/(nm('4')*ks*(nm('1')-q\verb|^|(nm('2')*ks)));\color{Green}
$\\$\color{BrickRed}c2\-\ =\-\ exp(-imag(tau)*nm('pi')*nm('2')-imag(z)*nm('pi')/omega\_1);\color{Green}
$\\$\color{BrickRed}c3\-\ =\-\ exp(-imag(tau)*nm('pi')*nm('2')+imag(z)*nm('pi')/omega\_1);\color{Green}
$\\$\color{BrickRed}c4\-\ =\-\ exp(-imag(tau)*nm('pi')*nm('2'));\color{Green}
$\\$
$\\$
$\\$\color{BrickRed}error\_bound\-\ =\-\ c1*(c2\verb|^|ks/(nm('1')-c2)+nm('2')*c4/(nm('1')-c4)+c3\verb|^|ks/(nm('1')-c3));\color{Green}
$\\$\color{BrickRed}\color{NavyBlue}\-\ while\-\ \color{BrickRed}\-\ sup(error\_bound)\-\ $>$\-\ AbsTol\color{Green}
$\\$\color{BrickRed}\-\ \-\ \-\ \-\ k\-\ =\-\ k+1;\color{Green}
$\\$\color{BrickRed}\-\ \-\ \-\ \-\ ks\-\ =\-\ nm(num2str(k));\color{Green}
$\\$\color{BrickRed}\-\ \-\ \-\ \-\ c1\-\ =\-\ nm('1')/(nm('4')*ks*(nm('1')-q\verb|^|(nm('2')*ks)));\color{Green}
$\\$\color{BrickRed}\-\ \-\ \-\ \-\ error\_bound\-\ =\-\ c1*(c2\verb|^|ks/(nm('1')-c2)+nm('2')*q\verb|^|(nm('2')*ks)/(nm('1')-c4)+c3\verb|^|ks/(nm('1')-c3));\color{Green}
$\\$\color{BrickRed}\color{NavyBlue}\-\ end\-\ \color{BrickRed}\color{Green}
$\\$
$\\$
$\\$\color{BrickRed}sum\_error\_bound\-\ =\-\ hull(-abs(error\_bound),abs(error\_bound))...\color{Green}
$\\$\color{BrickRed}\-\ \-\ \-\ \-\ \-\ \-\ \-\ \-\ \-\ \-\ \-\ \-\ \-\ \-\ +nm('1i')*hull(-abs(error\_bound),abs(error\_bound));\color{Green}
$\\$
$\\$
$\\$\color{BrickRed}sum\-\ =\-\ nm('0');\-\ \-\ \-\ \-\ \-\ \color{Green}
$\\$\color{BrickRed}\color{NavyBlue}\-\ for\-\ \color{BrickRed}\-\ n\-\ =\-\ 1:k-1\color{Green}
$\\$\color{BrickRed}\-\ \-\ \-\ \-\ ns\-\ =\-\ nm(num2str(n));\color{Green}
$\\$\color{BrickRed}\-\ \-\ \-\ \-\ sum\-\ =\-\ sum\-\ +\-\ (q\verb|^|(nm('2')*ns)/(ns*(nm('1')-...\color{Green}
$\\$\color{BrickRed}\-\ \-\ \-\ \-\ \-\ \-\ \-\ \-\ q\verb|^|(nm('2')*ns))))*sin(ns*nm('pi')*z/(nm('2')*omega\_1))\verb|^|nm('2');\color{Green}
$\\$\color{BrickRed}\color{NavyBlue}\-\ end\-\ \color{BrickRed}\color{Green}
$\\$
$\\$
$\\$\color{BrickRed}sum\-\ =\-\ sum\-\ +\-\ sum\_error\_bound;\color{Green}
$\\$
$\\$
$\\$\color{BrickRed}\color{NavyBlue}\-\ if\-\ \color{BrickRed}\-\ ~(exist('eta\_1'))==1\color{Green}
$\\$\color{BrickRed}\-\ \-\ \-\ \-\ eta\_1\-\ =\-\ weierstrass\_eta1(omega\_1,omega\_2);\color{Green}
$\\$\color{BrickRed}\color{NavyBlue}\-\ end\-\ \color{BrickRed}\color{Green}
$\\$
$\\$
$\\$\color{BrickRed}lg\-\ =\-\ eta\_1*z\verb|^|nm('2')/(nm('2')*omega\_1)+nm('4')*sum;\color{Green}
$\\$\color{BrickRed}out\-\ =\-\ (nm('2')*omega\_1/nm('pi'))*(sin(nm('pi')*z/(nm('2')*omega\_1)))*exp(lg);\color{Green}
$\\$
$\\$
$\\$\color{Green}$\%$\-\ use\-\ quasi-periodicity\-\ of\-\ the\-\ Weierstrass\-\ zeta\-\ function
$\\$\color{BrickRed}\color{NavyBlue}\-\ if\-\ \color{BrickRed}\-\ (m1==0)\&\&(m2==0)\color{Green}
$\\$\color{BrickRed}\-\ \-\ \-\ \-\ \color{NavyBlue}\-\ return\-\ \color{BrickRed}\color{Green}
$\\$\color{BrickRed}\color{NavyBlue}\-\ else\-\ \color{BrickRed}\color{Green}
$\\$\color{BrickRed}\-\ \-\ \-\ \-\ m1s\-\ =\-\ nm(num2str(m1));\color{Green}
$\\$\color{BrickRed}\-\ \-\ \-\ \-\ m2s\-\ =\-\ nm(num2str(m2));\color{Green}
$\\$\color{BrickRed}\-\ \-\ \-\ \-\ eta\_2\-\ =\-\ weierstrass\_zeta(omega\_2,omega\_1,omega\_2);\color{Green}
$\\$\color{BrickRed}\-\ \-\ \-\ \-\ out\-\ =\-\ out\-\ *nm('-1')\verb|^|(m1s*m2s+m1s+m2s)...\color{Green}
$\\$\color{BrickRed}\-\ \-\ \-\ \-\ \-\ \-\ \-\ \-\ *exp(nm('2')*(m1s*eta\_1+m2s*eta\_2)*(z+m1s*omega\_1+m2s*omega\_2));\color{Green}
$\\$\color{BrickRed}\color{NavyBlue}\-\ end\-\ \color{BrickRed}\color{Green}
$\\$
$\\$
$\\$
$\\$
$\\$
$\\$
$\\$
$\\$
$\\$
$\\$
$\\$
$\\$
$\\$
$\\$
$\\$
$\\$
$\\$
$\\$
$\\$
$\\$
$\\$
$\\$
$\\$
$\\$
$\\$
$\\$
$\\$
$\\$
$\\$\color{Black}\section{weierstrass\_zeta.m}

\color{Green}\color{BrickRed}\color{NavyBlue}\-\ function\-\ \color{BrickRed}\-\ out\-\ =\-\ weierstrass\_zeta(z,omega\_1,omega\_2,varargin)\color{Green}
$\\$
$\\$
$\\$\color{Green}$\%$\-\ use\-\ periodicity\-\ of\-\ the\-\ Weierstrass\-\ zeta\-\ function\-\ to\-\ bring\-\ argument
$\\$\color{Green}$\%$\-\ close\-\ to\-\ origin\-\ where\-\ the\-\ theta\_func\-\ program\-\ can\-\ succesffuly\-\ compute
$\\$\color{BrickRed}m\-\ =\-\ 0;\-\ \color{Green}
$\\$\color{BrickRed}n\-\ =\-\ 0;\color{Green}
$\\$\color{BrickRed}\color{NavyBlue}\-\ while\-\ \color{BrickRed}\-\ sup(real(z))\-\ $>$\-\ sup(real(omega\_1))\color{Green}
$\\$\color{BrickRed}\-\ \-\ \-\ \-\ z\-\ =\-\ z\-\ -\-\ nm('2')*omega\_1;\color{Green}
$\\$\color{BrickRed}\-\ \-\ \-\ \-\ m\-\ =\-\ m\-\ +\-\ 1;\color{Green}
$\\$\color{BrickRed}\color{NavyBlue}\-\ end\-\ \color{BrickRed}\color{Green}
$\\$\color{BrickRed}\color{NavyBlue}\-\ while\-\ \color{BrickRed}\-\ inf(real(z))\-\ $<$\-\ inf(-real(omega\_1))\color{Green}
$\\$\color{BrickRed}\-\ \-\ \-\ \-\ z\-\ =\-\ z\-\ +\-\ nm('2')*omega\_1;\color{Green}
$\\$\color{BrickRed}\-\ \-\ \-\ \-\ m\-\ =\-\ m\-\ -\-\ 1;\color{Green}
$\\$\color{BrickRed}\color{NavyBlue}\-\ end\-\ \color{BrickRed}\color{Green}
$\\$\color{BrickRed}\color{NavyBlue}\-\ while\-\ \color{BrickRed}\-\ sup(imag(z))\-\ $>$\-\ sup(imag(omega\_2))\color{Green}
$\\$\color{BrickRed}\-\ \-\ \-\ \-\ z\-\ =\-\ z\-\ -\-\ nm('2')*omega\_2;\color{Green}
$\\$\color{BrickRed}\-\ \-\ \-\ \-\ n\-\ =\-\ n\-\ +\-\ 1;\color{Green}
$\\$\color{BrickRed}\color{NavyBlue}\-\ end\-\ \color{BrickRed}\color{Green}
$\\$\color{BrickRed}\color{NavyBlue}\-\ while\-\ \color{BrickRed}\-\ inf(imag(z))\-\ $<$\-\ inf(-imag(omega\_2))\color{Green}
$\\$\color{BrickRed}\-\ \-\ \-\ \-\ z\-\ =\-\ z\-\ +nm('2')*omega\_2;\color{Green}
$\\$\color{BrickRed}\-\ \-\ \-\ \-\ n\-\ =\-\ n\-\ -\-\ 1;\color{Green}
$\\$\color{BrickRed}\color{NavyBlue}\-\ end\-\ \color{BrickRed}\color{Green}
$\\$
$\\$
$\\$\color{BrickRed}tau\-\ =\-\ omega\_2/omega\_1;\color{Green}
$\\$\color{Green}$\%$\-\ nome
$\\$\color{BrickRed}q\-\ =\-\ exp(nm('pi')*nm('1i')*tau);\color{Green}
$\\$
$\\$
$\\$\color{Green}$\%$
$\\$\color{Green}$\%$\-\ user\-\ error\-\ check
$\\$\color{Green}$\%$
$\\$
$\\$
$\\$\color{Green}$\%$\-\ make\-\ sure\-\ tau\-\ lies\-\ in\-\ the\-\ upper\-\ half\-\ of\-\ complex\-\ plane
$\\$\color{BrickRed}\color{NavyBlue}\-\ if\-\ \color{BrickRed}\-\ imag(tau)\-\ $<$=\-\ 0\color{Green}
$\\$\color{BrickRed}\-\ \-\ \-\ \-\ error('omega\_2/omega\_1\-\ must\-\ lie\-\ in\-\ the\-\ upper\-\ half\-\ of\-\ the\-\ complex\-\ plane')\color{Green}
$\\$\color{BrickRed}\color{NavyBlue}\-\ end\-\ \color{BrickRed}\color{Green}
$\\$
$\\$
$\\$\color{Green}$\%$\-\ make\-\ sure\-\ tau\-\ is\-\ purely\-\ imaginary
$\\$\color{BrickRed}\color{NavyBlue}\-\ if\-\ \color{BrickRed}\-\ real(tau)\-\ $>$\-\ 0\color{Green}
$\\$\color{BrickRed}\-\ \-\ \-\ \-\ error('tau\-\ must\-\ be\-\ purely\-\ imaginary');\color{Green}
$\\$\color{BrickRed}\color{NavyBlue}\-\ end\-\ \color{BrickRed}\color{Green}
$\\$
$\\$
$\\$\color{Green}$\%$
$\\$\color{Green}$\%$\-\ default\-\ values
$\\$\color{Green}$\%$
$\\$
$\\$
$\\$\color{Green}$\%$\-\ absolute\-\ error\-\ tolerance\-\ of\-\ truncated\-\ remainder
$\\$\color{BrickRed}abstol\-\ =\-\ 1e-16;\color{Green}
$\\$
$\\$
$\\$\color{Green}$\%$\-\ 
$\\$\color{Green}$\%$\-\ process\-\ optional\-\ user\-\ input
$\\$\color{Green}$\%$
$\\$
$\\$
$\\$\color{BrickRed}alen\-\ =\-\ length(varargin);\color{Green}
$\\$\color{BrickRed}pos\-\ =\-\ 0;\color{Green}
$\\$\color{BrickRed}\color{NavyBlue}\-\ while\-\ \color{BrickRed}\-\ pos\-\ $<$\-\ alen\color{Green}
$\\$\color{BrickRed}\-\ \-\ \-\ \-\ pos\-\ =\-\ pos+1;\color{Green}
$\\$\color{BrickRed}\-\ \-\ \-\ \-\ com\-\ =\-\ varargin{pos};\color{Green}
$\\$\color{BrickRed}\-\ \-\ \-\ \-\ \color{NavyBlue}\-\ if\-\ \color{BrickRed}\-\ isa(com,'char')\color{Green}
$\\$\color{BrickRed}\-\ \-\ \-\ \-\ \-\ \-\ \-\ \-\ \color{Green}
$\\$\color{BrickRed}\-\ \-\ \-\ \-\ \-\ \-\ \-\ \-\ switch\-\ \-\ lower(com)\color{Green}
$\\$\color{BrickRed}\-\ \-\ \-\ \-\ \-\ \-\ \-\ \-\ \-\ \-\ \-\ \-\ \color{Green}$\%$\-\ the\-\ smallest\-\ requirement\-\ on\-\ truncation\-\ error
$\\$\color{BrickRed}\-\ \-\ \-\ \-\ \-\ \-\ \-\ \-\ \-\ \-\ \-\ \-\ case\-\ 'abstol'\color{Green}
$\\$\color{BrickRed}\-\ \-\ \-\ \-\ \-\ \-\ \-\ \-\ \-\ \-\ \-\ \-\ \-\ \-\ \-\ \-\ pos\-\ =\-\ pos\-\ +\-\ 1;\color{Green}
$\\$\color{BrickRed}\-\ \-\ \-\ \-\ \-\ \-\ \-\ \-\ \-\ \-\ \-\ \-\ \-\ \-\ \-\ \-\ abstol\-\ =\-\ varargin{pos};\color{Green}
$\\$\color{BrickRed}\-\ \-\ \-\ \-\ \-\ \-\ \-\ \-\ \-\ \-\ \-\ \-\ \color{Green}$\%$\-\ Weierstrass\-\ eta\_1
$\\$\color{BrickRed}\-\ \-\ \-\ \-\ \-\ \-\ \-\ \-\ \-\ \-\ \-\ \-\ case\-\ 'eta1'\color{Green}
$\\$\color{BrickRed}\-\ \-\ \-\ \-\ \-\ \-\ \-\ \-\ \-\ \-\ \-\ \-\ \-\ \-\ \-\ \-\ pos\-\ =\-\ pos\-\ +\-\ 1;\color{Green}
$\\$\color{BrickRed}\-\ \-\ \-\ \-\ \-\ \-\ \-\ \-\ \-\ \-\ \-\ \-\ \-\ \-\ \-\ \-\ eta\_1\-\ =\-\ varargin{pos};\color{Green}
$\\$\color{BrickRed}\-\ \-\ \-\ \-\ \-\ \-\ \-\ \-\ \-\ \-\ \-\ \-\ otherwise\color{Green}
$\\$\color{BrickRed}\-\ \-\ \-\ \-\ \-\ \-\ \-\ \-\ \-\ \-\ \-\ \-\ \-\ \-\ \-\ \-\ error(['user\-\ input,\-\ ',com,'\-\ is\-\ not\-\ an\-\ option']);\color{Green}
$\\$\color{BrickRed}\-\ \-\ \-\ \-\ \-\ \-\ \-\ \-\ \color{NavyBlue}\-\ end\-\ \color{BrickRed}\color{Green}
$\\$\color{BrickRed}\-\ \-\ \-\ \-\ \-\ \-\ \-\ \-\ \color{Green}
$\\$\color{BrickRed}\-\ \-\ \-\ \-\ \color{NavyBlue}\-\ else\-\ \color{BrickRed}\color{Green}
$\\$\color{BrickRed}\-\ \-\ \-\ \-\ \-\ \-\ \-\ \-\ error('User\-\ error\-\ in\-\ optional\-\ arguments');\color{Green}
$\\$\color{BrickRed}\-\ \-\ \-\ \-\ \color{NavyBlue}\-\ end\-\ \color{BrickRed}\color{Green}
$\\$\color{BrickRed}\color{NavyBlue}\-\ end\-\ \color{BrickRed}\color{Green}
$\\$
$\\$
$\\$\color{BrickRed}\color{NavyBlue}\-\ if\-\ \color{BrickRed}\-\ ~(exist('eta\_1'))==1\color{Green}
$\\$\color{BrickRed}\-\ \-\ \-\ \-\ eta\_1\-\ =\-\ weierstrass\_eta1(omega\_1,omega\_2);\color{Green}
$\\$\color{BrickRed}\color{NavyBlue}\-\ end\-\ \color{BrickRed}\color{Green}
$\\$
$\\$
$\\$\color{Green}$\%$\-\ k\-\ so\-\ that\-\ the\-\ error\-\ bound\-\ is\-\ less\-\ than\-\ AbsTol
$\\$\color{BrickRed}k\-\ =\-\ 1;\color{Green}
$\\$\color{BrickRed}error\_bound\-\ =\-\ nm(abstol\-\ +\-\ 1);\color{Green}
$\\$\color{BrickRed}a\_plus\-\ =\-\ (nm('pi')/omega\_1)*(\-\ -nm('2')*imag(omega\_2)+abs(imag(z)));\color{Green}
$\\$\color{BrickRed}a\_minus\-\ =\-\ (nm('pi')/omega\_1)*(\-\ -nm('2')*imag(omega\_2)-abs(imag(z)));\color{Green}
$\\$\color{BrickRed}con\_plus\-\ =\-\ nm('1')-exp(a\_plus);\color{Green}
$\\$\color{BrickRed}con\_minus\-\ =\-\ nm('1')-exp(a\_minus);\color{Green}
$\\$\color{BrickRed}cnt\-\ =\-\ 0;\color{Green}
$\\$\color{BrickRed}\color{NavyBlue}\-\ while\-\ \color{BrickRed}\-\ sup(error\_bound)\-\ $>$\-\ abstol\color{Green}
$\\$\color{BrickRed}\-\ \-\ \-\ \-\ cnt\-\ =\-\ cnt\-\ +\-\ 1;\color{Green}
$\\$\color{BrickRed}\-\ \-\ \-\ \-\ \color{NavyBlue}\-\ if\-\ \color{BrickRed}\-\ cnt\-\ $>$\-\ 1000\color{Green}
$\\$\color{BrickRed}\-\ \-\ \-\ \-\ \-\ \-\ \-\ \-\ error('failure\-\ to\-\ find\-\ error\-\ bound');\color{Green}
$\\$\color{BrickRed}\-\ \-\ \-\ \-\ \color{NavyBlue}\-\ end\-\ \color{BrickRed}\color{Green}
$\\$\color{BrickRed}\-\ \-\ \-\ \-\ k\-\ =\-\ k+1;\color{Green}
$\\$\color{BrickRed}\-\ \-\ \-\ \-\ ks\-\ =\-\ nm(num2str(k));\color{Green}
$\\$\color{BrickRed}\-\ \-\ \-\ \-\ c\-\ =\-\ (nm('0.5')/(nm('1')-q\verb|^|(nm('2')*ks)));\color{Green}
$\\$\color{BrickRed}\-\ \-\ \-\ \-\ error\_bound\-\ =\-\ abs((c*(exp(a\_plus)\verb|^|ks/con\_plus\-\ +\-\ exp(a\_minus)\verb|^|ks/con\_minus)));\color{Green}
$\\$\color{BrickRed}\color{NavyBlue}\-\ end\-\ \color{BrickRed}\color{Green}
$\\$
$\\$
$\\$\color{Green}$\%$\-\ add\-\ the\-\ first\-\ k-1\-\ terms\-\ of\-\ the\-\ sum
$\\$\color{BrickRed}sum\-\ =\-\ nm('0');\color{Green}
$\\$\color{BrickRed}\color{NavyBlue}\-\ for\-\ \color{BrickRed}\-\ j\-\ =\-\ 1:k-1\color{Green}
$\\$\color{BrickRed}\-\ \-\ \-\ \-\ js\-\ =\-\ nm(num2str(j));\color{Green}
$\\$\color{BrickRed}\-\ \-\ \-\ \-\ sum\-\ =\-\ sum\-\ +\-\ (q\verb|^|(nm('2')*js)/(nm('1')-q\verb|^|(nm('2')*js)))*sin(js*nm('pi')*z/omega\_1);\color{Green}
$\\$\color{BrickRed}\color{NavyBlue}\-\ end\-\ \color{BrickRed}\color{Green}
$\\$
$\\$
$\\$\color{BrickRed}sum\-\ =\-\ sum\-\ +\-\ hull(-error\_bound,error\_bound)\-\ +\-\ nm('1i')*hull(-error\_bound,error\_bound);\color{Green}
$\\$
$\\$
$\\$\color{BrickRed}out\-\ =\-\ eta\_1*z/omega\_1+(nm('pi')/(nm('2')*omega\_1))*cot(nm('pi')*z/(nm('2')*omega\_1))+...\color{Green}
$\\$\color{BrickRed}\-\ \-\ \-\ \-\ (nm('2')*nm('pi')/omega\_1)*sum;\color{Green}
$\\$
$\\$
$\\$\color{Green}$\%$\-\ use\-\ quasi-periodicity\-\ of\-\ the\-\ Weierstrass\-\ zeta\-\ function
$\\$\color{BrickRed}\color{NavyBlue}\-\ if\-\ \color{BrickRed}\-\ (m==0)\&\&(n==0)\color{Green}
$\\$\color{BrickRed}\-\ \-\ \-\ \-\ \color{NavyBlue}\-\ return\-\ \color{BrickRed}\color{Green}
$\\$\color{BrickRed}\color{NavyBlue}\-\ else\-\ \color{BrickRed}\color{Green}
$\\$\color{BrickRed}\-\ \-\ \-\ \-\ out\-\ =\-\ out\-\ +\-\ nm('2')*nm(num2str(m))*eta\_1+nm('2')*nm(num2str(n))...\color{Green}
$\\$\color{BrickRed}\-\ \-\ \-\ \-\ \-\ \-\ \-\ \-\ *weierstrass\_zeta(omega\_2,omega\_1,omega\_2);\color{Green}
$\\$\color{BrickRed}\color{NavyBlue}\-\ end\-\ \color{BrickRed}\color{Green}
$\\$
$\\$
$\\$\color{Black}\section{xi\_der\_q\_psi.m}

\color{Green}\color{BrickRed}\color{NavyBlue}\-\ function\-\ \color{BrickRed}\-\ out\-\ =\-\ xi\_der\_q\_psi(q,psi,ntilde)\color{Green}
$\\$\color{Green}$\%$\-\ function\-\ out\-\ =\-\ xi\_der\_q\_psi(q,psi,ntilde)
$\\$\color{Green}$\%$
$\\$\color{Green}$\%$\-\ returns\-\ the\-\ derivative\-\ of\-\ \-\ omega*xi\-\ with\-\ respect\-\ to\-\ psi
$\\$
$\\$
$\\$
$\\$\color{Black}
Now 
\eq{
\pd{}{\psi}\omega \xi ( \omega + i\psi \omega')&= 2\omega \omega' \left(\left(\frac{\pi}{2\omega}\right)^2\sec^2\left(\frac{i\pi \psi \omega'}{2\omega}\right)-\frac{\pi^2}{\omega^2} \sum_{k=1}^{\infty}(-1)^k \frac{kq^{2k}}{1-q^{2k}}\left(q^{\psi k}+q^{-\psi k}\right) \right)\\
&= \frac{-\pi \log(q)}{2}\sec^2\left(\frac{\psi \log(q)i}{2}\right) + 2\pi \log(q)\sum_{k=1}^{\infty} (-1)^k \frac{kq^{2k}}{1-q^{2k}}\left(q^{\psi k}+q^{-\psi k}\right)
}{\notag}

$\\$
Note that 

$\\$
\eq{
\left|\sum_{k=N+1}^{\infty} (-1)^k \frac{kq^{2k}}{1-q^{2k}}\left(q^{\psi k}+q^{-\psi k}\right)\right|&\leq \sum_{k=N+1}^{\infty} \frac{kq^{2k}}{1-q^{2k}}\left(q^{\psi k}+q^{-\psi k}\right)\\
&\leq \frac{2}{1-q^{2(N+1)}}\sum_{k=0}^{\infty} (N+1+k)q^{(2-\psi)(N+1+k)}\\
&\leq \frac{2(N+1)q^{(2-\psi)(N+1)}}{1-q^{2(N+1)}}\sum_{k=0}^{\infty}q^{(2-\psi)k} + \frac{2q^{(2-\psi)N}}{1-q^{2(N+1)}}\sum_{k=0}^{\infty} k q^{(2-\psi)k}\\
&\leq  \frac{2(N+1)q^{(2-\psi)(N+1)}}{1-q^{2(N+1)}}\frac{1}{1-q^{(2-\psi)}} + \frac{2q^{(2-\psi)N}}{1-q^{2(N+1)}}\frac{q^{(2-\psi)}}{(1-q^{(2-\psi)})^2}
}{\notag}
\color{Green}
$\\$
$\\$\color{Green}$\%$$\%$
$\\$
$\\$
$\\$\color{Green}$\%$
$\\$\color{Green}$\%$\-\ constants
$\\$\color{Green}$\%$
$\\$
$\\$
$\\$\color{BrickRed}pie\-\ =\-\ nm('pi');\color{Green}
$\\$\color{BrickRed}psi0\-\ =\-\ max(sup(psi));\color{Green}
$\\$\color{BrickRed}q0\-\ =\-\ abs(q\verb|^|(2-psi0));\color{Green}
$\\$\color{BrickRed}qs0\-\ =\-\ q0\verb|^|(2-sup(abs(psi0)));\color{Green}
$\\$\color{BrickRed}q02\-\ =\-\ q0*q0;\color{Green}
$\\$
$\\$
$\\$\color{Green}$\%$\-\ 
$\\$\color{Green}$\%$\-\ find\-\ N\-\ for\-\ given\-\ error
$\\$\color{Green}$\%$
$\\$
$\\$
$\\$\color{BrickRed}N\-\ =\-\ 0;\color{Green}
$\\$\color{BrickRed}maxit\-\ =\-\ 1000;\color{Green}
$\\$\color{BrickRed}tol\-\ =\-\ 1e-17;\color{Green}
$\\$\color{BrickRed}err\-\ =\-\ tol\-\ +\-\ 1;\color{Green}
$\\$\color{BrickRed}\color{NavyBlue}\-\ while\-\ \color{BrickRed}\-\ err\-\ $>$\-\ tol\color{Green}
$\\$\color{BrickRed}\-\ \-\ \-\ \-\ N\-\ =\-\ N+1;\color{Green}
$\\$\color{BrickRed}\-\ \-\ \-\ \-\ \color{NavyBlue}\-\ if\-\ \color{BrickRed}\-\ N\-\ $>$\-\ maxit\color{Green}
$\\$\color{BrickRed}\-\ \-\ \-\ \-\ \-\ \-\ \-\ \-\ error('Maximum\-\ iterations\-\ exceeded');\color{Green}
$\\$\color{BrickRed}\-\ \-\ \-\ \-\ \color{NavyBlue}\-\ end\-\ \color{BrickRed}\color{Green}
$\\$\color{BrickRed}\-\ \-\ \-\ \-\ temp\-\ =\-\ 2*(N+1)*qs0\verb|^|(N+1)/((1-q02\verb|^|(N+1))*(1-qs0))\-\ +\-\ 2*qs0\verb|^|(N+1)/((1-q02\verb|^|(N+1))*(1-qs0)\verb|^|2);\color{Green}
$\\$\color{BrickRed}\-\ \-\ \-\ \-\ err\-\ =\-\ sup(temp);\color{Green}
$\\$\color{BrickRed}\color{NavyBlue}\-\ end\-\ \color{BrickRed}\color{Green}
$\\$
$\\$
$\\$\color{Green}$\%$\-\ initialize\-\ output
$\\$\color{BrickRed}out\-\ =\-\ nm(zeros(1,length(psi)));\color{Green}
$\\$
$\\$
$\\$\color{Green}$\%$\-\ find\-\ partial\-\ sum
$\\$\color{BrickRed}q2\-\ =\-\ q*q;\color{Green}
$\\$\color{BrickRed}\color{NavyBlue}\-\ for\-\ \color{BrickRed}\-\ k\-\ =\-\ 1:N\color{Green}
$\\$\color{BrickRed}\-\ \-\ \-\ \-\ q2k\-\ =\-\ q2\verb|^|k;\color{Green}
$\\$\color{BrickRed}\-\ \-\ \-\ \-\ out\-\ =\-\ out\-\ +\-\ ((-1)\verb|^|ntilde)\verb|^|k*(k*q2k/(1-q2k))*(q.\verb|^|(k*psi)+q.\verb|^|(-k*psi));\color{Green}
$\\$\color{BrickRed}\color{NavyBlue}\-\ end\-\ \color{BrickRed}\color{Green}
$\\$
$\\$
$\\$\color{Green}$\%$\-\ add\-\ error
$\\$\color{BrickRed}out\-\ =\-\ out\-\ +\-\ nm(-err,err)+1i*nm(-err,err);\color{Green}
$\\$
$\\$
$\\$\color{Green}$\%$\-\ add\-\ non\-\ infinite\-\ sum\-\ parts
$\\$\color{BrickRed}\color{NavyBlue}\-\ if\-\ \color{BrickRed}\-\ ntilde\-\ ==\-\ 0\color{Green}
$\\$\color{BrickRed}\-\ \-\ \-\ \-\ out\-\ =\-\ (-pie*log(q)/2)./(sin(1i*log(q)*psi/2).\verb|^|2)\-\ +\-\ 2*pie*log(q)*out;\color{Green}
$\\$\color{BrickRed}\color{NavyBlue}\-\ elseif\-\ \color{BrickRed}\-\ ntilde\-\ ==\-\ 1\color{Green}
$\\$\color{BrickRed}\-\ \-\ \-\ \-\ out\-\ =\-\ (-pie*log(q)/2)./(cos(1i*log(q)*psi/2).\verb|^|2)\-\ +\-\ 2*pie*log(q)*out;\color{Green}
$\\$\color{BrickRed}\color{NavyBlue}\-\ end\-\ \color{BrickRed}\color{Green}
$\\$
$\\$
$\\$
$\\$
$\\$
$\\$
$\\$
$\\$
$\\$
$\\$
$\\$
$\\$
$\\$
$\\$
$\\$
$\\$
$\\$
$\\$
$\\$
$\\$
$\\$
$\\$
$\\$
$\\$
$\\$
$\\$
$\\$
$\\$
$\\$
$\\$
$\\$
$\\$
$\\$
$\\$
$\\$
$\\$
$\\$
$\\$
$\\$
$\\$
$\\$
$\\$
$\\$
$\\$
$\\$\color{Black}\section{xi\_q\_psi.m}

\color{Green}\color{BrickRed}\color{NavyBlue}\-\ function\-\ \color{BrickRed}\-\ \-\ out\-\ =\-\ xi\_q\_psi(q,psi,ntilde)\color{Green}
$\\$\color{Green}$\%$\-\ computes\-\ omega*xi(ntilde*omega\-\ +\-\ 1i*psi*omega\_prime)
$\\$
$\\$
$\\$
$\\$\color{Black}
From 
\eq{
\vartheta_1(z) = 2\sum_{n=1}^{\infty} (-1)^{n+1}q^{(n-1/2)^2}\sin((2n-1)z),
}{\notag}

$\\$
we find that
\eq{
f(x)&:= \vartheta_1\left(\frac{\pi}{2\omega}(\omega x + i\omega' + \tilde n \omega + i\psi \omega')\right)
\\ &= -i\sum_{n=1}^{\infty}(-1)^{n+1}q^{(n-1/2)^2} \left(\hat v^{(2n-1)}-\hat v^{-(2n+1)}\right),
}{\notag}
where $\hat v := e^{i\pi(x+\tilde n)/2}q^{(1+\psi)/2}$. Hence

$\\$
\eq{
\pd{^m}{x^m} f(x)&=  -i(i\pi/2)^m\sum_{n=1}^{\infty}(-1)^{n+1}q^{(n-1/2)^2} (2n-1)\left(\hat v^{(2n-1)}-\hat v^{-(2n+1)}\right).
}{\notag}

$\\$
We find that
\eq{
Err &:= \left|-i(i\pi/2)^m\sum_{n=N+1}^{\infty}(-1)^{n+1}q^{(n-1/2)^2} (2n-1)\left(\hat v^{(2n-1)}-\hat v^{-(2n+1)}\right)\right| 
\\ &\leq 2q^{1/4}(\pi/2)^m\sum_{n = N+1}^{\infty}q^{n^2/2}\left((2n-1)^mq^{n^2/2-n-(2n-1)(1+\psi)/2}\right)\\
&\leq 2q^{1/4}(\pi/2)^m\sum_{n = N+1}^{\infty}q^{n^2/2}\\
&\leq2q^{1/4}(\pi/2)^mq^{(N+1)^2/2}\frac{1}{1-q}, 
}{\notag}
as long as we take $N$ large enough so that $g(x):= \left((2x-1)^mq^{n^2/2-x-(2x-1)(1+\psi)/2}\right)$ satisfies $g(N)< 1$, $g(x) < 0$ whenever $x> N$.

$\\$
\color{Green}
$\\$
$\\$\color{Green}$\%$$\%$
$\\$
$\\$
$\\$\color{Green}$\%$
$\\$\color{Green}$\%$\-\ constants
$\\$\color{Green}$\%$
$\\$
$\\$
$\\$\color{BrickRed}pie\-\ =\-\ nm('pi');\color{Green}
$\\$\color{BrickRed}psi0\-\ =\-\ max(sup(psi));\color{Green}
$\\$\color{BrickRed}q0\-\ =\-\ abs(q\verb|^|(2-psi0));\color{Green}
$\\$
$\\$
$\\$\color{Green}$\%$\-\ 
$\\$\color{Green}$\%$\-\ find\-\ N\-\ for\-\ given\-\ error
$\\$\color{Green}$\%$
$\\$
$\\$
$\\$\color{BrickRed}N\-\ =\-\ 0;\color{Green}
$\\$\color{BrickRed}maxit\-\ =\-\ 1000;\color{Green}
$\\$\color{BrickRed}tol\-\ =\-\ 1e-17;\color{Green}
$\\$\color{BrickRed}err\-\ =\-\ tol\-\ +\-\ 1;\color{Green}
$\\$\color{BrickRed}\color{NavyBlue}\-\ while\-\ \color{BrickRed}\-\ err\-\ $>$\-\ tol\color{Green}
$\\$\color{BrickRed}\-\ \-\ \-\ \-\ N\-\ =\-\ N+1;\color{Green}
$\\$\color{BrickRed}\-\ \-\ \-\ \-\ \color{NavyBlue}\-\ if\-\ \color{BrickRed}\-\ N\-\ $>$\-\ maxit\color{Green}
$\\$\color{BrickRed}\-\ \-\ \-\ \-\ \-\ \-\ \-\ \-\ error('Maximum\-\ iterations\-\ exceeded');\color{Green}
$\\$\color{BrickRed}\-\ \-\ \-\ \-\ \color{NavyBlue}\-\ end\-\ \color{BrickRed}\color{Green}
$\\$\color{BrickRed}\-\ \-\ \-\ \-\ temp\-\ =\-\ q0\verb|^|(N+1)/((1-q\verb|^|(2*(N+1)))*(1-q0));\color{Green}
$\\$\color{BrickRed}\-\ \-\ \-\ \-\ err\-\ =\-\ sup(temp);\color{Green}
$\\$\color{BrickRed}\color{NavyBlue}\-\ end\-\ \color{BrickRed}\color{Green}
$\\$
$\\$
$\\$\color{Green}$\%$\-\ initialize\-\ output
$\\$\color{BrickRed}out\-\ =\-\ nm(zeros(1,length(psi)));\color{Green}
$\\$
$\\$
$\\$\color{Green}$\%$\-\ add\-\ partial\-\ sum
$\\$\color{BrickRed}q2\-\ =\-\ q*q;\color{Green}
$\\$\color{BrickRed}logq\-\ =\-\ log(q);\color{Green}
$\\$\color{BrickRed}\color{NavyBlue}\-\ for\-\ \color{BrickRed}\-\ k\-\ =\-\ 1:N\color{Green}
$\\$\color{BrickRed}\-\ \-\ \-\ \-\ q2k\-\ =\-\ q2\verb|^|k;\color{Green}
$\\$\color{BrickRed}\-\ \-\ \-\ \-\ out\-\ =\-\ out\-\ +\-\ (q2k/(1-q2k))*sin(k*(pie*ntilde-1i*psi*logq));\color{Green}
$\\$\color{BrickRed}\color{NavyBlue}\-\ end\-\ \color{BrickRed}\color{Green}
$\\$
$\\$
$\\$\color{Green}$\%$\-\ add\-\ error
$\\$\color{BrickRed}out\-\ =\-\ out\-\ +\-\ nm(-err,err)\-\ +\-\ 1i*nm(-err,err);\color{Green}
$\\$
$\\$
$\\$\color{Green}$\%$\-\ multiply\-\ by\-\ constants
$\\$\color{BrickRed}out\-\ =\-\ 2*pi*out;\color{Green}
$\\$
$\\$
$\\$\color{Green}$\%$\-\ add\-\ non\-\ infinite\-\ sum\-\ parts
$\\$\color{BrickRed}out\-\ =\-\ out\-\ +\-\ (pie/2)*cot(pie*ntilde/2-1i*psi*logq/2);\color{Green}
$\\$
$\\$
$\\$\color{Green}$\%$\-\ multiply\-\ by\-\ constants\-\ to\-\ get\-\ omega*xi
$\\$\color{BrickRed}out\-\ =\-\ 2*1i*out;\color{Green}
$\\$
$\\$
$\\$
$\\$
$\\$
$\\$
$\\$
$\\$
$\\$
$\\$
$\\$
$\\$
$\\$
$\\$
$\\$
$\\$
$\\$
$\\$
$\\$
$\\$
$\\$
$\\$
$\\$